# Perspectives in Quantum Physics: Epistemological, Ontological and Pedagogical

## An investigation into student and expert perspectives on the physical interpretation of quantum mechanics, with implications for modern physics instruction.

Charles Raymond Baily

BA (1995), MS (2002) University of Colorado at Boulder

A thesis submitted to the
Faculty of the Graduate School of the
University of Colorado in partial fulfillment
of the requirements for the degree of
Doctor of Philosophy
Department of Physics
2011

This thesis entitled

Perspectives in Quantum Physics:
Epistemological, Ontological and Pedagogical

written by Charles Raymond Baily

has been approved by the Department of Physics

(Noah Finkelstein)

(Michael Dubson)

(5/2/2011)

The final copy of this thesis has been examined by the signatories, and we find that both the content and form meet acceptable presentation standards of scholarly work in the above mentioned discipline.

HRC Protocol # 0205.21

# ABSTRACT

Baily, Charles Raymond (Ph.D, Physics)
Title: Perspectives in Quantum Physics: Epistemological, Ontological and Pedagogical
Thesis directed by Associate Professor Noah D. Finkelstein

A common learning goal for modern physics instructors is for students to recognize a difference between the experimental uncertainty of classical physics and the fundamental uncertainty of quantum mechanics. Our studies suggest this notoriously difficult task may be frustrated by the intuitively *realist* perspectives of introductory students, and a lack of *ontological flexibility* in their conceptions of light and matter. We have developed a framework for understanding and characterizing student perspectives on the physical interpretation of quantum mechanics, and demonstrate the differential impact on student thinking of the myriad ways instructors approach interpretive themes in their introductory courses. Like expert physicists, students interpret quantum phenomena differently, and these interpretations are significantly influenced by their overall stances on questions central to the so-called *measurement problem*: Is the wave function physically real, or simply a mathematical tool? Is the *collapse of the wave function* an *ad hoc* rule, or a physical transition not described by any equation? Does an electron, being a form of matter, exist as a localized particle at all times? These questions, which are of personal and academic interest to our students, are largely only superficially addressed in our introductory courses, often for fear of opening a *Pandora's Box* of student questions, none of which have easy answers. We show how a transformed modern physics curriculum (recently implemented at the University of Colorado) may positively impact student perspectives on indeterminacy and wave-particle duality, by making questions of classical and quantum reality a central theme of our course, but also by making the beliefs of our students, and not just those of scientists, an explicit topic of discussion.

# ACKNOWLEDGEMENTS

This work has been supported in part by NSF CAREER Grant No. 0448176 and the University of Colorado. I wish to express my most sincere and unending gratitude for the cooperation of all the modern physics instructors and students who made these studies possible, but particularly our own students from the Fall 2010 semester, who were involuntary but enthusiastic participants in this dissertation project. I am also indebted to all the members of the Physics Education Research group at CU, for their continued insight and interest over the span of four years. Most importantly, I am grateful for the enduring support of my advisor and mentor, Noah Finkelstein.

# TABLE OF CONTENTS

# CHAPTER 1

## Perspectives in Quantum Physics

"Why do some textbooks not mention *complementarity*? Because it will not help in quantum mechanical calculations or in setting up experiments. Bohr's considerations are extremely relevant, however, to the scientist who occasionally likes to reflect on the meaning of what she or he is doing." – Abraham Pais [1]

### I. Introduction

### I.A. Notions of Classical and Quantum Reality

Albert Einstein considered the aim of physics to be "the complete description of any (individual) real situation (as it supposedly exists irrespective of any act of observation or substantiation)." [2] His statement on the *purpose* of science speaks also of his predisposition toward thinking of the universe itself in terms of *realist* expectations: there *is* an objective reality that exists independent of any human observation. In other words, a *complete* description (or theory) of that objective reality is minimally comprised of elements in one-to-one correspondence with physical quantities (such as position or momentum) that are assumed to have definite, objectively real values at all times. [3]

Such assumptions about the nature of reality are built into the equations of classical mechanics – in describing the position of a free electron with a given momentum at some later time, it is already assumed the electron was initially located at some definite, single point in space ($x_0$), and that its specific momentum ($p$) *predetermines* its definite location ($x$) at all later times ($t$):

(1.1) $$x(t) = x_0 + \frac{p}{m_e} \cdot t$$

Just as with the assumption of *universal time* for all observers in Galilean relativity, these classical assumptions are based on intuitive notions grounded in everyday experience, and it may not occur to classical thinking that these are even assumptions to begin with, or anything other than axiomatic.

Contrast this with an expression from quantum mechanics for the wave function ($\Psi$) describing a free electron with definite momentum (and therefore definite energy, $E$) as a function of position and time:

(1.2) $$\Psi(x,t) = A\exp\left[\frac{i}{\hbar}(p \cdot x - E \cdot t)\right].$$

Although the electron's momentum is well defined (Einstein would say it has *reality*), its location may only be described in terms of the probability for where it might be found when observed, which (according to Born's probabilistic interpretation of the wave function) is given by the modulus squared of this complex exponential:

$$\rho[x] = |\Psi(x)|^2 = \left(A^* \exp\left[-\frac{i}{\hbar} p \cdot x\right]\right) \cdot \left(A \exp\left[\frac{i}{\hbar} p \cdot x\right]\right) = |A|^2 = \text{constant} \tag{1.3}$$

(where the energy/time term has been suppressed). When its momentum is certain, the probability density for its location is constant, and the electron has an equal likelihood of being found anywhere in space – in the mathematics of quantum physics, the location of this free electron is not well defined, and the outcome of a position measurement cannot be predicted with any certainty. If, as Einstein assumed, the electron always exists as a localized particle and is indeed located at a specific point in space at all times (its position also has reality), then a probabilistic (statistical) description of the true state of that electron must be considered *incomplete*. [3] A physical quantity that has some definite value, but is not described by a theory, is known as a *hidden variable*.

According to quantum mechanics, the observables **p** and **x** are *incompatible* (their mathematical operator representations do not commute):

$$[\hat{p}, \hat{x}] = \hat{p} \cdot \hat{x} - \hat{x} \cdot \hat{p} = \frac{\hbar}{i} \qquad \leftrightarrow \qquad \Delta x \cdot \Delta p \geq \frac{\hbar}{2} \tag{1.4}$$

and so the position and momentum of a particle cannot be simultaneously described with arbitrary precision. If, in this scenario, the electron does not actually exist at some single, definite location until observed, then the theory of quantum mechanics is *not necessarily* an incomplete description of that reality.

In 1935, Einstein (along with Boris Podolsky and Nathan Rosen, collectively known as EPR) posited a second assumption about the nature of reality (which they considered to be "reasonable"): "If, at the time of measurement, two systems no longer interact, no real change can take place in the second system in consequence of anything that may be done in the first system." [3] This intuitive *assumption of locality* says that the outcome of a measurement performed on some System A can have no influence (or dependence) on any measurement performed on some other System B that is sufficiently isolated from the first. With their *condition of completeness* and the *assumption of locality* in hand, EPR argued that the position and momentum of a particle can be logically demonstrated to have simultaneous reality, and that the quantum mechanical description is therefore incomplete.

Originally formulated in terms of position and momentum measurements, the EPR argument has been reframed [4] in terms of spin measurements performed on systems of *entangled* particles. We imagine a pair of spin-1/2 fermions (Particles A & B) somehow formed in a state of zero total spin angular momentum and traveling in opposite directions.[1] Individual measurements of each particle's spin projection along any given axis will always yield one of two values (*up* or *down*, +1 or −1, however we choose to designate them). Moreover, spin measurements

[1] The argument does not depend on how this is done, but one method would involve preparing a positronium atom in a singlet state, and then dissociating the electron-positron bound state in such a way that the total linear and angular momentum of the system are conserved. [5]

performed on these entangled fermion pairs will always yield opposite values, so long as the measurements are performed along the same axis. In this way, a spin measurement performed along the z-axis for just one of the particles is sufficient for predicting with 100% certainty the outcome of a spin measurement performed on the second particle along that same axis; the actual measurement on the second particle need not be performed, but can be done so as to confirm the predicted outcome. The same is true for spin measurements performed along the x-axis, or any other axis we choose, so long as the axis of orientation is the same for both analyzers. Quantum mechanics says the operators for the x-component and the z-component of spin angular momentum are non-commuting, and therefore obey a similar uncertainty relation as with position and linear momentum (the components of spin angular momentum for a particle cannot be simultaneously specified along two different axes with arbitrary precision):

(1.5) $\left[\hat{S}_x,\hat{S}_z\right]\neq 0 \quad \leftrightarrow \quad \Delta S_x \cdot \Delta S_z \neq 0$

Now suppose the spin of Particle A is measured along the z-axis: an outcome of +1 for Particle A means that a similar measurement performed on Particle B will *always* yield the result of −1, *before* any such measurement on Particle B is actually made. The *assumption of locality* says that any measurement performed on Particle A can have no causal influence on the outcome of any measurement performed on Particle B.[2] EPR would then argue that the z-component of spin for Particle B must have had a definite (real) value at the time of its separation from Particle A, and that this value can be found without disturbing Particle B in any way. If the measurement on Particle B is instead performed along the x-axis, EPR would conclude that the spin projection for Particle B is now simultaneously specified along two different axes, both x (by the second measurement on Particle B) and z (by the first measurement on Particle A); they therefore have simultaneous reality, which is precluded in the quantum mechanical description. [Eq. 1.5] It follows that quantum mechanics offers an incomplete description of the objectively real state of Particle B.

In defense of the completeness of quantum physics, Niels Bohr took issue mainly with EPR's claim of *counterfactual definiteness* – there can be no definite statements (according to Bohr) regarding the outcomes of quantum measurements that haven't been performed. [6] He further insisted that no definitive line could be drawn between the measurement apparatus and the system being measured: "An independent reality in the ordinary [classical] physical sense can […] neither be ascribed to the phenomena nor to the agencies of observation." [7] Bohr ultimately went so far as to redefine the purpose of science: "It is wrong to think that the task of physics is to find out how nature is. Physics concerns what we can say about nature." [8]

[2] Assuming the two measurements are performed at space-like separations (the second measurement lies outside the light cone of the first), then special relativity precludes any cause-and-effect relationship between the two events.

**I.B. Philosophy or Science?**

It is generally agreed in the physics community that Bohr emerged triumphant in this debate, [8] though many physicists of today might feel hard-pressed to say exactly why. If anything, it has been argued that the positivistic[3] aspects of the *Copenhagen Interpretation* [9] (often referred to as the *orthodox* interpretation of quantum mechanics [10]) have contributed to its popularity over the years by allowing physicists to set aside questions of completeness and locality, and instead just use the wave function to "shut up and calculate!" [11] All the same, it was anyways widely believed that J. von Neumann had successfully ruled out the possibility for hidden quantum variables in 1932. [12]

Such beliefs went largely unchallenged [4] until the appearance in 1964 of a groundbreaking paper by J. S. Bell, who had come to realize that Einstein's assumptions were not just a matter of philosophical taste, and could be put to experimental test. [13] In his own discussion of the EPR argument, Bell maintained the assumption of locality in his demonstration that a more complete description of an entangled system of particles could never be specified in terms of hidden variables (a set of one or more unknown parameters, λ). If the result for Particle A is a function of the orientation of its Stern-Gerlach analyzer (unit vector ***a***) and the hidden parameters (λ); and if the outcome for Particle B is similarly a function of both the orientation of its Stern-Gerlach analyzer (***b***) and of λ, we may write this as

(1.6) $A(a,\lambda) = \pm 1 \quad \& \quad B(b,\lambda) = \pm 1$,

where A and B represent the measurement outcomes for Particles A & B, respectively. The assumption of locality may expressed as

(1.7) $A \neq A(a,b,\lambda) \quad \& \quad B \neq B(a,b,\lambda)$

which says merely that A cannot depend on how the other analyzer is oriented, and similarly for B. The anticorrelated nature of measurement outcomes along similar axes may be written as

(1.8) $A(a,\lambda) = -B(a,\lambda)$.

We may then find the expectation value in this local hidden variable (HV) theory for the *product* of the two measurements, by summing over all possible values for the hidden variables, weighted by some probability distribution for the hidden parameters (ρ):

(1.9) $E_{HV}(a,b) \equiv \left\langle \left(\vec{S}_1 \cdot a\right)\left(\vec{S}_2 \cdot b\right)\right\rangle_{HV} = \int d\lambda \; \rho(\lambda) A(a,\lambda) \, B(b,\lambda)$

We now show that the product of the hidden variable expectation values (where locality has been assumed) must obey an inequality that is violated by the predictions of quantum mechanics. We start by writing down the expression:

(1.10) $E_{HV}(a,b) - E_{HV}(a,c) = \int d\lambda \; \rho(\lambda)\left[A(a,\lambda)B(b,\lambda) - A(a,\lambda)B(c,\lambda)\right]$

where ***c*** is some other unit vector along which the spin projection might be measured. Using (1.8) this may be rewritten as

[3] In this context, we are referring to a refusal to speculate on that which can't be observed (measured).

(1.11) $E_{HV}(a,b) - E_{HV}(a,c) = -\int d\lambda\, \rho(\lambda)\left[A(a,\lambda)A(b,\lambda) - A(a,\lambda)A(c,\lambda)\right]$,

and then factored by recognizing that the square of any measurement outcome must be equal to +1, so that

(1.12) $E_{HV}(a,b) - E_{HV}(a,c) = -\int d\lambda\, \rho(\lambda)A(a,\lambda)A(b,\lambda)\left[1 - A(b,\lambda)A(c,\lambda)\right]$.

We must also have that:

(1.13) $|A(a,\lambda)A(b,\lambda)| \le +1$

so that taking absolute values in Eq. 1.12, and using the fact that $\rho(\lambda)$ is normalized, gives what is now known as *Bell's inequality*:

(1.14) $|E_{HV}(a,b) - E_{HV}(a,c)| \le 1 + E_{HV}(b,c)$.

The quantum mechanical expectation value for the product of spin measurements is

(1.15) $E_{QM}(a,b) \equiv \left\langle (\vec{S}_1 \cdot a)(\vec{S}_2 \cdot b) \right\rangle_{QM} = -\, a \cdot b = -\cos(\phi)$,

where $\phi$ is the angle between the unit vectors a and b. The equivalent expression for (1.14) in terms of the quantum mechanical (QM) expectation values is

(1.16) $|E_{QM}(a,b) - E_{QM}(a,c)| \le 1 + E_{QM}(b,c)$.

There are a variety of angles for which this quantum mechanical inequality holds, but for the simple case where the three vectors are situated at $60^0$ to each other, so that $\hat{a} \cdot \hat{b} = \cos(60^0)$, $\hat{b} \cdot \hat{c} = \cos(60^0)$ & $\hat{a} \cdot \hat{c} = \cos(120^0)$ we find:

(1.17) $1 \;=\; \left|\frac{1}{2} - \left(-\frac{1}{2}\right)\right| \;\le\; 1 - \frac{1}{2} \;=\; \frac{1}{2}$

which clearly violates Bell's inequality. Because quantum mechanics correctly predicts the observed expectation values (see below), it follows that at least one of EPR's assumptions (realism and/or locality) is not valid when describing quantum phenomena. If locality is instead *not* assumed in the above argument:

(1.18) $A = A(a,b,\lambda) \quad \& \quad B = B(a,b,\lambda)$

(the outcome for each measurement depends on the orientation of *both* analyzers), there are many functions (A & B) for which the quantum mechanical expectation value (Eq. 1.15) is reproduced, [14] and so it is the assumption of locality that must be set aside, leaving open the possibility for *non-local* hidden variable theories [4]

In 1969, Clauser, et al. generalized Bell's theorem to realizable experiments by allowing for detector inefficiencies, and for the possibility that the measurement correlations are imperfect (less than 100%). [15, 16] From all of this we may conclude that: (A) No local hidden variable theory can reproduce all of the predictions of quantum mechanics; and (B) An experiment may now be devised to differentiate between the two. Various refinements have been made, and a number of loopholes closed over the years, [17-21] but the first definitive test of the assumptions of *Local Realism* was made in 1981 by Alain Aspect and colleagues, [22] when they measured the polarization correlation rate for entangled photon pairs emitted in a radiative atomic cascade. [Fig. 1.1]

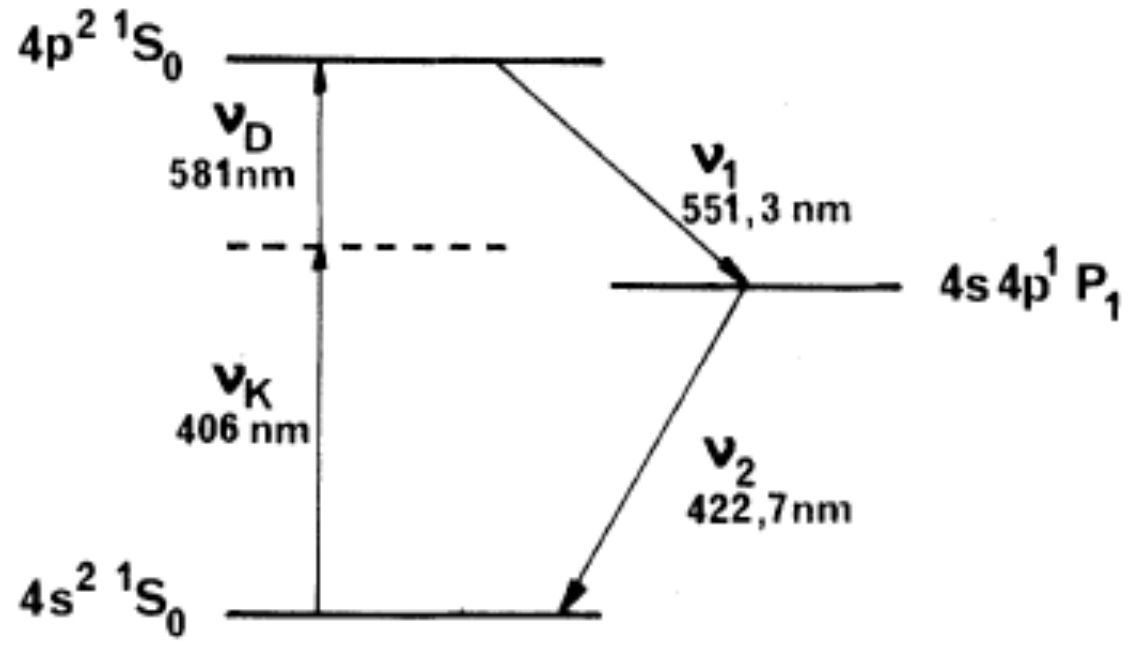

**FIG. 1.1.** Relevant energy levels of calcium. The atoms are excited by a two-photon absorption process ($\nu_K$ and $\nu_D$), and then decay by the emission of two visible photons ($\nu_1$ and $\nu_2$) that are correlated in polarization. [22]

In this experiment, entangled photon pairs were created using a calcium 40 cascade that yields two visible photons ($\upsilon_1$ and $\upsilon_2$). The calcium atoms were pumped to the upper level of the cascade from the ground state by two-photon absorption; the average decay lifetime of the intermediate decay state is $\tau$ = 4.7 ns. An atomic beam of calcium (with $\rho = 3\times10^{10}$ atoms/cm$^3$) was irradiated at $90^0$ by two laser beams with parallel polarizations, the first a krypton ion laser ($\lambda_K$ = 406.7 nm), then with a Rhodamine 6G dye laser tuned to resonance for the two-photon process ($\lambda_D$ = 581 nm). With each laser operating at 40mW, the typical cascade rate was $\sim 4\times10^7$ per second. [22]

In its ground state, calcium 40 has two valence electrons outside a closed shell; with their spins oppositely aligned, the total angular momentum (spin plus orbital) of this state is J = 0. The upper level of the cascade is also a J = 0 state, and the intermediate state is J = 1, so that the excited atom has two possible decay paths (m = +1 or m = −1) on its way to the ground state. By conservation of angular momentum, any photon pair ($\upsilon_1$ & $\upsilon_2$) that *happen to be emitted back-to-back* in this process must therefore have the same circular polarization: either both right-handed (R) or both left-handed (L). The entangled state of the two photons may then be written as:

(1.19) $|\Psi_{12}\rangle = |R_1\rangle|R_2\rangle + |L_1\rangle|L_2\rangle$.

Einstein would argue that each atom always decays by either one path or the other, so that each photon pair is produced in just one of the two polarization states with equal probability (determined by some hidden parameter), but that we cannot know which one until the photon pair is observed. He would say that the superposition state describing each photon pair is a reflection of *classical ignorance* (a lack of knowledge regarding the true state of the photon pair). Bohr would argue that the superposition state is a reflection of a more *fundamental uncertainty*, and that each photon pair exists in an indeterminate superposition state until measured. Observing only one of the two photons instantly *collapses* the superposition at random into just one of the two definite states with equal probability. The collapse must be instantaneous if the two measurements occur at space-like separation, since there would be no time for a signal to travel between the two photons regarding how they should behave when they encounter a polarizer.

Aspect measured the rate of coincidental detection of back-to-back photon pairs with the same type of polarization along a variety of relative angles, and found that these measurements violated the generalized Bell's inequality by more than 13 standard deviations, providing strong evidence against *any* local hidden-variable theory. [Fig. 1.2]

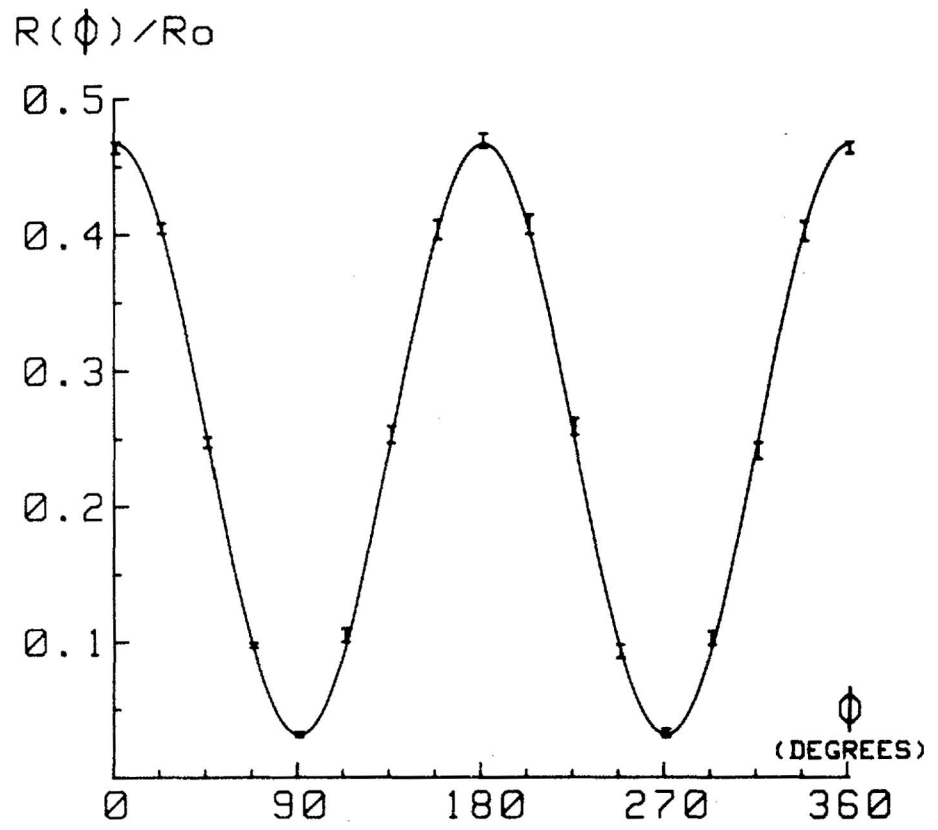

**FIG. 1.2.** Results of the first Aspect experiment testing Bell's inequality – normalized coincidence rate as a function of relative polarizer orientation. Error bars represent one standard deviation, and the curve drawn through the data points is not a best-fit curve, but rather what is predicted by quantum mechanics. [22]

## I.C. Wave-Particle Duality and Ontological Flexibility

In arguing for the incompleteness of quantum mechanics, Einstein was essentially questioning whether quantum mechanics could be used to describe the real state of individual particles, or merely a statistical distribution of measurement outcomes for an ensemble of similarly prepared systems (e.g., a coherent beam of single photons or electrons), where the final distribution of results is determined by some set of unknown, hidden parameters (initial position and/or momentum, for example). Does the *instantaneous collapse of the wave function* represent a change in knowledge of the observer regarding the true state of an individual system, or does it represent a physical transition for that system from an indeterminate state to one that is definite? Erwin Schrödinger famously questioned exactly when this so-called *collapse* is supposed to take place, when he ironically proposed a thought-experiment in which a macroscopic object (in this case, a cat in a box) is imagined to be in a superposition of two states (dead or alive) right up until the moment it is observed (when we open the box). [23] By 1950, Einstein had few allies remaining in the assault on realism in physics, as he expressed in a letter to Schrödinger from that time:

> "You are the only contemporary physicist, besides Laue, who sees that one cannot get around the assumption of reality – if only one is honest. Most of them simply do not see what sort of risky game they are playing with reality – reality as something independent of what is experimentally established. They somehow believe that the quantum theory provides a description of reality, and even a complete description; this interpretation is, however, refuted most elegantly by your system of radioactive atom [plus] cat in a box, in which the Ψ-function of the system contains the cat both alive and [dead]. Is the state of the cat to be created only when a physicist investigates the situation at some definite time? Nobody really doubts that the presence or absence of a [dead] cat is something independent of observation. But then the description by means of the Ψ-function is certainly incomplete, and there must be a more complete description." [24]

The practical significance of EPR's argument (and its refutation via Bell's Theorem) was not truly realized until the mid-to-late 1970's – as reflected in how their paper had a total of only 36 citations in *Physical Review* before 1980, but added 456 more citations in the period from 1980 to June 2003. [25] A similar trend can be seen [Fig. 1.3] in the belated, sudden increase in citations of Bell's paper, "On the Einstein-Podolsky-Rosen Paradox." [26]

It was the development during the 1970's and onward of experimental techniques for isolating and observing single quantum objects like photons, electrons, and atoms that caused physicists to take ideas about "quantum weirdness" seriously. According to Aspect: "I think it is not an exaggeration to say that the realization of the importance of entanglement and the clarification of the quantum description of single objects have been at the root of a *second quantum revolution*, and that John Bell was its prophet." [27]

Long before any such experiments were possible, physicists were already arguing for their preferred interpretations of quantum mechanics in terms of the individual behavior of quanta. In his own book on quantum mechanics, Dirac considers a thought experiment wherein individual photons are directed toward a beam splitter, and have equal probability of being transmitted or reflected. The quantum mechanical wave describing the probability for detecting the photon coherently splits at the beam splitter (it is both reflected and transmitted), but the result of any detection "must be either the whole photon or nothing at all. Thus the photon must suddenly change from being partly in one beam and partly in the other to being entirely in one of the beams." [28] Dirac argued this as a point of principle, despite there being no specific experimental evidence at the time for this assertion.

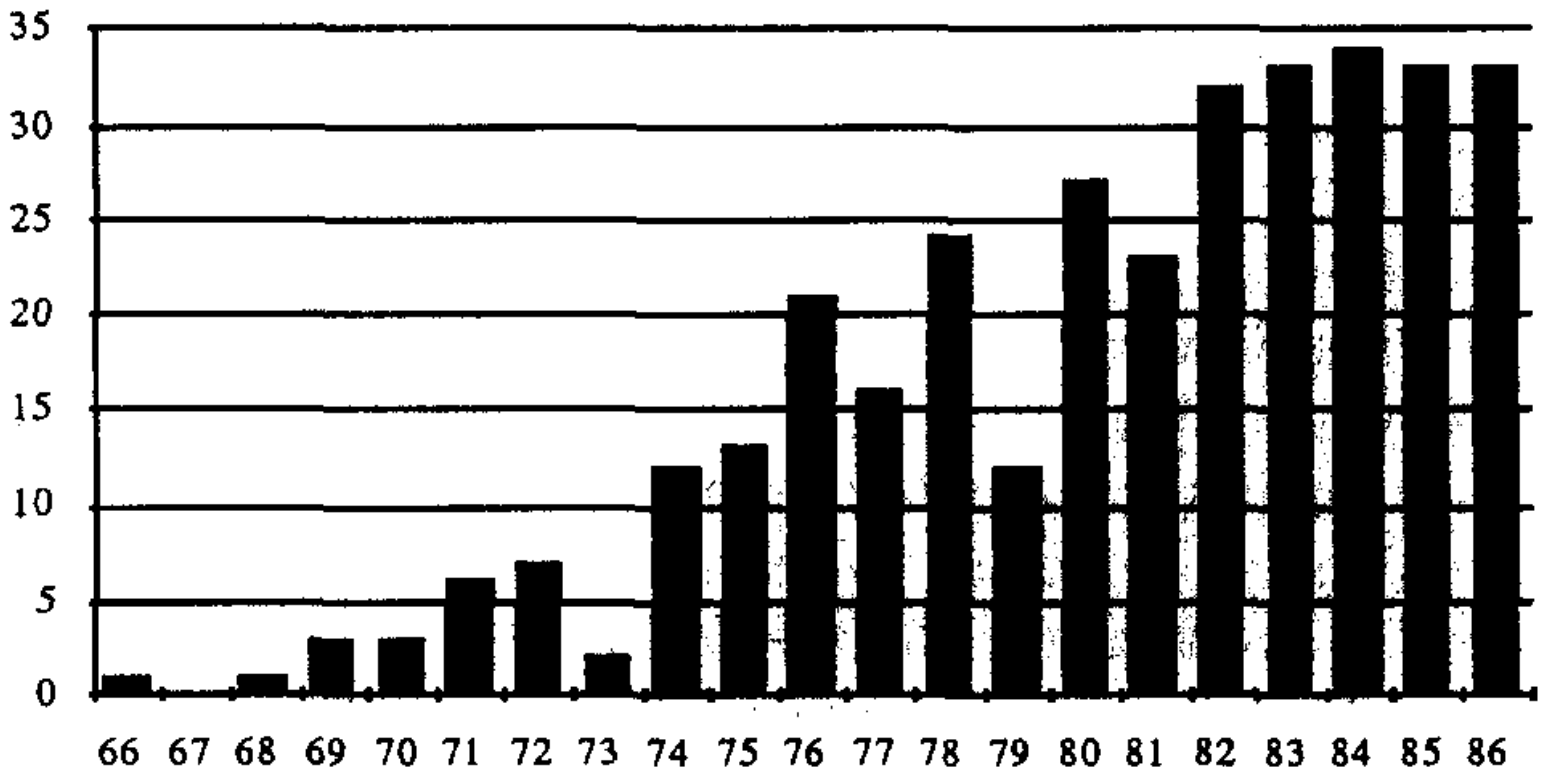

**FIG. 1.3.** Number of annual citations [1966-1986] of "On the Einstein-Podolsky-Rosen Paradox," by J. S. Bell, *Physics* **1**, 195 (1964). [26]

Definitive evidence for such behavior was most elegantly demonstrated by Grangier, Roger and Aspect in 1986 [29] using the same calcium 40 cascade photon source used in Aspect's first experiments testing the assumptions of Local Realism. [Fig. 1.4] Their first experiment was designed to demonstrate the particle-like behavior of photons; the second was meant to demonstrate the wave-like behavior of photons in a nearly identical situation. The experimental setup was along the lines proposed by Dirac in the thought experiment described above.

In each of these two experiments, the first photon ($\nu_1$) emitted in the calcium cascade serves as a trigger when detected in PM1, and the electronics opens a gate for a time equal to twice the lifetime of the intermediate state ($2\tau \sim 10$ ns), telling counters $N_A$ & $N_B$ to expect a second photon ($\nu_2$); a coincidence counter ($N_C$) is triggered if both photomultipliers fire during the short time the gate is open. The path to the beam splitter (BS1) from the source is collimated such that the second photon must have been one that was emitted back-to-back with the first, which greatly reduces the luminosity of this "single-photon" source. A set of mirrors ($M_A$ & $M_B$) direct the second photon toward either PMA (it was reflected at BS1) or PMB (it was transmitted at BS1).

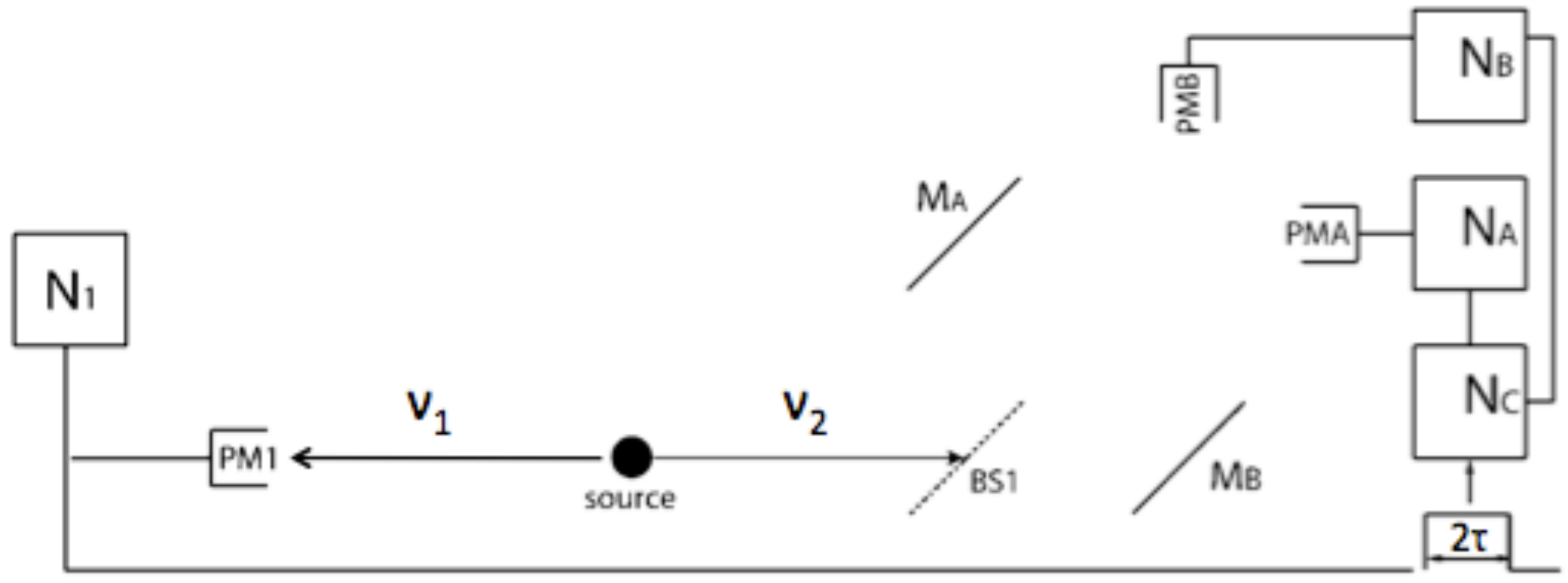

**FIG. 1.4.** Schematic diagram for the first anticoincidence experiment by Grangier, et al. PM1, PMA & PMB are photomultipliers; $N_1$, $N_A$, $N_B$ &$N_C$ are counters; BS1 is a beam splitter; $M_A$ and $M_B$ are mirrors. [29]

If the photon energy were coherently split at the beam splitter (wave-like behavior) it would be expected that energy would be deposited into the photomultipliers coincidentally, and that they would therefore fire together more often than separately. If the photon were instead either transmitted or reflected at the beam splitter (but not both; particle-like behavior) we expect the photomultipliers to always be triggered separately, so long as only one photon is in the apparatus at a time. We can quantify how often this is happening by defining an *anticorrelation parameter* (α):

(1.20) $$\alpha \equiv \frac{P_C}{P_A \cdot P_B}$$

where $P_A$ is the probability for PMA to fire, $P_B$ is the same for PMB, and $P_C$ the probability for both to fire during the time the gate is open.

- If individual photons are always detected in only one photomultiplier or the other (particle-like behavior), then $\alpha = 0$ since $P_C$ must be zero (there is zero probability that the two detectors click together during the time the gate is open).
- If the detectors are firing randomly and independently, then $\alpha = 1$, since $P_C$ is just the product of $P_A$ and $P_B$. This would be consistent with either many photons being present in the apparatus at once, or with waves depositing energy over time and randomly triggering the detectors.
- If there is a clustering of counts (higher than random probability that both detectors click together; consistent with wave-like behavior), then $\alpha > 1$ (i.e. $P_C$ is greater than just the product of $P_A$ and $P_B$).

The results for this first experiment show that, more often than not, photons are being detected in either one photomultiplier or the other during the time the gate is open, which is consistent with the predictions for particle-like behavior (α ≥ 0), while being inconsistent with the predictions for wave-like behavior (α ≥ 1). [Fig. 1.5 – the solid curve represents the predictions of quantum mechanics; error bars represent one standard deviation. It is necessary to extrapolate the measurements to "single-photon" intensity (α = 0) since the apparatus has a *dark rate* of ~300 counts/second.] We *interpret* these results as meaning that each photon must always take one path or the other on its way to detection – it is either reflected at the beam splitter or transmitted (but not both).

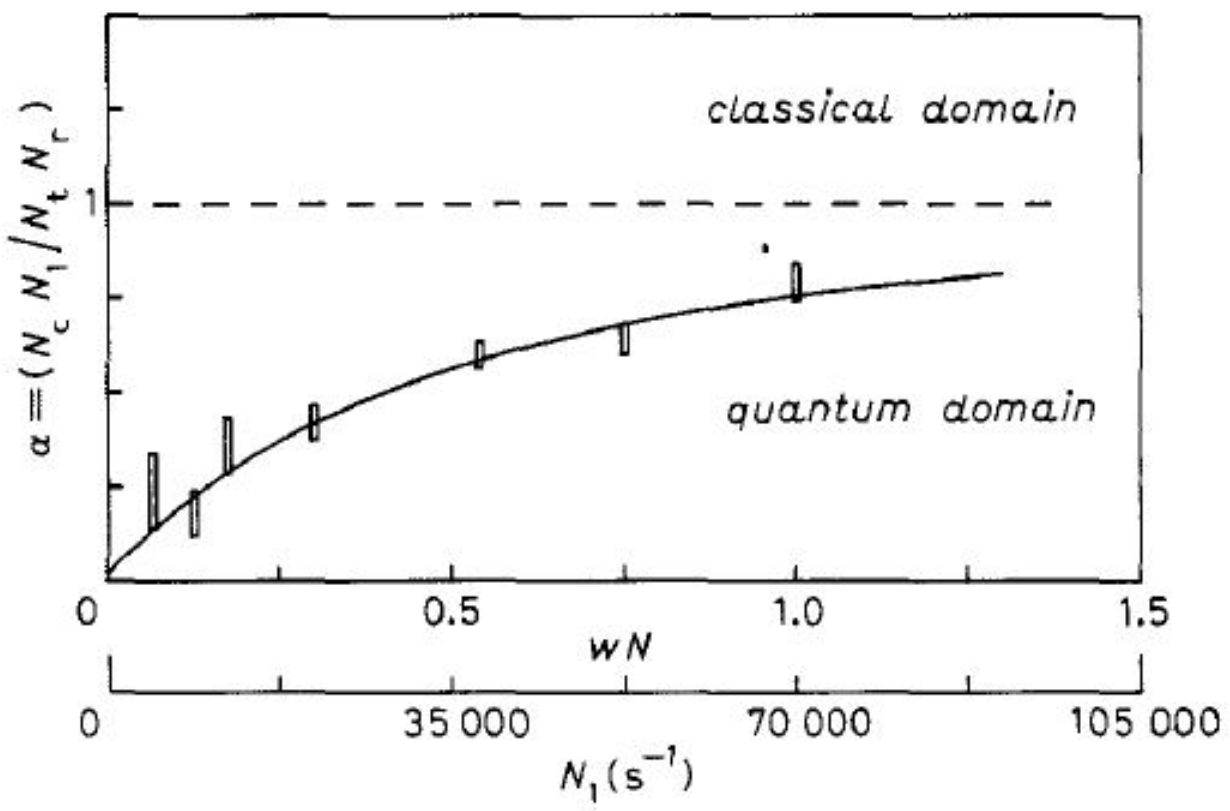

**FIG. 1.5.** Results from the first photon anticoincidence experiment performed by Grangier, et al. The anticorrelation parameter plotted as a function of the counting rate in PM1 (equivalently, the luminosity of the "single-photon" source). [29]

The experiment can be run a second time after a slight modification is made: inserting a second beam splitter into the paths taken by the photons (BS2). [Fig. 1.6] With BS2 in place, a photon might reach PMA by transmission at BS2 (Path A – it was reflected at BS1) or by reflection at BS2 (Path B – it was transmitted at BS1). Either way, a detection in PMA or PMB yields no information about the path taken by a photon to get there. According to quantum mechanics, the probabilities for photon detection in either PMA or PMB are oppositely modulated, as a function of the pathlength difference between Paths A & B. This means that, for certain pathlength differences (δ), *all* of the photons are detected in PMA and *none* are detected in PMB; and there are intermediate phases where detection in either photomultiplier is equally likely. [Fig. 1.7]

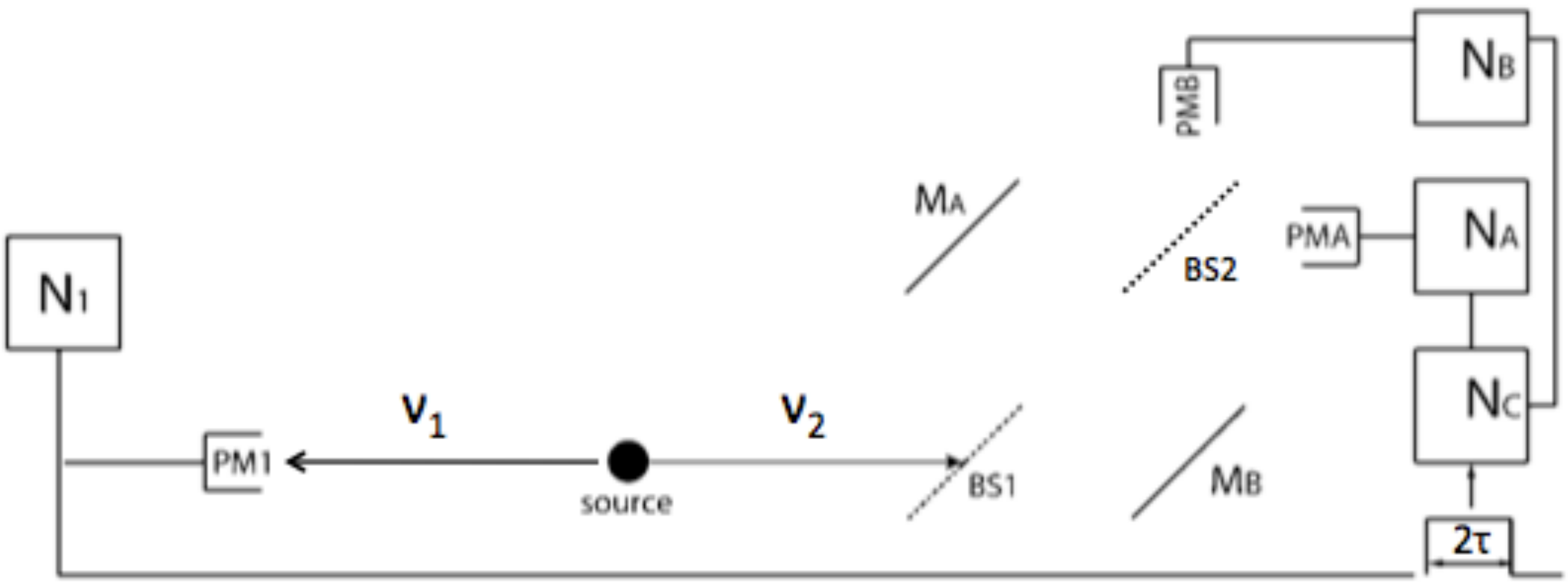

**FIG. 1.6.** Schematic diagram for the second anticoincidence experiment by Gragier, et al. PMA, PMB & PM1 are photomultipliers; $N_1$, $N_A$, $N_B$ &$N_C$ are counters; BS1 and BS2 are beam splitters; $M_A$ and $M_B$ are mirrors. [29]

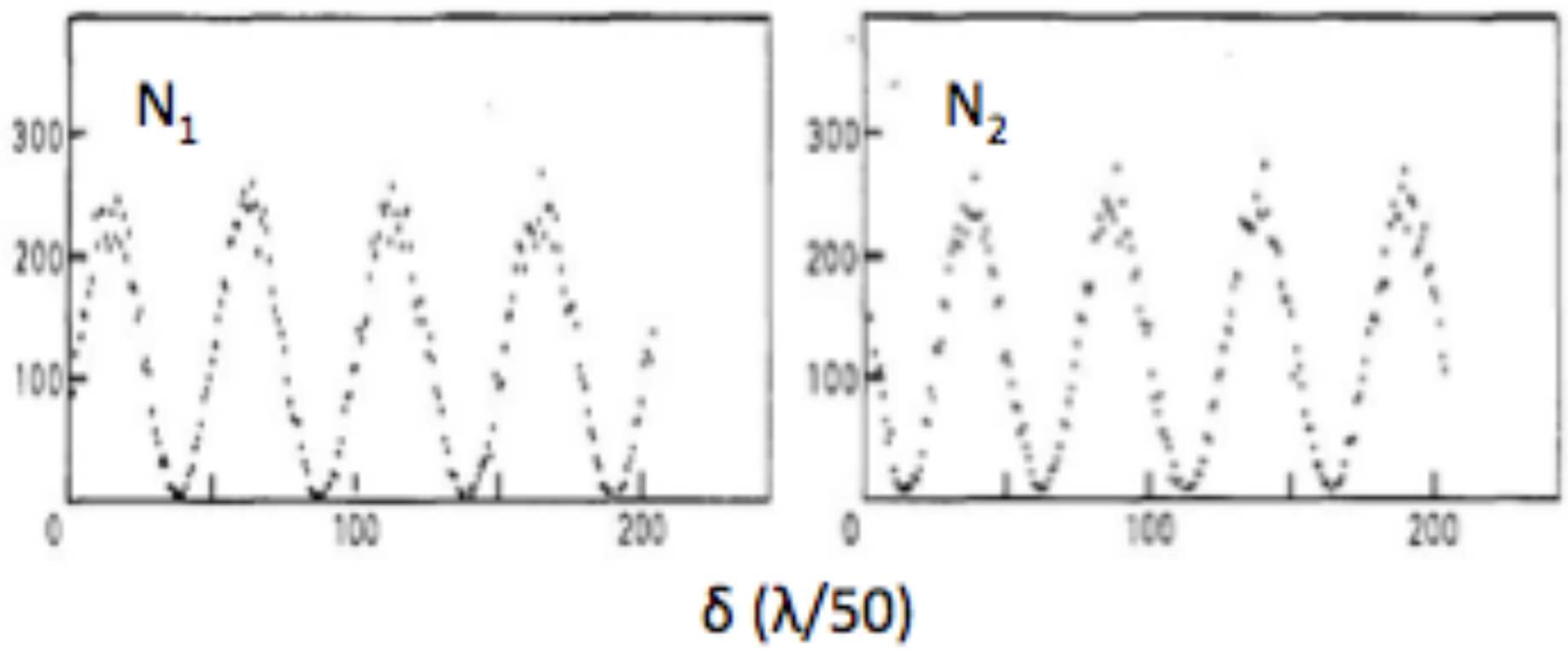

**FIG. 1.7.** Results from the second photon anticoincidence experiment performed by Grangier, et al. Counting rates at 15-second intervals for each of the two counters $N_1$ (left) and $N_2$ (right) as a function of path length difference (δ - in units of λ/50). For this experiment, α = 0.18. [29]

We *interpret* these results as meaning that each photon is coherently split at each beam splitter – it is both reflected *and* transmitted at BS1 (wave-like behavior, in contradiction with our conclusions from the first experiment) for, as the argument goes, how else could changing something about Path B affect the behavior of the photons that were supposed to have only taken Path A? For this second experiment, the anticorrelation parameter was small ($\alpha$ = 0.18), and so we must conclude that each photon is interfering with itself along the two paths (as opposed to many photons interfering with each other).

How are we to make sense of these two experiments, when the results seem to indicate contradictory behavior for the photons at BS1? How does each photon know whether BS2 is in place or not (whether we are conducting the first experiment or the second) when it first encounters BS1? Dirac would argue that every photon is coherently split as a delocalized wave at each beam splitter in *both experiments*, and that in each case the wave instantly collapsed down to a point when interacting with a detector.[4] Bohr would argue (more philosophically) that each photon is, from the very beginning, interacting with the entire apparatus as a whole, and that it behaves as it does at the first beam splitter (particle-like or wave-like) according to which type of behavior is allowed for which type of experiment.

In the end, these are all questions of *ontological category attribution* – it is clear that photons sometimes exhibit particle-like behavior, and sometimes exhibit wave-like behavior, depending on the experiment. Is it possible for photons to simultaneously behave as both particle and wave, for them to simultaneously *straddle* two (classically) distinct ontological categories?[5] A famous thought-experiment was proposed by Wheeler in 1978 [30] (and realized by Hellmuth, et al. in 1987 [31]) to test for this possibility. Imagine a photon entering the apparatus when only one path is available (the photon must take a single, definite path) from source to detector, but then a second path is opened up at the last moment (suddenly, two paths are available). If the photon had already "chosen" to take a single path at the first beam splitter, there should be no opportunity for the photon to interfere with itself, and no interference should be visible in the detectors.

In the actual experiment, [Fig. 1.8] a short-pulsed laser (less than a billionth of a second, with an average of one photon per pulse) was directed at a beam splitter, and the light then passed through 10-meter long optical fibers (in order to increase the transit time by ~30 ns). A Pockels cell (PC-A) in conjunction with a Glans prism was used to effectively insert and remove a path. When a voltage is applied to the Pockels cell, it rotates the plane of polarization of the light within five nanoseconds; the Glans prism then deflects away photons whose polarization has been altered, while transmitting unrotated photons. Therefore, when a voltage is applied to the Pockels cell, there is only one path by which a photon could reach the second beam splitter; with no voltage applied, both paths are possible. By randomly applying and removing voltages to the Pockels cell at the required frequency, it was possible to change the nature of the experimental setup after each photon had encountered the first beam splitter.

---

[4] Dirac did not take this "mental model" as a literal description of what was happening, but instead considered it to be a picture that helps to make sense of the situation: "One may extend the meaning of the word '*picture*' to include any way of looking at the fundamental laws which makes their self-consistency obvious." [28]

[5] Ontology concerns itself with the categorization of concepts, physical entities and processes according to their fundamental properties. Entities with similar characteristics belong to similar categories or sub-categories.

They found that when the experiment was run with initially only one path open (voltage applied), but then switched to both paths open after the photon had already encountered the first beam splitter (voltage removed), the photon still behaved as though two paths had been available all along, and interference was observed in the detectors. [Fig. 1.9] Wheeler argued that the photon's "choice" as to how to behave at BS1 (like a particle or a wave) must have been made *after the fact* (hence the term *delayed-choice experiment*, which may also refer to the delayed choice made by the observer of which experiment to conduct). [30]

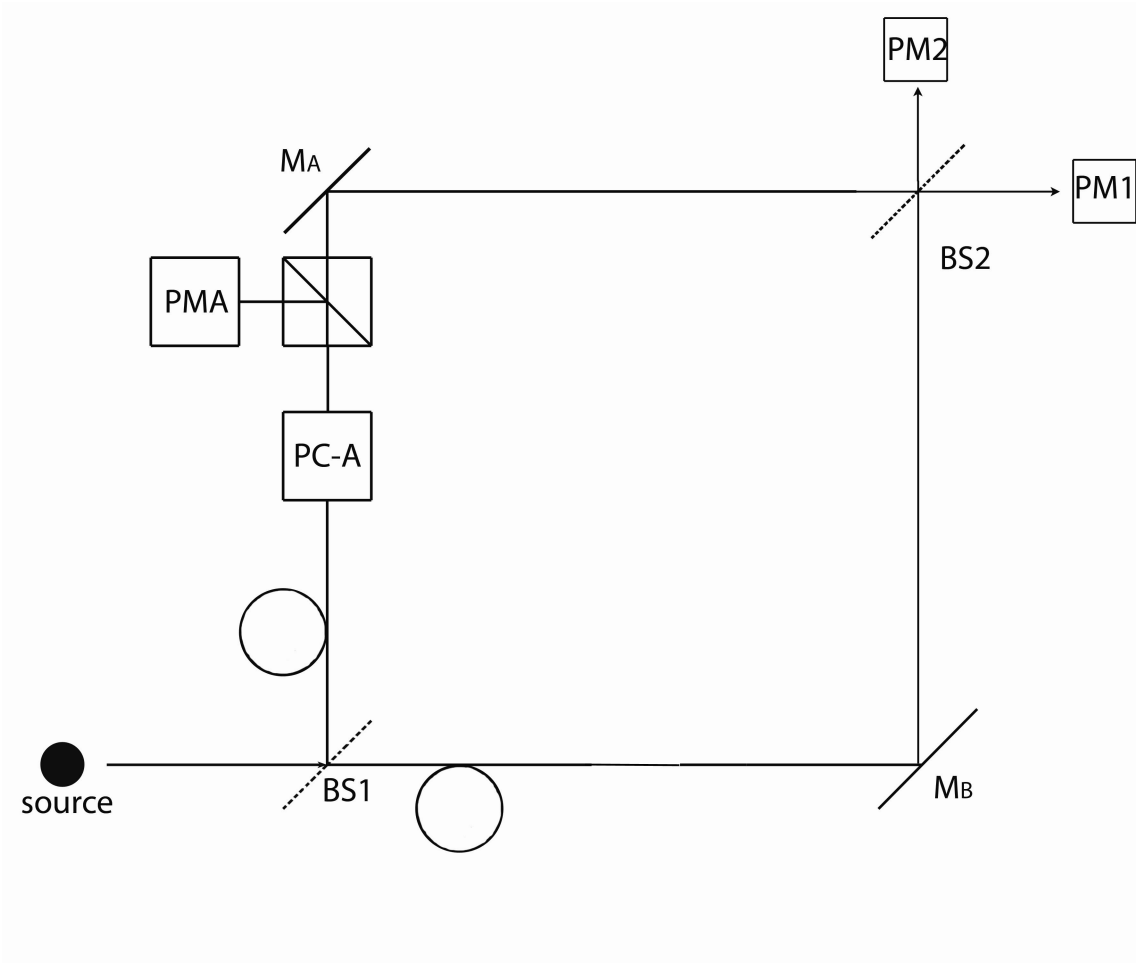

**FIG. 1.8.** Schematic diagram of the delayed-choice experiment conducted by Hellmuth, et al. PC-A is a Pockels cell used to rotate the plane of photon polarization when a voltage is applied; a Glans prism is used to pass unrotated photons, and to reflect away rotated photons. Only one path to BS2 is available with the voltage applied. [31]

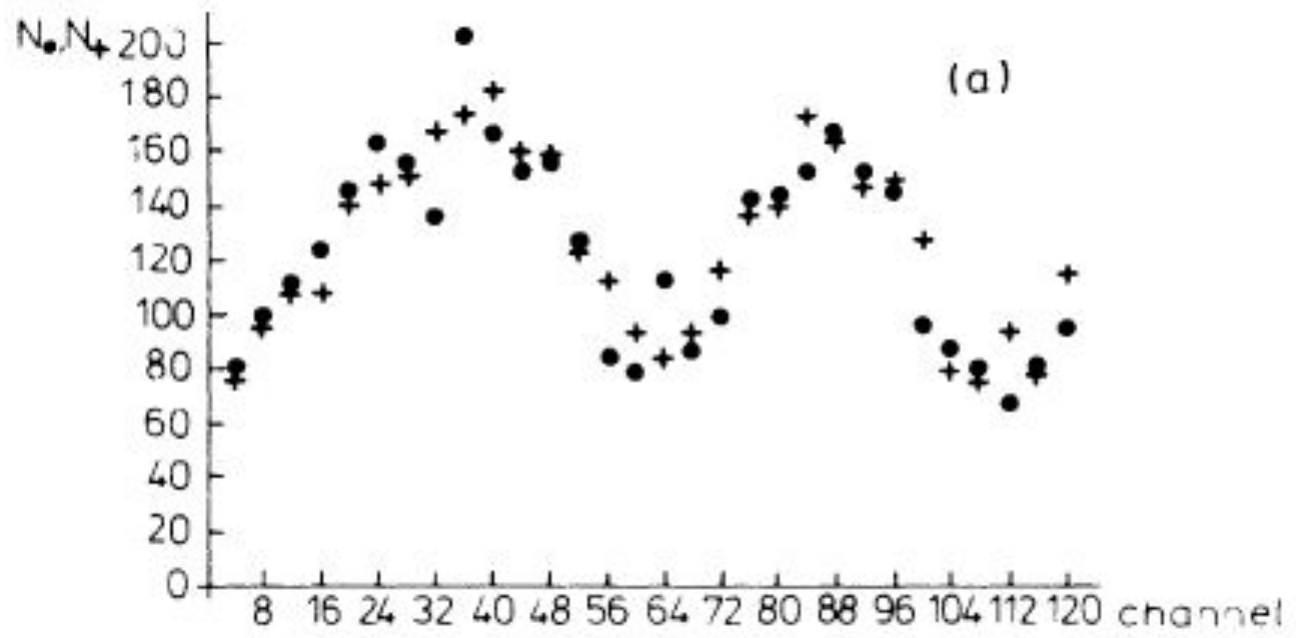

**FIG. 1.9.** Counting rates in "normal" mode [dots; no voltage applied throughout] and in "delayed-choice" mode [crosses; second path is unblocked after photon encounters BS1] as a function of path length difference. A clear interference pattern is observed in both data sets. [31]

It seems that no matter how an experiment is devised, we observe the behavior of quanta to be particle-like in some circumstances, and consistent with our expectations for classical waves in others, but we cannot demonstrate both types of behaviors simultaneously. Dirac has preemptively offered his interpretation of these experiments: each photon coherently divides at each beam splitter as a delocalized wave, interferes with itself when more than one path is available, and then instantly collapses to a point when interacting with a detector. Niels Bohr would characterize this *dual wave-particle* behavior as *complementary* (but exclusive) features of our ultimately classical understanding of an abstract quantum world. No single classical ontological category (particle or wave) can account for all the results of quantum experiments, but the union of these two complementary concepts allows for a generalized description of the whole. Like the *Yin* and *Yang* of Chinese philosophy, Bohr saw *Complementarity* as an epistemological[6] tool with broader implications; for example, he considered *truth* and *brevity* to be complementary concepts (the more you have of one, the less you have of the other). [1]

This complementary wave-particle duality is not limited to massless photons, but can be seen in the behavior of all kinds of matter. [32-34] A double-slit experiment performed with single electrons [35] is isomorphic to the experiments described above involving single photons. In this experiment, single electrons are passed through two slits and detected one at a time at seemingly random places, yet an interference pattern still builds up over time. [Fig. 1.10] A *matter-wave* interpretation of this result would insist that each electron propagates as a delocalized wave and is coherently split at both slits, interferes with itself, then becomes instantly localized in its interaction with the detecting screen. The *Copenhagen Interpretation* would say each electron's behavior at the two slits must be understood in terms of classical waves, and the nature of the detecting apparatus reveals a *complementary* electron behavior that can only be understood in terms of classical particles. Changing the nature of the experimental setup (e.g., blocking one of the slits, removing one of the paths) changes how the behavior of each electron is to be described over the course of the experiment.

[6] Epistemology concerns itself with the nature of knowledge, and how it is acquired. In simplest terms, it addresses the question: How do we know what we know?

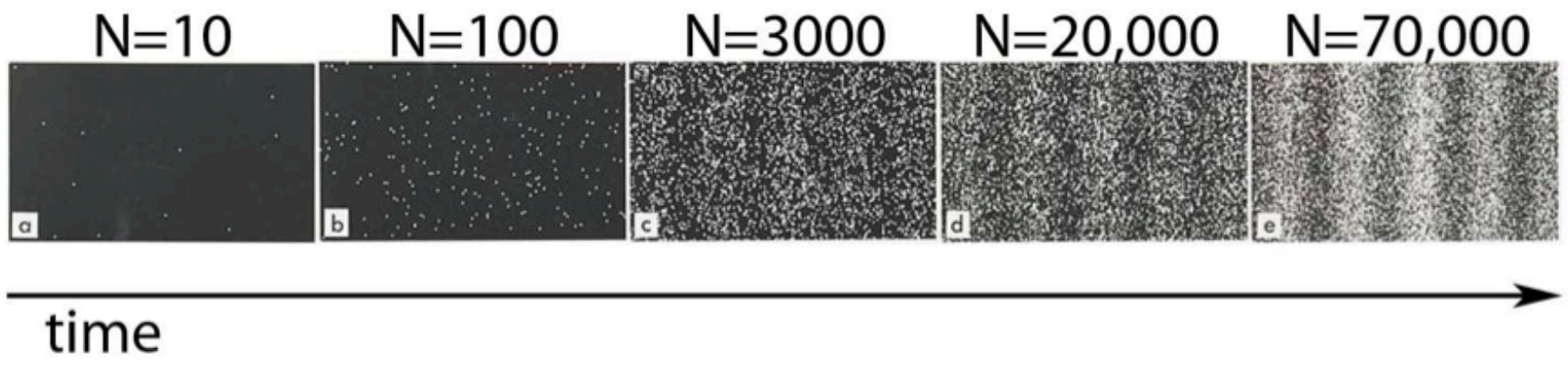

**FIG. 1.10.** Buildup of a single electron interference pattern. Single electrons are initially detected at seemingly random places, yet an interference pattern is still observed to build up after detecting many electrons. [35]

As epistemological tools, both interpretations are of similar use, in that we may employ either to decide which type of behavior will be observed in a given situation, without actually conducting the experiment: there should be no interference effects when only one path from source to detector is available; when two (or more) paths are allowed, interference will be observed. The two interpretations differ, however, in the physical meaning behind any switch between ontological descriptions of the behavior of quantum entities. In a *matter-wave* interpretation, each electron is viewed as a quantized excitation of a matter-field that (randomly) deposits its energy at a single point in its interaction with a detector. This *instantaneous collapse of the wave function* is viewed as a physical process by which these quantized excitations transition from a delocalized state (wave category) to one that is localized in space (particle category). *Complementarity* views this collapse as a moment when new information is available to the observer regarding the state of the quantum system in its interaction with the measurement apparatus (the line between which is arbitrarily drawn). The experiment reveals two sides of a more abstract quantum whole, each in analogy to classical behavior (particle- or wave-like), but any switch between ontological categories occurs only in the mind of the observer describing the system. Dividing the behavior of quantum entities into (classically) separate ontological categories is seen as a method for making sense of the decidedly nonclassical behavior of quantum entities, in terms of classical concepts intuitively associated with particles and waves.

However you choose to look at it, it should be clear that a proper understanding of quantum physics requires some degree of flexibility in the assignment of ontological categories when describing the behavior of quantum systems, and epistemological tools must be developed for understanding when each type of assignment is (or is not) appropriate for a given situation. A variety of formal interpretations of quantum theory may then be regarded in terms of coherent *epistemological and ontological framings* that guide the process of category assignment according to context. In this way, many difficulties in the conceptual understanding of quantum mechanics may be understood as stemming from varying degrees of commitment to *epistemological and ontological resources* that are in themselves neither right nor wrong, but which lead to incorrect or paradoxical conclusions when inappropriately applied to the description of quantum phenomena. We will see how this view has implications for the teaching and learning of quantum mechanics among introductory modern physics students.

## II. Epistemology and Ontology in Physics Instruction

Research into student learning has shown that, in contrast to the straight-forward acquisition of facts or skills, there are particular topics in science that are notoriously difficult for students, and where traditional modes of instruction have been demonstrated to be ineffective. Such difficulties in student learning are most generally thought of as stemming from any number of *prior ideas* held by students, which mediate the learning process, and which in some way or other must *change* before a proper (scientifically normative) understanding can be achieved. Precisely what it is that must change during this process of learning, whether it be concepts, beliefs, epistemological framings, or ontologies, is where education researchers primarily diverge. [36]

One line of research posits that many of the conceptual barriers faced by students in learning classical physics can be traced to unproductive or inappropriate degrees of commitment to *ontological category assignments*, and issues of *category inheritance*. It has been noted, for example, that *emergent processes* (such as electric current, resulting from the net motion of individual charged particles) are often alternatively conceptualized by students as *material substances* (electric current as a fluid that can be stored and consumed). [37] The general idea is that, whenever learners encounter some unfamiliar concept, they engage in a (conscious or unconscious) process of ontological categorization, whereby they sort the concept according to whatever information is available at the time. This information may include (but is not limited to) the context in which the concept is introduced, its similarity or co-occurrence with other concepts, or language patterns that give indications to its ontological nature. Once an ontological category (or sub-category) for that concept has been decided upon, it is believed that learners will then automatically associate with that concept the attributes of other concepts that fall within that same category – the new concept *inherits* the characteristics of other concepts that are ontologically similar in the mind of the learner. Many student difficulties in understanding emergent processes in classical physics can then be viewed as arising from the misattribution of properties intuitively associated with material substances. According to Chi, when the category assignment held by the learner is sufficiently distinct from the targeted (scientifically accepted) category, the process of reassignment cannot come about in gradual steps, and the learner must set aside their initial conceptualization in favor of a new conceptualization with other attributes. This *incompatibility hypothesis* motivates Chi's description of *radical conceptual change* in novice learners. [38]

A key question surrounding Chi's hypothesis is: What happens with the original ontology that is to be replaced? In their empirical work, Slotta and Chi make no real assertions regarding the ultimate fate of the original ontologies that are to be ignored by novice students, [39] though they have mentioned that

> "…physics experts do maintain substance-based conceptualizations in parallel with their more normative *process-like* views. In their everyday reasoning, physics experts often use substance-like models of heat, light, and electricity, although they are well aware of the limitations of such models, including when the models should be abandoned. Thus, if the early *substance-like* conceptions are not actually removed or replaced, we can interpret conceptual change as a matter of developing new conceptualizations alongside existing ones and understanding how and when to differentiate between alternatives." [40]

Slotta and Chi are therefore not only allowing for the possibility of *parallel ontologies* in student and expert thinking, they are insisting that productive use can be made of them by experts with a certain amount of sophistication in the flexible use of multiple ontological attributions for a single concept. [39]

Gupta, et al. have recently taken issue with the views of Slotta and Chi on ontologies in student and expert thinking, [41] most specifically with their delineation of ontologies into distinct, normative categories that remain *static*. [42] Gupta, et al. assert that not only do experts and novices often bridge between parallel ontologies, but that in many situations, clear distinctions between ontological categories don't even apply. Their view on *dynamic ontologies* claims that delineations between ontological categories and their associated attributes are not necessarily *rigid* in the minds of both experts and novices, and that they often *blend* material and process conceptualizations in their reasoning. They further take issue with the assumption that any one "scientific concept *correctly* belongs to a single ontological category." [42]

The differences in these two models of learning and cognition can be seen as analogous to the differences between material substances and emergent processes as ontological categories. A view of ontologies as distinct and stable *structures* (which is one way of accounting for the observed robustness of common student misconceptions) is contrasted with a dynamic view of flexible and adaptive ontologies that emerge in real time through the *coordinated activation* of cognitive resources (that are in themselves neither right nor wrong). In this way, the stability of misconceptions observed by Reiner, et al. [37] may be understood as resulting from contextually stable and coherent patterns of resource activation. [43] It is therefore the pattern of resource activation within a given context that must change in the minds of learners, and Gupta, et al. argue this may come about in gradual steps, so that matter-based reasoning can slowly lead to process-based reasoning. [41, 42]

It is possible these two perspectives are not entirely incompatible in the context of classical physics instruction; they may disagree on questions of meta-ontology (ontological attributions as stable cognitive structures versus emergent cognitive processes), but both agree that the learning of new concepts is mediated (and sometimes hindered) by prior knowledge (students do not enter the learning environment as blank slates), and that conceptual difficulties in learning physics often arise from the misattribution of ontological characteristics to unfamiliar concepts. And both agree that a degree of flexibility in switching between ontological attributions is not only possible, but also a *desirable* aspect of expert-like thinking. In the context of quantum physics, however, the wave-particle duality in the behavior of light and matter makes this flexibility *necessary* for a proper understanding of quantum mechanics.

We wish to extend these views on learning to the context of quantum physics in a way that would similarly address difficulties students have with changing their classical conceptions of light and matter. We first hypothesize that the intuitively *realist* perspectives of introductory physics students are reinforced by classical physics instruction, and that instruction in quantum physics can lead to measureable changes in student thinking. [Chapter 2] We will find that the highly contextual nature of student conceptions of light and matter are differentially influenced by the myriad ways in which instructors may choose (or choose not) to address interpretive themes in quantum mechanics, and that these instructional choices manifest themselves both explicitly and implicitly in the classroom. [Chapter 3] We further hypothesize that *realist* expectations among novices and experts in quantum physics are a manifestation of *classical ontological attribute inheritance*; in other words, quantum particles (at least initially, and despite evidence to the contrary) *inherit* many of their classical attributes, which can lead to incorrect or contradictory interpretations of quantum phenomena. We will demonstrate that novice quantum physics students exhibit varying degrees of flexibility in the ontological categorization of the behavior of quanta, and present evidence of students not only switching between ontological attributions both within and across contexts, but also creating a *blended* ontological category for quantum entities, simultaneously classifying them as both particle and wave (most consistent with a *pilot-wave* interpretation of quantum mechanics [4]). Moreover, it will be seen that ontological category reassignment among students can occur piecewise, context by context (particularly in cases where instruction is explicit), and that our findings are not reflective of some sudden, wholesale change in student perspectives on the ontological nature of quanta. [Chapter 4]

## III. Motivation and Overview of Dissertation Project

A detailed exploration of student perspectives on the physical interpretation of quantum mechanics is necessary, since these perspectives are an aspect of understanding physics, and have implications for how traditional content might be taught. Introductory modern physics courses are of particular interest since they often represent a first opportunity to transition students away from classical epistemologies and ontologies, to ones that are more aligned with those of practicing physicists.

In terms of assessing student difficulties in quantum mechanics, several conceptual surveys have been developed, [44-50] though most are appropriate for advanced undergraduate and beginning graduate students, since they address such advanced topics as the calculation of expectation values, or the time-evolution of quantum states. Because there does not seem to be a canonical curriculum for modern physics courses, the applicability of assessment instruments designed specifically for this kind of student population must be evaluated course-by-course. The Quantum Physics Conceptual Survey (QPCS) [51] is a recent example of an assessment instrument developed for introductory modern physics students. The authors of the QPCS found that students had the most difficulty with six questions they had classified as *interpretive*; for example, the two survey items with the lowest percentage of correct responses (~20% for each) ask whether, "according to the standard (Copenhagen) interpretation of quantum mechanics," light (or an electron) is behaving like a wave or a particle when traveling from source to detector. These authors also found that not only do a significant number of students perform reasonably well on non-interpretative questions while still scoring low on the interpretative items, there were no students who scored high on the interpretative questions but scored low on the non-interpretative ones. As the authors note, this parallels findings from Mazur [52] when comparing student performance on conventional classical physics problems versus ones requiring a solid conceptual understanding. Their results suggest that many introductory modern physics students may grasp how to use the computational tools of quantum mechanics, without a corresponding facility with notions (such as wave-particle duality) that are at odds with their classical intuitions.

Mannila, et al. [53] have previously explored student perspectives on particle-wave duality and the probabilistic nature of quantum mechanics within the context of a double-slit experiment, where a low intensity beam of quanta passes through a two-slit system and gradually forms a fringe pattern on a detecting screen. Their analysis of open-ended written student responses to a series of questions found they were dominated by "semi-classical" or "trajectory-based" ontologies, and that very few students expressed perspectives that were aligned with expert models, or even productive transitional models.[7] These authors also reported many instances of *mixed* student ontologies within that single context of a double-slit experiment, yet the design of their study provided no opportunity to further question students on any apparent inconsistencies. Our studies have demonstrated that student perspectives on quantum phenomena can vary significantly by context, [54–56, Chapters 2-4] so that it may not always be possible to make generalizations about student beliefs based on investigations within a single context.

[7] Non-local and/or statistical (probabilistic) perspectives, by their standards.

This dissertation concerns itself with a detailed exploration and characterization of student perspectives on the physical interpretation of quantum mechanics, and how these perspectives develop within the context of an introductory modern physics course. [Chapter 2] In doing so, we identify variations in teaching approaches with respect to interpretation, and their associated impacts on student thinking. [Chapters 3 & 4] These studies serve to inform the development of instructional materials designed to positively influence student perspectives on quantum physics. [Chapter 5] Further research conducted during the implementation of these materials in a modern physics course for engineering majors allow for an assessment of their effectiveness in influencing student perspectives, and inform their refinement for future use. [Chapter 6]

### Chapter 2: Development of Quantum Perspectives – Initial Studies

The first indication that student perspectives are being significantly influenced through formal instruction came from an analysis of student responses to a particular statement on the Colorado Learning Attitudes about Science Survey (CLASS) [57]: *It is possible for physicists to carefully perform the same experiment and get two very different results that are both correct.* There is a clear trend in how student responses to this statement change over the course of a three-semester introductory sequence of physics courses. In a cross-sectional study of student responses (PHYS1 [classical mechanics], N=2200; PHYS2 [classical electrodynamics], N=1650; PHYS3 [modern physics], N=730) we see a shift first from agreement to disagreement, and then back to agreement with this statement. [Fig. 1.11] At the beginning of instruction in classical mechanics (A), more students will agree (40%) with this statement than disagree (26%); yet the number in agreement decreases significantly (B) following instruction in classical physics (to 30%, $p<0.001$), while an increasing number of students disagree (to 39%, $p<0.001$). This trend then reverses itself over a single semester of modern physics (C), at the end of which a greater percentage of students agree with this statement (46%) than at the beginning of classical physics instruction.

We then analyzed the reasoning provided by approximately 600 students in an optional text box following the multiple choice response, in order to establish if their reasons for agreeing or disagreeing had changed. We find that, among students of introductory classical physics, those who disagree with this statement primarily concern themselves with the idea that there can be only one correct result for any physical measurement, while those in agreement are more conscious of the possibility for random, hidden variables to influence the outcomes of two otherwise identical experiments. Few students invoke quantum phenomena when responding before any formal instruction in modern physics; however, a single semester of modern physics instruction results in a significant increase in the percentage of students who believe that quantum phenomena would allow for two valid (but different) experimental results.

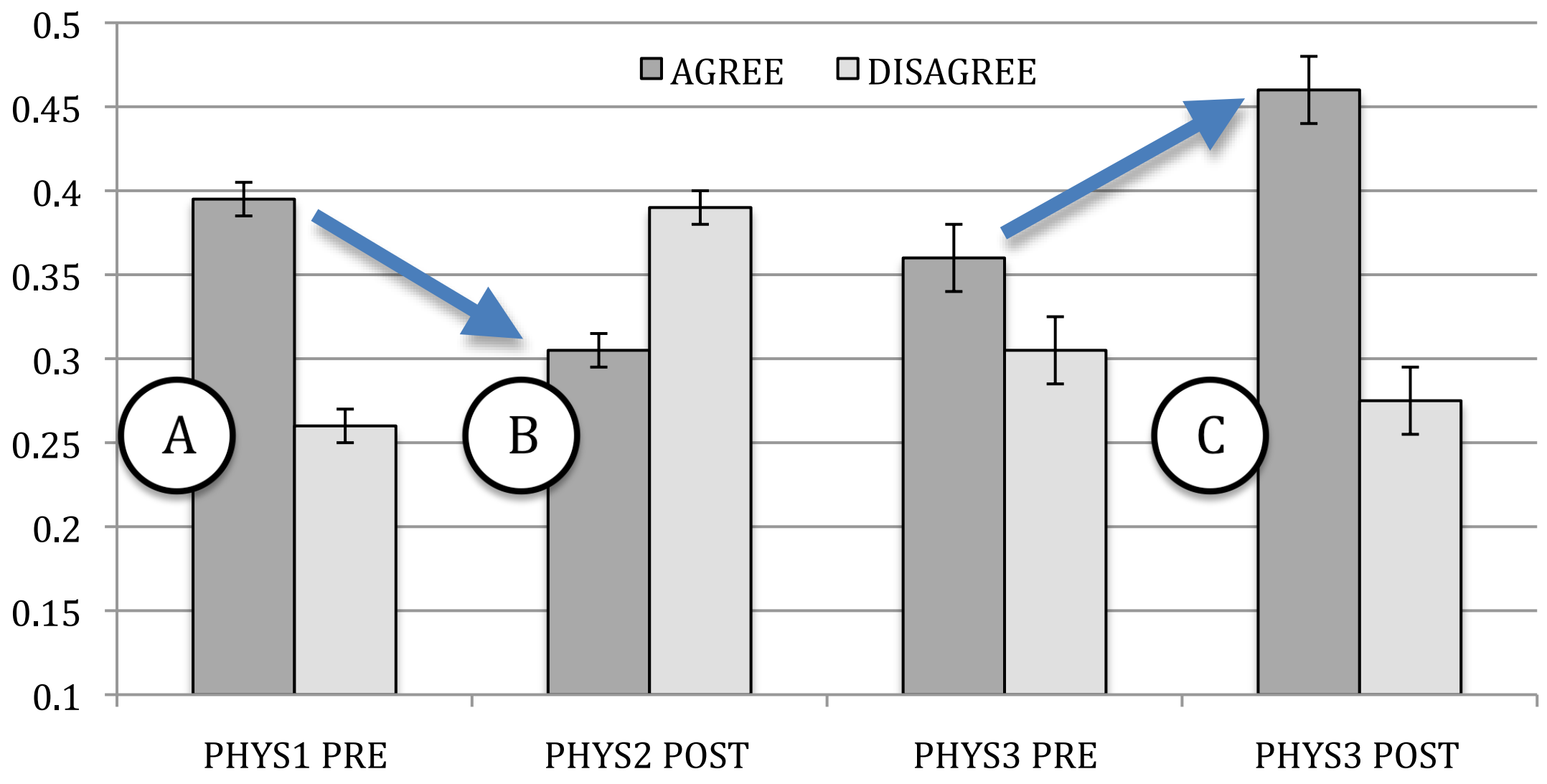

**FIG. 1.11.** Cross-sectional analysis of student responses to the statement: *It is possible for physicists to carefully perform the same experiment and get two very different results that are both correct* (expressed as a fraction of total responses: PHYS1, N=2200; PHYS2, N=1650; PHYS3, N=730). Error bars represent the standard error on the proportion.

## Chapter 3: Quantum Interpretation as Hidden Curriculum – Variations in Instructional Approaches and Associated Student Outcomes

Our efforts to characterize student perspectives on quantum physics were initially limited to the application of coarse labels (discussed below) to student responses to a post-instruction online essay question on interpretations of the double-slit experiment, coupled with responses to a survey statement concerning the existence of an electron's position within an atom. Students from courses that emphasized a *matter-wave* interpretation overwhelmingly preferred a wave description of electrons in the double-slit experiment (each electron passes through both slits and interferes with itself), while responses from courses taught from a *realist/statistical* perspective were dominated by realist interpretations (each electron goes through either one slit or the other, but not both).

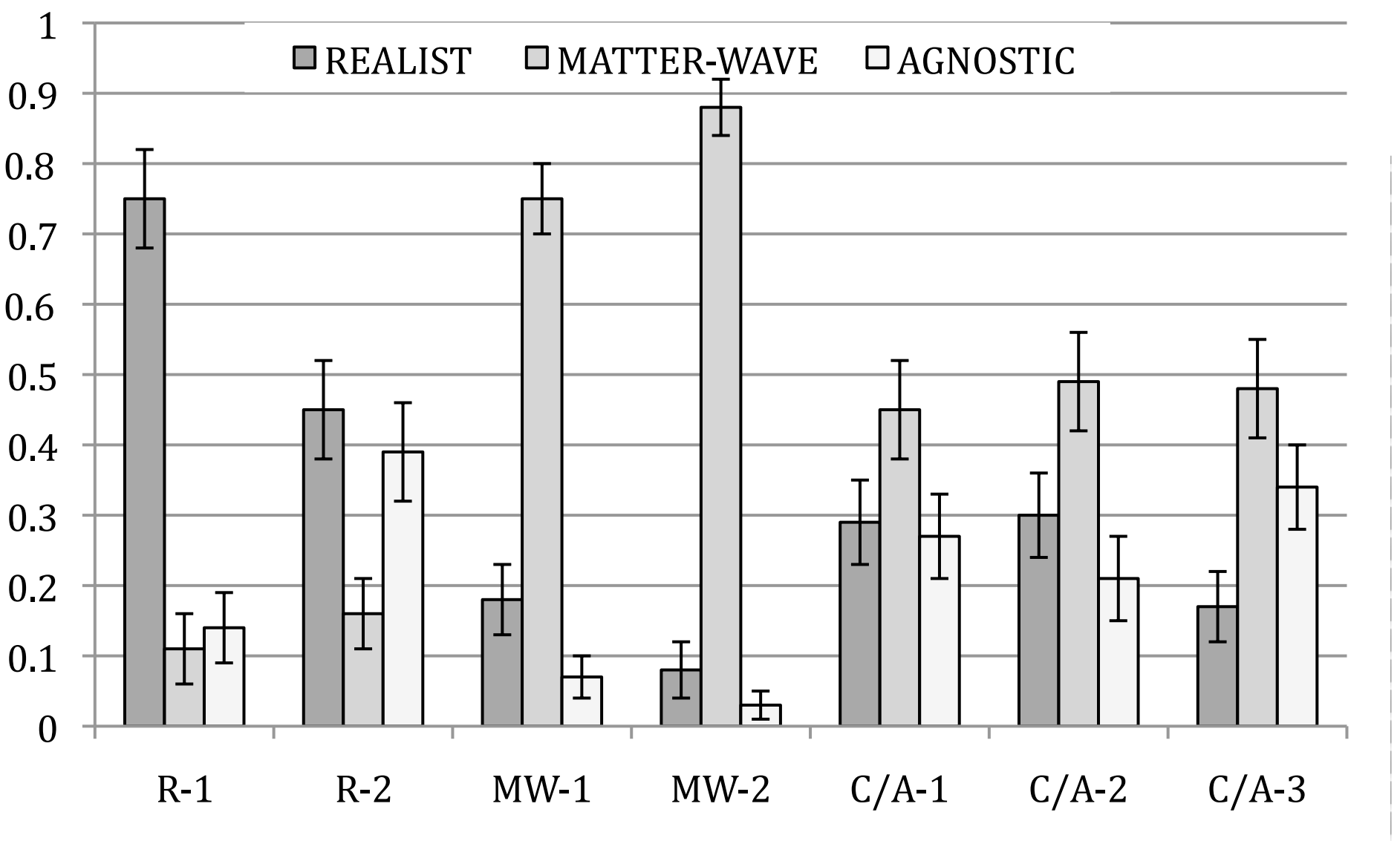

**FIG. 1.12.** Post-instruction student responses to the double-slit essay question, from seven different modern physics offerings of various instructional approaches [R = *Realist*; MW = *Matter-Wave*; C/A = *Copenhagen/Agnostic*]. Error bars represent the standard error on the proportion; N ~ 50-100 for each course.

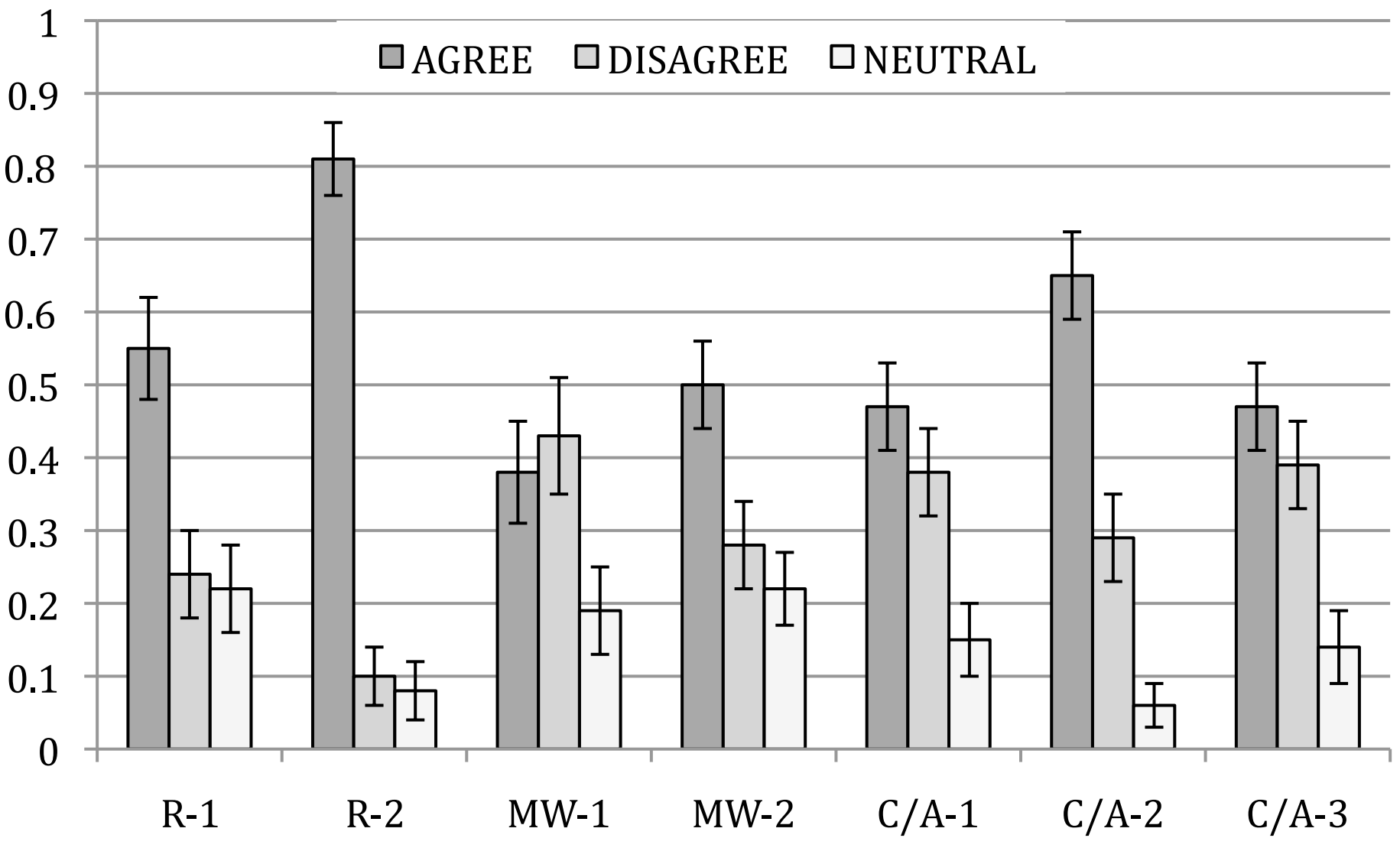

**FIG. 1.13.** Post-instruction student responses to the statement: *An electron in an atom exists at a definite (but unknown) position at each moment in time*, from seven different modern physics offerings of various instructional approaches [R = *Realist*; MW = *Matter-Wave*; C/A = *Copenhagen/Agnostic*]. Error bars represent the standard error on the proportion; N ~ 50-100 for each course.

Students from courses taught from a *Copenhagen* perspective (or ones that de-emphasized interpretation) offered more varied responses. These latter students were not only more likely to prefer an agnostic stance (quantum mechanics is about predicting the interference pattern, not discussing what happens in between), they were also more likely to align themselves with a realist interpretation. [Fig. 1.12] Of particular interest is how these same students responded to the statement: *An electron in an atom has a definite but unknown position at each moment in time*; [Fig. 1.13] Agreement with this statement would be most consistent with a realist perspective. Students from all of these types of modern physics courses were generally most likely to agree with this statement, including students from courses emphasizing a matter-wave interpretation.

When aggregate student responses from four modern physics offerings are combined so that responses to this statement on atomic electrons are grouped by how those same students responded to the essay question on the double-slit experiment, [Fig. 1.14] we see that students in the (double-slit) *Realist* category were the most consistent, with most preferring realist interpretations in both contexts. However, nearly half of the students who preferred a wave-packet description of electrons in the double-slit experiment would still agree that electrons in atoms exist as localized particles. Only those students who preferred an agnostic stance on the double-slit question were more likely to disagree with the statement than agree, and none of these students felt neutrally about whether atomic electrons are always localized. In addition, a small number of students from all courses (~5%, not shown) chose to agree with both *Matter-Wave* and *Realist* interpretations of the double-slit experiment. These findings indicate a need for more detailed characterizations of student perspectives on quantum phenomena.

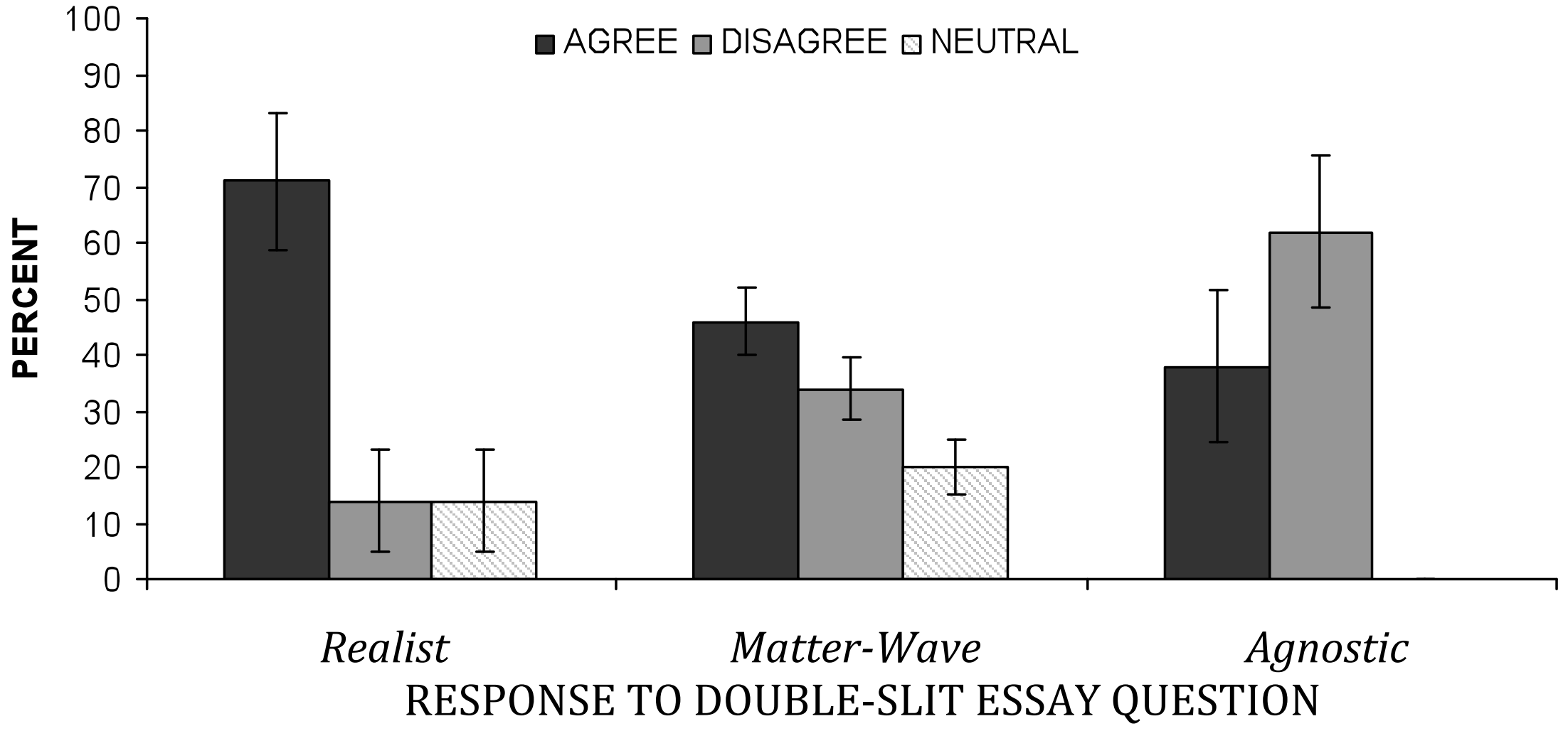

**FIG. 1.14.** Combined student responses from both PHYS3 courses to the statement: *An electron in an atom has a definite but unknown position at each moment of time*, grouped by how those students responded to the double-slit essay question. Error bars represent the standard error on the proportion (N~60).

## Chapter 4: Refined Characterization of Student Perspectives on Quantum Physics

A total of nineteen post-instruction interviews with students from four recent introductory modern physics courses taught at the University of Colorado have demonstrated that, though they may not employ the same formal language as expert physicists, students often invoke concepts and beliefs that parallel those invoked by expert physicists when arguing for their preferred interpretations of quantum mechanics. These parallels allow us to characterize student perspectives on quantum physics in terms of some of the same themes that distinguish these formal interpretations from each other. Of particular significance is the finding that students develop attitudes and opinions regarding these various themes of interpretation, regardless of whether these themes had been explicitly addressed by their instructors in class.

Results from these interviews show that, even when modern physics students have learned about "correct" responses from their instructors (or elsewhere), their classical intuitions may still influence their responses. Similar findings among classical physics students [58, 59] have shown that students most often explained differences between their *personal* and *public* perspectives in terms of responses that made intuitive sense to them (*personal*), versus ones based on their perceptions of scientists' beliefs (*public*), having not yet reconciled that knowledge with their own intuition. The inconsistent responses of some modern physics students may be similarly understood in terms of competing *personal* and *public* perspectives on quantum physics – when responding in interviews or surveys, some students frequently vacillated between what they personally believed and the answer they felt an expert physicist would give, without always explicitly distinguishing between the two.

A significant number of students from our interviews (ten of nineteen) demonstrated a preference for realist interpretations of quantum phenomena; however, only three of these students expressed personal confidence in the correctness of their perspectives, whereas four others differentiated between what made intuitive sense to them (*Realist*) and what they perceived to be correct responses (*Matter-Wave*). In addition to splits between intuition and authority, some of the seemingly contradictory responses from students may also be explained by their preferences for a *mixed* wave-particle ontology (a *pilot-wave* interpretation, wherein quanta are simultaneously *both* particle *and* wave). The realist and nonlocal beliefs of these three students were at odds with how wave-particle duality was addressed in class by their instructors (i.e. quanta are sometimes described by waves, and sometimes as particles, but never both simultaneously). The remaining nine of nineteen students seemed to express fairly consistent views that could be seen as in agreement with the instructional goals of their instructors. In other words, these students seemed to have successfully incorporated probabilistic and nonlocal views of quanta and quantum measurements into their personal perspectives.

## Chapter 5: Teaching Quantum Interpretations – Curriculum Development and Implementation

In exploring student perspectives on quantum physics, it seems natural that students should have attitudes regarding some themes of interpretation, in that these stances are reflections of each student's ideas about the very nature of reality, and the role of science in describing it. Is the universe deterministic or inherently probabilistic? When is a particle a particle, and when is it a wave? Is it unscientific to talk about that which can't be measured? A modern physics curriculum aimed at positively influencing student perspectives should provide students with the tools to formulate answers to such questions for themselves, since simply telling students about "scientifically accepted" answers does not seem to impact students at more than a superficial level.

The question remains: In what ways can student perspectives be addressed at a level appropriate for introductory modern physics students, without sacrificing traditional course content and learning goals? Although many instructors may feel that introductory students do not have the requisite sophistication to appreciate matters of interpretation in quantum mechanics, several authors have developed discussions of EPR correlations and Bell inequalities that are appropriate for the introductory level; [60, 61] relevant experimental tests of the foundations of quantum theory [5, 20, 22, 29, 31-35] may be addressed in a non-technical way. [17-19, 21, 61-64] Questions of interpretation may also be framed in terms of *scientific modeling*, an aspect of epistemological sophistication that is often emphasized in physics education research as a goal of instruction. [65] Moreover, a common lament among physics education researchers is that we are losing physics majors in the first years of their studies by only teaching them 19th-century physics in our introductory courses. Similar issues may arise when modern physics instructors limit course content mostly to the state of knowledge at the first half of the last century, or are reluctant to address questions that are clearly of personal and academic interest to students.

A modern physics course that specifically addresses student perspectives might do so within the following topics (among others):

**EPR Correlations/Entanglement:** Make explicit the assumptions of determinism and locality in the context of classical physics. The notion of atomic spin may be built up from a semi-classical (Bohr-like) atomic model; the limitations of this deterministic model become evident as it leads to predictions in conflict with experimental observation. Issues of measurement, quantum states and state preparation, and interpretation arise naturally. Indeterminacy and non-local aspects of quantum phenomena are demonstrated with simple probability arguments (thought experiments) [60, 61] and experimental evidence. [5, 17-22, 29, 31-36, 61-64] Address implications for quantum information theory (cryptography, computing, etc…). [5]

**Single-Quanta and Delayed-Choice Experiments:** The experiments of Aspect et al. demonstrate the complementary particle- and wave-like behavior of quanta, [29] providing opportunities to address various aspects of student perspectives on quantum mechanics enumerated in previous studies. [54-56, 65, 69, 70] Delayed-choice experiments [31] demonstrate the limitations of realist/statistical and pilot-wave interpretations. The basics of these experiments requires a simple understanding of atomic spectra and lasers, polarization and polarizers, beam-splitters [interferometry experiments] and photon detectors [photoelectric effect]. Discussion of these experiments can be facilitated by pointing students to non-technical articles. [17-19, 21, 61-64] Address complementarity as a general principle; help students develop an intuition for when interference effects should be visible, and when not.

**The Uncertainty Principle:** Discussions of the Uncertainty Principle (UP) follow naturally as a mathematical expression of complementarity. The UP can be framed in terms of Fourier decomposition and the properties of wave-packets. It may also be framed in terms of explicit formal interpretations. A realist/statistical interpretation is embodied in Heisenberg's Microscope. [71] A statistical interpretation concerns separate measurements performed on an ensemble of identically prepared system. [72, 73] Matter-wave and Copenhagen interpretations confront issues of indeterminacy in quantum measurement. Order-of-magnitude estimates can be made using simple models and assumptions, indicating a deeper physical meaning behind the UP beyond simple peculiarities of the measurement process. [5]

Such a curriculum has been implemented in the form of an introductory modern physics course for engineers in the Fall 2010 semester at the University of Colorado. Quantitative and qualitative data have been collected in the form of student responses to questions from previously validated instruments such as the CLASS [57], QMCS [49] and QPCS [51], as well as the same survey items and essay questions employed in our previous studies. [54-56] In this chapter, we discuss the guiding principles behind the development of this curriculum, and provide a detailed examination of specific, newly developed course materials designed to meet these goals. [A broader selection of relevant course materials can be found in Appendix C.] In doing so, we address the appropriateness and effectiveness of this curriculum by considering aggregate student responses to a subset of homework, exam, and survey items, as well as actual responses from four select students. We may employ the framework developed in Chapter 4 to characterize the perspectives of these four students as they progress through the course, and compare their incoming reasoning with how they responded at the end of the semester.

# Chapter 6: Teaching Quantum Interpretations – Comparative Outcomes and Curriculum Refinement

Results from these data collections may then be compared with previous incarnations of modern physics courses at the University of Colorado where similar data are available. We also examine student responses to specific exam questions and post-instruction content survey items, in an effort to identify which aspects of the new curriculum were most challenging for students, and propose refinements for the sake of potential future implementations and studies. Course materials specific to interpretation will be compiled and archived in a way that allows future instructors to incorporate them into their own curricula.

**References (Chapter 1)**

**1.** A. Pais, *Niels Bohr's Times, in Physics, Philosophy and Polity* (Clarendon Press, Oxford, 1991), p. 23.

**2.** A. Einstein, "Autobiographical Notes" in *Albert Einstein: Philosopher-Scientist*, P. A. Schillp (Ed.) (Open Court Publishing, Peru, Illinois 1949).

**3.** A. Einstein, B. Podolsky and N. Rosen, Can quantum mechanical description of physical reality be considered complete? *Phys. Rev.* **47**, 777 (1935).

**4.** D. Bohm and Y. Aharonov, Discussion of Experimental Proof for the Paradox of Einstein, Rosen and Podolsky, *Phys. Rev.* **108**, 1070 (1957).

**5.** G. Greenstein and A. J. Zajonc, The Quantum Challenge: Modern Research on the Foundations of Quantum Mechanics, 2$^{nd}$ ed. (Jones & Bartlett, Sudbury, MA, 2006).

**6.** N. Bohr, Can quantum mechanical description of physical reality be considered complete? *Phys. Rev.* **48**, 696 (1935).

**7.** Quoted in: A. Pais, *Niels Bohr's Times, in Physics, Philosophy and Polity* (Clarendon Press, Oxford, 1991), p. 314.

**8.** R. G. Newton, *How Physics Confronts Reality: Einstein was correct, but Bohr won the game.* (World Scientific, Singapore, 2009).

**9.** W. H. Stapp, The Copenhagen Interpretation, *Am. J. Phys.* **40**, 1098 (1972).

**10.** See, for example: D. J. Griffiths, *Introduction to Quantum Mechanics, 2nd Ed.* (Prentice Hall, Upper Saddle River, New Jersey, 2004), p. 3.

**11.** N. D. Mermin, What's Wrong with this Pillow? *Phys. Today* **42**, 9 (1989).

**12.** J. von Neumann, "Measurement and Reversibility" & "The Measuring Process" in *Mathematical Foundations of Quantum Mechanics* (Princeton University Press, Princeton, New Jersey, 1955), pp. 347-445.

**13.** J. S. Bell, On the Einstein-Podolsky-Rosen paradox, reprinted in *Speakable and Unspeakable in Quantum Mechanics* (Cambridge University Press, Cambridge, 1987), pp. 14-21.

**14.** J. S. Bell, "Locality in quantum mechanics: reply to critics," reprinted in *Speakable and Unspeakable in Quantum Mechanics* (Cambridge University Press, Cambridge, 1987), pp. 63-66.

**15.** J. F. Clauser, M. A. Horne, A. Shimony and R. A. Holt, Proposed Experiment to Test Local Hidden-Variable Theories, *Phys. Rev. Lett.* **23**, 880 (1969).

**16.** S. J. Freedman and J. F. Clauser, Experimental test of local hidden-variable theories, *Phys. Rev. Lett.* **28**, 938 (1972).

**17.** A. L. Robinson, Quantum Mechanics Passes Another Test, *Science* **217**, 435 (1982).

**18.** A. L. Robinson, Loophole Closed in Quantum Mechanics Test, *Science* **219**, 40 (1983).

**19.** A. Shimony, The Reality of the Quantum World, *Scientific American* (January 1988, pp. 46-53).

**20.** W. Tittel, J. Brendel, B. Gisin, T. Herzog, H Zbinden and N. Gisin, Experimental demonstration of quantum correlations over more than 10 km, *Phys. Rev. A* **57** (5), 3229 (1998).

**21.** A. Watson, Quantum Spookiness Wins, Einstein Loses in Photon Test, *Science* **277**, 481 (1997).

**22.** A. Aspect, P. Grangier and G. Roger, Experimental Tests of Realistic Local Theories via Bell's Theorem, *Phys. Rev. Letters* **47**, 460 (1981).

**23.** E. Schrödinger, "The Present Situation in Quantum Mechanics," in *Quantum Theory and Measurement*, J. A. Wheeler and W. H. Zurek (Eds.) (Princeton, NJ, 1983) p. 152. Originally published under the title, "Die gegenwärtige Situation in der Quantenmechanik," *Die Naturwissenschaften*, **23**, 807 (1935); translation into English by J. D. Trimmer, 1980.

**24.** A. Einstein, in *Letters on Wave Mechanics*, K. Przibram (Ed.) (Philosophical Library, New York, NY, 1986), p. 39.

**25.** S. Redner, Citation Statistics from 110 Years of Physical Review, *Phys. Today* **58**, 52 (2005).

**26.** L. E. Ballentine, Resource letter IQM-2: Foundations of quantum mechanics since the Bell inequalities, *Am. J. Phys.* **55**, 785 (1987).

**27.** A. Aspect, "John Bell and the second quantum revolution," in *Speakable and Unspeakable in Quantum Mechanics* (Cambridge University Press, Cambridge, 1987), p. xix.

**28.** P. A. M. Dirac, *The Principles of Quantum Mechanics, 3rd ed.*, (Clarendon Press, Oxford, 1947), p. 9.

**29.** P. Grangier, G. Roger, and A. Aspect, Experimental Evidence for a Photon Anticorrelation Effect on a Beam Splitter: A New Light on Single-Photon Interferences, *Europhysics Letters* **1**, 173 (1986).

**30.** J. A. Wheeler, in *Mathematical Foundations of Quantum Mechanics*, A. R. Marlow (Ed.) (Academic Press, New York, NY, 1978), pp. 9-48.

**31.** T. Hellmuth, H. Walther, A. Zajonc and W. Schleich, Delayed-choice experiments in quantum interference, *Phys. Rev. A* **35**, 2532 (1987).

**32.** R. Gaehler and A. Zeilinger, Wave-optical experiments with very cold neutrons, *Am. J. Phys.* **59** (4), 316 (1991).

**33.** O. Carnal and J. Mlynek, Young's Double-Slit Experiment with Atoms: A Simple Atom Interferometer, *Phys. Rev. Letters* **66** (21), 2689 (1991).

**34.** S. Frabboni, G. C. Gazzadi, and G. Pozzi, Nanofabrication and the realization of Feynman's two-slit experiment, *App. Phys. Letters* **93**, 073108 (2008).

**35.** A. Tonomura, J. Endo, T. Matsuda, T. Kawasaki and H. Exawa, Demonstration of single-electron buildup of an interference pattern, *Am. J. Phys.* **57**, 117 (1989).

**36.** A. A. diSessa, "A history of conceptual change research: Threads and fault lines," in *Cambridge Handbook of the Learning Sciences*, K. Sawyer (Ed.) (Cambridge University Press, Cambridge, 2006), pp. 265-281.

**37.** M. Reiner, J. D. Slotta, M. T. H. Chi and L. B. Resnick, Naïve physics reasoning: A commitment to substance-based conceptions, *Cognition and Instruction* **18**, 1 (2000).

**38.** M. T. H. Chi, Common sense misconceptions of emergent processes: Why some misconceptions are robust, *Journal of the Learning Sciences* **14**, 161 (2005).

**39.** J. D. Slotta, In defense of Chi's Ontological Incompatibility Hypothesis, *Journal of the Learning Sciences* **20**, 151 (2011).

**40.** J. D. Slotta and M. T. H. Chi, The impact of ontology training on conceptual change: Helping students understand the challenging topics in science, *Cognition and Instruction* **24,** 261 (2006).

**41.** A. Gupta, D. Hammer and E. F. Redish, The case for dynamic models of learners' ontologies in physics, *J. Learning Sciences* **19**, 285 (2010).

**42.** D. Hammer, A. Gupta and E. F. Redish, On Static and Dynamic Intuitive Ontologies, *J. Learning Sciences* **20**, 163 (2011).

**43.** D. Hammer, A. Elby, R. E. Scherr and E. F. Redish, "Resources, Framing and Transfer" in *Transfer of Learning*, J. Mestre (Ed.) (Information Age Publishing, 2005) pp. 89-119.

**44.** E. Cataloglu and R. Robinett, Testing the development of student conceptual and visualization understanding in quantum mechanics through the undergraduate career, *Am. J. Phys.* **70**, 238 (2002).

**45.** C. Singh, Student understanding of quantum mechanics, *Am. J. Phys.* **69**, 8 (2001).

**46.** C. Singh, Assessing and improving student understanding of quantum mechanics, *PERC Proceedings 2006* (AIP Press, Melville, NY, 2006).

**47.** C. Singh, Student understanding of quantum mechanics at the beginning of graduate instruction, *Am. J. Phys.* **76**, 3 (2008).

**48.** J. Falk, Developing a quantum mechanics concept inventory, unpublished master thesis, Uppsala University, Uppsala, Sweden (2005). Available at: http://johanfalk.net/node/87 (Retrieved January, 2011).

**49.** S. B. McKagan and C. E. Wieman, Exploring Student Understanding of Energy Through the Quantum Mechanics Conceptual Survey, *PERC Proceedings 2005* (AIP Press, Melville, NY, 2006).

**50.** S. Goldhaber, S. Pollock, M. Dubson, P. Beale and K. Perkins, Transforming Upper-Division Quantum Mechanics: Learning Goals and Assessment, *PERC Proceedings 2009* (AIP, Melville, NY, 2009).

**51.** S. Wuttiprom, M. D. Sharma, I. D. Johnston, R. Chitaree and C. Chernchok, Development and Use of a Conceptual Survey in Introductory Quantum Physics, *Int. J. Sci. Educ.* **31 (5)**, 631 (2009).

**52.** E. Mazur, *Peer instruction: A user's manual* (Prentice Hall, New York, NY, 1997).

**53.** K. Mannila, I. T. Koponen and J. A. Niskanen, Building a picture of students' conceptions of wave- and particle-like properties of quantum entities, *Euro. J. Phys.* **23**, 45-53 (2002).

**54.** C. Baily and N. D. Finkelstein, Development of quantum perspectives in modern physics, *Phys. Rev. ST: Physics Education Research* **5**, 010106 (2009).

**55.** C. Baily and N. D. Finkelstein, Teaching and understanding of quantum interpretations in modern physics courses, *Phys. Rev. ST: Physics Education Research* **6**, 010101 (2010).

**56.** C. Baily and N. D. Finkelstein, Refined characterization of student perspectives on quantum physics, *Phys. Rev. ST: Physics Education Research* **6**, 020113 (2010).

**57.** W. K. Adams, K. K. Perkins, N. Podolefsky, M. Dubson, N. D. Finkelstein and C. E. Wieman, A new instrument for measuring student beliefs about physics and learning physics: the Colorado Learning Attitudes about Science Survey, *Phys. Rev. ST: Physics Education Research* **2**, 1, 010101 (2006).

**58.** T. L. McCaskey, M. H. Dancy and A. Elby, Effects on assessment caused by splits between belief and understanding, *PERC Proceedings 2003* **720**, 37 (AIP, Melville, NY, 2004).

**59.** T. L. McCaskey and A. Elby, Probing Students' Epistemologies Using Split Tasks, *PERC Proceedings 2004* **790**, 57 (AIP, Melville, NY, 2005).

**60.** D. F. Styer, *The Strange World of Quantum Mechanics* (Cambridge University Press, Cambridge, 2000).

**61.** N. D. Mermin, Is the moon there when nobody looks? Reality and the quantum theory, *Phys. Today* **38** (4), 38 (1985).

**62.** M. Tegmark and J. A. Wheeler, 100 Years of Quantum Mysteries, *Scientific American* (February 2001, pp. 68-75).

**63.** D. Z. Albert and R. Galchen, A Quantum Threat to Special Relativity, *Scientific American* (March 2009, pp. 32-39).

**64.** B.-G. Englert, M. O. Scully and H. Walther, The Duality in Matter and Light, *Scientific American* (December 1994, pp. 86-92).

**65.** S. B. McKagan, K. K. Perkins and C. E. Wieman, Why we should teach the Bohr model and how to teach it effectively, *Phys. Rev. ST: Physics Education Research* **4**, 010103 (2008).

**66.** E. Redish, J. Saul and R. Steinberg, Student expectations in introductory physics, *Am. J. Phys.* **66**, 212 (1998).

**67.** R. Hake, Interactive-Engagement Versus Traditional Methods: A Six-Thousand-Student Survey of Mechanics Test Data for Introductory Physics Courses, *Am. J. Phys.* **66** (1), 64 (1998).

**68.** L. C. McDermott, Oersted Medal Lecture 2001: "Physics Education Research – The Key to Student Learning," *Am. J. Phys.* **69**, 1127 (2001).

**69.** S. B. McKagan, K. K. Perkins and C. E. Wieman, Reforming a large lecture modern physics course for engineering majors using a PER-based design, *PERC Proceedings 2006* (AIP Press, Melville, NY, 2006).

**70.** S. B. McKagan, K. K. Perkins, M. Dubson, C. Malley, S. Reid, R. LeMaster and C. E. Wieman, Developing and Researching PhET simulations for Teaching Quantum Mechanics, *Am. J. Phys.* **76**, 406 (2008).

**71.** W. Heisenberg, "The Physical Content of Quantum Kinematics and Mechanics," in *Quantum Theory and Measurement*, J. A. Wheeler and W. H. Zurek (Eds.) (Princeton, 1983) p. 62. Originally published under the title, "Uber den anschaulichen Inhalt der quantentheoretischen Kinematik und Mechanik," Zeitschrift fur Physik, **43**, 172 (1927); translation into English by J. A. Wheeler and W. H. Zurek, 1981.

**72.** L. E. Ballentine, The Statistical Interpretation of Quantum Mechanics, *Rev. Mod. Phys.* **42**, 358 (1970).

**73.** L. E. Ballentine, Quantum Mechanics: A Modern Development (World Scientific Publishing, Singapore, 1998).

# CHAPTER 2

## Development of Student Perspectives - Initial Studies

### I. Introduction

Our initial investigations into student perspectives seek to document and better understand the changes students undergo as they make the transition from learning classical physics to learning about quantum mechanics. We first analyze student responses to pre- and post-instruction surveys at various stages of an introductory physics sequence in order to demonstrate the development and reinforcement of deterministic perspectives during classical physics instruction, as well as the emergence of probabilistic and nondeterministic perspectives following instruction in modern physics. We also find that a modern physics instructor's choice of learning goals can significantly influence student responses: they are more likely to prefer either a *Realist* or *Quantum* (*matter-wave)* perspective in a context where such a perspective has been explicitly taught. Furthermore, a student's degree of commitment to any particular perspective is not necessarily robust across contexts: students may invoke both *Realist* and *Quantum* perspectives, without always knowing when either of these epistemological and ontological frames is appropriate. These studies serve as motivation for a more detailed exploration of variations in learning goals among modern physics instructors, and the associated impacts on student perspectives. [Chapter 3]

### II. Studies

The University of Colorado offers a three-semester sequence of calculus-based introductory physics courses: PHYS1 and PHYS2 are large-lecture courses [1] (N~300-600) in classical mechanics and electrodynamics, respectively; PHYS3 covers a variety of topics from modern physics, and is offered in two sections (N~50-100, each). At the beginning and end of each semester, students from several offerings of each of the above courses were asked to respond to a series of survey questions designed to probe their epistemic and ontological perspectives on physics. The first of these surveys was an online version of the Colorado Learning Attitudes about Science Survey (CLASS), [2] wherein students responded using a 5-point Likert-scale (ranging from strong disagreement to strong agreement) to a series of 42 statements, including:

> **#41:** It is possible for physicists to carefully perform the same experiment and get two very different results that are both correct.

CLASS researchers do not score student responses to this statement as favorable or unfavorable [2] due to a lack of consensus among expert responses[1]. The myriad ambiguities contained in this statement allow for a number of legitimate (but different) interpretations by expert physicists: they may disagree on what it means to conduct the *same* experiment, what qualify as *very different* results, or even what it means for an experimental result to be considered *correct*.

## II.A. Student ideas about measurement change over time.

There is a clear trend in how student responses to CLASS #41 change over the course of this introductory sequence. In a cross-sectional study of student responses from the three introductory physics courses (PHYS1, N=2200; PHYS2, N=1650; PHYS3, N=730) we see a shift first from agreement to disagreement, and then back to agreement with this statement. [Fig. 2.1] At the beginning of instruction in classical mechanics (A), more students will agree (40%) with this statement than disagree (26%); yet the number in agreement decreases significantly (B) following instruction in classical physics (to 30%, $p<0.001$), while an increasing number of students disagree (to 39%, $p<0.001$). This trend then reverses itself over a single semester of modern physics (C), at the end of which a greater percentage of students agree with this statement (46%) than prior to classical physics instruction.

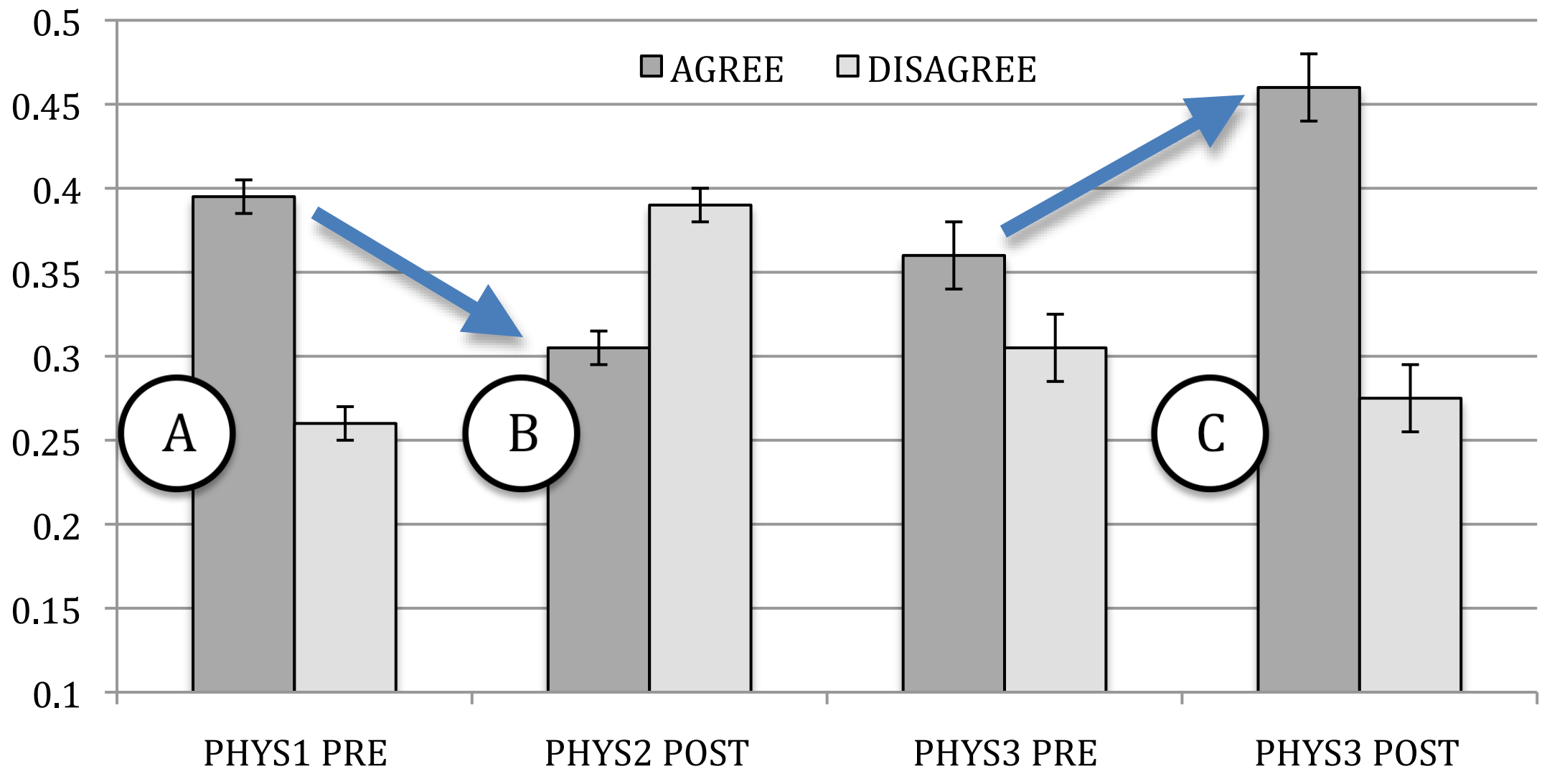

**FIG. 2.1.** Cross-sectional analysis of student responses to the statement: *It is possible for physicists to carefully perform the same experiment and get two very different results that are both correct* (expressed as a fraction of total responses: PHYS1, N=2200; PHYS2, N=1650; PHYS3, N=730). Error bars represent the standard error on the proportion.

[1] In informal interviews, physics faculty members at the University of Colorado responded approximately 35% Agree, 60% Disagree, and 5% Neutral.

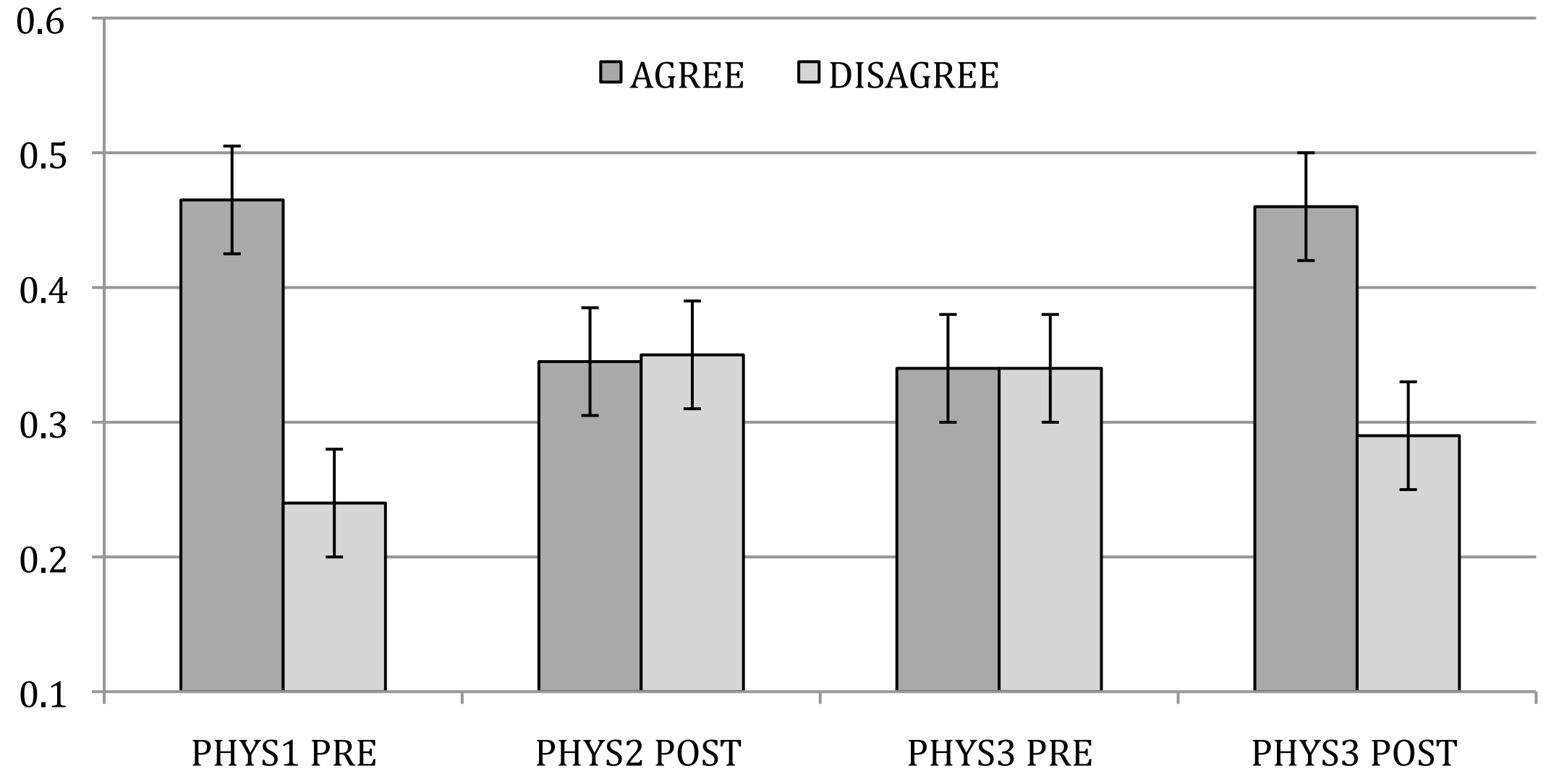

**FIG. 2.2.** Longitudinal study of student responses to the statement: *It is possible for physicists to carefully perform the same experiment and get two very different results that are both correct* (expressed as a fraction of total responses: N=124). Error bars represent the standard error on the proportion.

In a longitudinal study of 124 students over three semesters, we observe the same trends. [Fig. 2.2]

The distribution of student responses at the end of this introductory sequence is similar to that at the beginning (in terms of agreement versus disagreement); we are naturally interested then in finding out if and how the reasoning invoked by students in defense of their responses changes. We analyzed the reasoning provided by approximately 600 students in an optional text box appended to an online version of the CLASS. These open-ended responses were coded into five categories through an emergent coding scheme. [3] [Table 2.I] The types of reasons offered by modern physics students at the start of instruction was similar to that from students in classical physics courses (pre- and post-instruction), and so the data for both have been combined into a single, pre-quantum instruction group. [Table 2.II]

**TABLE 2.I.** Categorization of reasoning provided by students in response to the statement: *It is possible for physicists to carefully perform the same experiment and get two very different results that are both correct.*

| | |
|---|---|
| **A** | Quantum theory/phenomena |
| **B** | Relativity/different frames of reference |
| **C** | There can be more than one correct answer to a physics problem.<br>Experimental results are open to interpretation. |
| **D** | Experimental/random/human error<br>Hidden variables, chaotic systems |
| **E** | There can be only one correct answer to a physics problem.<br>Experimental results should be repeatable. |

**TABLE 2.II.** Distribution of reasoning provided by students before and after instruction in modern physics, in response to the statement: *It is possible for physicists to carefully perform the same experiment and get two very different results that are both correct.* Categories are as given in Table 2.I. Errors are the standard error on the proportion.

| CATEGORY | PRE-QM INSTRUCTION (+/-2%) | | POST-QM INSTRUCTION (+/-5%) | |
|---|---|---|---|---|
| | AGREE (N=231) | DISAGREE (N=199) | AGREE (N=41) | DISAGREE (N=26) |
| **A** | 10% | 5% | 32% | 27% |
| **B** | 3% | 0% | 17% | 4% |
| **C** | 28% | 6% | 10% | 8% |
| **D** | 59% | 20% | 41% | 19% |
| **E** | 0 | 69% | 0 | 42% |

Our analysis shows that, prior to instruction in modern physics, 59% of those who agreed with the statement offered Category **D** explanations (experimental error, hidden variables); Category E explanations (physics problems have only one correct answer) were preferred by those who disagreed (69%). These results (in conjunction with other studies [4]) allow us to conclude that most introductory classical physics students who disagree with this statement interpret the results of experimental measurements as an approximation of the true (real) value of the quantity being measured; whereas most of those who agree with the statement allow for the possibility of random, hidden factors to influence the outcome of two otherwise identical experiments.

We find that before any formal instruction in modern physics, few students invoke quantum phenomena, despite the fact that a majority of them reported having heard about quantum mechanics in popular venues before enrolling in the course (e.g., books by Greene [5] and Hawking, [6]). However, a single semester of modern physics instruction results in a significant increase in the number of students who believe that quantum physics could allow for two valid, but different, experimental results. Students shift from 13% to 49% in referencing quantum or relativistic reasons for agreeing with the statement. [Table 2.II] Responses from each population were compared with a Chi-Square test and were found to be statistically different ($p<0.001$).

## II.B. Instructional choices influence student perspectives.

To see if different types of instruction and learning goals can significantly influence student commitments to any particular perspective, we examined data from two PHYS3 offerings intended for physics majors. Course PHYS3A was taught by a PER instructor who employed in-class, research-based reforms [7], including interactive engagement and computer simulations [8] designed to provide students with a visualization of quantum processes; course PHYS3B was taught the following semester in the form of more traditional lectures. Both modern physics offerings were similar in devoting roughly one-third of the course to special relativity, with the remaining lectures covering the foundations of quantum mechanics and simple applications (as is typical at the University of Colorado). Notable differences in these two courses included the instructional approaches and learning goals of the instructors. Through informal end-of-term interviews and an analysis of course materials, it is clear that each of the instructors held different beliefs about incorporating interpretive aspects of quantum mechanics into a modern physics curriculum. In the context of a double-slit experiment performed with electrons, the instructor for PHYS3A ("Instructor A") explicitly taught that each electron propagates as a delocalized wave while passing through both slits, interferes with itself, and then becomes localized upon detection. Instructor B preferred a more agnostic stance on the physical interpretation of this experiment, and generally did not address such issues:

> "It seems like there's a new book about different interpretations of quantum mechanics coming out every other week, so I see this as something that is still up for debate among physicists. When I talked about the double-slit experiment in class, I used it to show students the need to think beyond F=ma, but I didn't talk about any of that other stuff. […] We did talk a little about [quantum weirdness] at the very end of the semester, but it was only because we had some time left over and I wanted to give the students something fun to talk about."

Despite Instructor B's self-reported *Agnostic* stance on quantum interpretations, his instructional practices differed in that he explicitly told students that each electron in a double-slit experiment passes through either one slit or the other, but that it is fundamentally impossible to determine which one without destroying the interference pattern (he characterized this *Realist* perspective as the one with which he was "least dissatisfied").

Students from both of these courses were given an end-of-term essay question asking them to argue for or against statements made by three fictional students discussing the Quantum Wave Interference (QWI) PhET simulation's [9] representation of a double-slit experiment with single electrons. [Fig. 2.3] In this simulation, a large circular spot (representing the magnitude of the wave function for a single electron, equivalent to the probability density) (A) emerges from a gun, (B) passes through two slits, and (C) a small dot appears on a detection screen; after a long time (many electrons) an interference pattern develops (not shown).

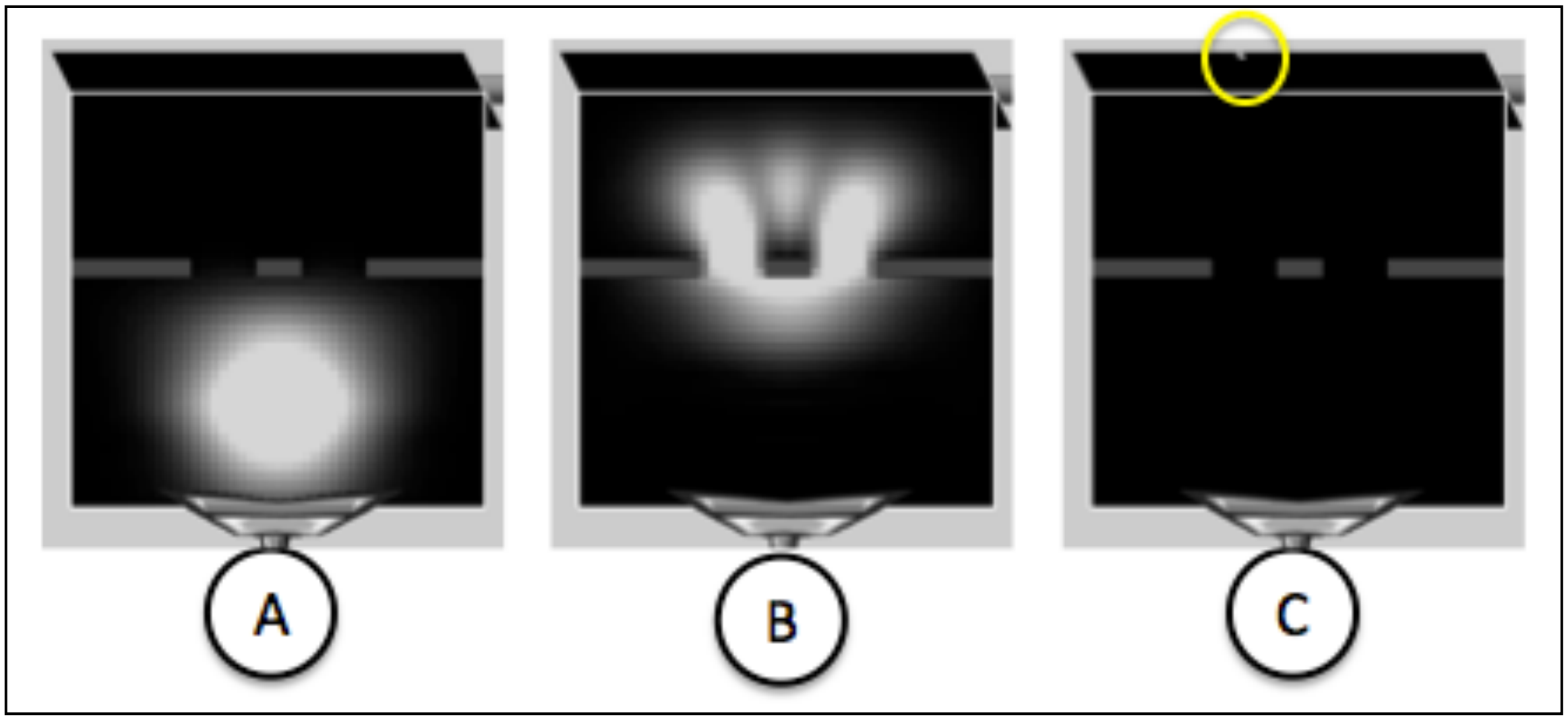

**FIG. 2.3.** Representation of a double-slit experiment with single electrons in the Quantum Wave Interference PhET simulation; used in the end-of-term essay question.

Each of the following statements (made by a *fictional* student) is meant to represent a potential perspective on how to think of an electron between the time it is emitted and when it is detected at the screen:

> **Student 1**: *That blob represents the probability density, so it tells you the probability of where the electron could have been before it hit the screen. We don't know where it was in that blob, but it must have actually been a tiny particle that was traveling in the direction it ended up, somewhere within that blob.*
>
> **Student 2**: *No, the electron isn't inside the blob, the blob represents the electron! It's not just that we don't know where it is, but that it isn't in any one place. It's really spread out over that large area up until it hits the screen*.
>
> **Student 3**: *Quantum mechanics says we'll never know for certain, so you can't ever say anything at all about where the electron is before it hits the screen.*

In this end-of-term survey question, students were asked to agree or disagree with any (or all) of the fictional students, and to provide evidence in support of their responses, which were then coded according to whether students preferred a *Realist* or a *Matter-Wave* perspective in their argumentation. A random sample of 20 student responses were re-coded by a PER researcher unaffiliated with this project as a test for inter-rater reliability; following discussion of the coding scheme, the two codings were in 100% agreement. The following sample of two student responses is illustrative of the types of responses seen:

> **Student Response (*Realist*):** "We just can't know EXACTLY where the electron is and thus the blob actually represents the probability density of that electron. In the end, only a single dot appears on the screen, thus the electron, wherever it was in the probability density cloud, traveled in its own direction to where it ended up."
>
> **Student Response (*Matter-Wave*)**: "The blob is the electron and an electron is a wave packet that will spread out over time. The electron acts as a wave and will go through both slits and interfere with itself. This is why a distinct interference pattern will show up on the screen after shooting out electrons for a period of time."

The distribution of all responses for the two courses is summarized in Table 2.III (columns do not add to 100% because some students provided a mixed or otherwise unclassifiable response; almost none of the responses favored Student 3). For this essay question, there is a strong bias towards a *Matter-Wave* perspective among PHYS3A students, while students from PHYS3B highly preferred a *Realist* perspective. Virtually no student agreed with fictional Student 3 (which might be consistent with an *Agnostic* perspective); among those who explicitly disagreed with Student 3, most felt that knowing about the probability density was a sufficient form of knowledge about this quantum system.

**TABLE 2.III.** Student responses to the Quantum Wave Interference essay question from two offerings of PHYS3. Numbers in parentheses represent the standard error on the proportion.

| CATEGORY | PHYS3A (%) (N=72) | PHYS3B (%) (N=44) |
|---|---|---|
| *Realist* | 18 (5) | 75 (7) |
| *Matter-Wave* | 78 (5) | 11 (5) |

Students from both PHYS3 courses also responded at the beginning and end of the semester to additional statements appended to an online version of the CLASS for modern physics students, including:

> **QA#2**: An electron in an atom has a definite but unknown position at each moment in time.

It might be expected that students who have learned to view an electron as delocalized until detected in the context of a double-slit experiment would also view it as such in other contexts, such as atoms. Disagreement with this statement on atomic electrons could be consistent with either a *Matter-Wave* or *Copenhagen/Agnostic* perspective, whereas agreement would be more consistent with a *Realist* perspective. While we again observe differences in student responses between the two PHYS3 course offerings [Table 2.IV] there is not the same strong bias toward a single perspective as seen in Table 2.III. Disagreement with this statement among PHYS3A students increased by 22%, and by 13% for PHYS3B students; agreement with this statement decreased by 5% in PHYS3A, while the number of PHYS3B students agreeing with this statement increased by a comparably small amount.

**TABLE 2.IV.** Student responses to the statement: *An electron in an atom has a definite but unknown position at each moment in time.* Numbers in parentheses represent the standard error on the proportion.

| RESPONSE | PHYS3A (%) (N=41) | | PHYS 3B (%) (N=36) | |
|---|---|---|---|---|
| | PRE | POST | PRE | POST |
| AGREE | 44 (8) | 39 (8) | 48 (8) | 54 (8) |
| NEUTRAL | 32 (7) | 17 (6) | 39 (8) | 21 (7) |
| DISAGREE | 22 (6) | 44 (8) | 10 (5) | 23 (7) |

## II.C Consistency of student perspectives

An important question remains: are there consistencies in student perspectives across domains? The differences in responses from PHYS3A and PHYS3B students are less significant for QA#2 [Table 2.IV] than those seen for the QWI essay question [Table 2.III], but together indicate a possible lack of consistency in their preferred perspectives in different contexts. This inconsistency can be better illustrated by combining matching data for both questions, and then grouping together students from both courses according to how they responded to the QWI essay question. [Table 2.V] In doing so, we see that students who preferred a *Matter-Wave* perspective in the essay question tended to disagree with the notion that atomic electrons exist as localized particles; and the majority of students who preferred a *Realist* perspective in the first case also took a *Realist* stance on the question of atomic electrons. Of particular interest, however, are the students who were not consistent in their responses: 18% of those who disagreed with QA#2, and 33% of those who agreed, offered a response that was inconsistent with their response to the QWI essay question. That is, 18% of students disagreed with the statement on atomic electrons, yet gave a *Realist* response on the interference question; 33% of students were the reverse: taking a *Realist* stance on atomic electrons, but preferring a *Matter-Wave* perspective on the question of electron interference.

**TABLE 2.IV.** Student responses to the statement: *An electron in an atom has a definite but unknown position at each moment in time,* grouped according to how they responded to the QWI essay question. Numbers in parentheses represent the standard error on the proportion.

| QA#2 - POST / QWI | DISAGREE (%) | NEUTRAL (%) | AGREE (%) |
|---|---|---|---|
| *Matter-Wave* (N=66) | 56 (6) | 11 (4) | 33 (6) |
| *Realist* (N=46) | 18 (6) | 18 (6) | 64 (7) |

## III. Summary and Discussion

The data presented in this chapter serve as evidence in support of three key findings. First, student perspectives with respect to measurement and determinism in the contexts of classical physics and quantum mechanics evolve over time. The distribution of reasoning provided by students in response to the CLASS survey statement indicate that the majority of those who disagree with this statement believe that experimental results should be repeatable, or that there can be only one correct answer to a physics problem. One could easily imagine that students begin their study of classical physics at the university level with a far more deterministic view of science than is evidenced by their initial responses (after all, most students do arrive with some training in classical science). We take the first significant shift in student responses (a decrease in agreement and an increase in disagreement with this statement, as shown in Fig. 2.1) to be indicative of the promotion and reinforcement of a deterministic perspective in students as a result of instruction in classical physics. After a course in modern physics, student responses shift a second time (an increase in agreement and a decrease in disagreement with the survey statement), although the reasoning behind their responses changed. Students of modern physics are instructed that different frames of reference could lead to different experimental results, both of which are correct (special relativity); they are also taught that the quantum-mechanical description of nature is probabilistic, and that the determinism assumed by Newtonian mechanics is no longer valid at the atomic scale. The impact of this type of instruction is reflected in the significant increase in the number of students who invoke relativistic or quantum phenomena as a reason for agreeing with the survey statement.

Second, we observe that how students develop and apply a particular perspective can depend upon the learning goals of their instructors. The results for the Quantum Wave Interference essay question indicate that how students view an electron within the context of a double-slit experiment can be significantly influenced by instruction. Instructor A explicitly taught students that each electron passes through both slits and interferes with itself, and provided students with an in-class visualization of this process via the QWI PhET simulation. The positivistic aspects of the *Copenhagen Interpretation* [10] insist that questions of which slit any particular electron passed through are (at best) ill-posed, and that quantum mechanics concerns itself only with the probabilistic prediction of experimental results. An *Agnostic* stance might say that the question of which slit an electron passed through is irrelevant to the proper application of the mathematical formalism. Although Instructor B reported personally holding an *Agnostic* stance on questions of interpretation in quantum mechanics, he did not teach this perspective explicitly, but rather was explicit in teaching a *Realist* interpretation of the double-slit experiment; this instructional approach is partly reflected in how the majority of PHYS3B students preferred a *Realist* stance on electrons in this context.

Third, we find that many students do not exhibit a consistent perspective on questions of ontology and epistemology across multiple contexts. While the data shown in Table 2.IV do demonstrate some amount of consistency in responses regarding the question of an electron's location, a significant number of students who preferred a *Matter-Wave* interpretation of an electron diffraction experiment would still agree that an electron in an atom has a definite (but unknown) position. We conclude that students will not necessarily develop robust concepts regarding the nature of quanta, which would be consistent with a resources view of student epistemologies and ontologies in physics. [16-19]

Without passing judgment on any particular set of instructional goals, it is worth acknowledging that significant differences in the teaching of modern physics courses do exist (as with upper-division courses in quantum mechanics[11]), and that these learning goals manifest themselves both explicitly and implicitly (intentionally, or not) during the course of instruction. It is in itself a significant finding that, at least in this regard, students are open to adopting their instructor's explicit interpretations of quantum phenomena (though it may be argued in the case of Instructor B that his explicit instruction was already in alignment with the *realist* expectations of his students); there is substantial evidence that students do not necessarily adopt an instructor's views in other contexts. Previous studies of introductory classical physics courses have shown that, with notably few exceptions, [12-14] students tend to shift to more unfavorable (novice-like) beliefs about physics and about the learning of physics [12, 15]. It has been demonstrated, however, that making epistemology an explicit aspect of instruction in introductory physics courses can positively influence this negative trend. [14] The studies presented in this chapter provide further indication that instructors should not take for granted that students will adopt their perspectives on quantum physics unless such learning goals are made explicit in their teaching.

In the end, it seems that a reasonable instructional objective would be for students to apply a particular perspective (deterministic or probabilistic, local or nonlocal) at the appropriate time. If we are to include these goals for our classes, it is important to understand how these messages are sent to our students, and what instructional practices may promote such understandings. [Chapter 3]

**References (Chapter 2)**

**1.** S. Pollock and N. D. Finkelstein, Sustaining Change: Instructor Effects in Transformed Large Lecture Courses, *PERC Proceedings 2006* (AIP Press, Melville, NY, 2006).

**2.** W. K. Adams, K. K. Perkins, N. Podolefsky, M. Dubson, N. D. Finkelstein and C. E. Wieman, A new instrument for measuring student beliefs about physics and learning physics: the Colorado Learning Attitudes about Science Survey, *Phys. Rev. ST: Physics Education Research* **2**, 1, 010101 (2006).

**3.** J. W. Creswell, *Education Research*, 2nd Ed. (Prentice Hall, Englewood Cliffs, NJ, 2005), pp. 397-398.

**4.** A. Buffler, S. Allie, F. Lubben & B. Campbell, The development of first year physics students' ideas about measurement in terms of point and set paradigms, *Int. J. Sci. Educ.* **23,** 11 (2001).

**5.** B. R. Greene, *The Elegant Universe* (Norton, New York, NY, 2003).

**6.** S. W. Hawking, *A Brief History of Time* (Bantam, New York, NY, 1988).

**7.** S. B. McKagan, K. K. Perkins and C. E. Wieman, Reforming a large lecture modern physics course for engineering majors using a PER-based design, *PERC Proceedings 2006* (AIP Press, Melville, NY, 2006).

**8.** http://phet.colorado.edu

**9.** http://phet.colorado.edu/simulations/sims.php?sim=QWI

**10.** W. H. Stapp, The Copenhagen Interpretation, *Am. J. Phys.* **40**, 1098 (1972).

**11.** S. Goldhaber, S. Pollock, M. Dubson, P. Beale, and K. Perkins, Transforming Upper-Division Quantum Mechanics: Learning Goals and Assessment, *PERC Proceedings 2009* (AIP, Melville, NY, 2009).

**12.** E. Redish, J. Saul and R. Steinberg, Student expectations in introductory physics, *Am. J. Phys.* **66**, 212 (1998).

**13.** V. K. Otero and K. E. Gray, "Attitudinal gains across multiple universities using the Physics and Everyday Thinking curriculum," *Phys. Rev. ST: Physics Education Research* **4** (1), 020104 (2008).

**14.** D. Hammer, Student resources for learning introductory physics, *Am. J. Phys.* **68**, S52 (2000).

**15.** S. Pollock, "No Single Cause: Learning Gains, Student Attitudes, and the Impacts of Multiple Effective Reforms" *PERC Proceedings 2004* (AIP Press, Melville, NY, 2005).

**16.** D. Hammer, Student resources for learning introductory physics, *Am. J. Phys.* **68**, S52 (2000).

**17.** D. Hammer, A. Elby, R. E. Scherr and E. F. Redish, "Resources, Framing and Transfer" in *Transfer of Learning*, edited by J. Mestre (Information Age Publishing, 2005) pp. 89-119.

**18.** A. Gupta, D. Hammer and E. F. Redish, The case for dynamic models of learners' ontologies in physics, *J. Learning Sciences* **19**, 285 (2010).

**19.** D. Hammer, A. Gupta and E. F. Redish, On Static and Dynamic Intuitive Ontologies, *J. Learning Sciences* **20**, 163 (2011).

# CHAPTER 3

## Quantum Interpretation as Hidden Curriculum - Variations in Instructor Practices and Associated Student Outcomes

### I. Introduction

In physics education research, the term *hidden curriculum* generally refers to aspects of science and learning about which students develop attitudes and opinions over the course of instruction, but which are primarily only implicitly addressed by instructors. [1] Students may hold varying beliefs regarding the relevance of course content to real-world problems, the coherence of scientific knowledge, or even the purpose of science itself, depending (in part) on the choices and actions of their instructors. Education research has demonstrated that student attitudes regarding such matters tend to remain or become less expert-like when instructors are not explicit in addressing them. [1] In this chapter we present similar findings: the less explicit an instructor is in addressing student perspectives within a given topic area, the greater the likelihood for students (within that specific context) to favor an intuitive, *realist* perspective. In other words, the less the interpretive aspects of quantum mechanics are explicitly addressed by instructors, the more they become part of a hidden curriculum. We explore here how modern physics instructors may (or may not) address this hidden curriculum, and examine the impact of specific instructional approaches on student thinking. Figs. 3.1 & 3.2 (where letters refer to specific instructors and their particular approaches, to be discussed below) illustrate how instructional choices can lead to significantly different student outcomes, as well as the mixed nature of student responses across contexts.

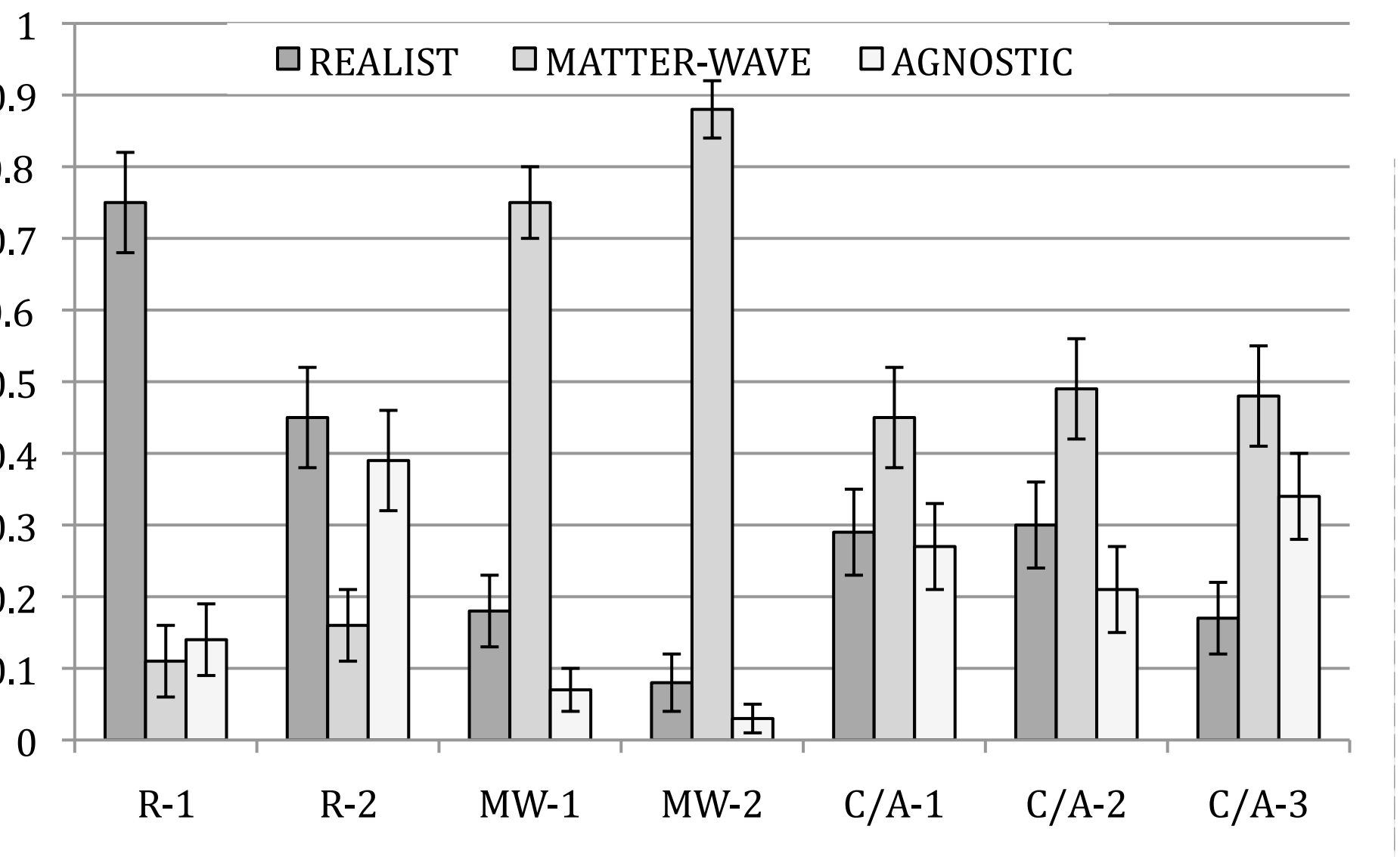

**FIG. 3.1.** Post-instruction student responses to the double-slit essay question, from seven different modern physics offerings of various instructional approaches [R = *Realist*; MW = *Matter-Wave*; C/A = *Copenhagen/Agnostic*]. Error bars represent the standard error on the proportion; N ~ 50-100 for each course.

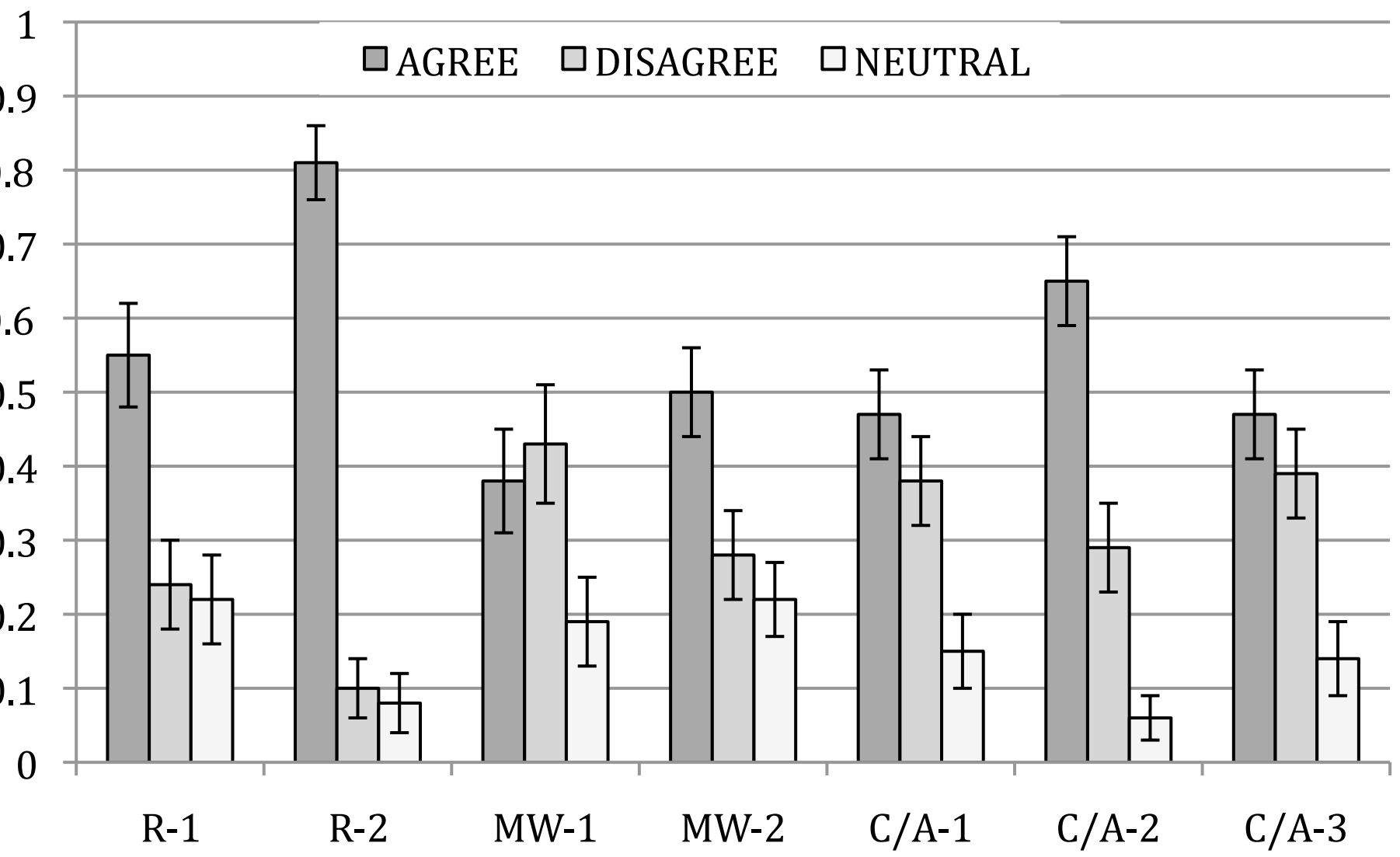

**FIG. 3.2.** Post-instruction student responses to the statement: *An electron in an atom exists at a definite (but unknown) position at each moment in time*, from seven different modern physics offerings of various instructional approaches [R = *Realist*; MW = *Matter-Wave*; C/A = *Copenhagen/Agnostic*]. Error bars represent the standard error on the proportion; N ~ 50-100 for each course.

## II. Instructors approach quantum interpretation differently

This section describes four specific approaches to addressing quantum interpretation in four different modern physics courses recently taught at the University of Colorado, each resulting in significant differences in student thinking by the end of the semester. All four courses were large-lecture (N~100), utilized interactive engagement in class, and devoted the usual proportions of lecture time to special relativity and quantum mechanics. Student responses to the double-slit essay question and statement on atomic electrons described in Chapter 2 are shown in Figs. 3.3 & 3.4, where letters refer to the specific instructors discussed in this section (and their particular approaches to instruction). With respect to the double-slit experiment with electrons, each of these instructors had been explicit in teaching one particular interpretation (*though not explicitly as an interpretation*); student responses in this context were generally reflective of the teaching approaches for each course. [Fig. 3.3]

In two of the four courses (B1 & C) instructors paid considerably less attention to interpretive themes at later stages of the course, as when students learned about the Schrödinger model of hydrogen. Students from all four courses were more likely to agree than disagree with the statement: *An electron in an atom has a definite (but unknown) position at each moment in time*. [Fig. 3.4] What follows is a more detailed discussion of the specific instructional approaches employed in the courses described above, where letters refer to specific instructors, as given in the figure captions.

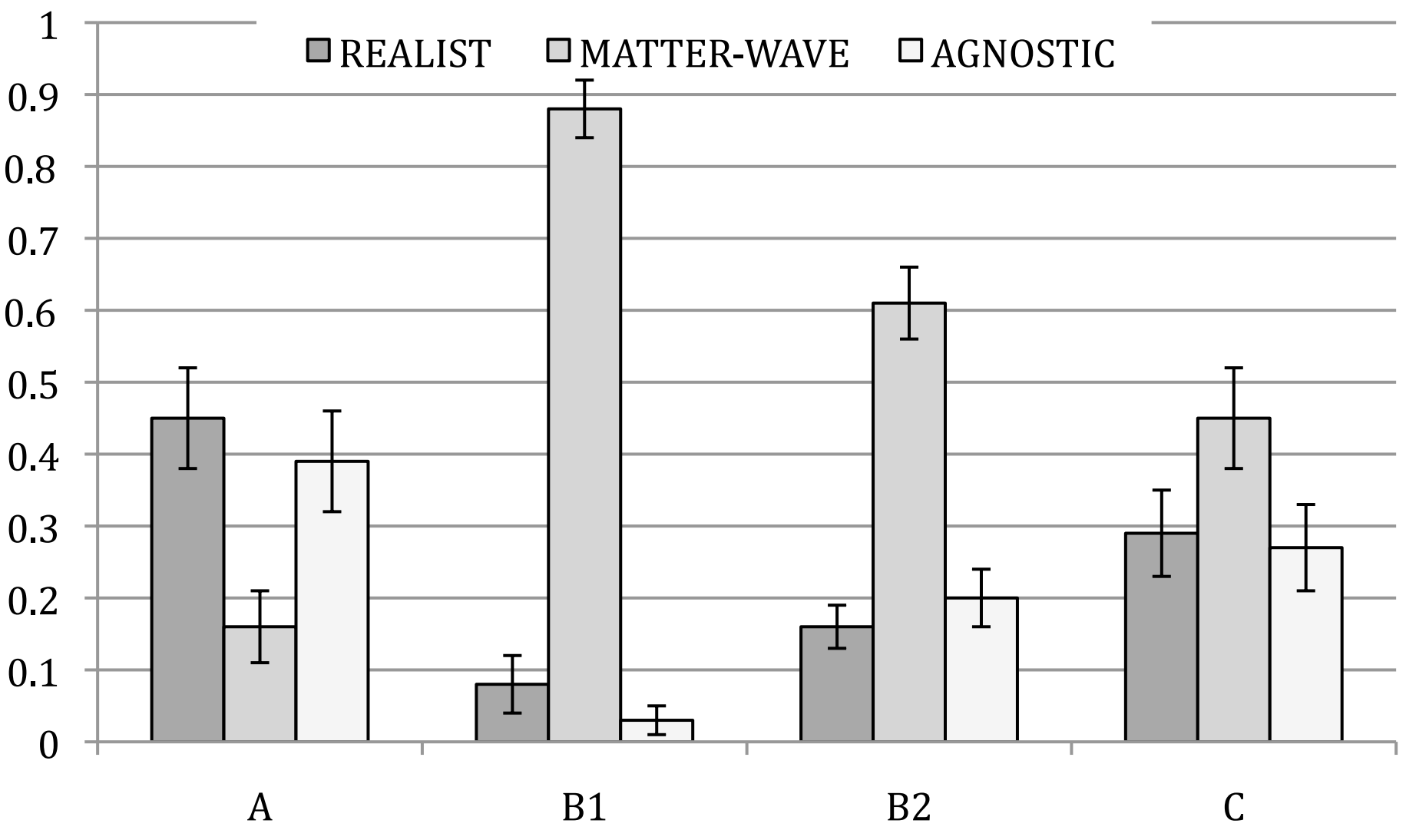

**FIG. 3.3.** Post-instruction student responses to the double-slit essay question, from four different modern physics offerings of various instructional approaches [A = *Realist/Statistical*; B1 & B2 = *Matter-Wave*; C = *Copenhagen/Agnostic*]. Error bars represent the standard error on the proportion; N ~ 100 for each course.

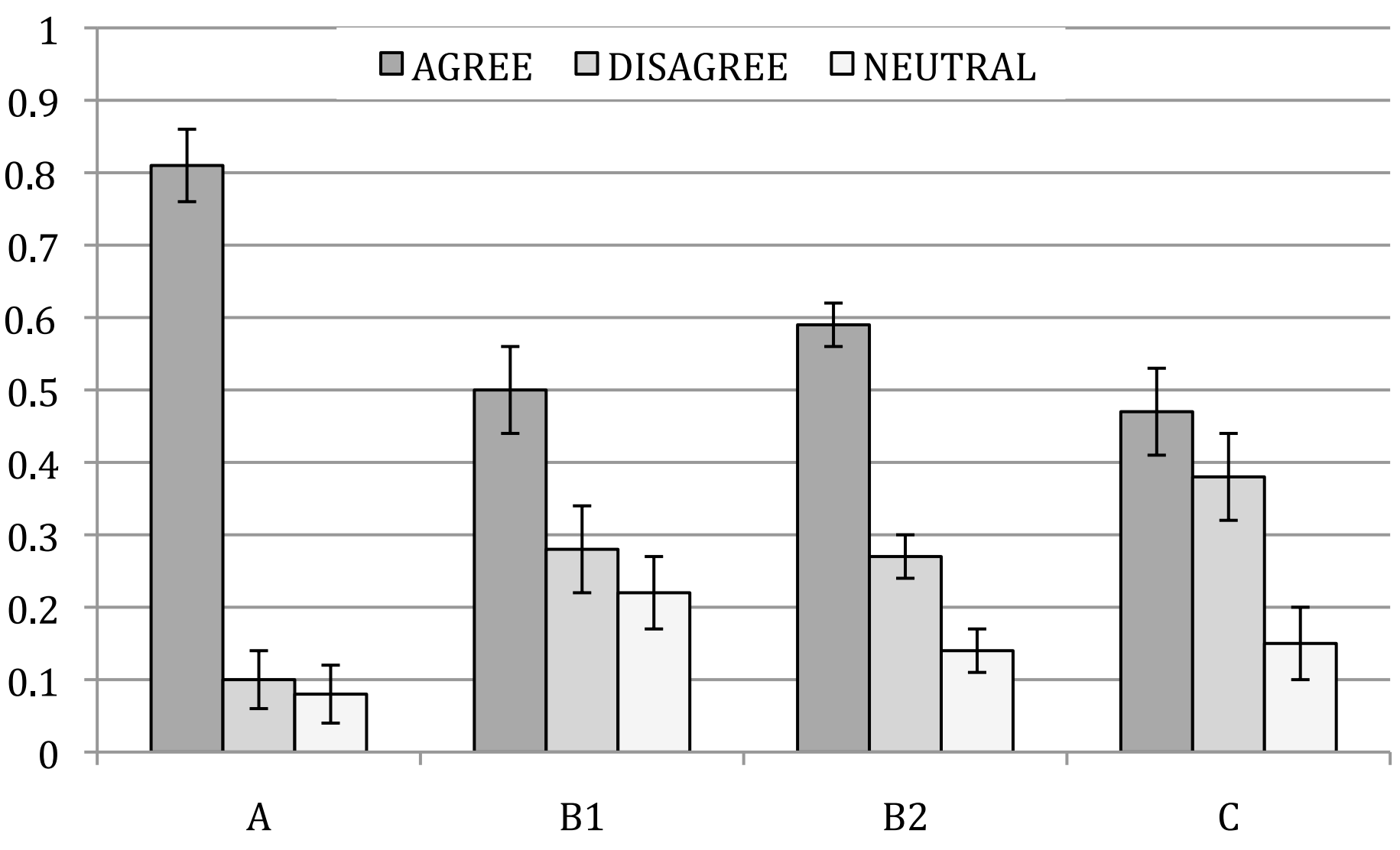

**FIG. 3.4.** Post-instruction student responses to the statement: *An electron in an atom exists at a definite (but unknown) position at each moment in time*, from four different modern physics offerings of various instructional approaches [A = *Realist/Statistical*; B1 & B2 = *Matter-Wave*; C = *Copenhagen/Agnostic*]. Error bars represent the standard error on the proportion; N ~ 100 for each course.

**A. Explicitly teach an interpretation that aligns with student intuition, without discussing alternatives:** Instructor A taught this course for engineering majors from a *Realist/Statistical* perspective (though he did not call it such), and explicitly referred to this in class as his own interpretation of quantum phenomena, one that other physicists would not necessarily agree with. Beyond his *Realist* stance on the double-slit experiment, students were explicitly instructed to think of atomic electrons as localized particles, and that energy quantization is the result of their average behavior; there was no discussion of alternatives to the perspective being promoted in class. Student responses from this course in both contexts were in alignment with Instructor A's explicit learning goals: they were the most likely to prefer a *Realist* interpretation of the double-slit experiment [each electron goes through either one slit or the other, but not both], as well as the most likely to agree that atomic electrons exist as localized particles. We believe student responses from this course are reflective not only of this instructor's explicit instruction, but also that this particular kind of interpretation of quantum mechanics is in agreement with intuitively *realist* expectations.

**B1. Teach one interpretation (though not explicitly as an interpretation) in some topic areas (particularly at the beginning of the course) and expect students to generalize to other contexts on their own:** When first teaching this modern physics course for engineering majors, Instructor B was explicit in modeling single quanta in the double-slit experiment as delocalized waves that pass through both slits simultaneously. He did not frame this discussion in terms of modeling or interpretation, but rather made what he saw as sufficient arguments in favor of this particular interpretation, as he stated in an informal post-instruction interview:

> "This image that [students] have of this [probability] cloud where the electron is localized, it doesn't work in the double-slit experiment. You wouldn't get diffraction. If you don't take into account both slits and the electron as a delocalized particle, then you will not come up with the right observation, and I think that's what counts. The theory should describe the observation appropriately. […] It really shouldn't be a philosophical question just because there are different ways of describing the same thing [i.e. as a wave or a particle]. They seem to disagree, but in the end they actually come up with the right answer."

Students from this *Matter-Wave* course overwhelmingly preferred a wave-packet description of individual electrons [each electron passes through both slits simultaneously and interferes with itself]. However, these students did not seem to generalize this notion of particles as delocalized waves to the context of atoms, where Instructor B was not explicit regarding the ontological nature of electrons, and where a majority still agreed that atomic electrons exist as localized particles. Students were more likely to prefer *Realist* notions in a topic area where Instructor B was not explicit regarding interpretation.

**B2. Teach one interpretation (though not explicitly as an interpretation) in some topic areas, combined with a more general discussion of interpretative themes towards the end of the course:** Instructor B later taught a second modern physics course for engineering majors in a similar manner, but this time devoted two lectures near the end of the course to interpretive themes in quantum mechanics, including a discussion of the interpretive aspects of the double-slit experiment (but without reference to atomic systems). Student responses were similar to the previous *Matter-Wave* course (B1) on interpretations of the double-slit experiment, but a majority of students still preferred a *Realist* stance on atomic electrons.

**C. Teach a Copenhagen/Agnostic perspective, or de-emphasize questions of interpretation:** In this modern physics course for physics majors, Instructor C did touch on some interpretive themes during the course, though he ultimately emphasized a perspective that was more pragmatic than philosophical, as when faced with the in-class question of whether particles have a definite but unknown position, or have no definite position until measured:

> "Newton's Laws presume that particles have a well defined position and momentum at all times. Einstein said we can't know the position. Bohr said, philosophically, it has no position. *Most physicists today say: We don't go there. I don't care as long as I can calculate what I need.*" [Emphasis added]

In an end-of-term interview, Instructor C clarified his attitude toward teaching any particular perspective to students in a sophomore-level course:

> "In my opinion, until you have a pretty firm grip on how QM actually works, and how to use the machine to make predictions, so that you can confront the physical measurements with pairs of theories that conflict with each other, there's no basis for ragging on the students about, 'Oh no, the electron, it's all in your head until you measure it.' They don't have the machinery at this point, and so anybody who wants to stand in front of [the class] and pound on the table and say some party line about what's really going on, nevertheless has to recognize that the students have no basis for buying it or not buying it, other than because they're being yelled at."

Student responses from this course to the double-slit essay question were more varied than with the other courses – students were not only more likely to prefer an *Agnostic* stance [quantum mechanics is about predicting the interference pattern, not discussing what happens between], a significant number of students (30%) preferred a *Realist* interpretation – more than with the *Matter-Wave* courses, but less so than with the *Realist/Statistical* course. Nearly half of all students from this course also preferred a *Realist* stance on atomic electrons.

## III. Comparing Instructor Practices (A Closer Look)

The goal of understanding the interplay between instructor practices and student perspectives calls for a more detailed comparison of two modern physics courses with similar content and presentation, but different in their approach to interpretive themes in quantum mechanics (Courses B1 & C from Section II, both of which took place in the semester immediately following the studies described in Chapter 2).

### III.A. Background on course materials and curriculum similarities.

Each semester, the University of Colorado (CU) offers two versions of its introductory modern physics course; one section is intended for engineering majors (e.g., Course B1), and the other for physics majors (Course C). The curricula for both versions of the course have traditionally been essentially the same, with variations from semester to semester according to instructor preferences. In the fall of 2005, a team from the physics education research (PER) group at CU introduced a transformed curriculum for the engineering course incorporating research-based principles. [2] This included interactive engagement techniques (in-class concept questions, peer instruction, and computer simulations [3]), as well as revised content intended to emphasize reasoning development, model building, and connections to real-world problems. These course transformations, implemented during the FA05-SP06 academic year, were continued in FA06-SP07 by another physics education researcher at CU, who then collaborated in the FA07 semester with a non-PER faculty member to adapt the course materials into a curriculum appropriate for physics majors (by including topics from special relativity).

The course materials [4] for all five of these semesters (which included lecture slides and concept tests) were made available to Instructors B & C, who both reported changing a majority of the lecture slides to some extent (as well as creating new ones). By examining the course syllabi and categorizing the lecture material for each course into ten standard introductory quantum physics topics, we find the general progression of topics in both classes to be essentially the same (the presentation of content was many times practically identical), with slight differences in emphasis. [Table 3.I]

**TABLE 3.I.** Progression of topics and number of lectures devoted to each topic from the quantum physics portion of both modern physics courses B1 & C.

| CODE | TOPIC | # OF LECTURES | |
|---|---|---|---|
| | | B1 | C |
| A | Introduction to quantum physics | 2 | 1 |
| B | Photoelectric effect, photons | 5 | 4 |
| C | Atomic spectra, Bohr model | 5 | 3 |
| D | DeBroglie waves/atomic model | 1 | 1 |
| E | Matter waves, interference/diffraction | 3 | 2 |
| F | Wave functions, Schrödinger equation | 2 | 5 |
| G | Potential energy, infinite/finite square well | 3 | 3 |
| H | Tunneling, alpha-decay, STM's | 2 | 4 |
| I | 3-D Schrödinger equation, hydrogen atom | 4 | 2 |
| J | Multi-electron atoms, periodic table, solids | 3 | 3 |

### III.B. Differences in instructional approaches.

While the learning environments and progression of topics for both modern physics courses were essentially the same, the two courses differed in sometimes obvious, other times more subtle ways with respect to how each instructor addressed student perspectives and themes of interpretation. An analysis of the instructional materials used in each of the two courses offers a first-pass comparison of the two approaches. When comparing the homework assignments for each course, there were no (or very minimal) opportunities for students to reflect on physical interpretations of quantum phenomena. Similarly, an examination of the midterms and finals from both courses revealed no emphasis on questions of interpretation. The one place that afforded the most faculty/student interaction with respect to interpretation was in the lecture portions of each course, and so we examine how these two instructors specifically addressed interpretation during lecture.

A first analysis of lecture materials entails a coding of lecture slides (which were later posted on the course website). We employ a simple counting scheme by which each slide is assigned a point value of zero or one in each of three categories, according to its relevance to three interpretive themes. [Table 3.II] These three categories (denoted as *Light*, *Matter,* & *Contrasting Perspectives*) were chosen to highlight key lecture slides that were explicit in promoting non-classical perspectives. Since light is classically described as a wave, slides that emphasized its particle-like nature, or explicitly addressed its dual wave-particle characteristics, were assigned a point in the *Light* category; similarly, slides that emphasized the wave nature of matter, or its dual wave/particle characteristics, were given a point in the *Matter* category. Other key slides (*Contrasting Perspectives* category) were those that addressed randomness, indeterminacy, or the probabilistic nature of quantum mechanics; or those that made explicit contrast between quantum results and what would be expected in a classical system. While most of the slides in Table 3.II received only one point in a single category, many slides were relevant to multiple categories, and so the point totals do not represent the total number of relevant slides from each course.

**TABLE 3.II.** Categorization of lecture slides relevant to promoting non-classical perspectives, with a point total for each category.

| THEME | DESCRIPTION OF LECTURE SLIDE | B1 | C |
|---|---|---|---|
| *Light* | Relevant to the dual wave/particle nature of light, or emphasizing its particle-like characteristics | 15 | 9 |
| *Matter* | Relevant to the dual wave/particle nature of matter, or emphasizing its wave-like characteristics | 15 | 16 |
| *Contrasting Perspectives* | Relevant to randomness, indeterminacy, or the probabilistic nature of quantum mechanics; explicit contrast between quantum & classical descriptions. | 28 | 22 |

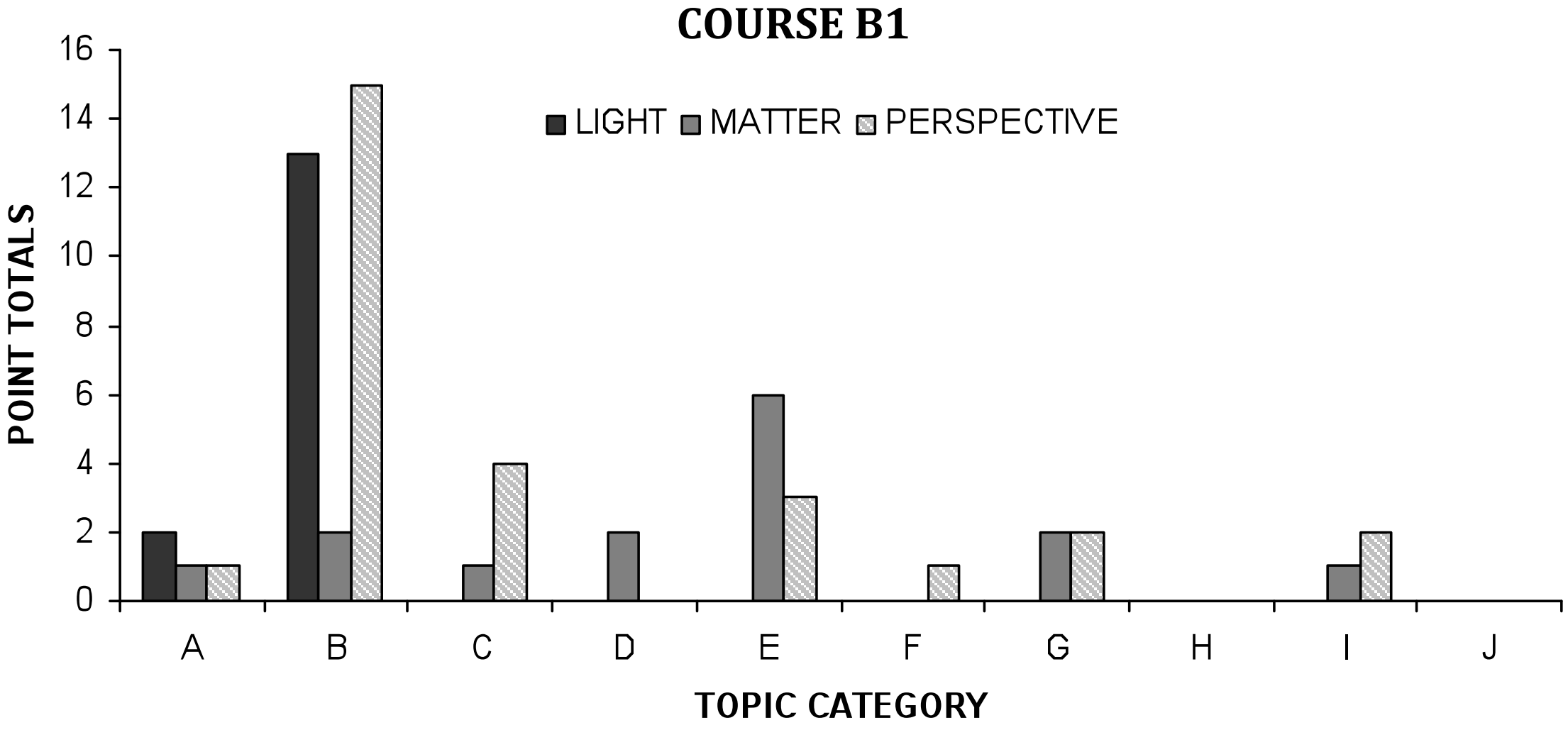

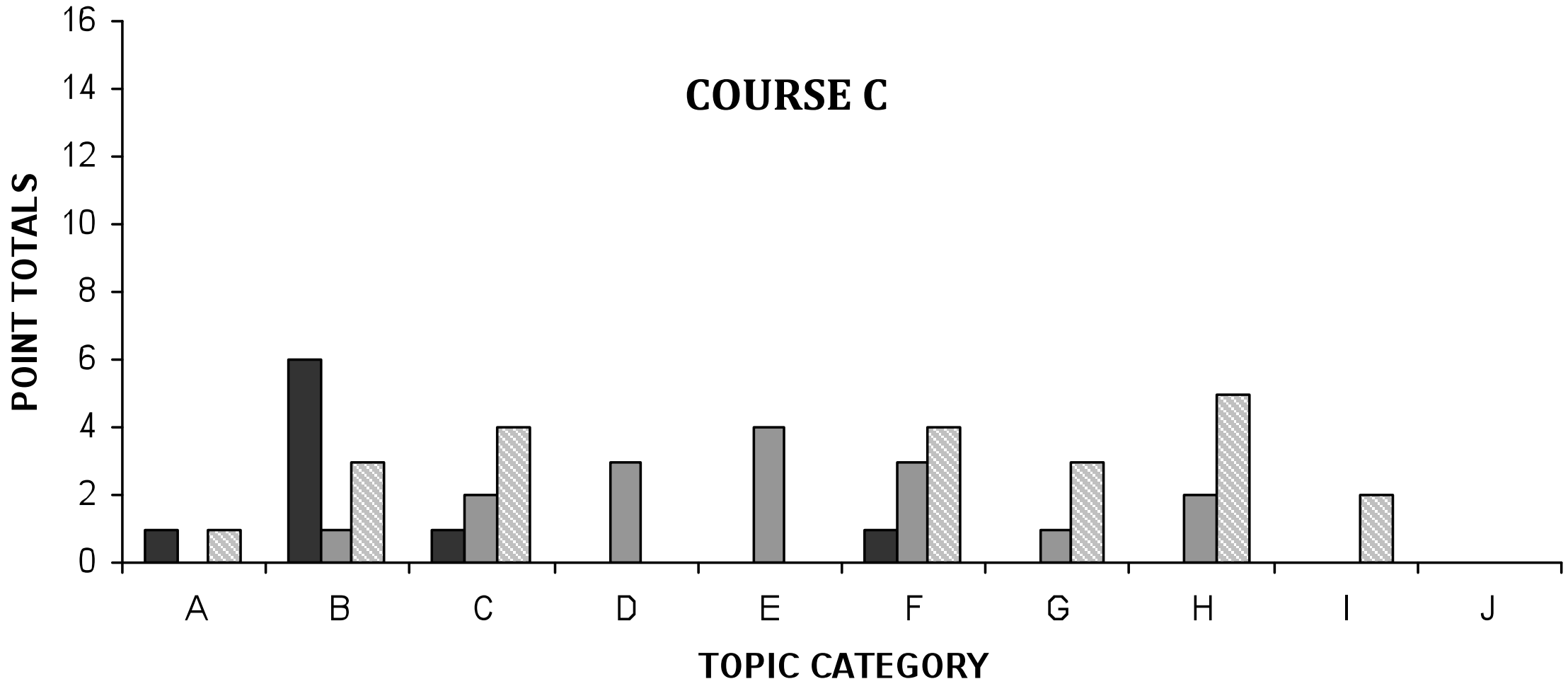

**FIG. 3.5.** The occurrence of lecture slides for both PHYS3 courses by topic (as describe in Table 3.I), for each of the themes described in Table 3.II.

Course B1 had a greater number of slides that scored in the *Light* and *Contrasting Perspectives* categories, though the graphs in Fig. 3.5 (which group the point totals for each course by topic area, as listed in Table 3.I) show that this difference can be largely attributed to instructor choices at the outset of the quantum physics sections of the two courses, in topic category B (photoelectric effect and photons). That this topic area should stand out in this analysis seems natural if one considers that: i) The photoelectric effect requires a particle description of light; ii) The double-slit experiment with single photons requires both a wave and a particle description of light in order to fully account for experimental observations; and iii) Being the first specific topic beyond the introductory quantum physics lecture(s), it represents an opportunity to frame the content of the course in terms of the need to think beyond classical physics. While both modern physics courses had the greatest point totals in this topic category, B1 devoted a greater portion of lecture time here to addressing themes of indeterminacy and probability (B1 also totaled more points in the *Light* category, though this difference can be largely attributed to Instructor B's brief coverage of lasers, a topic not covered in Course C).

Fig. 3.6 shows the ratio of the point totals for each of the three interpretive themes (from topic area B only) to the total number of slides used during these lectures; the differences between the two courses in terms of the amount of lecture time spent contrasting perspectives is statistically significant ($p=0.001$, by a one-tailed t-test). We note, finally, that in both courses all three of these interpretive themes received considerably less attention at later stages of the course.

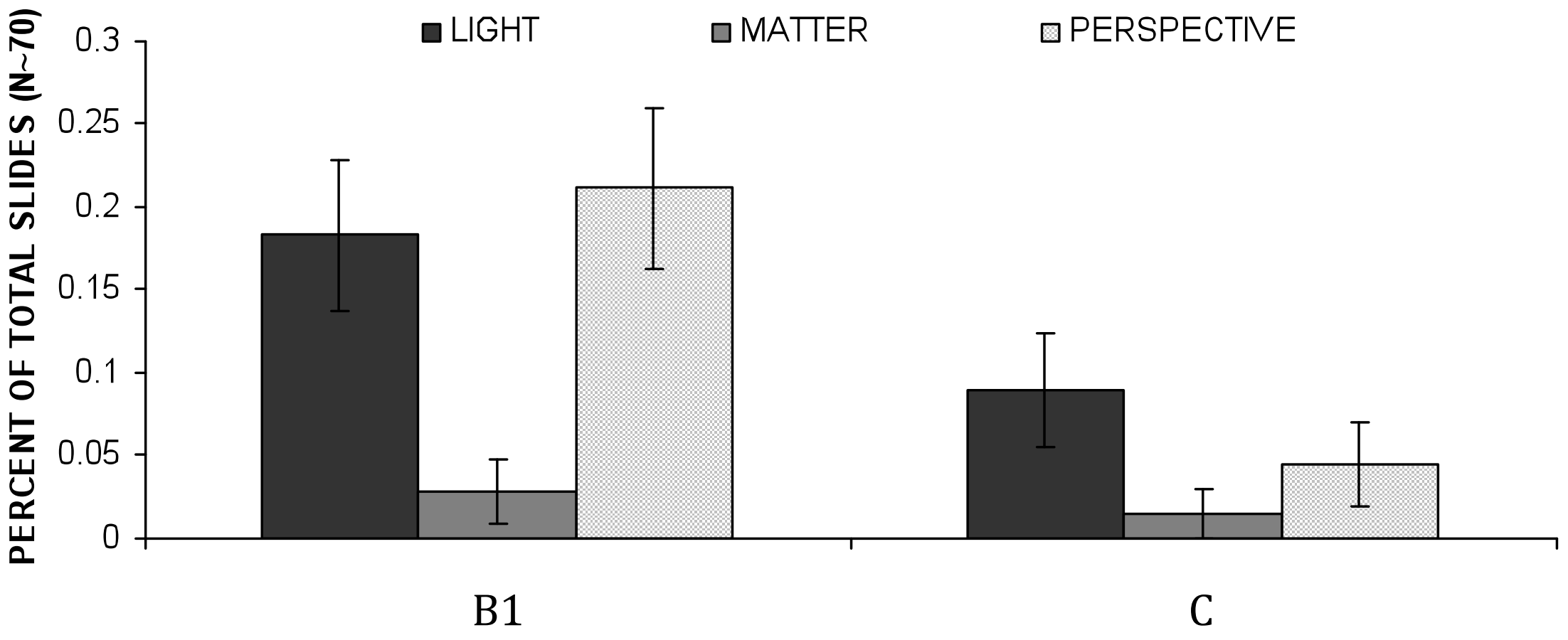

**FIG. 3.6.** Ratio of point totals from topic area B for each interpretive theme to the total number of slides used during these lectures. Error bars represent the standard error on the proportion.

The lecture slide shown in Fig. 3.7 is one example of how Course B1 differed from Course C in attending to student perspectives during the discussion of photons, by explicitly addressing the likelihood for students to think of quanta as being spatially localized. There were no comparable slides from Course C from this topic category, though this should not be taken to mean that Instructor C failed to address such issues at other times during the semester, or one-on-one with students. We note simply that there were no such explicit messages as part of the artifacts of the course in this topic area (which reflects a value judgment on the part of Instructor C regarding content), and students from Course C who accessed the lecture slides as posted online would have no indication that such ideas were deserving of any particular emphasis.

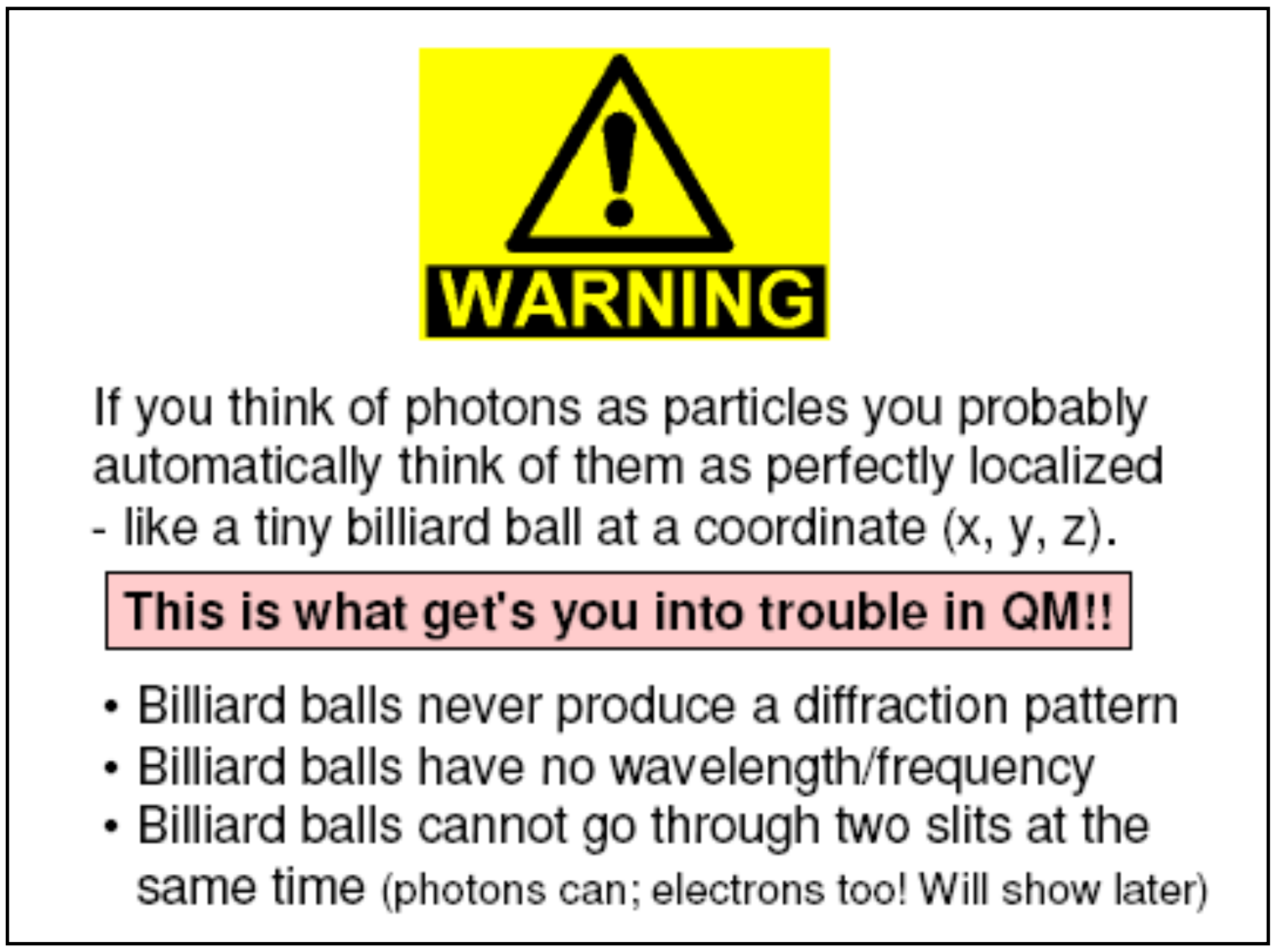

**FIG. 3.7.** A lecture slide used in Course B1 during the discussion of photons.

While there are coarse differences in how the instructors addressed student perspectives in some topic areas, the instructional approaches sometimes differed in more subtle ways. The two slides shown in Fig. 3.8 are illustrative of how the differences between the two courses could sometimes be less obvious, though still of potential significance. Both slides summarize the results for a system referred to in Course B1 as the *Infinite Square Well*, and by Instructor C as the *Particle in a Box*. At first glance, the two slides are almost identical: each depicts the first-excited state wave function of an electron in a potential well, as well as listing the normalized wave functions and quantized energy levels for this system. Both slides make an explicit contrast between the quantum mechanical description of this system and what would be expected classically, each pointing out that a classical particle can have any energy, whereas an electron confined in a potential well can only have specific energies.

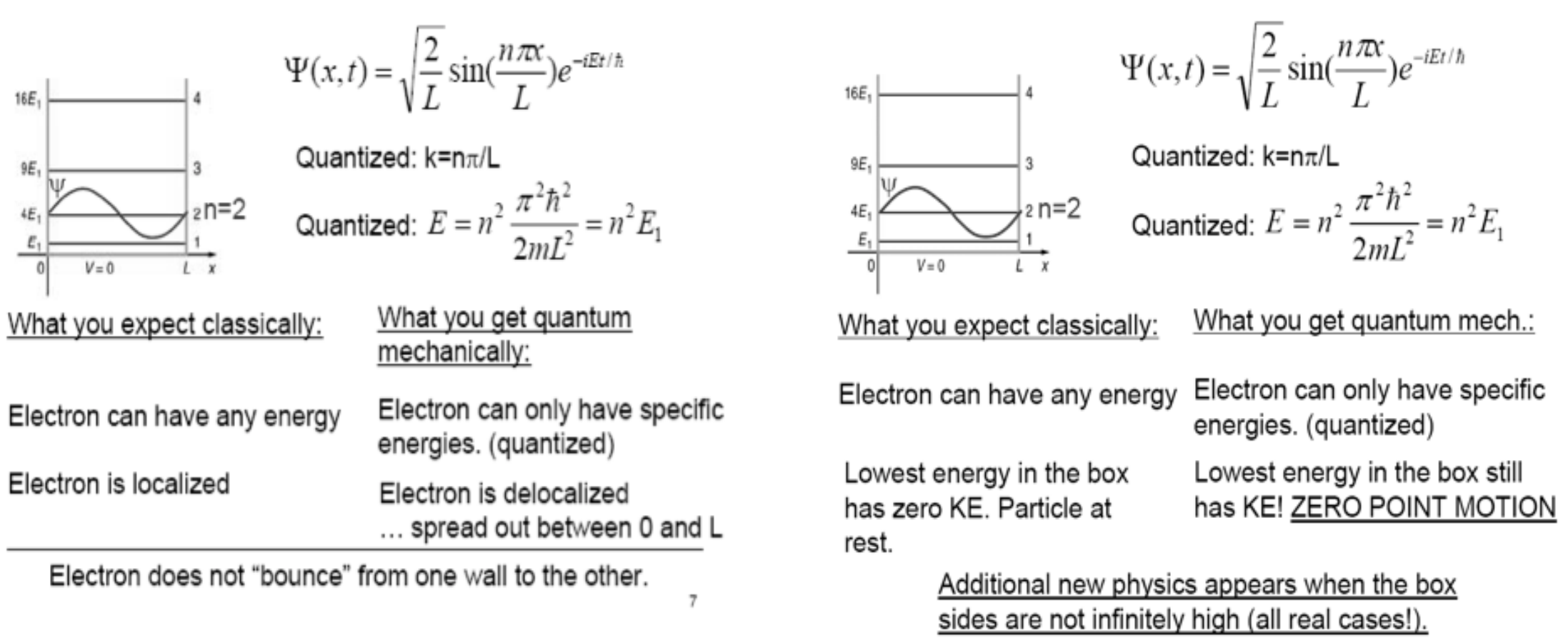

**FIG. 3.8.** Lecture slide from Course B1 (left, *Infinite Square Well*) and a nearly identical one from Course C (right, *Particle in a Box*).

However, Course B1 differed from Course C by emphasizing a wave model of the electron, delocalized and spread out, stating explicitly that the electron should not be thought of as bouncing back and forth between the two walls of the potential well. Instructor C focused instead on the kinetic energy of the system, pointing out that a classical particle can be at rest, whereas the quantum system has a non-zero ground state energy. It is arguable that Instructor C's choice of language, to speak of a *particle* in a box exhibiting zero-point motion, could implicitly reinforce in students the *realist* notion that in this system a localized electron is bouncing back and forth between two potential barriers. Both of these slides received a point in the *Contrasting Perspectives* category, but only the slide from PHYS3A received a point in the *Matter* category for its emphasis on the wave-like properties of an electron in a potential well.

### III.C. The double-slit experiment with single quanta.

As taught in these two courses, the double-slit experiment [Fig. 3.9] consists of a monochromatic beam of quanta that: (1) impinges on two closely spaced slits and diffracts; (2) wavelets spread out behind the slits and (3) interfere in the regions where they overlap; (4) bright fringes appear on the detection screen where the anti-nodal lines intersect.

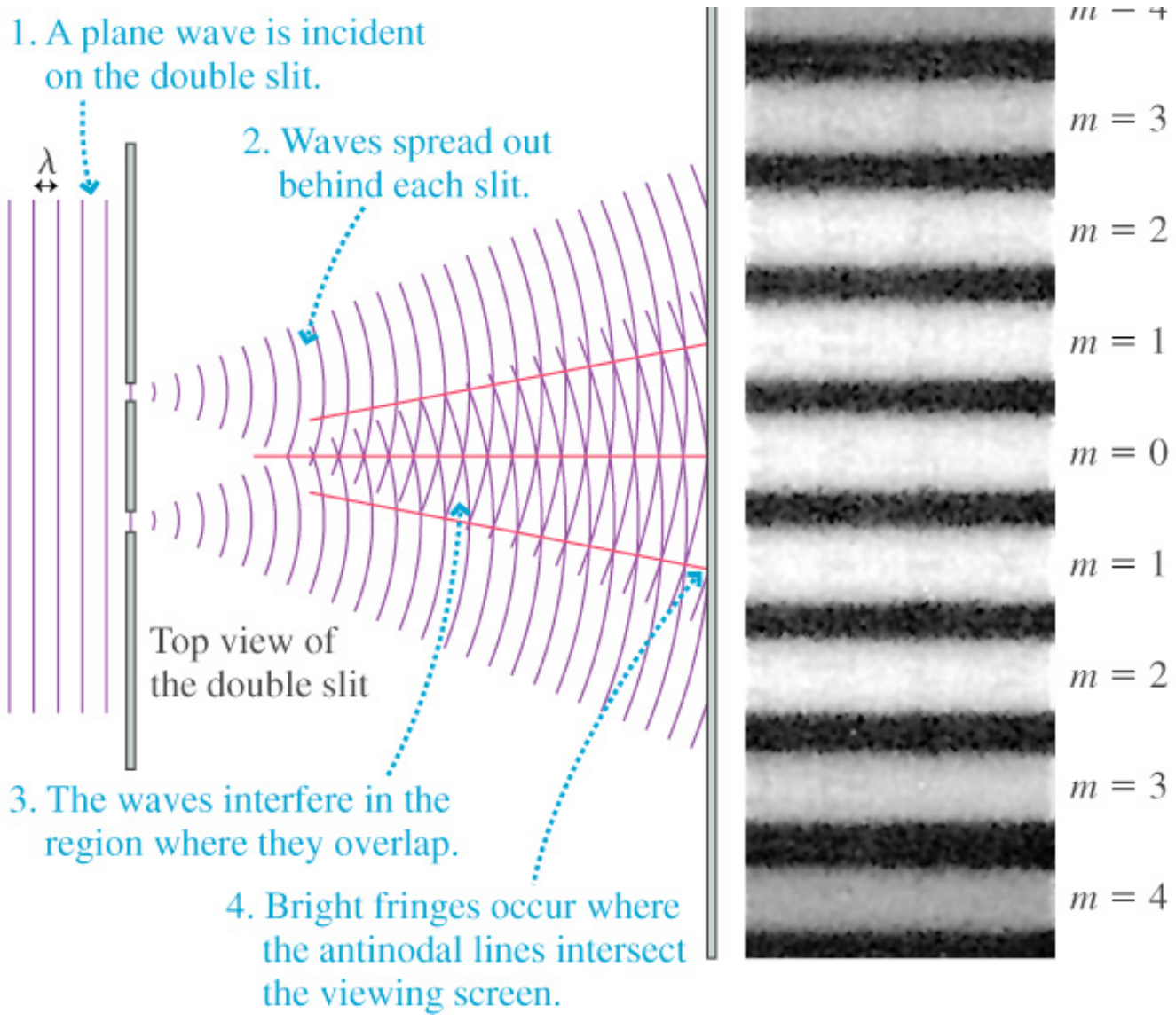

**FIG. 3.9.** Lecture slide used in both PHYS3 courses describing the double-slit experiment in terms of wave interference.

Both PHYS3 courses also instructed students that the intensity of the beam can be turned down to the point where only single quanta pass through the apparatus at a time; individual quanta are detected as localized particles on the screen, yet an interference pattern still develops over time. A wave description of quanta explains the interference pattern on the detection screen, while a particle description addresses the fact that individual quanta are detected as localized particles; in other words, a single ontological categorization of quanta (particle or wave) is inadequate for explaining all of what's observed in the double-slit experiment. Both instructors addressed during lecture a mathematical description of the interference pattern (how to relate the distance between the slits and the wavelength of the beam to the locations of fringe maxima and minima), and both used the Quantum Wave Interference simulation [5] in class to provide students with a visualization of the process. The approaches taken by the two instructors (B1 & C) with respect to quantum interpretation were as described in Section II; in brief, Instructor B took a *Matter-Wave* approach, while Instructor C was more *Agnostic* in his learning goals.

In the last week of the semester, students from both PHYS3 courses responded to an online survey designed to probe their ontological and epistemological beliefs about quantum mechanics. Students received homework credit for responding to the survey (equivalent to the number of points given for a typical homework problem), and the response rate for both courses was approximately 90%. Students were also told they would only receive full credit for providing thoughtful answers, and the text of the survey itself emphasized in bold type that there were no *right* or *wrong* answers to the questions being asked, but that we were particularly interested in what the students personally believed.

Instructors for both courses vetted the wording of the items on the survey, and interviews conducted after the end of the semester [Chapter 4] indicate that students interpreted the meaning of the questions in a way that was consistent with our intent. [See Appendix A for the evolution of the survey items (SP08-FA10).]

At the time of this study, the wording of the fictional student statements in the double-slit essay question had been changed in order to better reflect the language and argumentation of actual students (crafted in part from actual student responses from the study described in Chapter 2):

> **Student One:** *The probability density is so large because we don't know the true position of the electron. Since only a single dot at a time appears on the detecting screen, the electron must have been a tiny particle, traveling somewhere inside that blob, so that the electron went through one slit or the other on its way to the point where it was detected.*
>
> **Student Two:** *The blob represents the electron itself, since an electron is described by a wave packet that will spread out over time. The electron acts as a wave and will go through both slits and interfere with itself. That's why a distinct interference pattern will show up on the screen after shooting many electrons.*
>
> **Student Three:** *Quantum mechanics is only about predicting the outcomes of measurements, so we really can't know anything about what the electron is doing between being emitted from the gun and being detected on the screen.*

The results for both PHYS3 courses (B1 and C) are shown in Fig. 3.3, where responses are categorized according to which fictional student(s) the respondents agreed with (*Realist*, *Matter-Wave*, or *Agnostic*). While most students chose to agree with only a single statement, there were a few respondents from both courses who chose to agree with both the fictional *Realist* and *Agnostic* students, or with both the *Matter-Wave* and *Agnostic* students; we feel the *Realist* and *Matter-Wave* statements are not individually incompatible with the *Agnostic* statement, since simultaneously agreeing with the latter allowed students to acknowledge that they had no way of actually knowing if their preferred interpretations were correct. The relatively few students (~5%) who responded in this way are grouped together with the other students in the *Realist* or *Matter-Wave* categories, as appropriate.

As might be predicted based on the specific practices of Instructor B, most of his students chose to agree with the *Matter-Wave* statement (the electron is a delocalized wave packet that interferes with itself). The responses from Course C students were more varied: they were nearly four times more likely than B1 students to prefer a *Realist* interpretation; similarly, they were half as likely to favor the wave-packet description. More specifically, 29% of Course C students chose to agree with the *Realist* statement of Student One, and 27% of them agreed with the *Copenhagen/Agnostic* stance of Student Three, while only a combined 11% of students from Course B chose either of these responses.

### III.D. (In)consistency of student responses.

As seen in Fig. 3.5, both PHYS3 courses paid less explicit attention to student perspectives at later stages of instruction, as when covering the Schrödinger model of hydrogen. In lecture slides, both courses described an electron in the Schrödinger atomic model as a "cloud of probability surrounding the nucleus whose wave function is a solution of the Schrodinger equation," without further elaboration with respect to interpretation. We are interested in knowing if how students came to think of quanta in the context of the double-slit experiment would be relevant to how they thought of atomic electrons, particularly when they hadn't been given the same kind of explicit instruction in this topic area as with the double-slit experiment or the infinite square well.

In addition to the essay question, students responded (and provided reasoning) to the pre/post online survey statement regarding the position of atomic electrons; the following student quotes are illustrative of the reasoning offered by students in support of their responses:

> **AGREE**: "The probability cloud is like a graph method. It tells us where we are most likely to find the electron, but the electron is always a point-particle somewhere in the cloud."
>
> **DISAGREE**: "The electron is delocalized until we interfere with the system. It is distributed throughout the region where its wave function is non-zero. An electron only has a definite position when we make a measurement and collapse the wave function."

At the end of instruction, B1 students were just as likely to agree with the statement on atomic electrons as students from Course C, [Fig. 3.10] despite the emphasis given in Course B1 to thinking of an electron as delocalized in other contexts. Both courses showed a modest (and statistically insignificant) decrease in *Realist* responses to this statement between pre- and post-instruction, yet students from both courses were still more likely to agree than disagree with this statement in the end.

If responses from both courses to the statement on atomic electrons are grouped by how those same students responded to the double-slit essay question [Fig. 3.11] we see that 70% of students who preferred a *Realist* interpretation in the essay question took a stance on atomic electrons that would also be consistent with *realist* expectations. And while students who preferred a wave-packet description in the essay question were more likely than *Realist* category students to disagree with the statement on atomic electrons, 46% of those students still agreed that an electron in an atom has a definite position at all times. Only in the case of students who preferred the *Agnostic* statement did a majority disagree with this statement, and no students from this group responded neutrally.

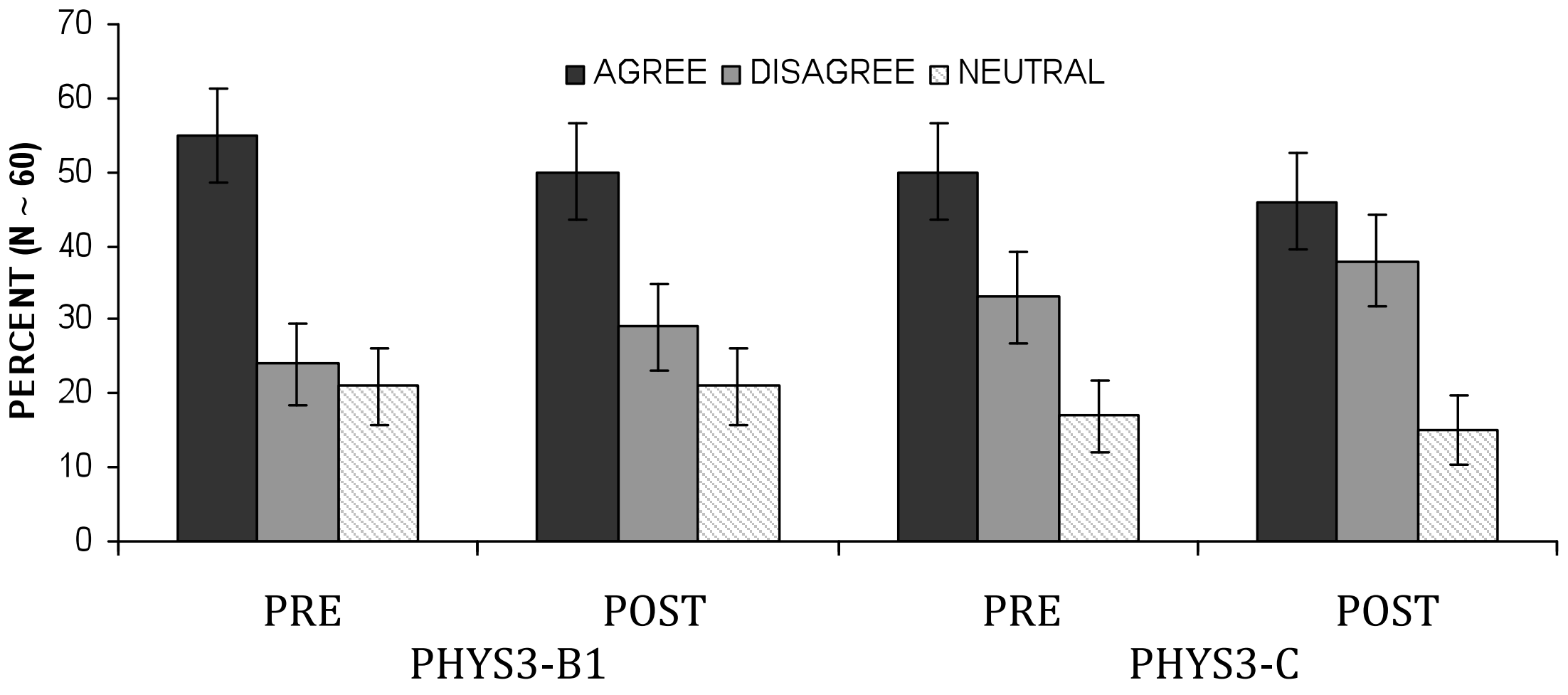

**FIG. 3.10.** Pre/post student responses from both PHYS3 courses to the statement: *An electron in an atom has a definite but unknown position at each moment of time.* Error bars represent the standard error on the proportion (N~60).

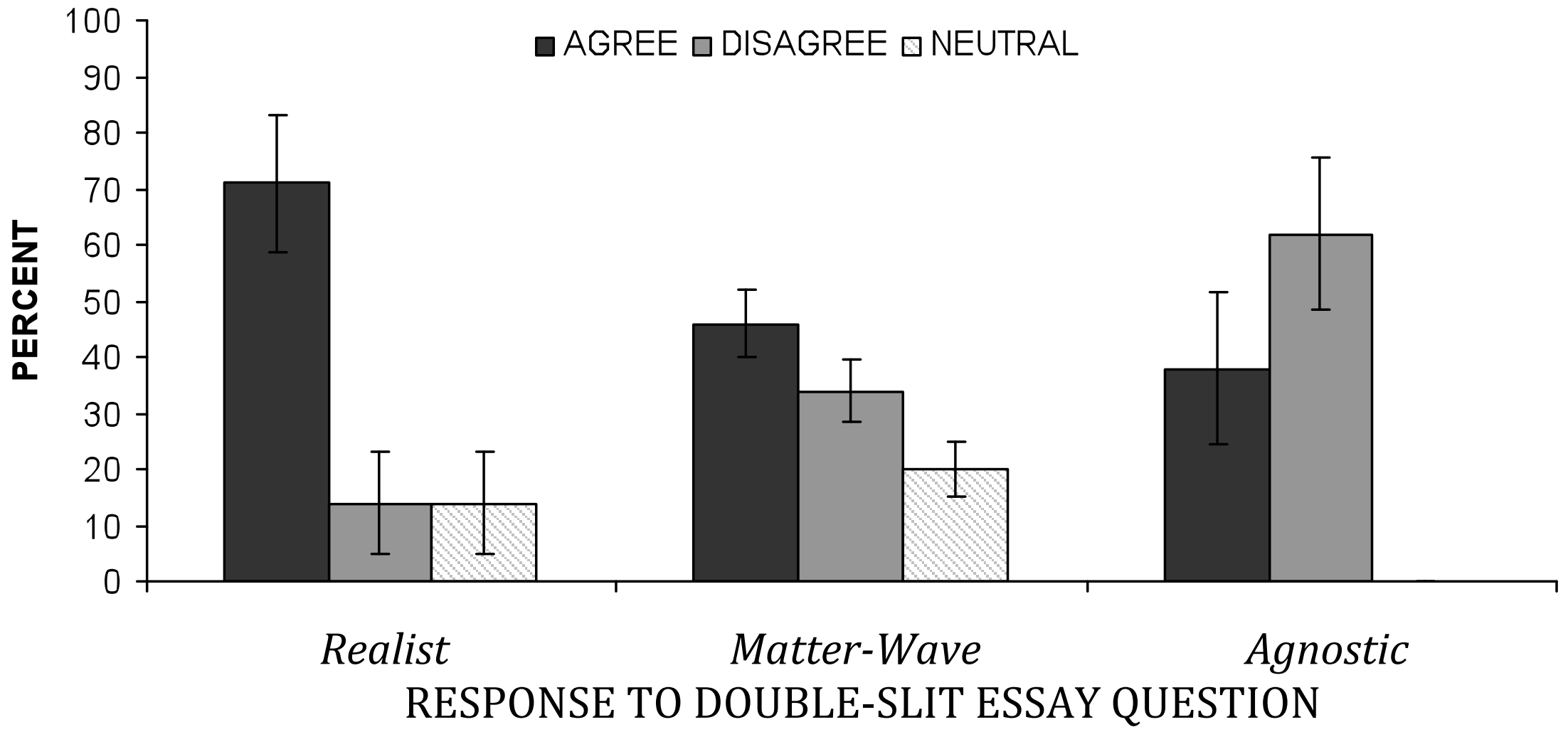

**FIG. 3.11.** Combined student responses from both PHYS3 courses to the statement: *An electron in an atom has a definite but unknown position at each moment of time*, grouped by how those students responded to the double-slit essay question. Error bars represent the standard error on the proportion (N~60).

**IV. Summary and Discussion**

Modern physics instructors differ not only in their personal perspectives regarding the physical interpretation of quantum mechanics, but also in their decisions to teach (or not teach) about quantum interpretations in their introductory courses. In this chapter, we have documented significant instructor effects in terms of how students respond to post-instruction surveys; we have also examined in detail two different approaches to addressing interpretative themes in two introductory modern physics courses with similar content.

When comparing these two courses in detail, Instructor B's more explicit approach to teaching a *matter-wave* interpretation of the double-slit experiment had a significant impact on how students said they thought of electrons within that specific context. Instructor C's less explicit and more *Agnostic* instructional approach is reflected in the greater variation of student responses to the essay question; not only were Course C students more likely than B1 students to prefer an *Agnostic* stance (which would be in alignment with Instructor C's instructional approach), these students were also more likely to align themselves with a *Realist* interpretation. In addition, the emphasis given in Course B toward thinking of electrons as delocalized in the double-slit experiment and the infinite square well had no discernible impact on student responses in areas where instruction was less explicit. Both courses were similar in their treatment of the Schrödinger atomic model, and student responses from both courses regarding the existence of an electron's position in an atom were not significantly different, with the majority of students from both courses favoring a *Realist* perspective in this specific context.

We may investigate the consistency in how students apply perspectives across contexts by comparing responses to the double-slit essay question with a statement regarding the position of an electron in an atom. We find that most every student who preferred a *Realist* interpretation of the double-slit experiment also took a *Realist* stance on the question of whether an electron in an atom has a definite position. On the other hand, almost half of the students who preferred the wave-packet description of a single electron in the double-slit experiment would still agree with particle-like descriptions of atomic electrons. Such responses evidence the greater likelihood for students to favor *Realist* perspectives in topic areas where instruction is less explicit, and suggest that instructors who wish to promote any particular perspective in quantum physics should do so explicitly across a range of topics, rather than assuming it to be sufficient to address student perspectives primarily at the outset.

These findings also indicate that, just as with topics in classical physics, [6-14] naïve intuition (being congruent with *realist* expectations) can serve as a barrier to conceptual understanding in quantum physics. A major difference between the intuitive barriers in classical physics and in quantum physics lies in the nature of the questions, both ontological (when is a particle a particle, and when is it a wave?) and epistemic (what is the difference between classical ignorance and fundamental uncertainty?). End-of-semester comments from Instructor C support the notion that students who preferred a *Realist* interpretation of the double-slit experiment were not doing so from a simple lack of understanding:

> "Some of the students who I considered to be the most engaged went with [the *Realist* statement. They said]: '…the electron is a real thing; it's got to be in there somehow. I know that's not what you told us, but that's what I'm thinking…' I thought that was just great; it was sort of honest. They were willing to recognize that that's not what we're saying, but they're grappling with that's how it's got to be anyways."

Furthermore, one-on-one interviews conducted with students from these two courses following the end of the semester [Chapter 4] showed that those who had favored a *Realist* perspective in the interference essay question were still able to correctly describe from memory the particulars of the double-slit experiment.

It is also worth noting that the two instructors considered in our detailed comparative study, while sometimes explicit in teaching *an* interpretation of quantum mechanics, were not explicit in teaching these interpretations *as interpretations*. In other words, they did not teach quantum mechanics from an axiomatic standpoint, did not explicitly teach the *Copenhagen Interpretation* (or any other formal interpretation); nor did they frame their interpretations in terms of *modeling,* or *nature of science* (NOS) issues. Instead, instructors for both courses addressed questions of interpretation as they arose within the contexts of specific topics, without making the physical interpretation of the wave function (beyond its probabilistic interpretation, à la Born [15]) into a major topic unto itself. The sense in which quantum interpretation is *hidden* in modern physics curricula becomes apparent when considering how students may default to intuitive *realist* expectations in topic areas where instructors are less explicit; and in recognizing that interpretive aspects of quantum physics tend to remain unaddressed in a way that is meaningful to students.

The studies considered in this chapter suggest that instructors should be aware of the potential impact they may have on student thinking as a consequence of their instructional choices – instructors who spend less time explicitly attending to student knowledge and intuition are less likely to transition students away from inappropriately *realist* perspectives. These studies have also indicated that students may favor a variety of perspectives in a way that may seem contradictory to expert physicists, indicating the need for a deeper exploration into the contextual aspects of student perspectives in quantum physics. [Chapter 4]

**References (Chapter 3)**

**1.** E. Redish, J. Saul and R. Steinberg, Student expectations in introductory physics, *Am. J. Phys.* **66**, 212 (1998).

**2.** S. B. McKagan, K. K. Perkins and C. E. Wieman, Reforming a large lecture modern physics course for engineering majors using a PER-based design, *PERC Proceedings 2006* (AIP Press, Melville, NY, 2006).

**3.** http://phet.colorado.edu

**4.** Available at http://www.colorado.edu/physics/EducationIssues/ modern/ (password protected, retrieved January, 2011).

**5.** http://phet.colorado.edu/simulations/sims.php?sim=QWI

**6.** A. A. diSessa, "A history of conceptual change research: Threads and fault lines," in *Cambridge Handbook of the Learning Sciences*, K. Sawyer (Ed.) (Cambridge University Press, Cambridge, 2006), pp. 265-281.

**7.** M. Reiner, J. D. Slotta, M. T. H. Chi and L. B. Resnick, Naïve physics reasoning: A commitment to substance-based conceptions, *Cognition and Instruction* **18**, 1 (2000).

**8.** M. T. H. Chi, Common sense misconceptions of emergent processes: Why some misconceptions are robust, *Journal of the Learning Sciences* **14**, 161 (2005).

**9.** J. D. Slotta, In defense of Chi's Ontological Incompatibility Hypothesis, *Journal of the Learning Sciences* **20**, 151 (2011).

**10.** J. D. Slotta and M. T. H. Chi, The impact of ontology training on conceptual change: Helping students understand the challenging topics in science, *Cognition and Instruction* **24,** 261 (2006).

**11.** D. Hammer, Student resources for learning introductory physics, *Am. J. Phys.* **68**, S52 (2000).

**12.** D. Hammer, A. Elby, R. E. Scherr and E. F. Redish, "Resources, Framing and Transfer" in *Transfer of Learning*, edited by J. Mestre (Information Age Publishing, 2005) pp. 89-119.

**13.** A. Gupta, D. Hammer and E. F. Redish, The case for dynamic models of learners' ontologies in physics, *J. Learning Sciences* **19**, 285 (2010).

**14.** D. Hammer, A. Gupta and E. F. Redish, On Static and Dynamic Intuitive Ontologies, *J. Learning Sciences* **20**, 163 (2011).

**10.** W. H. Stapp, The Copenhagen Interpretation, *Am. J. Phys.* **40**, 1098 (1972).

# CHAPTER 4

## Refined Characterizations of Student Perspectives on Quantum Physics

### I. Introduction

We have thus far seen how *realist* perspectives among modern physics students may translate into specific beliefs about quantum phenomena; e.g., particles are always localized in space, or that probabilistic descriptions of quantum measurements are the result of classical ignorance. We engage here in a more detailed exploration of student perspectives on quantum physics through a number of one-on-one student interviews. The resulting implications for modern physics instruction are particularly significant in that the learning goals for such courses typically include transitioning students away from classical epistemologies and ontologies, to ones that are more aligned with the beliefs of practicing physicists.

Still, it is not always clear exactly what expert physicists believe regarding the physical interpretation of quantum mechanics. [1] A recent survey [2] of quantum physics instructors at the University of Colorado and elsewhere (all of whom use quantum mechanics in their research) found that 30% of them interpreted the wave function as being physically real, while nearly half considered it to contain information only. The remaining respondents held some kind of mixed view on the physical interpretation of the wave function, or saw little distinction between the two choices. And only half of those who expressed a clear preference (matter-wave or information-wave) did so with confidence, being of the opinion that the other view was probably wrong. We find that students also develop attitudes and opinions regarding the reality of the wave function, as well as other interpretive themes from quantum mechanics.

The efforts described in Chapters 2 & 3 at characterizing student perspectives on quantum physics were limited to the application of three coarse labels (*Realist*, *Matter-Wave*, *Agnostic*) which are useful, but in light of the results of these studies, seem limited in terms of capturing the many nuances of student responses, and in particular understanding why students seem to exhibit contradictory perspectives between and within contexts. In this chapter, we therefore address the following:

1) How might our classification scheme be refined to better describe the nuances of student perspectives on interpretive themes in quantum physics?

2) For what reasons do students exhibit mixed perspectives within and across contexts?

From a total of 19 post-instruction interviews with students from four recent introductory modern physics courses at the University of Colorado we find that, though they may not employ the same formal language as expert physicists, students often invoke concepts and beliefs that parallel those invoked by expert physicists when arguing for their preferred interpretations of quantum mechanics. These parallels allow us to characterize student perspectives on quantum physics in terms of some of the same themes that distinguish these formal interpretations from each other. Of particular significance is the finding that students do indeed develop attitudes and opinions regarding a variety of interpretive themes in quantum mechanics, regardless of whether these themes had been explicitly addressed by their instructors. The mixed or seemingly contradictory student responses may be better understood in that: (A) some students prefer a mixed wave-particle ontology (a *pilot-wave* interpretation, wherein quanta are simultaneously *both* particle *and* wave); and (B) students are most likely to vacillate in their responses when what makes intuitive sense to them is not in agreement with what they perceive as a scientifically accepted response.

## II. Interview participants and course characteristics

We sought to recruit five students from each of the four modern physics offerings at the University of Colorado from a single academic year (immediately following the studies described in Chapter 2) to participate in an hour-long post-instruction interview. A mass email was sent to all students enrolled in these courses, offering a nominal sum of fifteen dollars in exchange for their participation; students were not informed ahead of time about the nature of the interview questions, only that we would be discussing some ideas from modern physics. There was no real opportunity to select among students since volunteers were sometimes scarce, and so there was no attempt to make the cohort representative of all students from those courses. A total of 19 students were interviewed from these four courses [Table 4.I], either in the last week of the semester or after the course had ended. Interview participants from the courses for physics majors were all physics or engineering physics majors, plus one astronomy major; those from the courses for engineers were all engineering majors (but not engineering physics), plus one mathematics major. The average final course grade for all 19 students was 3.4 (out of 4.0, where overall course averages fall in the 2.0–3.0 range), indicating that participants were generally better than average students, as might be expected for a group of volunteers. Interviews followed the protocol as given in Appendix B. It should be emphasized that our characterizations of instructional approaches in Table 4.I and elsewhere in this chapter come from analyses of course materials and practices, and are not necessarily reflective of each instructor's personal perspective on quantum mechanics, but rather of how that instructor addressed interpretive themes in class.

**TABLE 4.I** Summary of four courses from which students were recruited for interviews, including a characterization of each instructor's stance on interpretive themes, as taught in that course; instructor labels correspond to those given in Figs. 3.1 & 3.2.

| INSTRUCTOR | STUDENT POPULATION | INTERPRETIVE APPROACH | STUDENTS INTERVIEWED |
|---|---|---|---|
| MW-1 | Engineering | *Matter-Wave* | 3 |
| C/A-2 | | *Copenhagen* | 5 |
| C/A-1 | Physics | *Copenhagen/Agnostic* | 6 |
| C/A-3 | | | 5 |

The instructor labels given in Table 4.I correspond to those given in Figs. 3.1 & 3.2 (here, the labels MW-1 and C/A-1 correspond to Instructors/Courses B1 & C, respectively, as described in Chapter 3). The labels used for describing instructional approaches have been described earlier, but can be best illustrated by how each instructor addressed the double-slit experiment with single quanta. Instructor MW-1 (B1 in Chapter 3) was explicit in promoting a wave model of individual quanta as they propagate through both slits, interfere with themselves, and then become localized upon detection. Instructor C/A-2 told students that a *quantum mechanical wave of probability* passes through both slits, but that which-path questions change the circumstances of the experiment, making them ill-posed at best. While similar to C/A-2, Instructors C/A-1 (Instructor C in Chapter 3) and C/A-3 ultimately placed more emphasis on calculation (predicting features of the interference pattern) than matters of interpretation.

For the 19 students interviewed for the present studies, there were no discernible connections between a specific instructional approach and the preferred perspectives of the students interviewed from that course, likely due to the limited number of participants. Therefore, discussion in this chapter of specific instructional approaches will be limited to the brief characterizations given above, and a few specific statements below concerning the influence of an instructional approach on that student's individual responses.

## III. Refined characterizations of student perspectives

As will be demonstrated below, we find it useful to consider student perspectives in quantum physics in terms of concepts associated with some of the more common (i.e., less exotic) formal interpretations of quantum mechanics. In doing so, we do not mean to imply that student perspectives are as coherent or sophisticated as any formal interpretation (although other research [2] suggests that expert perspectives on quantum physics may be similarly tentative). In fact, our results can best be understood within a theoretical framework that views student perspectives (including the process of *ontological attribution*) as cognitive frameworks that are dynamic emergent processes (as opposed to fixed or static cognitive structures), that are contextually sensitive, and that sometimes simultaneously blend ontological attributions that belong to classically distinct categories. [See Refs. 3-5, as well as Chapter 1, Section II.] Nor do we assume that any one label is necessarily sufficient for describing the nuanced and sometimes inconsistent perspectives exhibited by any particular student; or even that the development of student perspectives on quantum physics follows along the lines of historical developments.

We do, however, find that some formal interpretations of quantum mechanics can be distinguished from each other in terms of a few key themes, and that students do have beliefs or ideas concerning these themes of interpretation, regardless of whether these themes had been explicitly addressed by their instructors. In other words, we have observed that many introductory modern physics students, when formulating a stance on these interpretive themes, employ some of the same epistemological tools used by expert physicists, and will sometimes invoke similar experimental results and intuitive notions of particles and waves as motivation for their preferred interpretations of quantum phenomena. An analysis of all 19 interview transcripts revealed student beliefs and attitudes (of varying degrees of sophistication) concerning the following three interpretive questions:

1) Is the position of a particle objectively real, or indeterminate and observation dependent? [Existence or non-existence of certain hidden variables.]

2) Is the wave function a mathematical tool that encodes probabilities [information-wave], or is it physically real [matter-wave]?

3) Does the *collapse of the wave function* (or *reduction of the state*) represent a physical process, or simply a change in knowledge of the observer?

## III.A. Discussion of formal interpretations

We present here a brief summary of some key features of several formal interpretations of quantum mechanics, in terms of the three interpretive themes given in Section II. [Table 4.II] Many aspects of these formal interpretations have been previously discussed in greater detail, [Chapter 1] and it should be emphasized that it would be impossible for these short summaries to be comprehensive, but are offered as working definitions for the sake of clarity when associating these labels with the expressed beliefs of individual students.

***Realist/Statistical***: From either a *Realist* or S*tatistical* perspective, the physical properties of a system are objectively real and independent of experimental observation (observations reveal reality, not create it). The state vector encodes probabilities for the outcomes of measurements performed on an ensemble of similarly prepared systems, but cannot provide a complete description of individual systems. The wave function is not physically real; the collapse of the wave function represents a change in the observer's knowledge of the system, and not a physical change brought about by the act of measurement.

***Copenhagen***: The probabilistic nature of quantum measurements is a reflection of the inherently probabilistic behavior of quantum entities; in general, the properties of a system are indeterminate until measured. The wave function is not a literal representation of a physical system, and the *collapse of the wave function* corresponds to a change in knowledge of the observer, though it does represent a physical transition from an indeterminate state to one where certain properties of the state become well defined.

***Matter-Wave***: Similar to the *Copenhagen Interpretation* with respect to indeterminacy and the non-existence of hidden variables, but also ascribes physical reality to the wave function. Though not described by the Schrödinger equation, the *collapse of the wave function* represents a physical process induced by measurement.

***Pilot-Wave***: From this perspective, quanta are simultaneously both particle and wave: localized particles follow trajectories determined by a physically real quantum wave. In the double-slit experiment, an electron is all at once both a particle that goes through only one slit, and a wave that passes through both slits and interferes with itself. In this context[1], the position of a particle is objectively real and predetermined based on unknowable initial conditions, so that the reduction of the state represents a change in knowledge of the observer.

[1] Nonlocal features come into play when other quantum effects (e.g., entanglement) are to be accounted for, in which case the *collapse of the wave function* must be seen as a (non-local) physical process.

**TABLE 4.II.** Summary of our characterizations of four formal interpretations of quantum mechanics, in terms of three interpretive themes (described in Section II). The *Agnostic* perspective is not a formal interpretation in itself, but is included for completeness.

| INTERPRETATION | HIDDEN VARIABLES? | INFO- OR MATTER-WAVE? | COLLAPSING WAVE FUNCTION? |
|---|---|---|---|
| *Realist/Statistical* | YES/AGNOSTIC | INFO | KNOWLEDGE |
| *Copenhagen* | NO | INFO | PHYSICAL |
| *Matter-Wave* | NO | MATTER | PHYSICAL |
| *Pilot-Wave* | YES | MATTER | KNOWLEDGE |
| *Agnostic* | AGNOSTIC | AGNOSTIC | AGNOSTIC |

***Agnostic:*** Though not a formal interpretation in itself, we distinguish between this stance and the positivistic aspects of the *Copenhagen Interpretation* (declining to speculate on the unobservable). The *Agnostic* perspective accounts for multiple interpretations of quantum mechanics and their ontological implications, but takes no definite stance on which might correspond to the best description of reality. The utility of quantum mechanics is generally favored over questions of interpretation.

### III.B. Students express beliefs that parallel those of expert proponents

We have hypothesized that the perspectives of many modern physics students on quantum phenomena are significantly influenced by the commonplace (and intuitive) notion of particles as localized in space. In classical physics, as in colloquial usage, the word particle generally connotes some small object, so it should not be surprising that students who have learned about particles primarily within the context of classical physics should persist in thinking of them as microscopic analogs to macroscopic objects when learning about quantum physics. This would be an example of *classical attribute inheritance*, in the sense that students may explicitly attribute to quantum particles *all* of their classical analogs, including a localized position (student codes are as given below in Table 4.III):

> "I guess an electron has to [always be at] a definite point. It is a particle, we've found it has mass and it has these intrinsic qualities, like the charge it has, so it will have a definite position, but due to uncertainty it will be a position that is unknown." [STUDENT QR2]

This statement reveals not only one student's belief in localized massive particles, it also suggests a stance on the uncertainty associated with a particle's position: its objectively real value will be unknown until revealed by measurement.

This student (and others with similar attitudes) reported interpreting the probability density for an atomic electron as strictly a mathematical tool used only for describing the probable locations for where that electron might be found once measured; probabilistic descriptions of such measurements were therefore seen as a reflection of *classical ignorance* concerning the true state of that particle just prior to measurement. We thus see how an intuitive notion of particles as localized objects can influence what physical meaning students ascribe to both the wave function and the probabilistic nature of quantum mechanics.

In a similar vein, another student explicitly objected to the idea that wave-packets could represent single particles. Here, this student is discussing the Quantum Wave Interference [6] (QWI) simulation's depiction of a wave-packet's propagation through both of two slits on its way to detection:

> "One electron can't go through both slits at the same time because electrons have mass. Wouldn't it violate conservation of mass and charge if [the electron] were split into two like it shows in the [QWI] simulation?" [STUDENT R1]

Such objections are reminiscent of those made by L. Ballentine (a major proponent of the *Statistical Interpretation* of quantum mechanics [7, 8]) when discussing a thought experiment in which an incident wave packet is divided by a semi-reflecting barrier into two distinct transmitted and reflected wave packets. The reflected and transmitted waves are then directed toward a pair of detectors connected to a coincidence counter. Ballentine argues:

> "Suppose that the wave packet *is* the particle. Then since each packet is divided in half, [...] the two detectors will always be simultaneously triggered by the two portions of the divided wave packet." [Ref. 8, p. 101, emphasis in original]

In this thought experiment (and in practice [9]), single quanta trigger either one detector or the other (and not both simultaneously); Ballentine therefore concludes that, while the wave function may have nonzero amplitude in two spatially separated regions, it cannot be interpreted as describing individual particles, since individual particles are never found in two places at once. In making this argument, Ballentine has implicitly assumed that the *collapse of the wave function* (or *reduction of the state*) represents a change in knowledge of the observer, and not an actual physical process induced by measurement.

In his own book on quantum mechanics, Dirac [10] considers the same type of thought experiment as Ballentine, but provides a radically different explanation:

> "The result of [the detection] must be either the whole photon or nothing at all. Thus the photon must change suddenly from being partly in one beam and partly in the other to being entirely in one of the beams." [Ref. 10, p. 9]

As counterintuitive as this interpretation may be, we find that a number of modern physics students report having accepted such ideas, and have incorporated them into their descriptions of quanta:

> "[T]he electron, until it's measured, until you try to figure out where it is, the electron is playing out all the possibilities of where it could go. Once you measure where it is, that collapses its wave function [and it] loses its properties as a wave and becomes particle in nature." [STUDENT Q3]

Students within this category all explicitly exhibited this kind of flexibility in their ontological descriptions of the behavior of electrons. Other students, like Ballentine, find these types of explanations unsatisfying:

> "[A] single electron is detected at the far screen, and I feel like that really can't be explained for the wave-packet, by one specific detection in a small place like that, if you say [the wave-packet] is the electron. That's really the only discrepancy I have with that: What happens when it hits the screen?" [STUDENT QR2]

Indeed, the question of what happens when individual quanta are detected in a double-slit experiment has played a significant role for some physicists in motivating their perspectives on quantum phenomena, as with Ballentine:

> "[I]t is possible to detect the arrival of individual electrons, and to see the diffraction pattern emerge as a statistical pattern made up of many small spots. *Evidently, quantum particles are indeed particles*, but particles whose behavior is very different from what classical physics would have led us to expect." [Ref. 8, p. 4, emphasis added]

This statement exemplifies a degree of ontological inflexibility in expert thinking: Ballentine is assuming that the detection of electrons as localized particles implies they exist as localized particles at all times. J. S. Bell has also invoked the double-slit experiment when discussing interpretation, but in this particular case as motivation for a pilot-wave interpretation, as proposed by Bohm and others [11]:

> "Is it not clear from the smallness of the scintillation on the screen that we have to do with a particle? And is it not clear, from the diffraction and interference patterns, that the motion of the particle is directed by a wave?" [Ref. 12, p. 191]

This student's discussion of the double-slit experiment echoes sentiments expressed by both Ballentine and Bell – by employing similar argumentation, he reaches similar conclusions:

> "For me, saying that the [wave] represents the electron isn't accurate because an electron, after it's measured on that screen, is a point-particle, you see a distinct interference pattern after shooting many electrons, but you still see one electron hit the screen individually. […] I do agree that the electron acts as a wave because that's obviously what causes the pattern; if it didn't interfere with itself, or create a wavelike function, then you wouldn't see the patterns on the screen also." [STUDENT R3]

Historically, and in our classrooms today, different physicists have offered different interpretations of quantum diffraction experiments. For Ballentine, diffraction patterns form as a consequence of the quantized momentum transfer between localized particles and the diffracting medium. [Ref. 8, p. 136] These patterns are more commonly explained in terms of wave interference, but for some, the wave is guiding the trajectory of a localized particle, while others would claim that each particle interferes with itself as a delocalized wave until becoming localized upon detection. At the same time, a number of *both* expert and student physicists find it unscientific to speculate on that which cannot be experimentally observed:

> "I understand why people would think [the electron] has to exist between here and where it impacts, and it does, but the necessity of [thinking of it] between here and where it impacts as an actual concept like a particle or a wave, I don't see much of the point. We're not going to observe what it is between here and there, so it doesn't seem like a statement for science to make. It seems right now to be entirely unobservable." [STUDENT C2]

The refusal to speculate on unobservable processes is a key feature of the orthodox *Copenhagen Interpretation* of quantum mechanics, which seems to be favored by a majority of practicing physicists, if only for the fact that it allows them to apply the mathematical tools of the theory without having to worry about what's "really" going on (as embodied in the popular phrase: *Shut Up and Calculate!* [12], and the sentiments expressed by Instructor C's [C/A-1, in this chapter] in-class comments from Chapter 3).

We also find it necessary to distinguish between the agnostic or positivistic aspects of an instructional approach, and the agnosticism of those who are aware of multiple interpretations, but are unsure as to which offers the best description of reality:

> "For now, for me, the electron is the wave function. But whether the electron is distributed among the wave function, and when you do an experiment, it sucks into one point, or whether it is indeed one particle at a point, statistically the average, I don't know." [STUDENT QA1]

### III.C. Categorization and summary of student responses

We summarize here a categorization of individual students in terms of the interpretive themes discussed above, grouped by overall perspective, as discussed in Section III.A; [Table 4.III] in this section, the label *Quantum* (Q) is used as shorthand for a *Matter-Wave* perspective, for brevity and for consistency with prior published research. [13] A discussion of key findings and commonalities among students within individual categories follows.

**TABLE 4.III.** Summary of individual student interview responses with respect to three interpretive themes (as described in Section III), grouped by overall perspective. The label *Quantum* (Q) is used as shorthand for a *Matter-Wave* perspective.

| STUDENT PERSPECTIVE | CODE | HIDDEN VARIABLES? | INFO OR MATTER WAVE? | COLLAPSING WAVE FUNCTION? |
|---|---|---|---|---|
| *Realist* | **R1** | YES | INFO | KNOWLEDGE |
| | **R2** | YES | INFO | KNOWLEDGE |
| | **R3** | YES | INFO | KNOWLEDGE |
| *Split Quantum/ Realist* | **QR1** | NO/YES | MATTER/INFO | PHYSICAL |
| | **QR2** | NO/YES | MATTER/INFO | KNOWLEDGE |
| | **QR3** | NO/YES | MATTER/INFO | KNOWLEDGE |
| | **QR4** | NO/YES | MATTER/INFO | AGNOSTIC |
| *Pilot-Wave* | **P1** | YES | MATTER | KNOWLEDGE |
| | **P2** | YES/ AGNOSTIC | MATTER/ AGNOSTIC | KNOWLEDGE/ AGNOSTIC |
| | **P3** | YES | MATTER | KNOWLEDGE |
| *Quantum (Matter-Wave)* | **Q1** | NO | MATTER | KNOWLEDGE |
| | **Q2** | NO | MATTER | PHYSICAL/AGNOSTIC |
| | **Q3** | NO | MATTER | PHYSICAL |
| | **Q4** | NO | MATTER | PHYSICAL |
| | **Q5** | NO | MATTER | PHYSICAL/AGNOSTIC |
| *Quantum/ Agnostic* | **QA1** | NO/ AGNOSTIC | MATTER/ AGNOSTIC | PHYSICAL/ AGNOSTIC |
| *Copenhagen* | **C1** | NO | INFO | PHYSICAL |
| | **C2** | NO | INFO/AGNOSTIC | PHYSICAL/AGNOSTIC |
| | **C3** | NO/ AGNOSTIC | INFO | AGNOSTIC |

We first note that many student responses agreed well with our characterizations of the formal interpretations, while other students provided one or more responses that were not entirely consistent with those characterizations; in the few such cases, a category was assigned based on what would be most consistent with the overall responses from that student. A second, independent physics education researcher coded a subset of five transcribed interviews (all students who were not quoted in this chapter), both by interpretive theme and by overall interpretation, with an initial inter-rater reliability of 93% on individual stances on the interpretive themes, and 100% on overall perspective; following discussion, there was 100% agreement between both coders.

All of the students in the *Split* category were explicit in distinguishing between what made intuitive sense to them (*Realist*) and what they perceived to be a correct response (*Quantum*). Other students offered opinions on specific themes when asked to take a stance, but chose to ultimately remain agnostic for lack of sufficient information (as indicated by the *XX/Agnostic* entries in the interpretive themes columns of Table 4.III). This agnostic characterization of individual responses differs from the overall stance of Student QA1, who said he preferred a *Quantum* interpretation, but expressed a sophisticated overall agnosticism on the legitimacy of a contrasting *Statistical* interpretation.

**Realist Category:** All three of these students considered probability waves to be mathematical tools used only to describe the probable outcomes of measurements. These students all objected to the idea that a wave packet could represent a single particle, and said they always consider an electron to be a localized object traveling somewhere inside the probability wave describing the system. These students were not classified as holding a *Statistical* perspective because they were explicit in their stance on electrons as localized particles (as opposed to agnostic), and did not have sufficient content knowledge (e.g., consequences of Bell's Theorem [14]) to appreciate why an agnostic stance on hidden variables might be necessary. All three of these students specifically objected to the notion of *wave function collapse*, calling it too counterintuitive or too unphysical to be a correct description of reality. These students all claimed to be aware of at least one alternative to their *Realist* interpretations, but said they hadn't yet been convinced by instructor arguments that their preferred perspective was incorrect.

**Split Quantum/Realist Category:** While the *Realist Category* students all expressed a measure of confidence in their perspectives on quantum physics (even when those perspectives differed from what they had heard in class), the four students in this *Split Quantum-Realist* category were, by the end of the interview, explicit in differentiating between what made intuitive sense to them, and what they considered to be a correct response. In example, Student QR1 first agreed that an electron in an atom always exists at a definite point, and continued with this line of thinking, both when first describing the double-slit experiment, and again as he began reading the *Realist* statement of Student One from the double-slit essay question:

**STUDENT QR1:** I would agree with what Student One is saying, that the electron is traveling somewhere inside that probability density blob, and it is a tiny particle. The problem here that I see is that the electron went through one slit or the other. [PAUSE] So, now I'm disagreeing with myself. OK, my intuition is fighting me right now. I said earlier that there should be one point in here that is the electron, and it goes through here and hits the screen, but I also know that I've been told that the electron goes through both slits and that's what gives you the interference pattern. Interesting. [LONG PAUSE] OK, somehow I feel like the answer is going to be that this probability density, it is the electron, and that can go through both slits, and then when it's observed with this screen, the probability density wave collapses, and then only exists at one point. But at the same time I feel that there should be a single particle, and that somehow a single, finite particle exists in this wave, and will either travel through one slit or the other. Why would a single particle be affected by a slit? That I don't have an answer to, other than that it's the wave that's actually being propagated, the wave is the electron.

**INTERVIEWER:** OK. It seems like you're talking about two different ideas. One is that the electron is a point somewhere inside this wave, and the other is saying the electron is the wave. Do you feel those two ideas conflict in any way?

**QR1:** Yeah, they do, because one says there is a finite particle at all times, and the other says that there's not, there is just this probability density, and I think the answer will turn out to be that the electron is the probability density, and that's contrary to what I said earlier. But I don't see how it could be the other way, with a finite particle. I don't see how you could get an interference pattern here with the electron being a finite particle the whole time.

**INT:** OK. What about [the *Matter-Wave*] statement?

**QR1:** [BEGINS READING] So, that goes off of what I was just saying. [READS] So, I agree with everything up to here, the electron acts as a wave and will go through both slits and interfere with itself, I believe that's true. And that's why an interference pattern develops after shooting many electrons; I guess I agree with that too, because when the blob gets to the screen, it can't just still have a probability density that would look like an interference pattern by itself. It's going to have one finite location. But after multiple electrons, multiple blobs have passed through, they will collectively form an interference pattern. So I would agree with Student Two.

**INT:** So you're agreeing with Student Two. And did you say that you disagree with Student One, or do you just have reservations about what they're saying?

**QR1:** Intuitively, I kind of agree with Student One, but I think I have reservations. I don't think, Student One, that they're right.

**INT:** But it appeals to you, what they're saying?

**QR1:** Based upon lecture, and upon those who have greater knowledge of physics than me, I would say that this [second] statement agrees more with that than the initial situation.

**INT:** So you say Student One's statement disagrees more with what you've heard in class?

**QR1:** Yes. But not more with what I envision. This [first] one kind of depicts more of my rational depiction, all that I can wrap myself around and understand, and the second one is more of what I've been told, but don't completely understand. I've been told it's right, so…

This excerpt serves two purposes. It first explicitly demonstrates how students may change as needed between ontological attributions in their descriptions of electrons in order to explain observed phenomena (electrons as particles in order to explain localized detections, electrons as waves in order to explain interference). It also underscores the need to distinguish between the *personal* and the *public* [15] perspectives of students on quantum physics: these students differed from their *Realist* category counterparts in that they explicitly differentiated between what made intuitive sense to them, and responses they perceived as being correct. This finding parallels studies by McCaskey et al., [16, 17] where students were asked to respond twice to the Force Concept Inventory, [18] first as they personally believed, and then as they felt a scientist would respond. These authors found that most every student *split* on at least one survey item, indicating a difference between their personal beliefs and their perceptions of scientists' beliefs. Following a series of validation interviews, these authors reported that students most often explained their personal responses in terms of what made intuitive sense to them, and that split responses reflected how students had learned a correct response from instruction, without having reconciled that knowledge with their own intuition. Similar studies probing the attitudes and beliefs of introductory classical physics students have demonstrated similar results. [19]

Regarding the public perspectives of modern physics students, we would also point out that students will not necessarily identify an authoritative stance based on specific knowledge of what expert physicists believe. Not only may their perceptions of what scientists believe be inaccurate, students may also employ undesirable epistemological strategies learned from their experiences in the classroom:

> "This [*Quantum* statement of Student Two] is more of a complex definition, I think. [...] Probably initially I would be confused by this statement if I hadn't taken this course, but I might be like the public and think the most complicated answer, that must be the right one. Because a lot of times—it's even happened with the [concept] questions in class—where I think: *That's got to be the answer*. But then I'll be like: No, that would be too easy, it's got to be something else. Sometimes that [strategy] can prove correct or incorrect." [STUDENT QR4]

**Pilot-Wave Category:** The responses from these three students indicated an ontology that blends attributes from both classical particles and waves. These students indicated a belief that wave-particle duality implies that quanta must be thought of as simultaneously *both* particle *and* wave. The following student explained the fringe pattern in the double-slit experiment in terms of constructive and destructive interference, and acknowledged that the experiment had been used in class to demonstrate the wave characteristics of quanta, but had his own ideas about the source of interference for localized particles:

> "It seems like the probable paths for the electron to follow interact with themselves, but the electron itself follows just one of those paths. It's like the electron rides on a track, like a train rides on a rail, but those rails or tracks go through both slits, and the possible paths for the electrons to follow interfere with themselves, create the interference pattern, but the physical electron just rides on the tracks, it picks one. Or maybe switches paths, if two of them cross. I don't know, it seems that the electron has to be on one of those tracks, but the tracks themselves cause the interference pattern." [STUDENT P3]

Of particular interest is the way in which this same student demonstrated how his realist (albeit nonlocal) perspective can be employed as an epistemological tool:

> "As [the electron is] traveling it's going to be somewhere in this [probability density] as it moves along until it's actually detected. And if it was here [INDICATES POINT NEAR DETECTING SCREEN] then it must have been here at one point in time [INDICATES SECOND POINT NEAR THE FIRST] and if it was here, then it had to be here at one point in time, all the way back to here [TRACES LINE BACK TO NEAR BOTH SLITS] in which case there's only two places it could be. So yes, I think it went through one slit or the other." [STUDENT P3]

As another example of the ontological flexibility exhibited in novice thinking, one of these students explained that, while it is necessary to think of an electron in the double-slit experiment as both wave and particle, it was unnecessary to employ a wave description for atomic electrons since, in his mind, there were no wave effects to be accounted for:

> "When I was thinking about [an electron] in an atom, there's really no reason that you have to think about it as a wave, in the fact that it's not really interacting with anything. In [the double-slit] experiment, yes I like to think of it as also a wave, because this is kind of the key experiment of quantum mechanics, to describe this [wave] phenomenon, and so for that reason it is more effective to think of it as both." [STUDENT P1]

With this excerpt, we call attention to the fact that sometimes students employ different models (ontological attribution assignments) in different contexts, without necessarily looking for or requiring internal consistency among them.

**Quantum** (**Matter-Wave**) **Category:** These five students were consistent in providing responses that indicated a *matter-wave* ontology:

> "I don't think of [the electron] as orbiting the nucleus because it doesn't, it just exists in that region of space. It exists in a volume element that defines the probability of finding the electron in that space […] and that's really what the electron is: a smeared out volume of charge." [STUDENT Q2]

All of these students described unobserved quanta strictly in terms of waves, and discussed the *collapse of the wave function* as a physical process where wave-like quanta suddenly exhibit particle-like properties. According to these students, their personal perspectives on quantum mechanics were in complete agreement with their perceptions of expert beliefs.

**Quantum/Agnostic Category:** We find it necessary to distinguish this one student from those in the strictly *Quantum* category because, while the *Quantum* category students had all expressed confidence in their *matter-wave* interpretations, this student expressed a degree of sophisticated uncertainty in his own views:

> "The way I think of an electron, I cannot ascribe to it any definite position, definite but unknown position. I mean, it may be that way, but I think that somehow the electron is represented by the wave function, which is just a probability, and if we want to localize it then we lose some of the information. So whether this is true or not is something of a philosophical question. I wish I knew, or understood it, but I don't. For now, for me, the electron is the wave function, so whether the electron is distributed among the wave function, and when you do an experiment, it sucks into one point, or whether it is indeed one particle at a point, statistically the average, I don't know." [STUDENT QA1]

**Copenhagen Category:** These three students were similar to the *Quantum Category* students in terms of the nonexistence of hidden variables, but saw probability waves as containing information only, rather than representing the actual physical state of a particle. As with student C2 (quoted previously in Section III.B) each of these students stated explicitly that it is unscientific to discuss that which can't be measured or observed. These three students said they considered electrons to be neither wave nor particle; that such concepts were in fact different models for describing the behavior of quanta under different circumstances. These students expressed what we consider to be a moderately sophisticated perspective on both the necessity and the desirability of switching between ontological categorizations.

It should also be noted in our studies that formal instruction is not the only source of information or influence for students regarding quantum physics, as with this student, who explained how his own personal solipsistic philosophy influenced his beliefs about quantum mechanics, and vice-versa:

> **STUDENT C3:** This is more of a philosophical point for me, but if we can't know something, there's no difference between it not existing and us not knowing it. So, for our purposes, it's more useful to say, if we can't know it, where the electron is, then it doesn't have a definite position. […] I believe, so long as we don't measure it, then an electron doesn't have a definite position.
>
> **INTERVIEWER:** What happens when we measure it?
>
> **C3:** Well, we find a position then**…** Then it does.
>
> **INT:** The position we find, is that where the particle was the moment before we measured it?
>
> **C3:** No. We can't know that. So, when we make a measurement, there's the particle. When we look away, the particle goes away. And I sort of felt this way before having learned about quantum mechanics. And it just solidified in my mind that there's no difference between me not knowing it, and it not existing.

In the class-wide online surveys, a majority of students from all of the four courses discussed here reported having previously heard about quantum mechanics in popular venues (e.g., books by Hawking [20] or Greene [21]) before enrolling in the course.

## IV. Summary and Discussion

Our more detailed characterization of the perspectives of modern physics students improves upon our previous efforts by addressing the contextual sensitivity of those perspectives, through an exploration of their expressed beliefs about quantum physics across three key interpretive themes. We find that, as a form of sense making, students develop a variety of ideas and opinions regarding the physical interpretation of quantum mechanics, in spite of how their instructors explicitly addressed matters of interpretation in class.

As with past studies, we find that a significant number of students from our interviews (10 of 19) expressed a preference for *realist* interpretations of quantum phenomena. However, the nature of these students' *realist* perspectives were not necessarily of the character we had anticipated from the results of earlier studies. Only three of these students consistently preferred *realist* interpretations of quantum phenomena, while simultaneously expressing confidence in the correctness of their perspectives; whereas four others differentiated between what made intuitive sense to them, and what they perceived to be correct responses. Their particular kind of switching between ontological framings may be best understood in terms of their competing *personal* and *public* perspectives [15] on quantum physics – when responding during interviews, these students frequently vacillated between what they personally believed and the answer they felt an expert physicist would give, without always articulating a difference between the two without prompting. This finding has implications for future research into the ontologies of quantum physics students, who may not always respond to such questions as they actually believe, but rather provide the responses that best mimic their instructors. Such issues are of particular significance with regard to matters of interpretation in quantum mechanics, where the beliefs of practicing physicists are at such variance with each other, which may confuse student perceptions.

The *Realist* beliefs of three other students were of a decidedly nonlocal character: localized quantum entities follow trajectories determined by the interaction of nonlocal quantum waves with the environment. None of these three students claimed to be aware of any formal *pilot-wave* interpretation, and their beliefs in quanta as simultaneously wave and particle were at odds with how wave-particle duality was addressed in class by their instructors (i.e., quanta are sometimes described by waves, and sometimes as particles, but never both simultaneously). The remaining nine of 19 students expressed fairly consistent views that could be seen as in agreement with the (implicit) learning goals of their instructors, whether Quantum or Copenhagen. In other words, these students seemed to have successfully incorporated probabilistic and nonlocal views of quanta and quantum measurements into their personal perspectives, and/or agreed that scientists should restrict discussions to that which can be measured and verified. While these findings are somewhat at odds with previous research into quantum ontologies, which have concluded that student perspectives are rarely in alignment with expert or productive transitional models, we emphasize that the relatively few students who participated in our interviews were generally better-than-average students, and were not representative of an entire class. Ultimately, we believe the value of these findings lies in the demonstration and documentation of a variety of student beliefs regarding quantum phenomena, and not a determination of the relative prevalence of any specific beliefs.

Of equal importance is the demonstration of students employing multiple parallel ontologies, or dynamic ontologies that are flexible and adaptive, each according to their immediate cognitive needs. In one specific case, Student QR1 initially described an electron as a particle localized in space, but wavered in his commitment to this description when he encountered the notion that each electron must have travelled through only one slit on its way to the detecting screen. After a moment of introspection, he concluded that a wave description was necessary in order to explain the observed interference pattern, for he had no explanation as to why a localized particle would be affected by the presence of a slit. He then explicitly stated that the "correct" way of looking at the situation is to equate the electron with the wave itself, which necessitated a corresponding belief in a physically collapsing wave function. Student QR1 was aware of the logical inconsistency in his two competing perspectives, but was able to articulate a need for maintaining both, one in correspondence with his intuition, and one in congruence with what he perceived as an authoritative stance, and which also led to an interpretation of the double-slit experiment that was consistent with observations. We can easily imagine this student's reasoning during the interview briefly recapitulated some of the thought processes he engaged in when first encountering this topic, as he initially seemed unaware of any need to think of electrons as anything other than localized particles, but immediately reconsidered his stance when confronted with an observation that he could only explain in terms of wave interference.

We have also seen how *classical attribute inheritance* will guide the thinking of both experts and novices, through the explicit statements of Ballentine, along with those from Students R1 and QR2: localized detections imply a continuously localized existence, particles are *by definition* localized in space; laws of mass and charge conservation preclude the possibility for particles to be spatially delocalized. These types of epistemological and ontological *resources* are not necessarily wrong in and of themselves, and may be of productive use in classical descriptions of matter, but have enormous implications for what kind of physical meaning students attach to the otherwise mathematically algorithmic process of deriving wave functions and calculating expectation values; and their activation in the context of quantum phenomena may lead students to interpretations that seem paradoxical or are inconsistent with observations.

The demonstration of student flexibility in assigning ontological attributes, switching back and forth (and sometimes blending) them as needed, does more than just explain the contextual sensitivity of student responses; it provides strong evidence of the dynamical nature of the ontologies employed by students when reasoning about quantum phenomena. The students falling into the strictly *Realist* category were the ones showing the least flexibility in their use of ontologies (and even these students were aware of alternative explanations, but hadn't yet bought into them). All of the other students demonstrated varying degrees of flexibility in their use of parallel ontologies: some distinguished between intuitive and normative ontologies; some perceived switches between ontological attributes as reflective of physical transitions; others blended attributes from classically distinct categories, or assigned them separately, all according to their cognitive needs of the moment.

These results are most consistent with the dynamic view of novice and expert ontologies discussed here and in Chapter 1, and are difficult to reconcile with the static, parallel ontologies promoted by Slotta and Chi. First, quantum mechanics describes the *behavior* of light and matter in terms of classically distinct ontological characteristics, and so a rigid (robust) assignment of ontological attributes is not possible for a complete description of electrons and photons. Nor do scientists agree on a normative view of the ontological nature of quanta, and instructors understandably vary in their choices of how to broach this topic in their introductory courses, sometimes fearful of opening a *Pandora's Box* of student questions with no easy answers.

Second, we observe that students frequently modify their patterns of ontological attribution assignment piecewise, both within and across multiple contexts. This type of gradual transition in student thinking cannot be plausibly explained in terms of rigid, parallel ontologies that are developed over the course of instruction, and which then replace the original, intuitive ontologies, unless one were to believe that students develop a whole multitude of parallel ontologies, each specific to the variety of situations they've encountered. In the end, Slotta has conceded that the disagreement between these two opposing views may ultimately be a matter of the degree of ontological flexibility and blending exhibited in both novice and expert thinking, [22] and both sides have made strong arguments in favor of their views on learning and cognition in the context of classical physics; their disparities become all the more apparent, however, in the context of quantum mechanics.

We also find it significant that most every student expressed distaste for deterministic ideas in the context of quantum phenomena, although it had been anticipated that *Realist Category* students might favor such notions. Not only did most every student say they were unfamiliar with the word *determinism* within the context of physics, practically every student believed either that any description of the behavior of quantum particles should be inherently probabilistic, or that the Heisenberg uncertainty principle places a fundamental limit on human knowledge of quantum systems, or a combination of both stances. A superficial analysis showed that the *Realist* and the *Split Quantum/Realist* students were more likely than other students to invoke the uncertainty principle when discussing notions of determinism; the remaining students were more likely to state that the behavior of quantum particles (or the nature of the universe) is inherently probabilistic. These responses indicate a need for a more detailed exploration of the uncertainty principle as an epistemological tool for quantum physics students.

These interviews have demonstrated how matters of interpretation are of both personal and academic interest to students, and modern physics instructors should recognize the potential impact on student thinking when choosing to de-emphasize interpretation in an introductory course. Not only do students develop their own ideas regarding the physical meaning behind quantum mechanics, they also develop attitudes (right or wrong) about the positivistic or agnostic stances of

their instructors:

> "It seems that there's this dogma among physicists, that you can't ask that question: *What is it doing between point A and point B? You can't ask that!* And I think that the only way we'll be able to make profound progress is by asking those questions. It doesn't make sense that somebody would say, don't ask that, or you can't ask that. I think somehow they're shutting down free seeking of knowledge. But I don't know enough about quantum mechanics. Maybe when I get more understanding of quantum mechanics, I too will be saying: *You can't ask that!* But as a naïve student it sounds like a bad attitude to have about physics." [STUDENT P3]

Although many instructors may argue that introductory students do not have the requisite sophistication to appreciate matters of interpretation in quantum mechanics, we note that several authors have already developed discussions of EPR correlations and Bell inequalities that are appropriate for the introductory level. [23, 24] Questions of interpretation may also be addressed in terms of *scientific modeling*, an aspect of epistemological sophistication that is often emphasized in physics education research as a goal of instruction, as well as in terms of *nature of science* issues. [25] In the end, we argue that modern physics instructors should concern themselves with matters of interpretation, if only because their students concern themselves with these matters, and as educators we should be concerned with what our students believe about physics and the nature of practicing physics. Modern physics instructors who aim to transition students away from classical epistemologies and ontologies may employ our framework for understanding and interpreting the myriad combinations of student ideas concerning the nature of quantum mechanics and its description of the natural world. Such insight may allow us to target instructional interventions that will positively influence student perspectives, and strengthen their abilities to make interpretations of physical phenomena, and to understand the limitations and bounds of these interpretations.

**References [Chapter 4]**

**1.** N. G. van Kampen, The Scandal of Quantum Mechanics, Am. J. Phys **76**, 989 (2008).

**2.** M. Dubson, S. Goldhaber, S. Pollock and K. Perkins, Faculty Disagreement about the Teaching of Quantum Mechanics, *PERC Proceedings 2009* (AIP, Melville, NY, 2009).

**3.** G. J. Posner, K. A. Strike, P. W. Hewson and W. A. Gertzog, Accommodation of a scientific conception: Toward a theory of conceptual change, *Sci. Educ.* **66**, 211 (1982).

**4.** A. Gupta, D. Hammer and E. F. Redish, The case for dynamic models of learners' ontologies in physics, *J. Learning Sciences* **19**, 285 (2010).

**5.** D. Hammer, A. Elby, R. E. Scherr and E. F. Redish, "Resources, Framing and Transfer" in *Transfer of Learning*, edited by J. Mestre (Information Age Publishing, 2005) pp. 89-119.

**6.** http://phet.colorado.edu/simulations/sims.php?sim=QWI

**7.** L. E. Ballentine, The Statistical Interpretation of Quantum Mechanics, *Rev. Mod. Phys.* **42**, 358 (1970).

**8.** L. E. Ballentine, Quantum Mechanics: A Modern Development (World Scientific Publishing, Singapore, 1998).

**9.** P. Grangier, G. Roger, and A. Aspect, Experimental Evidence for a Photon Anticorrelation Effect on a Beam Splitter: A New Light on Single-Photon Interferences, *Europhysics Letters* **1**, 173 (1986).

**10.** P. A. M. Dirac, *The Principles of Quantum Mechanics, 3rd ed.*, (Clarendon Press, Oxford, 1947).

**11.** D. Bohm and B. J. Hiley, *The Undivided Universe* (Routledge, New York, NY, 1995).

**12.** N. D. Mermin, What's Wrong with this Pillow? *Phys. Today* **42**, 9 (1989).

**13.** C. Baily and N. D. Finkelstein, Refined characterization of student perspectives on quantum physics, *Phys. Rev. ST: Physics Education Research* **5**, 020113 (2010).

**14.** J. S. Bell, Speakable and Unspeakable in Quantum Mechanics (Cambridge University Press, Cambridge, 1987).

**15.** L. Lising and A. Elby, The impact of epistemology on learning: A case study from introductory physics, *Am. J. Phys.* **73**, 372 (2005).

**16.** T. L. McCaskey, M. H. Dancy and A. Elby, Effects on assessment caused by splits between belief and understanding, *PERC Proceedings 2003* **720**, 37 (AIP, Melville, NY, 2004).

**17.** T. L. McCaskey and A. Elby, Probing Students' Epistemologies Using Split Tasks, *PERC Proceedings 2004* **790**, 57 (AIP, Melville, NY, 2005).

**18.** D. Hestenes, M. Wills, and G. Swackhamer, *Physics Teacher* **30**, 141 (1992).

**19.** K. E. Gray, W. K. Adams, C. E. Wieman and K. K. Perkins, Students know what physicists believe, but they don't agree: A study using the CLASS survey, *Phys. Rev. ST: Physics Education Research* **4**, 020106 (2008).

**20.** S. W. Hawking, *A Brief History of Time* (Bantam, New York, NY, 1988).

**21.** B. R. Greene, *The Elegant Universe* (Norton, New York, NY, 2003).

**22.** J. D. Slotta, In defense of Chi's Ontological Incompatibility Hypothesis, *Journal of the Learning Sciences* **20**, 151 (2011).

**23.** D. F. Styer, *The Strange World of Quantum Mechanics* (Cambridge University Press, Cambridge, 2000).

**24.** http://www.colorado.edu/physics/EducationIssues/modern/

**25.** S. B. McKagan, K. K. Perkins and C. E. Wieman, Why we should teach the Bohr model and how to teach it effectively, *Phys. Rev. ST: Physics Education Research* **4**, 010103 (2008).

# CHAPTER 5

## Teaching Quantum Interpretations – Curriculum Development and Implementation

"The tao that can be told is not the eternal Tao. The name that can be named is not the eternal Name." – Lao-tzu, Tao Te Ching

### I. Introduction

We wish to address one final question: Can the interpretive aspects of quantum mechanics be addressed at a level that is appropriate and meaningful for introductory modern physics students, without sacrificing traditional course content and learning goals? In fact, it would be hoped that an additional focus on interpretive topics (indeterminacy, the uncertainty principle, wave-particle duality, and the superposition of quantum states) would provide students with tools that would augment their overall understanding of traditional topics (quantum tunneling, atomic models); that discussions of the application of quantum mechanics could subsequently be framed in terms of language that has previously been unavailable to past instructors; and that students may develop more internal consistency in their interpretation of quantum phenomena.

The remainder of this dissertation will concern itself with the development of a modern physics curriculum designed to target these aspects of student thinking, and its recent implementation (Fall 2010) at the University of Colorado in the form of an introductory course for engineering majors. In this chapter, we discuss the guiding principles behind the development of this curriculum, and provide a detailed examination of specific, newly developed course materials designed to meet these goals. [A broader selection of relevant course materials can be found in Appendix C.] In doing so, we address the appropriateness and effectiveness of this curriculum by considering aggregate student responses to a subset of homework, exam, and survey items, as well as actual responses from four select students. [Appendix D contains a larger subset of complete responses from these particular four students.]

## II. Curriculum Development and Implementation

It must be *strongly emphasized* from the outset that it is our aim to *improve* upon an already-existing body of work, which has seen contributions from over a dozen physics education researchers and modern physics instructors at the University of Colorado. As was the case for many of the modern physics offerings discussed in these studies, a substantial portion of the course materials we used should be credited to the original work of S. B. McKagan, K. K. Perkins, and C. E. Wieman. Their original course transformations, [1] which served as the basis for our course, incorporated a number of principles learned from physics education research, which include, but are not limited to:

1. Students' attitudes toward science tend to become less expert-like unless instructors are explicit in addressing student beliefs. [2, 3] The original course transformations were explicit in addressing scientific method and logical deduction; experimental evidence and real-world applications; and the uses and limitations of models. [4]

2. Interactive engagement during lecture can lead to higher learning gains than traditional lectures, [5] and can be useful in eliciting known student misconceptions. [6] Concept tests (clicker questions) provide real-time feedback from students, allowing instructors to gauge student understanding, as well as target common misconceptions. Peer discussion during concept tests gives students an opportunity to articulate their knowledge and engage in scientific argumentation in a low-stakes environment. Weekly collaborative homework sessions offer similar benefits for both students and instructors.

3. In order for students to best gain conceptual understanding and reasoning skills, all aspects of the course (including lecture, homework, and exams) should emphasize conceptual understanding alongside numerical problem solving. [1]

4. Interactive simulations used in and outside of the classroom can be useful in helping students to build models and intuition about quantum physics, by providing visual representations of abstract concepts and unobservable processes. [7]

We have argued [Chapter 3] that interpretive themes in quantum mechanics are an often *hidden* aspect of modern physics instruction, according to three criteria: A) These issues are frequently superficially addressed, and in a way that is not meaningful for students beyond the specific contexts in which they arise; B) Students often develop their own ideas regarding these interpretive themes, even when instructors do not adequately attend to them; and C) Those beliefs tend to be more novice-like (intuitively realist) in contexts where instruction is less explicit. We therefore chose to directly confront the kinds of realist beliefs and attitudes that are common to introductory modern physics students, as informed by our own research into quantum perspectives. Our aim was not only to make students consciously aware of their own (often intuitive and tacit) beliefs, but also for them to acquire the necessary language and conceptual inventory to identify and articulate those beliefs (we are reminded that, even at post-instruction, most of the students in our interviews were not familiar with the word *determinism* in the context of physics, though they had certainly developed opinions about it).

We also chose to make the interpretation of quantum physics a course topic unto itself, primarily framing our discussions in terms of the historical back-and-forth between Albert Einstein and Niels Bohr. And though we decided to be explicit in promoting a matter-wave interpretation of quantum mechanics, our ultimate goal was for students to be able to distinguish between competing perspectives, to have the requisite tools for evaluating their advantages and limitations, and to be able to apply this knowledge in novel situations. In short, instead of trying to tell students what they should and shouldn't believe about quantum physics, we chose to engage them in an explicit, extended argument (with us and amongst themselves) against *Local Realism*. This argument was *extended* in two senses: 1) We were able to augment a number of standard topics (e.g., the uncertainty principle, atomic models) with discussions of interpretive themes; and 2) We introduced several entirely new topics (e.g., delayed-choice experiments) that created additional opportunities for students to explore the sometimes fluid boundaries between scientific interpretation and theory.

The entirety of our research has indicated that wave-particle duality is a particularly challenging topic for students, and wholly relevant to their beliefs regarding the physical meaning of quantum mechanics. Whether emphasized or not, *every* modern physics instructor considered in these studies made mention of the fact that double-slit experiments could be performed with single quanta, which are detected as localized particles, but which together form an interference pattern over time. This phenomenon was often (though not universally) demonstrated in class using the Quantum Wave Interference PhET simulation, [8] as seen in the post-instruction attitude surveys. Due to the distance scales involved, a true double-slit experiment was until recently only a thought experiment, crafted as a demonstration of principle; actual experiments had demonstrated the diffraction of electrons through periodic lattices (essentially, a many-slit experiment). [9] We sought in this course to emphasize connections between theory, interpretation, and experimental evidence, and so augmented these discussions with presentations on experimental realizations of these *Gedanken* experiments. In 2008, Frabboni, et al. employed nanofabrication techniques in the creation of a double-slit opening on a scale of tens of nanometers, which they then used to demonstrate electron diffraction, as well as the absence of interference after covering just one of the two slits (they also present in their paper STM images of the double-slits, formed by an ion beam in a gold foil, with both slits open and with one slit covered). [10] Tonomura, et al. have produced a movie that literally demonstrates single-electron detection and the gradual buildup of a fringe pattern. [11, 12] Students from prior courses were often skeptical as to whether such an experiment (where only a single electron passes through the apparatus at a time) could be done in practice – in this way, they can observe the phenomenon with their own eyes.

In addressing the tendency for students to interpret wave-particle duality as implying that quanta may act simultaneously as both particle and wave, we devoted additional class time to a presentation of the single-photon experiments discussed in the first chapter, which are essentially isomorphic to the double-slit arrangement (the double-slit and the beam splitters play analogous roles). One of the guiding principles in the design of this curriculum was to avoid as much as possible the expectation for students to accept our assertions as a matter of faith. Rather than describing what the experimentalists had meant to demonstrate, and then simply asserting that they had been successful, we presented students with the actual reported data, which required the use of statistical arguments, and thereby afforded further opportunity to highlight the role of probability in quantum mechanics. These single-photon experiments demonstrate for students the dualistic nature of photons, and provide strong evidence against realist interpretations, but only if the details and results of the experiments are accessible to them, and so we omitted from our presentation extraneous technical details, while still focusing on the very process of designing the experiment and creating an adequate photon source. Devoting an entire class period to these experiments afforded us the time to walk students through each of the three experiments, and for them to debate the implications of each, while creating further opportunities to distinguish between a collection of data points, and an interpretation of their meaning.

Just as importantly, these experiments call for an explicit discussion of the need for ontological flexibility (without naming it as such) in the description of quanta, from which we may easily segue into a comparison of competing interpretations. Bohr has offered up *Complementarity* as a guide to making sense of this dualistic behavior (note that we refrain here from digressing into a full explication of the Copenhagen Interpretation for our students), but this interpretation can come across as more a philosophical sidestepping of the *measurement problem*, than its scientific resolution. Dirac's matter-wave interpretation allows for a consistent description of the behavior of photons at the beam splitters, but the physical collapse of the wave function is not described by any equation, and accepting it as physically real requires a fairly large leap of faith in itself. Moreover, these discussions allow for the explicit development of quantum epistemological tools [two paths = interference; one path = no interference] that may facilitate student understanding, and which may be applied to novel situations.

Before presenting and evaluating any newly developed course materials, some general comments on the structure of the course in which they were used are in order. As with other modern physics courses described here, our course spanned a 15-week academic semester, and consisted of large lectures (N ~ 100) meeting three times per week, together with weekly online and written homework assignments, and twice-weekly problem-solving sessions staffed by the instructors. Course transformations for this semester occurred primarily during Weeks 6-8, spanning a total of nine lectures. [13] Instruction was collaborative, with two lead co-instructors (one of them the author, the other a PER faculty member associated with our prior investigations into quantum perspectives), along with two undergraduate learning assistants, [14] who helped facilitate student discussion during lecture. As with the original course transformations, we omitted topics from special relativity in order to win time for the introduction of new material, without eating into the usual time at the end of the course devoted to applications.

We selected Knight's *Physics for Scientists and Engineers* [15] as a textbook (mostly for its readability), but the lectures did not follow the textbook very closely (if at all), and it was necessary to provide students with outside reading materials for many of the new topics (e.g., single-photon experiments [16] and Local Realism [17]); these *Scientific American* articles were chosen for their non-technical, but scientifically correct, treatment of interpretive ideas and foundational experiments in quantum mechanics. An online discussion board was created to provide students with a forum to anonymously ask questions about the readings, and to provide answers to each other; following these discussions granted us ample opportunity to assess how students were responding to many of the new ideas we were introducing.[1] A total of 13 weekly homework assignments consisted of online submissions and written, long-answer problems; there was a broad mixture of conceptual and calculation problems, both requiring short-essay, multiple-choice, and numerical answers. There were a total of three midterm exams (held outside of class) and the course ended with a cumulative final exam. In lieu of a long answer section on the final exam, students were asked to write a 2-3 page (minimum) final essay on a topic from quantum mechanics of their choosing, or to write a personal reflection on their experience of learning about quantum mechanics in our class (an option chosen by ~40% of students). As opposed to a formal term paper, this assignment was meant to give students the opportunity to explore an aspect of quantum mechanics that was of personal interest to them. The almost universally positive nature of the feedback provided by students in their personal reflections is evidence for the popularity and effectiveness of our transformed curriculum, and its practical implementation.

The progression of topics may be broken into three main parts: classical and semi-classical physics; the development of quantum theory; and its application to physical systems). A complete explication and analysis of the entirety of this new curriculum and associated course materials would be beyond the scope of this dissertation, and so we conclude this section with a summary overview of the progression of topics covered in this class. The remaining sections of this chapter will address specific lecture, homework and exam materials, alongside aggregate and individual student responses from the Fall 2010 semester.

**PART I – Classical and Semi-Classical Physics (Weeks 1-5, Lectures 1-12):** Introduction to the course and the philosophy behind its structure. Review relevant mathematics (complex exponentials, differential equations, wave equations); review classical electricity and magnetism, Maxwell's equations and how they lead to a wave description of light. [Lectures 1-3] Cover properties of waves (superposition, interference); address the wave properties of light through Young's double-slit experiment and Michelson interferometers. Introduce polarization and polarizing filters in anticipation of future topics concerning photon detection. [Lecture 4]

[1] Students were asked to make a contribution to the discussion board each week of the latter half of the course as part of their homework assignment, but no efforts were made to verify their participation, and students were free to put as little or as much effort as they liked into their postings.

Discuss photoelectric effect experiment in terms of classical wave predictions, contrasted with a particle description of light. Photomultiplier tubes are introduced as an application of the photoelectric effect, but also so as to not be unfamiliar to students when they arise in the future. An emphasis on the physical meaning of the work function foreshadows applications of the Schrödinger equation to square well potentials. [Lectures 4-5] Review potential energy curves and explicitly relate them to models of physical systems. Discuss modeling in physics, and lead discussions on the differences between observation, interpretation, and theory. [Lectures 6-7] Relate spectral lines (Balmer series) to atomic energy levels via the energy-frequency relationship established in the photoelectric effect, and use them to make inferences about quantized atomic energy levels. Emphasize the differences between photon absorption (an all-or-nothing process) and collisional excitation of atoms (discharge tubes). [Lectures 8-9] Apply knowledge of photon absorption and emission processes to the construction of lasers. Compare and contrast wave and particle descriptions of light, and address their ranges of applicability. Relate wave intensity to the probability for photon detection in the context of a single-photon double-slit experiment (simulated). [Lectures 10-11] Review for the first exam. [Lecture 12]

**PART II – Development of Quantum Theory (Weeks 5-8, Lectures 13-24):** Review potential and kinetic energy of electrons in a Coulomb potential, then introduce the semi-classical Bohr model of hydrogen. Discuss the ad-hoc mixture of classical and quantum rules, along with the strengths and weaknesses of the model. Introduce de Broglie waves and his atomic model as an explanation for quantized energy levels. [Lectures 13-14] Review the behavior of magnets in response to homogeneous and inhomogeneous magnetic fields; employ a Bohr-like model for atomic magnetic moments, and explicitly address classical expectations for their behavior in a Stern-Gerlach type apparatus.[2] [Lecture 15] Use repeated spin-projection measurements to introduce ideas of: quantization of atomic spin (two-state systems); definite versus indefinite states; state preparation; and probabilistic descriptions of measurement outcomes. Digress briefly to cover classical probability, statistical distributions, and the calculation of expectation values. [Lectures 16-17] Offer multiple interpretations of repeated spin measurements for future evaluation, and discuss the differences between classical ignorance and quantum uncertainty. Introduce *entanglement* in the context of distant, correlated atomic spin measurements, and relate to topics in quantum cryptography. Make explicit definitions of *hidden variables*, *locality*, *completeness* and *Local Realism*, followed by a discussion of the EPR argument and its implications for the nature of quantum superpositions. Use the notion of *instruction sets* as a first pass deterministic model, and reveal its limitations in the face of observation.[3] [Lectures 18-19] Use the single-photon experiments by Aspect, et al. as an argument against simultaneous wave and particle descriptions of photons. Invoke *Complementarity*

---

[2] Much of the lecture and homework material on magnetic moments and repeated spin measurements was inspired by D. F. Styer. [18]

[3] The "Local Reality Machine" argument is due to N. D. Mermin. [17]

and other interpretive stances in the establishment of quantum epistemological tools. [Lectures 20-21] Relate conclusions drawn from single-photon experiments to an understanding of the double-slit experiment performed with single electrons. Plane wave descriptions of single particles lead to more generalized notions of quantum wave functions and their probabilistic interpretation. Introduce the Heisenberg uncertainty principle, its mathematical expression, and various interpretations of its physical meaning. [Lectures 22-23] Review for second exam. [Lecture 24]

**PART III – Applications of Quantum Mechanics (Weeks 9-15, Lectures 25-44):** Motivate the Schrödinger equation through analogies with electromagnetic waves and solve for free particles in terms of plane waves. [Lectures 25-26] Introduce square well potentials (infinite and finite) and use them to model electrons in wires. [Lectures 27-28] Frame discussions of quantum tunneling as a consequence of the wave behavior of matter, then apply tunneling to scanning tunneling microscopes, and a description of alpha-decay. [Lectures 29-31] Apply the Schrödinger equation to an electron in a 3-D Coulomb potential and develop the Schrödinger model of hydrogen. Generalize to multi-electron atoms and account for the periodicity of elements. [Lectures 32-35] Review for the third exam. [Lecture 36] Explain molecular bonding and conduction banding in terms of the superposition of atomic potentials and electron wave functions. [Lectures 37-39] Apply these concepts to the theory of transistors and diodes. [Lecture 40] Finish with a foray into radioactivity, nuclear energy, and nuclear weapons (at student request) [Lectures 41-42] Review for the final exam. [Lectures 43-44]

## II.A. Assessing Incoming Student Perspectives and Conceptual Understanding

Developing pre-instruction content surveys for modern physics students is more difficult than assessing incoming student beliefs about classical physics, for several reasons. First, it is expected that introductory students with little knowledge of Newtonian mechanics will have already developed intuitions (right or wrong) through their everyday experiences about the motion of macroscopic objects; in contrast, our everyday experiences with applied quantum physics (e.g. computers) provide little insight into the rules governing the behavior of quantum entities. Second, many of the learning goals for modern physics courses concern topics, such as quantum tunneling, that are entirely foreign to introductory students; and so, for example, it is practically meaningless to discuss incoming student responses to questions regarding deBroglie wavelengths and transmission probabilities, since the distributions of responses are often statistically indistinguishable from guessing.[4] Third, the broad variation in learning goals among modern physics instructors indicates a lack of consensus in the physics community regarding canonical course content, making it difficult to develop general assessment instruments that would be appropriate for a range of course offerings and student populations.

[4] For example, an (unpublished) analysis by this author of pre-instruction QMCS scores from several modern physics courses showed them to be normally distributed about an average consistent with random guessing.

We therefore constructed a content survey (administered in the first week of the semester) that would be appropriate for the specific learning goals of this course, by culling questions from a variety of previously validated assessment instruments, [19-21] and then limiting pre-instruction items to ones where it could be reasonably expected that students would have specific reasons for responding as they do beyond random guessing (i.e., prior content knowledge or intuitive expectations). So, for example, even if students have never heard of a double-slit experiment performed with electrons, their intuitive notions of particles might still lead them expect a pattern that would be consistent with their expectations for macroscopic particles in an analogous situation (these questions taken from the QPCS; [21] student responses are given in Table 5.I):

**The following questions** refer to the following three experiments:

In one experiment electrons pass through a double-slit as they travel from a source to a detecting screen. In a second experiment light passes through a double-slit as it travels from a source to a photographic plate. In a third experiment marbles pass through two slit-like openings as they travel from a source to an array of collecting bins, side-by-side.

The right-hand figure diagrams the experimental setup, and the figures below show roughly the possible patterns that could be detected on the various screens.

**Top view of experimental set-up (not to scale)**

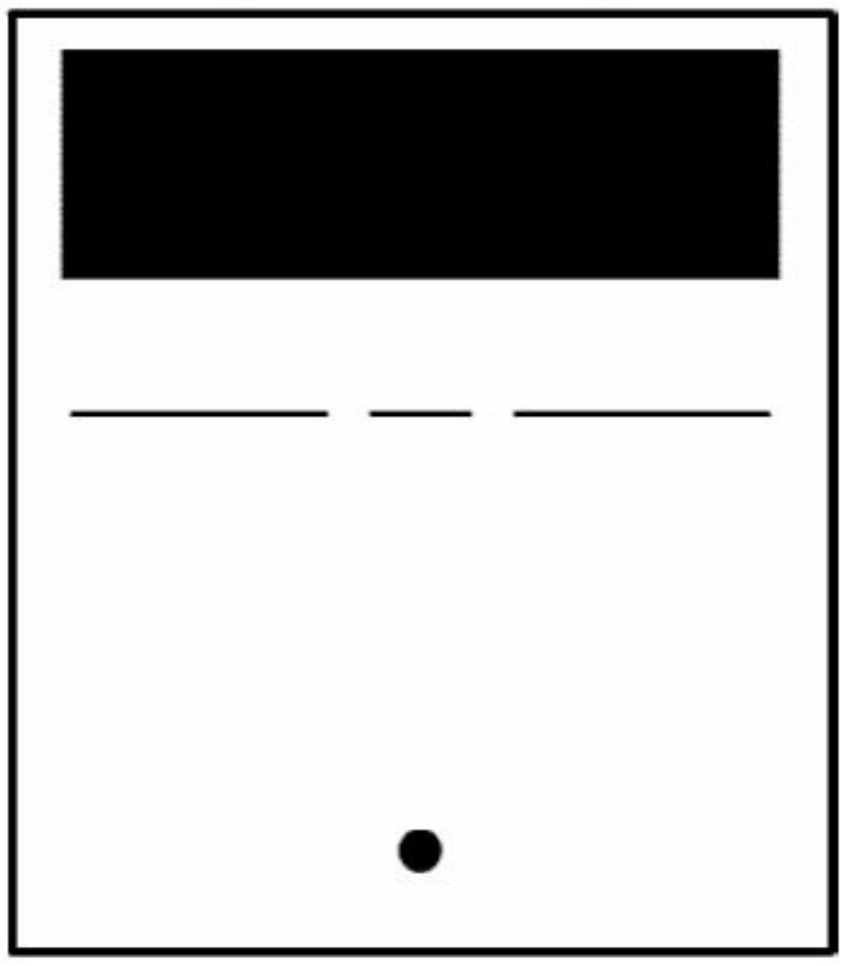

**Possible patterns (not to scale)**

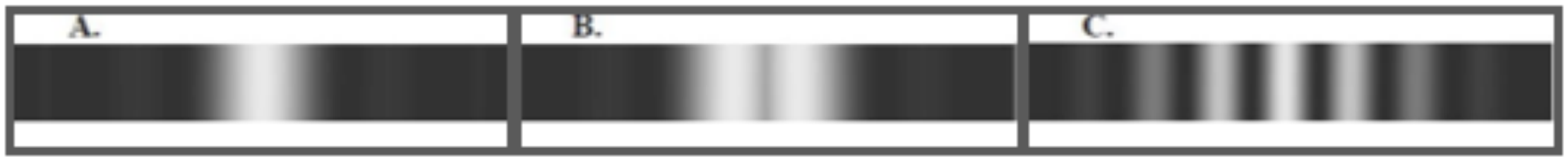

A through C represent some patterns which might be observed. If you think none is appropriate, answer D. Which pattern would you expect to observe when…

**6. …***marbles* pass through the double opening?

**7. …***electrons* pass through the double slit?

**TABLE 5.I.** Pre- and post-instruction student responses (in percent) to items 6 & 7 from the content survey used in the modern physics course from Fall 2010. The standard error on the proportion for all cases was ~5% (Pre: N=110; Post: N=88). Students shift from expecting *similar* behavior for marbles and electrons, to expecting *different* behavior.

| **PRE (N=110)** | **A** | **B** | **C** | **D** |
|---|---|---|---|---|
| Marbles | 15% | 60% | 21% | 5% |
| Electrons | 14% | 51% | 35% | 1% |
| **POST (N=88)** | **A** | **B** | **C** | **D** |
| Marbles | 9% | 86% | 2% | 2% |
| Electrons | 0% | 12% | 88% | 0% |

We note first that, prior to instruction, the most popular response to both items was the same (B), indicating that most students expected *similar* behavior for both electrons and marbles in similar situations. These responses are consistent with our hypothesis that incoming students have particle-like expectations for the behavior of all matter. These items saw dramatic shifts in post-instruction student responses, indicating that most students expected *different* behavior for macroscopic marbles and microscopic electrons by the end of the course. The class average on the pre-instruction content survey was 46% (+/- 2%), and the average for post-instruction items common to both surveys was 80% (+/- 3%), for a normalized gain of 0.63. [See Appendix C for a complete list of pre- and post-instruction items from the content survey, with an item-by-item summary of student responses.]

As part of their first homework assignment, students were also asked to complete the same online attitudes survey administered in other courses. We summarize below the distribution of pre-instruction student responses (in terms of agree/neutral/disagree) for the entire class, along with the full responses of four select students. These four students (denoted as A, B, C & D) were not selected in order to be representative of any one group of students; their responses instead serve to demonstrate typical pre/post differences in student reasoning, even when overall responses to survey items (agreement or disagreement) had not changed. Their specific homework submissions and exam responses will later serve to address the question of whether topics that are new to the curriculum are accessible to students. Closely following these four students also allows for a more detailed exploration of the curriculum's influence on some of the aspects of student thinking that had been targeted, without making unnecessary extrapolations to the entire class population. Together, these two types of pre-instruction data will allow us to establish a baseline on incoming student perspectives.

**1.** It is possible for physicists to carefully perform the same measurement and get two very different results that are both correct.

| PRE | Agree | Neutral | Disagree |
|---|---|---|---|
| **Class (N=94)** | **0.65** | **0.13** | **0.22** |

**Student A:** **(Agree)** I feel that no matter how much technology advances or how much we learn, we can never fully understand how the world works and in many cases, we use outcomes of experiments to look at phenomena in different ways that may or may not be entirely correct in the real world. For instance, looking at the behavior of light as both a particle and wave. So, yes, I believe that an experiment came be conducted twice with different outcomes.

**Student B:** **(Agree)** I don't know of any examples, but the fact that quantum physics has some things that seem counter-intuitive and contradict classical physics, it seems that this could be a possibility.

**Student C:** **(Strongly Agree)** What the two physicists are measuring could be highly unstable and sensitive to multiple external stimulus.

**Student D:** **(Strongly Agree)** It is possible for identical measurements to produce different results if that which is being measured can exist in more than one state at the same time. Thus, one would not know whether the subject of the measurement is the object in one state or the other. Interpreting this question differently, one could comment on the fact that the very act of measuring itself introduces new elements into a system, and thus actually changes the outcome of the measurement.

Overall class responses are consistent with prior results, with a strong majority of students agreeing with this statement, though it should be cautioned that students vary greatly in the reasoning behind their responses, as seen in Chapter 2. Students A, B & D have all invoked quantum phenomena in their agreement with this statement, with varying degrees of sophistication. Student D speaks of quantum superposition and the physical influence of observation; Student A notes that light may be described as both particle and wave; Student B simply states his impression that quantum mechanics will challenge his intuition, so perhaps this statement might be true. Student C's reasoning is more consistent with the idea that chaotic, hidden variables may randomly influence the outcomes of similar measurements – an attitude commonly seen in pre-instruction responses.

**2.** The probabilistic nature of quantum mechanics is mostly due to physical limitations of our measurement instruments.

| PRE | Agree | Neutral | Disagree |
|---|---|---|---|
| **Class (N=94)** | **0.46** | **0.32** | **0.22** |

**Student A:** **(Neutral)** I really don't know enough about quantum theory to make a guess on that. However, even our most basic assumptions about the world have sometimes proven to be incorrect and quantum seems to involve so much theory that we can never really be sure if it actually functions the way physicists think it does or if we are coming up with theories that just fit what we find without even seeing the entire picture.

**Student B:** **(Strongly Agree)** I believe that in the future, we would be able to make more accurate and exact assertions due to technological advances and would not need to rely on probability.

**Student C:** **(Neutral)** I don't know what quantum mechanics is yet.

**Student D:** **(Strongly Disagree)** The probabilistic nature of quantum mechanics is a fundamental property of the system. For example: it is impossible to define (not just measure) the position and momentum of an electron at the same instant in time (Heisenberg's uncertainty principle). Thus, the uncertainty exists outside of the instruments used to try to measure those properties. (I would really, really like to learn the math behind these statements!)

Responses here were more varied than with the first statement, though agreement amongst the class is moderately favored; the individual responses range from strong agreement to strong disagreement. The two neutral responses from Students A & C indicate a similar tentativeness due to a lack of knowledge about quantum mechanics; Students A & B both echo a common perception that knowledge in science is itself tentative, and that profound progress (technological or theoretical) often upends previously held beliefs. In contrast, Student D identifies quantum uncertainty as fundamentally different from experimental uncertainty, explicitly stating there are limits not only on the precision of simultaneous measurements, but also on simultaneous quantum descriptions of incompatible observables (position and momentum, specifically).

**3.** When not being observed, an electron in an atom still exists at a definite (but unknown) position at each moment in time.

| PRE | Agree | Neutral | Disagree |
|---|---|---|---|
| **Class (N=94)** | **0.72** | **0.09** | **0.19** |

**Student A:** **(Strongly Agree)** An electron is a fundamental piece of an atom, though it moves extremely fast, so at any point in time, yes it does occupy a position being that it is matter.

**Student B:** **(Strongly Agree)** An electron is a particle, and every particle has a definite position at each moment in time.

**Student C:** **(Agree)** Because I have been told this since 9th grade.

**Student D:** **(Agree)** An electron occupies a single definite position at any given point in time. It is only our measurement (and thus knowledge) of that position at any given point in time that is subject to the Heisenberg uncertainty principle, where either the position or the momentum of the electron may be measured to a high level of precision, but not both.

As expected, a strong majority of incoming students chose to respond in a manner that would be consistent with realist expectations; all four of our individual students were in agreement that atomic electrons should exist as localized particles. The reasoning invoked by Students A & B is consistent with our hypothesis of *classical attribute inheritance* – electrons, as a form of matter, have the same properties as macroscopic particles, including a localized position at all times; Student A further implies that the uncertainty in an electron's position can be attributed to its swift, chaotic motion about the nucleus – similar to the hidden-variable style reasoning of Student C in response to the first survey item. Here, Student C makes an appeal to authority: the idea of localized electrons conforms to what he has been told in school since (presumably) first learning about the structure of atoms. Most interestingly, Student D is explicit in asserting the realist belief that electrons always exist as localized particles; he claims it is our simultaneous *knowledge* of incompatible observables that is constrained by the uncertainty principle.

**4.** I think quantum mechanics is an interesting subject.

| PRE | Agree | Neutral | Disagree |
|---|---|---|---|
| Class (N=94) | 0.85 | 0.13 | 0.02 |

**Student A:** **(Strongly Agree)** From the examples I have heard and some of the theory, I think quantum mechanic is very interesting.

**Student B:** **(Strongly Agree)** I think that I'm going to learn that what I would think is correct is actually completely incorrect. Plus, it just sounds cool.

**Student C:** **(Neutral)** I don't know yet.

**Student D:** **(Strongly Agree)** Quantum mechanics fascinates me precisely because it is so counterintuitive. I want to challenge my perception of the world, and there are few better ways to do that than QM. It is also interesting to me because I am much more used to physics on very large, indeed cosmic scales. It is especially interesting to see how the world of the unimaginably tiny and the world of the unimaginably large interact...

**5.** I have heard about quantum mechanics through popular venues (books, films, websites, etc...)

| PRE | Agree | Neutral | Disagree |
|---|---|---|---|
| Class (N=94) | 0.61 | 0.19 | 0.20 |

**Student A:** **(Strongly Agree)** [BLANK]

**Student B:** **(Strongly Disagree)** I'm completely out of the "physics loop" and hope to get more into it in this class!

**Student C:** **(Agree)** I read part of the book In Search Of Schrodinger's Cat by John Gribbin

**Student D:** **(Agree)** In high school, I got a taster of quantum mechanics through generalized physics books, but nothing more in depth. Beyond that, my knowledge of quantum mechanics is limited, and comes primarily from several online lectures by MIT (through itunes U) and several from the University of Madras (posted on youtube).

The reported incoming interest in quantum mechanics for these students is somewhat higher (85%) than is usually seen in a course for engineering majors (~75%; and comparable with typical incoming attitudes among physics majors; see Chapter 6). Because we have no other reason to believe that students from this semester would be any different from previous populations for this course, we can only speculate that this is what resulted from all four members of the instruction team hyping the excitement of quantum physics on the first day of lecture. And as with previous introductory modern physics courses, a majority of students reported having heard *something* about quantum mechanics before enrolling in the course, which underscores the fact that incoming students are not entirely blank slates when it comes to quantum physics, and will certainly bring *some* preconceived notions into the course – incoming students will have impressions about the nature of quantum mechanics, positive or negative.

With these considerations in mind, it seems reasonable to conclude that this particular group of students held incoming attitudes and beliefs that were typical of similar student populations (as measured by these specific assessments), and to assert that any aggregate student outcomes associated with the implementation of this curriculum should not be attributed to there being anything unique about this particular class. We have no means of objectively assessing just how representative Students A – D are of the overall student population, but it is our subjective opinion (based on the experience of studying a wide variety of modern physics offerings over the span of several academic years) that Students A, B & C represent several points of view that are common among incoming engineering students. It is also our subjective assessment that Student D holds a relatively sophisticated view on quantum mechanics for an incoming student, but one that could be categorized as *Realist/Statistical* in light of his explicit belief in the localized nature of electrons, and his assertion that the uncertainty principle constrains simultaneous *knowledge* of incompatible observables.

### II.B. Lecture Materials

In their end-of-term reflective essays, the topics most frequently cited by students as having influenced their perspectives on quantum physics were the single-quanta experiments with light and/or matter, and so we focus our attention here on one lecture (#20) primarily devoted to the experiments performed by Aspect, et al. (as described in Chapter 1). Topics from immediately prior to this lecture included: hidden variables, Local Realism, and indeterminacy in quantum mechanics. [Lectures 18-19] Our primary objectives for this lecture were for students to understand how two similar experimental setups can lead to dramatically different observations; to highlight the differences between observation and inference (interpretation of experimental facts); and to provide experimental evidence that contradicts the simultaneous attribution of particle and wave characteristics to photons.

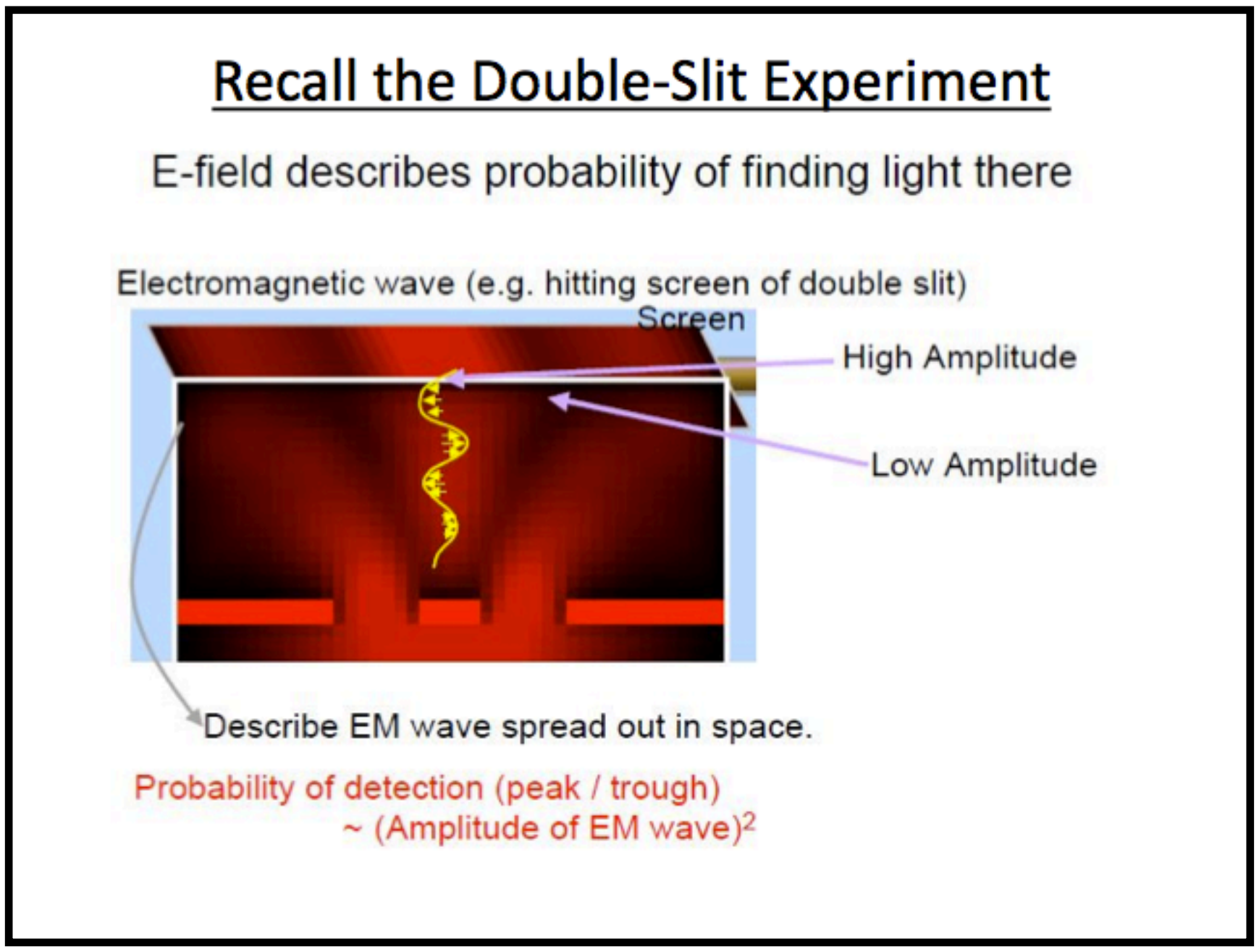

**L20.S01.** Students are reminded that the double-slit experiment can be performed with single photons, which are detected individually. Wave intensity is associated with the probability for detection, which is greater in locations where there is constructive interference.

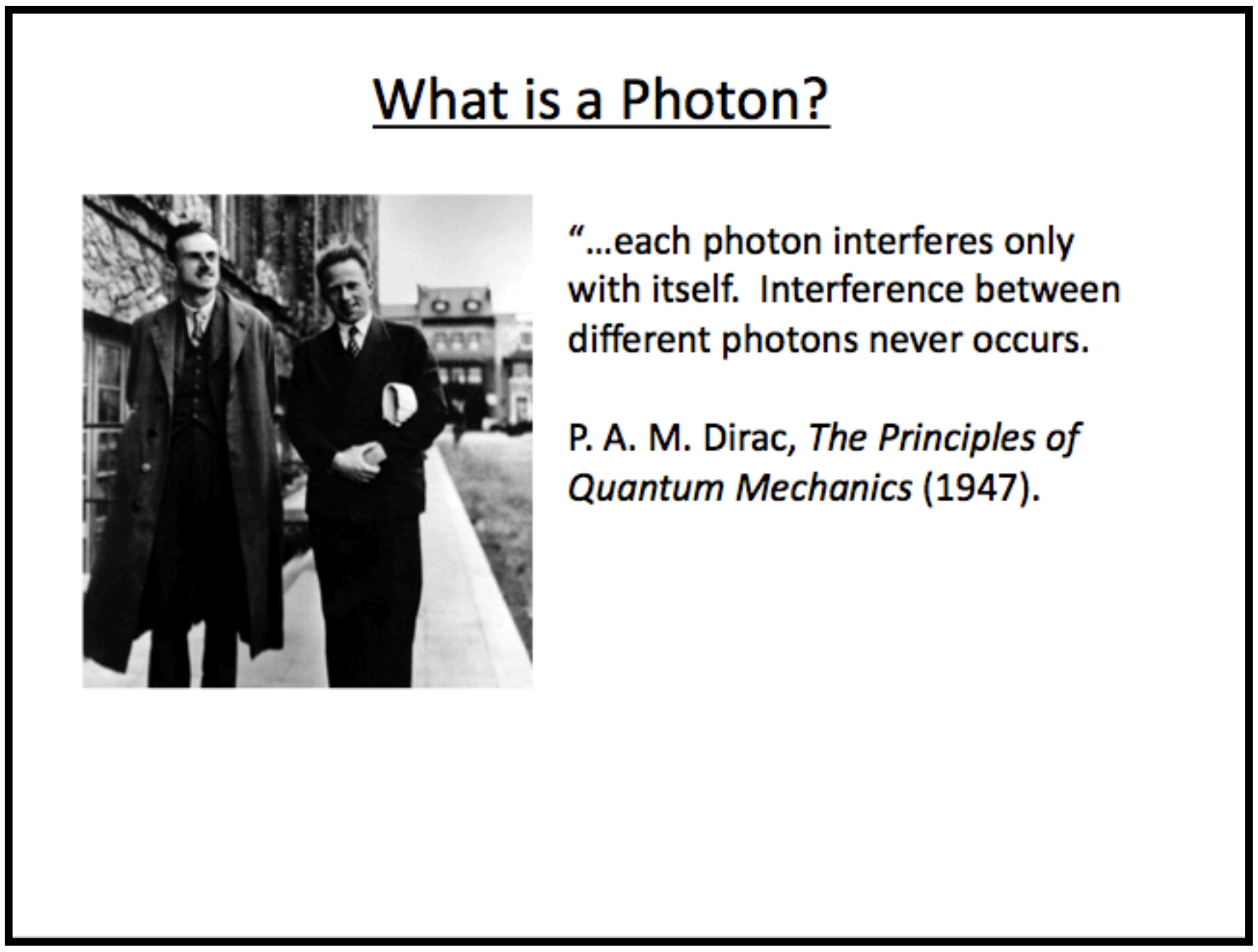

**L20.S02.** Dirac offered his interpretation of these kinds of experiments long before they could be realized: each photon must pass through both slits as a delocalized wave and interfere with itself; interference with other photons does not occur.

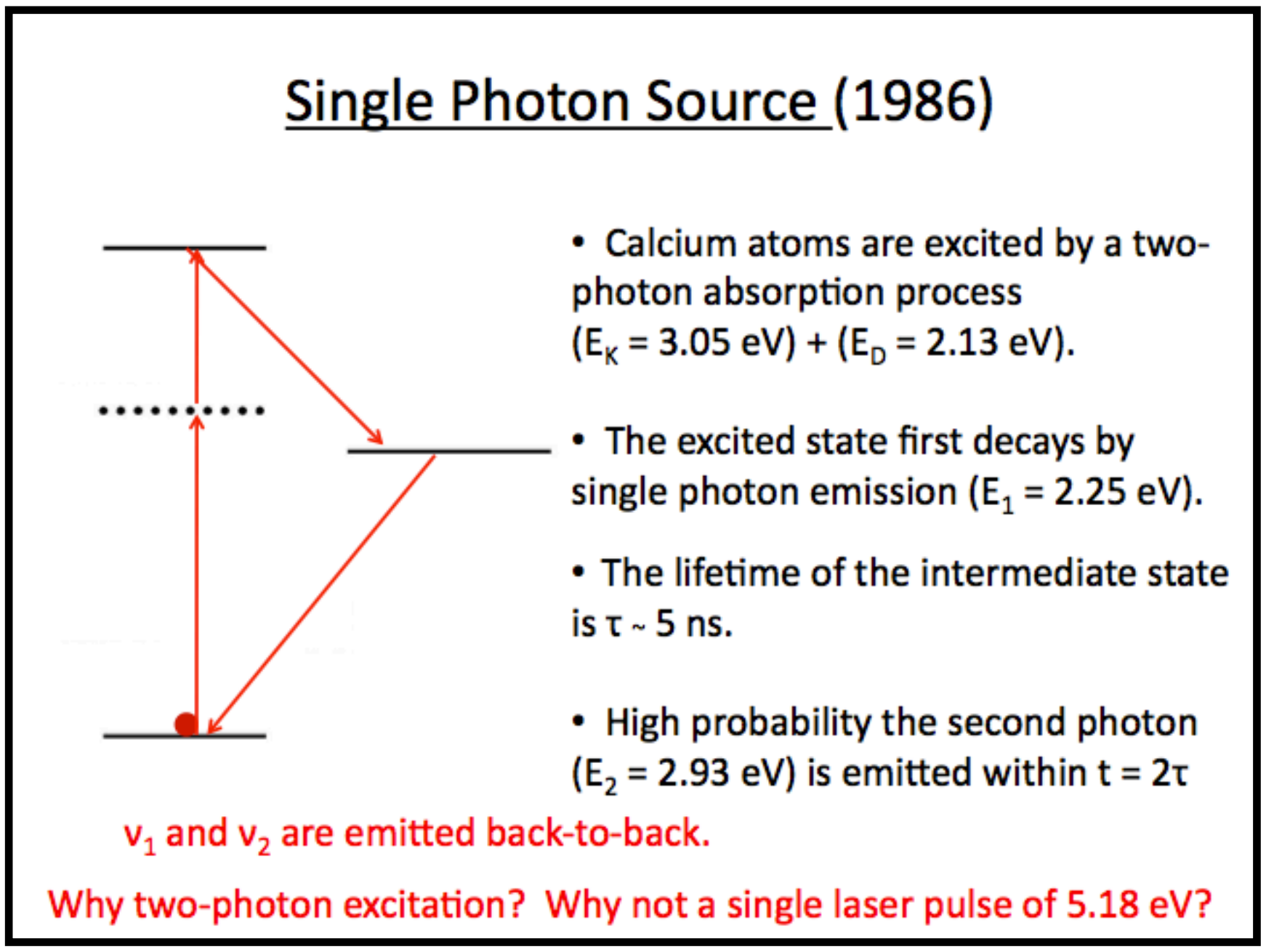

**L20.S03.** A "single-photon source" was employed by Aspect in 1986 to explore the wave-particle duality of photons. The two-step excitation process greatly reduces the intensity of the source, where the goal is to detect only specific photons: ones emitted in a two-step, back-to-back de-excitation process.

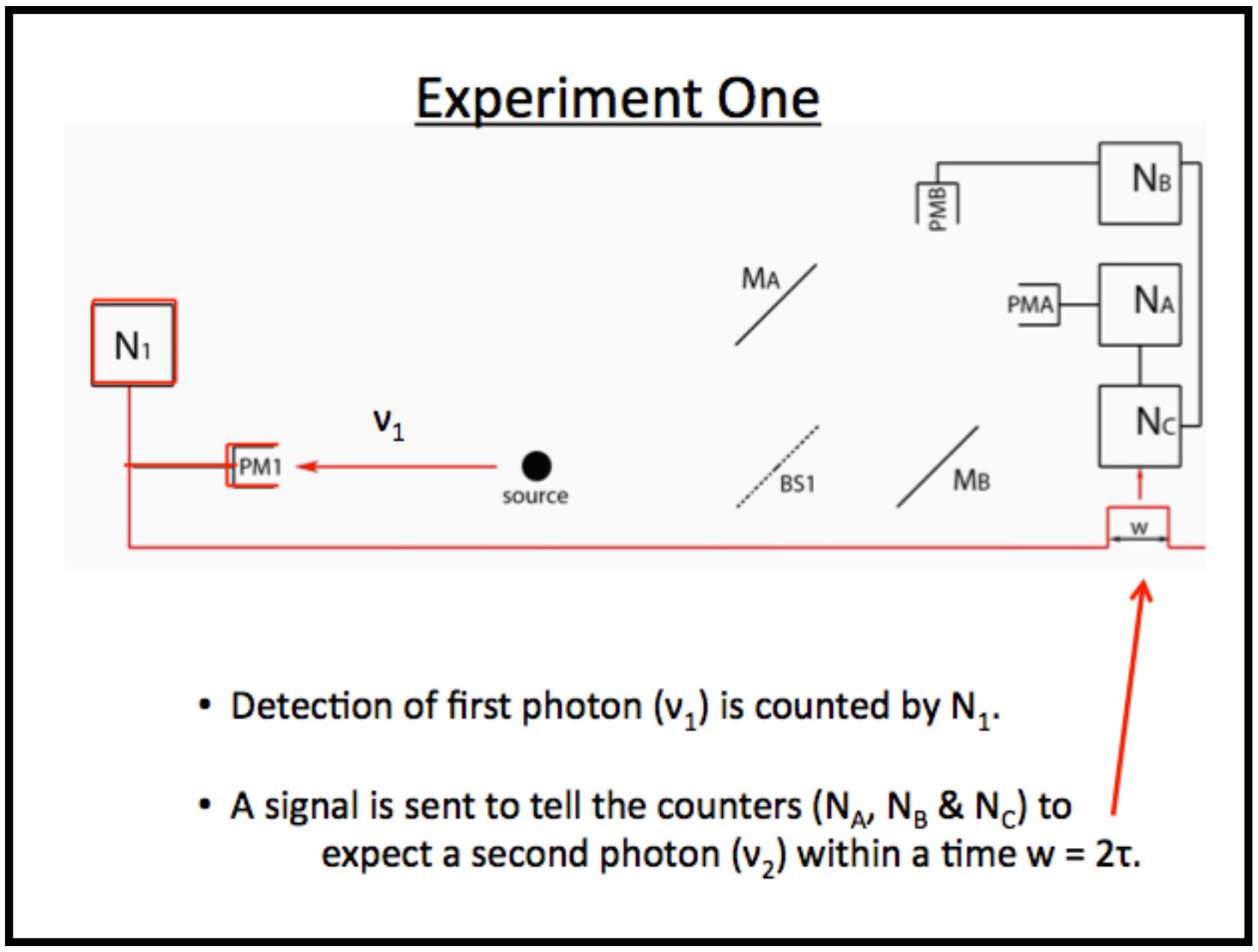

**L20.S04.** Detection of the first photon ($ν_1$) in PM1 signals the counters to await the detection of the second photon ($ν_2$). The gate is open for a time equal to twice the lifetime of the intermediate state, making it highly probable that a second photon was emitted during that time period.

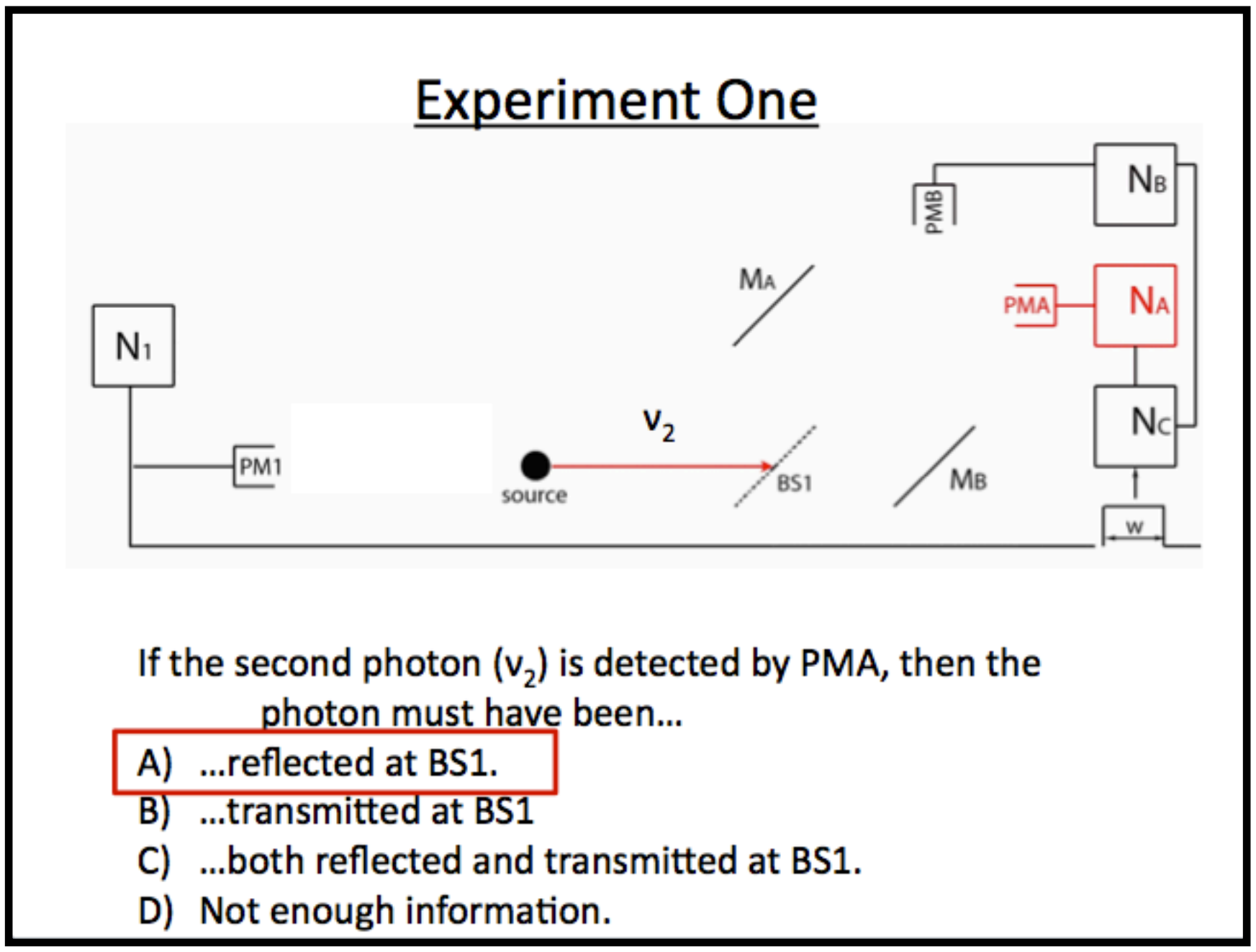

**L20.S06.** With a little discussion, students quickly converged on (A). The greatest student confusion arose from the schematic nature of the diagram, which implies there is open space between BS1 and the two photomultipliers, which might allow for a photon reflected at BS1 to reach PMB. This question helps check that students understand the purpose of each element of the experimental setup (beamsplitter, mirror, detector, counter).

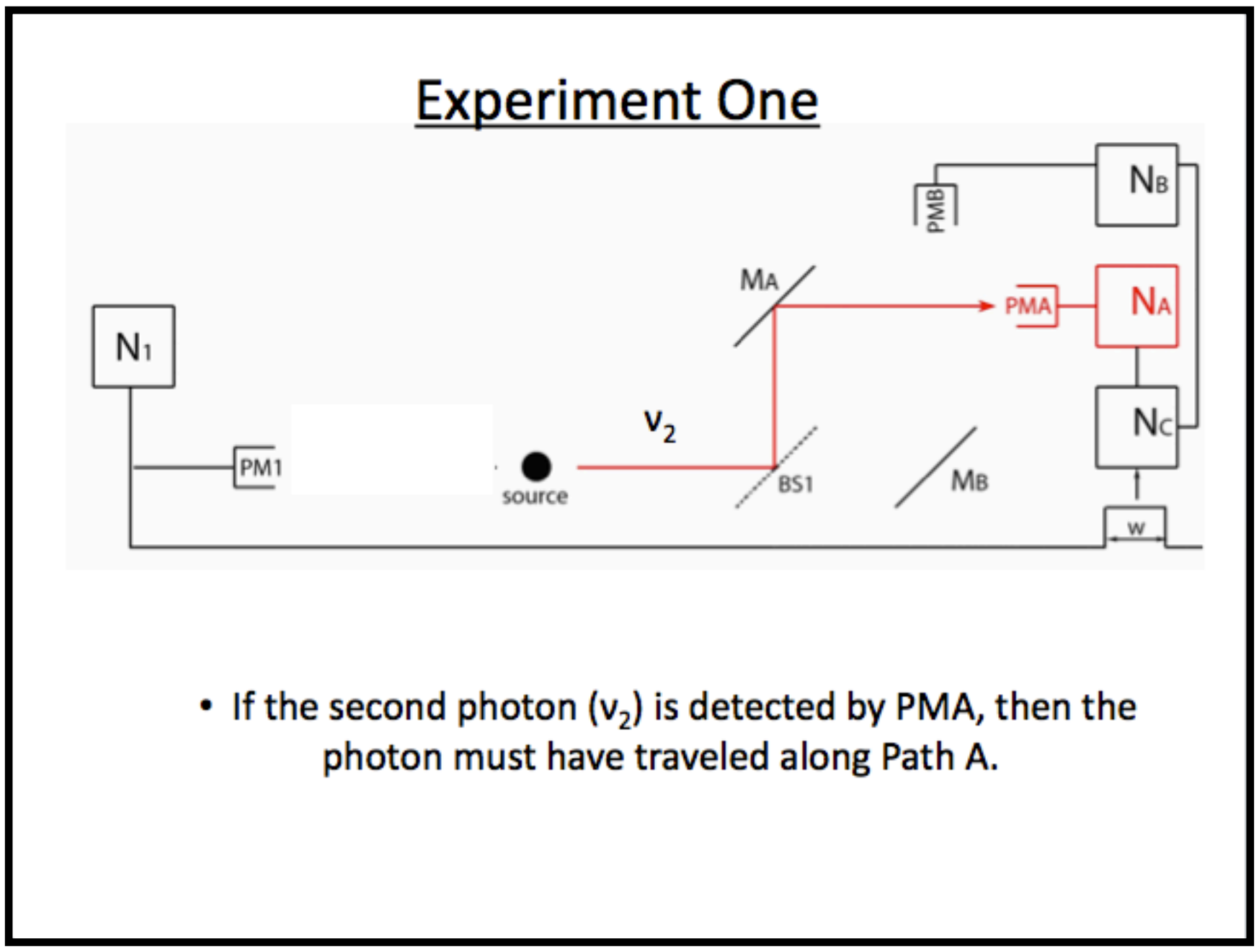

**L20.S08.** Following the previous concept test and subsequent discussion, it should now be clear there is only one path by which a photon might reach PMA: it must have traveled along Path A, by reflection at BS1, and reflection again at $M_A$.

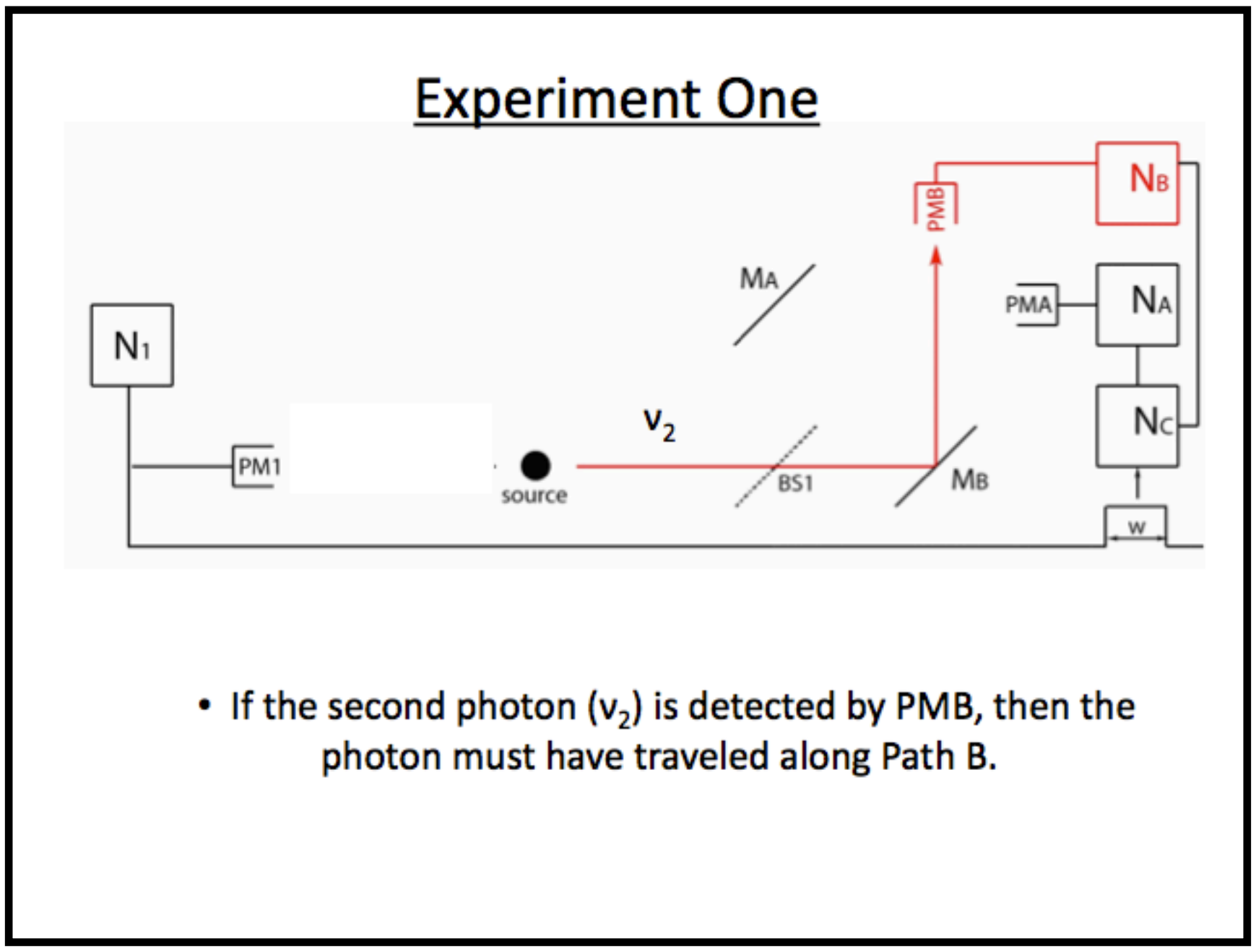

**L20.S09.** The same is true for a detection in PMB: the photon can only have traveled via Path B, by transmission at BS1, and reflection at $M_B$.

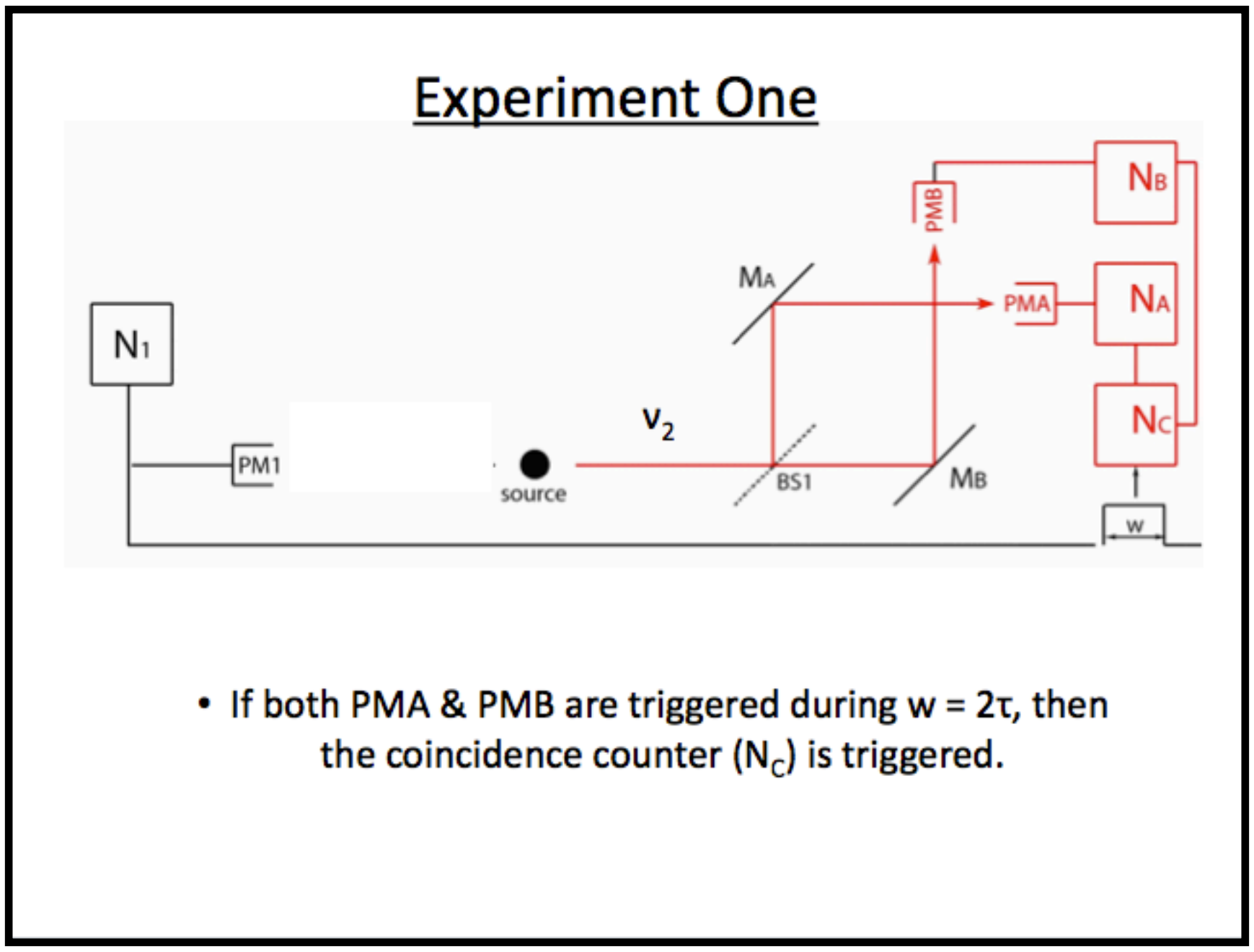

**L20.S10.** It is still possible to record a detection in both photomultipliers during the short time the gate is open – when this happens, the coincidence counter ($N_C$) is triggered. How often this happens has implications for how we interpret the behavior of photons.

## Anti-Correlation Parameter

- Need some kind of measure of how often PMA & PMB are being triggered at the same time.
- Let $\alpha \equiv \frac{P_C}{P_A P_B}$
- $P_A$ is the probability for $N_A$ to be triggered.
- $P_B$ is the probability for $N_B$ to be triggered.
- $P_C$ is the probability for the coincidence counter ($N_C$) to be triggered (both $N_A$ and $N_B$ during $t = 2\tau$).

**L20.S11.** We first require some kind of statistical measure of how often the two photomultipliers are firing together versus firing separately. This can be defined in terms of a ratio of the counting rates per unit time for each of the three counters, or equivalently, in terms of the probability for each of the counters to be triggered during the short time the gate is open.

## Anti-Correlation Parameter

$$\alpha \equiv \frac{P_C}{P_A P_B}$$

- If $N_A$ and $N_B$ are being triggered randomly and independently, then $\alpha = 1$.

  $P_C = P_A \times P_B$ which is consistent with:
  - Many photons present at once
  - EM waves triggering $N_A$ & $N_B$ at random.
- If photons act like particles, then $\alpha \geq 0$.

  $P_C = 0$ when particles are detected by PMA or by PMB, but not both simultaneously.
- If photons act like waves, then $\alpha \geq 1$.

  $P_C > P_A \times P_B$ means PMA and PMB are firing together more often than by themselves ("clustered").

**L20.S12.** If the detectors were to fire together more often than not (implying that the photon energy is coherently split at BS1 and deposited equally in both detectors – wave behavior), then α should be ≥ 1. It will be less than one if the detectors tend to fire independently (implying each detection corresponds to a single photon following a single path – particle behavior).

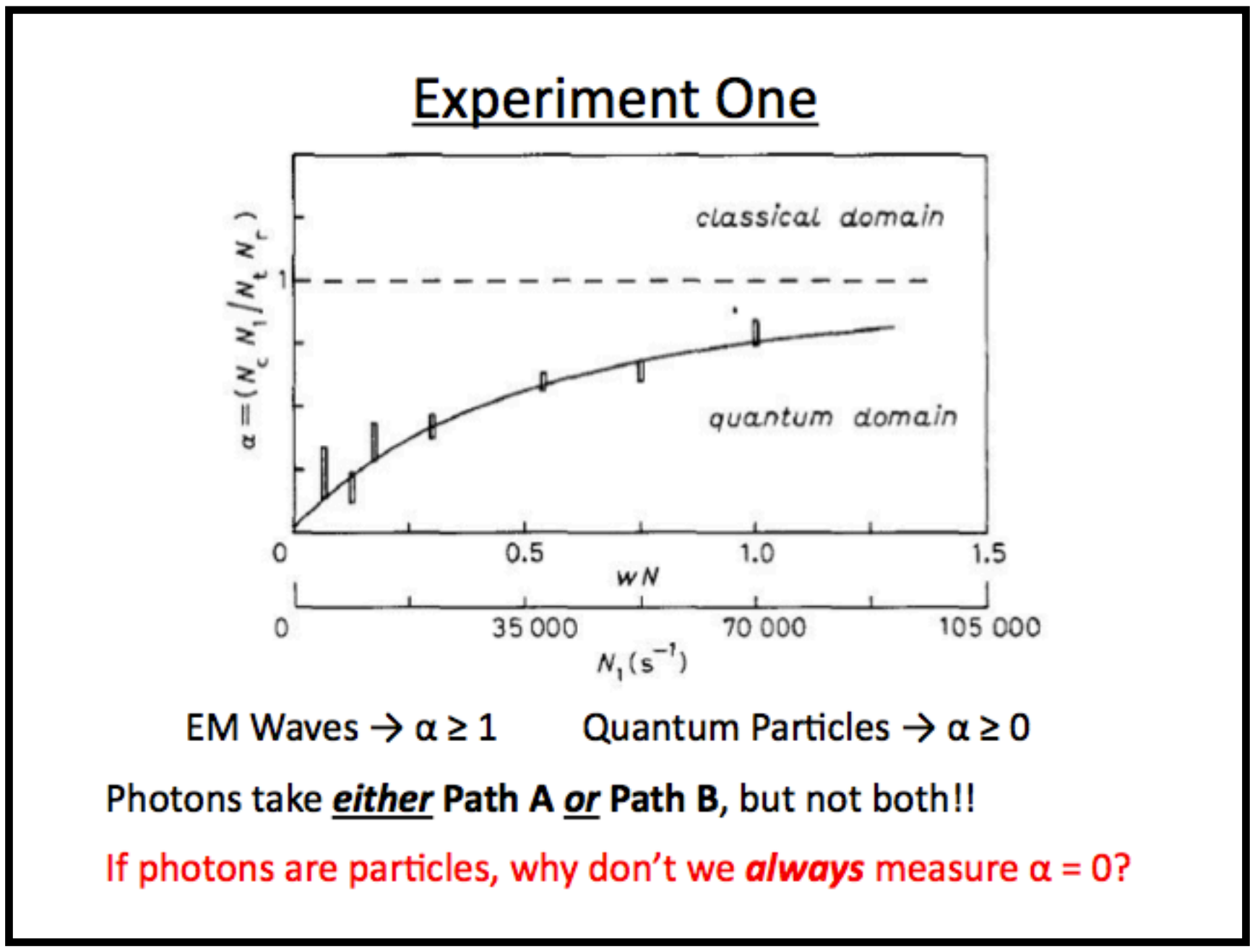

**L20.S13.** At all intensities (but particularly at low counting rates), the two photomultipliers fire independently more often than not. Since only a single path leads to either of the two detectors, we interpret these results as indicating that each photon is either reflected or transmitted at BS1, but not both.

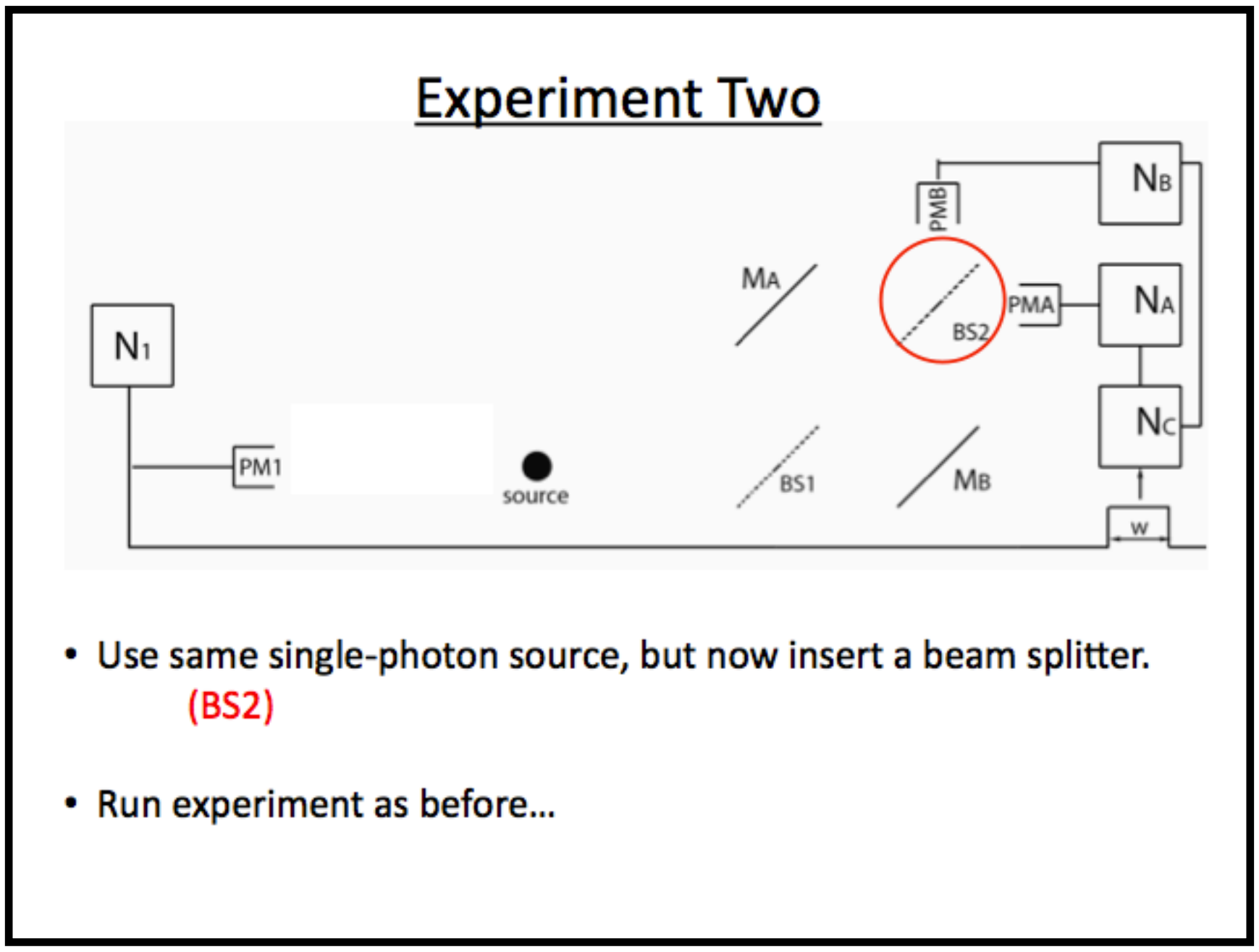

**L20.S14.** The experiment is run again as before, except that now a second beam splitter (BS2) is inserted into the path. It is impossible to determine which-path information through a detection in either one of the photomultipliers.

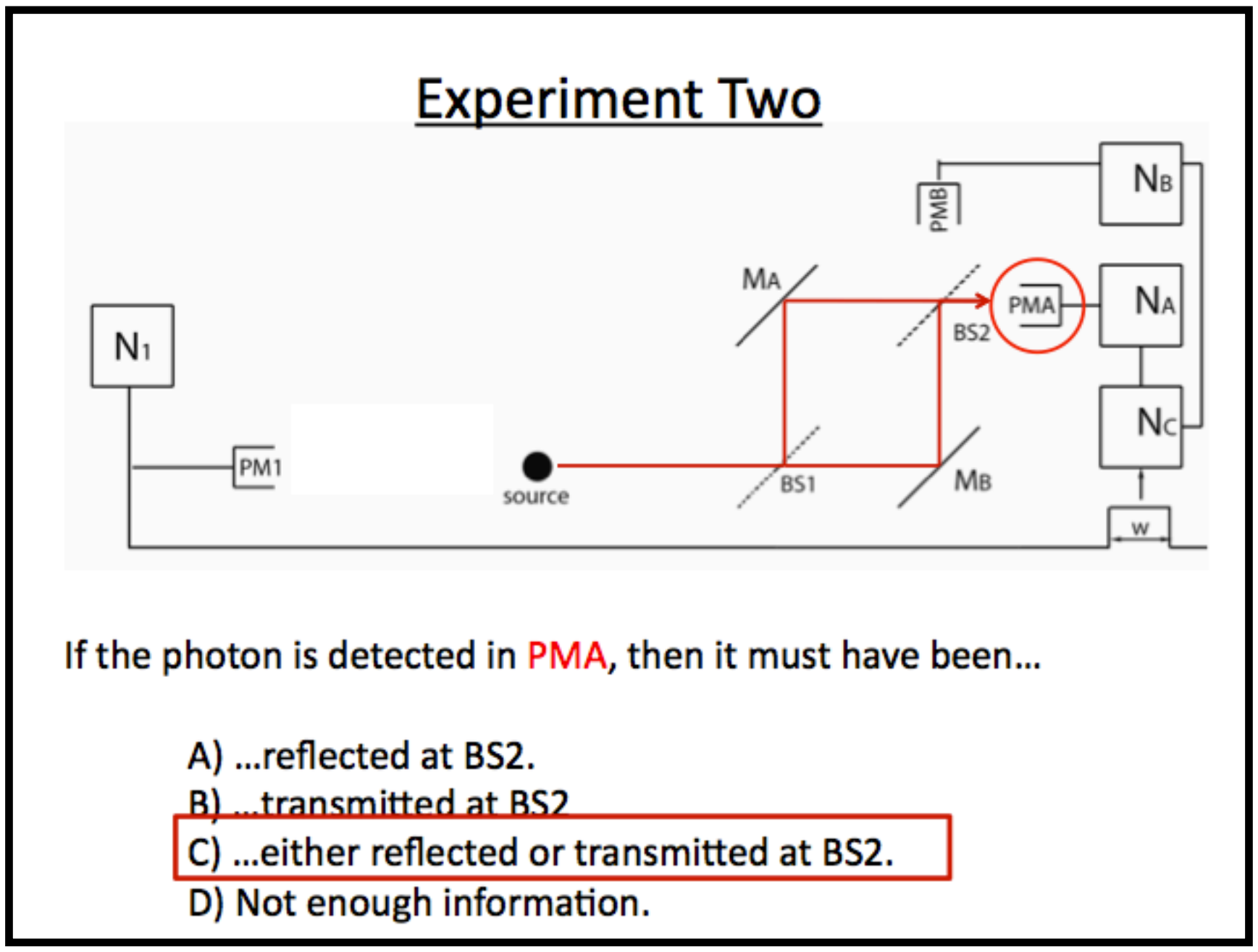

**L20.S15.** With the second beam splitter in place, there are now multiple paths a photon could take to be detected in a given photomultiplier. Students were quick to converge on (C) as the correct answer, with less discussion than was required for the first concept test.

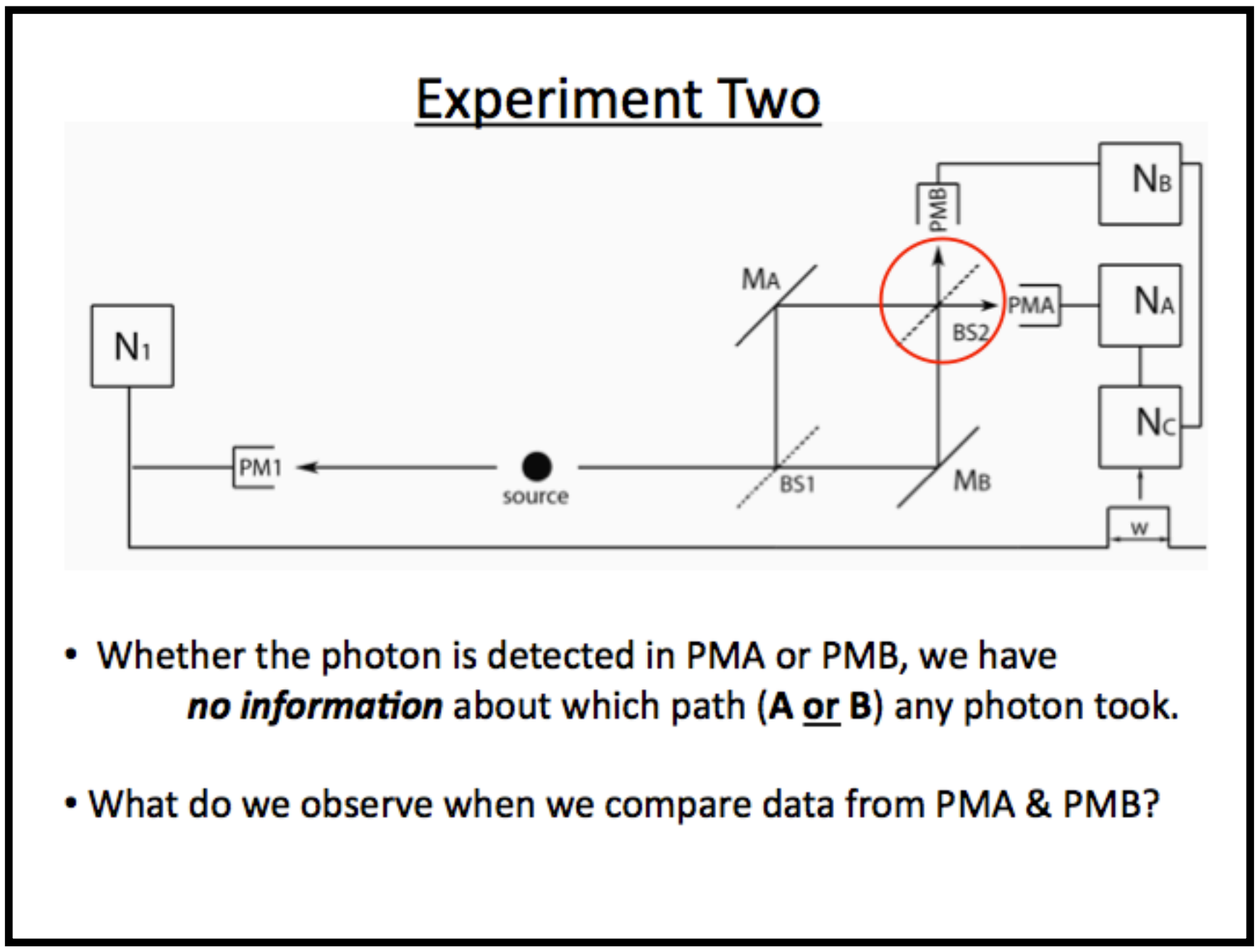

**L20.S16.** Detection in either of the photomultipliers yields no information about which path a photon must have taken to get there. With multiple possible paths, interference effects are expected, though not of a kind previously encountered by students. In this case, interference is observed by comparing the counting rates in the two detectors.

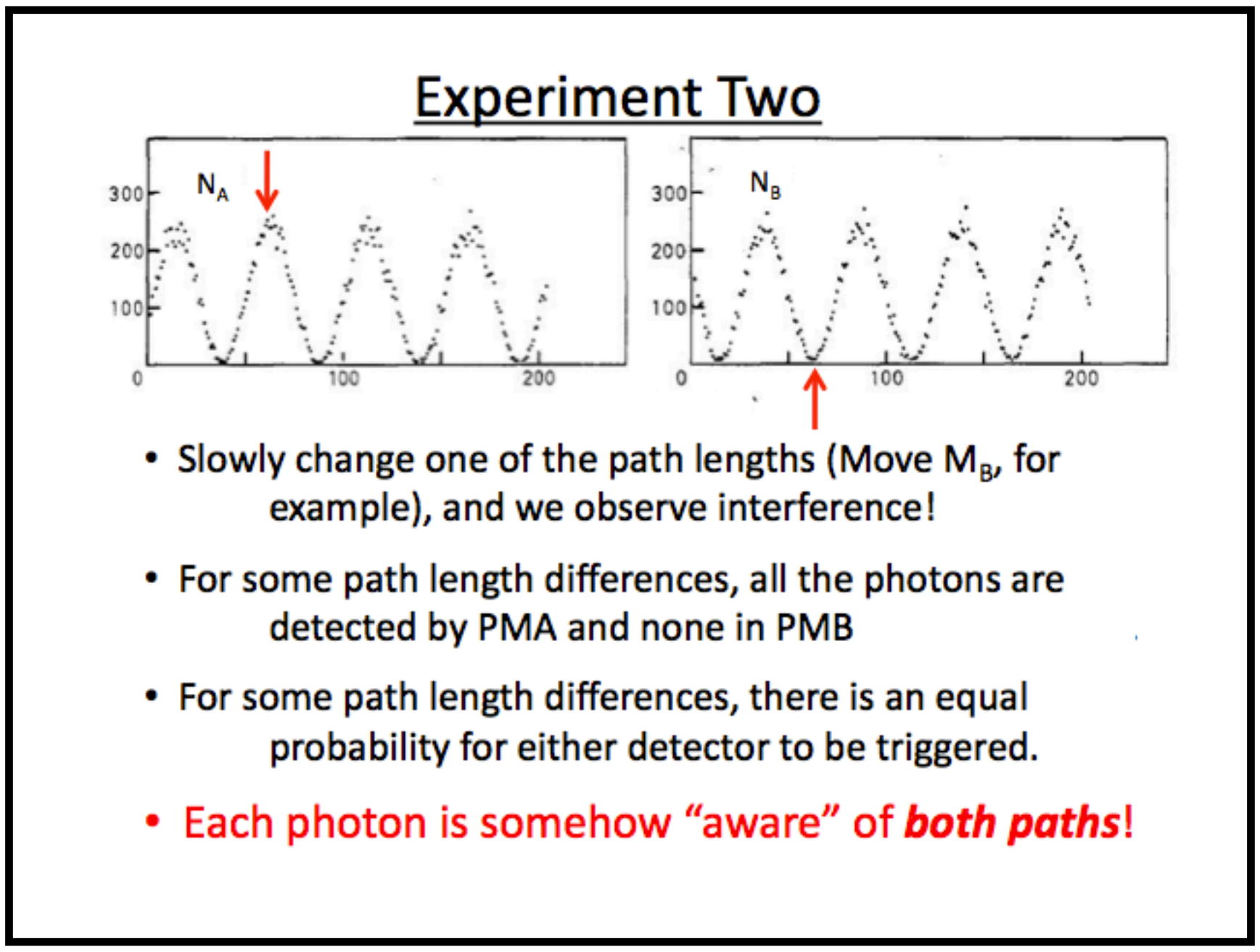

**L20.S17.** According to quantum mechanics, the counting rates in the two detectors are oppositely modulated according to the difference in path lengths between A & B. Photons that had only taken Path A should not be affected by any changes made to Path B, yet their behavior at BS2 is determined entirely by the relative lengths of **both paths**.

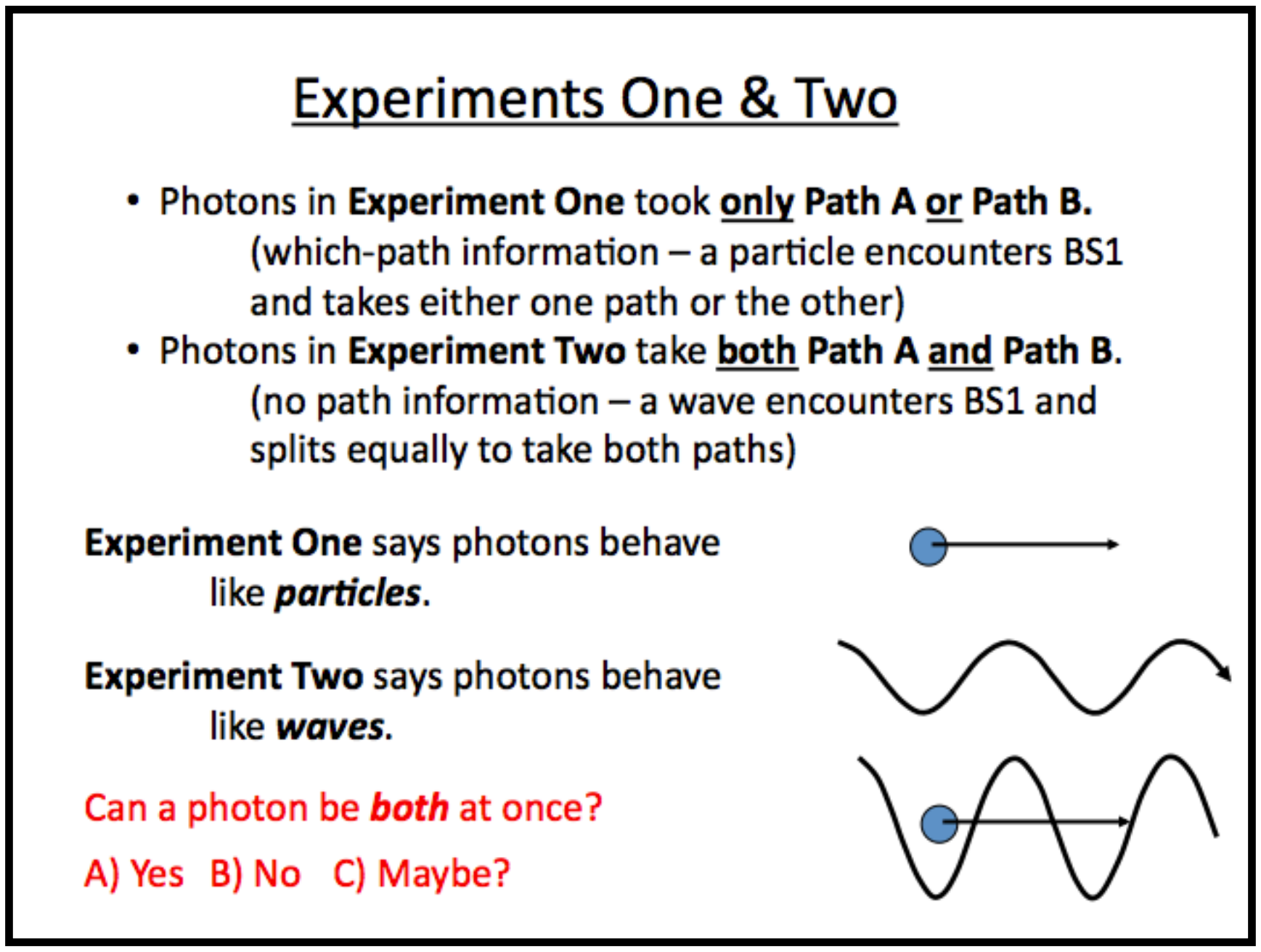

**L20.S18.** An explicit connection is made between the interpretation of a photon's behavior at BS1 and the which-path information available to the experimenter. There was no favored response to this moderately rhetorical clicker question, which was meant more to get students thinking and talking about the validity of our interpretations, and to prime them for the delayed-choice experiment.

## The "Conspiracy" Theory

How can the photon "know" whether we are conducting Experiment One or Experiment Two when it encounters BS1?

Perhaps each photon "senses" the entire experimental apparatus and always behaves accordingly.

Can we "trick" a photon into acting like a particle when it should act like a wave, or the other way around?

Suppose we let the photon enter the apparatus when the second beam splitter is absent (particles take one path or the other), but then insert the beam splitter at the last moment.

**L20.S19.** The question is now whether we can make a change in the experimental apparatus after the photon has encountered the first beam splitter; in such a way that we go from conducting Exp. 1 to Exp. 2 (or vice-versa) after the photon has already "decided" how to behave when it encounters BS1.

## Experiment Three

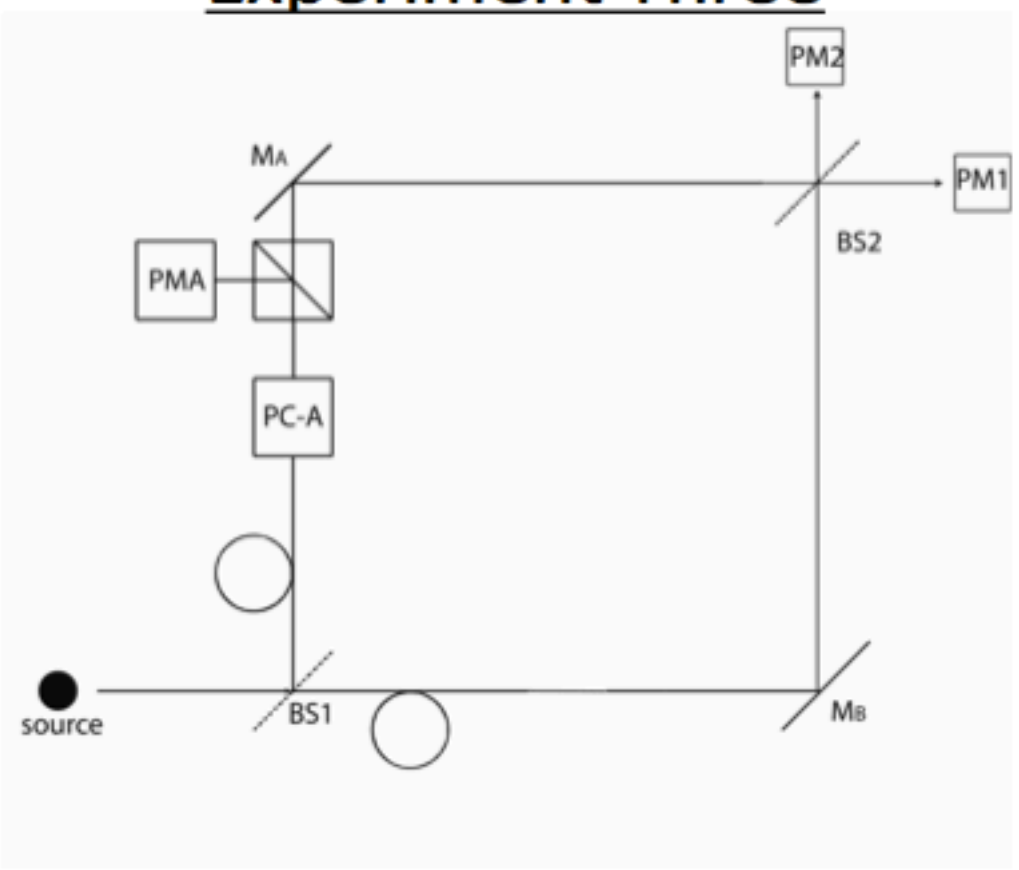

Impossible to physically remove actual beam splitter at the necessary speed, but this type of experimental setup is equivalent to what we just described.

**L20.S20.** While structurally similar to the first experiment, this one utilizes a laser tuned to such low intensity that there is, on average, only one photon per pulse.

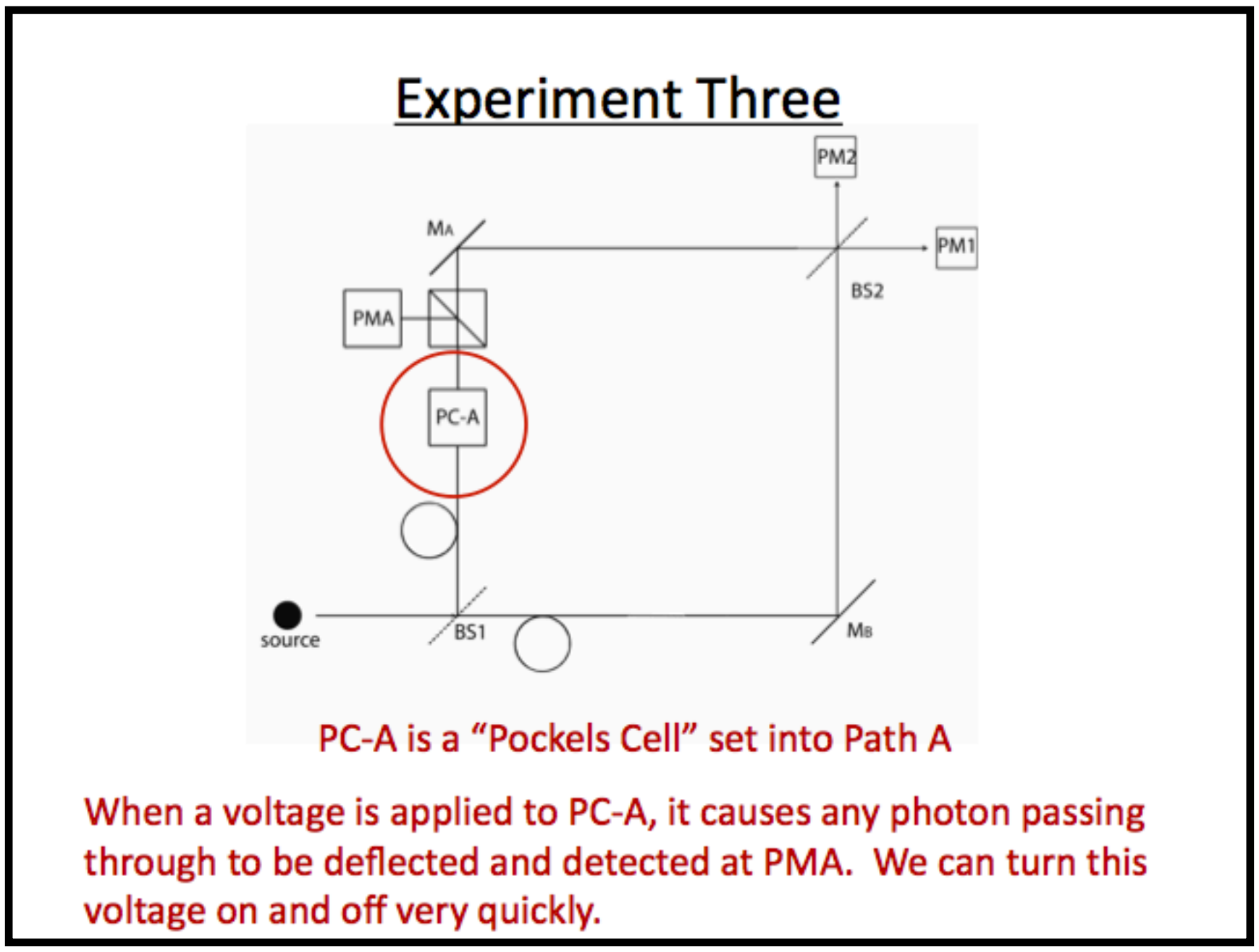

**L20.S21.** When a voltage is applied to the Pockels cell it rotates the plane of polarization of a photon such that it is always reflected by the Glans prism into PMA. This voltage can be turned on and off with a frequency that is sufficient for the time resolution of this experiment.

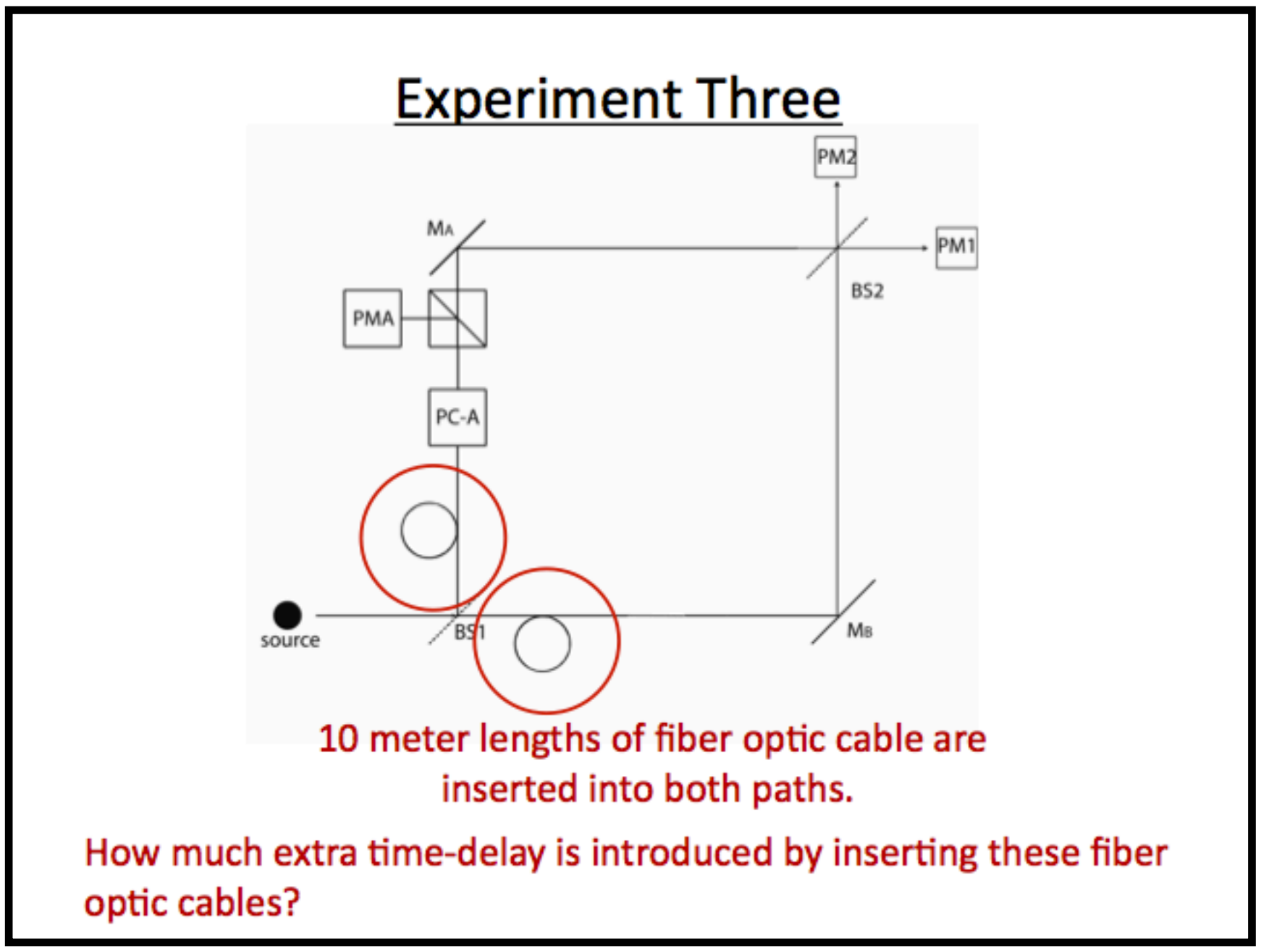

**L20.S22.** Two 10-meter lengths of fiber optic cable introduce a transit delay time of about 30 nanoseconds after the photon has encountered the first beam splitter.

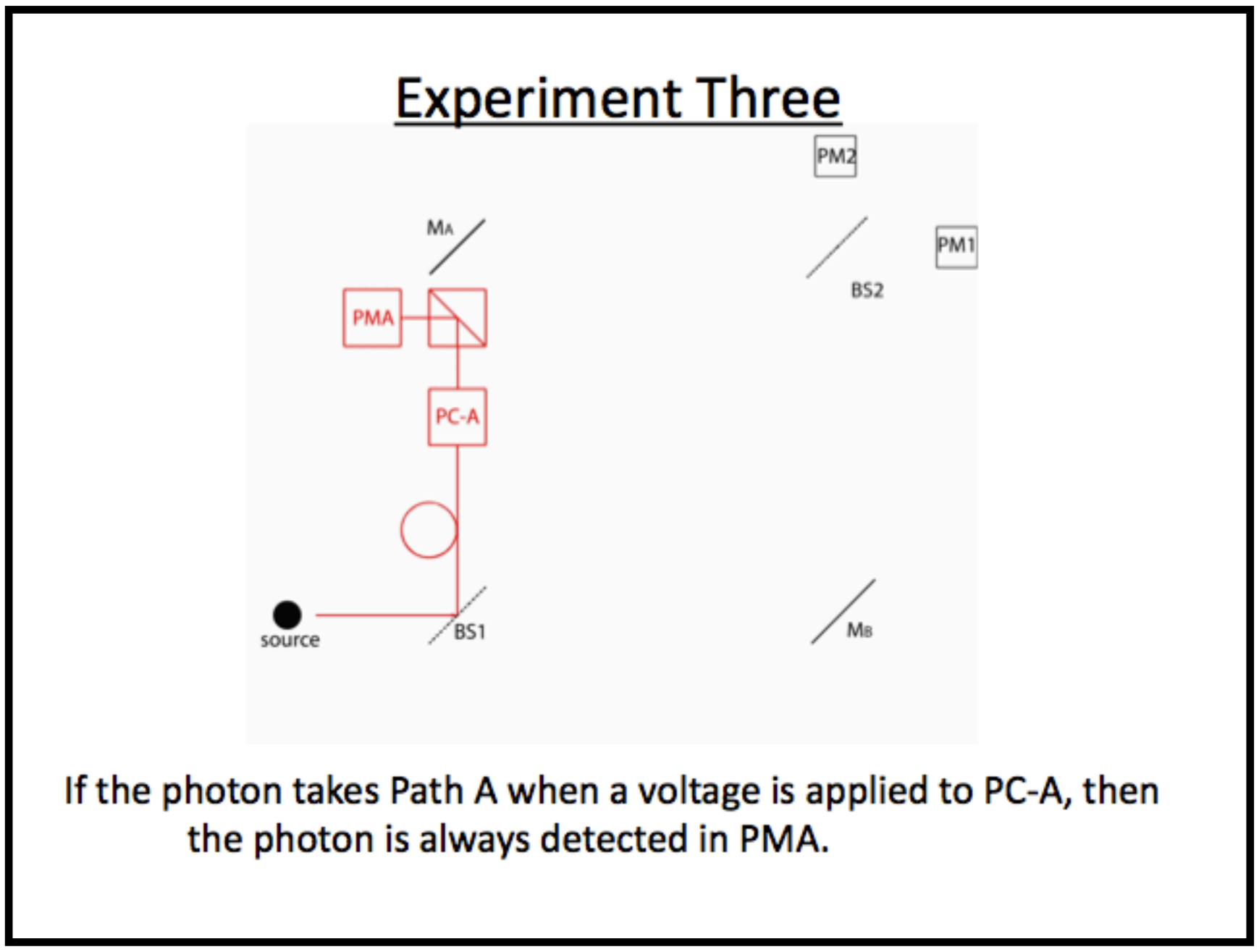

**L20.S22.** With a voltage applied to the Pockels cell (PC-A), any photon reflected at BS1 will be detected in PMA with 100% probability.

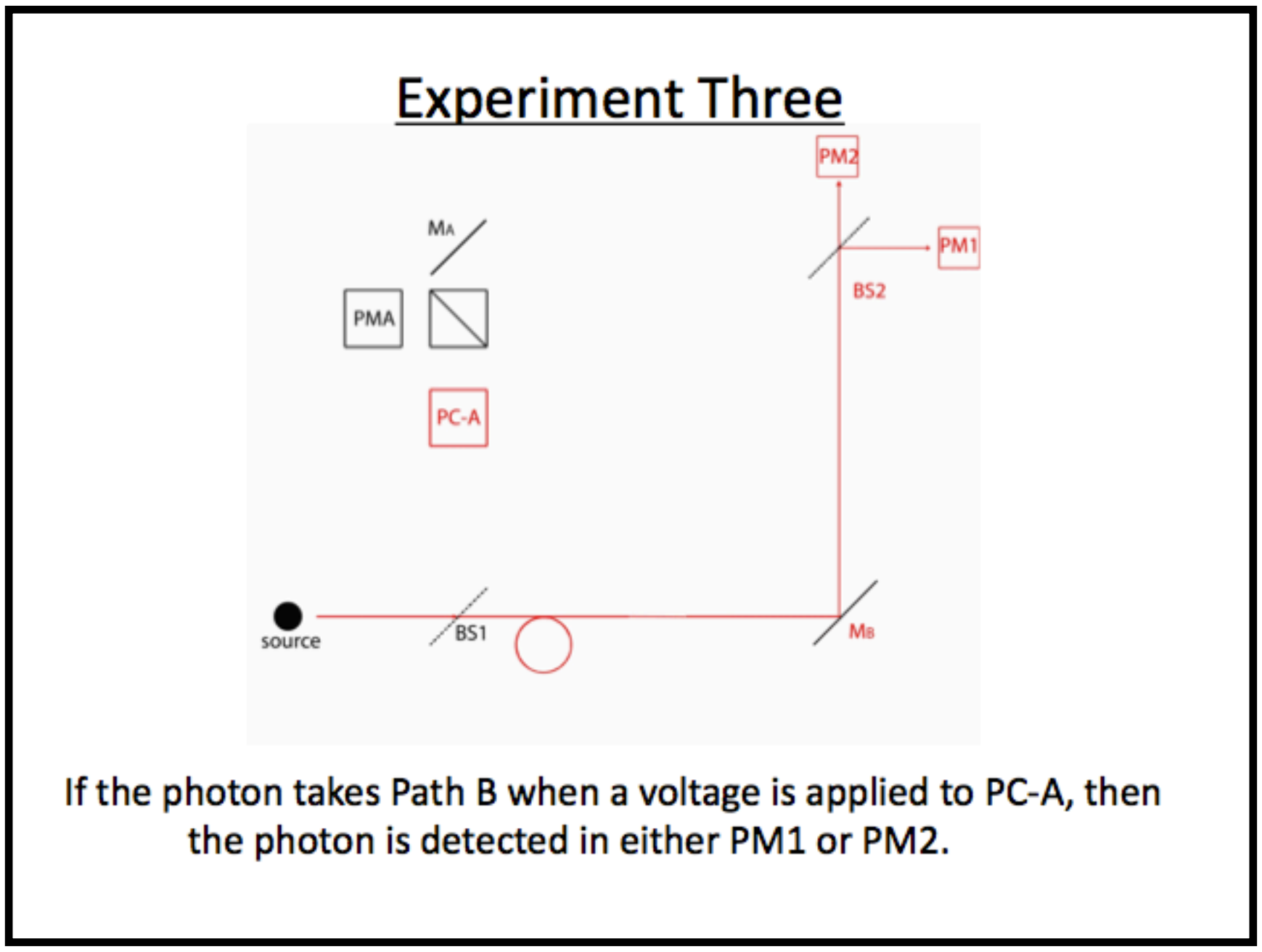

**L20.S24.** With a voltage applied to the Pockels cell, any photon transmitted at BS1 will have an equal likelihood of being detected in either PM1 or PM2.

## Experiment Three

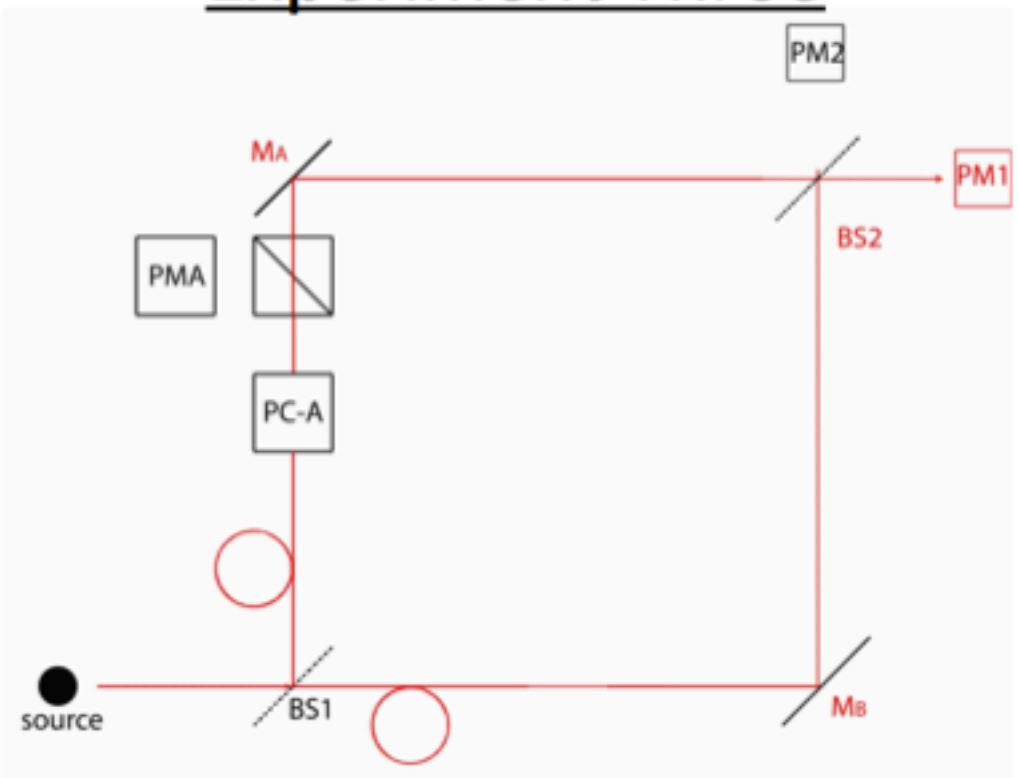

When ***NO*** voltage is applied to PC-A, then both Paths A & B are possible.

Fix the mirrors so that photons are always detected in PM1
(Interference)

**L20.S25.** With no voltage applied to the Pockels cell, both Path A and Path B are open to the photon. Since self-interference is possible in this case, we may fix the mirrors so that every photon is detected only in PM1 when no voltage is applied.

## Experiment Three

***No voltage*** applied to PC-A:

Both paths are possible and photon is detected in PM1 only.

TWO PATHS = INTERFERENCE

***Voltage applied*** to PC-A:

If photon detected in PMA <-> Photon took Path A

If photon detected in PM1 or PM2 <-> Photon took Path B

ONE PATH = NO INTERFERENCE.

**L20.S26.** This may form the basis of a quantum epistemological tool for students. With only one path possible, no interference effects should be seen (photons behave like particles); two (or more) paths means interference should be visible (photons behave like waves).

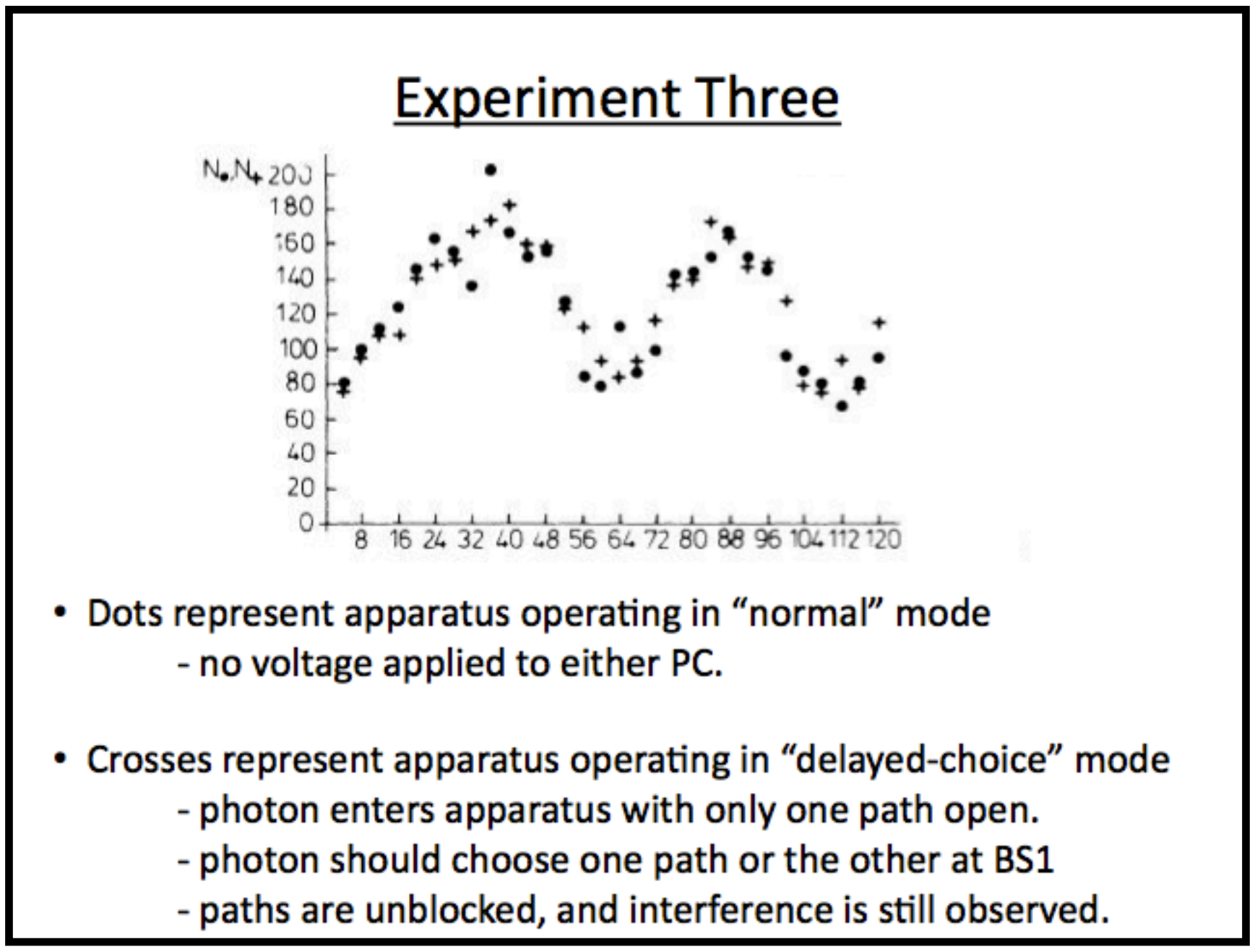

**L20.S27.** When the experiment is run, interference is seen whenever two paths were open to the photon, and absent when only one path was open, regardless of which was the case at the time the photon encountered the first beam splitter.

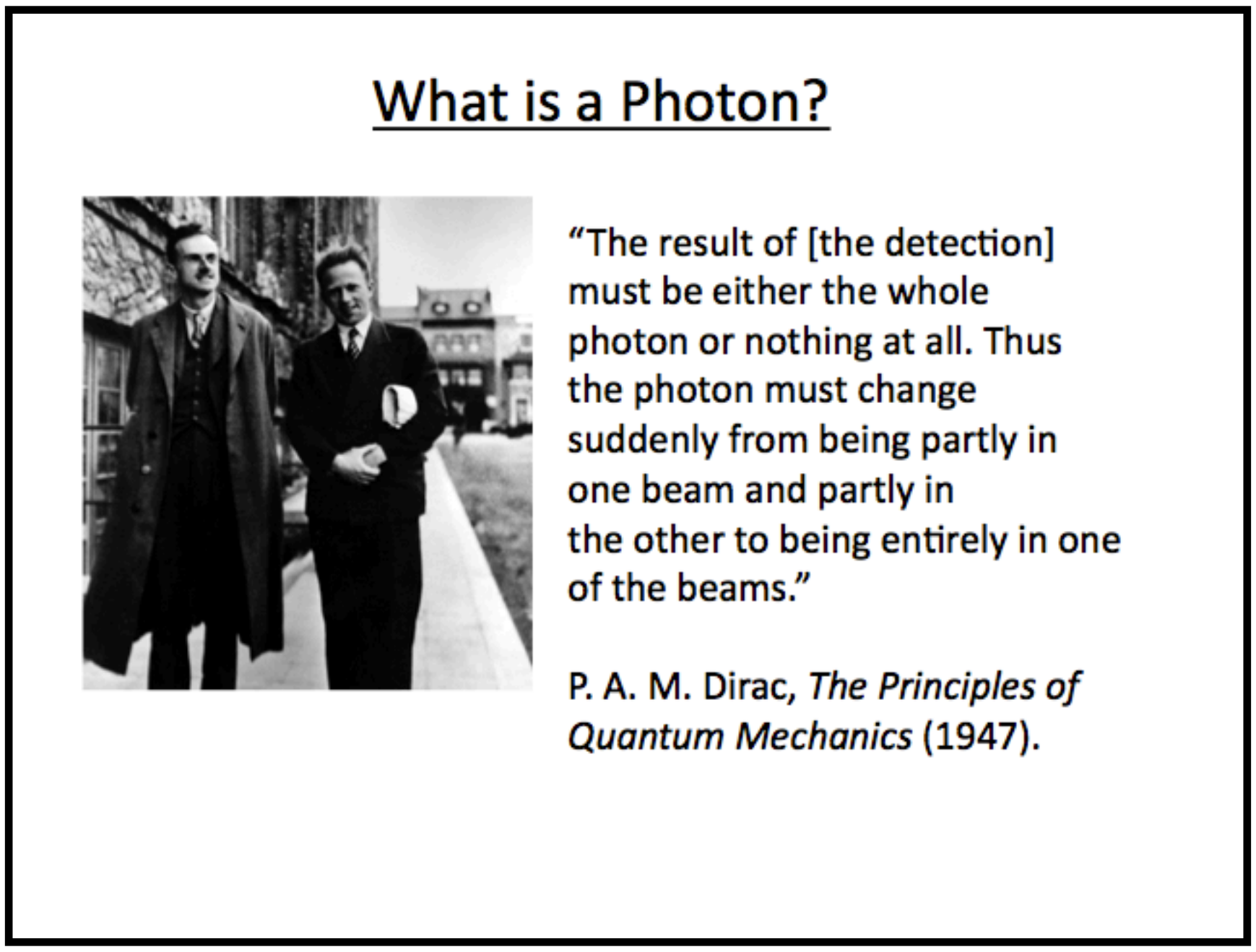

**L20.S28.** Dirac's interpretation suggests the photon is coherently split into a superposition state at the first beam splitter in all three experiments, and then collapses to a point when (randomly) interacting with a detector.

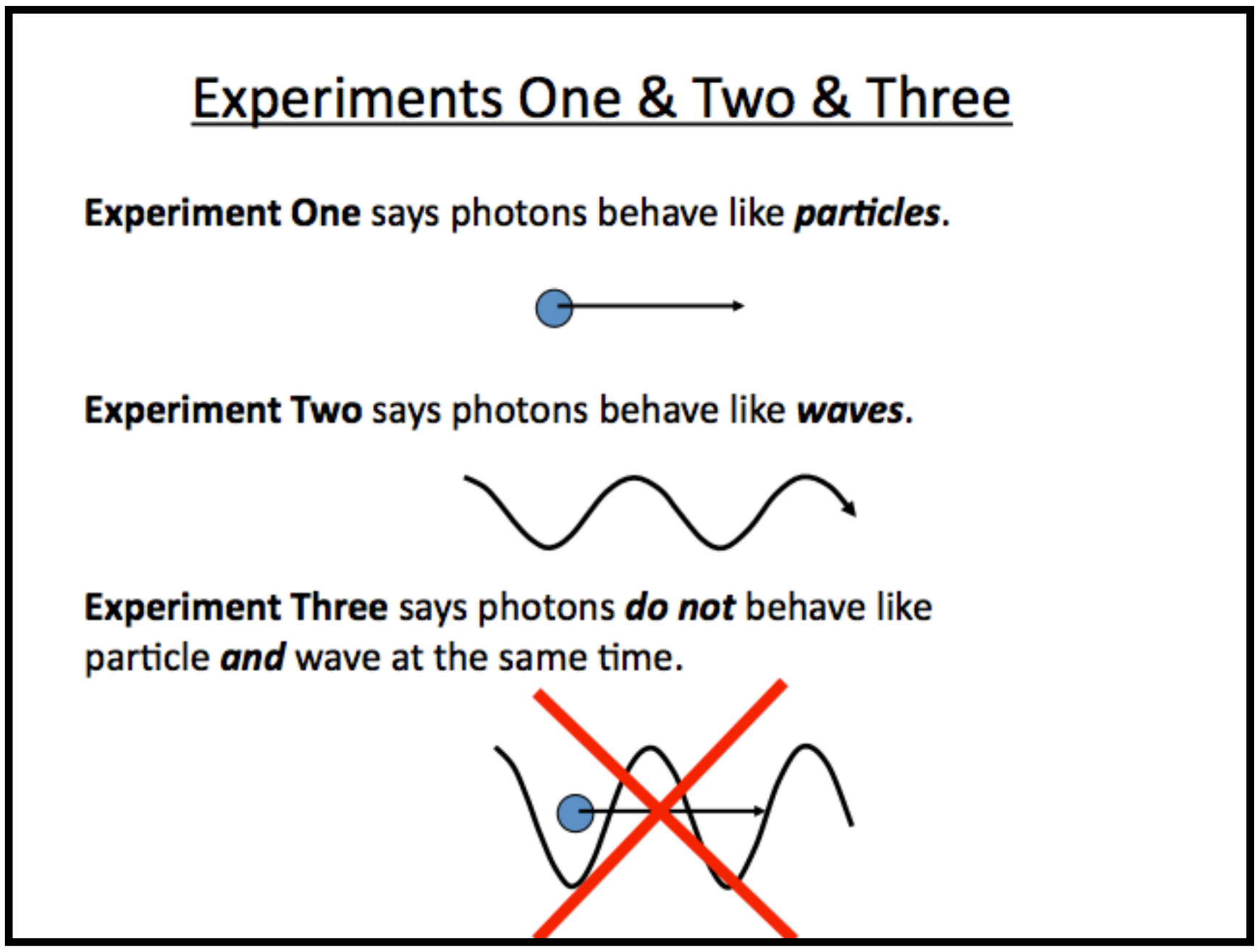

**L20.S29.** It is hoped that, by this point, students will not just accept, but conclude for themselves that photons never exhibit both types of behaviors simultaneously.

## II.C. Homework

Informal interviews with modern physics instructors have revealed a common concern that a proper treatment of the interpretive aspects of quantum theory requires an understanding and knowledge base that is beyond the reach of most introductory students, and may only open a Pandora's Box of unanswerable questions that could ultimately lead to more confusion. We believe, however, that this end result is more likely in a course where students are not given the requisite tools, including language, to fully appreciate the arguments against classical thinking in quantum contexts; and that it is precisely these kinds of open questions in physics that inspire the excitement and imagination of our students. We also believe that realist preferences are common, and so intuitive to students that many are simply lacking a name for beliefs they had already articulated in their pre-instruction survey responses. The full implications of nonlocality in quantum phenomena might not be appreciated by every student, but most will readily agree that a measurement performed on one of two physically separated systems should have no influence on the outcome of a measurement performed on the second. We wish to address here just how accessible some of the formal definitions of concepts associated with *Local Realism* are to students, following their discussion in class and in the assigned reading. [16]

One of the homework essay questions from Week 7 asks students to articulate their own understanding of the terms *realism*, *locality*, and *completeness*, and to provide some examples of *hidden variables*:

**Student A:** To me, realism can be described as the idea that things happen whether someone is there to witness it. For example, if a tree falls in the middle of the woods and there is nothing around to hear it, does it still make a sound? Locality represents an intuition that objects around us can only be directly influenced by other objects in its immediate surrounding. Completeness is a description of the world that is represented by the smallest physical attributes such as particles, electrons, waves, atoms, etc. Completeness describes the complete world as one. A great example of hidden variables is the example referred to in class about 2 socks being put into different boxes, mixed up and sent to opposite sides of the universe. Once you discover the color of one sock, you know the color of the other one... entanglement. These socks are hidden variables until one sock's color is discovered.

**Student B:** Realism is a property in which every measurable quantity exists. In other words, everything is definite, and there is no superposition. The only thing that keeps us from knowing what all the quantities are is our ignorance. Completeness refers to a theory that can describe everything without leaving anything unknown. By this definition, quantum physics is not complete because when we measure a certain quantity such as the projection of the atom in the Z direction, then we can't know its projection in the X direction.

Locality is the concept of being able to relate all actions to actions that occurred before them. For example, locality can describe a car accident – all the events that lead up to the car accident are clear and relate to one another. Bohr's interpretation of entanglement is not local, because we have no way of explaining how the observation of one atom collapses the wave such that the other atom (which would be miles apart) instantaneously is affected.

**Student C:** Locality: Locality of the two particles that are being separated and measured means that in some way the particles are linked to each other. These two linked particles are then able to influence each other with out traveling faster than the speed of light.

Realism: Realism suggests that no quantum superposition exists. If I see a red sock in the classic two socks in box experiment, the sock was red all along and the other sock was blue all along.

Completeness: If the sum total parts of any experiment is known, the outcome can be predicted. There is completeness to an experiment that can always be predicted. Quantum mechanics suggests otherwise.

Hidden Variables: A hidden variable could influence the outcome of an experiment and explain the non-locality of entangled particles. A tachyon is an example of a hidden variable, it is something that can travel faster than the speed of light.

**Student D:** Realism states that a quantity in a measured system has an objectively real value, even if it isn't known. For example, under a realist interpretation, an atom always has a particular spin, we are simply unable to know that spin before we measure it (it is "hidden"). Locality is the concept that there must always be a causative chain in the real world linking two events, in other words, that one object may only effect another by causing a change in its local surroundings that may eventually propagate to cause a change in the second object through its local surroundings. Entanglement appears to violate this principle by allowing two particles to influence the state of each other regardless of their physical separation or the material in-between them. For a physical theory to be "Complete" according to the guidelines set by EPR, it must be able to explain the nature and behavior of everything in physical reality. In this sense, quantum mechanics is not complete; if locality is not to be violated quantum mechanics cannot explain all of the physical properties of a system at the most basic level.

Not surprisingly, the coherence of Student D's overall response indicates a solid understanding of each of these terms, not only individually, but also in how they relate to each other in making up EPR's argument for the incompleteness of quantum mechanics. Student B's responses are also satisfactory, and a careful reading reveals his continued preference for realist notions: his specific choice of language implies that an atom can indeed have a definite spin projection along multiple axes, and that our quantum mechanical knowledge of the system is therefore incomplete. Student A's definition of *completeness* seems not far off the mark, though his last statement on the matter is somewhat vague – does he mean that a complete theory consists of a complete description of everything in the universe, or that a complete theory describes everything as a complete and undivided whole? Student C's ideas about *completeness* are linked with determinism: knowing all of the relevant variables would make the outcomes of measurements predictable. In defining *locality*, Student C actually describes a state of *entanglement*, though he later correctly refers to entanglement as being *non-local* in his description of hidden variables. He is also correct in asserting that, should tachyons exist, their unknown presence may have some hidden influence on the outcome of measurements, but we consider it preferable that students focus their attention on more concrete examples of hidden variables (such as position or momentum), as opposed to exotic, hypothetical phenomena.

Fortunately, this was not the last opportunity for students to wrestle with the meaning of these terms, and all that they imply. During Weeks 6-8, students responded each week to an online reading quiz, which merely asked them to pose (at least) one question about something (anything) from the reading assignments for that week. These questions were then compiled and used as seeds for an online class discussion forum. For each of the subsequent five weeks, students were asked to make a contribution to the discussion board as part of their weekly homework assignments, but no efforts were made to verify their participation, and students were free to put as little or as much effort as they liked into their postings. Student postings were anonymous (even to the instructors), though we could verify at the end of the semester how many postings a student had made. Figure 5.1 shows how a large majority (> 75%) of students made at least four contributions to the discussion board during the course of the semester (the few students who made zero contributions are not shown).

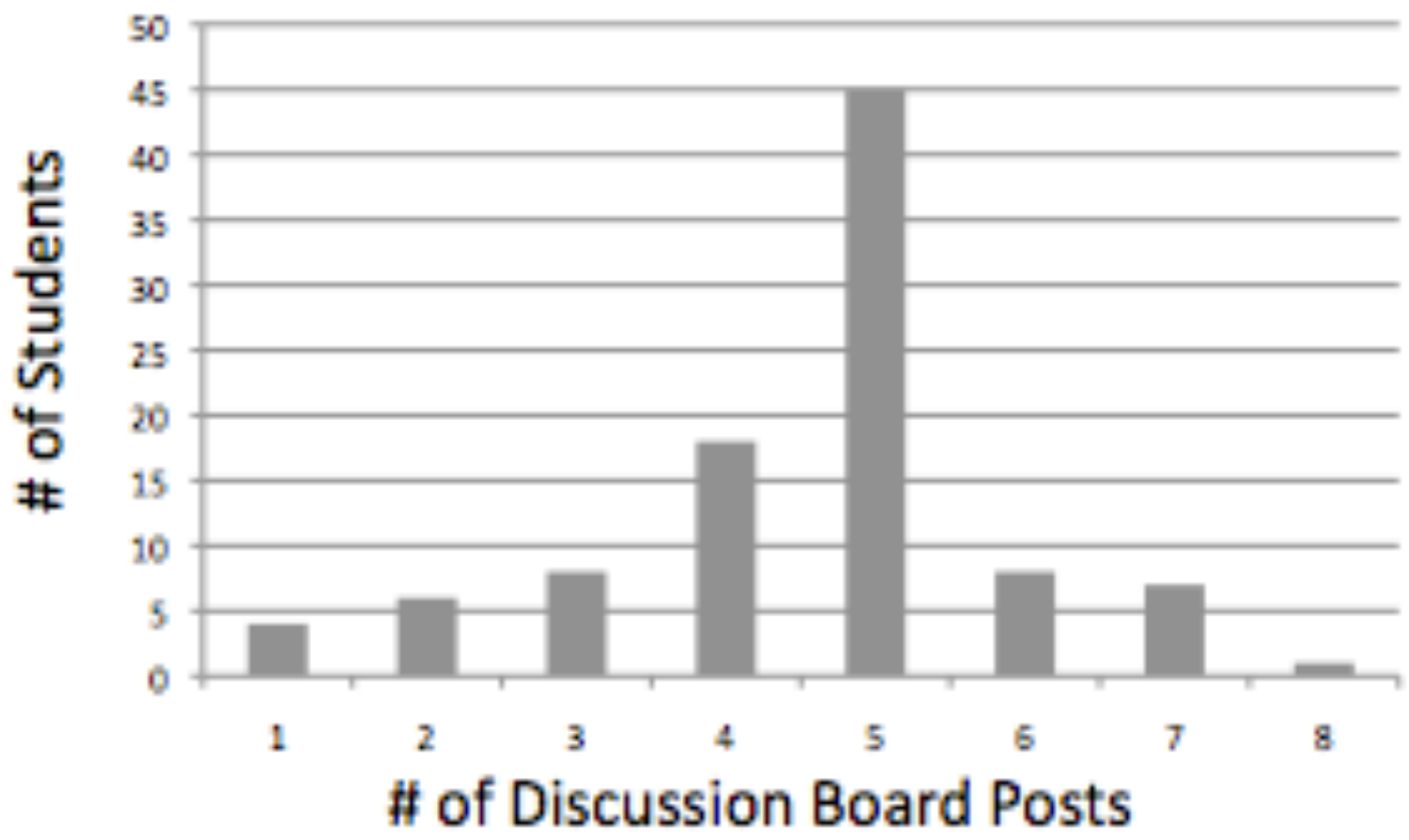

**FIG. 5.1.** Total number of postings made by students by the end of the Fall 2010 semester. Well over 3/4 of the enrolled students made at least four contributions to the discussion board over the course of the semester.

Our overall assessment would be that students engaged each other in a thoughtful and creative exchange of ideas, sometimes within topics that were fairly removed from our immediate focus (tachyons, time travel, warped space, and the like…). Many of the discussion threads centered on students clarifying their understanding of specific concepts (with the *occasional* intervention of an instructor, in order to stem the propagation of misconceptions), but a good deal more showed how many of the students didn't struggle so much with understanding what the interpretations were about; they struggled more with what they implied about the nature of science and reality. In just one excerpt from a discussion thread, [see Appendix F for a larger selection] we see how students are troubled by the idea of collapsing wave functions – is it some ad hoc rule invented to make the theory conform with observation? We see opposing views on questions of ontology: a literal switch between categories, or a switch between descriptions, or do photons belong to a category all their own? What are our everyday experiences with quantum phenomena, and where do we draw the line between the classical and the quantum world?

Subject: Delayed-Choice Experiments
Date: October 12, 2010 10:53 PM

[…] It seems that what's important for the argument is what's going on at the first beamsplitter. I think Dirac is saying that we can think of each photon always taking both paths and then the collapse of the wavefunction forces the photon to suddenly go from being in both paths to being in just one?

> Date: October 17, 2010 7:46 PM
>
> I got the same message from Dirac's statement that "each photon interferes only with itself" and that the photon is wavelike until observed as a particle. Or innocent until proven guilty if you will ;)
>
> Still, riddle me this, how can a propagating wave suddenly switch to particle like behavior?
>
> And the weirdness of quantum mechanics persists.
>
>> Date: October 19, 2010 3:34 AM
>>
>> That has been tough to grasp for me as well, how do we understand that there is some mechanism for the wave to switch to particle behavior?
>>
>> We have only the wave equation collapse and probability which seem like the algorithms we discarded earlier in the semester for the "farmer and the seed". I know there isn't an answer yet of the process its what me have to accept for now since the math coincides with experimentation so perfectly. (My observations thus far)
>
>> Date: October 19, 2010 9:56 AM
>>
>> I've been thinking about the nature of photons and the like, and I've decided that "behaving like a particle/wave" doesn't say anything about what the photon actually is. These comparisons just give us something to relate them to, at certain times. Photons are in a category all their own, and behave like nothing we know classically.
>>
>>> Date: November 3, 2010 12:39 AM
>>>
>>> Like so much in our world: words can never suffice.
>>>
>>> It's just so very perturbing to me: the idea a wave acts like a wave when we want it to and vice versa with the particle. Why is the measurement so important? Have particles such as photons always acted this way even when we were ignorant of things not just at the quantum level, but at simply the cellular level? I sometimes wonder if the world behaves in a quantum manner just because we are observing it behave in a quantum manner, like the whole of existence is just a hypothetical wave in someone's photon experiment and there's a whole other particle-side out there which we don't know about. Is it just a question of making an effort to find it?
>>
>>> Date: November 9, 2010 7:14 PM
>>>
>>> I wholeheartedly agree. Light quanta is a concept used to explain certain phenomena we perceive in certain experiments, not the absolute truth. What the photon actually is can only be described in partially complete terms "wave or particle" that end up confusing the people.
>>>
>>> But light behaves in a so called "classical" manner, does it not? You perceive light all the time. As you are reading this light is stimulating nerves in your eyes. You know the effects of light well. So, do photons truly behave like nothing we know classically?
>>>
>>>> Date: November 15, 2010 9:49 PM
>>>>
>>>> We've discussed plenty of times that objects that were previously believed to have only "classical" properties behave in a quantum manner. Bucky balls for instance are quite "large" especially compared to an electron or photon and in general I would say that we would think of the Bucky ball behaving "classically." That said, we've seen interference patterns from them which is strictly a quantum behavior. What is your justification for light behaving "classically"? Remember that your retina is a measurement device and will destructively alter the quantum state of a photon.

**II.D. Exam Materials**

One learning goal for this section of the course was for students to be able to identify a perspective as being realist, and to have some facility with the arguments in favor or against any particular interpretation. Since our usual post-instruction essay question on the double-slit experiment had proven useful in our interviews (in terms of eliciting students' attitudes toward some interpretive themes), we thought it appropriate to adapt this question for the second midterm exam. The problem statement for the exam question was identical to its presentation in the post-instruction online survey, but here students were asked first to identify and characterize the assumptions of Student One in terms of the interpretations of quantum mechanics we had discussed in class:

**Student A:** Student One interprets this sequence of screen shots classically, he obviously is thinking of this problem not quantum mechanically because if he did he would think the electron is going through both slits at the same time although he is thinking of this in terms of the Bohr model a bit. I think this is because he knows that we don't know the true position of the electron which means he is also thinking of it in terms of the uncertainty principle too. He thinks classically because he thinks it can't go through 2 slits at the same time.

**Student B:** Student One believes that the electron is indeed just a particle the whole time, but is moving around so fast in a random way that we can't detect it. He does not believe in wave-particle duality of electrons. He does believe that there are hidden variables (i.e., position). He also does not believe that there is a superposition. Overall, he has a realist point of view that the electron has a specific path but we just don't know it.

**Student C:** Student 1 is taking a somewhat realist perspective. They are assuming the electron traveled through one slit or the other. They claim the reality of the situation is the particle-like electron existed in a cloud of probability, and passes through one slit or the other as the cloud moved through the double slits. This explanation does not mention the probability density predicted by the wave equation.

**Student D:** Student 1's statement is consistent with that of someone who holds realism to be true. He/she assumes that: 1) The electron was always a particle with a fixed position in space and time; and 2) The only reason that the probability field is so large is because we are unable to determine its position (a "hidden variable") prior to it striking the screen. Thus, he believes that the properties of the electron are always the same, but we (the observer) are only able to observe those properties under a given set of circumstances (when the particle hits the screen).

Like Student A, there were some students who didn't utilize the specific terminology we had developed in class (e.g., distinguishing only between *classical* and *quantum* thinking, or *particle* and *wave* perspectives, without employing terms like *realism*); virtually every single student was regardless able to recognize that Student One's belief in localized electrons was an assumption. The second part of the essay question asks students to list any rationale or evidence that favors or refutes the first two statements; and to explain whether the third statement is claiming the first two are wrong, and why such a stance might or might not be favored by practicing physicists:

**Student A:** For Student 1, I agree that the prob. density is large because we don't know position of the electron – we never do. I disagree that this can't be represented quantum mechanically. From experiments in the past it is proven that we get fringes (pattern).

For Student 2, I disagree that the electron is the blob because in the brighter part of the blob there is a higher probability that an electron will be detected than in the dimmer part. Although I agree the electron acts as a wave, I disagree that a single electron can be described as a wave packet.

The third student isn't saying the first 2 are wrong. All he is saying is that the interference patterns are a result of probability not classical physics and that both are right. We don't know how we get the results we do so we work with probabilities.

**Student B:** Since Student One believes that the electron was traveling within the blob and went through only one slit, he believes that electrons act as particles. This would mean that he would never observe interference. This is not true though because the experiment shows that over a long time, interference is observed. (Even the nickel atoms in a crystal lattice experiment shows this too.) Since Student 2 believes that the electron acts as a wave packet, he suggests that we have a small uncertainty in its position (and large uncertainty in its momentum). However, if we had a small uncertainty in its position, then we could later predict where it would show up on the screen. The double-slit experiment shows this. In other words, the blob doesn't represent the electron, but rather the probability density of the electron to be detected. Experiments show that we don't really know what the electron is doing before we detect it. Student 3 is indeed disagreeing with Students 1 & 2 by saying that Students 1 & 2 can't make some of their claims, as we really just can't tell what the electron is doing between being emitted from the gun and being detected on the screen. He might not be stating that Students 1 & 2 are necessarily wrong, but he says that quantum mechanics can't conclude their conclusions. A practicing physicist would most likely agree with Student 3 because it is consistent with the Aspect experiment for photons.

**Student C:** Student 2 describes the electron as a wave packet. When a double slit experiment is performed, the interference pattern that is observed corresponds to a probability density that can be described by a wave-packet equation. A packet of waves would interfere with itself, creating a probability of the electron to pass through both slits. Also, which slit the electron went through cannot be measured without altering the uncertainty in the momentum.

**Student D:** **Rationale/Evidence for Student 1 (aka EPR):**
Realism argument: all objects must have definite properties within the system regardless of observation. Location is real but hidden variable. Makes intuitive sense.

**Against Student 1:**
Idea of definite quantities for all states (Local Realism) does not hold to experiment. Probabilistic provides correct explanation, deterministic does not. Single-photon interference experiments.

**Rationale/Evidence for Student 2 (aka Bohr):**
Electron is a wave function that collapses to a determinate state at plate. Consistent with matter waves argument put forward by deBroglie. Allows for interference with only one electron.

**Against Student 2:**
Fails when applied quantitatively; no mechanism for wave collapse yet developed.

No, Student Three is simply stating the theory behind the interpretations put forth by the first two students. In other words, he is limiting his assessment of the experiment to what can be predicted and explained through existing QM theory. A practicing physicist would tend to agree with Student 3 because his description requires the least assumptions and adheres to what we know as opposed to what we postulate.

Once again, Student D offers a near textbook response. Student B employs standard arguments against a strictly particle view of electrons, and in favor of a wave representation, but is explicit in saying that the wave corresponds to the probability for where an electron might be found, and not the electron itself. He is also cognizant of the incompatibility of the two statements – it is not possible for both of the fictional students to be correct. Not every student saw these two views as contradictory, in the sense that they reduced the two statements down to simply representing either a particle view or a wave view, without considering how each statement makes an explicit assertion regarding the behavior of the electron at the slits – it either goes through one slit or it goes through both. In other words, not every student took a definitive stance on the question of whether an electron always passes through one slit or both, focusing more on the legitimacy of particle or wave views in this context.

Interestingly, Student A's response is an almost exact recapitulation of Student R3's reasoning in Chapter 4: they both agree the electron is somehow behaving like a wave in this experiment, but object to the idea that a wave packet can describe an individual particle. Student A also indicates a belief that we can never know the true position of an electron, hence the large probability density. At this stage, it seems that Student A is not yet *split* in his beliefs – he hasn't conceded that an authoritative stance trumps his intuitive views, and indeed implies that scientists might believe that Students One & Two are both right, and that we can't really know why we observe what we do. Student C is not explicit in arguing against Student One, but instead explains why Student Two's description conforms to observation. As we shall see in the final portion of this exam question, Student C still believes in a continuously localized existence for electrons in this experiment:

**(Part III)** Which student(s) (if any) do you ***personally*** agree with? If you have a different interpretation of what is happening in this experiment, then say what that is. Would it be reasonable or not to agree with ***both*** Student 1 & Student 2? This question is about your personal beliefs, and so there is no "correct" or "incorrect" answer, but you will be graded on making a reasonable effort in explaining why you believe what you do.

**Student A:** I think from what I have learned in this class that Student 3 is correct. Probability can show us patterns but we really don't know what's going on before. It is reasonable to agree with both Student One who thinks classically and Student 2 who thinks quantum mechanically because that allows you to form your own ideas about what is going on but the truth is that we don't know what's going on between emission and the screen.

**Student B:** I personally believe that the electron acts like a wave until we observe it. This is Dirac's interpretation. Student 1 & Student 2 can't both be right because that would suggest that the electron acts like a wave and particle at the same time, and there is experimental evidence that refutes this.

**Student C:** Since electrons show both wave and particle like behavior, it would be reasonable to side with either Student 1 or 2. Student 2 used a more wave-like interpretation, Student 1 used a more particle like interpretation.

I personally visualize the situation as a flow of some fluid that travels through the two slits in waves. It appears through all space as soon as the electron is fired. The electron then rides this chaotic fluid toward the screen and strikes in a location that is somewhat determined by the interference patterns of the fluid. Trying to measure this fluid flow collapses the waves created.

**Student D:** I personally agree with Student 3. I see no reason to jump to a conclusion regarding the electron's behavior without a quantitative mechanism to explain its behavior between source and the plate. We know from this experiment that an electron exhibits behavior consistent with that of a wave, but we do not know exactly <u>why</u> or <u>how</u> that is so. That being said, I find Student 2's statement a more convenient way to think about the electron's behavior.

Student A merely restates his earlier stance: we require probabilistic descriptions because we can't really know what is going on between source and detection, and so either point of view might be equally legitimate. In the end, it seems this student is asserting his right to believe as he chooses when science has no definitive answer. At this point, we would characterize Student A as *Agnostic* – he recognizes the implications of competing perspectives, but is unwilling to take a stance on which might best describe reality.

Student B does not explicitly say which student he agrees with, but reports his belief in Dirac's matter-wave interpretation. Notice, however, that he says the electron *acts* like a wave, and not that an electron *is* a wave. Without further information from Student B, his views at this point might be consistent with either a *Quantum* or a *Copenhagen* perspective, since his stance on the reality of the wave function, and the nature of its collapse, is unclear.

We may easily place Student C within the *Pilot-Wave* category; indeed, his response sounds eerily similar to Student P3 (from Chapter 4) – the interference of nonlocal quantum waves determines the trajectories of localized particles. These two students arrived at the same conclusions independently; we made only cursory mention of Bohm's interpretation in our class, and it was not discussed at all in Student P3's class. This suggests that such ideas may be more prevalent among students than it seemed at first glance.

Student D's sentiments are not so different from Student A – it isn't known why quanta behave as they do, and so being agnostic requires the fewest assumptions (though he does mention that he finds it useful to employ a wave description in this situation). It seems reasonable to characterize Student D as subscribing to a *Copenhagen/Agnostic* perspective at this stage of the course.

The class as a whole performed well on this exam question: ~75% of students received full credit for their responses; the remaining students primarily lost one or more points (usually not more than three, from a total of ten points) for providing incomplete responses (very few students made any assertions that were unequivocally false). Overall, we would say that several of our learning goals surrounding this material were met by the majority of our students: they were able to identify the realist assumptions of the first fictional student, and to contrast them with an alternative perspective; they could provide evidence that favors or refutes competing points of view; and they were able to articulate their own beliefs regarding the interpretation of this quantum experiment. All of this regardless of whether they actually employed the exact terminology that had been developed in class (though most students did indeed use terms like *realism* and *hidden variables* in their argumentation). 18% of students chose to explicitly agree with Student One, though only one of them agreed with this statement exclusively; the remaining students were split between agreeing with both of the first two statements, or agreeing with all three. 46% of students said they agree with Student Two, or with both of the last two statements, while 36% preferred Student Three's statement exclusively.

### II.E. Assessing Outgoing Perspectives

As part of their final homework assignment, students were asked to respond to the same post-instruction attitudes survey that had been administered in other courses. We report here the final class wide responses to each survey item, juxtaposed with how they responded at the beginning of the semester. We similarly offer complete responses from Students A, B & C. Student D did not respond to this final survey, but we shall hear from him again in our discussion of the final essay assignment below. [Section II.F]

**1.** It is possible for physicists to carefully perform the same measurement and get two very different results that are both correct.

| | Agree | Neutral | Disagree |
|---|---|---|---|
| **POST (N=90)** | **0.78** | **0.06** | **0.17** |
| **PRE (N=94)** | **0.65** | **0.13** | **0.22** |

**Student A:** **(Disagree)** Take the example of hidden variables. If you put one red sock and one blue sock into identical boxes and both socks are identical beside their color, and you send them across the universe, then your technically performing the same measurement. When you open one box you find out what color the sock is in that box and it can be either red or blue, two different results. At the same time you also know what is in the other box every time you perform the experiment, in that respect, you are kinda getting the same result.
**(PRE: Agree)**

**Student B:** **(Strongly Agree)** This is possible especially when it comes to measuring the position of an electron. This is because there is no definite position to begin with. All we can know is the probability of finding the electron in a particular position, but probability does not determine where the electron will be when we measure it.
**(PRE: Agree)**

**Student C:** **(Strongly Agree)** Two very different results could confirm the same fact. Being correct is nothing more than confirming a fact.
**(PRE: Strongly Agree)**

Students shifted towards more agreement with this question (and less neutrality), but drawing conclusions from overall agreement or disagreement should be done with caution, for there are quantum mechanical reasons for disagreeing with this statement. For example, it has been argued by students that, in practice, scientists perform a number of measurements in any given experiment, and it is the statistical distribution of data that is the final result, which should be always be the same for similar experiments:

> "…if we are measuring the position of an electron, we will measure a different position each time. But if we compile all our results we will find positions that correspond to the wave function. I strongly disagree with the above statement because if an experiment is performed correctly it should produce the same results!"

The distribution in Table 5.II of the kinds of reasoning invoked by students at pre- and post-instruction (by the same categorization scheme employed in Chapter 2) shows that students shifted dramatically in their preferences for deterministic and hidden-variable style thinking (Categories D & E). Students shifted from 47% to 17% in providing Category D & E responses (whether in agreement or disagreement). And while only 17% of students invoked quantum phenomena (Category A) at the outset of the course, 65% of post-instruction responses made reference to quantum systems. Most students agreed with this statement before and after instruction, but learning about quantum mechanics caused most of them to consider it in a new light. For example, Student B has confirmed his pre-instruction suspicion that quantum mechanics might allow for this statement to be true. Student A originally agreed because of wave-particle duality, but now disagrees through an example of hidden variables and classical ignorance. Student C strongly agreed in both cases, first providing a Category D response, and then one more consistent with Category C.

**TABLE 5.II.** Categorization (as in Chapter 2) and distribution of reasoning provided at pre- and post-instruction, in agreement or disagreement with the statement: *It is possible for physicists to carefully perform the same measurement and get two very different results that are both correct*; standard error on the proportion ≤ 5% in each case.

| **CATEGORY DESCRIPTION** | |
|---|---|
| **A** | Quantum theory/phenomena |
| **B** | Relativity/different frames of reference |
| **C** | There can be more than one correct answer to a physics problem.<br>Experimental results are open to interpretation. |
| **D** | Experimental/random/human error<br>Hidden variables, chaotic systems |
| **E** | There can be only one correct answer to a physics problem.<br>Experimental results should be repeatable. |

| **CATEGORY** | **PRE-INSTRUCTION (N=94)** | | **POST-INSTRUCTION (N=90)** | |
|---|---|---|---|---|
| | **AGREE** | **DISAGREE** | **AGREE** | **DISAGREE** |
| **A** | 15% | 2% | 58% | 7% |
| **B** | 4% | 0 | 0 | 0 |
| **C** | 13% | 0 | 10% | 0 |
| **D** | 29% | 3% | 9% | 4% |
| **E** | 1% | 14% | 0 | 4% |
| **TOTAL** | 62% | 19% | 77% | 15% |

**2.** The probabilistic nature of quantum mechanics is mostly due to physical limitations of our measurement instruments.

| | Agree | Neutral | Disagree |
|---|---|---|---|
| **POST (N=90)** | **0.18** | **0.21** | **0.61** |
| **PRE (N=94)** | **0.46** | **0.32** | **0.22** |

**Student A:** **(Strongly Agree)** The probabilistic nature of quantum mechanics comes from the fact that there are aspects of quantum mechanics that can't be measured due to physical limitations of our measurement instruments. For instance how the uncertainty principle interacts with electrons orbiting a nucleus. Electrons are too small and move too fast for humans to know exactly where an electron is at a certain moment, so we can only perform one measurement at a time. Position and momentum of a particle can't be known at the same time, we can only calculate the probability of finding them there.
**(PRE: Neutral)**

**Student B:** **(Strongly Disagree)** It seems that the probabilistic nature of quantum mechanics is mostly due to the nature of sub-atomic particles rather than the limitations of our measurement instruments. If the particles were in definite states and definite positions to begin with, or even if there were a wave function that could define the exact state of the particles at any time, then one could argue that the problem is our measurement instruments. Perhaps such a formula will exist in the future, but that would mean that the limitation is our knowledge, not our instruments.
**(PRE: Strongly Agree)**

**Student C:** **(Neutral)** I have no idea.
**(PRE: Neutral)**

There was a strong shift away from agreement and in favor of disagreement by the end of the class; without passing judgment on students who feel neutrally towards this statement (after all, we do not consider agnosticism to be unsophisticated), we would at least like for our student to *not agree* with the notion that technology might one day reduce the need for probabilistic descriptions of quantum phenomena. Student B's response is desirable, in that he identifies uncertainty in quantum mechanics as fundamental, and not a consequence of experimental uncertainty. Student A's response is consistent with his reasoning on atomic electrons at the beginning of the course: their chaotic, rapid motion precludes knowledge of their true positions. We placed Student A in the *Agnostic* category at the time of the second exam, but we shall now see his explicit preference for realism:

**3.** When not being observed, an electron in an atom still exists at a definite (but unknown) position at each moment in time.

| | Agree | Neutral | Disagree |
|---|---|---|---|
| **POST (N=90)** | **0.26** | **0.18** | **0.57** |
| **PRE (N=94)** | **0.72** | **0.09** | **0.19** |

**Student A:** **(Strongly Agree)** Every physical thing exists whether it is being observed or not. This is the idea of realism, and I completely agree with it. An electron is a particle therefore I believe that it has a physical manifestation. An electron will definitely still exist at a definite position at every moment in time. This correlates with my answer above.
**(PRE: Strongly Agree)**

**Student B:** **(Disagree)** This thought process only makes sense if one were to view electrons as particles (like billiard balls). However, we know from experimentation that the electron has wave-like properties and can be described in the form of an electron cloud (Schrodinger's model). Thus, we can have an idea of where we are likely to find the electron if we make a measurement, but when we don't make a measurement, the electron should not be acting like a particle. But then again, we can't be 100% sure of what's happening when we aren't measuring...
**(PRE: Strongly Agree)**

**Student C:** **(Neutral)** If an electron orbits a nucleus in a forest and no physicist is there to observe it, does it obey the uncertainty principle?
**(PRE: Agree)**

As with the second survey item, we would have liked for our students to *not* choose to agree with this statement, and only 26% of them did by the end of the semester. We may not infer too much from Student C's tongue-in-cheek response, except to suggest his neutral attitude implies this question may now have as little (or as much) meaning to him as considering the sound of one hand clapping – at a minimum, his response has shifted away from agreement. In his disagreement, Student B explicitly addresses the wave-like properties of atomic electrons, though he also expresses a modicum of tentativeness in his beliefs.

Even though Student A has come through this course with explicitly realist notions intact (perhaps even reinforced), we would still consider his response to be in keeping with at least some of our learning goals: he has given conscious consideration to his intuitive beliefs and confirmed them to himself, and he can now articulate those beliefs in terms of language that been previously unavailable to him. At the very least, he did not use such language in his pre-instruction responses, which focused more on the tentativeness of scientific knowledge. Let us consider these students' last thoughts on the double-slit experiment before drawing any final conclusions on their overall outgoing perspectives:

**Student A:** I agree with Student 1 mostly except for the fact that the electron could be going through both slits at the same time for all we know. I also agree with student 2 because I think that the electron is acting as a wave and again possibly go through both slits at the same time. Therefore I agree more with student 3 because we really don't know what is happening between the moment the electron is shot from the gun and it hits the detection screen.

**Student B:** I agree with student three because it seems that the electron can act as a wave until we observe it. Even if this isn't the reality, there's nothing we can know about it from when the electron is emitted to when it is detected. However, student one and student two cannot be both correct because the electron cannot act like a wave (student 2) and a particle (student 1) at the same time, because there is experimental evidence that refutes this.

**Student C:** Student One is assuming the electron is always a particle. Student Two is assuming that the electron is pretty much a wave until it gets smooshed by the screen. Student three is sticking to the fact that the electron has a probability of going in certain places on the screen. I think there will always be a more accurate description of observations and quantum mechanics is, for now, an accurate description of reality.

And so it would have been premature to consider Student A to be a confirmed *Realist*, seeing how he maintains an explicit tentativeness regarding what can actually be known in this experiment, and so we might best characterize his overall final responses as *Realist/Agnostic.* Student B's earlier exam responses placed him somewhere between the *Quantum* and *Copenhagen* categories, but his overall language has consistently referred to the *behavior* of quanta, and he has explicitly refused to equate the wave with the particle it describes. Considering his final agreement with Student Three, and his concession that a wave description of quanta may ultimately not conform to reality, Student B's outgoing perspective on quantum mechanics is most consistent with the *Copenhagen* category. Student C's final response requires some thought: we believe he is suggesting there will one day be a *more accurate* description of reality, but that quantum mechanics is currently a *sufficiently accurate* description of that reality, and so we don't interpret his response as implying that quantum mechanics is necessarily incomplete. Student C expressed beliefs in non-local realism at mid-semester, and we did not ask him for his own interpretation of the double-slit experiment in the post-instruction survey, but his overall final response indicate he would be best described as being in the *Agnostic* category.

A final look at the overall class responses to this post-instruction essay question, in conjunction with their responses on atomic electrons, provides some insight into the consistency of student perspectives, which was part of our original motivations for our investigations. [Chapter 2] Only five of the 87 students who provided clear responses to this survey item explicitly agreed with Student One, and three of them did so in their expression of agreement with all three statements. Of these five students, three of them agreed with the statement on atomic electrons, one was neutral, and the other replied in disagreement. This means that 23% of students who chose to *not agree* with Student One in the double-slit experiment essay question offered a response to the statement on atomic electrons that would be consistent with realist expectations. Even though we are only considering five students here (meaning there is significant statistical error), we note that this distribution of responses on atomic electrons for students who had expressed realist preferences in the double-slit experiment matches our findings in Chapter 2 exactly. We also note that this 23% (±4%) of students evidencing inconsistent thinking across these two contexts is significantly less than the 33% (±6%) found in our initial studies ($p<0.001$, by a one-tailed t-test). We believe these results allow us to conclude that another of our learning goals had been achieved for a majority of our students – the consistency of student perspectives between these two contexts has been significantly increased over prior incarnations of modern physics courses.

We conclude this section by considering the level of personal interest in quantum mechanics expressed by students at the end of the semester:

**4.** I think quantum mechanics is an interesting subject.

| | Agree | Neutral | Disagree |
|---|---|---|---|
| POST (N=90) | 0.98 | 0.02 | 0.0 |
| PRE (N=94) | 0.85 | 0.13 | 0.02 |

**Student A:** **(Strongly Agree)** I found quantum mechanics to be an interesting subject because the concepts around it are not proven. A lot of what is behind quantum mechanics is qualitative which is very different than most physics classes which are quantitative. It is nice to look at a complex subject such as physics from a qualitative manner because for the past two years I've been taking all engineering classes which are all involving math significantly.
**(PRE: Strongly Agree)**

**Student B:** **(Strongly Agree)** The fact that there are truths associated with quantum mechanics that still can't be explained is a very interesting concept. I have never been taught something in school that is proven in experiments but still lacks a proper reasoning (such as entanglement). I also think it's very interesting to learn how sub-atomic particles behave so differently than macroscopic particles.
**(PRE: Strongly Agree)**

**Student C:** **(Strongly Agree)** Quantum mechanics is strange and interesting and mind stretching. This has been a great course.
**(PRE: Neutral)**

We find it remarkable that virtually every student expressed an interest in quantum mechanics by the end of the course, and that only two students responded neutrally – these final numbers are contrary to the usual decrease in interest among engineering students, and are on par with what is typically seen in a course populated with physics majors, where it is fairly safe to assume that nearly every student is already interested in learning about quantum mechanics coming into the course. [Chapter 6.] Still, considering the relatively high rate of incoming interest in quantum mechanics for students from our course, it is not entirely clear how effective we were in influencing student attitudes without considering a more detailed breakdown of their responses. In all other cases, *agreement* and *strong agreement* had been collapsed into a single category, and similarly for *disagreement* and *strong disagreement*; we therefore consider the number of students who became *more emphatic* in their agreement. Initially, 32% of students merely agreed that quantum mechanics is an interesting subject, and 53% were in strong agreement – these numbers shifted by the end of the course to 20% and 78%, respectively. We may therefore conclude that this curriculum, as implemented, was successful in not only maintaining student interest in physics, but in promoting it as well. As a final comment, we note that Students A, B & C all express a strong interest in the subject, and their responses suggest that it is precisely the still-open questions in quantum mechanics that inspire their fascination – Pandora's Box has been opened, and we don't have to be afraid!

## II.E. Final Essay

In lieu of a long answer section on the final exam, students were asked to write a 2-3 page (minimum) final essay on a topic from quantum mechanics of their choosing, or to write a personal reflection on their experience of learning about quantum mechanics in our class (an option chosen by ~40% of students). As opposed to a formal term paper, this assignment was meant to give students the opportunity to explore an aspect of quantum mechanics that was of personal interest to them. Topics selected by students for their final essays (ones that were not personal reflections) included: quantum cryptography; quantum computing; enzymatic quantum tunneling; bosons and fermions; the Quantum Zeno Effect; string theory; atomic transistors; quantum mechanics in science fiction; and more… The nearly universally positive nature of the feedback provided by students in their personal reflections is evidence for the popularity and effectiveness of our transformed curriculum, and its practical implementation. [Excerpts from *each* of the submitted personal reflections from the Fall 2010 semester are collected in Appendix E.]

We recall from earlier in this chapter that Student D had entered this course with a relatively sophisticated view on quantum mechanics, but one that was explicitly realist/statistical. We are interested, of course, in whether this curriculum has something new to offer students with a high degree of background knowledge coming into the semester. Though he did not complete the end-of-term attitudes survey, we may still draw some conclusions regarding the effectiveness of this curriculum at influencing Student D's interpretive stances:

> "Upon entering the class, I was most excited to learn about the various interpretations put forth to explain quantum mechanical phenomena. I already had a fairly strong footing in the actual mathematics of the material, both from my own independent studies and from an exceptional AP Physics course I had taken in my senior year in high school. However, neither of those pursuits had given me a strong grounding in the overarching theoretical principles behind the material, especially when it came to interpreting the experimental data in the more recent work such as Aspect's single photon experiments and electron diffraction. I came in understanding the results of those experiments, but not their implications for the nature of light and matter. This class did a fantastic job of patching those holes in my understanding. […] Although this class has not significantly changed my ideas about physics and the practice of science, it has been one of the few courses I have taken that accurately portrays the scientific method of careful observation. The course was exceptional in how it handled conclusions drawn from experimental results, the most memorable example being the refutation of the "hidden variable" interpretation. The class was at its best when discussing the interpretations of experiments and the implications of their results; Aspect's single photon experiments were explained with particular clarity and care."

We may not know precisely how Student D would have responded to the post-instruction survey, but we may infer from his statements that he no longer personally subscribes to the notion of *hidden variables*. We assert that Student D successfully transitioned from a *Realist*/*Statistical* perspective on quantum mechanics, to one that is more aligned with the beliefs of practicing physicists (*Copenhagen*).

**References (Chapter 5):**

**1.** S. B. McKagan, K. K. Perkins and C. E. Wieman, Reforming a large lecture modern physics course for engineering majors using a PER-based design, *PERC Proceedings 2006* (AIP Press, Melville, NY, 2006).

**2.** W. K. Adams, K. K. Perkins, N. Podolefsky, M. Dubson, N. D. Finkelstein and C. E. Wieman, A new instrument for measuring student beliefs about physics and learning physics: the Colorado Learning Attitudes about Science Survey, *Phys. Rev. ST: Physics Education Research* **2**, 1, 010101 (2006).

**3.** E. Redish, J. Saul and R. Steinberg, Student expectations in introductory physics, *Am. J. Phys.* **66**, 212 (1998).

**4.** S. B. McKagan, K. K. Perkins and C. E. Wieman, Why we should teach the Bohr model and how to teach it effectively, *Phys. Rev. ST: Physics Education Research* **4**, 010103 (2008).

**5.** R. Hake, Interactive-Engagement Versus Traditional Methods: A Six-Thousand-Student Survey of Mechanics Test Data for Introductory Physics Courses, *Am. J. Phys.* **66** (1), 64 (1998).

**6.** L. C. McDermott, Oersted Medal Lecture 2001: "Physics Education Research – The Key to Student Learning," *Am. J. Phys.* **69**, 1127 (2001).

**7.** S. B. McKagan, K. K. Perkins, M. Dubson, C. Malley, S. Reid, R. LeMaster and C. E. Wieman, Developing and Researching PhET simulations for Teaching Quantum Mechanics, *Am. J. Phys.* **76**, 406 (2008).

**8.** http://phet.colorado.edu/simulations/sims.php?sim=QWI

**9.** C. Davisson and L. H. Germer, Diffraction of electrons by a crystal of nickel, *Phys. Rev.* **30**, 705 (1927).

**10.** S. Frabboni, G. C. Gazzadi, and G. Pozzi, Nanofabrication and the realization of Feynman's two-slit experiment, *App. Phys. Letters* **93**, 073108 (2008).

**11.** A. Tonomura, J. Endo, T. Matsuda, T. Kawasaki and H. Exawa, Demonstration of single-electron buildup of an interference pattern, *Am. J. Phys.* **57**, 117 (1989).

**12.** http://rdg.ext.hitachi.co.jp/rd/moviee/doubleslite.wmv

**13.** New lecture materials and summaries are available online at:
http://www.colorado.edu/physics/EducationIssues/baily/dissertation/

**14.** V. Otero, S. Pollock and N. Finkelstein, A Physics Department's Role in Preparing Physics Teachers: The Colorado Learning Assistant Model, *Amer. J. Phys.* **78**, 1218 (2010).

**15.** R. D. Knight, *Physics for Scientists and Engineers: A strategic approach*, 2nd Ed. (Addison Wesley, San Francisco, CA, 2004).

**16.** A. Shimony, The Reality of the Quantum World, *Scientific American* (January 1988, pp. 46-53).

**17.** N. D. Mermin, Is the moon there when nobody looks? Reality and the quantum theory, *Phys. Today* **38** (4), 38 (1985).

**18.** D. F. Styer, *The Strange World of Quantum Mechanics* (Cambridge University Press, Cambridge, 2000).

**19.** E. Cataloglu and R. Robinett, Testing the development of student conceptual and visualization understanding in quantum mechanics through the undergraduate career, *Am. J. Phys.* **70**, 238 (2002).

**20.** S. B. McKagan and C. E. Wieman, Exploring Student Understanding of Energy Through the Quantum Mechanics Conceptual Survey, *PERC Proceedings 2005* (AIP Press, Melville, NY, 2006).

**21.** S. Wuttiprom, M. D. Sharma, I. D. Johnston, R. Chitaree and C. Chernchok, Development and Use of a Conceptual Survey in Introductory Quantum Physics, *Int. J. Sci. Educ.* **31 (5)**, 631 (2009).

# CHAPTER 6

## Teaching Quantum Interpretations – Comparative Outcomes and Curriculum Refinement

### I. Introduction

In the previous chapter, we considered the design and implementation of a transformed modern physics curriculum for engineers, taught at the University of Colorado in the Fall 2010 semester. The accessibility and effectiveness of this new curriculum was discussed in terms of some measures that were entirely new and specific to that course – student responses to homework and exam questions relevant to the physical interpretation of quantum mechanics. But we have also gauged learning outcomes according to measures that had been employed in prior studies, (Chapter 3) and so we shall address in this chapter how some of the outcomes for this transformed course compare with three previous modern physics offerings.

Naturally, the outcomes from this course would be less significant if our learning goals had not represented a challenge for our students, or for ourselves as instructors and curriculum designers. Any newly implemented curriculum will certainly have need for refinement, requiring first the identification of specific student difficulties with the new material, which may then inform our suggestions for improvement. In light of our focus throughout this dissertation, it seems most appropriate to discuss problems students had in understanding the single-photon experiments, as revealed through their responses to another long-answer exam question from the second midterm. At the same time, we may also assess their use of some of the epistemological tools we had worked to establish in lecture. We will also consider aggregate and individual student responses to several of the multiple choice questions from our exams and the post-instruction content survey, which may indicate other student difficulties requiring future study.

### II. Comparative Outcomes

We have already seen how certain instructional approaches with respect to interpretation can be associated with specific student outcomes (e.g., there is a greater prevalence of realist beliefs in contexts where instruction has been less explicit in promoting an alternative perspective, or in topic areas where realist/statistical interpretations were deliberately promoted). There are many similarities between our course from Fall 2010 and the four courses discussed in detail in Chapter 3, Section 3.II - they were all large-lecture courses (N > 60) where interactive engagement was employed during class, and covered roughly the same progression of topics from quantum mechanics and its applications. And all but the course taught from a realist/statistical perspective utilized many of the same lecture materials that had been developed during the first round of course transformations in 2005-2007. Yet they all differed in their instructional approaches to interpretation, though we would say that Course B2 (as denoted in Section 3.II) was most similar to our own, in that the instructor was explicit in promoting a matter-wave interpretation of the double-slit experiment, and significant lecture time was given toward the very end of the semester to discussions of measurement and interpretation in quantum mechanics (but without specific reference to atomic systems). There were no significant differences in the wording or presentation of the online attitudes survey administered in each course. Before making direct comparisons of student outcomes, we first (briefly) remind ourselves of our characterizations of the courses with which we'll be making our comparisons, and establish how they will be denoted in this chapter. [Table 6.I]

**TABLE 6.I** Summary of the four courses to be compared in this section, including a characterization of each instructor's approach to interpretive themes. For reference, how each course was denoted in Chapter 3 is also included [n/a = not applicable].

| STUDENT POPULATION | COURSE | INSTRUCTIONAL APPROACH | CH. 3 DENOTION |
|---|---|---|---|
| Engineering | ENG-FA10 | *Matter-Wave* | n/a |
| | ENG-R/S | *Realist/Statistical* | A |
| | ENG-MW | *Matter-Wave* | B2 |
| Physics | PHYS-C/A | *Copenhagen/Agnostic* | C |

ENG-R/S is the only engineering class considered in our studies that was taught from a realist/statistical perspective. ENG-MW is the engineering course most similar to ours (ENG-FA10), in that similar lecture materials were used, a matter-wave perspective was promoted, and interpretive themes were discussed near the end. PHYS-C/A is a class for physics majors that also used many of the same lecture materials, but with less emphasis on interpretation.

### II.A. Student Interest in Quantum Mechanics

It is now well known in physics education research that student attitudes toward physics have a tendency to become less positive after instruction in introductory courses of all kinds, including ones where specific attention had been paid to student attitudes and beliefs. [3, 4] Similar effects have been seen in modern physics courses; one study showed that traditional modern physics instruction typically led to significant negative shifts in student attitudes (as measured by the CLASS [4]), while a curriculum transformed using principles from PER saw no significant pre/post-instruction shifts, meaning overall student attitudes at least did not get worse. [1] By combining pre-instruction survey responses on their reported interest in quantum mechanics from six semesters of engineering courses (including the Fall 2010 semester), we see that incoming interest for engineers is on average between 75-80% favorable. [Fig. 6.1] The average post-instruction interest among engineering students from five of these course offerings dropped to below 70%, while negative responses increased significantly ($p<0.001$) – approximately 1/3 of engineering students would not agree that quantum mechanics is an interesting subject after having learned about it in our modern physics courses! Students from the Fall 2010 semester were nearly unanimous (98%) in their reported interest in quantum physics, and not one student responded with a negative opinion. [Relative to the number of students who completed the final exam, the response rate for our post-survey was ~90%.]

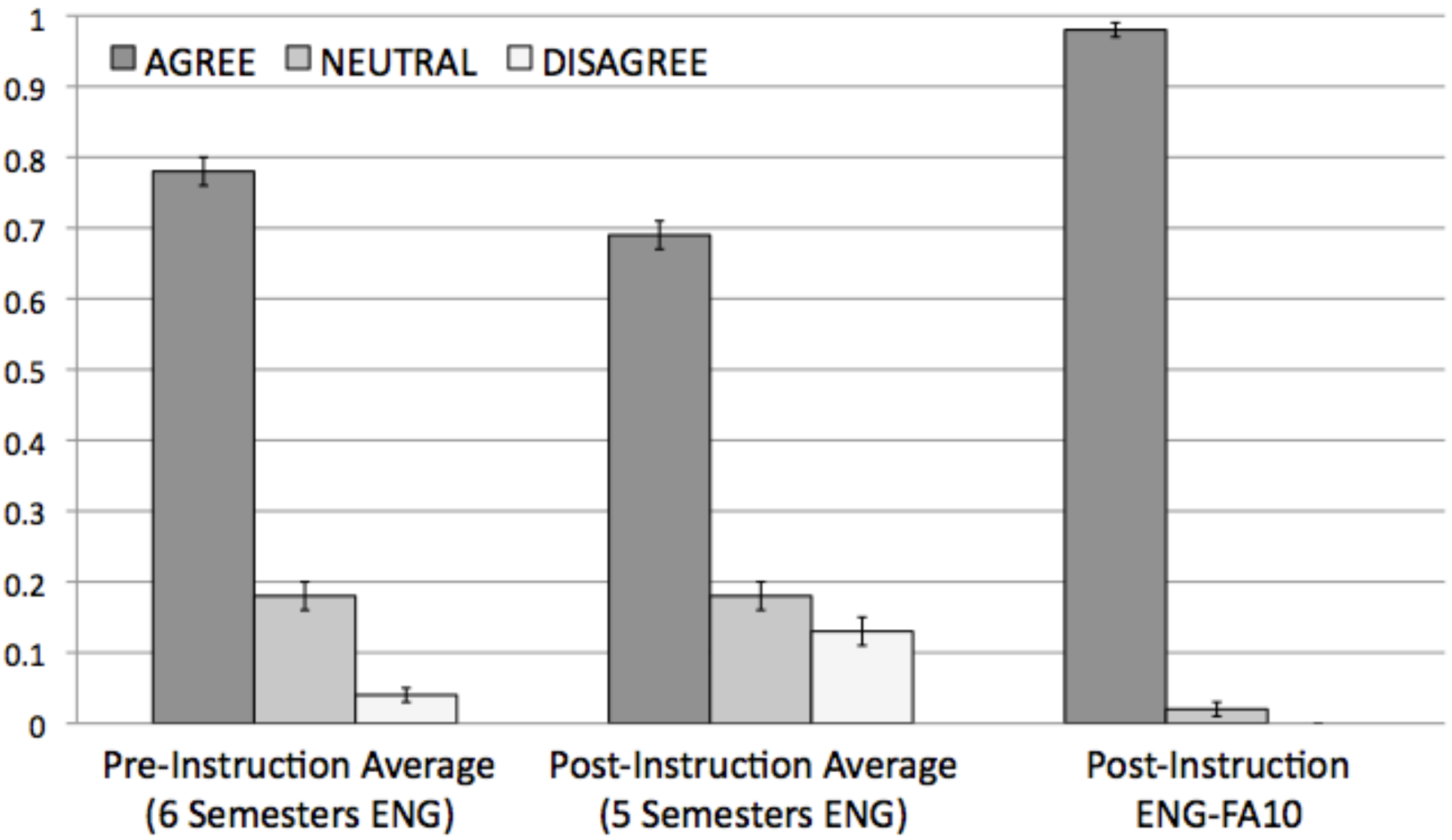

**FIG. 6.1.** Average pre- and post-instruction student responses to the statement: *I think quantum mechanics is an interesting subject*, from five modern physics courses for engineers, plus the FA10 semester. [Error bars represent the standard error on the proportion; N ~ 50-100 for each course].

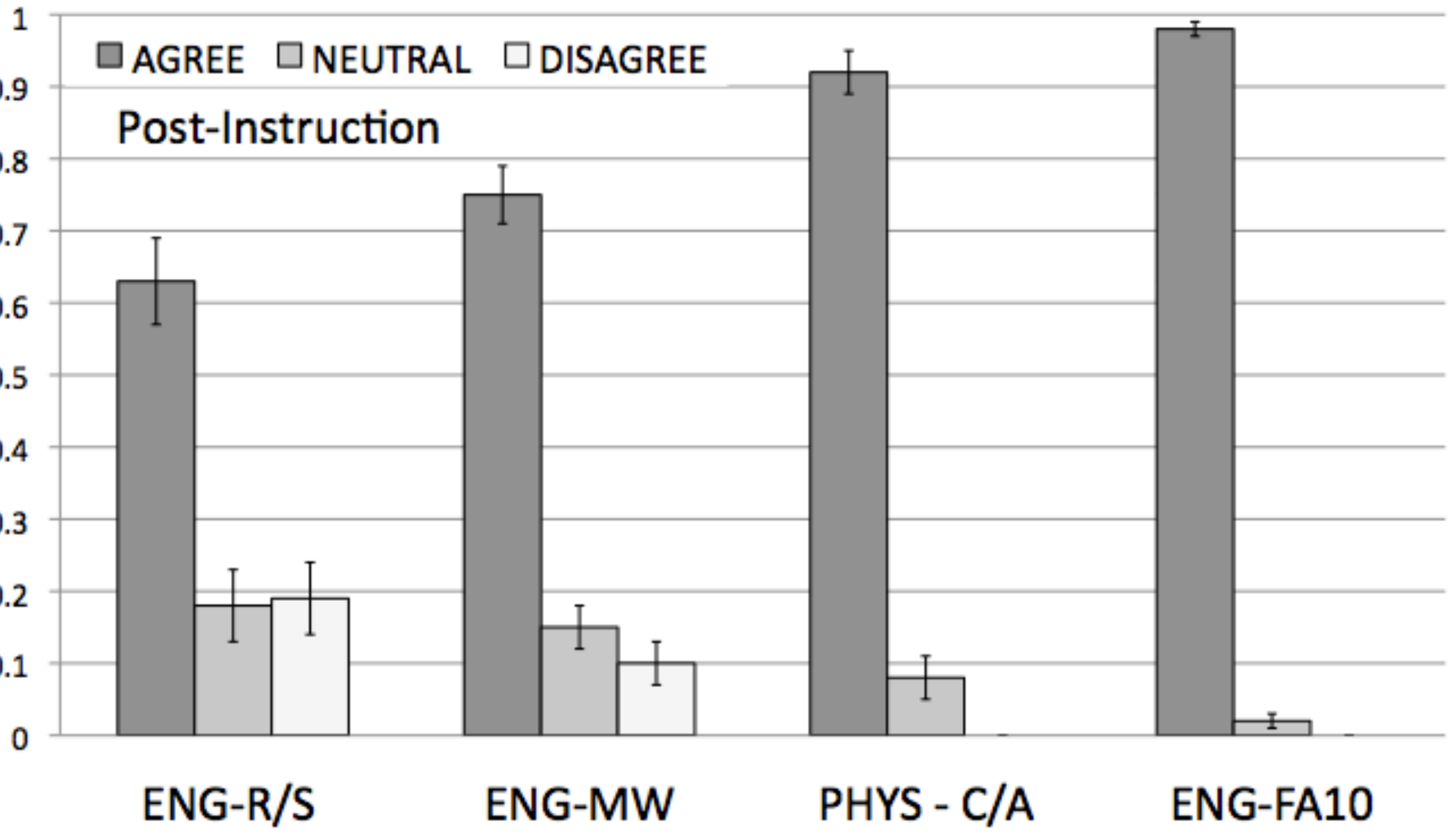

**FIG. 6.2.** Post-instruction student responses to the statement: *I think quantum mechanics is an interesting subject*, from four modern physics offerings, as denoted in Table 6.I. [Error bars represent the standard error on the proportion; N ~ 50-100 for each course]

It would be too great an assumption to conclude that shifts in student interest are necessarily directly correlated with the interpretive approach of the instructor, or with the student population. There are surely myriad other affective considerations, such as instructor popularity or choice of textbook, and we have seen courses for physics majors where overall interest in quantum mechanics declined. Nonetheless, we note that the Fall 2010 course had the greatest proportion of students reporting positive post-instruction attitudes towards quantum mechanics, including the course for physics majors; [Fig. 3.2] and that end-of-term student evaluations from ENG-MW, PHYS-C/A and ENG-FA10 ranked all of those instructors in the top 25%, relative to departmental averages (the instructor for ENG-R/S was ranked lower, at 32%). Different results were achieved by instructors of comparable popularity, and the responses from students to the newly introduced topics were overwhelmingly positive, which leads us to conclude that the new curriculum was at least partly responsible for the increased popularity of the course.

### II.B. Interpretive Attitudes

We may assess the relative impact of our transformed curriculum on student perspectives by further considering their post-instruction survey responses in relation to outcomes from previous modern physics offerings. The overall distribution of student responses from our course to the double-slit essay question is consistent with prior results, which had shown them to be generally reflective of each instructor's specific approach to that particular topic, whether *Realist*, *Quantum* or *Agnostic*. [Fig. 6.3] Considering this question had been adapted for use on the second exam, and that exam solutions detailing "acceptable" responses were later available online, it might be reasonably argued that the near absence of student preference for a realist interpretation of this experiment is mere confirmation of the effect of explicit instruction in that context.

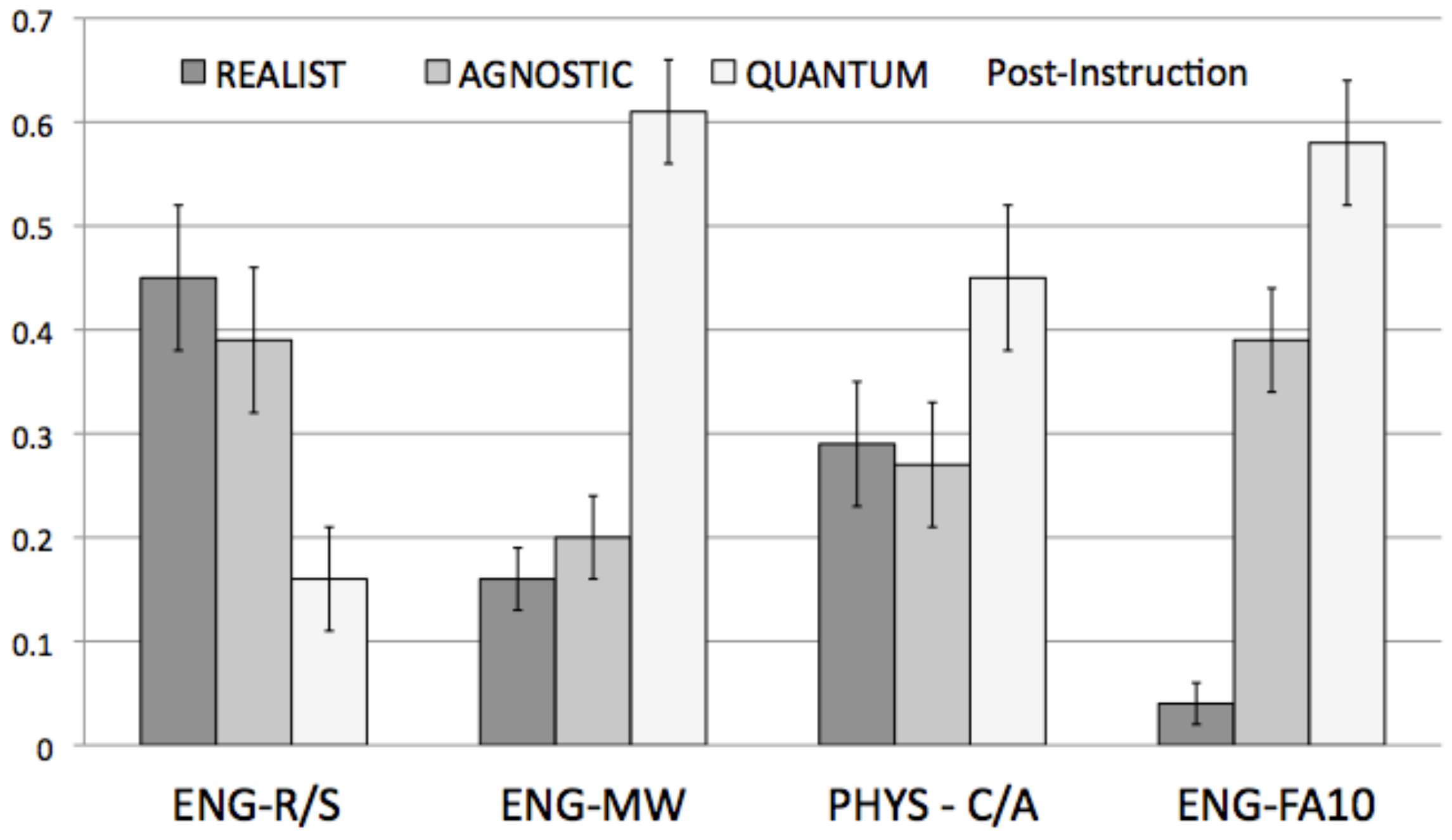

**FIG. 6.3.** Post-instruction student responses to the double-slit essay question, from four different modern physics courses, as denoted in Table 3.I. [Error bars represent the standard error on the proportion; N ~ 50-100 for each course.]

However, we made no mention during the entire semester of student responses to the pre-instruction attitudes survey, and did not give students any indication they would be revisiting these questions at the end of the course. We offered no explicit instruction as to what kinds of responses would be considered "acceptable", and repeatedly emphasized in the survey and in the homework assignments that we were most interested in what students actually believed. The lecture materials used during our treatment of the Schrödinger model of hydrogen were essentially the same as those used in ENG-MW and PHYS-C/A, with a few notable exceptions. Like the instructors for those two courses, we showed students how the Schrödinger model predicts zero orbital angular momentum for an electron in the ground state, and contrasted this result with the predictions of Bohr and de Broglie. But we continued by explicitly arguing how this result has implications for the physical interpretation of the wave function – for how could conservation of angular momentum allow for a localized particle to exist in a state of zero angular momentum in its orbit about the nucleus? This difficulty is removed when we choose to view atomic electrons as delocalized standing waves in quantized modes of vibration. More importantly, having already established language and concepts specific to interpretive themes, we were able to explicitly identify the position of an atomic electron as yet another example of a hidden variable, which we had argued couldn't exist as a matter of principle. Ours is the only course among these four where a significant majority of students chose to disagree with the idea of localized atomic electrons at the end of the semester. [Fig. 6.4]

The instructor for ENG-R/S told students during lecture that they *should* think of atomic electrons as localized, and overall responses from his course reflect this instruction. More specifically, he explained that quantized energy levels represent the average behavior of electrons over a time scale that is long relative to their orbital frequency, and that atomic electrons may be found to have a continuous range of energies when the time scale of the energy measurement is short (as enforced by the time-energy uncertainty relation); hence the broadening of spectral lines. This kind of reasoning is not unique among physicists, [5] and has therefore likely been utilized by modern physics instructors elsewhere.

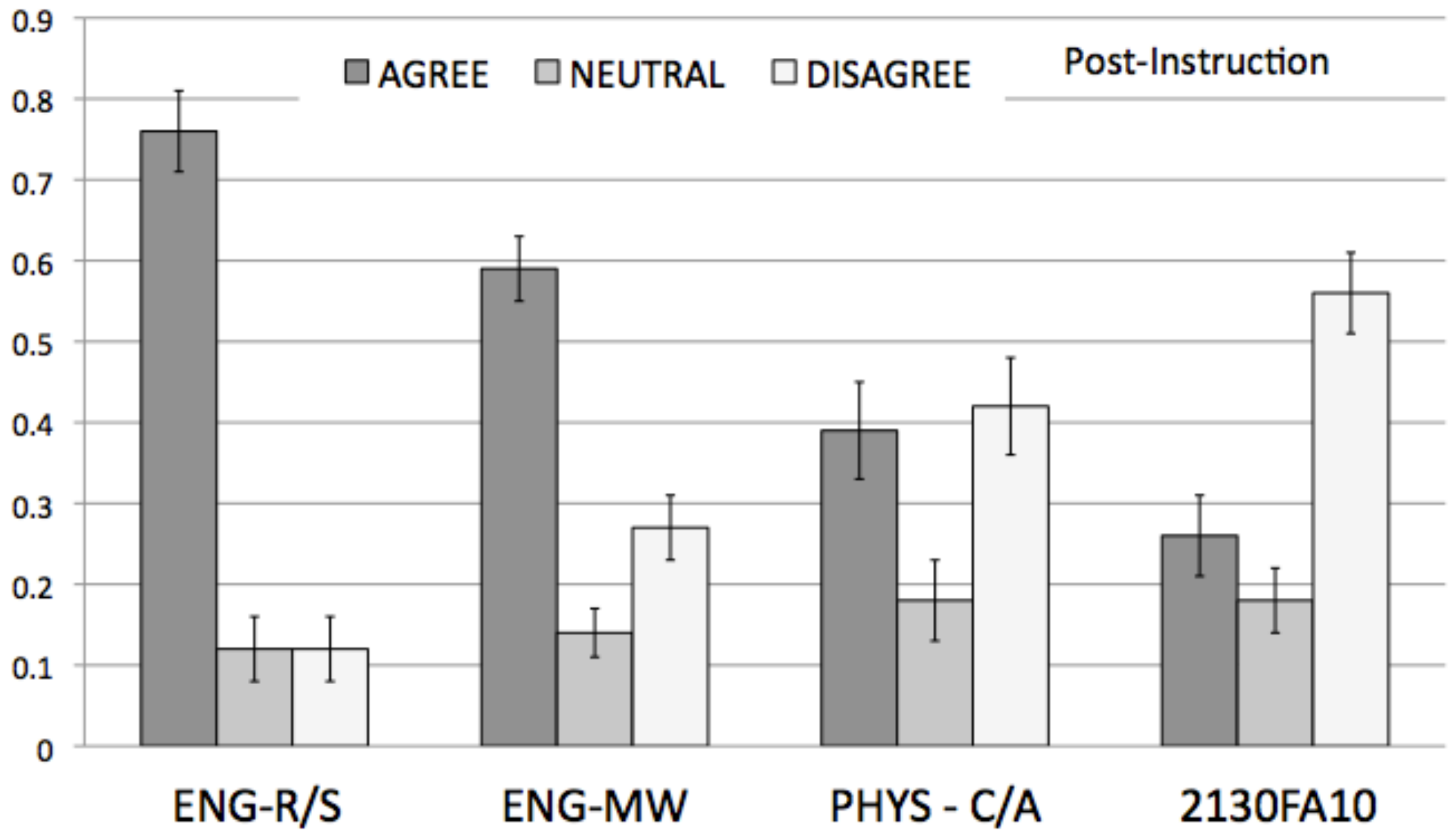

**FIG. 6.4.** Post-instruction student responses to the statement: *When not being observed, an electron in an atom still exists at a definite (but unknown) position at each moment in time*, from four different modern physics courses, as denoted in Table 3.I. [Error bars represent the standard error on the proportion; N ~ 50-100 for each course.]

We have previously characterized the other two courses as having de-emphasized matters of interpretation in the latter parts of the course, [Chapter 3] and heard from one instructor [PHYS-C/A; Instructor C in Chapter 3] about what had influenced his instructional choices – he felt that giving students a facility with the mathematical tools of quantum mechanics should take precedence over a detailed exploration of its physical interpretation, which might anyways be beyond the sophistication of introductory students. Though different in his overall interpretive approach, it turns out the other instructor [ENG-MW; Instructor B2 in Chapter 3] offered similar reasoning for having made a similar choice, and so it is worthwhile to consider one last time in detail what we consider to be a common motivation for the de-emphasis of interpretive themes in introductory modern physics courses, according to the instructor for ENG-MW:

> "This [probabilistic] aspect of quantum mechanics I feel is very important, but I don't expect undergraduate students to grasp it after two months. So that's why I can understand why [the statement on atomic electrons] was not answered to my satisfaction, but that was not my primary goal of this course – not at this level. We don't spend much time on this introduction to quantum mechanics, and there are many aspects of it that are significant enough at this level – it is really great for students to understand how solids work, how does conductivity work, how does a semiconductor work, and these things you can understand after this class. If all of the students would understand how a semiconductor works, that would be a great outcome. I feel that probably at this level – especially with many non-physics majors – I think that's more important at this point. But still, they have to understand the probabilistic nature of quantum mechanics, and I hope, for instance, that this is done with the hydrogen atom orbitals, not that everyone would understand that, but if the majority gets it that would be nice. These are very hard concepts. At this level, I feel it should still have enough connections to what they already understand, and what they want to know. They want to know how a semiconductor works probably much more than where is an electron in a hydrogen atom. […] I don't think the [engineering] students will be more successful in their scientific endeavors, whether it's a personal interest or career, by giving them lots and lots of information about how to think of the wave function. The really important concept I feel is to see that there is some sort of uncertainty involved, which is new, which is different from classical mechanics. […] At the undergraduate level, I feel it is important to make the students curious to learn more about it – and so even if they don't understand everything from this course, if they are curious about it, that's more important than to know where the electron really is, I think."

We see the instructor for ENG-MW *would have liked* for his students to disagree with this statement, and yet 75% of them chose to *not disagree*. Recall that this instructor made his own modifications to the first modern physics curriculum, to include an entire lecture on quantum measurement and interpretation towards the end of the course (but without specific reference to atomic systems).

At the end of the introduction to matter waves, our transformed course and ENG-MW both utilized a lecture slide similar to the one shown in Fig. 6.5 – note that both courses offered similar explicit guidance, albeit decontextualized, on how to think of electrons when not being observed: as delocalized waves. We believe this kind of general guidance is not by itself sufficient to cause most students to reconsider their conceptions of atomic electrons, as evidenced by the distribution of responses from a course that did not apply more specific guidance in the context of atoms. [Fig. 6.4] But specifically telling students to think of atomic electrons as delocalized would also not by itself be sufficient for significantly influencing students' overall perceptions of uncertainty in quantum mechanics.

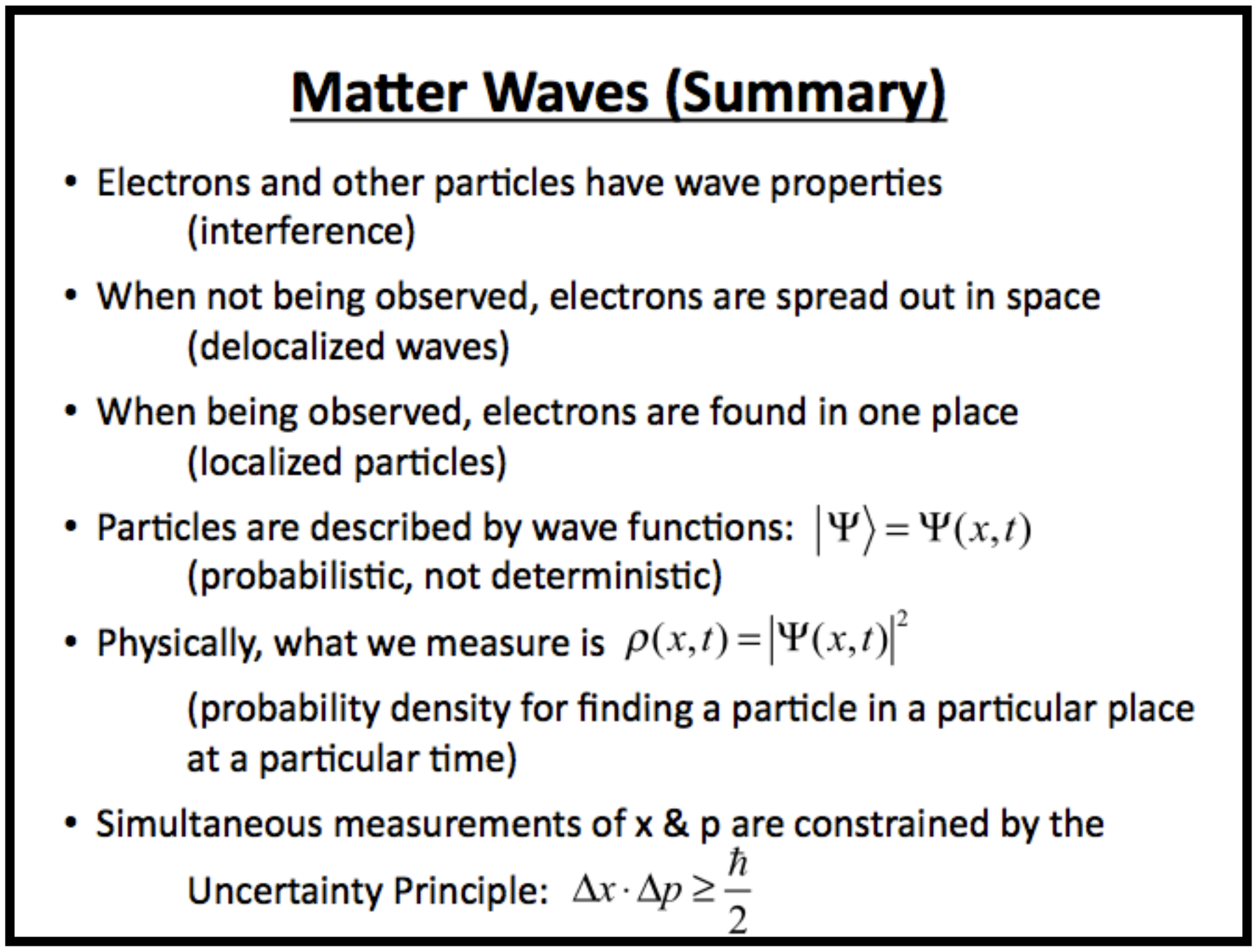

**FIG. 6.5.** A lecture slide equivalent to one used in each of two modern physics courses for engineers, ENG-MW & ENG-FA10. This slide offers explicit, but decontextualized, guidance on how to think of matter when not being observed.

We may conclude this from our observation that explicit instruction in one context does not necessarily influence student perspectives in other contexts, but also by other considerations. Even if the physical interpretation of atomic wave functions is not a primary learning goal for every instructor, we may safely say that our course shared with ENG-MW and PHYS-C/A a common learning goal that *was* primary: recognizing a difference between the experimental uncertainty of classical mechanics and the fundamental uncertainty of quantum physics. How do these four courses compare with respect to student responses to our last attitudes statement on the probabilistic nature of quantum mechanics? Realist expectations might lead incoming students to favor agreement with the statement: *The probabilistic nature of quantum mechanics is mostly due to the limitations of our measurement instruments.* We find that the incoming percentage of students from all three of the engineering courses agreeing with this statement is nearly identical, [ENG-R/S: 45%; ENG-MW: 48%; ENG-FA10: 46%] but that incoming attitudes for physics majors were significantly more favorable (with only a quarter of them agreeing, and over half disagreeing before instruction).

It would seem from his explanation of atomic energy quantization that the *Realist/Statistical* instructor would consider the uncertainty in quantum mechanics as being introduced by the *measurement process*, which is not the same as asserting that quantum uncertainty is experimental in origin, or that technology might one day find a way around these fundamental limits on observation. Regardless, there was a mild uptick in students from his course agreeing with this survey statement at the end of the semester. [Fig. 6.6] The instructor for PHYS-C/A had the greatest proportion of favorable responses at post-instruction – despite a de-emphasis on interpretive themes, he was successful in positively influencing student perspectives on uncertainty in quantum mechanics, though we must keep in mind the student population of his course, and the already relatively favorable incoming attitudes of his students.

In fact, the differential impact on student responses from these four modern physics courses is most dramatically illustrated by normalizing (post – pre) shifts in student agreement with this survey statement, according to their rate of agreement at the start of the course.[1] [Fig. 6.7] By this measure, our course had the greatest positive impact on student attitudes regarding the relationship between fundamental uncertainty in quantum mechanics and classical experimental uncertainty.

We conclude this section with some comments on the statements of the instructor for ENG-MW, regarding what we might like for students to take away from our introductory courses. First, if the aim of instruction is not necessarily a universal understanding of concepts, but for students to come away with a continued interest in modern physics, then we would claim that our course was the more successful of the two: student interest in quantum mechanics increased from 70% to 75% for his course (with 10% responding negatively at post-instruction), but the reported interest among students from our course increased from 85% to 98%, which we have argued must be in part attributable to the transformed curriculum itself. Second, we shouldn't presume to know exactly where the interests of our engineering students lie. The results from our curriculum implementation would suggest that students are in fact *just as interested*, if not more so, in questions about the nature of reality, as they are in learning about the theory of semiconductors.

---

[1] We define *favorable gain* as the negative of this, since we consider a *decrease* in agreement with this statement as favorable. This definition is equivalent to the definition of *normalized gain* = (post – pre)/(1 – pre), except that the target response rate is zero, and not 100%.

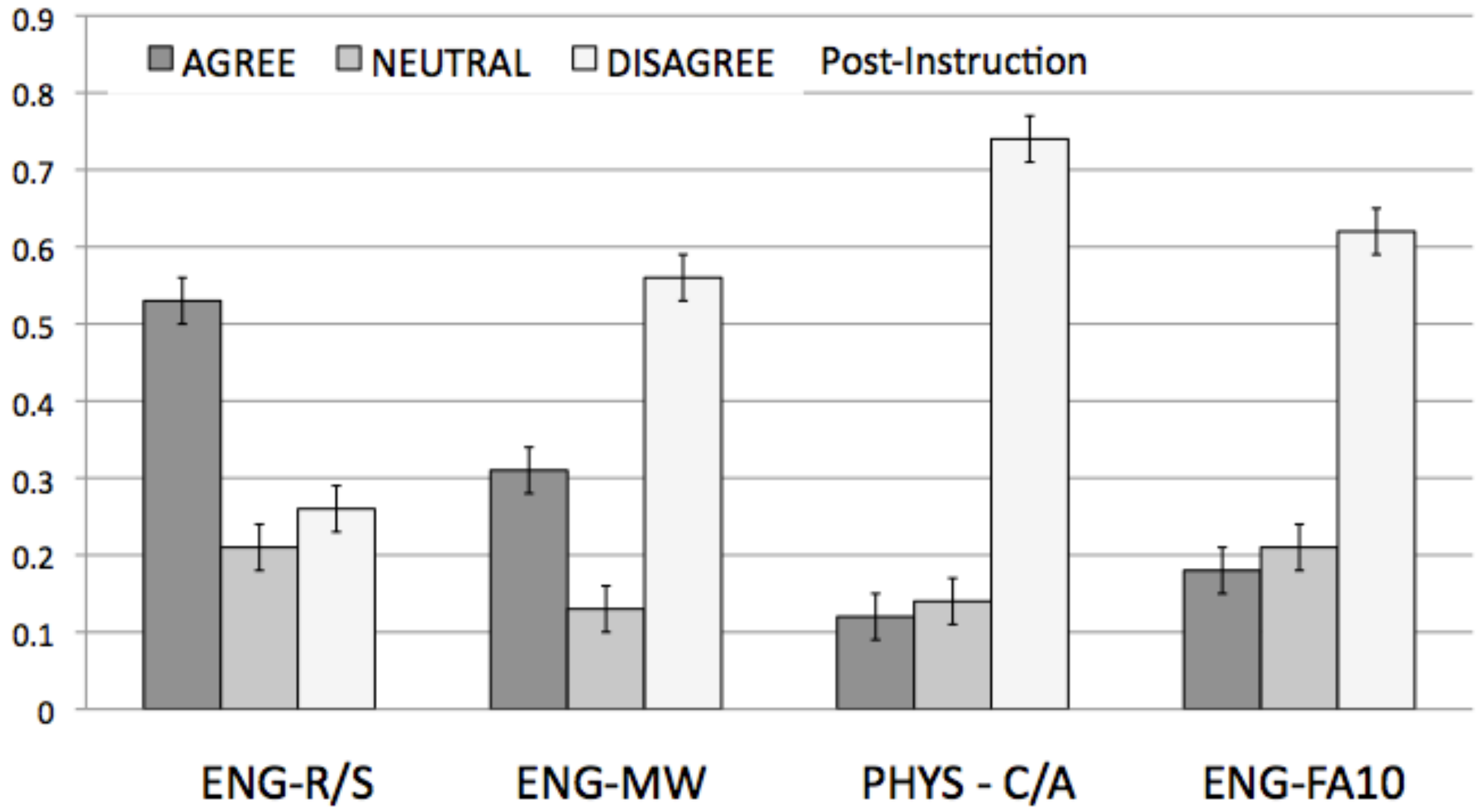

**FIG. 6.6.** Post-instruction student responses to the statement: *The probabilistic nature of quantum mechanics is mostly due to the limitations of our measurement instruments,* from four different modern physics courses, as denoted in Table 3.I. [Error bars represent the standard error on the proportion; N ~ 50-100 for each course.]

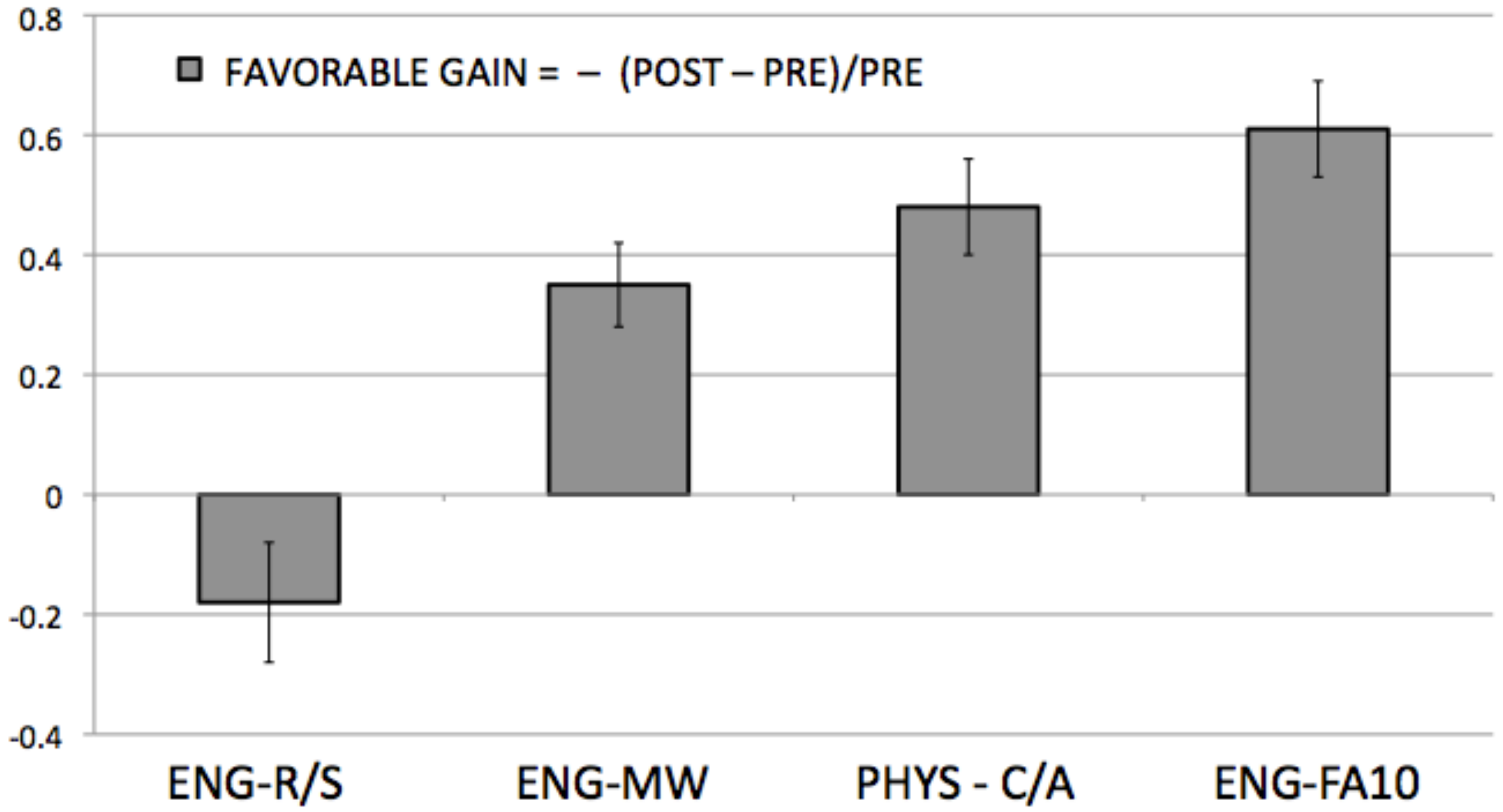

**FIG. 6.7.** Normalized favorable gain in the (post – pre) rate of student agreement with the statement: *The probabilistic nature of quantum mechanics is mostly due to the limitations of our measurement instruments,* from four different modern physics courses, as denoted in Table 3.I. A positive favorable gain is defined as a *decrease* in agreement with this statement. [N ~ 50-100 for each course.]

And finally, we didn't just give our students lots and lots of information about how to think of the wave function – we also gave them lots and lots of information about molecular bonding, conduction banding, semiconductors, transistors and diodes; as well as lasers, scanning tunneling microscopes, and nuclear energy; not to mention applications of nonlocality to quantum cryptography and computing. We had time for this because our course omitted topics from special relativity, which have generally cost other modern physics courses a minimum of three weeks out of a 15-week semester. We wouldn't claim that special relativity is not a relevant and worthy topic for engineering students, but the original decision to omit special relativity was in part a response to an overall consensus among engineering faculty at the University of Colorado, that their students would be better served by a curriculum that emphasized the quantum origins of material structures, and other real-world applications. [1] Every modern physics instructor at CU has had the option of removing special relativity from the engineering curriculum, and its re-emergence following the first round of course transformations is symbolic of a deep sense of tradition surrounding the topic, and stands in recognition of the profound influence its development has had on modern scientific thinking.

Our students had ample opportunity to contemplate the myriad contributions of Einstein's genius to the twentieth-century, but many of them were even more fascinated by the idea that Einstein could have been wrong about *anything*! And his glory was in no way diminished by telling our students this story of his confusion; for as we wove this tale of classical and quantum reality, he became a champion for those who expressed a deep commitment to their intuitions, which had become all the more apparent to them when we made their own beliefs (and not just our own) a topic of discussion. In the end, it is a question for each instructor of the pedagogical costs and benefits when deciding which story from the history of physics to tell our students, but we have made our best argument that the benefits may far outweigh any costs when we make the physical interpretation of quantum mechanics a central theme of our modern physics courses.

## III. Curriculum Refinement and Other Future Directions

For the sake of future implementations of this curriculum, efforts should be made to assess where students had the most difficulty, so that suggestions for improvement can be made. Given the volumes of data collected in this dissertation project, we must confine our discussion here to specific examples of potentially fruitful changes, and suggestions for future studies.

### III.A. Single-Photon Experiments

We begin by examining student responses to another essay question from the second midterm exam, designed to test student understanding of the single-photon experiments; we focus on this specific topic area for several reasons. First, single-quanta experiments with electrons and photons were the topics most commonly cited by students in their personal reflections as having influenced their perspectives on quantum physics, indicating this to be a key component of this curriculum's successful implementation. Second, the content of this lecture is fairly self-contained, and might easily be adapted by instructors who wish to augment their own courses without adopting the entire curriculum, and is therefore worthy of extra attention. Third, we are unaware of any instructional materials having yet been developed for introductory modern physics students concerning such experiments, and so have had no basis for judging ahead of time whether their implications for the meaning of wave-particle duality would be fully appreciated by our students.

For this midterm, students were required to answer the first essay question on interpretations of the double-slit experiment, but were given the option of answering just one of the remaining two essay questions; if students chose to answer both of the remaining two problems, they received credit for the higher of the two scores. Naturally, we will have no insight into the difficulties faced by students who opted out of answering this question, but 75% of the 103 students who took the exam did respond, which should represent a fair sampling of overall student understanding of this topic. Generally speaking, students performed well on this question: the average total score was 6.75 out of 8 points, and 85% of responses received a total score of 6 or better. We shall first give the problem statement below, and then consider individual responses of our four students (A–D) from Chapter 5. Their individual answers will help to illustrate the coding scheme that emerged in our analysis of aggregate student responses, but also the quality of responses from students with whom we are already somewhat familiar.

The beginning of each of the first two parts asks students to identify for which experimental setup, X or Y, (see below) they would expect photons to exhibit particle-like behavior, and which for wave-like behavior. Calling these two experiments X and Y (instead of 1 and 2, as in the lecture slides; see Chapter 5), and reversing their order of presentation seemed to have no impact on student responses, since all students but one were correct in their identification for each case. We felt a key step in assessing student understanding of the implications of these experiments would be to determine whether they could describe in what sense the photon is behaving like a particle or wave in each setup. We were also interested in finding out which kinds of epistemological tools would be favored by students in justifying why each type of behavior could be expected in a given situation. The final part of the this essay question concerns a delayed-choice experiment that is the reverse of the situation described during lecture: here, the second beam splitter is in place at the time a photon encounters the first beam splitter. If the second beam splitter were to be quickly removed before the photon had passed through the apparatus, there would be no opportunity for the photon to interfere with itself, meaning there is an equal likelihood for it to be detected in either photomultiplier.

**E3. (OPTION TWO – 3 PARTS, 8 POINTS TOTAL)** For the diagrams below depicting Experiments X & Y, M = Mirror, BS=Beam Splitter, PM = Photomultiplier, N = Counter. In each experiment a single-photon source sends photons to the right through the apparatus one at a time.

# EXPERIMENT X

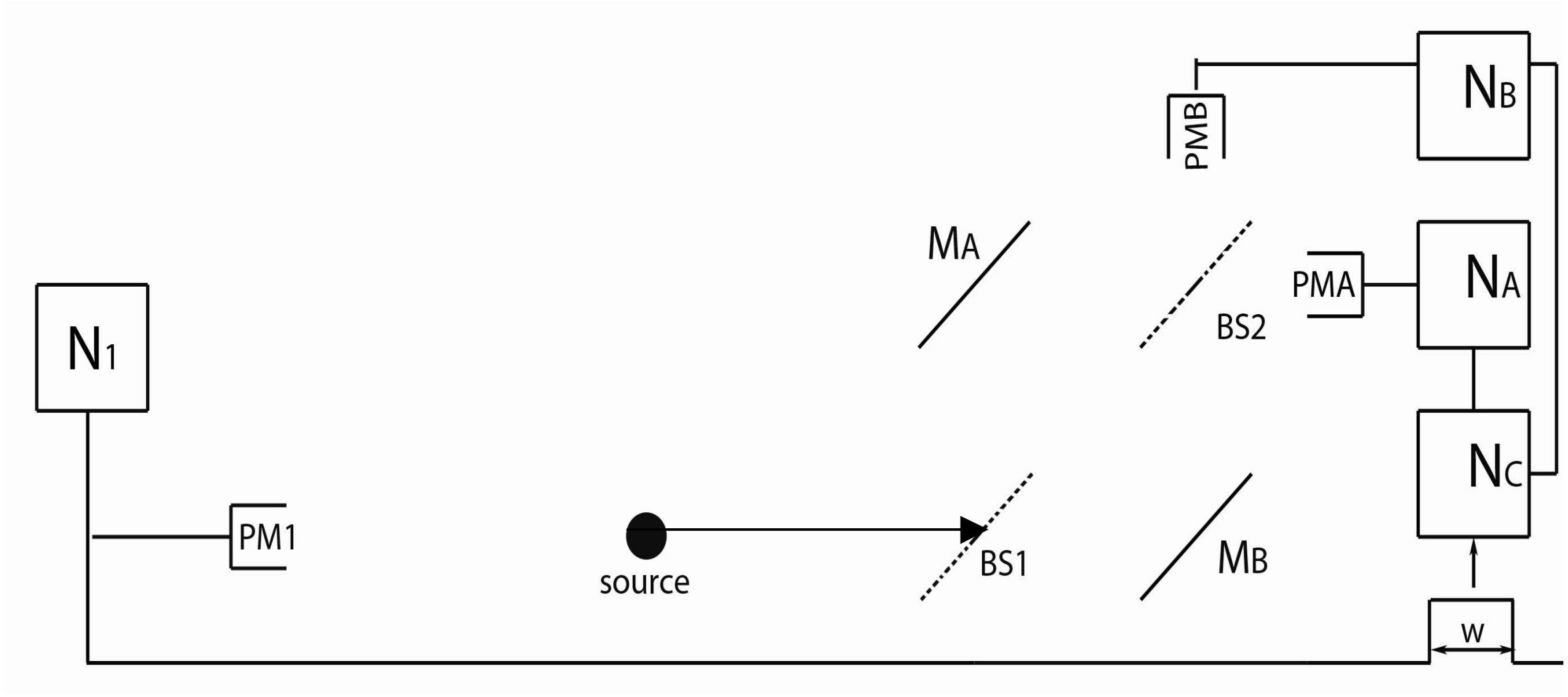

# EXPERIMENT Y

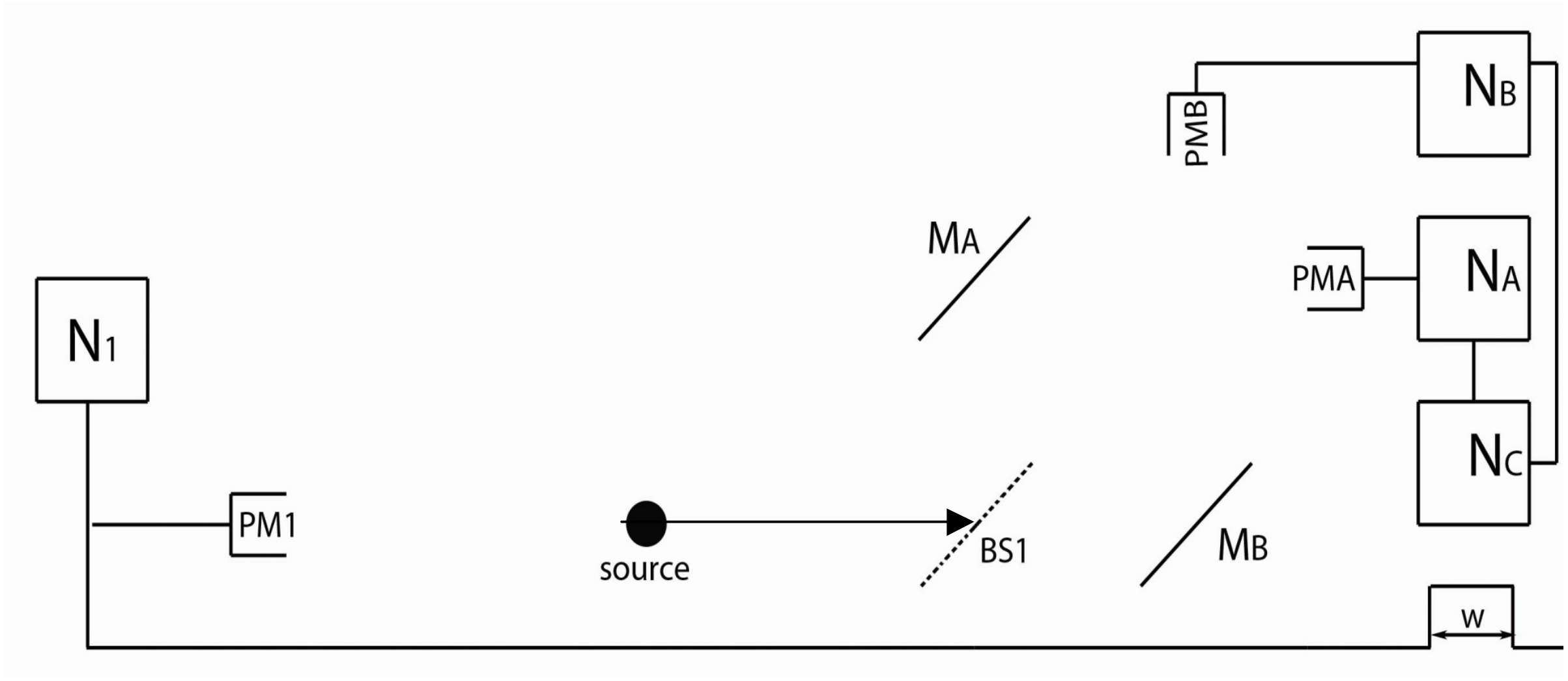

**E3.A (3 Points)** For which experimental setup (X or Y) would you expect photons to exhibit particle-like behavior? Describe in what sense the photon is behaving like a particle during this experiment. What features of the experimental setup allow you to draw this conclusion without actually conducting the experiment?

**Student A:** In setup Y, the photons exhibit particle like behavior because the photon can only have one path to get to a particular photomultiplier. I know this because beamsplitter one will either allow the photon through or reflect it. If it reflects it it will go to PMA, if it is let through it will go to PMB. It can't take Path A to get to PMB thus there is one path to take, it acts as a particle.

**Student B:** Particle-like behavior expected in setup Y. Photon's path is predictable depending on the detector in which it was detected. It either gets reflected or transmitted at BS1, thus if detected at PMA, it must have been reflected and if detected at PMB, it must have been transmitted. We also know that $\alpha = P_C/(P_A P_B) = 0$ if there is only one photon in the apparatus during the time constant. This implies that $P_C = 0$ and no wave like behavior, acts like a particle. There is only one BS, so it will act like a particle (we know this even before conducting exp.)

**Student C:** Experiment Y should show photons acting like a particle. This is due to the fact that which path the photon takes can be determined by which photomultiplier is triggered. If the photon struck mirror B, PMB will fire, if the photon struck mirror A, PMA will fire. If there was truly only a single photon in the source only one of the photomultipliers will fire, and each would fire with a 50/50 chance.

**Student D:** Experiment Y (Aspect's 1st Experiment)
The photon may take one of 2 paths, but not both, and thus travels along a defined path consistent with the behavior of a particle. The way the experiment is set up, a photon may only take one of:
source – beamsplitter – mirror A – photomultiplier A
source – beamsplitter – mirror B – photomultiplier B
If a photon is to be detected in PM1, its pair must have exited the source in exactly the opposite direction, and by geometry can only take one of the two paths listed above.

Of the three parts to this essay question, this one presented the least problems for students, and 95% of them received full credit for their responses. Students were fairly uniform in the types of argumentation and reasoning they employed, and a simple coding scheme was almost immediately apparent. Many students offered multiple justifications for their answers, and so we ranked each type of argument according to its prominence in the student's response, or by which appeared first if they seemed to carry equal weight; we report here statistics only on students' primary responses.

In describing the behavior of a photon in Experiment Y, 58% of students said that, as a particle, it is only taking one path or other on its way from source to detector (Students A & D); and 40% said its particle nature is demonstrated by being detected in either one PMT or the other, but not both (Students B & C). It seems significant that the majority of students associated particle behavior with definite trajectories (taking a single path), while fewer students associated particles with localized detections. This focus is also reflected in their identification of which features of the setup would allow them to predict particle-like behavior: 66% cited the fact that only a single path existed between source and each detector; 14% claimed the ability to determine which path a given photon had taken was sufficient for predicting this specific behavior. [16% focused on the literal difference between the two experiments – the absence of a second beam splitter.] So, a relatively small number of students relied on the new and more abstract epistemological tool developed in lecture, the availability which-path *information* as a determiner of behavior (as opposed to the existence of a single path). Not only did fewer students associate particles with localized detections, only 10% of all students made mention of measuring the anticorrelation parameter, or referred to counting rates and coincidence detections, even though these had been significant aspects of our presentation. This suggests that students are not entirely comfortable with the statistical nature of the argument for interpreting particle-like behavior in this experiment, which likely has implications for why students had greater difficulties with the flip side to this question:

**E3.B (3 Points)** For which experimental setup (X or Y) would you expect photons to exhibit wave-like behavior? Describe in what sense the photon is behaving like a wave during this experiment. What features of the experimental setup allow you to draw this conclusion without actually conducting the experiment?

**Student A:** In setup X, the photons exhibit wave like behavior because the photon can take either Path A or Path B and still get to PMA or PMB, we don't know which path it took, thus since it is unpredictable, it acts like a wave. Since it can take either path and still get to either photomultiplier, I know it can be represented as a wave.

**Student B:** Wave-like behavior expected in Setup X. Photon behaves like a wave because there is interference if we change the path length (move BS2). Thus it seems to interfere with itself. In this experiment, we can't know which path the photon takes due to the existence of BS2 (it could be detected by either PM, and have taken either path). We can also change BS2's location such that all the photons are detected in PMA or PMB. Throughout the experiment, it seems that the photon somehow "knows" that there are both paths. The BS2 lets us conclude this before starting the experiment (that it can behave like a wave).

**Student C:** Experiment X should show photons acting like waves. The path the photon took is undeterminable. Mirror B could have been hit with a photon and either PMB or PMA could fire. This implies a wave is being propagated through both possible paths. The wave then describes an equal probability of triggering each photomultiplier provided each path is the same length. Interference can happen if the paths are different length and cause only one photomultiplier to trigger.

**Student D:** Experiment X (Aspect's 2nd Experiment)
The exact path taken by the photon is rendered indeterminate by the second beamsplitter; we can't know which path the photon actually took to PMA or PMB. If we vary the path length of A or B, and observe interference as a result in the detectors, a logical explanation is that the wave that represents the photon split at beamsplitter 1, and then (due to the difference in phase created by the changed path length) interfered with itself to produce the observed results. The presence of the 2nd beamsplitter essentially randomizes whether a photon travelling along path A or B ends up in PMA or PMB (50% chance of either for fixed path length), thus rendering the path of the photon indeterminate, which allows for the above conclusions to be drawn.

Only 51% of students received full credit for their responses to this part of the question, but a total of 90% were given a score of 2/3 or better. 43% said that photons manifest their wave behavior in the form of interference (Students B, C & D), and 35% claimed that wave-like photons take both paths in this experiment. This is not precisely what Student A said – he mentions that photons are *capable* of taking both paths, but not that photons *are* taking both paths. In fact, his responses to this part of the question and the last suggest that he associates wave-like behavior with indeterminacy – photons are still presumed to take only one of two paths – it is our knowledge of which that is indefinite.

Most significant was the finding that 21% of students mistakenly believed, in Experiment X, that photons would be detected in both PMT's simultaneously; 5% explicitly stated that measuring the anticorrelation parameter as greater than one (coincidental detection) would be evidence of the photon's wave behavior in this case. In fact, for the data run presented in class demonstrating interference through path length modulation, the anticorrelation parameter was calculated to be 0.18 (less than unity, as it should be). We believe this confusion may be likely attributed to two factors. First, we only implied individual photon detections in our comparison of counting rates, but did not explicitly point out that the anticorrelation parameter had been found here to also be less than one. The specific wording of Slide 12 from this lecture [Fig. 6.8] could also be confusing for students. We want them to associate wave-like behavior in this experiment with what each photon does at the beam splitter, yet this slide could lead them to believe that wave behavior should be universally associated with coincidental detections. This misunderstanding could be directly addressed by placing greater emphasis on the connection between wave behavior and self-interference, or indefinite trajectories; and by placing greater emphasis on the continued particle-like *detection* of photons, focusing student attention instead on the behavior of the photons at each beam splitter.

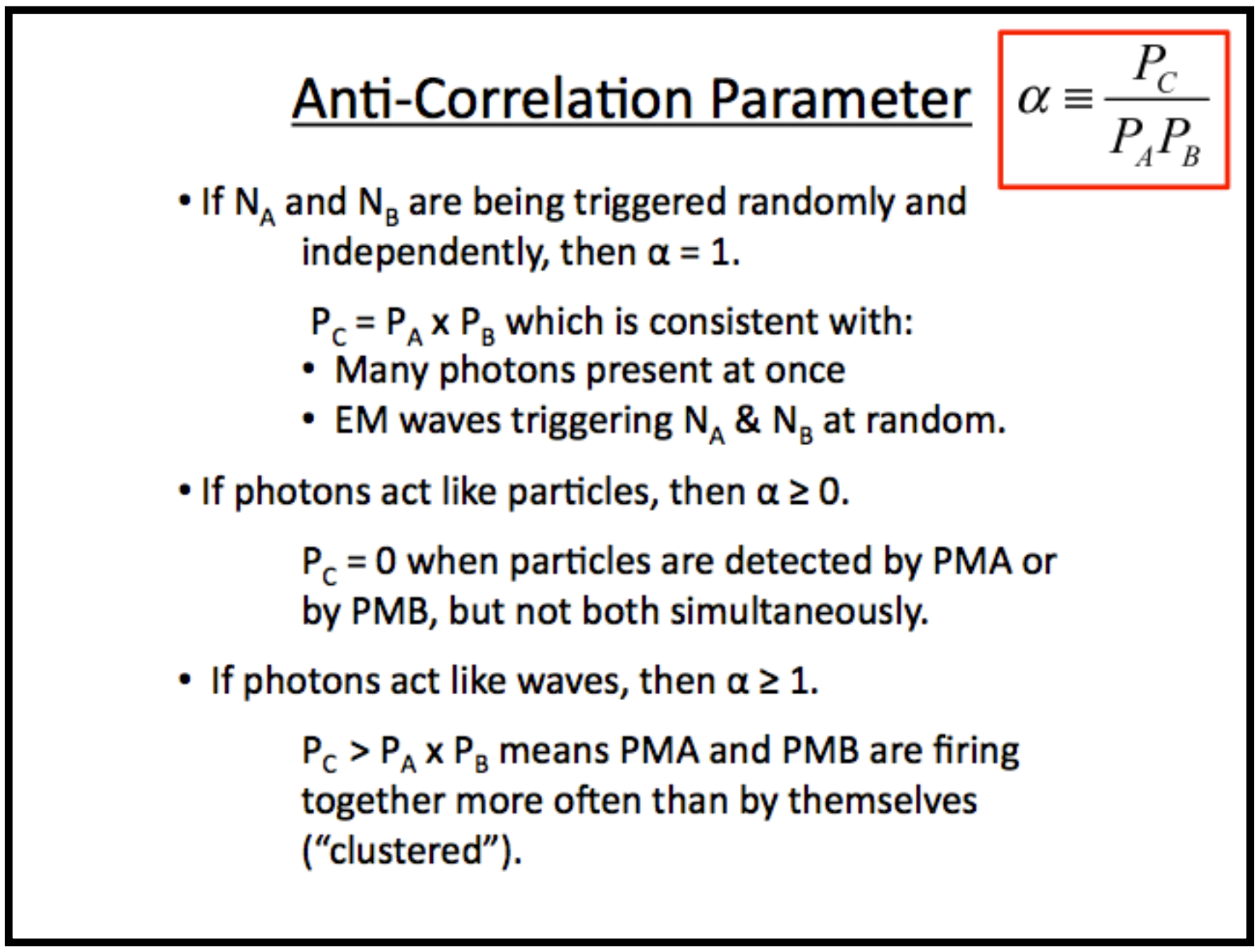

**FIG. 6.8.** Slide 12 from Lecture 20 (Single-Photon Experiments, see Chapter 5). In the first experiment, wave behavior is associated with coincidental detection; it is associated with indefinite trajectories and self-interference in the second.

We have further indication that students are uncomfortable with how wave interference is manifested in this experiment, which is different from directly observing a fringe pattern. Of all the students who mentioned interference as evidence of wave behavior, only half specifically said that it would be observed by making changes to the relative path lengths; the other half only commented that interference would be observed. Moreover, only 26% correctly spoke of interference in terms of modulated detection rates in the two photomultipliers, and 5% incorrectly believed that fixing the mirrors would cause every photon to take just one of the paths (as opposed to being detected in just one of the PMT's). Whereas only 16% of students had cited the absence of the second beam splitter as being the key feature of Experiment Y, a full 38% of students focused on its presence in Experiment X as being key to determining what kind of behavior would be observed. 36% employed an epistemological tool developed in class: no *which-path information* would be available; and 23% said the availability of two paths for the photon was key to predicting wave-like behavior in this experiment.

These results, and those from the first part of the question, suggest that students attach greater significance to the question of which path a photon takes (strong associations with particle behavior), and focus less on its behavior at the beam splitter (weak associations with wave behavior). The argument for wave behavior presented in class centered on the behavior of the photon at the beam splitter, and so perhaps this emphasis was not properly communicated to students; but we may also consider exploiting the strength of student preference for which-path arguments by giving them greater prominence in our argumentation. After all, we had been trying to develop the concept of *which-path information* as an epistemological tool, which might be aided by placing less emphasis on the response of photons to beam splitters, where students are less likely to have had any exposure to in previous classes. We had discussed them earlier in the context of the Michelson-Morley experiment, but perhaps there was insufficient connection made between the coherent 50/50 splitting of a classical EM wave, and a 50/50 probability for transmission or reflection of a photon.

Responses to the third part of this question show that the subtlety of the delayed-choice experiments was not entirely lost on students, but also provide additional evidence of student difficulties with probabilistic descriptions of measurement outcomes:

**E3.C (2 Points)** Suppose we are conducting Experiment X (the second beam splitter (BS2) is present) when a photon enters the apparatus and encounters the first beam splitter (BS1). Afterwards, while the photon is still travelling through the apparatus (but before it encounters a detector), we suddenly remove the second beam splitter (switch to Experiment Y). Can we determine the probability for the photon to be detected in PMA? If not, why not? If so, what would be that probability? Explain your reasoning.

**Student A:** No, we could not because we don't know which path the photon took, it could have taken path A in which it which it would be detected by photomultiplier A or it could have taken path B and not been detected by PMA. Since it has not been detected yet we can't determine the probability it's already on a definite path.

**Student B:** This is the delayed-choice experiment. We can indeed predict the path that the photon took if BS2 is not present depending on the detector in which it was detected. Thus, the probability of being detected in PMA would be 50/50 (0.5). It would act just as if we ran experiment Y and behave like a particle. Put the beam splitter back and it acts like a wave again. There is no "tricking" the photon!

**Student C:** First assume that experiment X is set up so that interference occurs and only PMA is firing. If the photon is still traveling through the apparatus, and BS2 is then suddenly removed, the photon will switch to acting like a particle. The photon will no longer only fire in PMA due to interference, but will instead show particle-like behavior and trigger either PMB or PMA with a 50/50 probability. BS1 results in either path from BS1 being 50/50 probable. Because when BS2 is removed, the path the photon took is now better known and particle like behavior is observed. In other words, once BS2 is removed PMB firing means MB was hit by a photon, and PMA firing means MA was hit by a photon.

**Student D:** Yes, the probability will be 0.5 – same result as Experiment X with equal path lengths, but with a definite path for any given photon. A photon may exhibit either wave-like or particle-like properties, but not both in the same instant. Removing the 2nd beamsplitter "forces" the photon to exhibit particle-like behavior by making its path definite retroactively – example of a "delayed choice" experiment.

We note that each of the four students suggested that removing the second beam splitter forces a photon into taking a definite path (not that self-interference would no longer be possible), and only Student C's response makes explicit mention of the lack of interference. Again, student associations seem to be strongest between particles and definite paths. 81% of students said that the probability for detection in PMA could be known, but only 3/4 of those students explicitly stated that probability as being 50%. Student A seems to be close to drawing this conclusion, but there appears to be a disconnect between a completely indeterminate outcome and a 50/50 likelihood for either occurrence. Regardless of whether they felt the probability could be known, almost 40% of students did not state that the probability for detection in PMA is 0.5; this suggests that students require more practice with the use of probabilities, beyond the single lecture we devoted to classical probability and probability distributions.

**II.B. Entanglement and Correlated Measurements**

The need for future studies into student difficulties with our transformed curriculum is illustrated by responses to a multiple-choice exam question concerning distant, anticorrelated measurements performed on entangled atom pairs:

**6.** Suppose we have two "Local Reality Machines" (Stern-Gerlach analyzers capable of being oriented along three different axes: A, B, & C, each oriented at $120^0$ to each other, as shown) set up to detect atom-pairs emitted in an entangled state:

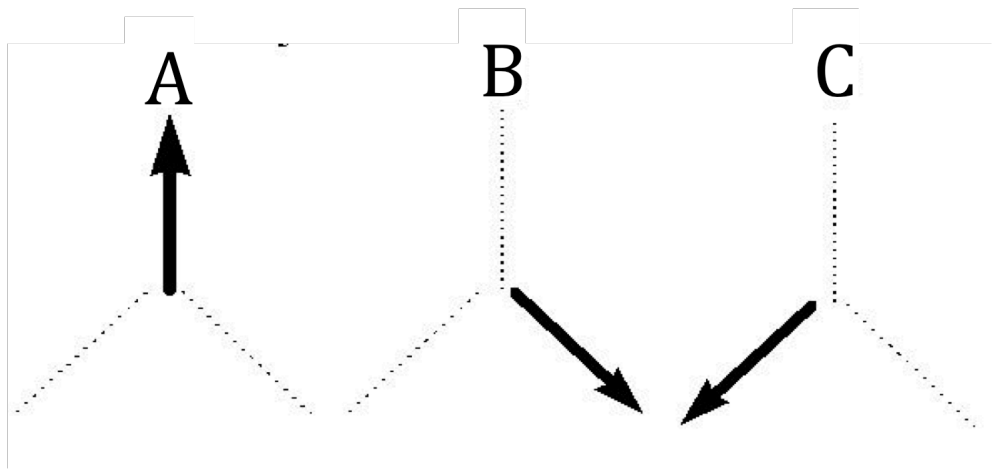

$$\left|\Psi_{12}\right\rangle = \left|\uparrow_1\right\rangle\left|\downarrow_2\right\rangle + \left|\downarrow_1\right\rangle\left|\uparrow_2\right\rangle$$

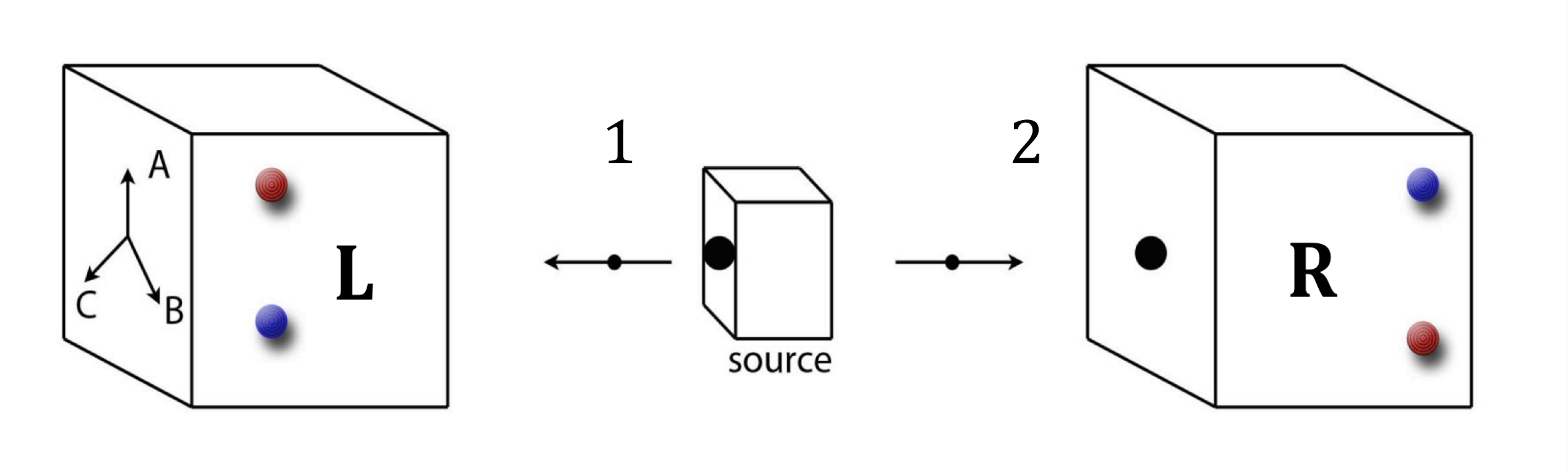

The leftward travelling atom (1) reaches the left analyzer (L) before the rightward traveling atom (2) reaches the right analyzer (R). The left analyzer is set on A and measures atom 1 to be "up" along the vertically oriented A-axis (it exited from the plus-channel). A short time later, atom 2 enters the right-side analyzer. If the right analyzer is set on B ($120^0$ from the vertical axis), what is the probability for atom 2 to exit from the plus-channel of the right analyzer?

**TABLE 6.II** Distribution of student responses to multiple-choice question #6 from the second midterm exam – correct response highlighted in bold.

| A) 0 | B) 1/4 | C) 1/2 | **D) 3/4** | E) 1 | RIGHT | WRONG |
|---|---|---|---|---|---|---|
| 14% | 34% | 12% | **42%** | 0 | 42% | 58% |

The majority of the class got this question wrong (58%, see Table 6.II), but we have little insight into the reasons for this, since each option is a significant distractor, with many potential sources of confusion. We note that no student selected option E (1), and so we may at least conclude that students did not believe that a measurement of "up" in the first detection requires an "up" measurement for the second. Option C (1/2) was chosen by 12% of students, which may indicate they did not recognize how the entangled state of the atom pair, and therefore the outcome of the first measurement, establishes a definite state of "down" for the second particle along Axis-A; this response would be correct if there were no influence of the first measurement on the second. Option B (1/4) was the most popular incorrect response, which comes from using ($120^0/2$) as the relevant angle in calculating the probability for an "up" measurement along Axis-B, when it is actually ($60^0/2$) – it varies according to the cosine squared of the half-angle between incoming state and axis of analyzer orientation. Students who correctly identified this angle may have forgotten to divide by two; or they may have correctly applied the formula, but thought the second atom would also be measured as "up" along Axis-A; or they may have simply been distracted by the prominence of the $120^0$ angle in the problem statement. Option A (0) was also a popular response (14%), which may imply these students felt that an "up" measurement for the first particle precluded an "up" measurement for the second particle along *any* axis. Adapting this specific question into a short-answer problem, where students would be required to provide their reasoning, would be a first step toward understanding some of the difficulties students have with entanglement and distant correlated measurements.

### II.C. Atomic Models and Probability

One of the questions adopted from the QMCS [1] for our post-instruction content survey was designed to elicit common student misconceptions regarding the outcome of a position measurement for an atomic electron in the ground state of hydrogen:

**30.** The electron in a hydrogen atom is in its ground state. You measure the distance of the electron from the nucleus. What will be the result of this measurement?

A. You will measure the distance to be the Bohr radius.
B. You could measure any distance between zero and infinity with equal probability.
C. You are most likely to measure the distance to be the Bohr radius, but there is a range of other distances that you could possibly measure.
D. There is a mostly equal probability of finding the electron at any distance within a range from a little bit less than the Bohr radius to a little bit more than the Bohr radius.

**TABLE 6.III** Distribution of student responses to multiple-choice question #30 from the post-instruction content survey – correct response highlighted in bold.

| A | B | **C** | D | E | %Correct | %Incorrect |
|---|---|---|---|---|---|---|
| 30% | 2% | **49%** | 18% | 0 | 49% | 51% |

An analysis of student responses to a midterm exam question on atomic models showed that only ~10% of students exclusively employed a planetary model in their descriptions of hydrogen, yet 30% of students incorrectly answered on the post-instruction survey that the electron would definitely be found at the Bohr radius, and 18% thought it was equally likely be found somewhere in that vicinity. This apparent disconnect may be explained by further difficulties students have in using probabilities to describe the outcome of quantum measurements, but it may also indicate realist commitments that were not revealed by the attitudes survey statement on atomic electrons. [See above, Section II.B.] Option D may have been popular among students that favor the de Broglie atomic model over a planetary description, but we must only speculate without the opportunity to further question students on the reasons for their responses, which is impossible in an end-of-term, multiple-choice format.

## IV. Concluding Remarks

Perhaps the most important take-home message from these studies is that students will develop their own attitudes (right or wrong, sophisticated or not) regarding the physical interpretation of quantum mechanics when we, as instructors, do not explicitly attend to the realist beliefs that are so common among our introductory modern physics students. We have frequently heard that a primary goal when introducing students to quantum mechanics is for them to recognize a fundamental difference between classical and quantum uncertainty. The notorious difficulty of this has lead many instructors to view this learning goal as superficially possible, but largely unachievable in a meaningful way for most introductory students. We believe our studies demonstrate otherwise. By addressing the physical interpretation of quantum phenomena across a variety of contexts, but also by making questions of classical and quantum reality a central theme of our course, we were able to positively influence student thinking across a variety of measures, both attitudinal and in content-specific topic areas.

We have developed a framework for understanding student interpretations of quantum mechanics, which show how their overall perspectives may be influenced by their specific attitudes toward several individual themes central to the question of probabilistic measurement outcomes. Is the wave function physically real, or a mathematical tool? Is the reduction of quantum superpositions to definite states an ad hoc rule established to make theory agree with observation, or does it represent some kind of physical transition not described by any equation? Is an electron, being a form of matter, strictly localized at all times? We have identified student attitudes regarding these questions as playing a key role when formulating their thoughts on quantum phenomena, and have seen how the myriad ways in which these attitudes may combine can lead to a variety of overall interpretive stances. If we wish to have significant influence on student perspectives, and if we are to take seriously the lessons learned from education research on the impact of *hidden curricula*, then we must choose to explicitly address these beliefs in our introductory courses.

We also believe that a *static* view of student and expert ontologies, however useful in addressing student difficulties in classical physics, is too limited to account for the contextually sensitive and highly dynamic thought processes of our students when it comes to ontological attributions. We have seen students blend attributes from the classically distinct categories of particles and waves; they may switch between views according to their cognitive needs of the moment; and they often distinguish between their intuitive perspectives, and what they have learned from authority. At the very least, we may conclude that ontological flexibility does not come easily to most students, and that the contextual sensitivity of their responses is most consistent with students engaging in a piecewise altering of their perspectives, rather than some wholesale shift (or replacement) in ontologies. Most importantly, many of our students demonstrated exactly the kind of ontological flexibility that is required for a proper understanding of quantum mechanics. We believe this learning goal is more easily achieved by placing greater emphasis on the meaning of wave-particle duality, and by providing experimental evidence that favors dualistic descriptions, but also by explicitly addressing in class the commonly held beliefs of students revealed by our studies. Among the many learning goals for our transformed curriculum was for students to be consciously aware of their own (often intuitive and tacit) beliefs, but also for them to acquire the necessary language and conceptual inventory to identify and articulate those beliefs. This was accomplished in part by presenting them with specific terminology relevant to perceptions of reality and locality, but also by making the beliefs of students (and not just the beliefs of scientists) a topic of discussion in our course.

It would be too simplistic to say that our aim was for students to consistently *not agree* with realist interpretations of quantum phenomena. After all, there are a variety of situations in quantum mechanics where the physical interpretation of the wave function has no relevance or bearing on the outcome of a calculation. It is not that a particle view of matter is entirely illegitimate in quantum mechanics; it is simply that its consistent application in all contexts is not adequate in accounting for all of what we observe in nature. We suggest that a significant amount of the confusion introductory students feel when learning about quantum mechanics results from the paradoxical conclusions that come as a consequence of realist expectations and ontological inflexibility.

Nor would we wish to connote too much negativity with the fact that students are relying on their intuition as a form of sense making. It is true we are telling them that their everyday thinking can be misleading in quantum physics, but that is not a sufficient argument for the wholesale abandonment of productive epistemological tools. Indeed, our approach to teaching quantum interpretations frequently required an appeal to student intuitions about the classical behavior of particles (they are transmitted or reflected; they are localized upon detection), and similarly with waves. A more important goal is for students to achieve more internal consistency in their thinking, which may be cultivated by developing epistemological tools that aid in deciding which type of behavior should be expected in which type of situation. Considering the observed strong associations students make between particles and definite paths, it seems that framing such tools in terms of *which-path information* [two paths = interference; one path = no interference] may be particularly useful for students.

Of the many potential studies that might be conducted as an improvement on those presented here, we believe that focusing on the thinking associated with *Agnostic* students would be particularly beneficial. We have never considered an agnostic perspective to be unsophisticated; in fact, our *Agnostic* category [as defined in Chapter 4] was meant to include both students *and* experts who acknowledge the potential legitimacy of competing perspectives, without taking a definitive stance. Agnosticism, by this definition, would therefore involve an acknowledgement of evidence that favors more than just a single interpretation, which is clearly different from students who exclusively assert the legitimacy of their realist intuitions *in spite of* evidence to the contrary. At the same time, an agnostic stance may be indicative of the perception that nothing can truly be known or understood in science, since many of the assumptions we make about the world turn out to be demonstrably false, and so much in quantum physics cannot be directly observed. Either way, an agnostic stance among students may be interpreted as an intermediary stage in the transition away from realism, but might also signal unfavorable perceptions on the *nature of science*. Negative perceptions about what can and can't be known in science might be an unintended consequence of our curriculum transformations, and require further detailed consideration.

## References (Chapter 6)

**1.** S. B. McKagan, K. K. Perkins and C. E. Wieman, Reforming a large lecture modern physics course for engineering majors using a PER-based design, *PERC Proceedings 2006* (AIP Press, Melville, NY, 2006).

**2.** S. B. McKagan, K. K. Perkins and C. E. Wieman, Why we should teach the Bohr model and how to teach it effectively, *Phys. Rev. ST: Physics Education Research* **4**, 010103 (2008).

**3.** E. Redish, J. Saul and R. Steinberg, Student expectations in introductory physics, *Am. J. Phys.* **66**, 212 (1998).

**4.** W. K. Adams, K. K. Perkins, N. Podolefsky, M. Dubson, N. D. Finkelstein and C. E. Wieman, A new instrument for measuring student beliefs about physics and learning physics: the Colorado Learning Attitudes about Science Survey, *Phys. Rev. ST: Physics Education Research* **2**, 1, 010101 (2006).

**5.** G. Greenstein and A. J. Zajonc, The Quantum Challenge: Modern Research on the Foundations of Quantum Mechanics, 2$^{nd}$ ed. (Jones & Bartlett, Sudbury, MA, 2006).

# BIBLIOGRAPHY

W. K. Adams, K. K. Perkins, N. Podolefsky, M. Dubson, N. D. Finkelstein and C. E. Wieman, A new instrument for measuring student beliefs about physics and learning physics: the Colorado Learning Attitudes about Science Survey, *Phys. Rev. ST: Physics Education Research* **2**, 1, 010101 (2006).

D. Z. Albert and R. Galchen, A Quantum Threat to Special Relativity, *Scientific American* (March 2009, pp. 32-39).

A. Aspect, P. Grangier and G. Roger, Experimental Tests of Realistic Local Theories via Bell's Theorem, *Phys. Rev. Letters* **47**, 460 (1981).

C. Baily and N. D. Finkelstein, Development of quantum perspectives in modern physics, *Phys. Rev. ST: Physics Education Research* **5**, 010106 (2009).

C. Baily and N. D. Finkelstein, Teaching and understanding of quantum interpretations in modern physics courses, *Phys. Rev. ST: Physics Education Research* **6**, 010101 (2010).

C. Baily and N. D. Finkelstein, Refined characterization of student perspectives on quantum physics, *Phys. Rev. ST: Physics Education Research* **6**, 020113 (2010).

L. E. Ballentine, The Statistical Interpretation of Quantum Mechanics, *Rev. Mod. Phys.* **42**, 358 (1970).

L. E. Ballentine, Resource letter IQM-2: Foundations of quantum mechanics since the Bell inequalities, *Am. J. Phys.* **55**, 785 (1987).

L. E. Ballentine, Quantum Mechanics: A Modern Development (World Scientific Publishing, Singapore, 1998).

J. S. Bell, *Speakable and Unspeakable in Quantum Mechanics* (Cambridge University Press, Cambridge, 1987).

D. Bohm and Y. Aharonov, Discussion of Experimental Proof for the Paradox of Einstein, Rosen, and Podolsky, *Phys. Rev.* **108**, 1070 (1957).

D. Bohm and B. J. Hiley, *The Undivided Universe* (Routledge, New York, NY, 1995).

N. Bohr, Can quantum mechanical description of physical reality be considered complete? *Phys. Rev.* **48**, 696 (1935).

A. Buffler, S. Allie, F. Lubben and B. Campbell, The development of first year physics students' ideas about measurement in terms of point and set paradigms, *Int. J. Sci. Educ.* **23,** 11 (2001).

O. Carnal and J. Mlynek, Young's Double-Slit Experiment with Atoms: A Simple Atom Interferometer, *Phys. Rev. Lett.* **66** (21), 2689 (1991).

E. Cataloglu and R. Robinett, Testing the development of student conceptual and visualization understanding in quantum mechanics through the undergraduate career, *Am. J. Phys.* **70**, 238 (2002).

M. T. H. Chi, Common sense misconceptions of emergent processes: Why some misconceptions are robust, *Journal of the Learning Sciences* **14**, 161 (2005).

J. F. Clauser, M. A. Horne, A. Shimony and R. A. Holt, Proposed Experiment to Test Local Hidden-Variable Theories, *Phys. Rev. Lett.* **23**, 880 (1969).

J. G. Cramer, The Transactional Interpretation of Quantum Mechanics, *Rev. Mod. Phys.* **58**, 647 (1986).

J. W. Creswell, *Education Research*, 2nd Ed. (Prentice Hall, Englewood Cliffs, NJ, 2005), pp. 397-398.

C. Davisson and L. H. Germer, Diffraction of electrons by a crystal of nickel, *Phys. Rev.* **30**, 705 (1927).

P. A. M. Dirac, *The Principles of Quantum Mechanics, 3rd ed.*, (Clarendon Press, Oxford, 1947).

A. A. diSessa, "A history of conceptual change research: Threads and fault lines," in *Cambridge Handbook of the Learning Sciences*, K. Sawyer (Ed.) (Cambridge University Press, Cambridge, 2006), pp. 265-281.

M. Dubson, S. Goldhaber, S. Pollock and K. Perkins, Faculty Disagreement about the Teaching of Quantum Mechanics, *PERC Proceedings 2009* (AIP, Melville, NY, 2009).

A. Einstein, "Autobiographical Notes" in *Albert Einstein: Philosopher-Scientist*, P. A. Schillp (Ed.) (Open Court Publishing, Peru, Illinois 1949).

A. Einstein, *Letters on Wave Mechanics*, K. Przibram (Ed.) (Philosophical Library, New York, NY, 1986).

A. Einstein, B. Podolsky and N. Rosen, Can quantum mechanical description of physical reality be considered complete? *Phys. Rev.* **47**, 777 (1935).

B.-G. Englert, M. O. Scully and H. Walther, The Duality in Matter and Light, *Scientific American* (December 1994, pp. 86-92).

S. J. Freedman and J. F. Clauser, Experimental test of local hidden-variable theories, *Phys. Rev. Lett.* **28**, 938 (1972).

J. Falk, Developing a quantum mechanics concept inventory, unpublished master thesis, Uppsala University, Uppsala, Sweden (2005). Available at: http://johanfalk.net/node/87 (Retrieved January, 2011).

S. Frabboni, G. C. Gazzadi, and G. Pozzi, Nanofabrication and the realization of Feynman's two-slit experiment, *App. Phys. Letters* **93**, 073108 (2008).

R. Gaehler and A. Zeilinger, Wave-optical experiments with very cold neutrons, *Am. J. Phys.* **59** (4), 316 (1991).

S. Goldhaber, S. Pollock, M. Dubson, P. Beale and K. Perkins, Transforming Upper-Division Quantum Mechanics: Learning Goals and Assessment, *PERC Proceedings 2009* (AIP, Melville, NY, 2009).

P. Grangier, G. Roger, and A. Aspect, Experimental Evidence for a Photon Anticorrelation Effect on a Beam Splitter: A New Light on Single-Photon Interferences, *Europhysics Letters* **1**, 173 (1986).

K. E. Gray, W. K. Adams, C. E. Wieman and K. K. Perkins, Students know what physicists believe, but they don't agree: A study using the CLASS survey, *Phys. Rev. ST: Physics Education Research* **4**, 020106 (2008).

B. R. Greene, *The Elegant Universe* (Norton, New York, NY, 2003).

G. Greenstein and A. J. Zajonc, The Quantum Challenge: Modern Research on the Foundations of Quantum Mechanics, 2nd ed. (Jones & Bartlett, Sudbury, MA, 2006).

D. J. Griffiths, *Introduction to Quantum Mechanics, 2nd Ed.* (Prentice Hall, Upper Saddle River, New Jersey, 2004).

A. Gupta, D. Hammer and E. F. Redish, The case for dynamic models of learners' ontologies in physics, *J. Learning Sciences* **19**, 285 (2010).

R. Hake, Interactive-Engagement Versus Traditional Methods: A Six-Thousand-Student Survey of Mechanics Test Data for Introductory Physics Courses, *Am. J. Phys.* **66** (1), 64 (1998).

D. Hammer, Student resources for learning introductory physics, *Am. J. Phys.* **68**, S52 (2000).

D. Hammer, A. Elby, R. E. Scherr and E. F. Redish, "Resources, Framing and Transfer" in *Transfer of Learning*, J. Mestre (Ed.) (Information Age Publishing, 2005) pp. 89-119.

D. Hammer, A. Gupta and E. F. Redish, On Static and Dynamic Intuitive Ontologies, *J. Learning Sciences* **20**, 163 (2011).

S. W. Hawking, *A Brief History of Time* (Bantam, New York, NY, 1988).

W. Heisenberg, The Physical Content of Quantum Kinematics and Mechanics, in *Quantum Theory and Measurement*, J. A. Wheeler and W. H. Zurek (Eds.) (Princeton, NJ, 1983) p. 62. Originally published under the title, "Uber den anschaulichen Inhalt der quantentheoretischen Kinematik und Mechanik," *Zeitschrift fur Physik*, **43**, 172 (1927); translation into English by J. A. Wheeler and W. H. Zurek, 1981.

T. Hellmuth, H. Walther, A. Zajonc and W. Schleich, Delayed-choice experiments in quantum interference, *Phys. Rev. A* **35**, 2532 (1987).

D. Hestenes, M. Wills and G. Swackhamer, *Physics Teacher* **30**, 141 (1992).

R. D. Knight, *Physics for Scientists and Engineers: A strategic approach*, 2nd Ed. (Addison Wesley, San Francisco, CA, 2004)

L. Lising and A. Elby, The impact of epistemology on learning: A case study from introductory physics, *Am. J. Phys.* **73**, 372 (2005).

K. Mannila, I. T. Koponen and J. A. Niskanen, Building a picture of students' conceptions of wave- and particle-like properties of quantum entities, *Euro. J. Phys.* **23**, 45-53 (2002).

A. R. Marlow (Ed.), *Mathematical Foundations of Quantum Mechanics* (Academic Press, New York, NY, 1978).

E. Mazur, *Peer instruction: A user's manual* (Prentice Hall, New York, NY, 1997).

T. L. McCaskey, M. H. Dancy and A. Elby, Effects on assessment caused by splits between belief and understanding, *PERC Proceedings 2003* **720**, 37 (AIP, Melville, NY, 2004).

T. L. McCaskey and A. Elby, Probing Students' Epistemologies Using Split Tasks, *PERC Proceedings 2004* **790**, 57 (AIP, Melville, NY, 2005).

L. C. McDermott, Oersted Medal Lecture 2001: 'Physics Education Research – The Key to Student Learning', *Am. J. Phys.* **69**, 1127 (2001).

S. B. McKagan and C. E. Wieman, Exploring Student Understanding of Energy Through the Quantum Mechanics Conceptual Survey, *PERC Proceedings 2005* (AIP Press, Melville, NY, 2006).

S. B. McKagan, K. K. Perkins and C. E. Wieman, Reforming a large lecture modern physics course for engineering majors using a PER-based design, *PERC Proceedings 2006* (AIP Press, Melville, NY, 2006).

S. B. McKagan, K. K. Perkins, M. Dubson, C. Malley, S. Reid, R. LeMaster and C. E. Wieman, Developing and Researching PhET simulations for Teaching Quantum Mechanics, *Am. J. Phys.* **76**, 406 (2008).

S. B. McKagan, K. K. Perkins and C. E. Wieman, Why we should teach the Bohr model and how to teach it effectively, *Phys. Rev. ST: Physics Education Research* **4**, 010103 (2008).

N. D. Mermin, Is the moon there when nobody looks? Reality and the quantum theory, *Phys. Today* **38** (4), 38 (1985).

N. D. Mermin, What's Wrong with this Pillow? *Phys. Today* **42**, 9 (1989).

R. G. Newton, *How Physics Confronts Reality: Einstein was correct, but Bohr won the game* (World Scientific, Singapore, 2009).

V. K. Otero and K. E. Gray, Attitudinal gains across multiple universities using the Physics and Everyday Thinking curriculum, *Phys. Rev. ST: Physics Education Research* **4** (1), 020104 (2008).

V. Otero, S. Pollock and N. Finkelstein, A Physics Department's Role in Preparing Physics Teachers: The Colorado Learning Assistant Model, *Amer. J. Phys.* **78**, 1218 (2010).

A. Pais, *Niels Bohr's Times, in Physics, Philosophy and Polity* (Clarendon Press, Oxford, 1991).

G. J. Posner, K. A. Strike, P. W. Hewson and W. A. Gertzog, Accommodation of a scientific conception: Toward a theory of conceptual change, *Sci. Educ.* **66**, 211 (1982).

S. Pollock, No Single Cause: Learning Gains, Student Attitudes, and the Impacts of Multiple Effective Reforms, *PERC Proceedings 2004* (AIP Press, Melville, NY, 2005).

S. Pollock and N. D. Finkelstein, Sustaining Change: Instructor Effects in Transformed Large Lecture Courses, *PERC Proceedings 2006* (AIP Press, Melville, NY, 2006).

E. Redish, J. Saul and R. Steinberg, Student expectations in introductory physics, *Am. J. Phys.* **66**, 212 (1998).

S. Redner, Citation Statistics from 110 Years of Physical Review, *Phys. Today* **58**, 52 (2005).

M. Reiner, J. D. Slotta, M. T. H. Chi and L. B. Resnick, Naïve physics reasoning: A commitment to substance-based conceptions, *Cognition and Instruction* **18**, 1 (2000).

A. L. Robinson, Quantum Mechanics Passes Another Test, *Science* **217**, 435 (1982).

A. L. Robinson, Loophole Closed in Quantum Mechanics Test, *Science* **219**, 40 (1983).

E. Schrödinger, "The Present Situation in Quantum Mechanics," in *Quantum Theory and Measurement*, J. A. Wheeler and W. H. Zurek (Eds.) (Princeton, NJ, 1983) p. 152. Originally published under the title, "Die gegenwärtige Situation in der Quantenmechanik," *Die Naturwissenschaften*, **23**, 807 (1935); translation into English by J. D. Trimmer, 1980.

A. Shimony, The Reality of the Quantum World, *Scientific American* (January 1988, pp. 46-53).

C. Singh, Student understanding of quantum mechanics, *Am. J. Phys.* **69**, 8 (2001).

C. Singh, Assessing and improving student understanding of quantum mechanics, *PERC Proceedings 2006* (AIP Press, Melville, NY, 2006).

C. Singh, Student understanding of quantum mechanics at the beginning of graduate instruction, *Am. J. Phys.* **76**, 3 (2008).

J. D. Slotta, In defense of Chi's Ontological Incompatibility Hypothesis, *Journal of the Learning Sciences* **20**, 151 (2011).

J. D. Slotta and M. T. H. Chi, The impact of ontology training on conceptual change: Helping students understand the challenging topics in science, *Cognition and Instruction* **24,** 261 (2006).

W. H. Stapp, The Copenhagen Interpretation, *Am. J. Phys.* **40**, 1098 (1972).

D. F. Styer, *The Strange World of Quantum Mechanics* (Cambridge University Press, Cambridge, 2000).

M. Tegmark and J. A. Wheeler, 100 Years of Quantum Mysteries, *Scientific American* (February 2001, pp. 68-75).

W. Tittel, J. Brendel, B. Gisin, T. Herzog, H Zbinden and N. Gisin, Experimental demonstration of quantum correlations over more than 10 km, *Phys. Rev. A* **57** (5), 3229 (1998).

A. Tonomura, J. Endo, T. Matsuda, T. Kawasaki and H. Exawa, Demonstration of single-electron buildup of an interference pattern, *Am. J. Phys.* **57**, 117 (1989).

N. G. van Kampen, The Scandal of Quantum Mechanics, Am. J. Phys **76**, 989 (2008).

J. von Neumann, "Measurement and Reversibility" and "The Measuring Process" in *Mathematical Foundations of Quantum Mechanics* (Princeton University Press, Princeton, New Jersey, 1955), pp. 347-445.

A. Watson, Quantum Spookiness Wins, Einstein Loses in Photon Test, *Science* **277**, 481 (1997).

A. Whitaker, *Einstein, Bohr and the Quantum Dilemma: From Quantum Theory to Quantum Information* (Cambridge University Press, Cambridge, 1996).

S. Wuttiprom, M. D. Sharma, I. D. Johnston, R. Chitaree and C. Chernchok, Development and Use of a Conceptual Survey in Introductory Quantum Physics, *Int. J. Sci. Educ.* **31 (5)**, 631 (2009).

# APPENDIX A

## Evolution of Online Survey Items

**SURVEY INSTRUCTIONS (FA08, SP09, FA09, SP10, FA10):**

Below are several statements that may or may not describe your beliefs or opinions.

You are asked to rate each statement by selecting a number between 1 and 5 where the numbers mean the following:

1. Strongly Disagree
2. Disagree
3. Neutral
4. Agree
5. Strongly Agree

Choose one of the above five choices that best expresses your feeling about the statement. If you have no strong opinion either way, choose 3.

A textbox follows each statement, asking you to explain the reasoning behind your answer (e.g. what is going through your mind as you formulate a response). Please note that, for some of the statements, the wording is deliberately ambiguous; we are particularly interested in how you interpret each statement, and any argumentation in support of your response.

WE ARE ASKING THAT YOU EXPRESS YOUR OWN BELIEFS. YOUR SPECIFIC ANSWERS WILL NOT AFFECT ANY EVALUATION OF YOU AS A STUDENT.

**SURVEY STATEMENTS:**

**1.** It is possible for physicists to carefully perform the same experiment and get two very different results that are both correct. **(FA08, SP09, FA09, SP10)**

**1.** It is possible for physicists to carefully perform the same measurement and get two very different results that are both correct. **(FA10)**

**2.** Uncertainty in quantum mechanics is mostly due to the limited accuracy of our measurement instruments. **(FA08)**

**2.** The probabilistic nature of quantum mechanics is mostly due to physical limitations of our measurement instruments. **(SP09, FA09, SP10)**

**2.** The probabilistic nature of quantum mechanics is mostly due to the limitations of our measurement instruments. **(FA10)**

**3.** An electron in an atom has a definite but unknown position at each moment in time. **(FA08, SP09)**

**3.** When not being observed, an electron in an atom still exists at a definite (but unknown) position at each moment in time. **(FA09, SP10, FA10)**

**4.** I think quantum mechanics is an interesting subject. **(FA08, SP09, FA09, SP10, FA10)**

**5.** I have heard about quantum mechanics through popular venues (books, films, websites, etc...) **(FA08, SP09, FA09, SP10, FA10)**

**<u>DOUBLE-SLIT ESSAY QUESTION</u>:**

PLEASE NOTE: As before, there are no "right" or "wrong" answers to the following question. We are interested in what you actually believe. Play with the Quantum Wave Interference simulation, at:
http://phet.colorado.edu/new/simulations/sims.php?sim=QWI

In particular, select the "Single Particles" tab at the top of the screen, and then click on the "Double Slits" button at the right of the display. Select "Electrons", and then fire the gun... Now consider the following:

**1.** A blob (representing the probability density for a single electron) emerges from an "electron gun".

**2.** The blob passes through the two slits

**3.** A single electron is detected on the far screen. After many electrons, an interference pattern develops.

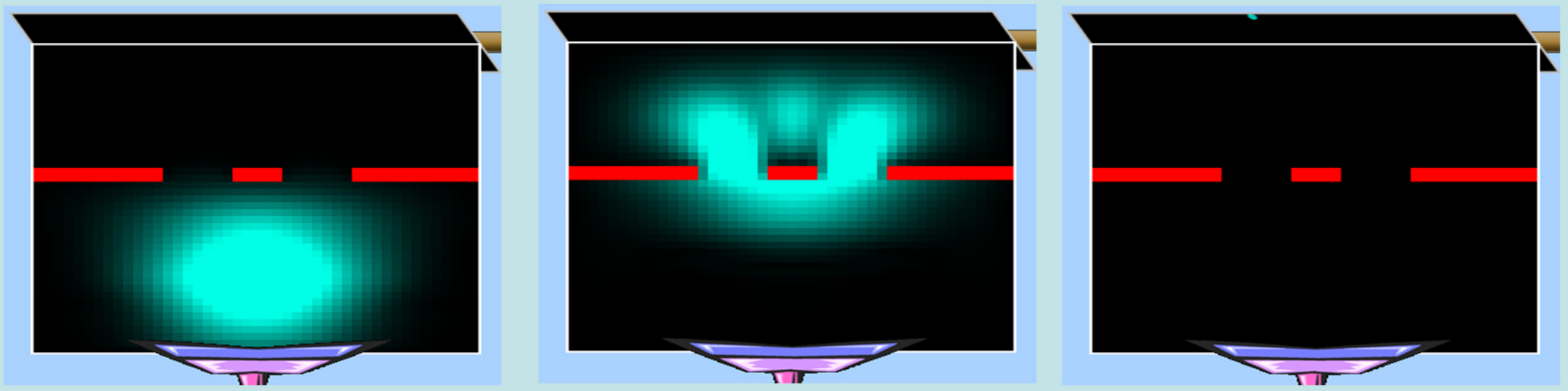

Three students discuss this Quantum Wave Interference simulation:

**<u>Student 1</u>:**

That blob represents the probability density, so it tells you the probability of where the electron could have been before it hit the screen. We don't know where it was in that blob, but it must have actually been a tiny particle that was traveling in the direction it ended up, somewhere within that blob. **(SP08)**

The probability density is so large because we don't know the true position of the electron. Since only a single dot at a time appears on the detecting screen, the electron must have been a tiny particle, traveling somewhere inside that blob, so that the electron went through one slit or the other on its way to the point where it was detected. **(FA08, SP09, FA09, SP10, FA10)**

**Student 2**:

No, the electron isn't inside the blob, the blob represents the electron! It's not just that we don't know where it is, but that it isn't in any one place. It's really spread out over that large area up until it hits the screen. **(SP08)**

The blob represents the electron itself, since an electron is described by a wave packet that will spread out over time. The electron acts as a wave and will go through both slits and interfere with itself. That's why a distinct interference pattern will show up on the screen after shooting many electrons. **(FA08, SP09, SP10, FA10)**

I think the blob represents the electron itself, since a free electron is properly described by a wave packet. The electron acts as a wave and will go through both slits and interfere with itself. That's why a distinct interference pattern will show up on the screen after shooting many electrons. **(FA09)**

**Student 3**:

Quantum mechanics says we'll never know for certain, so you can't ever say anything at all about where the electron is before it hits the screen. **(SP08)**

Quantum mechanics is only about predicting the outcomes of measurements, so we really can't know anything about what the electron is doing between being emitted from the gun and being detected on the screen. **(FA08, SP09)**

All we can really know is the probability for where the electron will be detected. Quantum mechanics may predict the likelihood for a measurement outcome, but it really doesn't tell us what the electron is doing between being emitted from the gun and being detected at the screen. **(FA09, FA10)**

Quantum mechanics lets us predict the interference pattern, but I think we really can't know or say anything about what each electron was doing between being emitted by the gun and being detected on the screen. **(SP10)**

# APPENDIX B

## Interview Protocol

BACKGROUND INFORMATION (name, declared major, previous physics and mathematics courses, both at CU and in high school, motivations for enrolling in the course)

ASK STUDENTS TO DESCRIBE AN ELECTRON

DESCRIBE AN ELECTRON IN AN ATOM
(Do students use a planetary model as a first-pass description? Are students aware of the limitations of the Bohr model? Do electrons move in orbits as localized particles? Does the student describe the electron in terms of an *electron cloud*, or a *cloud of probability*? What is this cloud? Does it represent something physical, or is it a mathematical tool? If the electron is described as a wave, what is it that's waving? Is there something moving up and down in space?)

RESPOND TO THE STATEMENT:
*An electron in an atom has a definite but unknown position at each moment in time.*
IN AGREEMENT OR DISAGREEMENT AND EXPLAIN REASONING
(Is the student's response consistent with their earlier descriptions of atomic electrons?

DESCRIBE THE SETUP FOR THE DOUBLE-SLIT EXPERIMENT
(What is observed? Can the experiment be run with both light and electrons? What is observed when only single quanta pass through the apparatus at a time? What happens if you block one of the slits? What happens if you place a detector at one of the slits to see which slit individual quanta passed through? How do students explain the fringe pattern? If they explain the experiment in terms of localized particles, what is the source of interference? If they prefer a wave-description of quanta, how did the student explain why single quanta are detected as localized points? Is it possible for a particle to pass through two slits simultaneously? Does a wave packet description of individual particles reflect an ignorance of that particle's true position or momentum?

RESPOND TO THE ONLINE SURVEY QUESTION ON THE DOUBLE-SLIT EXPERIMENT:

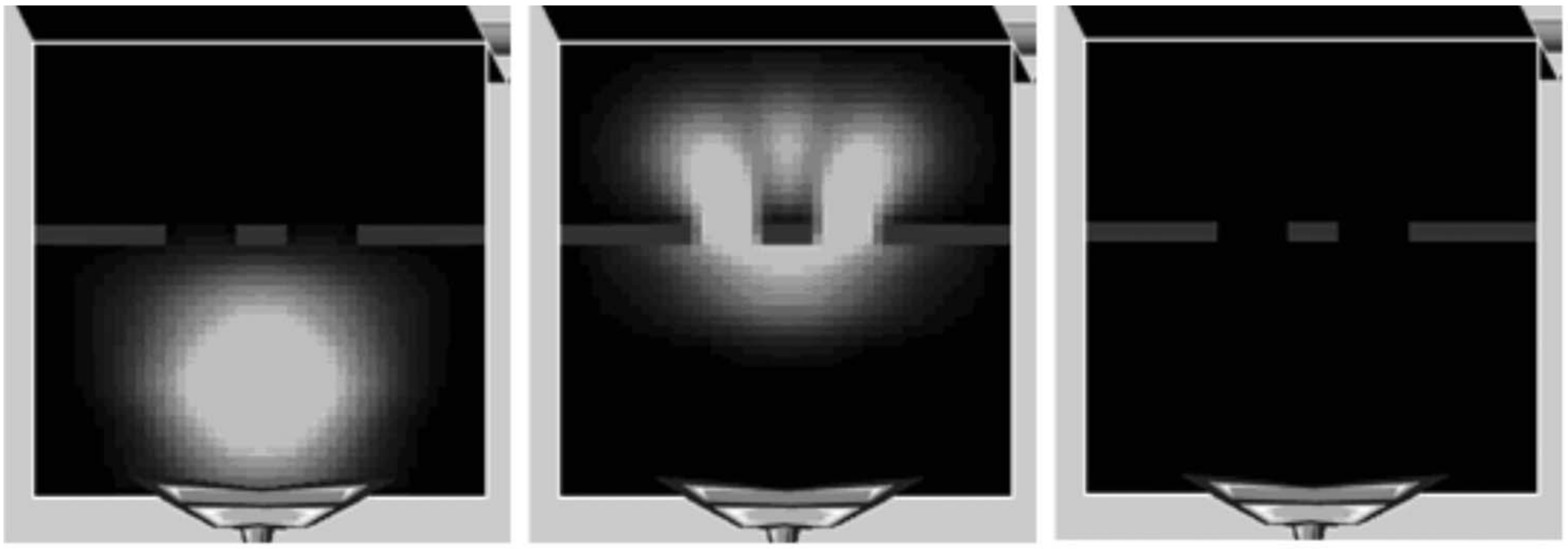

(A) (B) (C)

A sequence of screen shots from the Quantum Wave Interference simulation. A bright spot (representing the probability density for a single electron) emerges from an electron gun (A), passes through both slits (B), and a single electron is detected on the far screen (C). After many electrons, a fringe pattern develops (not shown).

**Three students discuss the Quantum Wave Interference simulation:**

**Student 1**: The probability density is so large because we don't know the true position of the electron. Since only a single dot at a time appears on the detecting screen, the electron must have been a tiny particle, traveling somewhere inside that blob, so that the electron went through one slit or the other on its way to the point where it was detected.

**Student 2**: The blob represents the electron itself, since an electron is described by a wave packet that will spread out over time. The electron acts as a wave and will go through both slits and interfere with itself. That's why a distinct interference pattern will show up on the screen after shooting many electrons.

**Student 3**: Quantum mechanics is only about predicting the outcomes of measurements, so we really can't know anything about what the electron is doing between being emitted from the gun and being detected on the screen.

(Ask students to read each statement one at a time, and respond before moving on to the next statement. Are student responses to the essay question consistent with their earlier descriptions of the experiment? Are student responses consistent with their earlier descriptions of atomic electrons? If not, why not? Is the student aware of inconsistencies?)

QUESTIONS REGARDING INTERPRETATIONS OF QUANTUM MECHANICS
(Is the student aware there are multiple interpretations of quantum mechanics? Can they name any of them or describe their features? Has the student heard of the Copenhagen Interpretation, and can they describe what it entails? Does the student know what the word *determinism* means within the context of physics? Did they have an opinion as to how they think their instructor would have wanted them to respond to earlier interview questions?

RESPOND TO THE STATEMENT:
*It is possible for physicists to carefully perform the same experiment and get two very different results that are both correct.*
IN AGREEMENT OR DISAGREEMENT AND EXPLAIN REASONING.

RESPOND TO THE STATEMENT:
*The probabilistic nature of quantum mechanics is mostly due to the limitations of our measurement instruments.*
IN AGREEMENT OR DISAGREEMENT AND EXPLAIN REASONING.

# APPENDIX C

## Selected Modern Physics Course Materials (Fall 2010)

**Lecture slides for Weeks 6-8 [New Material] are available online at:**

http://www.colorado.edu/physics/EducationIssues/baily/dissertation/

**Course materials from previous modern physics offerings are available online at:**

http://www.colorado.edu/physics/EducationIssues/modern/

# Course Calendar
# Physics 2130, Fall 2010

**Week 1 (8/23 – 8/27):**

**L01: Introduction**

**L02: Review Math & EM Waves**

**L03: EM Waves (cont.)**

**Week 2 (8/30 – 9/3):**

**L04: Waves and Superposition**

**L05: Photoelectric Effect 1**

**L06: Photoelectric Effect 2**

**Week 3 (9/6 – 9/10):**

**L07: Modeling in Physics**

**L08: Atomic Spectra**

**Week 4 (9/13 – 9/17):**

**L09: Atomic Spectra & Discharge Lamps**

**L10: Atomic Spectra (cont.) & Lasers**

**L11: Finish Lasers & Double-Slit Experiment with Light**

**Week 5 (9/20 – 9/24):**

**L12: Exam 1 Review**

**L13: Balmer Series & Bohr Model**

**L14: Bohr Model (cont.)**

**Week 6 (9/27 – 10/1) [NEW MATERIAL]:**

**L15: Atomic Spin**

**L16: Stern-Gerlach Experiments**

**L17: Probability**

**Week 7 (10/4 – 10/8) [NEW MATERIAL]:**

**L18: EPR/Entanglement**

**L19: Local Realism**

**L20: Single-Photon Experiments**

**Week 8 (10/11 – 10/15) [NEW MATERIAL]:**

**L21: Complementarity**

**L22: Electron Diffraction/Matter Waves**

**L23: Wave Function/Uncertainty Principle**

**Week 9 (10/18 – 10/22):**

**L24: Exam 2 Review**

**L25: Schrödinger Equation**

**L26: Potentials/Potential Energy**

**Week 10 (10/25 – 10/29):**

**L27: Infinite Square Well**

**L28: Finite Square Well/Tunneling**

**L29: Tunneling Tutorial (in-class)**

**Week 11 (11/1 – 11/5):**

**L30: Alpha-Decay/STM**

**L31: STM (cont.)/Measurement**

**L32: Hydrogen Atom 1**

**Week 12 (11/8 – 11/12):**

**L33: Hydrogen Atom 2**

**L34: Multi-Electron Atoms/Periodic Table**

**L35: Bonding & Color**

**Week 13 (11/15 – 11/19):**

**L36: Exam 3 Review**

**L37: Molecular Bonding/Solids**

**L38: Conductivity**

**Week 14 (11/29 – 12/3):**

**L39: Semi-Conductors/Diodes/Transistors**

**L40: Semi-Conductors/Diodes/Transistors (cont.)**

**L41: Nuclear Weapons & Nuclear Energy**

**Week 15:**

**L42: Nuclear Weapons & Nuclear Energy (cont.)**

**L43: Final Exam Review**

**L44: Final Exam Review/Last Day of Class**

# Course Syllabus
# Physics 2130, Fall 2010

**Course Materials:**

**Textbook: Knight, Physics for Scientists and Engineers – Volumes 3 & 5**
**Calculator:** Bring this to class.
**A Clicker** for use in class, available in bookstore.

**Instructors:**

Professor Noah Finkelstein & Charles Baily

**Graduate Graders:**

Yi-Ping Huang, Yu Ye, Xiao Yin

**Undergraduate Learning Assistants:**

Danny Rehn & Sam Milton

**Prerequisites:** Physics II (E&M), Introductory Lab

**Co-Requisite:** Calculus III

## Overall Course Goals:

**Reason for the course.**

The goal of this course is for you to understand the microscopic origin of the behavior of materials that you may encounter in the world around you or in technological applications. Engineers and scientists use simplified models to describe material properties, and most of the time these approximations work fine, but not always. This course will help you to understand why these models work and where they become unreliable and why. The latter issues become particularly important as one is working in the area of nanotechnology. A secondary goal is to increase your general knowledge through understanding the "new" (in past hundred years) way physicists have come to understand how the universe behaves, i.e. according to the laws of quantum physics.

**General Structure of the course.**

I. The key experimental results that gave rise to the ideas of quantum mechanics and how the ideas of quantum mechanics explain these experiments. In all three sections, the use and limitations of mathematics will be considered.

II. Deeper exploration of the ideas of QM and learning to apply these ideas in simple systems such as simple atoms and model situations.

III. Applying ideas to a number of real world situations.

IV. Understand how scientists think and work, as well as be able to make interpretations about physical experiments and models.

## Guiding principles of the instruction:

1. People understand concepts better by seeing them in action and thinking about them than by hearing them explained.
2. Understanding physics (and solving problems that test that understanding) is a learned skill, like cooking, or playing basketball or the violin. It takes time, effort, and practice.
3. People learn best by thinking about topics and discussing them with others.
4. Students learn most when they take the responsibility for what is learned.

In keeping with these principles, there will be a substantial number of homework problems each week. You will have considerable difficulty completing them if you follow non-expert problem solving approaches and/or you work alone. However, if you work with other students and develop an "expert" approach to problem solving, the homework problems should take you less time and effort, and you will learn a lot from doing them. Although you are encouraged to work out the solutions to problems together with other students, <u>you are required to write up the answers in **your own words**. So each student's wording should be unique to the student. ***I will fail any student who submits work that is not his/her own or permits another student to do so***</u>.

Typically you will need to spend between four and six hours outside of class to master the material. (Your homework will typically require 4+ hours and you should spend a couple hours each week reading and preparing for class.)

There will be several problem-solving sessions Monday and Tuesday where you will be able to conveniently get together with other students to work on homework. The instructors, Noah Finkelstein, Charles Baily, and the LAs will be present at these sessions to provide "coaching" on problem solving methods. You are encouraged to come to these to work with other students and get coaching in problem solving as necessary. The times and room numbers are listed above. The physics help room is also open 40 hours per week, and there are always students and TAs there, although they are not necessarily from 2130.

Students begin this class with a range of backgrounds in physics and math. As a result, it is impossible for each class to be perfectly matched to everyone's background. The primary purpose of office hours is to provide individual help to students that need it. We are anxious to provide whatever help is necessary for every student, regardless of their background, to do well in the course and achieve all of the learning goals. However, it is your responsibility to recognize that you need that help, and to take advantage of its availability by asking to meet with us.

## GRADES:

**Grading philosophy**: the amount you will learn depends on how much thought and practice you put in distributed sensibly over the term. Everyone who makes an honest attempt to do all the assigned work on time will pass, normally with a grade of A or B.

**Grade components**:

In-class activities and Online Participation (10% of grade):

- Participation points (not graded): 2 pts per class for participating in clicker questions. Occasionally additional participation points for more extended activities.

- Graded in-class points: 3 pts for each reading quiz (~once per week). Occasionally there will be a graded clicker questions during class (1 pt each).

- Drop lowest 3 participation days.

Homework (42.5% of grade): Weekly homework, various numbers of points each.

- Drop lowest homework grade.

- Online Participation/Feedback: + 1 extra-credit point each week towards HW grade

Exams (47.5% of grade): 3 x 1 -Hour Exams are 50pts each, Final Exam is 75 points.

**1. Homework**

- Homework will typically be due **Tues midnight** (11:59p Tues) on CU Learn & Long Answers due in the basement cabinet Wed **by the beginning of class**.
- Homework are web based and accessed through the course homepage table of contents. They are available directly on CU Learn. Many weeks there will be one long answer write up where you will need to show your work that will be due in class on Wednesday.
- **We encourage you to work together on the homework problems, but you must write up the answers in your own words.** There will be 2 Problem Solving Sessions each week: Mondays and Tues. The location is at reserved tables near the rear of the Physics Help Room. This is a great opportunity to come work on the homework with your classmates!

- Homework is a large part of your grade, so failing to turn in more than one assignment, and thereby getting a 0 will have a big impact on your grade. Talk to us, NOW, if you will have a scheduling problem during the term so that you will be unable to complete any of the assignments. Permission for exceptions from the normal class work schedule must be requested in advance.
- It is best if you print out the assignment early, so you see the problems before class.
- Homework solutions may be accessed through the Physics 2130 Home Page. The answers and solutions to the previous weeks homework will normally be available at noon on the Wed they are due.
- Homework grading will be done by the course graders. There will be essay questions on the homework. Each week, a sampling of the essays will be graded with the grading rubric. When answering the essays you should keep in mind the grading rubric:

**Criteria for grading non-mathematical, short-essay questions.**

Many of the homework questions ask you to use physical principles discussed in class to analyze a situation and reason an outcome. For each of the questions or parts of questions, the answer will be graded for correctness on a scale of 0 to 7 based on the following rubric:

1. Identifies physical principle or principles that are relevant to answering the question: (2 if correct, 0 if irrelevant principle, 1 if have both some relevant and some irrelevant principles.)
2. Explains how the principle(s) apply to the situation described in the problem: (1 if correct, 0 if not)
3. Employs proper reasoning to explain the logic in going from how principle applies to the situation to the answer to the question: (2, 1, 0 according to level of correctness. If #1 or #2 are 0, this should automatically be 0.)
4. Clarity of writing: (2 if good, 1 if difficult but can be figured out, 0 if incomprehensible. If 0 here, all the others will be 0.)

So if your answer is scored as 5/7 and the question (part) is worth 1 pt, you will receive 0.71 pts for that part.

**2. In-class questions, activities, and quizzes on reading:**

**Clickers:** You will need to buy transmitters (usually referred to as “clickers”) from the bookstore for answering questions in class.

**Reading Quizzes:** After each reading assignment there will be a very short quiz covering the material in class worth 2 to 3 points

**In-class clicker questions:** During class there will be many questions on which you enter your response using clickers. Your answers will be recorded and you will receive 2 points towards your in-class grade per class for submitting any answer to all of the questions, whether or not your answers are correct. There will be a few questions, typically 0 to 2 per classes, for which you will receive one point if you have the correct answer, and 0 if incorrect. Graded questions will usually be late in the class and ones that nearly all students get correct if they have been paying attention.

**In-class clicker activities:** Some weeks there will be a more extended in-class activity for which you will receive additional points for participating..

**Online Participation due Tues at midnight:** Each week you will be asked to fill out an online participation form to give us feedback on various aspects of the course, what you are learning, how you are thinking about the ideas, etc. You are not graded on your response, but we value this feedback and thoughts as it helps us better understand how to teach the class effectively. You will receive 1 point of extra-credit towards homework for submitting the form.

**3. Hour exams:**

- Each exam is worth 50 points towards your total exam grade.
- There will be no early or late exams given and no make-up exams.
- **Be sure to bring formula card(s) and calculator.** All exams will be closed book. You may make up a single 3 x 5 formula card for each exam and bring your previous cards with you to subsequent exams so you will have one card for first exam, two for the second and four for the final. You can write anything you want on your formula card, but you must write it by hand - no photocopying or printing allowed. You should bring a calculator to class and exam. Sharing of calculators during exams and quizzes will not be allowed.
- **Important:** To accommodate travel, illness, etc, 1-hour exam score can be dropped. You should not need to be excused for a second. Only in the rare instance of a severe medical or family emergency will an excuse for a second absence be considered. To be excused you must notify Noah Finkelstein by email or telephone before the exam, and you must provide a physician's note or other documentation to one of us within two weeks of the exam. If you failed to call before the exam, you must provide documentation why a medical condition made this impossible. For an excused absence, you will be given the class average on that exam.
- Exam grades and solutions will be posted after the exam on the course website.

**4. Final exam**: The final is worth 75 points towards your total exam grade.

**5. Extra Credit Points:**

- Approximately once per week there will be an online participation question - one point per week of extra credit towards your homework grade.
- The Online Participation forms are **due by Tues midnight with the HW.**

## General rules:

The rules in this list may seem rather harsh and arbitrary, but they are essential to maintaining the integrity of the course. There is a painful story behind every one of them. Although most of you will never come up against any of the rules, there are a handful of students each semester that just cannot seem to avoid them. These rules are primarily to prevent these students from obtaining an unfair advantage over the others in the class. If these rules are going to cramp your style, then this class is probably not for you.

No students will fail who makes a serious effort at all the assigned work. If you miss a homework assignment or do not take an exam, it becomes possible for you to fail the course.

Although you are encouraged to work together with other students, you must hand in your own work and put the explanation in your own words. Handing in a copy of another student's work is considered cheating. ***We will fail any student who submits for a grade work that is not their own or permits another student to do so***.

**Student Classroom and Course-Related Behavior:** Students and faculty each have responsibility for maintaining an appropriate learning environment. Students who fail to adhere to such behavioral standards may be subject to discipline. Faculty have the professional responsibility to treat all students with understanding, dignity and respect, to guide classroom discussion and to set reasonable limits on the manner in which they and their students express opinions. Professional courtesy and sensitivity are especially important with respect to individuals and topics dealing with differences of race, culture, religion, politics, sexual orientation, gender variance, and nationalities. Class rosters are provided to the instructor with the student's legal name. I will gladly honor your request to address you by an alternate name or gender pronoun. Please advise me of this preference early in the semester so that I may make appropriate changes to my records.

**Modern Physics Survey (Pre-Instruction)**
**Physics 2130, Fall 2010**

## INTRODUCTION

Below are several statements that may or may not describe your beliefs or opinions. You are asked to rate each statement by selecting a number between 1 and 5 where the numbers mean the following:

1. Strongly Disagree
2. Disagree
3. Neutral
4. Agree
5. Strongly Agree

Choose one of the above five choices that best expresses <u>your</u> feeling about the statement. If you have no strong opinion either way, choose 3.

A textbox follows each statement, asking you to briefly explain the reasoning behind your answer (e.g. what is going through your mind as you were formulating a response). Please note that, for some of the statements, the wording is deliberately ambiguous; we are particularly interested in how you interpret each statement, and any argumentation in support of your response.

**WE ARE ASKING THAT YOU EXPRESS YOUR OWN BELIEFS. YOUR SPECIFIC ANSWERS WILL <u>NOT</u> AFFECT ANY EVALUATION OF YOU AS A STUDENT.**

**1. It is possible for physicists to carefully perform the same measurement and get two <u>very different</u> results that are both correct.**

**2. The probabilistic nature of quantum mechanics is mostly due to physical limitations of our measurement instruments.**

**3. When not being observed, an electron in an atom still exists at a definite (but unknown) position at each moment in time.**

**4. I think quantum mechanics is an interesting subject.**

**5. I have heard about quantum mechanics through popular venues (books, films, websites, etc...)**

**Modern Physics Survey (Post-Instruction)**
**Physics 2130, Fall 2010**

## INTRODUCTION

Below are several statements that may or may not describe your beliefs or opinions. You are asked to rate each statement by selecting a number between 1 and 5 where the numbers mean the following:

1. Strongly Disagree
2. Disagree
3. Neutral
4. Agree
5. Strongly Agree

Choose one of the above five choices that best expresses <u>your</u> feeling about the statement. If you have no strong opinion either way, choose 3.

A textbox follows each statement, asking you to briefly explain the reasoning behind your answer (e.g. what is going through your mind as you were formulating a response). Please note that, for some of the statements, the wording is deliberately ambiguous; we are particularly interested in how you interpret each statement, and any argumentation in support of your response.

**WE ARE ASKING THAT YOU EXPRESS YOUR OWN BELIEFS. YOUR SPECIFIC ANSWERS WILL <u>NOT</u> AFFECT ANY EVALUATION OF YOU AS A STUDENT.**

## SURVEY

**1. It is possible for physicists to carefully perform the same measurement and get two <u>very different</u> results that are both correct.**

**2. The probabilistic nature of quantum mechanics is mostly due to physical limitations of our measurement instruments.**

**3. When not being observed, an electron in an atom still exists at a definite (but unknown) position at each moment in time.**

**4. I think quantum mechanics is an interesting subject.**

## ESSAY QUESTION:

**PLEASE NOTE: As before, there are no “right” or “wrong answers to the following question. We are interested in what you actually believe.**

Play with the Quantum Wave Interference simulation, at
http://phet.colorado.edu/new/simulations/sims.php?sim=QWI

In particular, select the “Single Particles” tab at the top of the screen, and then click on the “Double Slits” button at the right of the display. Select “Electrons” and then fire the gun…

Now consider the following:

**1.** A blob (representing the probability density for a single electron) emerges from an “electron gun”.

**2.** The blob passes through the two slits

**3.** A single electron is detected on the far screen. After many electrons, an interference pattern develops.

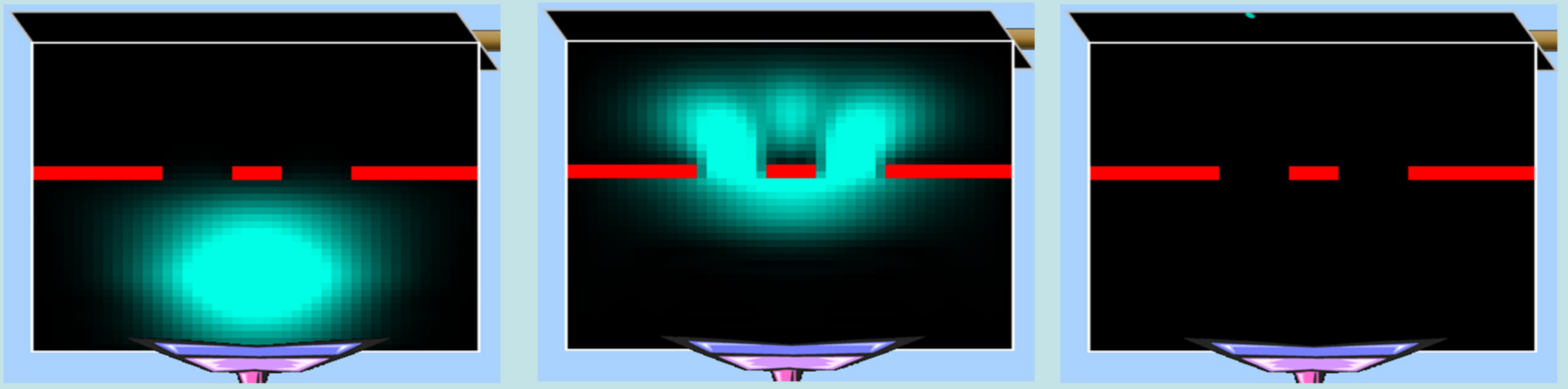

**Three students discuss this Quantum Wave Interference simulation:**

**STUDENT ONE:** The probability density is so large because we don't know the true position of the electron. Since only a single dot at a time appears on the detecting screen, the electron must have been a tiny particle traveling somewhere inside that blob, so that the electron went through one slit or the other on its way to the point where it was detected.

**STUDENT TWO:** I think the blob represents the electron itself, since a free electron is properly described by a wave packet. The electron acts as a wave and will go through both slits and interfere with itself. That's why a distinct interference pattern will show up on the screen after shooting many electrons.

**STUDENT THREE:** All we can really know is the probability for where the electron will be detected. Quantum mechanics may predict the likelihood for a measurement outcome, but it really doesn’t tell us what the electron is doing between being emitted from the gun and being detected at the screen.

**Which students (if any) do you agree with, and why? What’s wrong with the other students’ arguments? What is the evidence that supports your answer?**

# Physics 2130 Fall 2010
# Homework 06 Solutions

**M/C & Short Answer: 10 Points**
**Essays: 11 Points**
**Long Answer: 9 Points**

**3. (0.5 Points)** Which of the following is NOT a possible probability?

A. 25/100
B. 1.25
C. 1
D. 0

Answer: B

Remember that probabilities are normalized so that the sum of the probabilities for all possible outcomes is equal to one. Therefore, a probability expressed as a number that is not between 0 and 1 (inclusive) makes no sense – we can't say there is a 125% chance of an outcome occurring.

**4. (0.5 Points)** Among twenty-five items, nine are defective, six having only minor defects and three having major defects. Determine the probability that an item selected at random has major defects, given that it has defects.

A. 1/3
B. 0.25
C. 0.12
D. 0.08

Answer: A

The answer is not C [3/25 = 0.12] because that is the probability of selecting an item with a major defect from the collection of all 25 items. We are assuming that on this particular trial we have selected one of the nine items that have defects, of which only three of those nine have a major defect. The answer is therefore 3/9 = 1/3 = A.

**5. (1 Point)** *ABCD* is a square. *M* is the midpoint of *BC* and *N* is the midpoint of *CD*. A point is selected at random in the square. Calculate the probability that it lies in the triangle *MCN*.

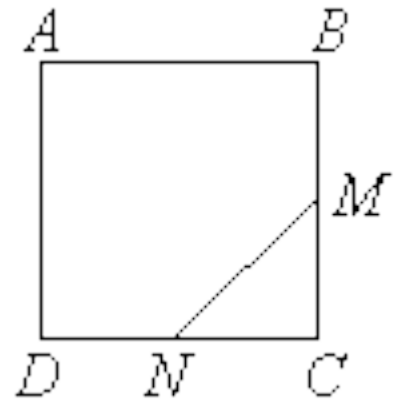

SOLUTION:

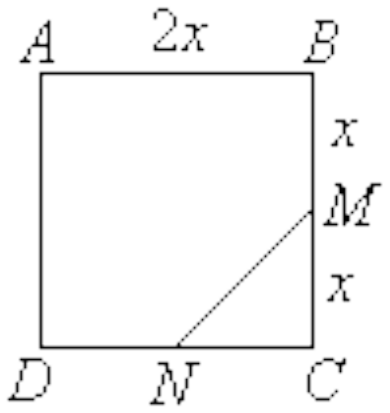

Let $2x$ be the length of the square.

Area of square = $2x \times 2x = 4x^2$.

Area of triangle $MCN = \frac{1}{2}x^2$.

P[Point in $MCN$] = $\dfrac{\text{Area of } MCN}{\text{Area of } ABCD} = \dfrac{\frac{1}{2}x^2}{4x^2} = \dfrac{1}{8}$

**6.** Two balanced dice are rolled. Let X be the sum of the two dice. Obtain the probability distribution of X (i.e. what are the possible values for X and the probability for obtaining each value?). Check that the probabilities sum to one.

**(1 Point)** What is the probability for obtaining X >= 8?

SOLUTION:

When the two balanced dice are rolled, there are 36 equally likely possible outcomes:

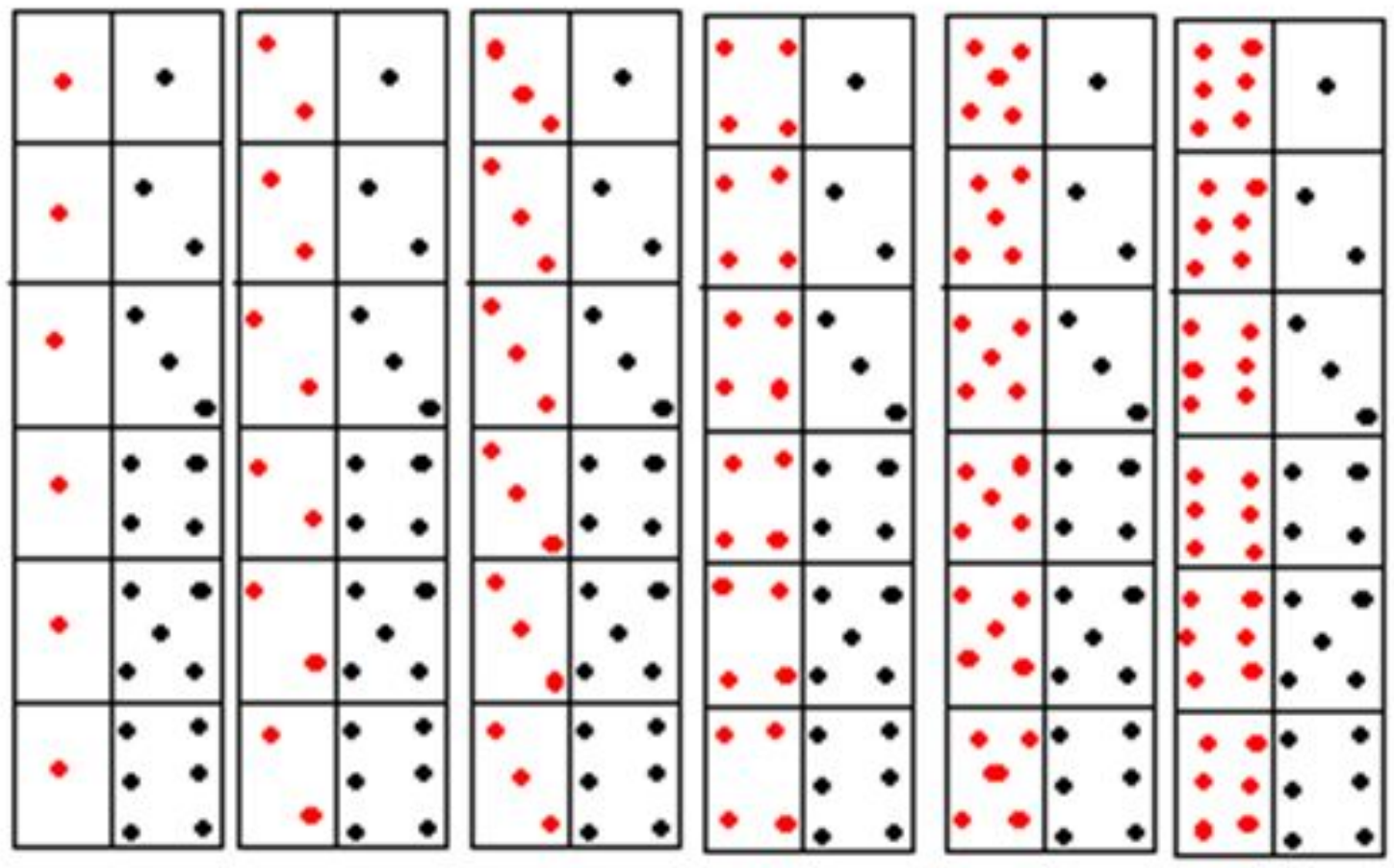

The possible values of X are: 2, 3, 4, 5, 6, 7, 8, 9, 10, 11 and 12.

The possible outcomes are equally likely, so the probabilities P[X] are given by:

P[2] = P1,1] = 1 / 36

P[3] = P[1,2] + P[2,1] = 2 / 36 = 1 / 18

P[4] = P[1,3] + P[2,2] + P[3,1] = 3 / 36 = 1 / 12

P[5] = P[1,4] + P[2,3] + P[3,2] + P[4,1] = 4 / 36 = 1 / 9

P[6] = P[1,5] + P[2,4] + P[3,3] + P[4,2] + P[5,1] = 5 / 36

P[7] = P[1,6] + P[2,5] + P[3,4] + P[4,3] + P[5,2] + P[6,1] = 6 / 36 = 1 / 6

P[8] = P[2,6] + P[3,5] + P[4,4] + P[5,3] + P[6,2] = 5 / 36

P[9] = P[3,6] + P[4,5] + P[5,4] + P[6,3] = 4 / 36 = 1 / 9

P[10] = P[4,6] + P[5,5] + P[6,4] = 3 / 36 = 1 / 12

P[11] = P[5,6] + P[6,5] = 2 / 36 = 1 / 18

P[12] = P[6,6] = 1 / 36

Therefore, P[X >= 8] = 5/36 + 4/36 + 3/36 + 2/36 + 1/36 = 15/36

**7. (1 Points)** What is the average value of X?

SOLUTION:

The average of X is given by

**<X>** = ∑ X P(X) = 2*(1/36) + 3*(1/18) + 4*(1/12) + 5*(1/9) + 6*(5/36)

+ 7*(1/6) + 8*(5/36) + 9*(1/9) + 10*(1/12) + 11*(1/18) + 12*(1/36) = 7

**8-9.** An atom in the state $|\uparrow_z\rangle$ is shot into the following line of three Stern-Gerlach analyzers. Analyzer A is tilted at an angle of $-\alpha$ from the vertical, Analyzer B at $+\beta$, and Analyzer C at $-\gamma$.

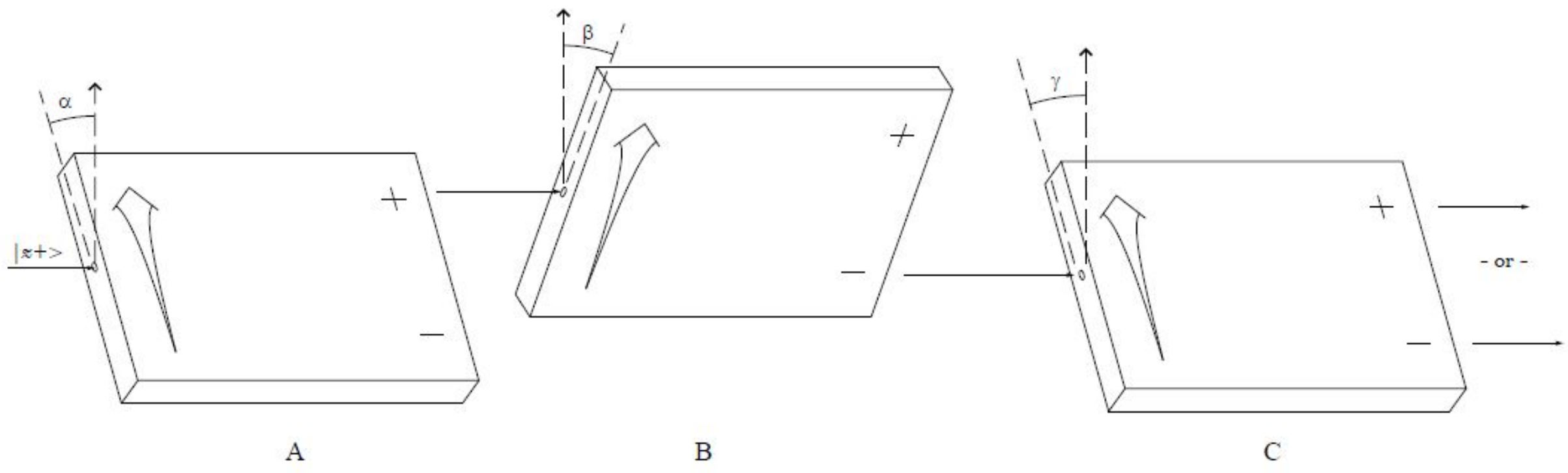

For these problems $\alpha = 15^0$, $\beta = 35^0$ & $\gamma = 20^0$.

**(1 Point)** What is the probability the atom exits from the plus-channel of Analyzer C?

**(1 Point)** What is the probability the atom exits from the minus-channel of Analyzer C?

For Questions 8 & 9, you may find it useful to play with the Stern-Gerlach Sim, but you should calculate the exact probabilities, and then check that your answer is close to what is measured in the simulation - a number that may or may not exactly match the correct answers for the below.

Probability of atom entering A emerging from the plus-channel is $\cos^2(\alpha/2)$.

Probability of atom entering B emerging from the minus-channel is $\sin^2((\alpha+\beta)/2)$.

Probability of atom entering C emerging from the plus-channel is $\cos^2\left(\frac{180^0-(\beta+\gamma)}{2}\right) = \sin^2((\beta+\gamma)/2)$. [Remember: We are interested in the relative angle between the incoming state and the orientation of the analyzer.]

Probability of atom entering C emerging from the minus-channel is $\cos^2((\beta+\gamma)/2)$.

We are asking about the probability for a specific outcome at Analyzer A AND a specific outcome at Analyzer B AND a specific outcome at Analyzer C – the AND is a signal that the total probability is the product of the individual probabilities. The probability of the atom entering Analyzer A emerging from the plus-channel of Analyzer C is therefore:

$$\cos^2(\alpha/2)\sin^2((\alpha+\beta)/2)\sin^2((\beta+\gamma)/2)$$

The probability of it emerging from the minus-channel of Analyzer C is:

$$\cos^2(\alpha/2)\sin^2((\alpha+\beta)/2)\cos^2((\beta+\gamma)/2)$$

If $\alpha = 15^0$, $\beta = 35^0$ & $\gamma = 20^0$, then the probability for emerging from the plus-channel of Analyzer C is:

$$\cos^2(7.5^0)\sin^2(25^0)\sin^2(27.5^0) = 0.037$$

…and the probability for emerging from the minus-channel is:

$$\cos^2(7.5^0)\sin^2(25^0)\cos^2(27.5^0) = 0.138$$

**10. (1 Point)** Why don't these two probabilities sum to 1?

These two probabilities don't sum to one because there's also some probability that the atom will leave through the minus-channel of Analyzer A or the plus-channel of Analyzer B. Note that the probability for an atom entering Analyzer C to leave from either the plus-channel OR the minus-channel of Analyzer C **is** equal to one (because we always get one of two answers at any analyzer) – but we've asked about the probability for an atom entering A to exit at C.

$$\sin^2((\beta+\gamma)/2) + \cos^2((\beta+\gamma)/2) = 1$$

**11.** A particular Stern-Gerlach analyzer has three settings, each oriented $120^0$ from the other. During lecture we found the probability for an atom that entered in a definite state of $|\uparrow_z\rangle$ to leave from the plus-channel if the detector setting is random. In this same situation:

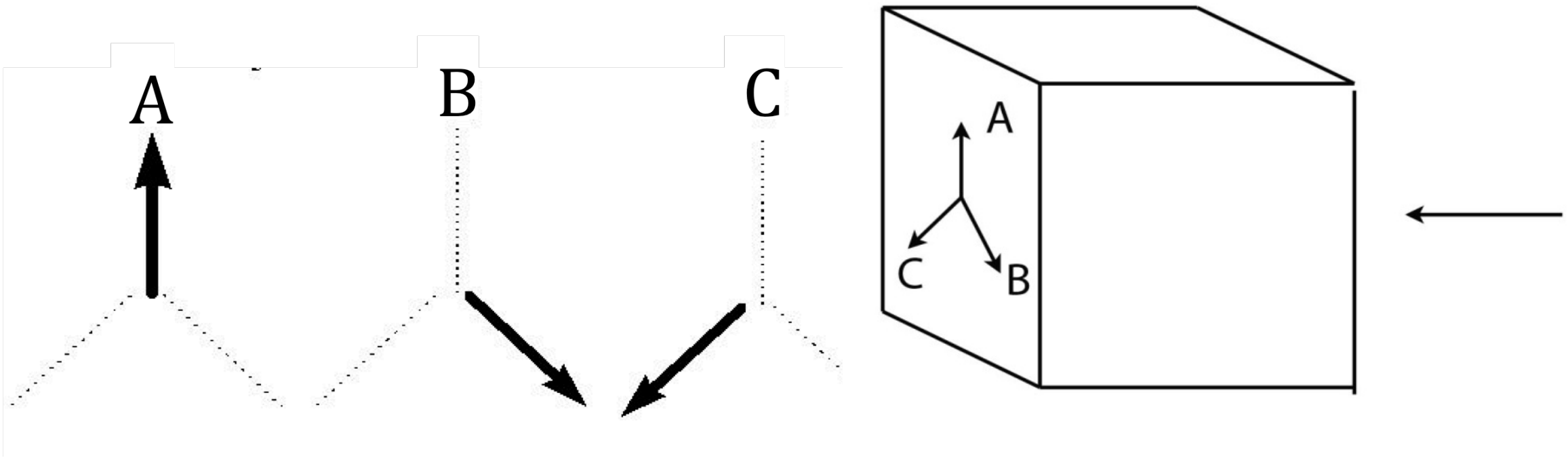

**(0.5 Points)** What is the probability for an atom to leave the minus-channel if the incoming atom is in the state $m_Z = +m_B$?

We solve this problem just as in lecture. The probability to leave the minus channel if the incoming atom is in the state $m_Z = +m_B$ is 0 if the analyzer is set to A, and 3/4 if the analyzer is set to either B or C. If the settings on the analyzer are random, then the total probability to exit from the minus channel is [1/3 x 0] + [1/3 x 3/4] + [1/3 x 3/4] = 2/4 = 50.0% [1/3 for each possible random orientation, and the 3/4 comes from $\sin^2(120^0/2)$].

**12. (0.5 Points)** What is the probability for an atom to leave the plus-channel if the incoming atom is in the state $m_{120^0} = +m_B$?

The probability of an atom entering in any state that is aligned (or anti-aligned) with ***any*** one of the three settings to exit from the plus-channel is [1/3 x 1] + [1/3 x 1/4] + [1/3 x 1/4] = 50.0%

**13. (0.5 Points)** What is the probability for an atom to leave the minus-channel if the incoming atom is in the state $m_{120^0} = +m_B$?

The probability to leave the minus channel if the incoming atom is in the state $m_Z = +m_B$ is 0 if the analyzer is set to A, and 3/4 if the analyzer is set to either B or C. The total probability to exit from the minus channel is [1/3 x 0] + [1/3 x 3/4] + [1/3 x 3/4] = 2/4 = 50.0%. We also see this if we recognize that the probability in #12 and the probability in this problem add up to 1.

**14.** Consider the same situation as in Questions 11, 12 & 13, **but now** settings B and C are oriented at +/- 110 degrees from the vertical (instead of 120 degrees).

**(1 Point)** What is the probability for an atom in the state of *spin up* along the +z-axis to leave from the plus-channel if the settings are random?

Answer: 55.3 %

Just as in the lecture problem, the probability of an atom entering with $|\uparrow_Z\rangle$ leaving through the plus-channel is [1/3 x 1] + [1/3 x 0.328] + [1/3 x 0.328] = 55.26% [1/3 for each possible random orientation, and the 0.328 comes from $\cos^2(110^0/2)$]. Comparing this result with previous problems, we should understand that the 1/2 probabilities in those situations are special circumstances for the axes oriented at 120 degrees, and isn't a general statement about two-state systems always having equal likelihoods for either outcome.

**(0.5 Points)** In the experiment depicted below, which of the following best describes the state of an atom that leaves the plus-channel of Analyzer B?

A) $|\uparrow_Z\rangle$

B) $|\uparrow_X\rangle$

C) $|\uparrow_Z\rangle|\uparrow_X\rangle$

D) $|\uparrow_X\rangle|\uparrow_Z\rangle$

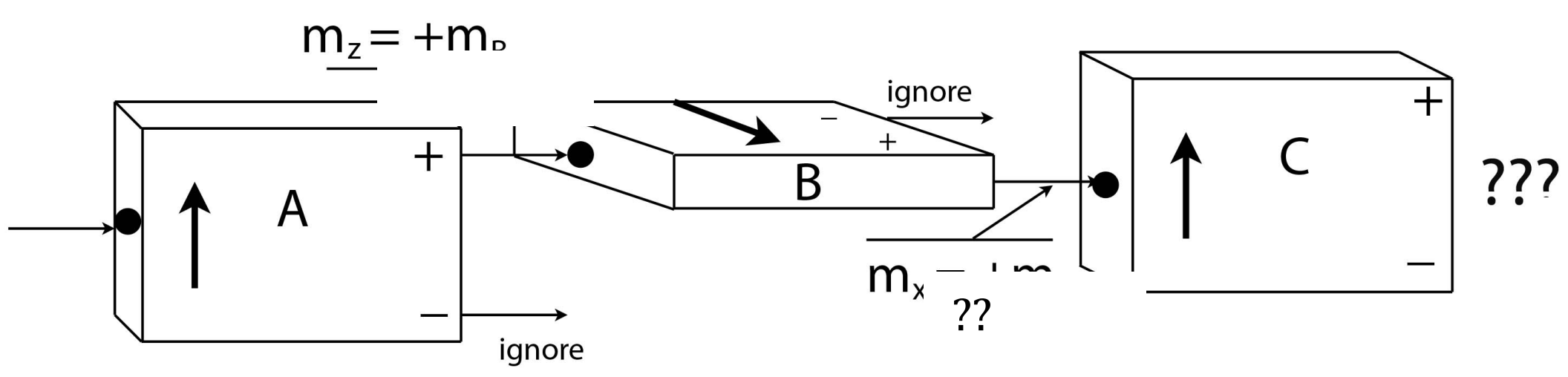

Answer: B

The atom is in the definite state $\left|\uparrow_X\right\rangle$ since it left from the plus-channel of the horizontal Analyzer B; $m_Z$ is indefinite: there is an equal probability for the atom to exit from either channel of Analyzer C. Either state $\left|\uparrow_Z\right\rangle\left|\uparrow_X\right\rangle$ or $\left|\uparrow_X\right\rangle\left|\uparrow_Z\right\rangle$ would say that the outcome at Analyzer C is pre-determined to be spin-up, when it is actually indeterminate until measured. It doesn't matter what order we write multiplicative terms in a quantum state, so $\left|\uparrow_Z\right\rangle\left|\uparrow_X\right\rangle = \left|\uparrow_X\right\rangle\left|\uparrow_Z\right\rangle$ [they both contain the same information, so are equivalent; the order does not reflect the order in which measurements are made].

Questions 16 – 18 refer to the reading "100 Years of Quantum Mysteries".

**16. (2 Points)** As discussed in this article, what were some of the problems in classical physics that led to the development of quantum theory?

1) Attempts to use classical physics to predict the spectrum of thermal radiation from a perfect absorber (and emitter – a "blackbody") found that that the power radiated by a blackbody would be infinite (the so-called "Ultraviolet Catastrophe"). In 1900, Planck was able to predict the correct blackbody spectrum, but only if he assumed (without physical justification) that the thermal radiation exists in discrete amounts (and not continuous). This idea was further applied by Einstein in explaining the photoelectric effect, who interpreted the results as implying that the energy of electromagnetic radiation comes in quantized bits (photons).

2) Rutherford had demonstrated from scattering experiments that the atom should consist of electrons orbiting about a compact nucleus, yet classical electromagnetic theory predicted that accelerating (orbiting) charges should radiate away their energy in the form of light. This would happen over a very short time, which contradicts the observation that most atoms are stable. Bohr side-stepped this problem by supposing that the laws of classical electromagnetism don't apply at the atomic level, and that a single photon is emitted only when an electron transitions from a higher to a lower energy state.

**17. (2 Points)** How are the terms *theory* and *interpretation* used in this article?

At the end of the article, the authors refer to *theories* as consisting of two components: mathematical equations, and words that explain how the equations are connected to what is observed in experiments. They also discuss how some theories are really subsets of a more general theory, but with a limited range of application – quantum mechanics is the low-energy version of the more fundamental quantum field theory; and similarly with Newtonian gravity and general relativity (Einstein's theory of gravity). Newtonian gravity is good enough to put satellites in orbit, and quantum mechanics is good enough to talk about the behavior of particles at low energies, but the more fundamental theories also describe nature at much higher energies than what we experience in everyday life (which is why they were discovered later).

In contrast, the interpretations discussed in this article don't relate to what is directly observable. They are the result of scientists trying to make physical sense of what the mathematical equations imply, but they generally have to do with "what's really going on" when we're not making a measurement (observation) [or rather, what's going on in the time just before we make a measurement or just after] – sort of like questions about whether a tree falling in the forest makes a sound when no one is around. Einstein favored deterministic interpretations, whereas Bohr decided that the probabilistic nature of quantum theory was reflective of the inherently probabilistic behavior of quantum systems. These kinds of interpretative themes were generally considered to be a matter of philosophical taste for many years; as we'll learn later in the course, physicists eventually began to realize that some of these questions could be put to experimental test.

**18. (2 Points)** Is there any experimental evidence in favor of any of the interpretations discussed in this article? In the cases where there is not, why would a scientist favor one interpretation over another?

As mentioned in Question #17, the interpretations discussed in this article can't be distinguished from each other by any physical measurement, since they mostly have to do with what's happening when we're ***not*** making measurements. Still, physicists may favor one type of interpretation over another for various reasons, usually having to do with the physical intuitions we build up through experience. Decoherence seems to be popular among physicists as a solution to the measurement problem, in part because it provides a plausible physical mechanism for why we can never observe quantum superpositions in macroscopic systems. The Many-Worlds interpretation may be equivalent to decoherence in the sense that both make identical predictions about what is observed (not superposition states), but it may also seem like nothing more than crazy science-fiction to many scientists. In this course we will evaluate the merits and drawbacks of some of these interpretations of quantum mechanics – in the end, you can decide for yourself what kind of interpretation works best for you, but you should always be conscious of when that interpretation is legitimate, and if/when it will lead you to wrong answers.

**19. This question refers to the reading assignment "Probability"**

**(2 Points)** According to this article, in what way(s) is quantum mechanics a probabilistic theory?

As described in this article, in quantum mechanics the position of a particle must be described with a probability density, which tells us about the likelihood of finding the electron within a region of space when we make a measurement. The outcome of any individual position measurement can't be predicted ahead of time, but if we make many position measurements on similar systems, a graph of the data for where we find the electron each time should follow the same pattern as the probability density.

**LONG ANSWER**

Suppose you have a classical particle in a 1-dimensional box, bouncing back and forth between the two walls without friction or other loss of energy. Since the particle is bouncing between the two walls at constant speed, there is an equal likelihood of finding it at any point in the box if we look at some random time. The probability density for the position of this classical particle is therefore constant (a flat line, meaning equal probability everywhere between the two walls):

$\rho(x) = A$ for $0 \le x \le L$; $\rho(x) = 0$ otherwise

…where L is the length of the box, and A is some constant with appropriate units.

**1) (1 Point)** What must A be equal to in order for $\rho(x)$ to be normalized? In other words, for what value of A is the normalization condition $\int_{-\infty}^{+\infty} \rho(x)\ dx = 1$ satisified?

SOLUTION:

$$1 = \int_{-\infty}^{+\infty} \rho(x)\ dx = \int_0^L A\ dx = A \cdot \left[x\right]_0^L = A \cdot L = 1 \quad \rightarrow \quad A = \frac{1}{L}$$

**2) (1 Point)** Make a sketch of the normalized $\rho(x)$ for $0 \le x \le L$.

SOLUTION:

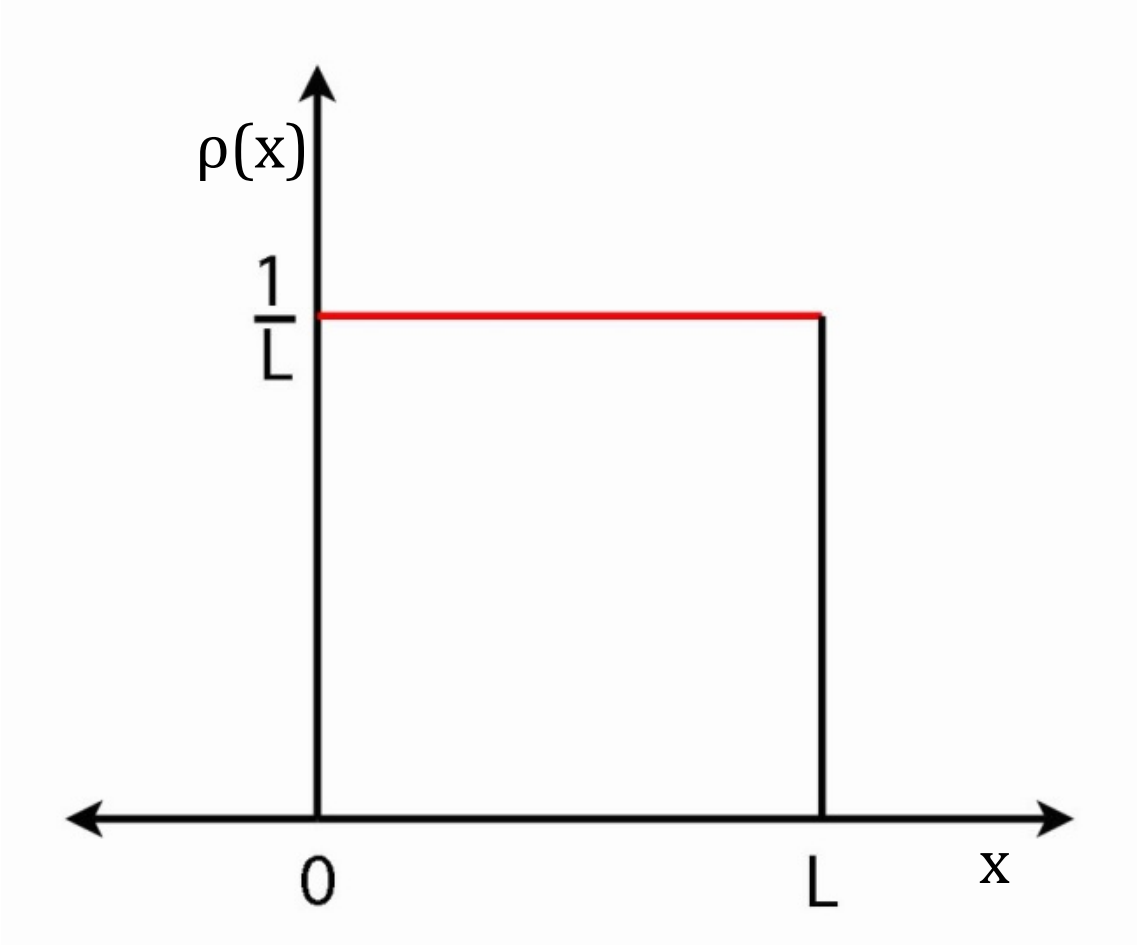

**3) (1 Point)** ***Show*** (using mathematics, not symmetry arguments) that the average value of x is equal to L/2. In other words, compute $\langle x \rangle = \int_{-\infty}^{+\infty} x\ \rho(x)\ dx$

SOLUTION:

$$\langle x \rangle = \int_{-\infty}^{+\infty} x\ \rho(x)\ dx = \frac{1}{L}\int_0^L x\ \ dx = \frac{1}{L}\left[\frac{1}{2}x^2\right]_0^L = \frac{L}{2}$$

Suppose instead the probability density for the position of a particle were given by:

$\rho(x) = A\sin^2\left(\frac{\pi x}{L}\right)$ for $0 \le x \le L$; $\rho(x) = 0$ otherwise

**4) (2 Points)** What must A be equal to in order for $\rho(x)$ to be normalized?

HINT: Use the trigonometric identity $\sin^2(u) = \frac{1-\cos(2u)}{2}$

SOLUTION:

$$1 = \int_{-\infty}^{+\infty} \rho(x)\ dx = A\cdot\int_0^L \sin^2\left(\frac{\pi x}{L}\right) dx = \frac{A}{2}\cdot\int_0^L 1-\cos\left(\frac{2\pi x}{L}\right) dx$$

$$= \frac{A}{2}\left[x - \frac{L}{2\pi}\sin\left(\frac{2\pi x}{L}\right)\right]_0^L = \frac{A}{2}\left[(L-0) - \frac{L}{2\pi}(\sin 2\pi - \sin 0)\right] = \frac{A}{2}\left[L - \frac{L}{2\pi}(0-0)\right]$$

$$\rightarrow \quad \frac{AL}{2} = 1 \quad \rightarrow \quad A = \frac{2}{L} \quad \rightarrow \quad \rho(x) = \frac{2}{L}\sin^2\left(\frac{\pi x}{L}\right)$$

**5) (1 Point)** Make a sketch of the normalized $\rho(x)$ for $0 \le x \le L$.

SOLUTION:

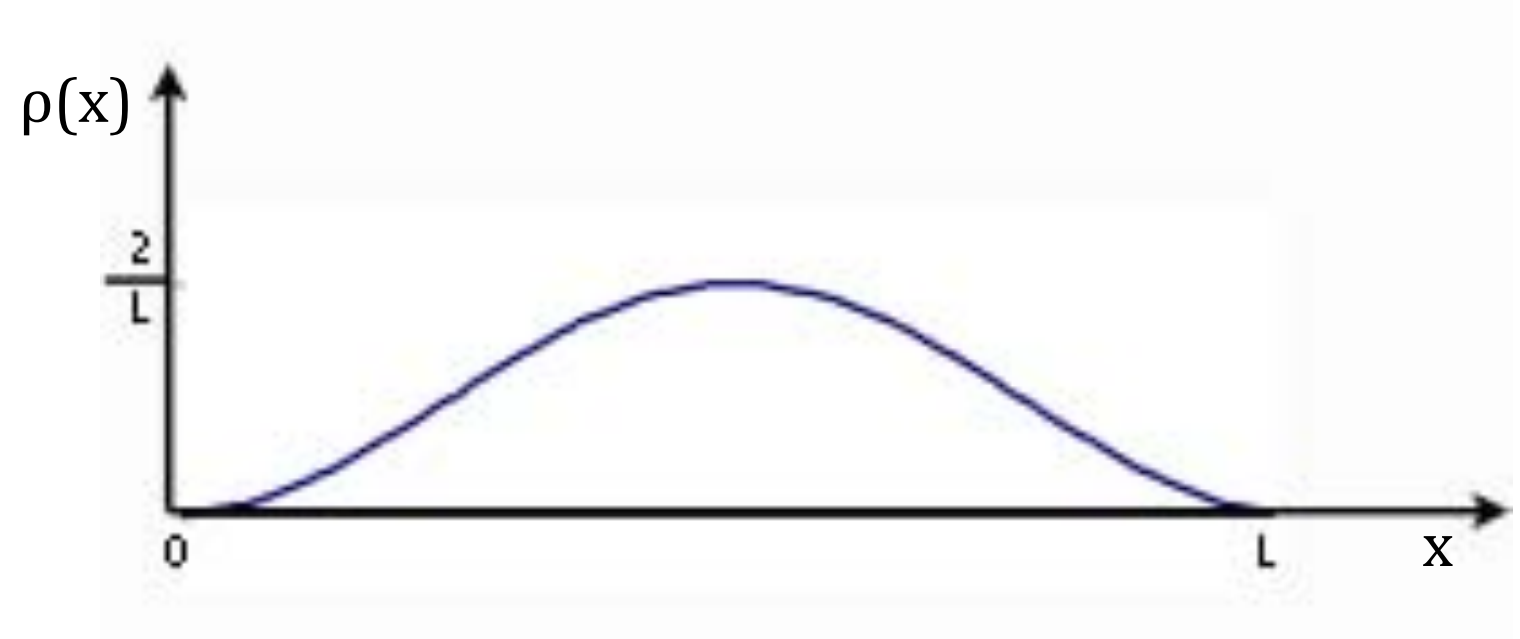

**6) (1 Point)** What is the probability of finding the particle in the left third of the box? (i.e., what is $P[0 \le x \le L/3]$?) How does this compare with finding a classical particle in the left third of the box when the probability density is constant?

SOLUTION:

Using our solution to Question #4, we now evaluate the integral between 0 and L/3:

$$P[0 \le x \le L/3] = \frac{A}{2}\left[x - \frac{L}{2\pi}\sin\left(\frac{2\pi x}{L}\right)\right]_0^{L/3} = \frac{1}{L}\left[\left(\frac{L}{3} - 0\right) - \frac{L}{2\pi}\left(\sin\frac{2\pi}{3} - \sin 0\right)\right]$$

$$= \frac{1}{3} - \frac{1}{2\pi}\frac{\sqrt{3}}{2} = 0.195 \sim 20\%$$

Compare this with 33.3% for a classical particle.

**7) (2 Points) *Show*** (using mathematics, not symmetry arguments) that the average value of x is equal to L/2. In other words, compute $\langle x \rangle = \int_{-\infty}^{+\infty} x\ \rho(x)\ dx$.

HINT: Use the same trigonometric identity as in #4, and then integrate by parts: $\int u\ dv = uv - \int v\ du$

SOLUTION:

$$\langle x \rangle = \int_{-\infty}^{+\infty} x\ \rho(x)\ dx = \frac{2}{L}\int_0^L x\sin^2\left(\frac{\pi x}{L}\right) dx = \frac{2}{L}\int_0^L x\frac{1}{2}\left(1 - \cos\left(\frac{2\pi x}{L}\right)\right) dx$$

$$= \frac{1}{L}\left\{\left[\frac{1}{2}x^2\right]_0^L - \int_0^L x\cos\left(\frac{2\pi x}{L}\right) dx\right\} = \frac{L}{2} - \frac{1}{L}\int_0^L x\cos\left(\frac{2\pi x}{L}\right) dx$$

To evaluate the second integral, let $u = x$ so that $du = dx$

...and let $v = \sin\left(\frac{2\pi x}{L}\right) \quad \rightarrow \quad \frac{L}{2\pi}\ dv = \cos\left(\frac{2\pi x}{L}\right) dx$

$$\int_0^L x\cos\left(\frac{2\pi x}{L}\right) dx = \frac{L}{2\pi}\int_{x=0}^{x=L} u\ dv = \frac{L}{2\pi}\left\{[uv]_{x=0}^{x=L} - \int_{x=0}^{x=L} v\ du\right\} = \frac{L}{2\pi}\left\{\left[x\sin\left(\frac{2\pi x}{L}\right)\right]_0^L - \int_0^L \sin\left(\frac{2\pi x}{L}\right) dx\right\}$$

$$= \frac{L}{2\pi}[L\sin 2\pi - 0\cdot\sin 0] - \frac{L}{2\pi}\left[-\frac{2\pi}{L}\cos\left(\frac{2\pi x}{L}\right)\right]_0^L = 0 + [\cos 2\pi - \cos 0] = [1-1] = 0$$

$$\rightarrow \quad \langle x \rangle = \frac{L}{2}$$

**Physics 2130 Fall 2010**
**Homework 07 Solutions**

**M/C & Short Answer: 5 Points**
**Essays: 14 Points**
**Long Answer: 10 Points**

**3. (1 Point)** The function ψ(x) is shown in the graph below:

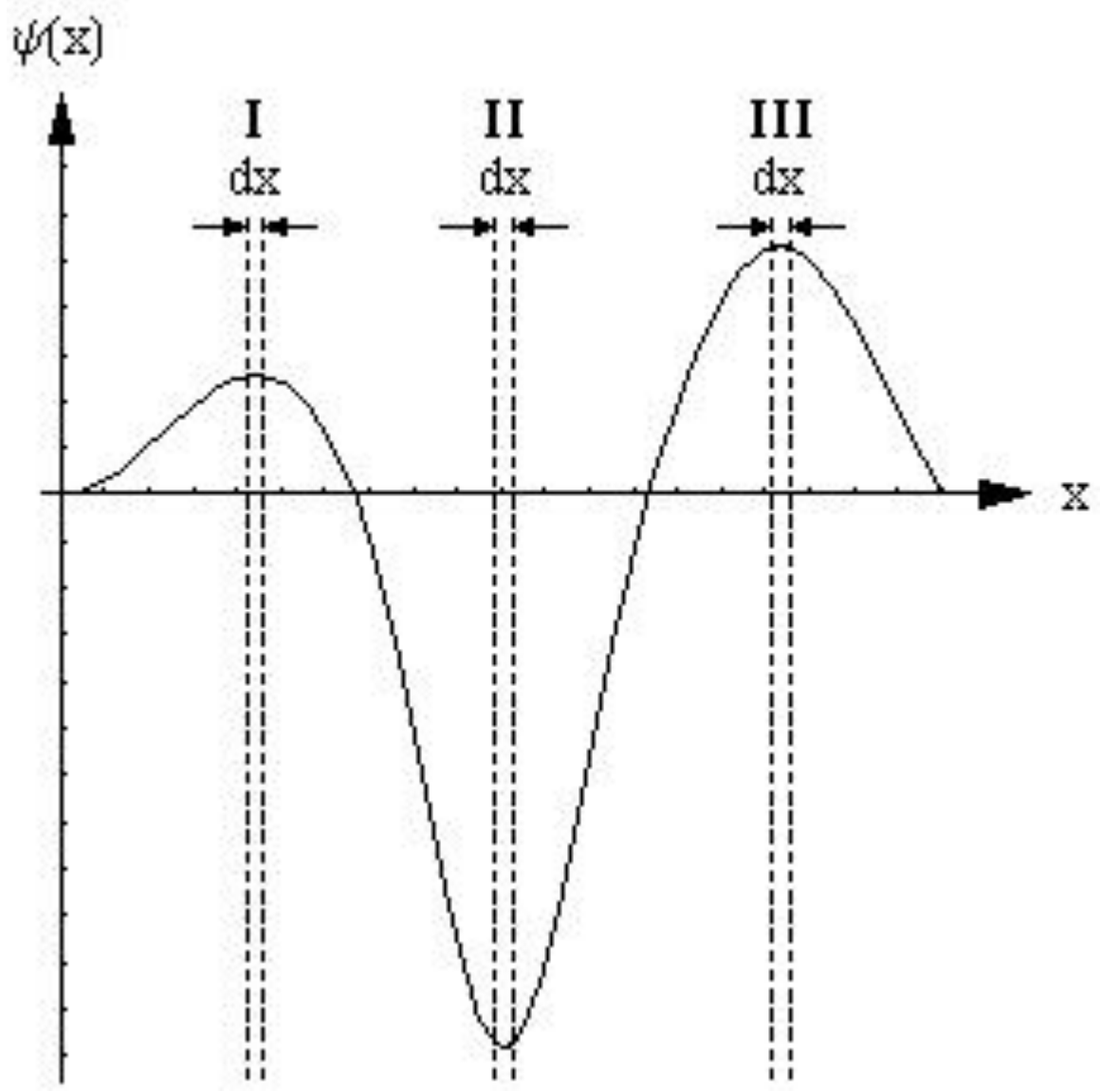

If a probability density function for the position of a particle is given by $\rho(x) = |\psi(x)|^2$ [i.e., the probability distribution is equal to the magnitude squared of the function ψ(x)], rank the probabilities of finding the particle in the regions shown.

A) P[III] > P[I] > P[II]
B) P[II] > P[I] > P[III]
C) P[III] > P[II] > P[I]
D) P[I] > P[II] > P[III]
E) P[II] > P[III] > P[I]

Answer: E

The most probable region is where the probability density is greatest, and the square of a real function is always positive. Since $\rho(x) = |\psi(x)|^2$, ρ(x) is greatest in Region II, then Region III, followed by Region I.

**4. (1 Point)** Atoms leaving the plus-channel of a vertically oriented Stern-Gerlach analyzer are fed into a second analyzer oriented in the +x-direction. With **Analyzer 2** oriented at $90^0$ to **Analyzer 1**, either result $\left|\uparrow_X\right\rangle$ or $\left|\uparrow_X\right\rangle$ is equally likely.

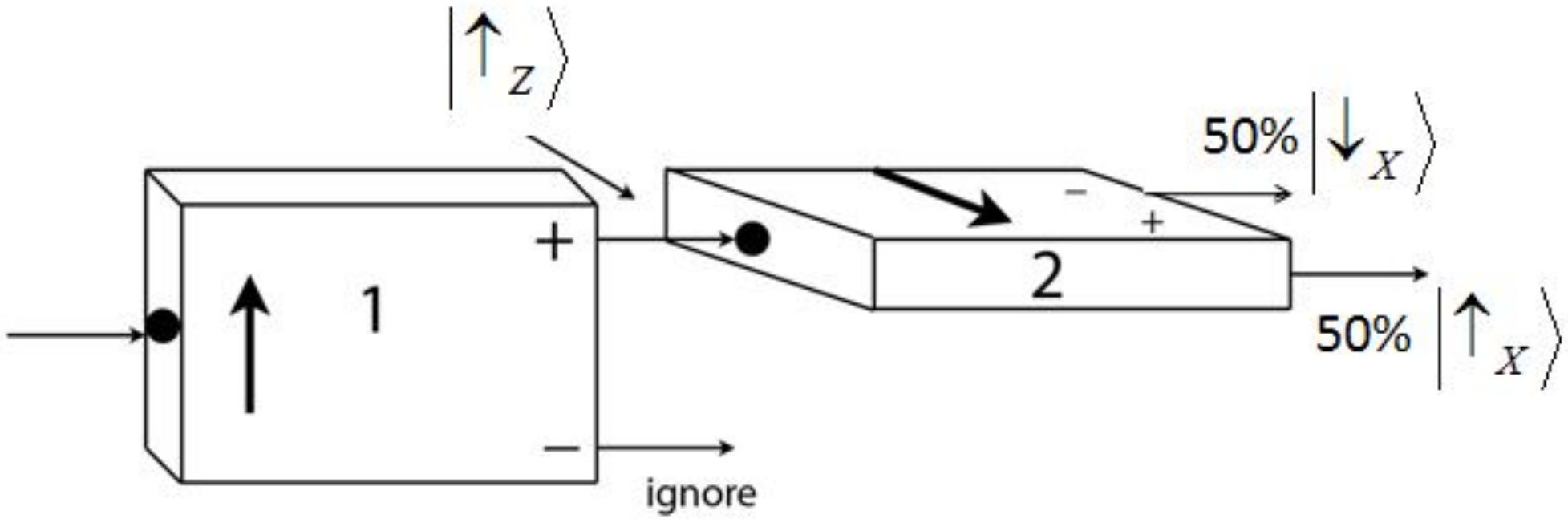

What is the average value of $m_X$? [$\langle m_X \rangle = ?$]

Answer: 0

$$\left\langle m_X \right\rangle = P\left[\left|\uparrow_X\right\rangle\right](+m_B) + P\left[\left|\downarrow_X\right\rangle\right](-m_B) = (0.50)(+m_B) + (0.50)(-m_B) = 0$$

**5-8.** In an Einstein-Podolsky-Rosen (EPR) experiment, an initial state of an atom pair is represented by initial and various hypothetical final states are shown below. What are the probabilities for observing each of the final states?

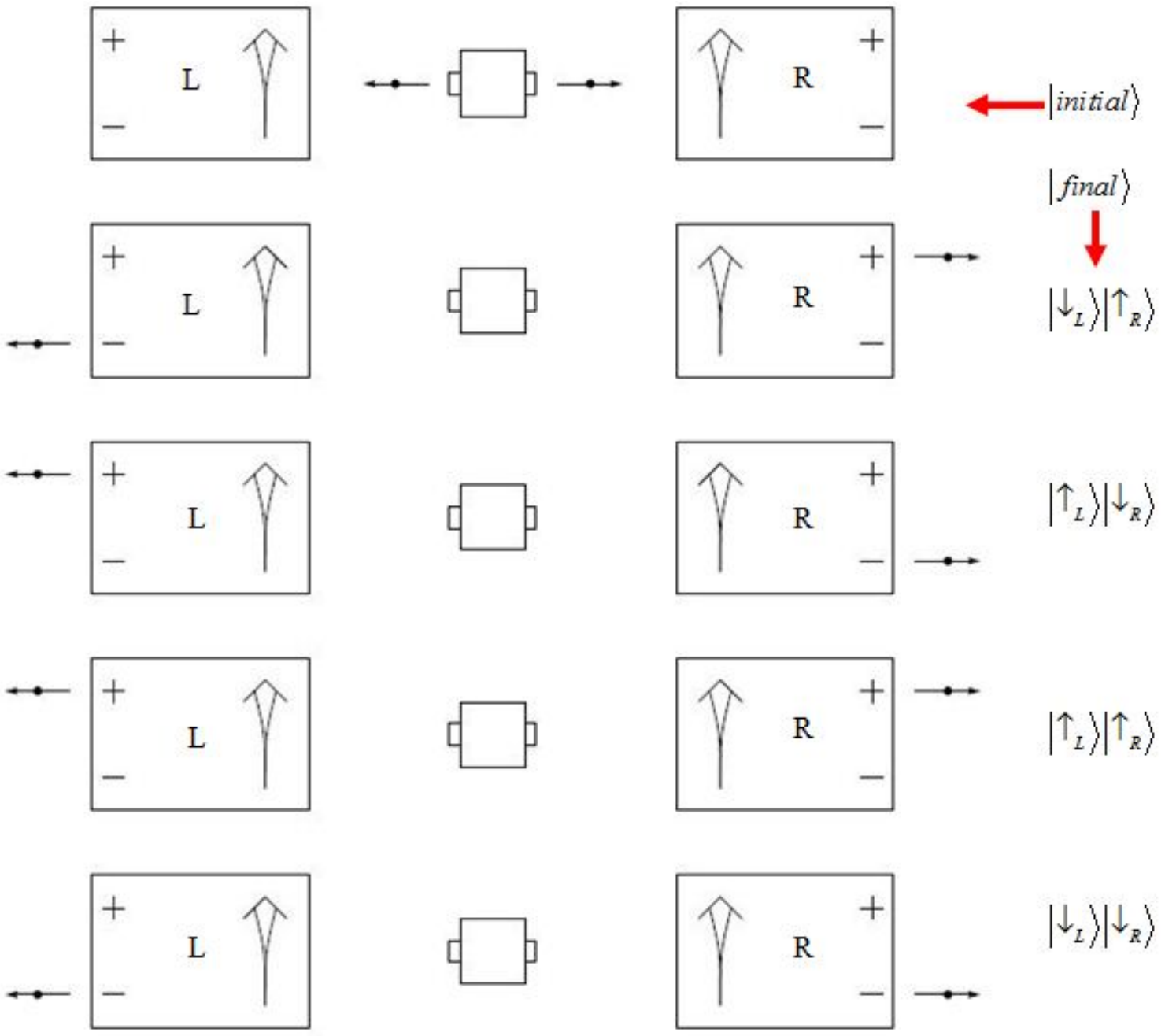

SOLUTION: (0.5 Points Each)

The entangled atom-pairs are in a state of zero total spin, meaning the results at the analyzers are perfectly anti-correlated – we always get opposite answers. The first two outcomes therefore occur with equal probability (50/50). The second two outcomes never occur (zero probability) since the total spin for the atom pairs must add up to zero.

(Notice that in this situation there is no such thing as a probability just for the right atom to exit from the plus-channel, because the probability for the right atom to exit from the plus-channel depends on whether the left atom exits from the plus-channel or minus-channel. The two atoms exist in an entangled state, but the right atom doesn't have a state unto itself.)

**Questions 9 & 10 refer to the readings: "A Quantum Threat to Special Relativity" & "Is the moon there when nobody looks?"**

**9. (4 Points – 1 Point Each)** What is meant by the terms ***realism, locality*** & ***completeness***? What are some examples of hidden variables?

**(1 Point)** ***Realism*** refers to a perspective where the properties of physical systems are considered to be objectively real (observation independent), in the sense that they exist and have definite values independent of any observer, human or not. Physical systems exist in definite states, whether we can completely know what that state is or not.

**(1 Point)** ***Locality*** refers to the intuitive notion that a measurement performed on one of two systems which are physically isolated from each other can't have any influence on the outcome of a measurement performed on the other, and vice-versa. [We find however, that correlated measurements can be performed on photon-pairs separated by more than 10 km.] Locality assumes there must be a physical mechanism for any interaction between two distant systems.

**(1 Point)** ***Completeness*** refers to whether a theory would be able (in principle) to describe (predict) all of the relevant properties of a physical system. If particles do indeed have simultaneously a definite position and momentum, then quantum mechanics would be considered incomplete, since it is unable to describe both of these quantities simultaneously. However, if these quantities do not have any definite value independent of measurement, then quantum mechanics might indeed be a complete description of reality.

**(1 Point)** The term ***hidden variable*** refers to any of these physical quantities not described by an incomplete theory. Some examples might be the position or momentum of a particle (things to do with its trajectory), the orientation of an atom's magnetic moment, the polarization state of a photon, etc…

**10. (2 Points)** Does ***entanglement*** allow for faster-than-light communication? If so, what kind of information can be communicated? If not, why not?

No. The entangled particles are traveling at a finite speed, even though the collapse of the wave function is assumed to be instantaneous. When distant correlated measurements are being performed, both observers will know ahead of time what the outcome of the other's measurement will be before they compare data, as long as the analyzers are oriented along the same axis. In this way, a **randomly** generated encryption key composed of strings of 1's and 0's ("up" and "down") could be transmitted between the two persons (each knows what the other measured), but it still takes a finite amount of time for the atom-pairs to travel the distance from source to detector. Most importantly, neither of the two observers (nor any person at the source) has any way to control the outcome of any particular measurement (it is random), which is what we'd need to be able to do in order to transmit a faster-than-light signal.

When we wrote this question, we were thinking about communicating information between two humans, but if you answered that entanglement allows two particles to communicate with each other at faster-than-light speeds, then that answer will also be accepted – but let's not be too liberal about using the word "communication" when we're talking about inanimate objects.

**11. (2 Points)** In the two Aspect experiments discussed in class, where the goal was to produce a "single-photon" source, the calcium atoms were excited to the upper level by a two-photon absorption process:

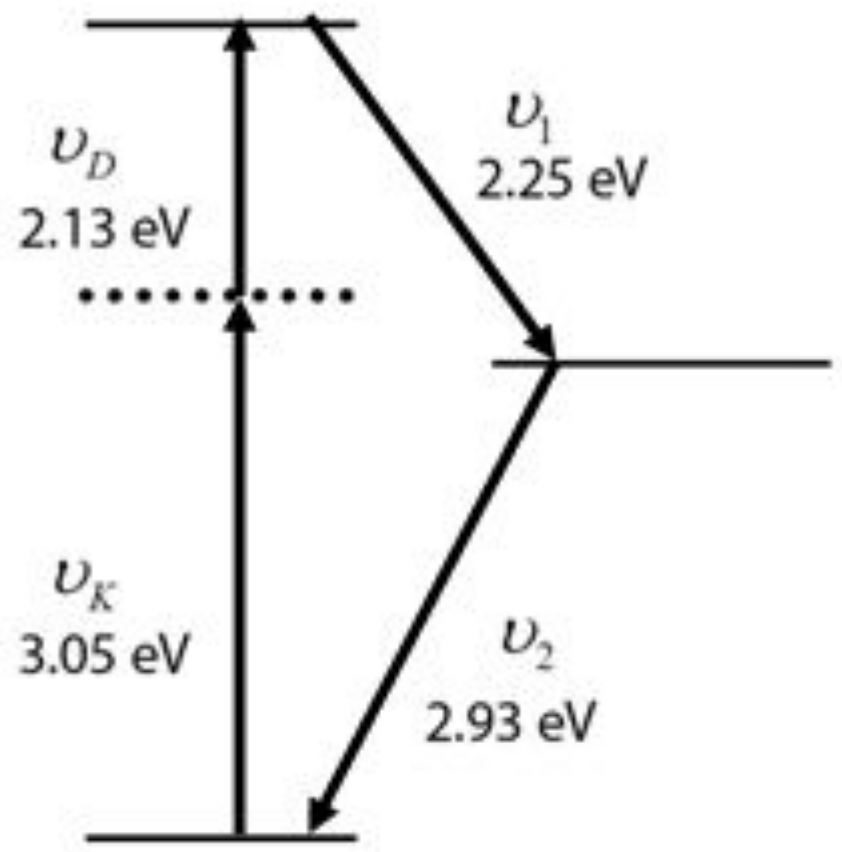

Why did the experimenters excite the calcium atoms with a laser pulse of 3.05 eV photons followed by a pulse of 2.13 eV photons, rather than with single photons with the same energy as the two original photons combined (single-photon excitation)?

In this course we have emphasized all along that two-photon absorption by an atom is a much rarer occurrence than any single-photon excitation. The experimenters exploited this fact in order to keep the intensity of the photon source as low as possible. If the experimenters had excited the calcium atoms to the upper level through single-photon absorption, then a large number of atoms would have been able to undergo de-excitation by back-to-back two-photon emission, and the intensity of the photon source would have been too high to expect to have single photons in the apparatus at one time. Since the two-photon absorption process is much rarer, the experimenters were able to reduce the intensity down to an acceptable level.

**12. (2 Points)** In your own words, explain what the anti-correlation parameter (**α**) is, both in terms of its mathematical definition, and in terms of what it physically tells us, in the context of single-photon experiments as performed by Aspect. Why didn't Aspect measure **α** = 0 if photons are supposed to be acting like particles?

In the first Aspect experiment, there were two PMT's connected to a coincidence counter, which clicked only when a photon was detected in both PMT's during the time the gate was open. The anti-correlation parameter **α** is defined as:

$\alpha = \frac{P_{12}}{P_1 P_2}$, where $P_1$ and $P_2$ are the probability for PMT1 and PMT2 to be triggered respectively, and $P_{12}$ is the probability for both PMT's to be triggered while the gate is open. When the intensity of the photon source is low, **α** >= 1 if photons are behaving like classical waves (since $P_{12}$ is greater than $P_1$ times $P_2$, meaning the PMT's are being triggered together more often than not) & **α** >= 0 if photons are behaving like particles (since both PMT's should not trigger simultaneously if there is only a single photon in the apparatus). **α** should equal zero only if there were always exactly one photon in the apparatus during the time the gate is open, but the intensity of the photon source is not quite that low. Aspect continually reduced the intensity of the photon source, but had to extrapolate the data down to single-photon luminosity.

**13. (2 Points)** Log onto CU Learn and click on the "Discussions" tab on the left-hand side. There you will find several topic threads that have been seeded with some of your questions from two of the reading assignments: "100 Years of Quantum Mysteries" & "A Quantum Threat to Special Relativity". Each week, you will receive two homework points for contributing to the discussion there by making ***at least one*** post per week. A posting worth full credit can be an answer or explanation to one of the questions already posted (or an elaboration or re-wording of an already posted explanation), an additional question relevant to the topic thread, a comment on a previous posting that contributes to the overall discussion, or even starting a new topic thread on something else relevant to the course. All of the postings will be anonymous to other students (but not to the instructors), so feel free to say what's really on your mind, but please remember to be respectful at all times.

## LONG ANSWER

**(5 Points)** In class we have discussed correlated measurements performed on systems of two entangled atoms. The assigned reading "The Reality of the Quantum World" discusses correlated measurements performed on entangled photon pairs. In what ways are these systems of entangled photon pairs similar or different from systems of entangled atom pairs? In what sense are the particles in each system entangled (i.e., what properties are correlated for each of the two types of systems)? What types of measurements are performed to determine these properties, and what are probabilities for the possible results of these measurements for both types of systems?

The systems are similar in the sense that for both, two quantum particles are produced in an entangled superposition state, where the outcome of a given set of measurements is indeterminate until the moment one or both of the particles are observed. Even when the two measurements are performed at a distance, the outcomes of the two measurements are correlated in some way when we are asking the same question.

In the case of atoms, the pair is produced in a superposition state of zero total spin; a set of Stern-Gerlach analyzers measure the projection of each atom's spin along some axis. There is a 50/50 probability that a given atom will be observed in either the "spin-up" state or the "spin-down" state along the axis of measurement, but the outcomes for each atom-pair are always anti-correlated, meaning we always get **opposite** answers at the two analyzers.

In the case of photons, the pair is produced in a superposition state of equal parts vertical and horizontal linear polarization. When each photon is passed through a set of linear polarizing films oriented along some axis, there is a 50/50 probability for a photon to either be transmitted or blocked, but the outcomes for each photon-pair are always strictly correlated, meaning either both are transmitted or both or blocked, so that we always get the **same** answers at the two analyzers.

**(5 Points)** As discussed in class and in the readings, what do the two single-photon experiments performed by Aspect tell us about the nature of photons? How were the two experiments designed to demonstrate the particle and the wave nature of photons? When answering, don't concern yourself with technical details (such as how the photons were produced); focus instead on how the design of each experimental setup determined which type of photon behavior would be observed. How are the elements of these two experiments combined in a delayed-choice experiment, and what do delayed-choice experiments (along with the two Aspect experiments) tell us about the nature of photons?

The two experiments performed by Aspect were designed to demonstrate either the particle behavior of photons, or their wave behavior. In each experiment, a photon was passed through a beam splitter, where it had a 50/50 probability of being either transmitted or reflected along one of two possible paths. In the first experiment, photons were detected either in one PMT or the other (but not both simultaneously), meaning each photon could have taken only one of the two possible paths on the way to being detected. This experiment demonstrated that photons behave like particles, always taking either one path or the other when they encounter the beam splitter.

The second experiment was similar to the first, except now a second beam splitter was inserted at the point where the two separate paths intersected. Now, each photon might be again either transmitted or reflected at the second beam splitter, and detected in one PMT or the other, but now we have no information about which path the photon took. By changing the distance travelled along just one of the paths (by moving one of the mirrors) each photon is observed to interfere with itself – we can arrange it so all the photons are detected in one PMT, or that all of them are detected in the other PMT, or anywhere in between (depending on the path-length difference). Since each photon is somehow "aware" of the length travelled along both paths, it must have acted like a wave, taking both paths simultaneously when it encounters the beam splitter.

In a delayed-choice experiment, it is now arranged so that we can open or block one of the paths after the photon has encountered the first beam splitter. If one of the paths is blocked, then any photon we detect must have travelled along just one of the two paths – the photon should behave like a particle when it encounters the first beam splitter and go one way or the other. If both paths are left open, then the photon can have travelled by either path and interference is observed – each photon was coherently split at the first beam splitter and took both paths, behaving like a wave. This setup allows us to switch from Experiment 1 (photon takes Path A or Path B at the first beamsplitter) to Experiment 2 (photon takes both Paths A & B at the first beamsplitter) after the photon has encountered the first beam splitter, yet the photon still acts as though we've been conducting Experiment 2 all along, which makes us question whether the photon could really have already "decided" whether to behave like a wave or a particle before the choice was made to block one of the paths or leave it open.

Results from these three experiments demonstrate that photons behave like waves when wavelike properties are measured (interference), and like particles when particle-like properties are measured (which-path information), but also that we should not think of a photon as simultaneously both particle and wave (a "wavicle"). We can take a Complementarity point of view and say no experiment can be designed to simultaneously demonstrate both particle and wave behavior. Dirac says we can always think about an unobserved photon acting like a wave, but we must accept the collapse of the wavefunction at the moment of detection to mean the photon goes from being in both paths, to suddenly being in just one.

# Exam II

**MULTIPLE CHOICE: 22 Questions (1 Point Each) = 22 Points**

**1.** If we halve the distance between an electron and the nucleus (with Z-protons) in a single electron atom, what happens to the potential energy of the electron?

**A) It becomes more negative by a factor of two.**
B) It becomes more negative by a factor of 2Z.
C) It becomes more positive by a factor of two.
D) It becomes more positive by a factor of 2Z
E) It changes by a factor of 4.

$$PE(r) = -\frac{kZe^2}{r} \quad \rightarrow \quad PE(r/2) = -\frac{kZe^2}{r/2} = -2\frac{kZe^2}{r} = -2\left|PE(r)\right|$$

**2.** Bohr's model predicts spectra for any atom if it is ionized such that there is only one electron orbiting. Consider a doubly-ionized Li++ atom (3 protons, 4 neutrons, 1 electron).

If the light curve (which is the same in each of the graphs below) represents the PE curve for H as a function of distance from the nucleus, which of the dark curves represents the PE curve for Li++?

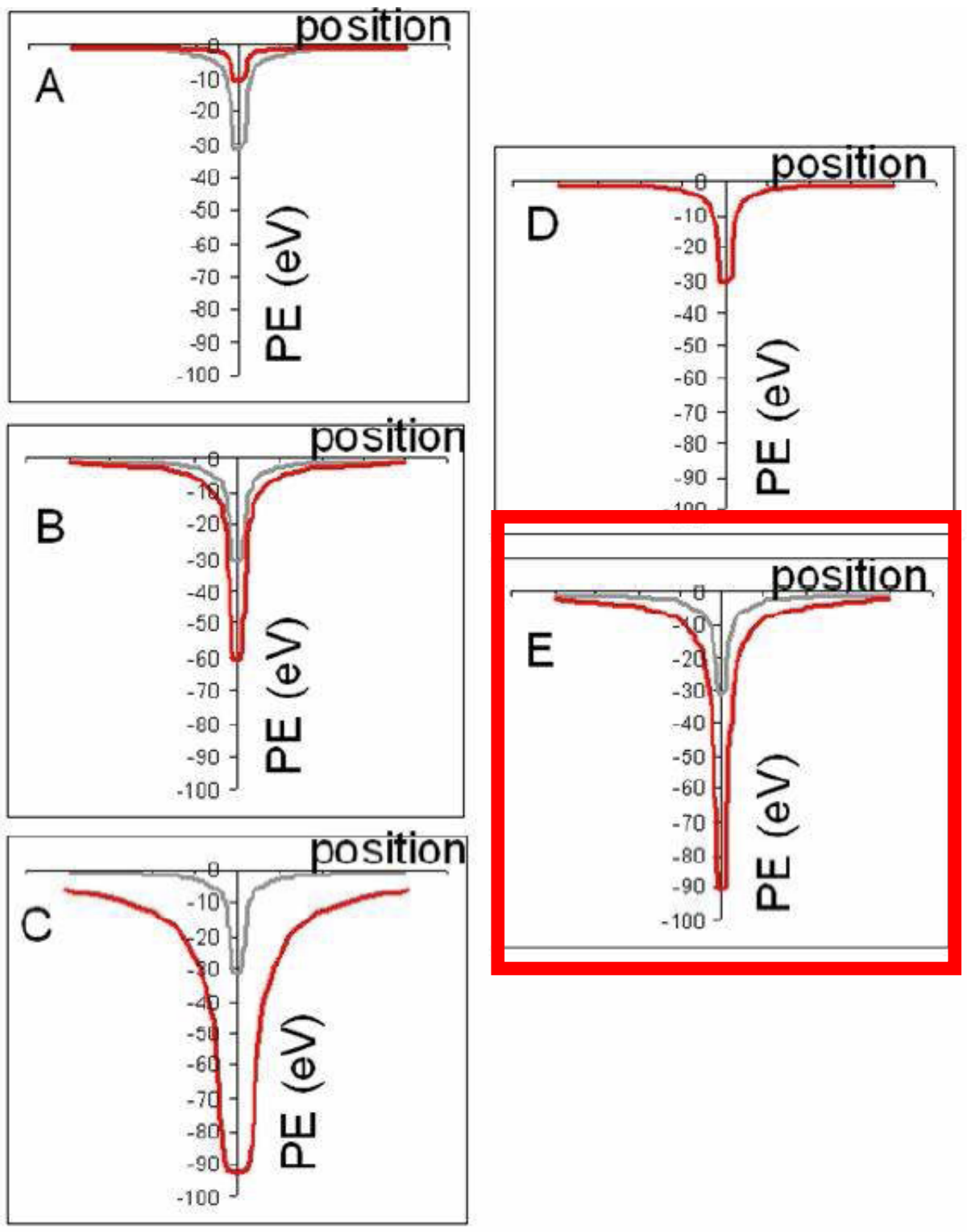

**3 & 4.** The energy level diagram at right roughly shows the lowest four electronic energy levels for a hydrogen atom (ground state $E_1$ = -13.6 eV and the first three excited states: $E_2$, $E_3$ & $E_4$ – **note that the diagram is not drawn to scale**).

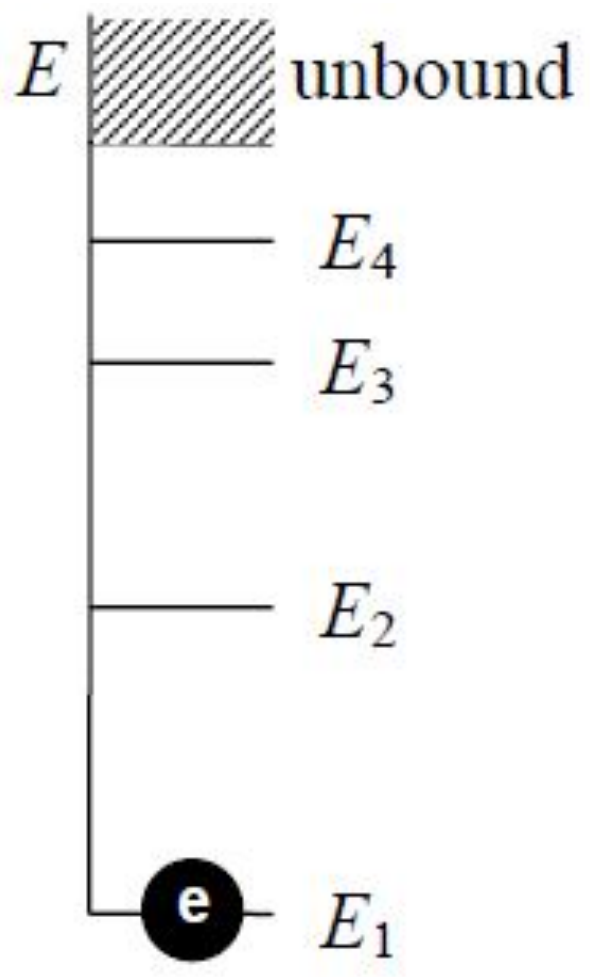

**3.** If the electron starts out in the ground state and is excited to level $E_3$ by an incoming photon, what was the wavelength of that photon (in nm)?

A. 95.4 nm
B. 102.5 nm
C. 121.5 nm
D. 136.7 nm
E. 182.3 nm

$$\Delta E = E_3 - E_1 = \frac{-13.6 \text{ eV}}{(3)^2} - \frac{-13.6 \text{ eV}}{(1)^2} = -1.51 \text{ eV} + 13.6 \text{ eV} = 12.1 \text{ eV}$$

$$\lambda = \frac{hc}{E_\gamma} = \frac{1240 \text{ eV} \cdot \text{nm}}{12.1 \text{ eV}} = 102.5 \text{ nm}$$

**4.** When the electron transitioned from $E_1$ to $E_3$ its orbital radius increased by a factor of:

A) 1 (It didn't change)
B) 2
C) 3
D) 4
E) 9

$$r_n = n^2 a_B \quad \rightarrow \quad \frac{r_3}{r_1} = \frac{(3)^2}{(1)^2} = 9$$

**5.** Atoms exiting from the plus-channel of Stern-Gerlach Analyzer 1 (oriented along the +z-axis) are fed into Analyzer 2 (oriented along the –z-axis). What is the probability for the atoms to exit from the minus-channel of Analyzer 2?

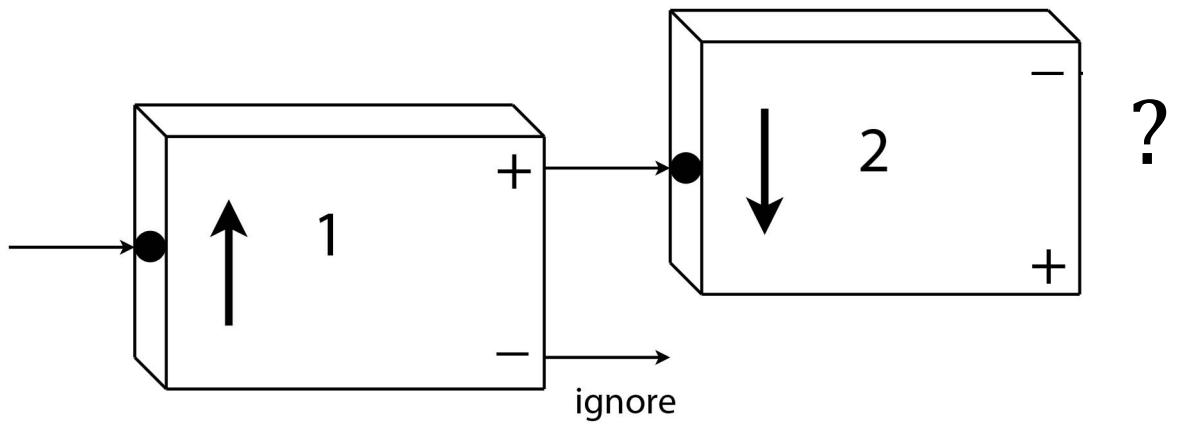

A) 0 B) 1/4 C) 1/2
D) 3/4 E) 1

**The atoms exiting Analyzer 1 are in the definite state:** $|\uparrow_z\rangle = |\downarrow_{-z}\rangle$

**Therefore, 100% of atoms exit from the minus-channel**

**6.** Suppose we have two "Local Reality Machines" (Stern-Gerlach analyzers capable of being oriented along three different axes: A, B, & C, each oriented at $120^0$ to each other, as shown) set up to detect atom-pairs emitted in an entangled state:

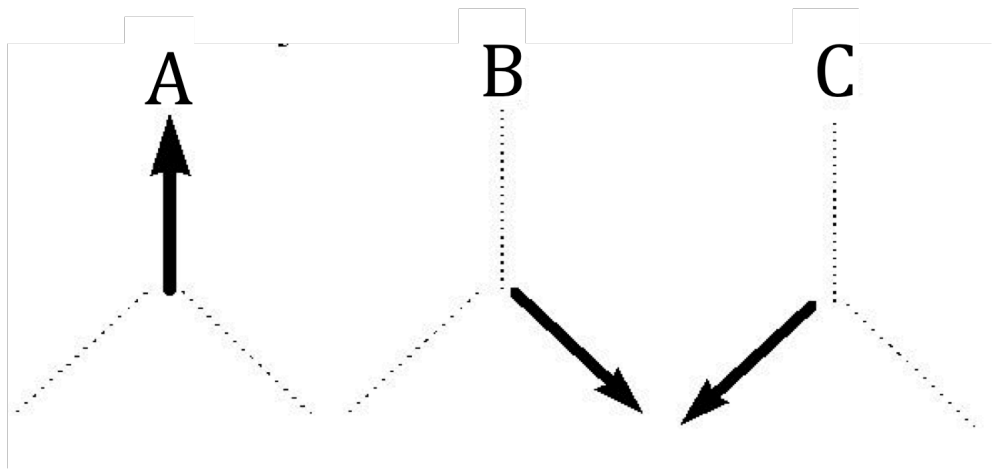

$$\left|\Psi_{12}\right\rangle = \left|\uparrow_1\right\rangle\left|\downarrow_2\right\rangle + \left|\downarrow_1\right\rangle\left|\uparrow_2\right\rangle$$

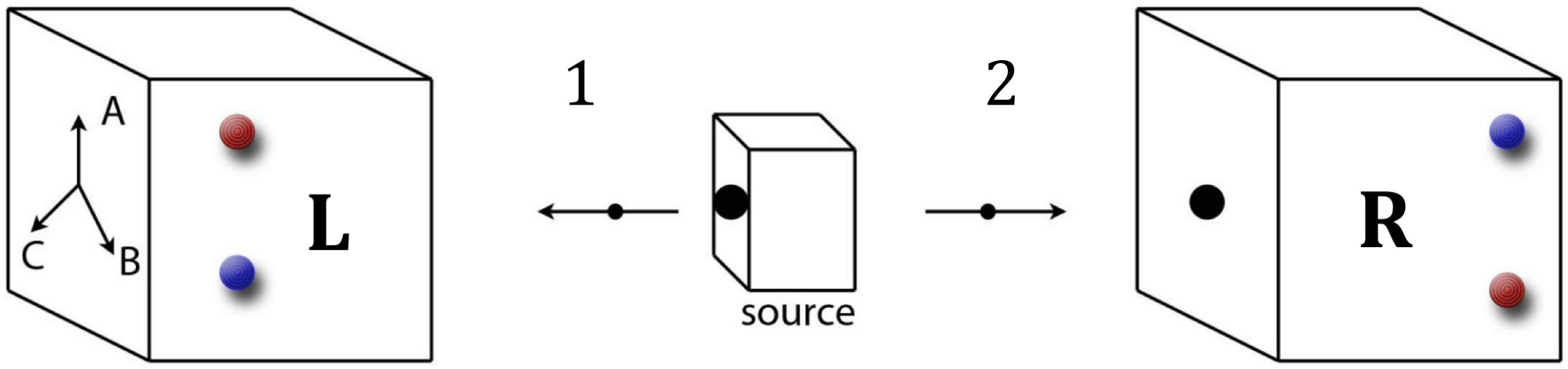

The leftward travelling atom (1) reaches the left analyzer (L) before the rightward traveling atom (2) reaches the right analyzer (R). The left analyzer is set on A and measures atom 1 to be "up" along the vertically oriented A-axis (it exited from the plus-channel). A short time later, atom 2 enters the right-side analyzer. If the right analyzer is set on B ($120^0$ from the vertical axis), what is the probability for atom 2 to exit from the plus-channel of the right analyzer?

A) 0
B) 1/4
C) 1/2
D) 3/4
E) 1

**Atom 2 is in the definite state** $\left|\downarrow_Z\right\rangle \quad \rightarrow \quad P\left[\left|\uparrow_{120^0}\right\rangle\right] = \cos^2\left(\frac{60^0}{2}\right) = \left(\frac{\sqrt{3}}{2}\right)^2 = 3/4$

**7 & 8.** Suppose you have an electron and a photon both moving through free space, each with a total energy of 9 eV.

**7.** What is the deBroglie wavelength in nm of the photon?
A) 0.41 nm  B) 138 nm  C) 276 nm  D) 410 nm
E) Photons do not have a deBroglie wavelength.

$$\lambda = \frac{hc}{E_\gamma} = \frac{1240 \text{ eV}\cdot\text{nm}}{9 \text{ eV}} = 138 \text{ nm}$$

**Same as the wavelength we've calculated all along.**

**8.** What is the deBroglie wavelength of the electron?

A) 0.41 nm  B) 0.58 nm  C) 138 nm  D) 276 nm  E) 726 nm

$$\lambda = \frac{h}{p} \quad \& \quad \frac{p^2}{2m} = E \quad \rightarrow \quad \lambda = \frac{h}{\sqrt{2mE}}$$

$$\lambda = \frac{6.6\times10^{-34} J\cdot s}{\left[2(9.11\times10^{-31} kg)(9 \text{ eV})(1.6\times10^{-19} J/s)\right]^{1/2}} = 4.1\times10^{-10} \text{ m} = 0.41 \text{ nm}$$

**9 & 10.** An electron's wave function between x = 0 and x = L is described by the following function:

$$\Psi(x) = \sqrt{\frac{2}{L}}\sin\left(\frac{2\pi x}{L}\right) \text{ for } 0 \le x \le L$$

$$\Psi(x) = 0 \text{ for } x < 0 \text{ and } x > L$$

**9.** What does the function $\Psi(x)$ look like?

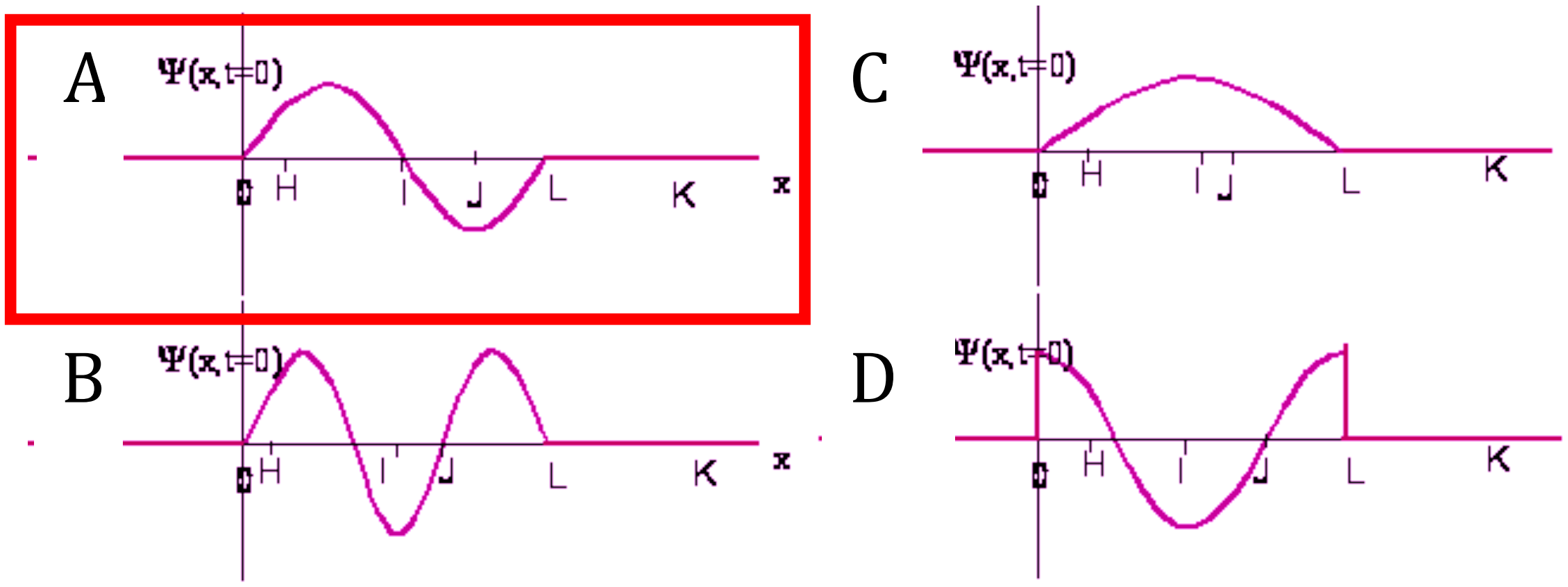

**10.** For an electron described by this same function:

The likelihood that you find the particle at x = L/2 is _____ the likelihood of finding it at L/4.

A) greater than  B) smaller than  C) the same as
D) Impossible to tell

$$\rho(x = L/2) = \frac{2}{L}\sin^2(\pi) = 0 < \rho(x = L/4) = \frac{2}{L}\sin^2\left(\frac{\pi}{2}\right) = \frac{2}{L}$$

**11.** For ***a different wave function***, plotted below on the right, how do the probabilities of finding the electron very close (within a very small distance dx) to x = G, H, I, J, and K compare? (G=Probability of finding the electron near point G, etc…):

A) G = H = I = J = K
B) H > J = G = K > I
C) I > H > G = J = K
D) J > H > I = G = K
E) H > I > J > G = K

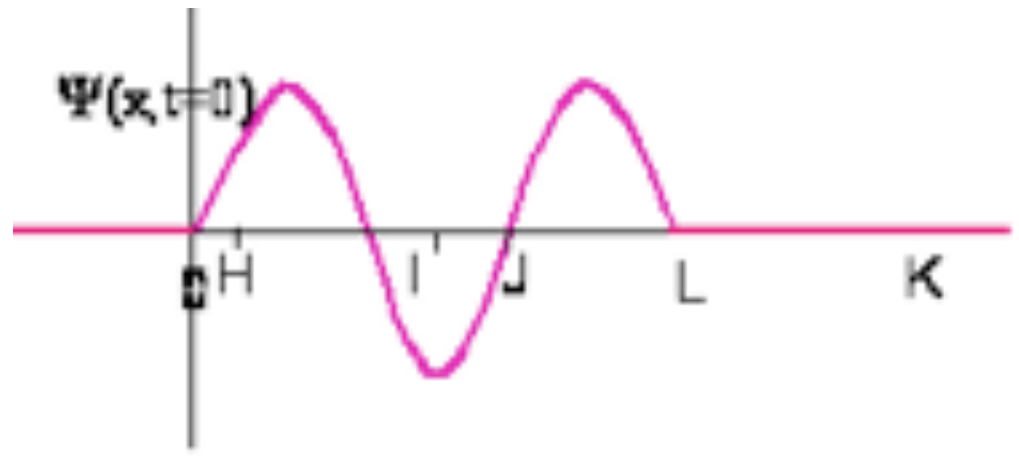

**12.** A free electron is generally described by the wave function $\Psi(x) = A\exp(ikx)$. According to Euler's formula, this is $\Psi(x) = A\exp(ikx) = A \cdot [\cos(kx) + i\sin(kx)]$

For this wave function, a bigger **k** …

A) means a bigger wavelength
B) has no effect on wavelength
C) means a smaller momentum
D) means a larger momentum
E) has no effect on momentum

$$p = \frac{h}{\lambda} \quad \& \quad k = \frac{2\pi}{\lambda} \quad \rightarrow \quad p = \hbar k$$

**13.** The Heisenberg Uncertainty Principle is generally applied to very small objects such as electrons and protons. Which of the following statements **best explains** why don't we use the uncertainty principle on larger objects such as cars and tennis balls?

A) The errors of measurement can always, in principle, be made smaller by using more sensitive equipment.
B) Large objects at any instant of time have an exact position and exact momentum and with sufficient care we can measure both precisely.
C) Large objects obey Newton's laws of motion, to which the uncertainty principle does not apply.
D) Because it does apply to large objects, but the uncertainties are so small that we don't notice them.

**14.** Three particles of equal mass are traveling in the same direction. The de Broglie waves of the three particles are as shown at right. Rank the speeds of the particles I, II and III:

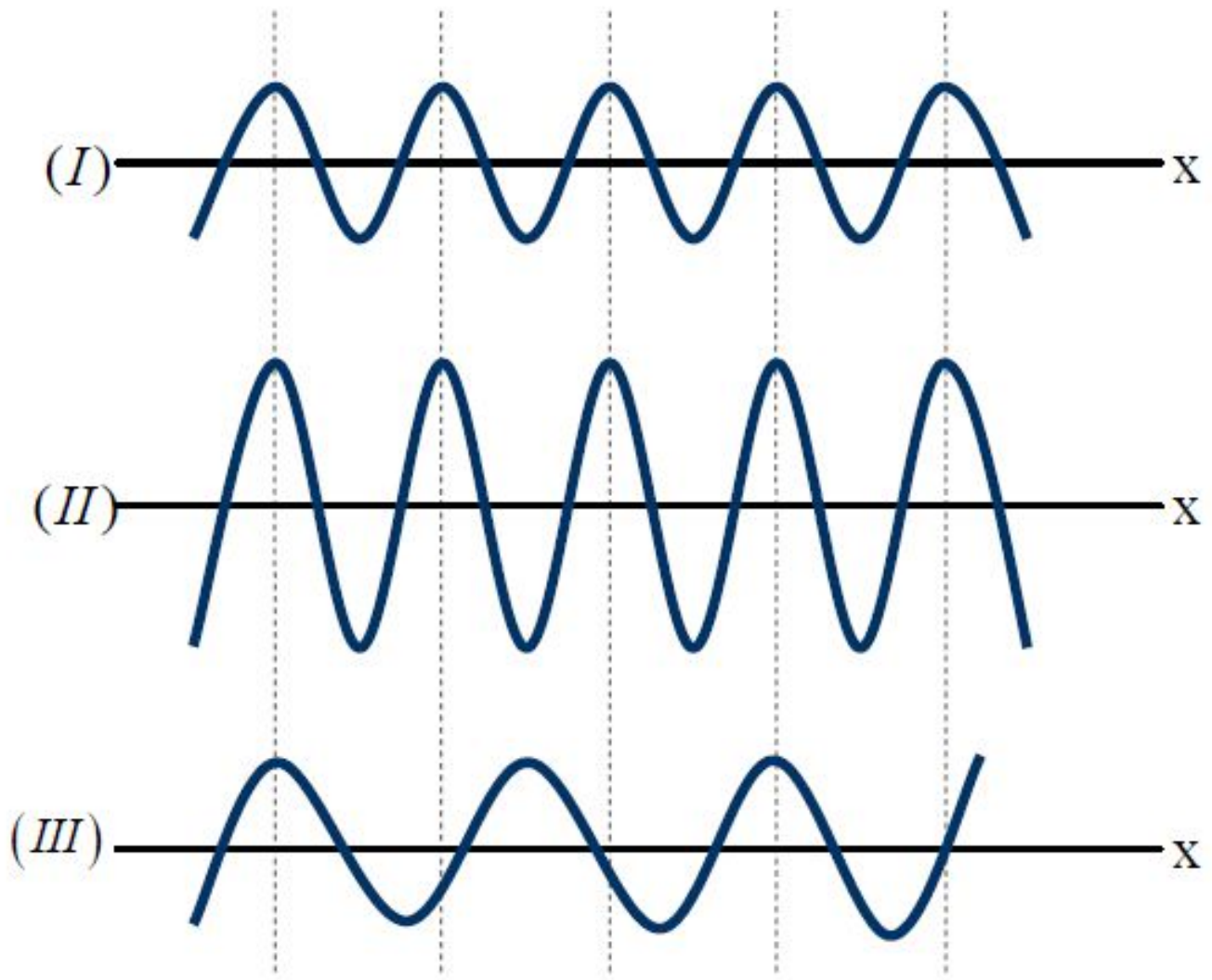

A) $V_{II} > V_I > V_{III}$
B) $V_{II} > V_{III} > V_I$
C) $V_I = V_{II} > V_{III}$
D) $V_{II} > V_I = V_{III}$

**Amplitude has no relation to wavelength, which is the same for I & II; these are shorter than III, and so III has the least speed, while I & II are the same.**

**15 & 16.** Consider the following two cases:

**Case 1**: At time t = 0, an electron is described by a plane wave, $\psi(x) = Ae^{ikx}$, where we have drawn the real part below (the wave keeps going forever off the edge of the paper):

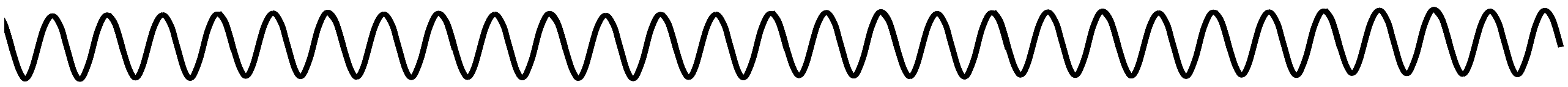

**Case 2**: At time t = 0, an electron is described by a wave packet as drawn below:

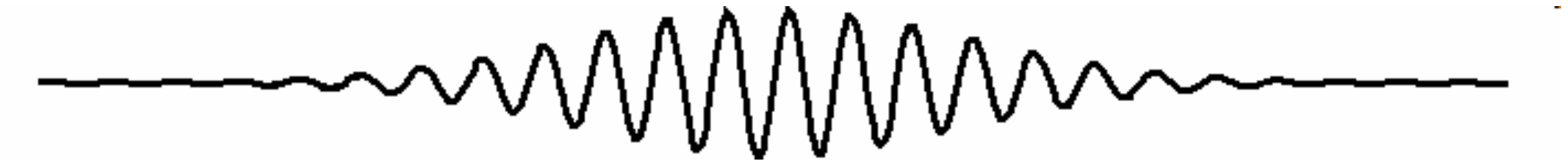

**15.** Are the following statements true or false?
**I.** There is no uncertainty in the momentum of the electron in Case 1.
**II.** The uncertainty in the position of the electron in Case 1 is less than the uncertainty in the position of the electron in Case 2.

A) I = true, II = true
B) I = true, II = false
C) I = false, II = true
D) I = false, II = false

**16.** Which of the following statements **best explains** what is happening with the uncertainty in momentum for Case 2 and why:

A) The uncertainty in momentum for case 2 is the same as for case 1, because the wavelength is the same.
B) The uncertainty in momentum for case 2 is less than for case 1, because the wave is less spread out.
C) The uncertainty in momentum for case 2 is greater than for case 1, because to create a localized wave packet requires the superposition of sine waves with a range of wavelengths and a range of momentum.
D) The uncertainty in momentum for case 2 is less than for case 1, because the wave is more spread out and uncertainty in momentum and position trade off.
E) The uncertainty in momentum for case 2 is greater than for case 1, because the wave packet is made up of a number of electrons with a range of momentum.

**Questions 17 through 21** refer to the following three experiments:

In one experiment ***electrons*** pass through a double-slit as they travel from a source to a detecting screen.

In a second experiment ***light*** passes through a double-slit as it travels from a source to a photographic plate.

In a third experiment ***marbles*** pass through two slit-like openings as they travel from a source to an array of collecting bins, side-by-side.

The right-hand figure diagrams the experimental setup, and the figures below show roughly the possible patterns which could be detected on the various screens.

Top view of experimental set-up (not to scale)

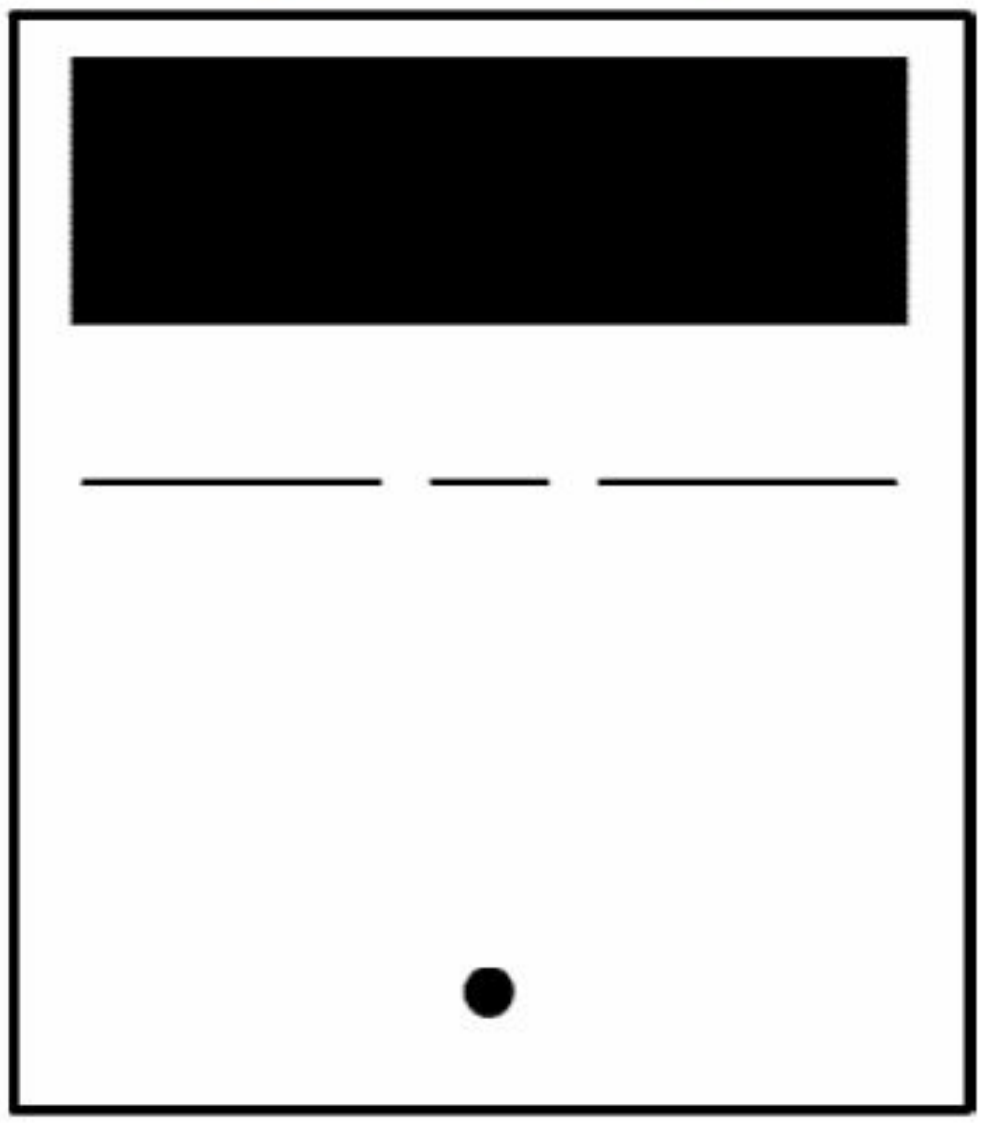

Possible patterns (not to scale)

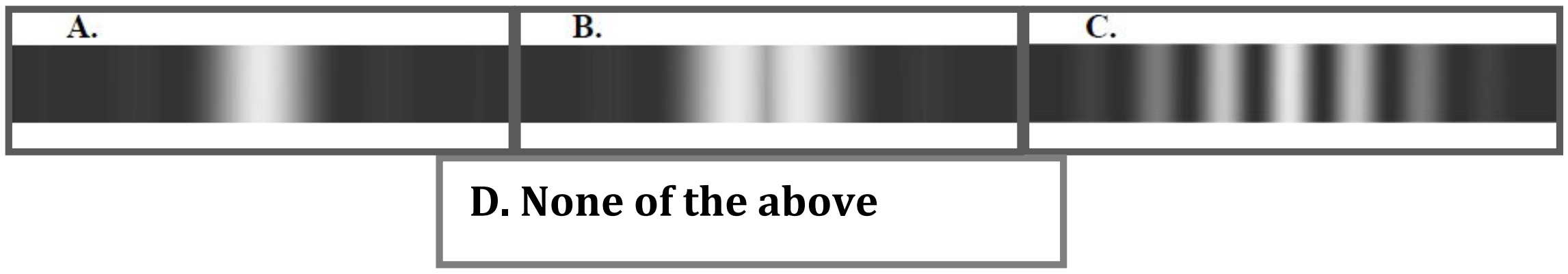

A, B & C represent some patterns which might be observed. If you think none is appropriate, answer D. Which type of pattern would you expect to observe when…

**17.** **…***light* passes through the double slit? **C**

**18.** **…***marbles* pass through the double opening? **B**

**19.** **…***electrons* pass through the double slit? **C**

**20.** **…***light* passes through the apparatus when *one of the slits is covered*? **A**

**21.** **…***electrons* pass through the apparatus when *one of the slits is covered*? **A**

**22.** Protons are accelerated from rest through a potential difference of 1000 Volts before passing through a double-slit apparatus and being detected on a screen. What would be the deBroglie wavelength for these protons in picometers? (1 pm = $10^{-12}$ m)

A) 0.90 pm B) 1.80 pm C) 38.7 pm D) 77.3 pm E) 1240 pm

$$\lambda = \frac{h}{\sqrt{2mE}} = \frac{(6.6\times10^{-34}\,J\cdot s)}{\left[2(1.67\times10^{-27}\,kg)(1000\text{ eV})(1.6\times10^{-19}\,J/\text{eV}\right]} = 9.0\times10^{-13}\,m = 0.90\,pm$$

**ESSAY QUESTIONS: PLEASE NOTE – YOU MUST ANSWER THE FIRST ESSAY QUESTION (E1 – 10 POINTS) AND THEN ONE OF THE TWO REMAINING ESSAY QUESTIONS (E2 OR E3 – 8 POINTS). IF YOU CHOOSE TO RESPOND TO BOTH OF THE LAST TWO ESSAY QUESTIONS, WE WILL GIVE YOU THE POINTS FOR THE HIGHER OF THE TWO SCORES.**

**E1. (REQUIRED – 3 PARTS, 10 POINTS TOTAL)** In the sequence of screenshots shown below (taken from the PhET Quantum Wave Interference simulation), we see: A) a bright spot (representing the probability density for a single electron) emerges from an electron gun; B) passes through both slits; and C) a single electron is detected on the far screen. After many electrons pass through and are detected, a fringe pattern develops (not shown).

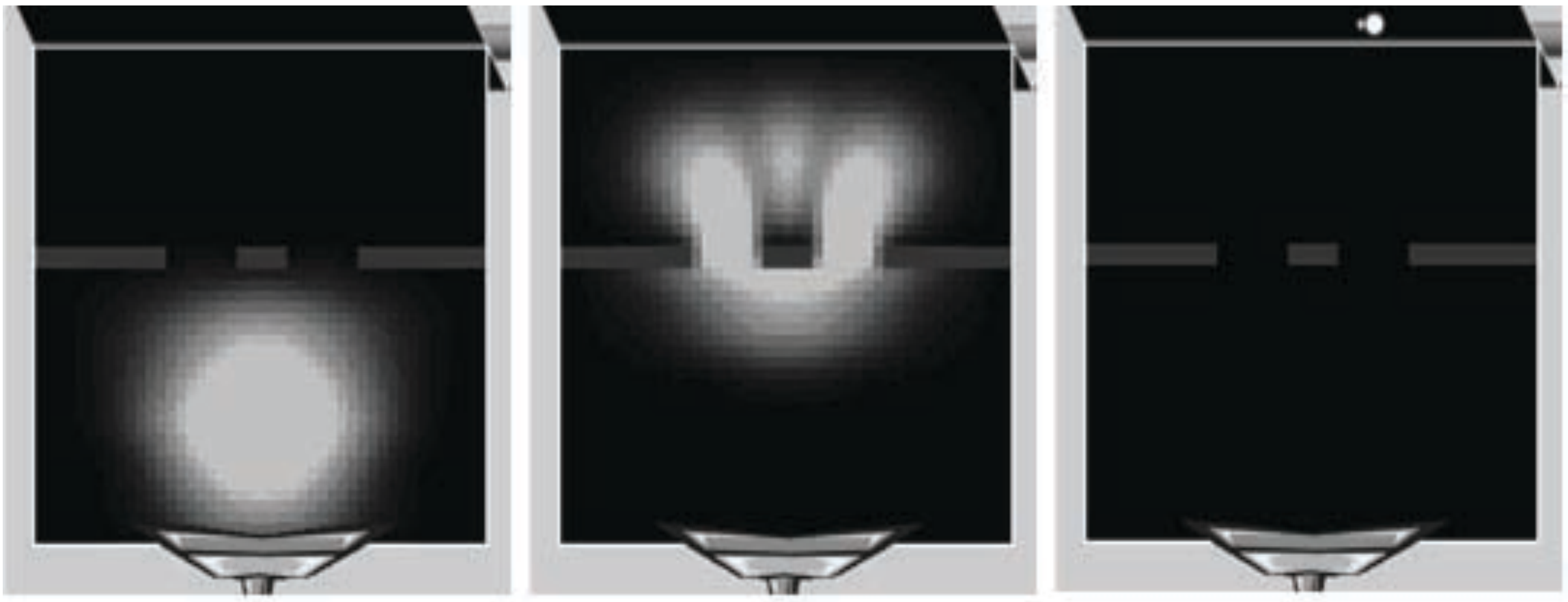

(A) (B) (C)

**Three students discuss the Quantum Wave Interference simulation:**

**Student 1**: The probability density is so large because we don't know the true position of the electron. Since only a single dot at a time appears on the detecting screen, the electron must have been a tiny particle traveling somewhere inside that blob, so that the electron went through one slit or the other on its way to the point where it was detected.

**Student 2**: The blob represents the electron itself, since a free electron is properly described by a wave packet. The electron acts as a wave and will go through both slits and interfere with itself. That's why a distinct interference pattern will show up on the screen after shooting many electrons.

**Student 3**: All we can really know is the probability for where the electron will be detected. Quantum mechanics may predict the likelihood for a measurement outcome, but it really doesn't tell us what the electron is doing between being emitted from the gun and being detected at the screen.

**E1.A (2 Points)** In terms of the interpretations of quantum phenomena we've discussed in class, how would you characterize the perspective represented by Student 1's statement? What assumptions are being made by Student 1 that allows you to identify their perspective on this double-slit experiment?

**Student One's statement would be consistent with a realist perspective (that of Albert Einstein). [Other key words we accepted: classical ignorance, hidden variables, "anti-Complementarity, and the like. Student One assumes the position of the electron is a real**

**but unknown quantity, and that the superposition wave packet is an expression of classical ignorance. Student One assumes that if the electron is detected at only one point in space, then it must have always existed at some one point in space as it was passing through the apparatus, and therefore must pass through one slit or the other on its way from source to detector. Student One considers the position of the electron to be a hidden variable.**

**E1.B (6 Points)** For each of the first two statements (made by Students 1 & 2), what rationale or evidence (experimental or otherwise, if any) exists that favors or refutes these two points of view? As for the third statement, is Student 3 saying that Students 1 & 2 are wrong? Why would a practicing physicist choose to agree or disagree with Student 3?

**In favor of Student One:**
**Intuition, Classical/Newtonian physics.**
**Particles are always localized upon detection.**
**If a detector is placed at either (or both) of the slits, electrons are detected at one slit or the other, but not both.**

**Against Student One:**
**Covering one slit shouldn't affect particles that were only passing through the other slit.**
**Localized particles shouldn't create an interference pattern.**
**Electrons show the same wave behavior as photons, which have been shown to be capable of taking multiple paths.**

**In favor of Student Two:**
**All arguments against Student One – particularly the formation of an interference pattern.**
**Bell's Theorem, along with entanglement and tests of local schemes, as well as single-photon experiments, tell us that quantum phenomena must be nonlocal; and that superposition states are real, and not a result of classical ignorance.**

**Against Student Two:**
**Doesn't explain why particles are only detected at a single point.**
**No known evidence that Student Two's perspective on the double-slit experiment is incorrect, but we must accept collapse of the wave function to explain the outcome of measurements..**
**Some might object to the student saying the wave packet "is" the electron. However, we are certainly modeling a single electron with a wave packet, and this model doesn't lead to incorrect predictions.**

**Student Three:**
**Not necessarily saying Students 1 & 2 are wrong, but arguing that quantum mechanics explains how to predict probabilities for measurements, and that Students 1 & 2 are speculating on what is going on between source and detection without really knowing for sure. Physicists may choose to go with Student 3 since problems about how to physically interpret the wave function don't have to get in the way of making the correct calculations. Physicists may choose to disagree with Student 3 since scientists regularly make physical interpretations of mathematical theories, which can lead to greater insight, or eventually new tests of old and new ideas. We also observe evidence of wave behavior, and so thinking of an electron as a wave packet is consistent with this.**

**E1.C (2 Points)** Which student(s) (if any) do you ***personally*** agree with? If you have a different interpretation of what is happening in this experiment, then say what that is. Would it be reasonable or not to agree with ***both*** Student 1 & Student 2? This question is about your personal beliefs, and so there is no "correct" or "incorrect" answer, but you will be graded on making a reasonable effort in explaining why you believe what you do.

**E2. (OPTION ONE – 4 PARTS, 8 POINTS TOTAL)** A double-slit experiment is performed with a low-intensity beam of electrons, so that only one electron at a time is passing through the apparatus; after a period of time a fringe pattern forms on the detection screen.

**E2.A (2 Points)** Discuss what aspects of this experiment are consistent with electrons acting like a particle.

**Particles are always detected one at a time, and always at one point in space. If we were to place a detector at one or both of the slits, we would always observe an electron at one slit or the other, but not both.**

**E2.B (2 Points)** Discuss what aspects of this experiment are consistent with electrons acting like a wave.

**An interference pattern develops over time, even though electrons pass through one at a time, indicating that each electron must be interfering with itself as a wave after passing through the double-slits.**

**E2.C (3 Points)** Suppose that the slits are separated by a distance D = 0.5 μm; the distance (H) between the center of the pattern and the ***second*** bright region is 1.0 mm; and the distance (L) between the screen and the slits is 2.0 m. Use the results of this experiment to determine the wavelength of the electrons. Explain how you arrived at your answer.

$$H = \frac{m\lambda L}{D} \quad \rightarrow \quad \lambda = \frac{DH}{mL} = \frac{\left(0.5\times10^{-6}m\right)\left(1\times10^{-3}m\right)}{(2)\left(2m\right)} = 0.125\times10^{-9}m = 0.125\ nm$$

**E2.D (1 Point)** Suppose the same experiment were conducted with protons having the same kinetic energy as the electrons in the previous experiment. Describe (qualitatively) how, if at all, the appearance of the fringe pattern would change. Explain your reasoning.

**Each proton has the same energy as the electrons in the previous case, but has greater mass, and so greater momentum than the electrons. Greater momentum means a smaller de Broglie wavelength, which corresponds to a smaller distance between bright fringes**

**E3. (OPTION TWO – 3 PARTS, 8 POINTS TOTAL)** For the diagrams below depicting Experiments X & Y, M = Mirror, BS=Beam Splitter, PM = Photomultiplier, N = Counter. In each experiment a single-photon source sends photons to the right through the apparatus one at a time.

# EXPERIMENT X

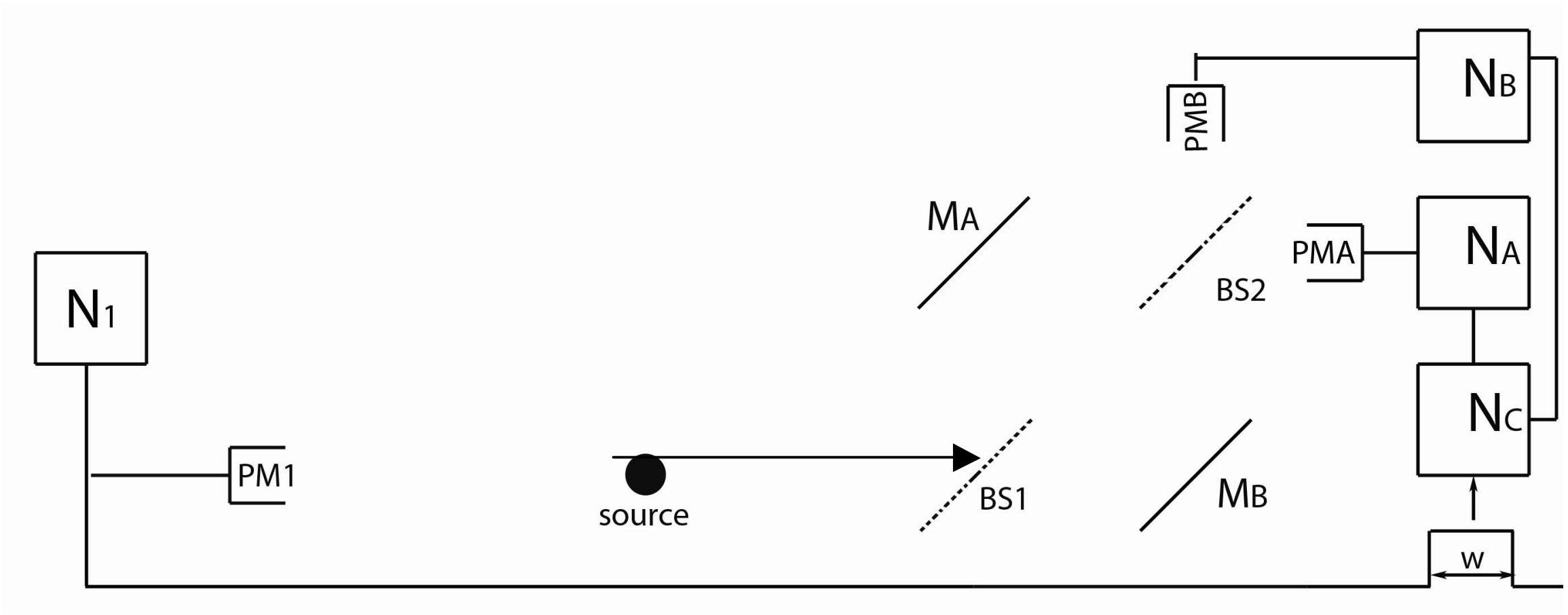

# EXPERIMENT Y

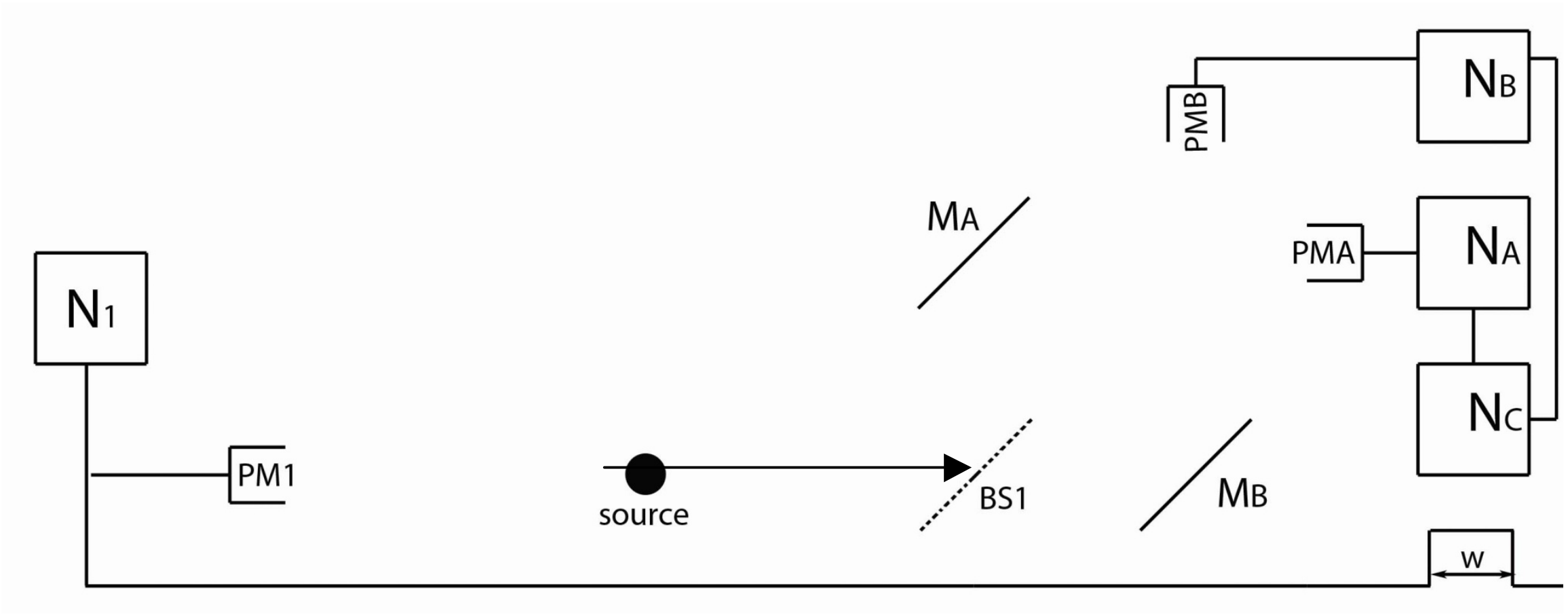

**E3.A (3 Points)** For which experimental setup (X or Y) would you expect photons to exhibit particle-like behavior? Describe in what sense the photon is behaving like a particle during this experiment. What features of the experimental setup allow you to draw this conclusion without actually conducting the experiment?

**In Experiment Y, photons detected in PMA must have travelled by Path A (via Mirror A) and photons detected in PMB must have travelled by Path B (via Mirror B). The photon acts like a particle by taking only one path or the other when it encounters BS1, but also when it is detected in one (but not both) of the photomultipliers. Since only one path is possible, there should be no interference effects.**

**E3.B (3 Points)** For which experimental setup (X or Y) would you expect photons to exhibit wave-like behavior? Describe in what sense the photon is behaving like a wave during this experiment. What features of the experimental setup allow you to draw this conclusion without actually conducting the experiment?

**In Experiment X, no information is known about the path taken by photons detected in either PMA or PMB since both paths are possible. When we alter the length of just one of the paths, we can affect the behavior of particles that were assumed in Experiment Y to take only the other path, so each photon must be aware of both paths, taking both paths at BS1 and interfering with itself. Since two paths are possible, interference should be observed.**

**E3.C (2 Points)** Suppose we are conducting Experiment X (the second beam splitter (BS2) is present) when a photon enters the apparatus and encounters the first beam splitter (BS1). Afterwards, while the photon is still travelling through the apparatus (but before it encounters a detector), we suddenly remove the second beam splitter (switch to Experiment Y). Can we determine the probability for the photon to be detected in PMA? If not, why not? If so, what would be that probability? Explain your reasoning.

**Although we start off conducting an experiment that should demonstrate interference (Experiment X), where the probability for being detected in PMA depends on the path length difference, we switch to an experiment that should show particle-like behavior (Experiment Y) after the photon has encountered BS1, but before it has reached a detector. No interference should be visible and the photon has a 50/50 chance of being detected in PMA – just as though we had been conducting Experiment Y all along.**

# Exam III H-Atom Essay Question

**ESSAY QUESTIONS: PLEASE NOTE – YOU <u>MUST</u> ANSWER THE FIRST ESSAY QUESTION (E1 – 10 POINTS) AND THEN <u>ONE OF THE TWO</u> REMAINING ESSAY QUESTIONS (E2 <u>OR</u> E3 – 10 POINTS). IF YOU CHOOSE TO RESPOND TO BOTH OF THE LAST TWO ESSAY QUESTIONS, WE WILL GIVE YOU THE POINTS FOR THE HIGHER OF THE TWO SCORES.**

**E1. (Required – 3 Parts – 10 Points Total)** A hydrogen atom is in its lowest energy state. Use words, graphs, and diagrams to describe the structure of the Hydrogen atom **in its lowest energy state (ground state).** Include in your description:

- **(4 Points)** At least two ideas important to any accurate description of a hydrogen atom.

- **(3 Points)** An electron energy level diagram of this atom, including numerical values for the first few energy levels, and indicating the level that the electron is in when it is in its ground state.

- **(3 Points)** A diagram illustrating how to accurately think about the distance of the electron from the nucleus for this atom.

(On these diagrams, be quantitative where possible. Label the axes and include any specific information that can help to characterize hydrogen and its electron in this ground state.)

The energy levels are quantized: $E_n = \dfrac{-13.6 \text{ eV}}{n^2}$ n = 1, 2, 3, ...

$E_1 = -13.6 \text{ eV}; \quad E_2 = -3.4 \text{ eV}; \quad E_3 = -1.5 \text{ eV}; \quad \text{etc...}$

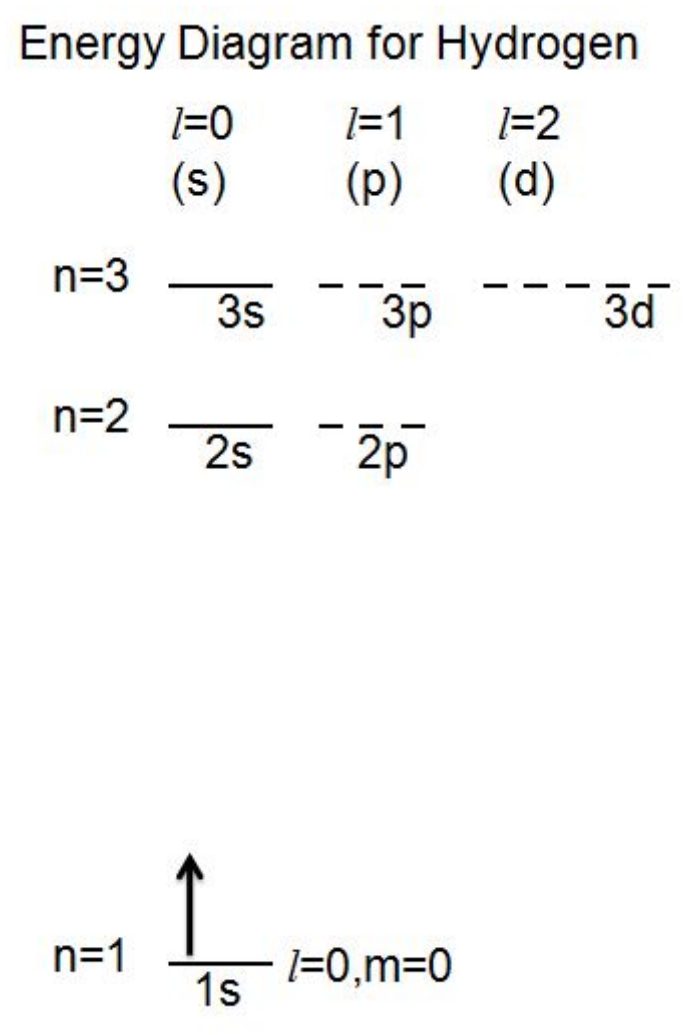

The orbital angular momentum of the electron is quantized: $|\vec{L}| = \sqrt{l(l+1)}\hbar \quad l = 0,1,2,\ldots,n-1$

The z-projection of the orbital angular momentum is quantized: $L_Z = m\hbar \quad m = -l,\ldots,+l$

The energy only depends on n; the quantum numbers *l* & m determine the shape of the wave function. When n=1, then *l*=0 and the probability density is spherically symmetric.

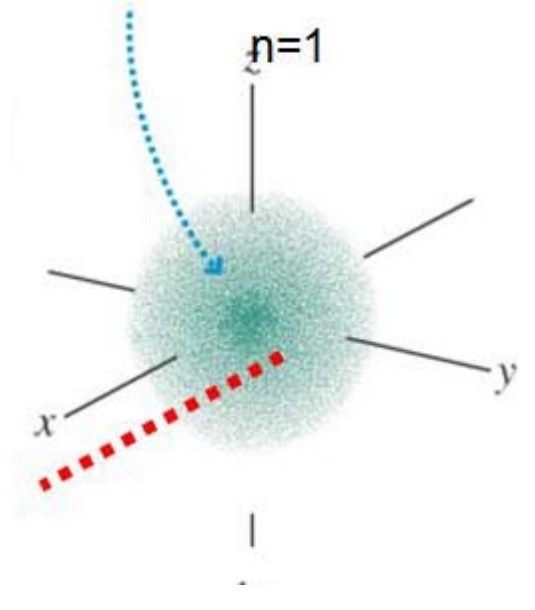

The principle quantum number in the ground state is one (n=1). The orbital angular momentum of the ground state is zero (*l* = 0). The z-projection of the orbital angular momentum in the ground state is zero (m=0).

The electron is described by a standing wave, with a range of possible positions for where the electron might be found (probability cloud). The probability density for the electron in the ground state is spherically symmetric.

The single most likely place to find the electron in the ground state of hydrogen is at the center (r=0); this is where the radial wave function peaks.

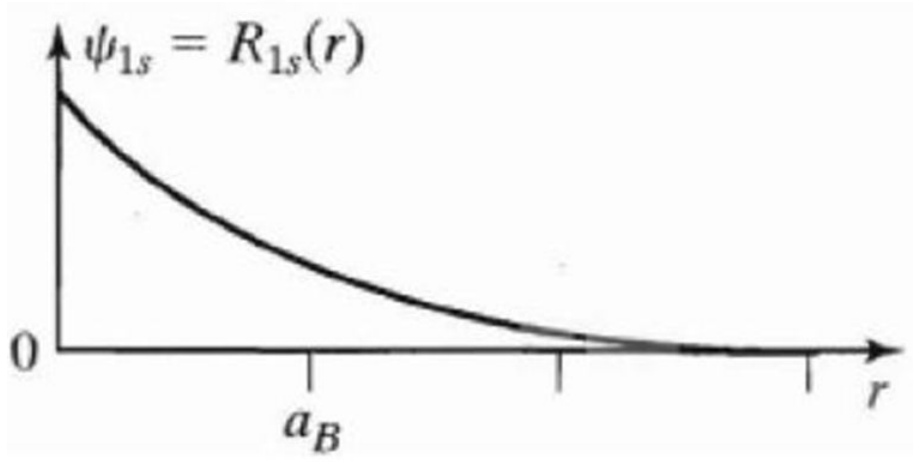

The radial probability distribution for hydrogen in the ground state peaks at the semi-classical Bohr radius ($a_B$) – meaning that the electron is most likely to be found on the surface of an imaginary sphere with $r = a_B$, though any single point on that sphere has less probability than for finding the electron at r=0.

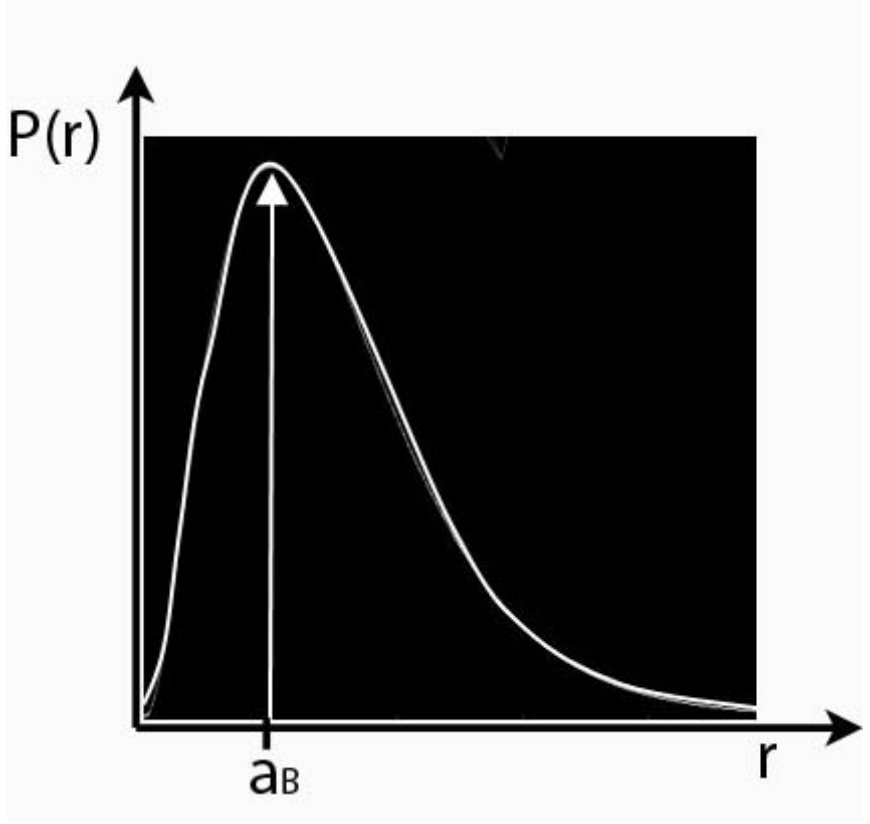

The Bohr and de Broglie models both give the correct energy levels for hydrogen, but are inaccurate in other regards:

The Bohr model and the de Broglie model both describe the electron as being located at a discrete distance from the nucleus, with $r_n = n^2 a_B$ , n = 1, 2, 3… where $a_B$ = Bohr radius.

The Bohr model describes the electron as a localized particle in a circular orbit about the nucleus; the orbital angular momentum is quantized (assumption), but incorrectly describes the ground state of hydrogen as L = ħ.

The de Broglie model describes the electron as a standing wave on a ring, with wavelength given by the de Broglie relation: $\lambda = h/p$. This quantizes the angular momentum, but also incorrectly describes the ground state as L = ħ.

# Quantum Tunneling Tutorial (New)

**PART A: CLASSICAL PARTICLE**
A ball of mass *m* rolls back and forth without any loss of energy between two very high walls (at x=0 & x=2L). There is a ramp centered at x=L that extends upward to a height less than ***h***. The ball has total energy $E_{TOTAL}=mgh$ (kinetic plus potential).

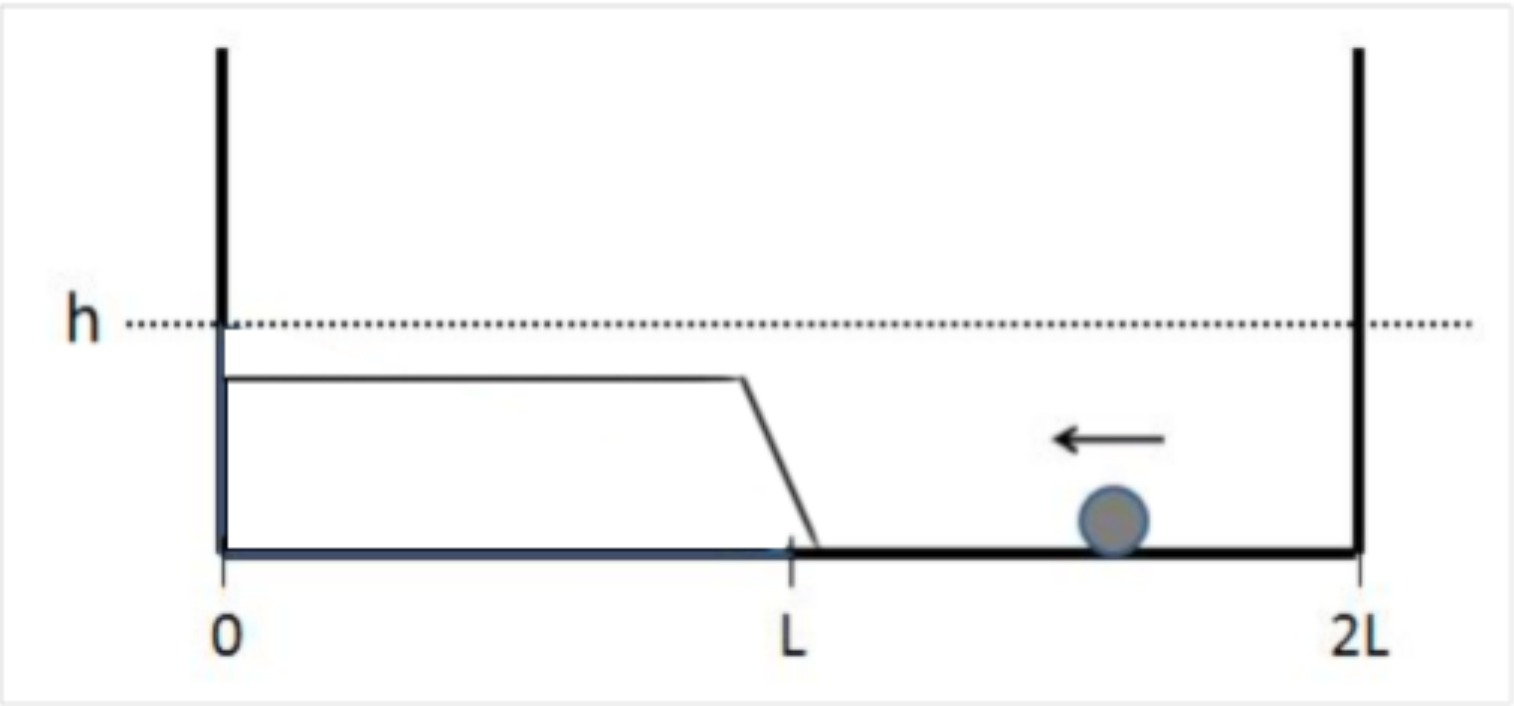

**1)** Is the total energy of the ball as it rolls from x=2L to x=0 increasing, decreasing or staying the same? (Explain your answer)

The total energy of the ball remains constant (E=mgh). The kinetic and potential energies of the ball change, but the sum of the two is constant (conservation of energy).

**2)** Sketch below (all on one graph) the kinetic energy, gravitational potential energy, and total energy of the ball between x=0 and x=2L.

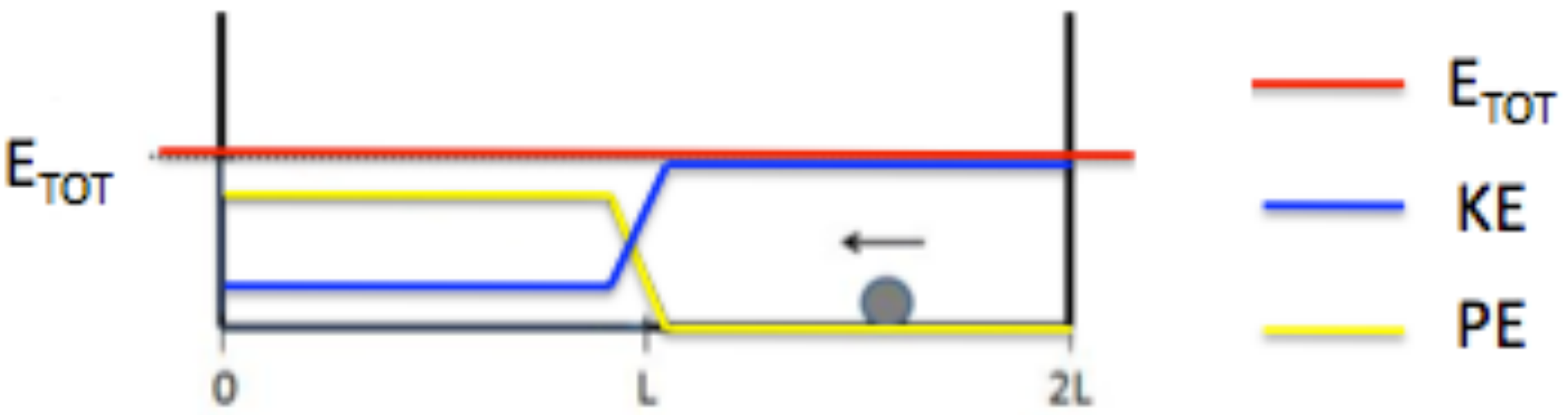

**3)** Is the amount of time the ball spends between x=0 and x=L greater than, less than or equal to the amount of time it spends between x=L and x=2L? (Ignore any time the ball spends on the ramp.) Why?

Because its kinetic energy is least in the region x=0 to x=L, the ball moves at a slower speed there, and therefore spends more time in that region than in the region where it is moving quickly (x=L to x=2L).

**4)** If someone were to take a photograph of the ball at some random time, would the probability of finding the ball in the first half (between x=0 & x=L) be greater than, less than, or equal to the probability of finding it in the second half (between x=L & x=2L)? Why?

The ball spends the most time in the left side, and so the probability for finding it there when looking at some random time is greater than where it spends the least time.

## PART B: SOLUTIONS TO SCHRÖDINGER'S EQUATION

The time-independent Schrödinger equation is given by:

$$\frac{-\hbar^2}{2m}\frac{d^2\psi(x)}{dx^2}+V(x)\psi(x)=E_{TOT}\psi(x)$$

This can be rewritten as:

$$\frac{d^2\psi}{dx^2}=-\frac{2m}{\hbar^2}\left(E-V\right)\psi=\frac{2m}{\hbar^2}\left(V-E\right)\psi$$

**1)** If $E < V$, will the solutions to Schrödinger's equation be real exponentials or complex exponentials? [Hint: Is the quantity on the right-hand side positive or negative in this case?]

If E < V, then the quantity on the right is positive, and the solutions to the equation are real exponentials (exponential growth or decay).

**2)** Write down the most general solution to Schrödinger's equation for the case when $E < V$ [in terms of the quantities given – you may define any new constants, as needed].

$\Psi_{E<V}(x)=A\exp(+\alpha x)+B\exp(-\alpha x)$ where $\alpha=\sqrt{\frac{2m}{\hbar^2}(V-E)}$

**3)** If $E > V$, will the solutions to Schrödinger's equation be real exponentials or complex exponentials? [Again, consider whether the right-hand side is positive or negative.]

If E > V, then the quantity on the right is negative, and the solutions to the equation are complex exponentials (oscillatory solutions).

**4)** Write down the most general solution to Schrödinger's equation for the case when $E > V$ [in terms of the quantities given – you may define any new constants, as needed].

$\Psi_{E>V}(x)=C\exp(+ikx)+D\exp(-ikx)$ where $k=\sqrt{\frac{2m}{\hbar^2}(E-V)}$

**PART C: ELECTRON IN A WIRE ($E > V$)**

Consider an electron with total energy ***E*** moving to the right through a very long smooth copper wire with a ***small*** air gap in the middle:

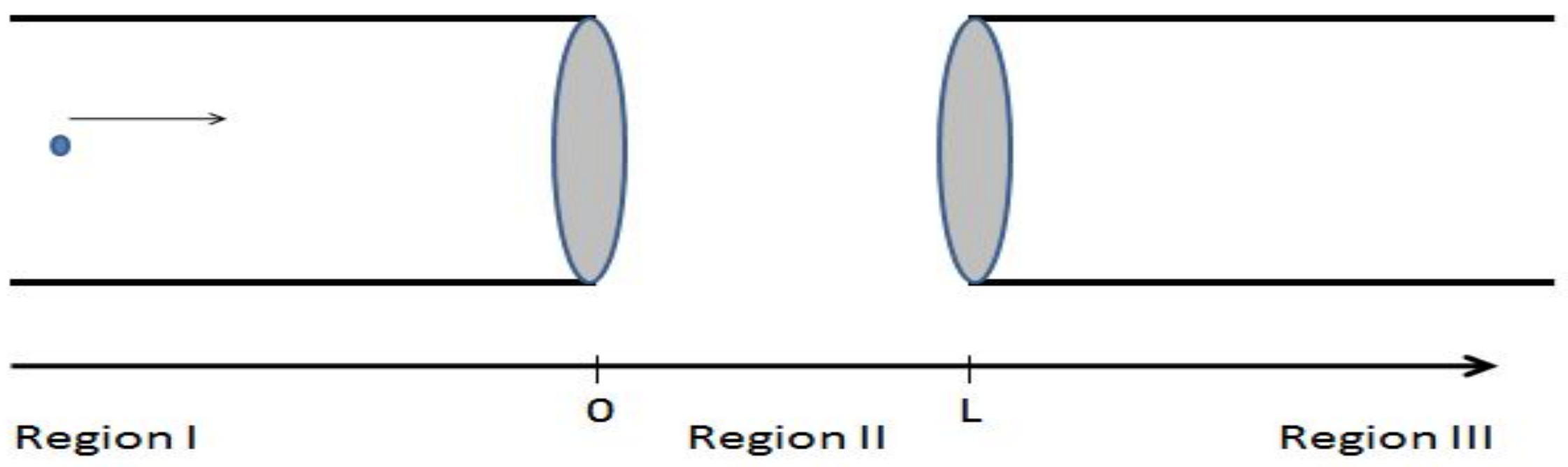

Assume that the work function of the wire is $V_0$ and that $V = 0$ inside the wire.

**1)** If **E > $V_0$**, draw a graph of the electron's potential and kinetic energy in all three regions. Also draw a dashed line indicating the total energy of the electron.

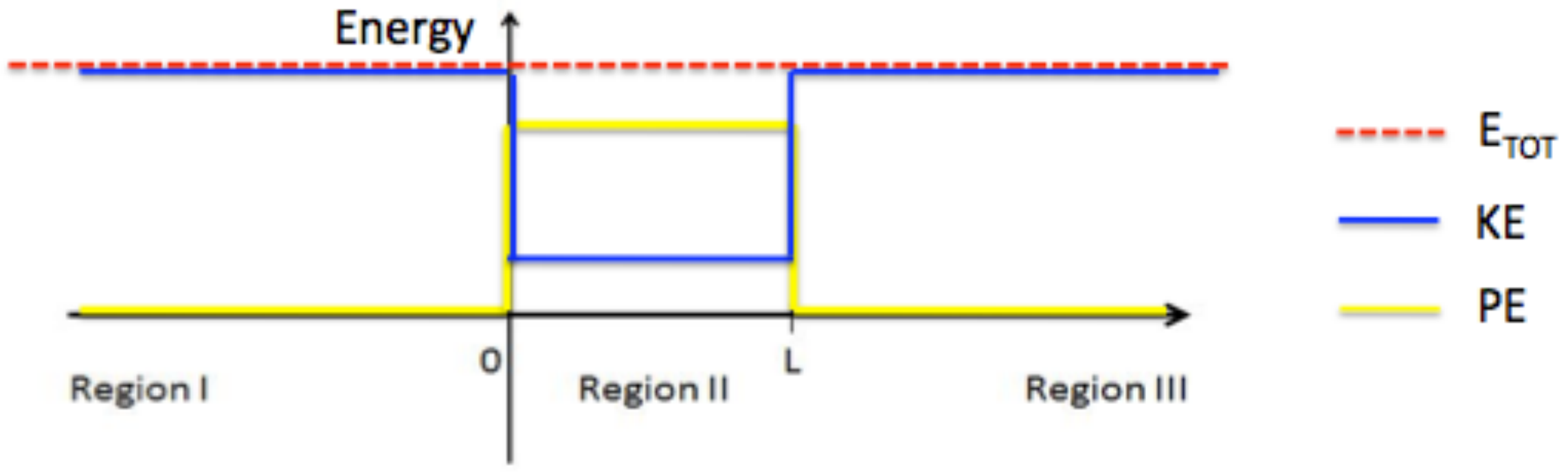

**2)** In each of the three regions, are the solutions to Schrödinger's equation real exponentials or complex exponentials? Write down a solution for each of the three regions corresponding to an electron traveling to the right.

**Region I:** **(E > V)** **Complex exponential:** $\Psi_I(x) = A\exp(+ik_I x)$

**Region II:** **(E > V)** **Complex exponential:** $\Psi_{II}(x) = B\exp(+ik_{II} x)$

**Region III:** **(E > V)** **Complex exponential:** $\Psi_{III}(x) = C\exp(+ik_{III} x)$

**3)** How does the deBroglie wavelength of the electron compare in each of the three regions? Rank the wavelengths in the three regions ($\lambda_1$, $\lambda_2$, $\lambda_3$) from largest to smallest. If the wavelength is not defined in a particular region, then say so.

The wavelengths in regions I & III are the same, since the kinetic energy (= total energy) is the same there. The kinetic energy in region II is smaller, so it has less momentum, and therefore a longer wavelength. $\lambda_2 > \lambda_1 = \lambda_3$.

**4)** How does the amplitude of the electron's wave function compare in each of the three regions? [Hint: think about $|\psi(x)|^2$ what tells you in terms of probabilities – remember your answers to Part A].

Just as with the ball in the gravitational field, if we look at some random time the electron will be most likely to be found in the region where its kinetic energy is least. Greater probability corresponds to greater amplitude, and so the amplitude should be largest in region II. If we consider also that there is some probability for the wave to be reflected at the barrier (even when E > V), then the amplitude on the right side should be smaller than the amplitude on the left side.

**5)** With this information in mind, sketch the ***real part*** of the electron's wave function in all three regions [keep in mind requirements on the continuity of the wave function]:

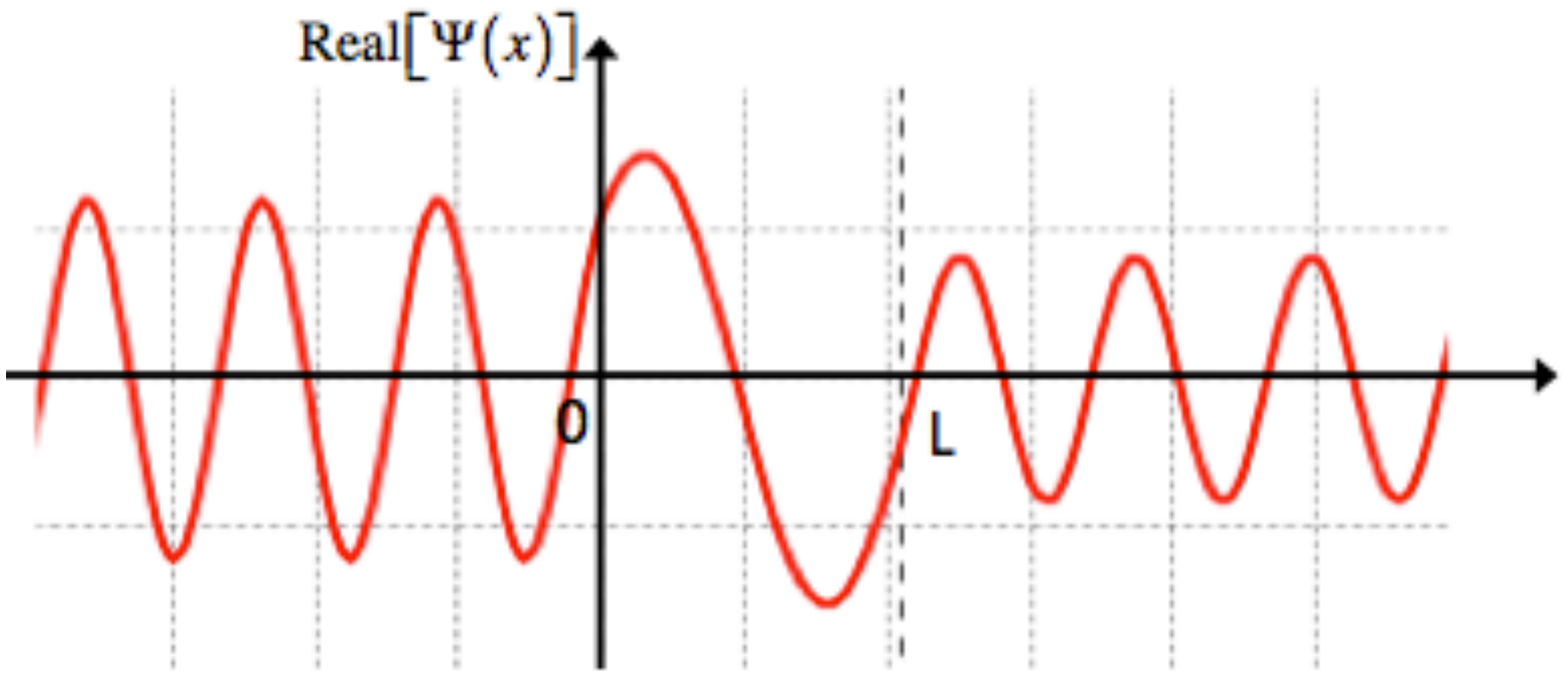

**PART D: ELECTRON IN A WIRE ($E < V$)**

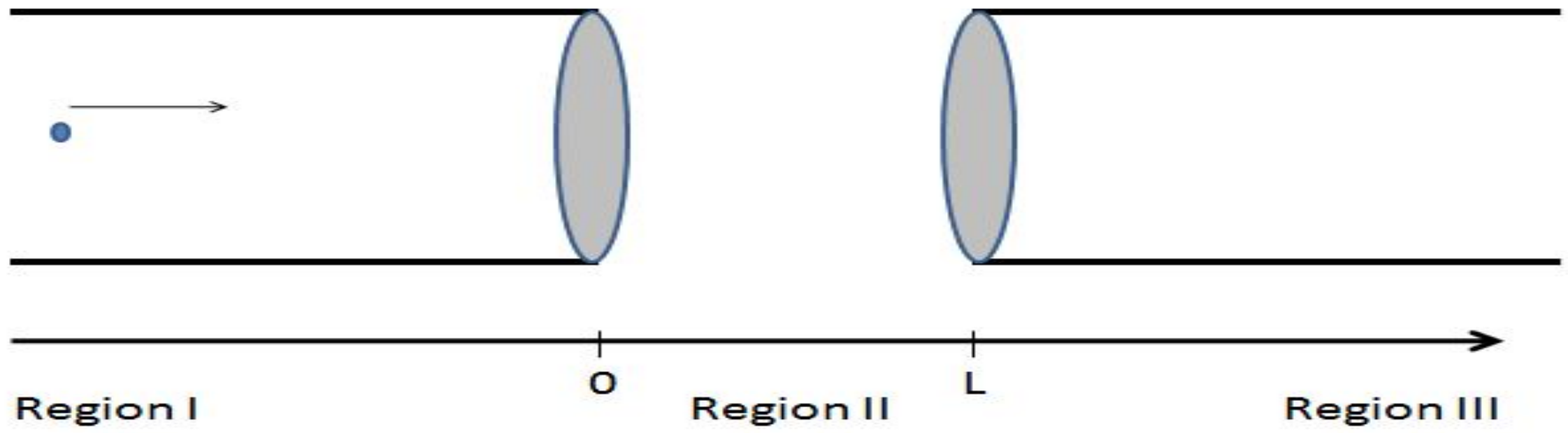

Consider the same situation as in **Part C**, but now the total energy $E$ of the electron is ***less than*** the work function $V_0$.

**1)** If **E < $V_0$**, draw a graph of the electron's potential and kinetic energy in all three regions. Also draw a dashed line indicating the total energy of the electron.

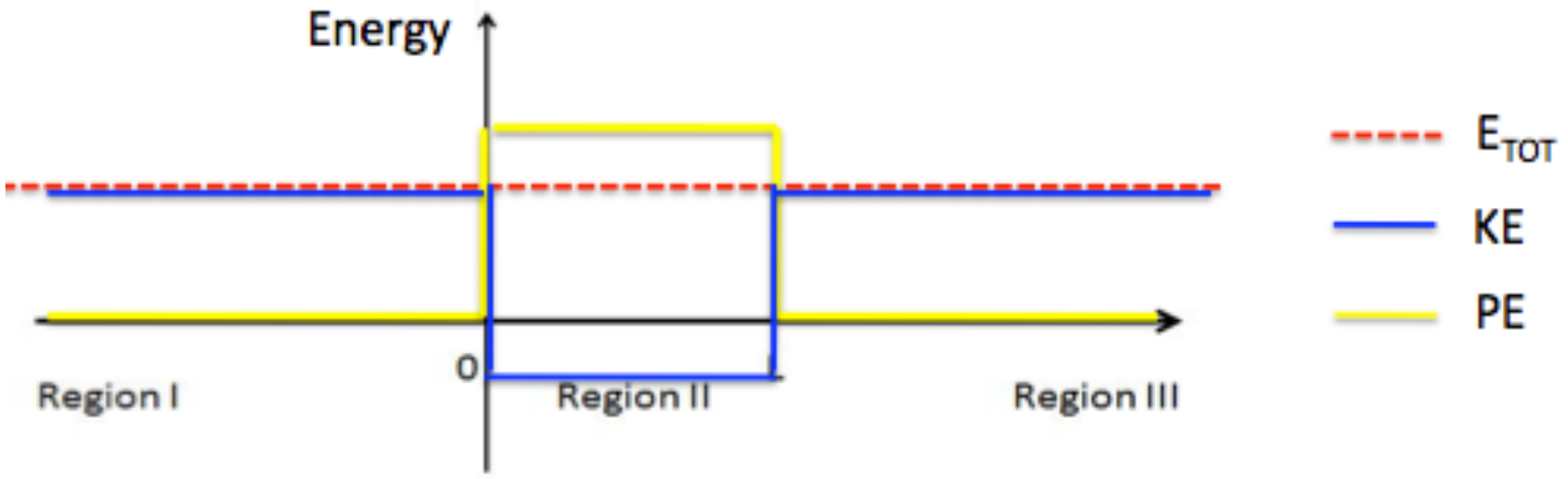

**2)** In each of the three regions, are the solutions to Schrödinger's equation real exponentials or complex exponentials? Write down a solution for each of the three regions corresponding to an electron traveling to the right.

**Region I:** **(E > V)** **Complex exponential:** $\Psi_I(x) = A\exp(+ik_I x)$

**Region II:** **(E < V)** **Decaying real exponential:** $\Psi_{II}(x) = B\exp(-\alpha x)$

**Region III:** **(E > V)** **Complex exponential:** $\Psi_{III}(x) = C\exp(+ik_{III} x)$

**3)** How does the deBroglie wavelength of the electron compare in each of the three regions? Rank the wavelengths in the three regions ($\lambda_1$, $\lambda_2$, $\lambda_3$) from largest to smallest. If the wavelength is not defined in a particular region, then say so.

Again, the wavelengths in regions I & III are the same, since the kinetic energy (= total energy) is the same there. The kinetic energy in region II is negative(!), and the resulting exponentially decaying wave function has no wavelength associated with it. $\lambda_2$ not defined; $\lambda_1 = \lambda_3$.

**4)** How does the amplitude of the electron's wave function compare in each of the three regions? [Hint: think about $|\psi(x)|^2$ what tells you in terms of probabilities]. Explain what physical meaning we can make from the shape of the wave function in Region II.

Here, the electron is much more likely to be reflected than transmitted, and so the amplitude should be greater in Region I than in Region III. The wave function is exponentially decaying in Region II, meaning there is a decreasing probability for transmission the thicker (or higher) the barrier is.

5**)** With this information in mind, sketch the ***real part*** of the wave function for this electron [keep in mind requirements on the continuity of the wave function]:

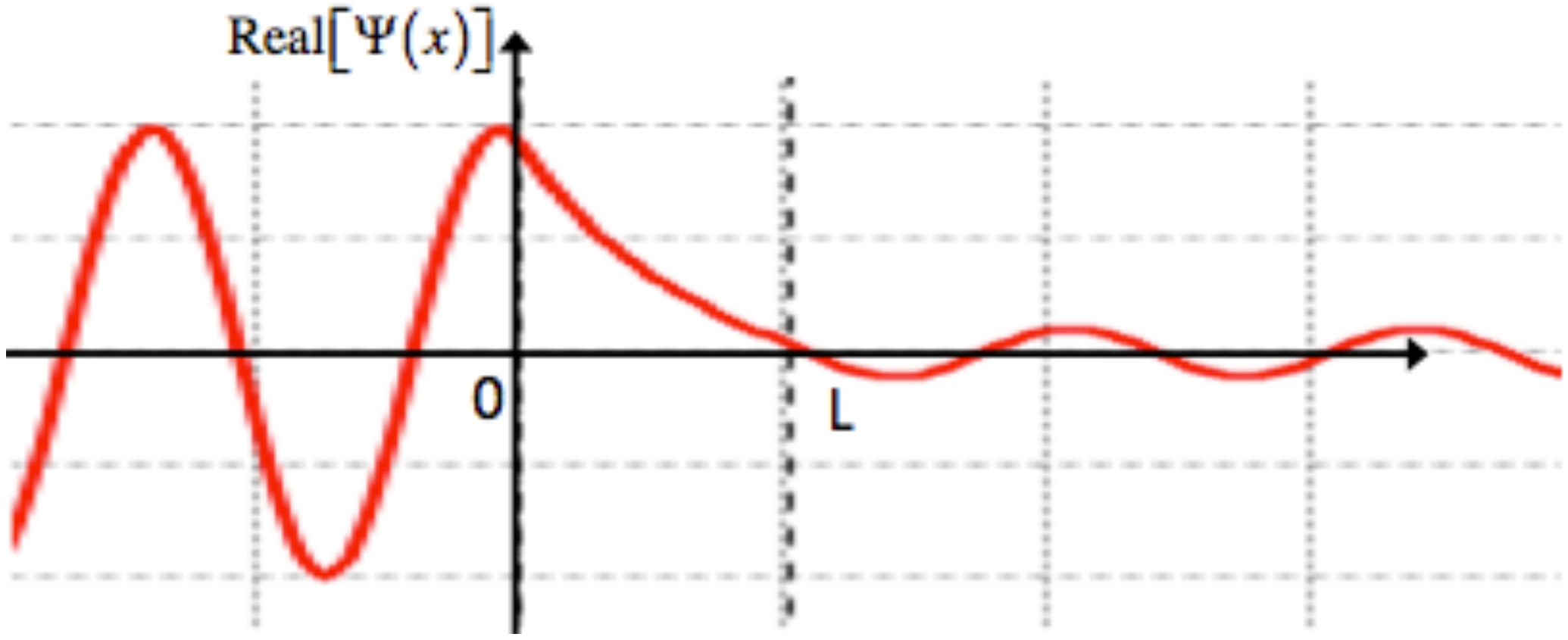

**6)** Using the solution to **#5**, what conclusions can you make about the possible position of the particle? How is this different than a classical particle in the same situation? Can you offer an explanation of why classical objects (people) don't exhibit the same property, called tunneling?

This means there is a (small) chance that the electron can be found on the other side of the potential barrier, since the wave function penetrates into the "classically forbidden" region. A classical (localized) particle should not be able to penetrate into any region where its total energy is less than its potential energy. The kinetic energy of the electron in the barrier region is negative(!), which doesn't make sense for a localized particle, but implies exponentially decaying solutions for a wave. Classical objects don't tunnel because they don't have the properties of waves.

# Tutorial: Quantum Tunneling with PhET Simulation (new)
## Daniel Rehn

**Part A: Classical Particle**

A ball of mass $m$ rolls to the right on a flat, frictionless surface with total energy $E = 3mgh$. The ball soon encounters a sloped surface and rolls up to height $2h$. After, the ball rolls back down the ramp, always staying in contact with the surface.

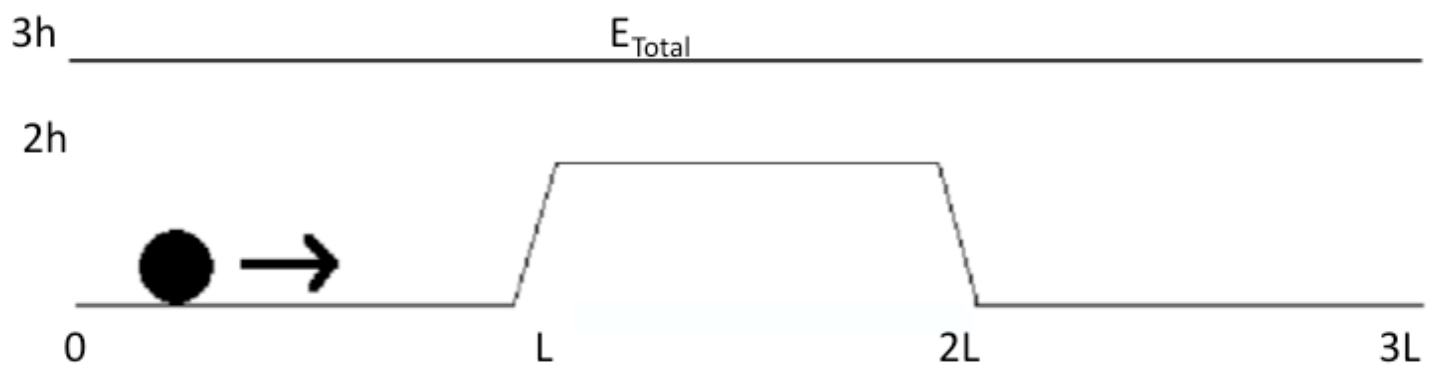

1) Is the total energy of the ball as it rolls from 0 to 3L increasing, decreasing, or staying the same?

2) Sketch the kinetic energy, gravitational potential energy, and total energy of the ball between 0 and 3L.

3) Is the amount of time the ball spends between L and 2L greater than, less than, or equal to the amount of time it spends between 0 and L? How does it compare to the amount of time it spends between 2L and 3L? (Ignore the time the ball spends on the ramp.)

4) Now imagine that we take a photograph of the ball at some random time BEFORE it reaches 2L. Is the probability of finding the ball between 0 and L greater than, less than or equal to the probability of finding it between L and 2L? Why?

**PART B: Quantum Particle with E > V (Using PhET sim)**

1) Observe the plot of the wave function for plane wave solutions for the case where E > V. Why is the wave function oscillating up and down?

2) Now widen the width of the wire gap (where V > 0) to 3.5 dashed-lines wide. How does the wavelength of the wave function in this region compare to the wavelength in the region to the left? How about to the region on the right? Lastly, how do the wavelengths in the regions on the left and right compare to each other?

3) What does your answer to (2) tell you about the kinetic energy of the particle in each of these three regions? Be sure to discuss this with your group members.

4) Now refer back to the classical particle case. How does your answer to (3) compare to your answer to PART A question 2? How is wavelength related to kinetic energy?

5) Now let's look at the amplitude of the wave function. What does the amplitude of $\psi$ (or $|\psi|^2$ ) tell you?

6) If we were to make a measurement of position of the wave function, would we be more likely to find it in the region to the left of the air gap or in the air gap (for now ignore the region to the right of the air gap, very much like you did in PART A)?

7) In this case of E > V, explain how measurements of position for a quantum particle compare to taking a photograph of a classical particle.

**PART C: Classical Particle with E < V**

Now imagine that the same ball from PART A has an initial total energy of E = 2mgh, while the height of the hill remains at 3h.

1) What happens to the ball as it starts to go up the hill? Is it possible for the ball to be found between L and 2L? How about between 2L and 3L?

**PART D: Quantum Particle with E < V**

1) Now, using the PhET sim, decrease the size of the wire gap to 1 dashed-line wide and increase the height of the potential energy line all the way to the top. What type of function do you see in region 1 and 2?

2) What type of function is shown inside the wire gap? Hint: It might be more obvious if you look at the wave function when the air gap is very wide… but return to 1 dashed-line wide for the next question!

3) How do the wavelengths of the wave function on the left and right of the air gap compare to each other? What does that tell you about the kinetic energy of the particle in each of those regions?

4) Now refer back to PART C with the classical particle. How does the kinetic energy of the classical particle in regions 2 and 3 compare to the kinetic energy of the quantum particle in regions 2 and 3?

5) What does the amplitude of $\psi$ (or $|\psi|^2$) tell you about finding a particle in regions 2 or 3?

6) If we were to make a measurement of the particle's position with E < V, which region would we be most likely to find it in? Compare this to the case of the classical particle.

# Tutorial: Quantum Tunneling (old)
## Sam McKagan, et al.

In this tutorial you will explore the physics of an electron traveling through an air gap in a wire – first in the case in which the electron has enough energy to get through the gap classically, and then in the case in which it does not. If you're paying attention, you should be surprised by some of the results.

Consider an electron initially moving to the right through a very long smooth copper wire with a small air gap in the middle. (See figure below.) The work function of copper is $V_0$.

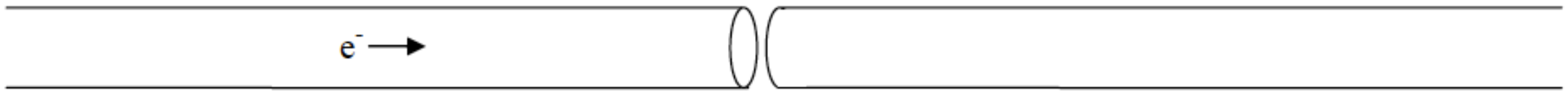

**PART I: E > $V_0$:** Suppose the electron shown above has an initial energy $E > V_0$.

**1.** In the space below, sketch a graph of the potential energy V of the electron as a function of horizontal position x. Define V = 0 <u>inside the wire</u>. Once you have V(x) sketched, use a dashed line to show the energy of an electron that satisfies the $E > V_0$ condition.

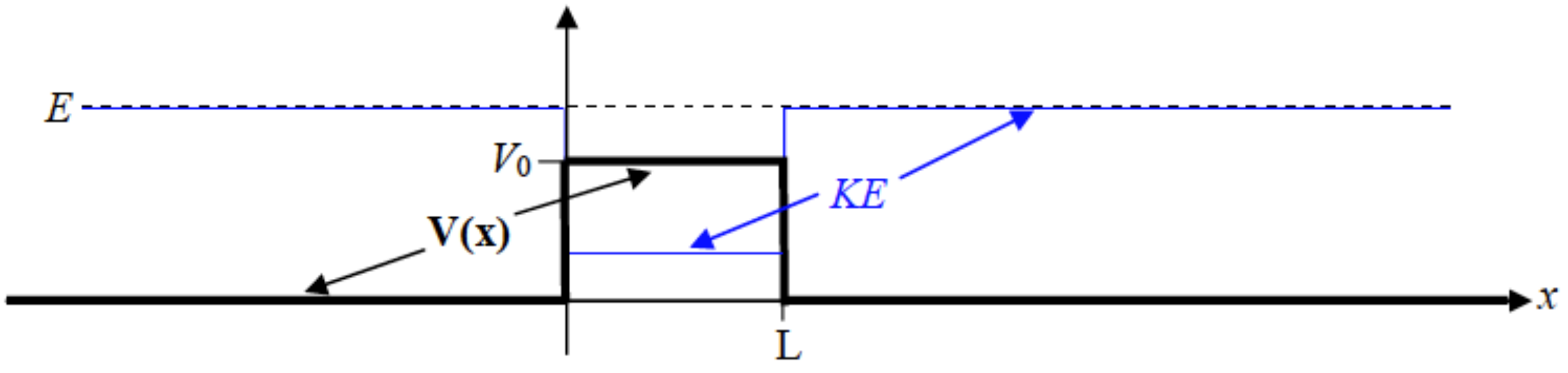

**2.** For the region in the copper wire to the left of the air gap, write down the general solution for Ψ(x,t). Plug it into the Schrodinger Equation to make sure it works and solve for the total energy E of the electron.

The general solution is: $\Psi(x,t) = \left(Ae^{ikx} + Be^{-ikx}\right)e^{-iEt/\hbar}$

Plugging into the Schrodinger equation: $-\frac{\hbar^2}{2m}\left(-k^2\right)\Psi(x,t) + 0 = i\hbar\left(-iE/\hbar\right)\Psi(x,t)$

This simplifies to: $\frac{\hbar^2 k^2}{2m} = E$

For the region in the air gap, write down the general solution for Ψ(x,t). Is the value of k here the same as the value of k in the previous region? Why or why not? If not, call it k' to distinguish it from the k above. Plug your solution into the Schrodinger Equation to make sure it works and solve for the total energy E of the electron.

The general solution is: $\Psi(x,t) = \left(Ae^{ik'x} + Be^{-ik'x}\right)e^{-iEt/\hbar}$

Plugging into the Schrodinger equation: $\left(-\frac{\hbar^2}{2m}\left(-k'^2\right) + V_0\right)\Psi(x,t) = i\hbar\left(-iE/\hbar\right)\Psi(x,t)$

This simplifies to: $\frac{\hbar^2 k'^2}{2m} + V_0 = E$

The k here is not the same as the k above because it is related to the total energy E in a different way than the k above. Another way to say this is that the kinetic energy in the gap is different than the kinetic energy in the wire.

For the region in the copper wire to the right of the air gap, write down the general solution for Ψ(x,t). Is the value of k here the same as either of the values of k above? Why or why not? If not, call it k" to distinguish it from the k's above. Plug your solution into the Schrodinger Equation to make sure it works and solve for the total energy E of the electron.

Because the potential here is the same as in the left wire, the solution is the same, the k is the same, and the energy is the same: $\frac{\hbar^2 k^2}{2m} = E$

**3.** In the plot below, sketch the shape of the real part of the wave function at t = 0 in each region for the case where E > $V_0$. The air gap starts at x = 0 and ends at x = L. Don't worry about the relative magnitudes of the waves in the different regions, but think carefully about the general shape of the graph in each region.

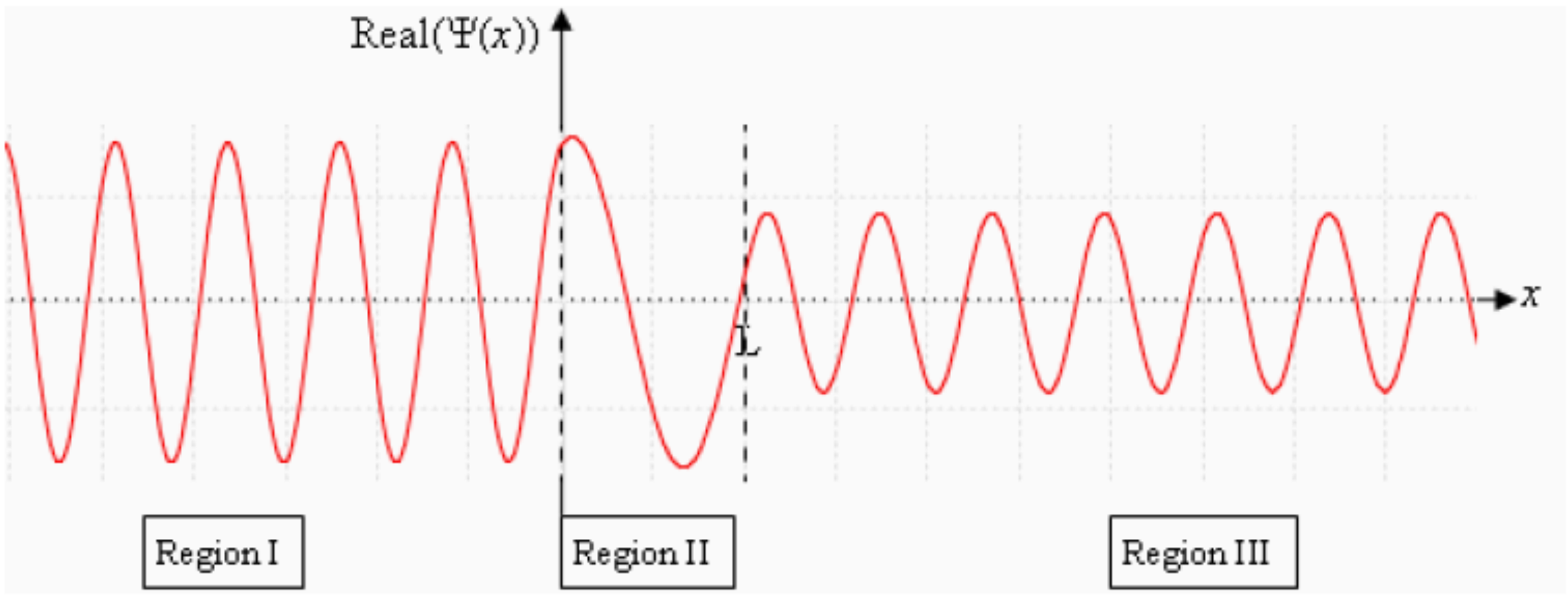

**4.** What is the basic shape of the real part of the wave function in each of the three regions? For example, is it linear, constant, quadratic, exponential, sinusoidal, or something else?

It's sinusoidal in all three regions.

**5.** Is the total energy of the electron to the right of the air gap greater than, less than, or equal to the energy of the electron to the left of the air gap? Explain how you arrived at your answer.

Equal to. Energy must be conserved, so the total energy can't change.

**6.** Fill in the values for the potential, kinetic, and total energy of the particle in each of the three regions in the table below. Your answers should be in terms of E and $V_0$.

| | Left Wire | Air Gap | Right Wire |
|---|---|---|---|
| Potential Energy | 0 | $V_0$ | 0 |
| Kinetic Energy | E | $E-V_0$ | E |
| Total Energy | E | E | E |

**7.** On top of the graph you drew in question 1, now sketch the kinetic energy KE of the electron as a function of position. Be sure to label each of the energies clearly.

**8.** Write an equation that relates the kinetic energy of a particle to its deBroglie wavelength.

$$KE = \frac{p^2}{2m} = \frac{h^2}{2m\lambda^2}$$

**9.** Are the wave functions you sketched in question 3 consistent with your equation in question 8 and your kinetic energies in question 6? Resolve any discrepancies.

My equation in question 8 tells me that the as the kinetic energy increases, the wavelength decreases, and vice versa. In question 6 I said that the kinetic energy in the right wire is equal to the kinetic energy in the left wire, and this is consistent with my drawing, which shows the same wavelength in these two regions. I also said that kinetic energy is smaller in the gap, and this is also consistent with my drawing, which shows a larger wavelength in this region.

**PART II: E < $V_0$:** Now suppose the electron has an initial kinetic energy E < $V_0$.

**10.** In the space below, sketch a graph of the potential energy V of the electron as a function of horizontal position x. Define where V = 0 inside the wire. Once you have V(x) sketched, use a dashed line to show the energy of an electron that satisfies the E < $V_0$ condition.

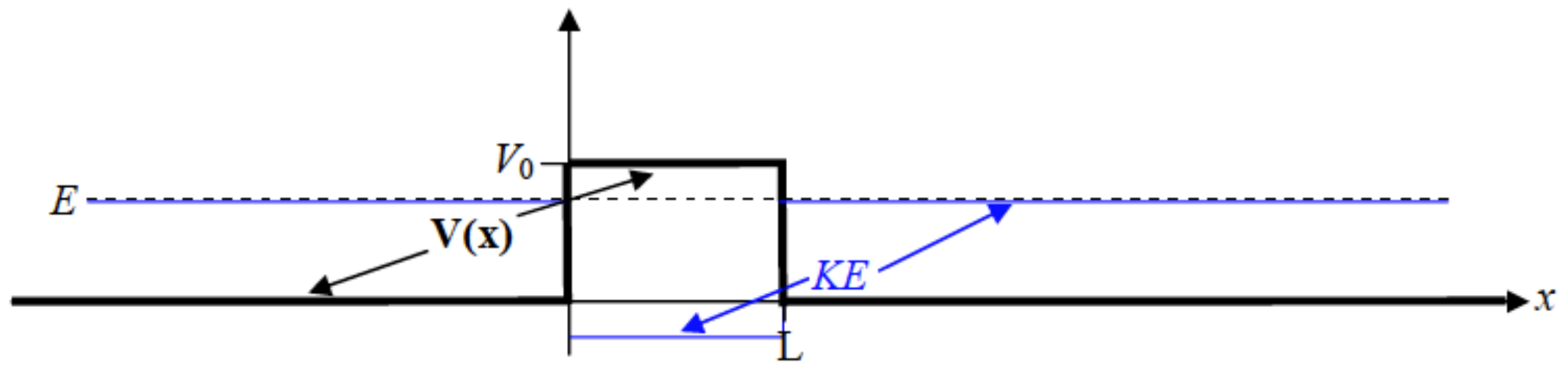

**11.** For the region in the copper wire to the left of the air gap, write down the general solution for Ψ(x,t). Plug it into the Schrodinger Equation to make sure it works and solve for the total energy E of the electron.

The general solution is: $\Psi(x,t)=\left(Ae^{ikx}+Be^{-ikx}\right)e^{-iEt/\hbar}$

Plugging into the Schrodinger equation: $-\frac{\hbar^2}{2m}\left(-k^2\right)\Psi(x,t)+0=i\hbar\left(-iE/\hbar\right)\Psi(x,t)$

This simplifies to: $\frac{\hbar^2k^2}{2m}=E$

For the region in the air gap, write down the general solution for .(x,t). Plug your solution into the Schrodinger Equation to make sure it works and solve for the total energy E of the electron.

The general solution is: $\Psi(x,t)=\left(Ae^{\alpha x}+Be^{-\alpha x}\right)e^{-iEt/\hbar}$

Plugging into the Schrodinger equation: $\left(-\frac{\hbar^2}{2m}\left(\alpha^2\right)+V_0\right)\Psi(x,t)=i\hbar\left(-iE/\hbar\right)\Psi(x,t)$

This simplifies to: $-\frac{\hbar^2\alpha^2}{2m}+V_0=E$

For the region in the copper wire to the right of the air gap, write down the general solution for Ψ(x,t). Plug your solution into the Schrodinger Equation to make sure it works and solve for the total energy E of the electron.

The general solution is: $\Psi(x,t)=\left(Ae^{ikx}+Be^{-ikx}\right)e^{-iEt/\hbar}$

Plugging into the Schrodinger equation: $-\frac{\hbar^2}{2m}\left(-k^2\right)\Psi(x,t)+0=i\hbar\left(-iE/\hbar\right)\Psi(x,t)$

This simplifies to: $\frac{\hbar^2k^2}{2m}=E$

**12.** In the plot below, sketch the shape of the real part of the wave function at t = 0 in each region for the case where $E < V_0$. The air gap starts at x = 0 and ends at x = L. Don't worry about the relative magnitudes of the waves in the different regions, but think carefully about the general shape of the graph in each region.

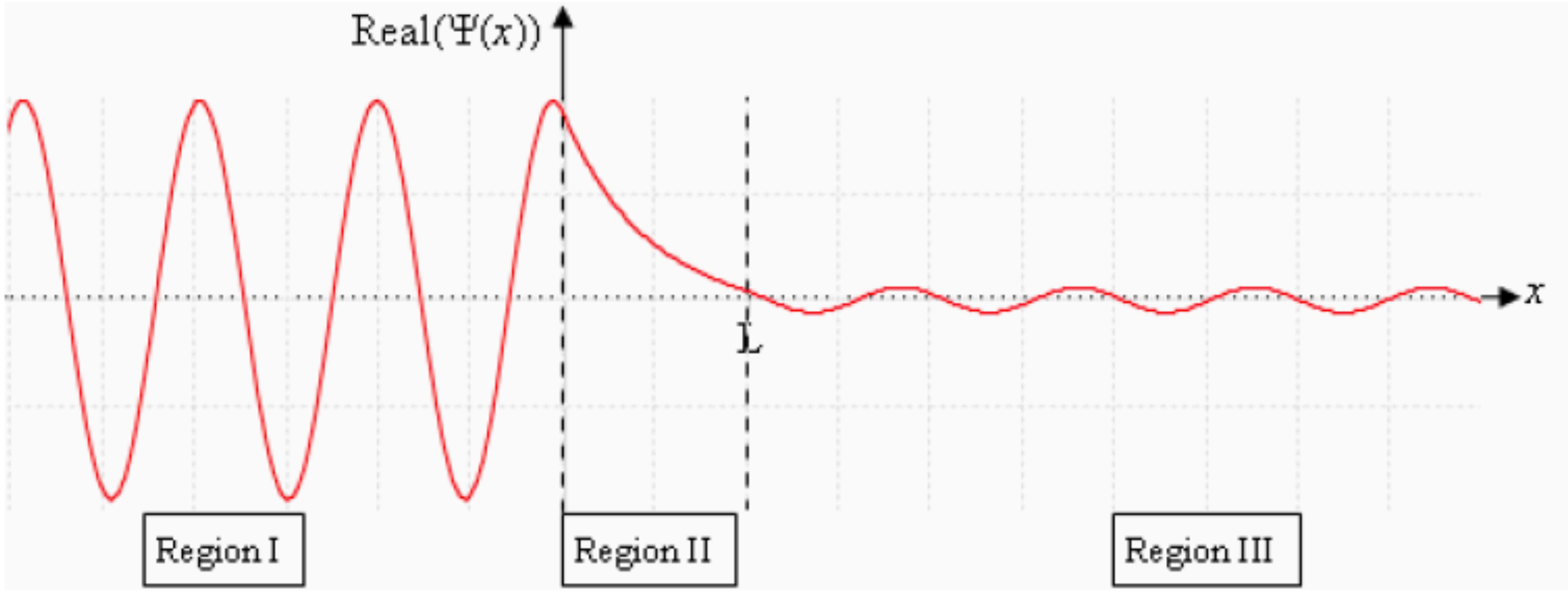

**13.** What is the basic shape of the real part of the wave function in each of the three regions? For example, is it linear, constant, quadratic, exponential, sinusoidal, or something else?

It's sinusoidal in regions I and III and it's a decaying exponential in region II.

**14.** It is often stated that a particle can quantum mechanically tunnel through a barrier. Explain what is meant by this.

In this example, although classically the electron does NOT have enough energy to get out of the metal and into the air gap, there is a solution to the Schrodinger equation in this region. This means there is a non-zero probability to find the electron in the air gap. The probability decays exponentially as you go farther into the gap, but if the gap is thin enough, the electron can "tunnel" through the gap into the copper wire on the right, wire it has enough energy to stay.

**15.** Consider an electron that has tunneled through the barrier. Is the energy of the electron to the right of the air gap greater than, less than, or equal to the energy of the electron to the left of the air gap? Explain how you arrived at your answer.

Equal to. Energy must be conserved, so the total energy can't change.

**16.** Fill in the values for the potential, kinetic, and total energy of the particle in each of the three regions in the table below. Your answers should be in terms of E and $V_0$.

| | Left Wire | Air Gap | Right Wire |
|---|---|---|---|
| Potential Energy | 0 | $V_0$ | 0 |
| Kinetic Energy | E | $E-V_0$ | E |
| Total Energy | E | E | E |

**17.** On top of the graph you drew in question 10, now sketch the kinetic energy KE of the electron as a function of position. Be sure to label each of the energies clearly.

**18.** Do you notice anything unusual about the kinetic energy?

The kinetic energy is negative in the air gap! This is because the particle doesn't have enough total energy to be in this region, but it is "borrowing" some kinetic energy to compensate for having a potential energy greater than its total energy. There is no classical analogue to the situation, and if you try to think of the electron as a classical particle moving around in this region, it won't work. But we can describe the behavior of the wave function in this region without any problem and it accurately predicts the results of experiments.

## Final Paper

In lieu of answering the essay questions on the final, students will instead write a short essay on a topic related to quantum mechanics, or on your experience of learning about quantum mechanics. Papers should be **at least** two (2) pages in length (minimum, single-spaced, and definitely **not more** than five (5) pages). Papers must be submitted electronically (in Word or PDF format) and will be due on the last day of class (Friday, 12/10/2010). Students must have their topics approved before the beginning of Fall Break (11/19/2010). We will offer suggestions for improvement to any student who turns in a draft of their paper at least one week before the due date. This is meant to be an exploration of a topic beyond our discussions in class, or in the readings, so students must cite **at least one source** that is not among the assigned readings from this course (NOTE: Wikipedia and the like do NOT count as legitimate sources). As a starting-off point for some of you, several additional articles on quantum information, cryptography & computing are available in the Readings folder on CU Learn. We are open to students expressing their knowledge and understanding creatively – youtube videos, podcasts, computer animations, interpretive dance and the like would be acceptable as a final project, as long as the instructors approve of your proposal ahead of time. We will make time on the last day of classes for students to make presentations to the class if they would like to do so.
Possible topics include:

- Heisenberg's Uncertainty Principle
- Wave-Particle Duality
- The Copenhagen Interpretation
- The Many-Worlds Interpretation
- Decoherence
- Schrödinger's Cat
- Objective Reality
- Indeterminacy in Quantum Mechanics
- Quantum Cryptography
- Quantum Computing
- Quantum Teleportation
- String Theory
- Quantum Gravity
- Bosons and Fermions
- Classical Physics vs. Quantum Physics
- Quantum physics in popular culture
- Any important experiment in the history of quantum physics
- **Or, you may choose to write a personal reflection on your experiences in this course by answering the following questions in as much detail as possible (plus anything else you would like to add):**

Describe your experience of learning about quantum mechanics in this course. What motivated you to take this course, what sort of questions did you have coming in, and were these questions or motivations addressed during the course? Has this course changed your ideas about physics and the practice of science in any way? If so, were there any particular ideas or discussions that led to this change in your perceptions? What topic(s) from this course were most interesting to you (and why)? Are there topics that you wish we had covered in this course, or ones that you wish we hadn't? What teaching techniques (lecture, peer-instruction, readings, concept tests, simulations, etc…) were helpful for you in learning about quantum mechanics, and in what way?

# Modern Physics Conceptual Survey

[Correct answers are highlighted in **BOLD**]

**Questions 1 through 4** refer to the following two experiments:

In one experiment electrons are traveling from a source to a detecting screen.
In a second experiment light is traveling from a source to a photographic plate.

For each question, choose from the options A through D below the most appropriate answer according to quantum physics.

A. It is behaving like a particle.
B. It is behaving like a wave.
C. It is behaving like both a particle and a wave.
D. You cannot tell if it is behaving like a particle or a wave.

How is the particle/wave behaving when…

**1. …**an electron is traveling from the source to the detecting screen?

POST

| A | **B** | C | D | %Correct | %Incorrect |
|---|---|---|---|---|---|
| 9.1% | **53.4%** | 21.6% | 15.9% | 53.4 | 46.6% |

PRE

| A | **B** | C | D | %Correct | %Incorrect |
|---|---|---|---|---|---|
| 61.3% | **17.1%** | 18.9% | 2.7% | 17.1% | 82.9% |

**2. …**the light is traveling from the source to the photographic plate?

POST

| A | **B** | C | D | %Correct | %Incorrect |
|---|---|---|---|---|---|
| 2.3% | **63.6%** | 21.6% | 12.5% | 63.6 | 36.4% |

PRE

| A | **B** | C | D | %Correct | %Incorrect |
|---|---|---|---|---|---|
| 13.5% | **52.3%** | 31.5% | 2.7% | 52.3% | 47.7% |

**3. …**an electron interacts with the detecting screen?

POST

| **A** | B | C | D | %Correct | %Incorrect |
|---|---|---|---|---|---|
| **93.2%** | 5.7% | 1.1% | 0 | 93.2% | 6.8% |

PRE

| **A** | B | C | D | %Correct | %Incorrect |
|---|---|---|---|---|---|
| **64.0%** | 9.0% | 21.6% | 5.4% | 64.0% | 36.0% |

**4. ...**the light interacts with the photographic plate?

POST

| **A** | B | C | D | %Correct | %Incorrect |
|---|---|---|---|---|---|
| **64.8%** | 30.7% | 4.5% | 0 | 64.8% | 35.2% |

PRE

| **A** | B | C | D | %Correct | %Incorrect |
|---|---|---|---|---|---|
| **31.5%** | 24.3% | 29.7% | 14.4% | 31.5% | 68.5% |

**Questions 5 through 9** refer to the following three experiments:

In one experiment electrons pass through a double-slit as they travel from a source to a detecting screen.

In a second experiment light passes through a double-slit as it travels from a source to a photographic plate.

In a third experiment marbles pass through two slit-like openings as they travel from a source to an array of collecting bins, side-by-side. The right-hand figure diagrams the experimental setup, and the figures below show roughly the possible patterns which could be detected on the various screens.

**Top view of experimental set-up (not to scale)**

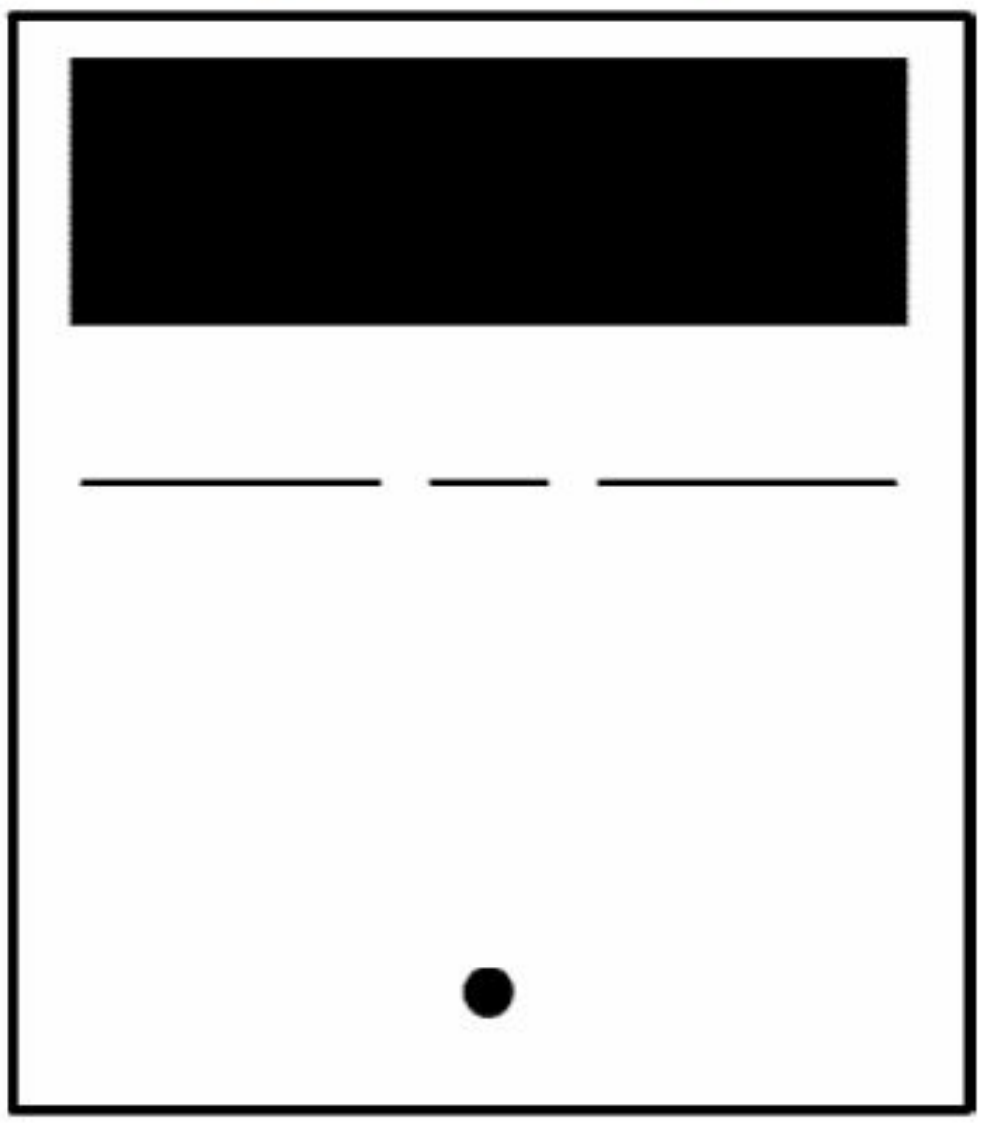

**Possible patterns (not to scale)**

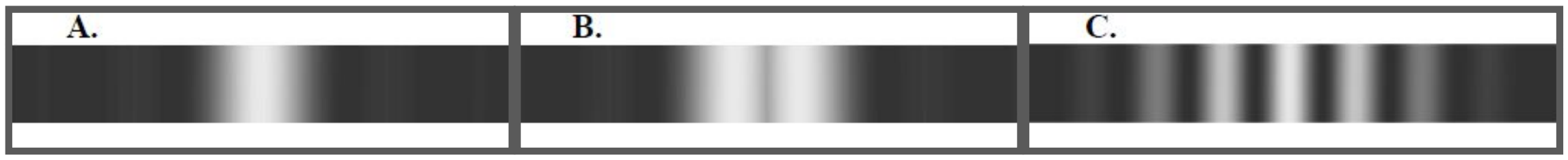

A through C represent some patterns which might be observed. If you think none is appropriate, answer D.

Which pattern would you expect to observe when…

**5. ...***light* passes through the double slit?

POST

| A | B | **C** | D | %Correct | %Incorrect |
|---|---|---|---|---|---|
| 0 | 1.1% | **98.9%** | 0 | 98.9% | 1.1% |

PRE

| A | B | **C** | D | %Correct | %Incorrect |
|---|---|---|---|---|---|
| 10.8% | 45.1% | **43.2%** | 0.9% | 43.2% | 56.8% |

**6. ...***marbles* pass through the double opening?

POST

| A | **B** | C | D | %Correct | %Incorrect |
|---|---|---|---|---|---|
| 0.9% | **86.4%** | 2.3% | 2.3% | 86.4% | 13.6% |

PRE

| A | **B** | C | D | %Correct | %Incorrect |
|---|---|---|---|---|---|
| 14.4% | **59.5%** | 20.7% | 5.4% | 59.5% | 40.5% |

**7. ...***electrons* pass through the double slit?

POST

| A | B | **C** | D | %Correct | %Incorrect |
|---|---|---|---|---|---|
| 0 | 12.5% | **87.5%** | 0 | 87.5% | 12.5% |

PRE

| A | B | **C** | D | %Correct | %Incorrect |
|---|---|---|---|---|---|
| 14.4% | 51.0% | **34.6%** | 0.9% | 34.6% | 65.4% |

**8. ...***light* passes through the apparatus when *one of the slits is covered*?

POST

| **A** | B | C | D | %Correct | %Incorrect |
|---|---|---|---|---|---|
| **92.0%** | 4.5% | 3.4% | 0 | 92.0% | 8.0% |

PRE

| **A** | B | C | D | %Correct | %Incorrect |
|---|---|---|---|---|---|
| **64.0%** | 12.6% | 20.7% | 2.7% | 64.0% | 36.0% |

**9. ...***electrons* pass through the apparatus when *one of the slits is covered*?

POST

| **A** | B | C | D | %Correct | %Incorrect |
|---|---|---|---|---|---|
| **97.7%** | 2.3% | 0 | 0 | 97.7% | 2.3% |

PRE

| **A** | B | C | D | %Correct | %Incorrect |
|---|---|---|---|---|---|
| **70.3%** | 8.1% | 18.0% | 3.6% | 70.3% | 29.7% |

**10.** According to the uncertainty principle, the more we know about an electron's position, the less we know about its...

A. …speed.
B. …momentum.
C. …kinetic energy.
D. All of these.

POST

| A | B | C | **D** | %Correct | %Incorrect |
|---|---|---|---|---|---|
| 2.3% | 45.5% | 1.1% | **51.1%** | 51.1% | 48.9% |

PRE

| A | B | C | **D** | %Correct | %Incorrect |
|---|---|---|---|---|---|
| 15.3% | 7.2% | 5.4% | **72.1%** | 72.1% | 27.9% |

**11.** Choose the answer A through D that is the most appropriate answer according to quantum physics. The Heisenberg Uncertainty Principle is mostly applied to very small objects such as electrons and protons. Why don't we use the uncertainty principle with larger objects such as cars and tennis balls?

A. The errors of measurement can always, in principle, be made smaller by using more sensitive equipment.
B. Large objects at any instant of time have an exact position and exact momentum, and with sufficient care we can measure both precisely.
C. Large objects obey Newton's laws of motion, to which the uncertainty principle does not apply.
D. Because it does apply to large objects, but the uncertainties are so small that we don't notice them.

POST

| A | B | C | **D** | %Correct | %Incorrect |
|---|---|---|---|---|---|
| 0 | 6.8% | 9.1% | **84.1%** | 84.1% | 15.9% |

PRE

| A | B | C | **D** | %Correct | %Incorrect |
|---|---|---|---|---|---|
| 3.6% | 18.0% | 23.4% | **55.0%** | 55.0% | 45.0% |

**12.** The diagram at right shows the electronic energy levels in an atom with an electron at energy level $E_m$. When this electron moves from energy level $E_m$ to $E_n$, light is emitted. The greater the energy difference between the electronic energy levels $E_m$ and $E_n$ ...

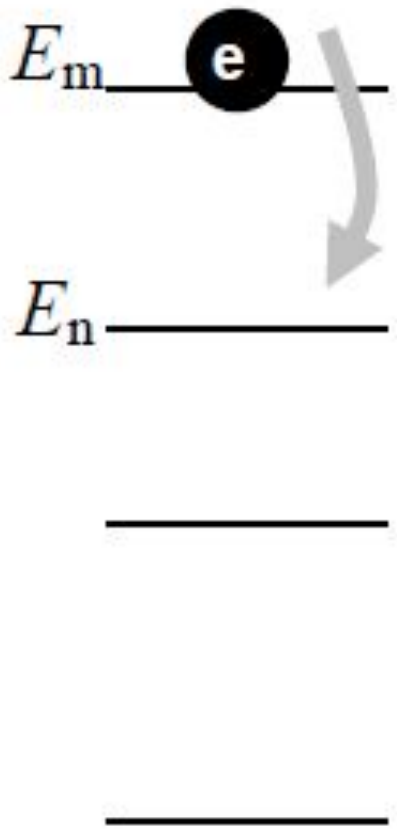

A. …the more photons emitted.
B. …the brighter (higher intensity) the light emitted.
C. …the longer the wavelength (the more red) of the light emitted.
D. …the shorter the wavelength (the more blue) of the light emitted.
E. More than one of the above answers is correct.

POST

| A | B | C | **D** | E | %Correct | %Incorrect |
|---|---|---|---|---|---|---|
| 0 | 2.3% | 12.5% | **73.9%** | 11.4% | 73.9% | 26.1% |

PRE

| A | B | C | **D** | E | %Correct | %Incorrect |
|---|---|---|---|---|---|---|
| 5.4% | 20.7% | 5.4% | **22.5%** | 45.9% | 22.5% | 77.5% |

**13.** An electron in an atom has the energy level diagram at right. The electron is in its lowest energy state, as shown in the diagram. What is the lowest energy photon that it can absorb?

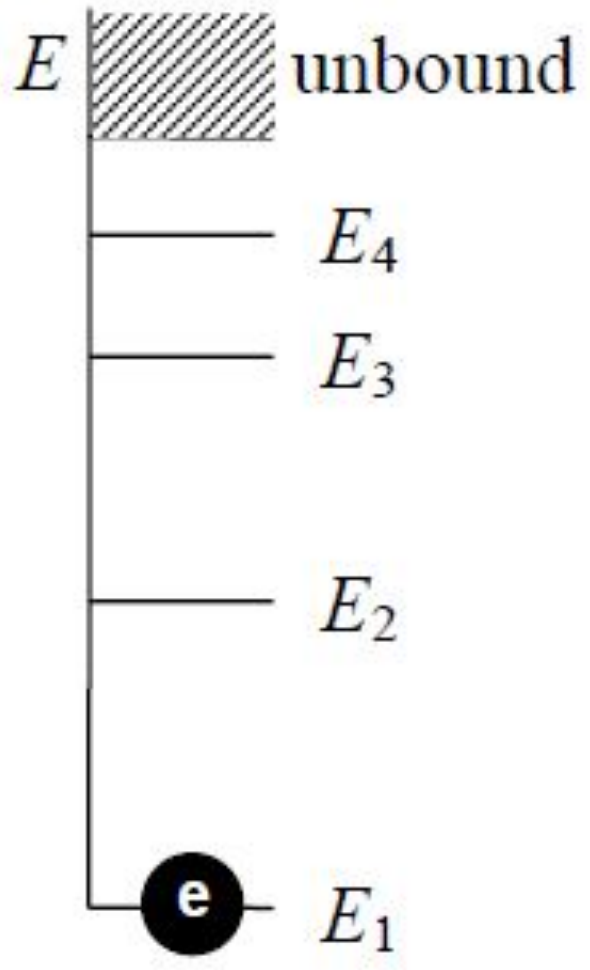

A. It can absorb a photon of any energy.
B. $E_1$
C. $E_2$
D. $E_2 - E_1$
E. $E_4 - E_3$

POST

| A | B | C | **D** | E | %Correct | %Incorrect |
|---|---|---|---|---|---|---|
| 3.4% | 2.3% | 14.8% | **78.4%** | 1.1% | 78.4% | 21.6% |

PRE

| A | B | C | **D** | E | %Correct | %Incorrect |
|---|---|---|---|---|---|---|
| 20.7% | 20.7% | 17.1% | **33.3%** | 8.1% | 33.3% | 66.7% |

**For each question 14 through 16**, choose the description from A through D which best describes the wave packet illustrated to the right.

A. Poorly defined position, well defined wavelength.
B. Well defined position, poorly defined wavelength.
C. Well defined position, well defined wavelength.
D. Poorly defined position, poorly defined wavelength.

**14.** ____

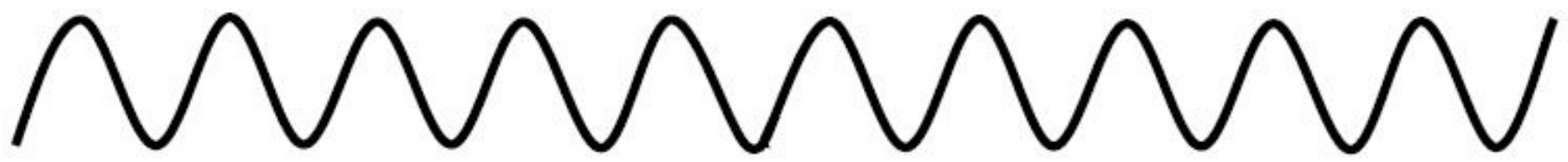

**15.** ____

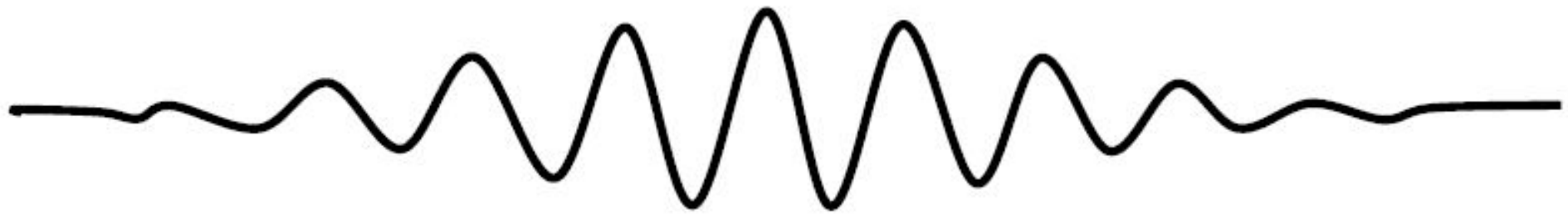

**16.** ____

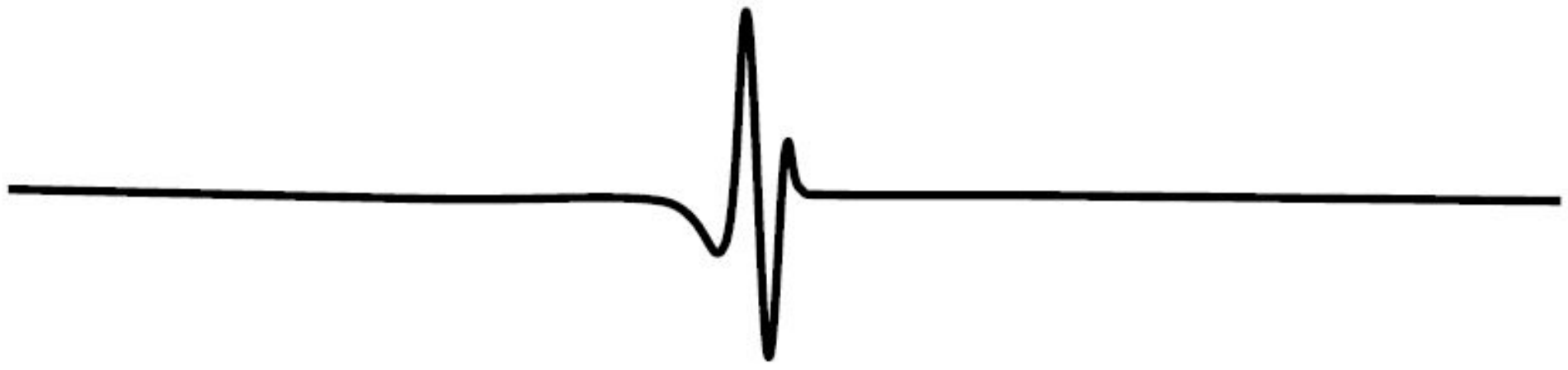

14.
POST

| A | B | C | D | E | %Correct | %Incorrect |
|---|---|---|---|---|---|---|
| **97.7%** | 0 | 1.1% | 1.1% | 0 | 97.7% | 2.3% |

PRE

| A | B | C | D | E | %Correct | %Incorrect |
|---|---|---|---|---|---|---|
| **44.1%** | 1.8% | 50.5% | 3.6% | 0 | 44.1% | 55.9% |

15.
POST

| A | B | C | D | E | %Correct | %Incorrect |
|---|---|---|---|---|---|---|
| 2.3% | 5.7% | 12.5% | **79.5%** | 0 | 79.5% | 20.5% |

PRE

| A | B | C | D | E | %Correct | %Incorrect |
|---|---|---|---|---|---|---|
| 27.9% | 24.3% | 23.4% | **24.3%** | 0 | 24.3% | 75.7% |

16.
POST

| A | B | C | D | E | %Correct | %Incorrect |
|---|---|---|---|---|---|---|
| 0 | **92.0%** | 2.3% | 5.7% | 0 | 92.0% | 8.0% |

PRE

| A | B | C | D | E | %Correct | %Incorrect |
|---|---|---|---|---|---|---|
| 6.4% | **50.4%** | 9.9% | 33.3% | 0 | 50.4% | 49.6% |

**17.** Three particles of equal mass are traveling in the same direction. The de Broglie waves of the three particles are as shown at right.

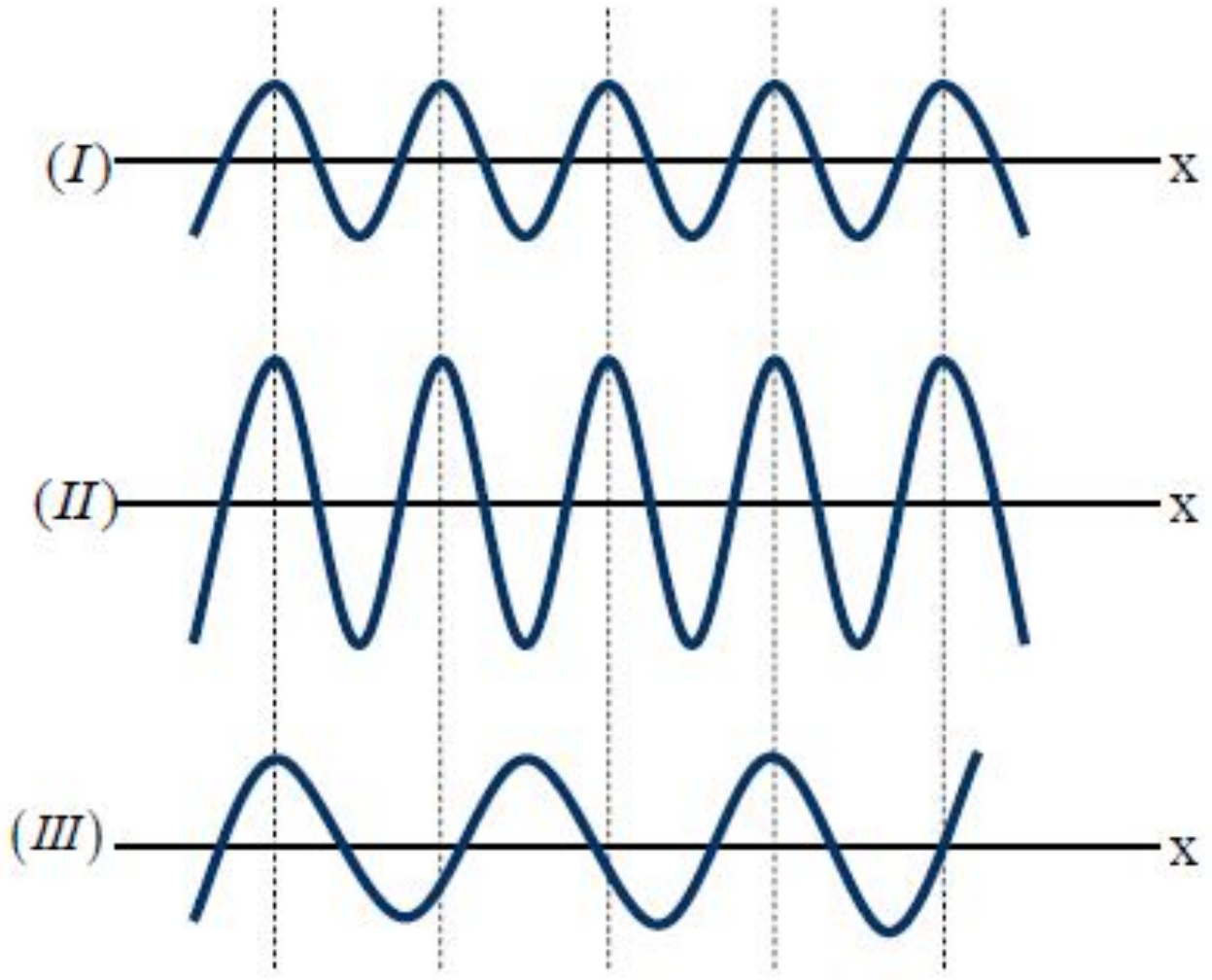

Rank the speeds of the particles I, II and III

A) $V_{II} > V_{I} > V_{III}$
E) $V_{II} > V_{III} > V_{I}$
F) $V_{I} = V_{II} > V_{III}$
G) $V_{II} > V_{I} = V_{III}$

POST

| A | B | C | D | E | %Correct | %Incorrect |
|---|---|---|---|---|---|---|
| 4.5% | 4.5% | **86.4%** | 4.6% | 0 | 86.4% | 13.6% |

**For each question 18 through 20,** choose the most appropriate answer from A through C.

A. The de Broglie wavelength of the particle will increase.
B. The de Broglie wavelength of the particle will decrease.
C. The de Broglie wavelength of the particle will remain the same.

What will happen when a quantum particle is traveling from left to right with constant total energy (dashed line: $E_0$), in a region in which the potential energy (solid line: V(x)) is…

**18.** ____ Constant?

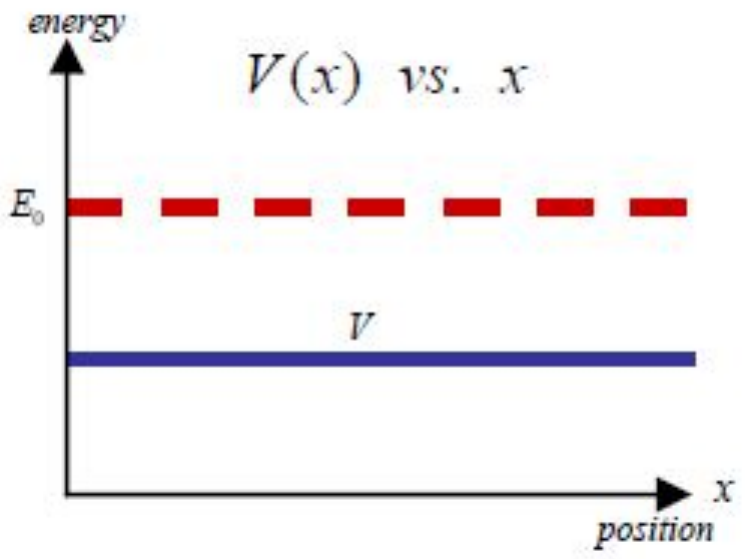

**19.** ____ Increasing?

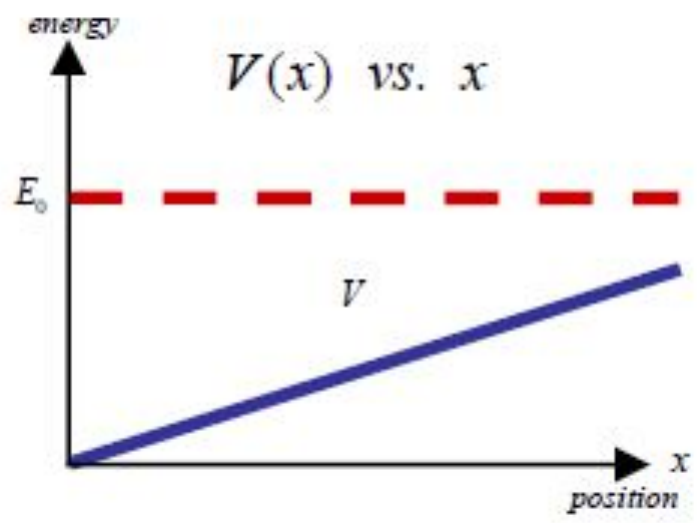

**20.** ____ Decreasing?

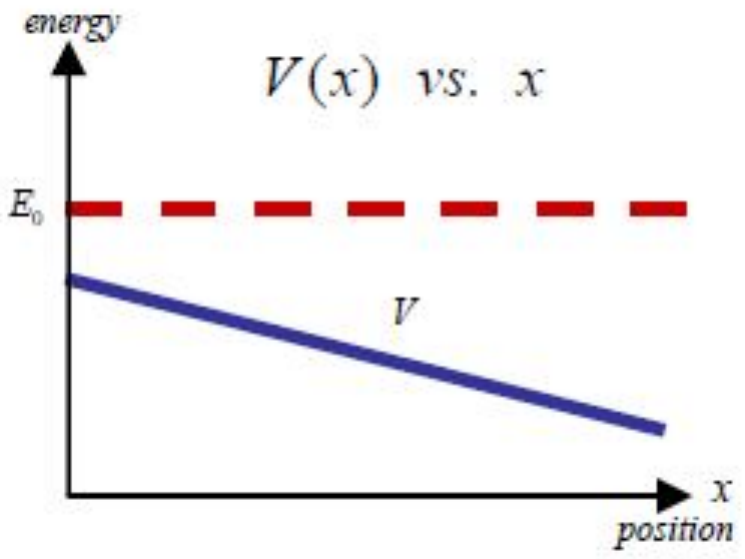

**18.**
POST

| A | B | **C** | D | E | %Correct | %Incorrect |
|---|---|---|---|---|---|---|
| 1.1% | 0 | **98.9%** | 0 | 0 | 98.9% | 1.1% |

**19.**
POST

| **A** | B | C | D | E | %Correct | %Incorrect |
|---|---|---|---|---|---|---|
| **59.1%** | 39.8% | 1.1% | 0 | 0 | 59.1% | 40.9% |

**20.**
POST

| A | **B** | C | D | E | %Correct | %Incorrect |
|---|---|---|---|---|---|---|
| 36.4% | **60.2%** | 3.4% | 0 | 0 | 60.2% | 39.8% |

**21.** You see an electron and a neutron moving by you at the <u>same speed</u>. How do their wavelengths λ compare?

A. $\lambda_{neutron} > \lambda_{electron}$
B. $\lambda_{neutron} < \lambda_{electron}$
C. $\lambda_{neutron} = \lambda_{electron}$

POST

| A | **B** | C | D | E | %Correct | %Incorrect |
|---|---|---|---|---|---|---|
| 18.2% | **54.5%** | 27.3% | 0 | 0 | 54.5% | 45.5% |

**For questions 22-23**, consider an electron with the potential energy:

$$U(x) = \begin{cases} 0 & 0 < x < L \\ \infty & x < 0 \quad or \quad x > L \end{cases}.$$

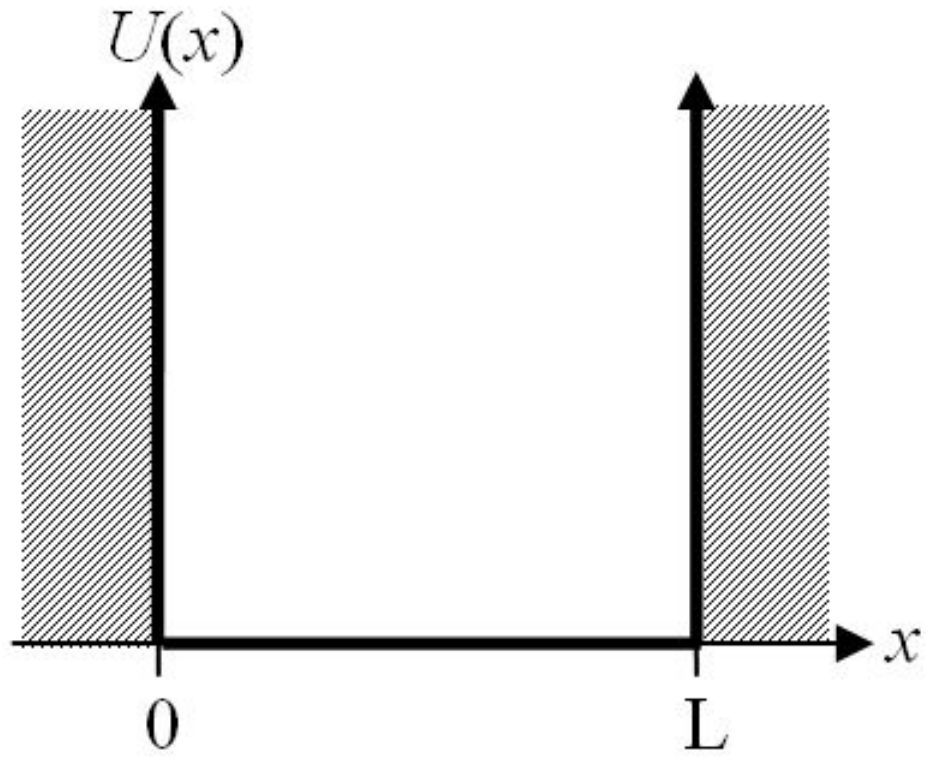

This potential energy function, plotted at right, is often referred to as an infinite square well or a rigid box. Your electron is in the lowest energy state of this potential energy, with a wave function $\psi(x) = \psi_1(x)$ and a corresponding energy $E_1$.

**22.** Suppose that you measure the position of this electron very precisely, without destroying the electron. Which of the following statements most accurately describes the result of this measurement?

A. You will find that the electron is spread out in space between 0 and L.
B. You will find the electron at the bottom of the well.
C. You will find the electron at $x$ = L/2.
D. You will find the electron at one particular location, somewhere between 0 and L. There is an equal probability of finding the electron anywhere between 0 and L
E. You will find the electron at one particular location, somewhere between 0 and L. It is most likely to be found at $x$ = L/2, but it could also be found elsewhere with some probability.

POST

| A | B | C | D | **E** | %Correct | %Incorrect |
|---|---|---|---|---|---|---|
| 1.1% | 1.1% | 1.1% | 14.8% | **81.9%** | 81.9% | 18.1% |

**23.** *After* measuring the position, you measure the *energy* of the same electron. Which of the following statements describes the result of this energy measurement?

A. The value that you measure will be $E_1$.
B. The value that you measure could possibly be $E_1$.
C. The value that you measure will *not* be $E_1$.

POST

| A | **B** | C | D | E | %Correct | %Incorrect |
|---|---|---|---|---|---|---|
| 22.7% | **48.9%** | 28.4% | 0 | 0 | 48.9% | 51.1% |

**24.** The figure at right shows a potential energy function $U(x)$, where the potential energy is infinite if $x$ is less than 0 or greater than L, and has a slanted bottom in between 0 and L, so that the potential well is deeper on the right than on the left. Which of the plots of $|\psi(x)|^2$ vs. $x$ is most likely to correspond to a stationary state of this potential well?

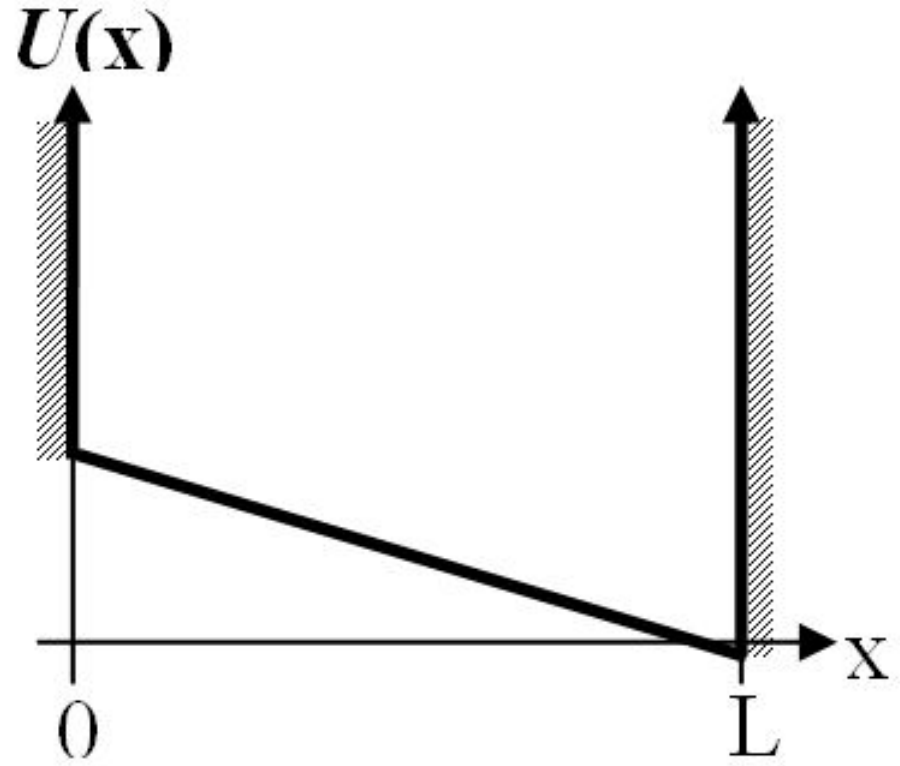

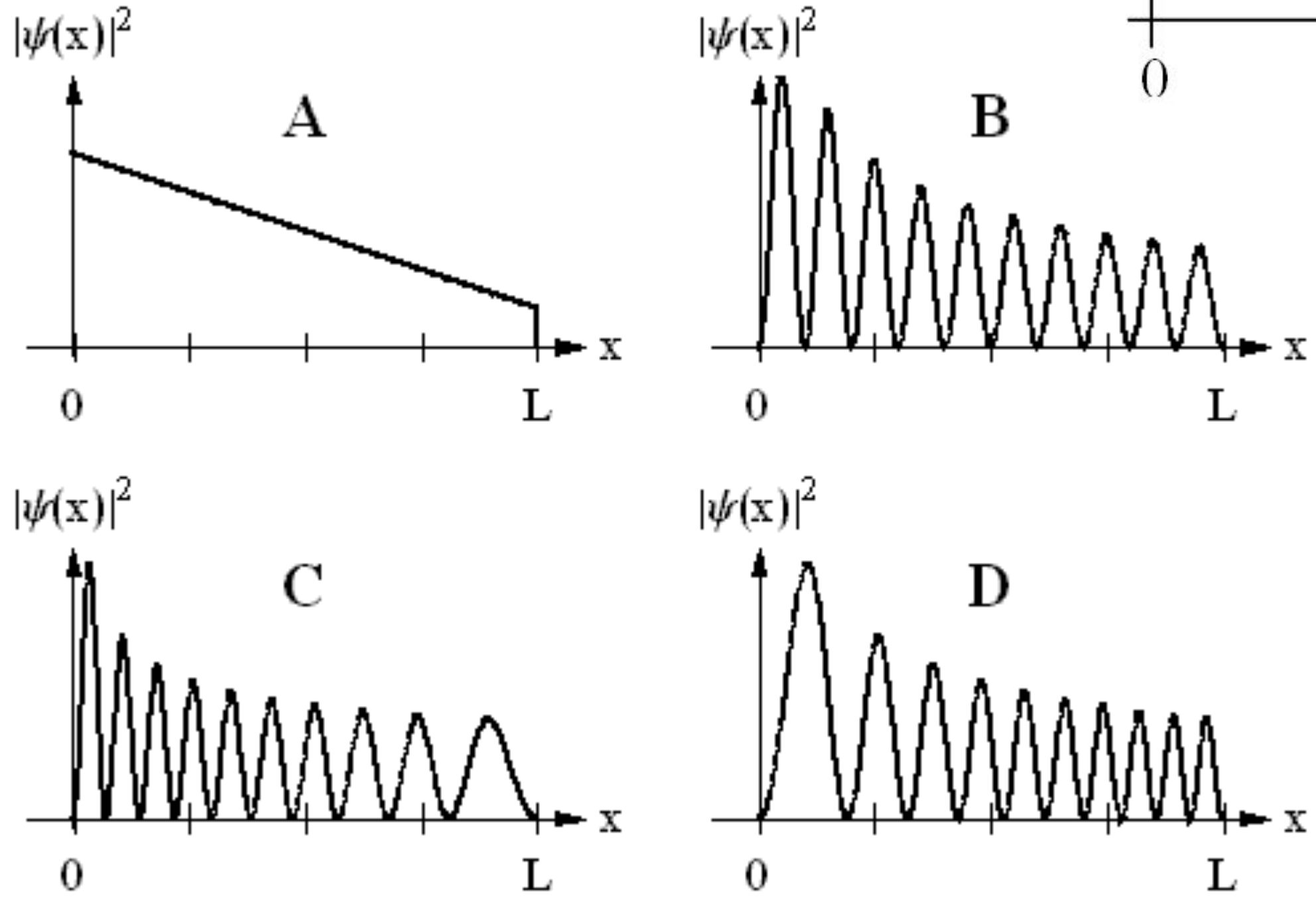

E. More than one of these is a possible stationary state.

POST

| A | B | C | **D** | E | %Correct | %Incorrect |
|---|---|---|---|---|---|---|
| 2.3% | 15.9% | 27.3% | **42.0%** | 12.5% | 42.0% | 58.0% |

**25.** The plot at right shows a snapshot of the spatial part of a one-dimensional wave function for a particle, ψ(x), versus x. ψ(x) is purely real. The labels, I, II, and III, indicate regions in which measurements of the position of the particle can be made. Order the probabilities, P, of finding the particle in regions I, II, and III, from biggest to smallest.

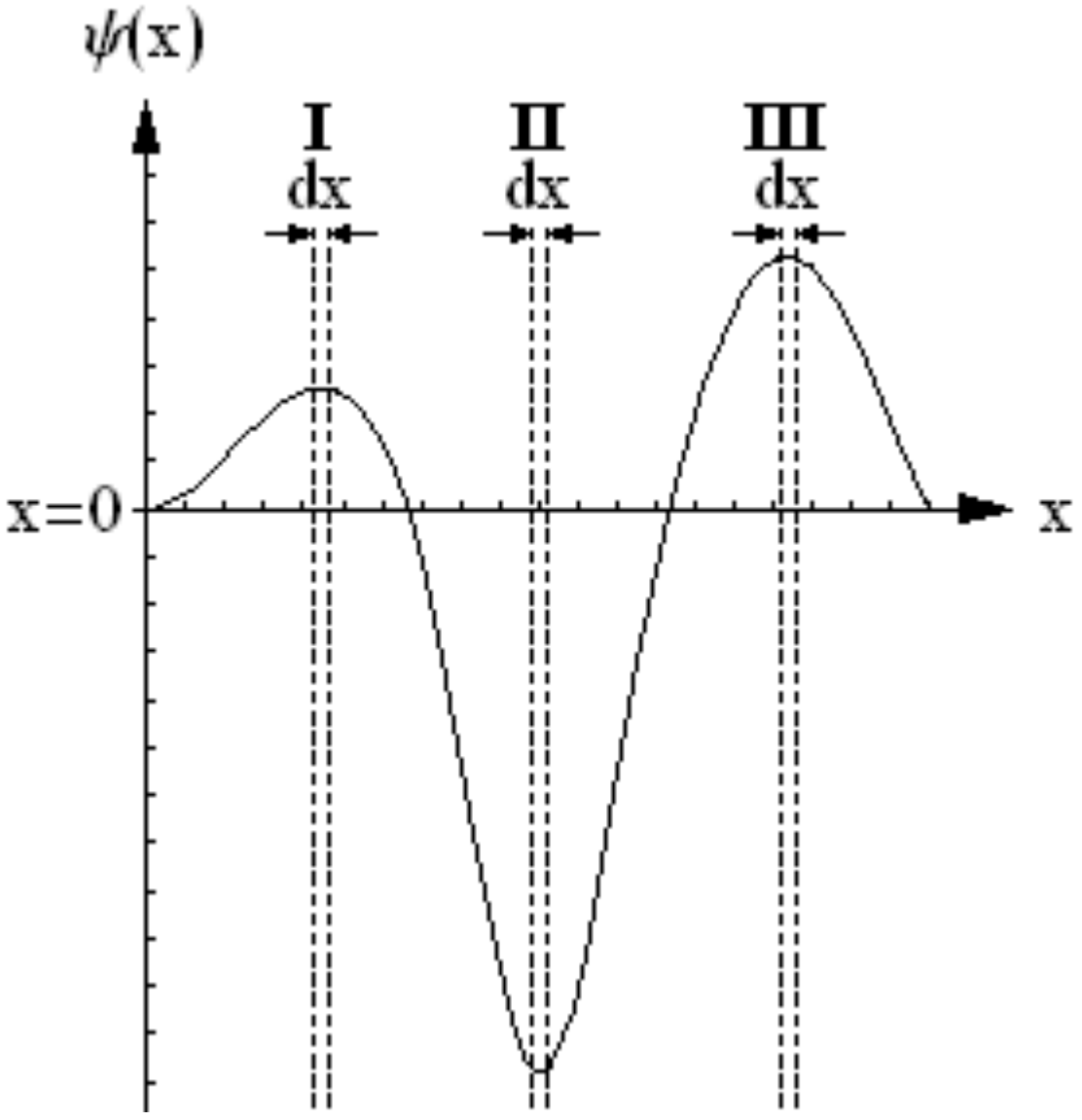

A. P(III) > P(I) > P(II)
B. P(II) > P(I) > P(III)
C. P(III) > P(II) > P(I)
D. P(I) > P(II) > P(III)
E. P(II) > P(III) > P(I)

POST

| A | B | C | D | **E** | %Correct | %Incorrect |
|---|---|---|---|---|---|---|
| 0 | 0 | 0 | 0 | **100%** | 100% | 0% |

**26.** If a particle is bound in a potential well, which statement best describes the values of total energy that particle can possibly have?

A. It can have only certain discrete values of total energy.
B. It can have any value of total energy as long as it is small enough that the particle remains bound.
C. It can have any value of total energy as long as it is large enough that the particle remains bound.

POST

| **A** | B | C | D | E | %Correct | %Incorrect |
|---|---|---|---|---|---|---|
| **71.6%** | 25.0% | 3.4% | 0 | 0 | 71.6% | 28.4% |

**27.** If a particle is *NOT* bound in a potential well, which statement best describes the values of total energy that particle can possibly have?

A. It can have only certain discrete values of total energy.
B. It can have any value of total energy as long as it is small enough that the particle remains unbound.
C. It can have any value of total energy as long as it is large enough that the particle remains unbound.

POST

| A | B | **C** | D | E | %Correct | %Incorrect |
|---|---|---|---|---|---|---|
| 17.0% | 6.8% | **76.2%** | 0 | 0 | 76.2% | 23.8% |

**28.** An electron with energy $E$ is traveling through a conducting wire when it encounters a small gap in the wire of width w. The potential energy of the electron as a function of position is given by the plot at right, where $U_0 > E$. Which of the following sketches most accurately describes a snapshot of the real part of the wave function of this electron?

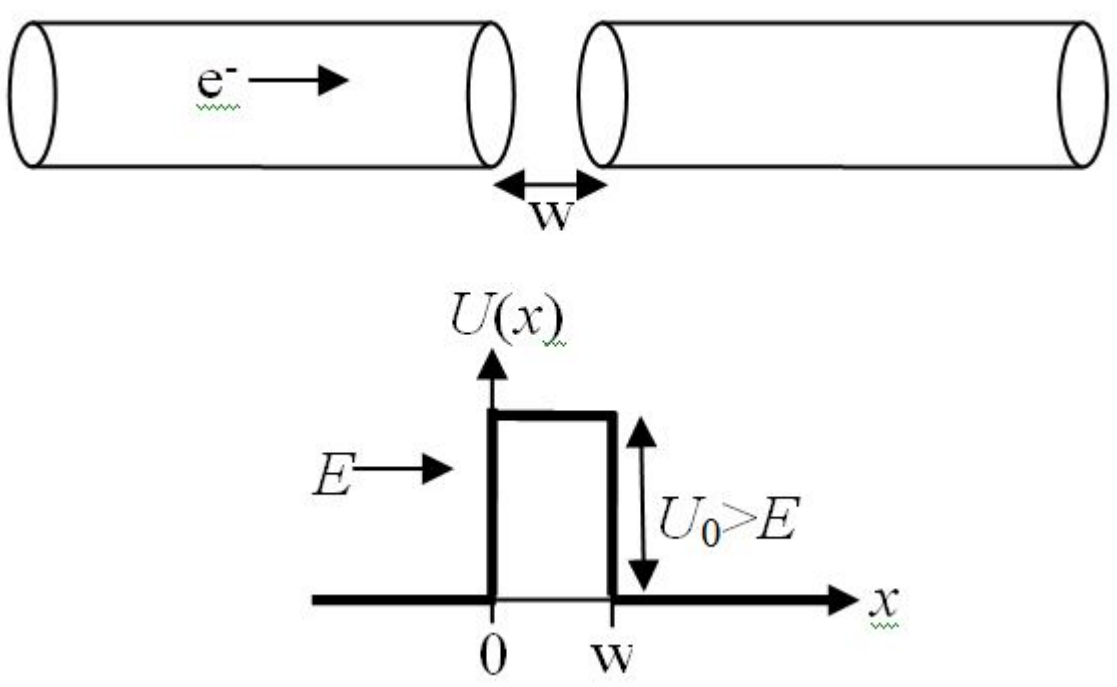

**A**

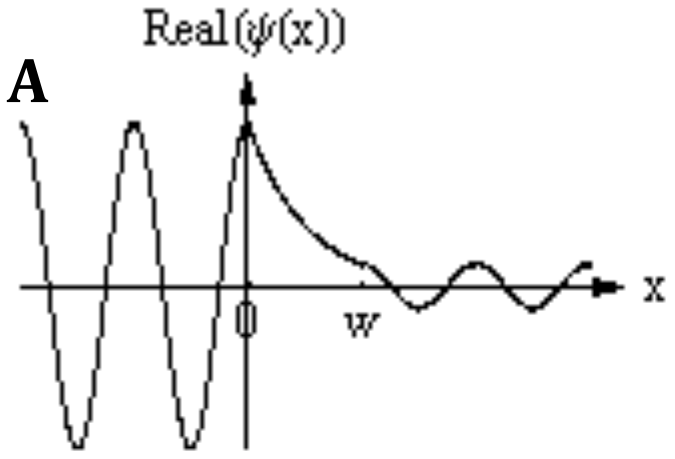

**B**

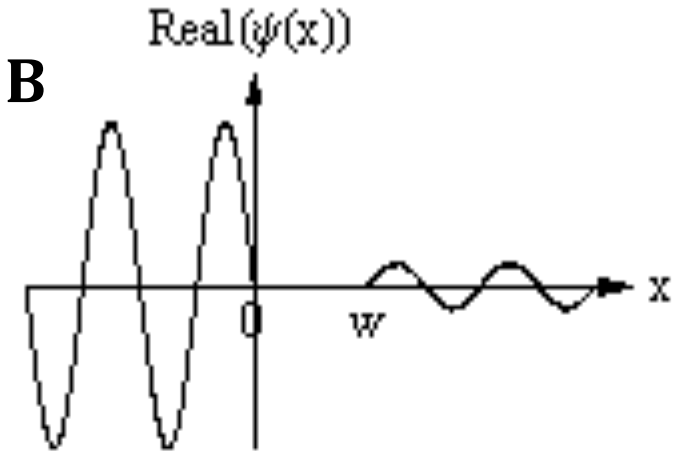

**C**

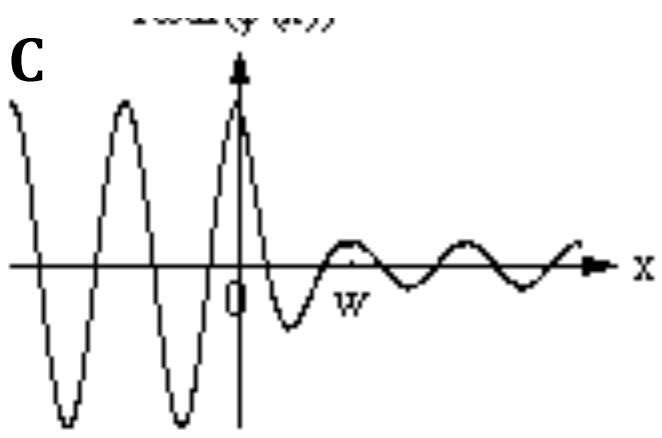

**D**

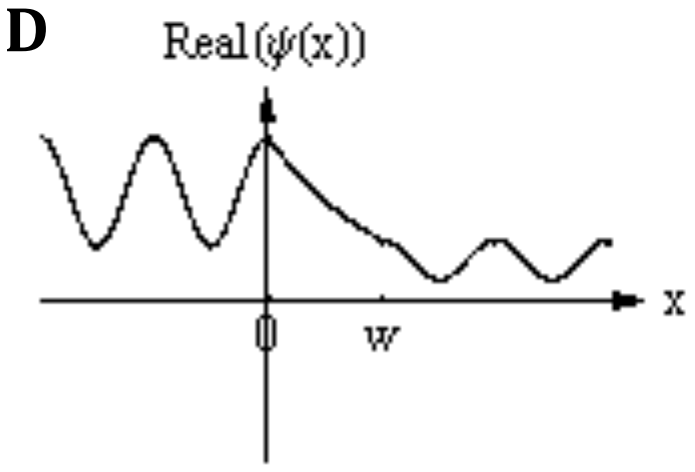

**E**

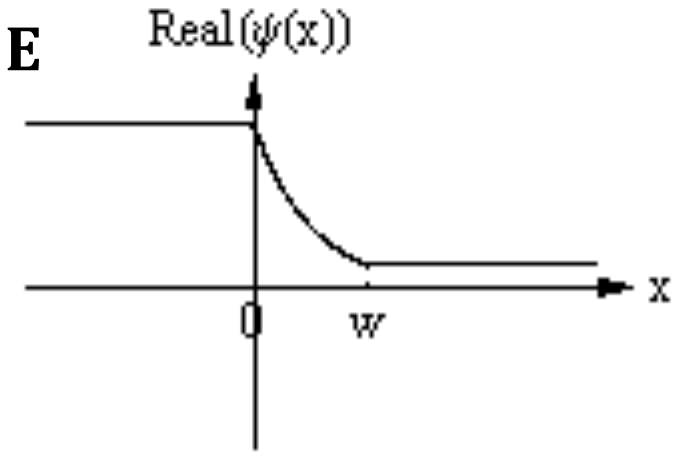

POST

| **A** | B | C | D | E | %Correct | %Incorrect |
|---|---|---|---|---|---|---|
| **71.6%** | 3.4% | 12.5% | 12.5% | 0 | 71.6% | 28.4% |

**29.** Suppose that in the experiment described **in question 28**, you would like to *decrease* the speed of the electron coming out on the right side. Which of the following changes to the experimental set-up would decrease this speed?

A. Increase the width w of the gap:

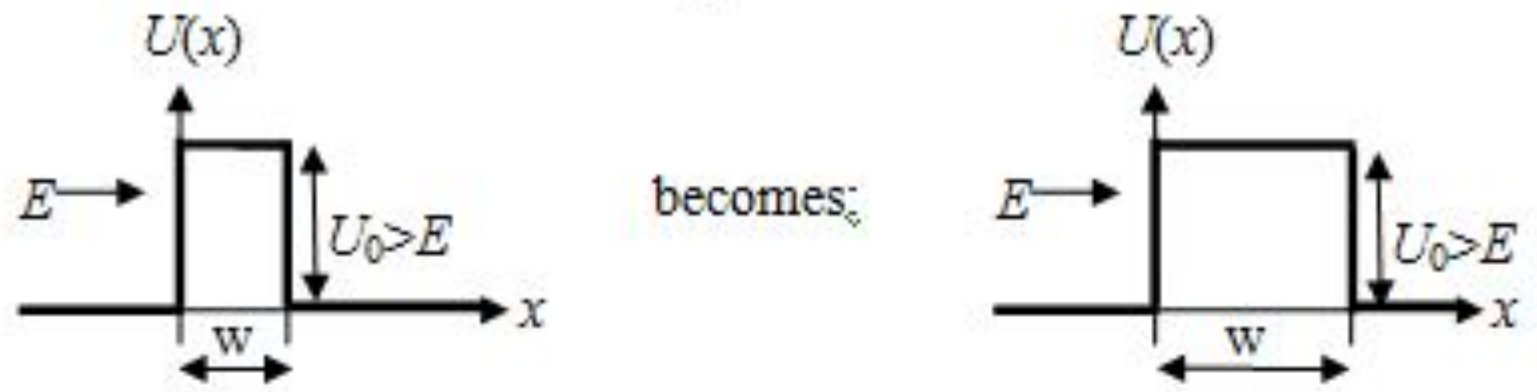

B. Increase $U_0$, the potential energy of the gap:

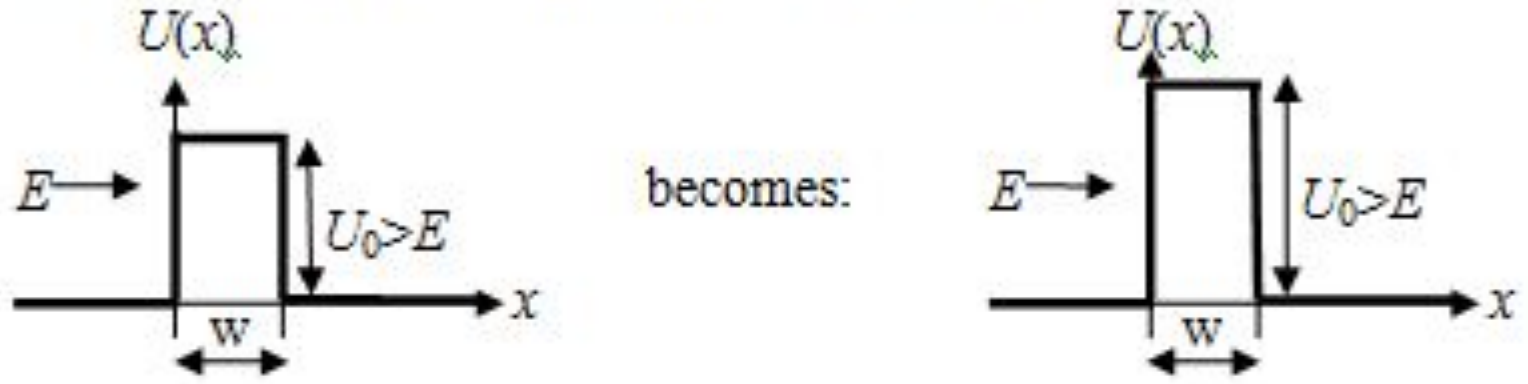

C. Increase the potential energy to the right of the gap:

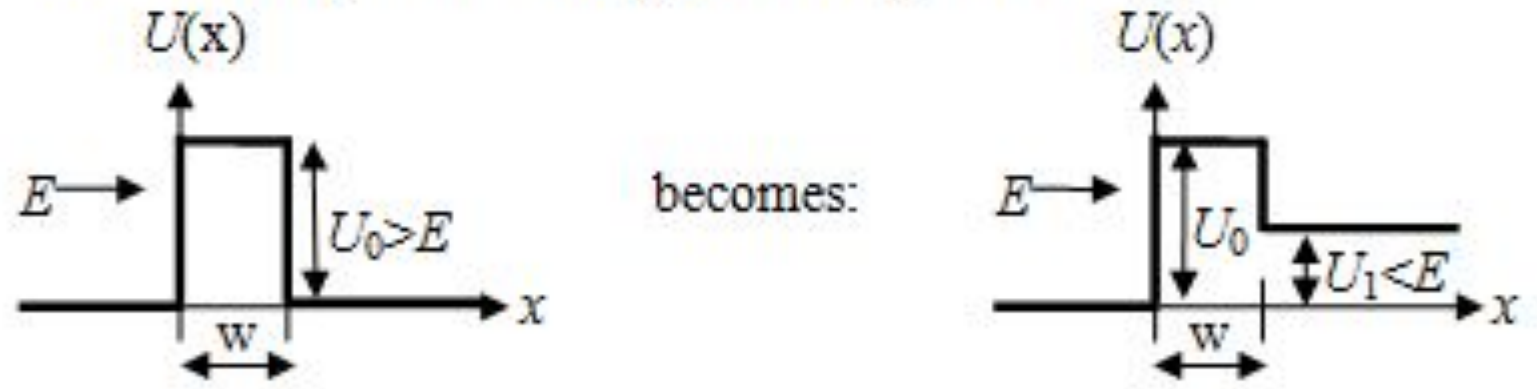

D. Decrease the potential energy to the right of the gap:

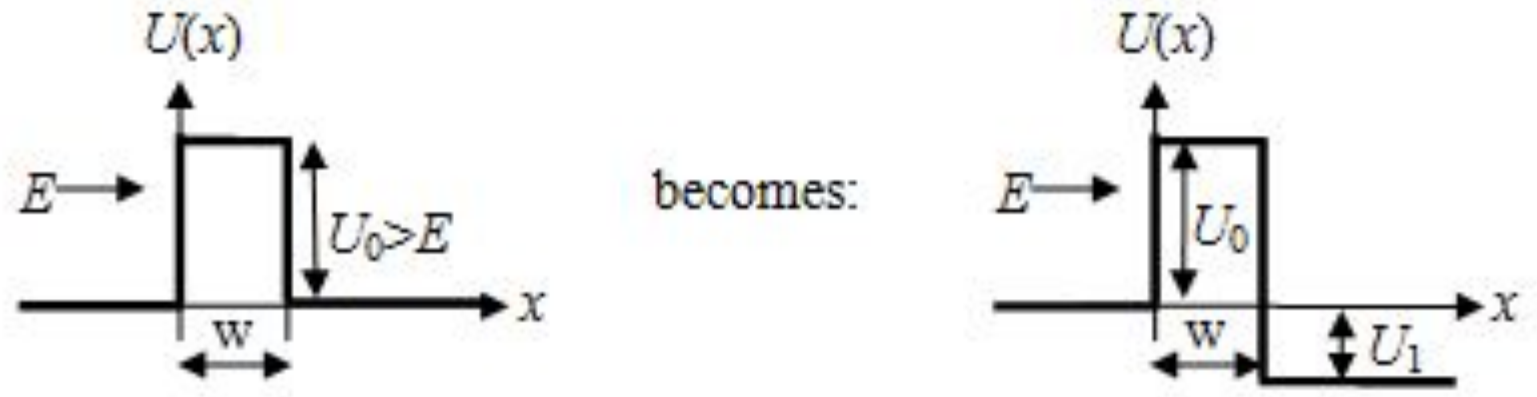

E. More than one of the changes above would decrease the speed of the electron.

POST

| A | B | **C** | D | E | %Correct | %Incorrect |
|---|---|---|---|---|---|---|
| 13.6% | 4.5% | **46.6%** | 10.2% | 22.7% | 46.6% | 53.4% |

**30.** The electron in a hydrogen atom is in its ground state. You measure the distance of the electron from the nucleus. What will be the result of this measurement?

A. You will measure the distance to be the Bohr radius.
B. You could measure any distance between zero and infinity with equal probability.
C. You are most likely to measure the distance to be the Bohr radius, but there is a range of other distances that you could possibly measure.
D. There is a mostly equal probability of finding the electron at any distance within a range from a little bit less than the Bohr radius to a little bit more than the Bohr radius.

POST

| A | B | **C** | D | E | %Correct | %Incorrect |
|---|---|---|---|---|---|---|
| 29.5% | 2.3% | **48.9%** | 18.1% | 0 | 48.9% | 51.1% |

# APPENDIX D

## Selected Homework, Exam, Survey and Final Essay Submissions from Four Students (Fall 2010)

## Student A

### Pre-Instruction Survey

**1.** It is possible for physicists to carefully perform the same measurement and get two very different results that are both correct.

**(Agree)** I feel that no matter how much technology advances or how much we learn, we can never fully understand how the world works and in many cases, we use outcomes of experiments to look at phenomena in different ways that may or may not be entirely correct in the real world. For instance, looking at the behavior of light as both a particle and wave. So, yes, I believe that an experiment came be conducted twice with different outcomes.

**2.** The probabilistic nature of quantum mechanics is mostly due to physical limitations of our measurement instruments.

**(Neutral)** I really don't know enough about quantum theory to make a guess on that. However, even our most basic assumptions about the world have sometimes proven to be incorrect and quantum seems to involve so much theory that we can never really be sure if it actually functions the way physicists think it does or if we are coming up with theories that just fit what we find without even seeing the entire picture.

**3.** When not being observed, an electron in an atom still exists at a definite (but unknown) position at each moment in time.

**(Strongly Agree)** An electron is a fundamental piece of an atom, though it moves extremely fast, so at any point in time, yes it does occupy a position being that it is matter.

**4.** I think quantum mechanics is an interesting subject.

**(Strongly Agree)** From the examples I have heard and some of the theory, I think quantum mechanic is very interesting.

**5.** I have heard about quantum mechanics through popular venues (books, films, websites, etc...)

**(Strongly Agree)** [Blank]

**Homework Problems**

**<u>HW01:</u>**

**13.** What are atoms made of, and how are the parts of the atom configured?

Atoms are made up of a particular configuration of electrons, protons and neutrons. The center of an atom is called its nucleus which is comprised of the protons and neutrons. Electrons surround the nucleus in different orbit levels.

**21.** The force (F) experienced by a charge (q) in an electric field (E) is given by the equation: $\vec{F} = q \cdot \vec{E}$

In fact, the electric field at any point in space is ***defined*** in terms of the force that would be experienced by a charge placed at that point in space:

$\vec{E} \equiv \frac{\vec{F}}{q}$ (Electric field defined as force per unit charge)

How do you (personally) ***interpret*** the concept of an electric field? Is the electric field something that is physically real, even though we can't observe it directly? Or is it a mathematical tool devised by physicists to explain our observations and to make calculations easier? A mix of both, or is it something else entirely? Remember, we are interested in what you actually think – there are no "correct" or "incorrect" answers to this question. Please provide the reasoning behind your response.

I think that electric fields are real things that cant be seen heard, etc. because they are comprised of negative charges that can be harnessed into electrical energy. This energy has been used to build the world that we live in today.

**<u>HW02:</u>**

**12.** It is said that the photoelectric effect demonstrates the particle-like nature of light. Explain how this conclusion is reached. That is, what *experimental evidence* is consistent with particle-like behavior for light but not with wave-like behavior? Cite at least two pieces of evidence.

Any light will give at least a few electrons enough energy to move like a partical because each light quantum delivers all of its energy to just one electron. Therefore, all the energy used to project an electron is used in one direction making it act like a particle and not like a wave. If we increase the intensity of the light, then more light quantum will deliver its energy to a single electron.

**13.** In explaining the observations from the photoelectric effect experiment, did Einstein propose a new theory, a new model, or a new interpretation of the data? In your mind, are there any differences between the three terms (theory, model, interpretation)? If so, what would they be? If not, why would it be OK to use the three terms interchangeably?

Remember, for these types of questions, we are interested in having you express your opinion, so there are no "right" or "wrong" answers. Your response will be graded on the effort you put into it.

Einstein proposed a new theory, a new model and a new interpretation of the data. To me there are not really any differences between the three in this case. Einstein came up with the model E=hf, which in turn made a whole new theory about light quantum, and it is obvious that this would create a new interpretation of the data.

**HW03:**

**3.** What was the point of the “Farmer and the Seeds” story (from lecture)?

To show an analogy of theory, model, and interpretation can all work together to form an abstract idea. The different schemes in this analogy also play a role in this abstract idea.

**4.** Review the discussion of the terms *theory, model & interpretation* in the solution set from HW02. How would these terms, as **you** understand them, apply to the different elements of the “Farmer and the Seeds” story, and to the schemes you came up with in class?

Remember, we are interested in what **you** think. There are no “right” or “wrong” answers to this question.

Theory: Schemes of why certain #s of sprouts came up.

Model: Way of showing the different combinations to produce different #s of sprouts.

Interpretation: How one interprets this analogy as a whole and why we get the results that we do.

**HW06:**

**Questions 16 – 18 refer to the reading “100 Years of Quantum Mysteries”.**

**16.** As discussed in this article, what were some of the problems in classical physics that led to the development of quantum theory?

Electromagnetic theory predicted that orbiting electrons would radiate away their energy, and would spiral into the nucleus continuously, but a hydrogen atom was known to be stable. This didn’t make any sense because it led to an underprediction of hydrogen's lifetime by 40 years. Bohr's theory used quanta to explain this discrepancy.

**17.** How are the terms *theory* and *interpretation* used in this article?

In this article, theories are used to describe ideas that scientists, and physicists came up with to explain why this happen the way that they do. Interpretations are used to describe the way that scientists think of a given theory. It represents the thought processes of a given person/scientist.

**18.** Is there any experimental evidence in favor of any of the interpretations discussed in this article? In the cases where there is not, why would a scientist favor one interpretation over another?

There is no experimental evidence in favor of one interpretation over another. A scientist would favor one interpretation over another because he/she feels or believes in that particular interpretation. It also may depend on what the scientist is trying to do with a specific interpretation.

**19. This question refers to the reading assignment "Probability"**

According to this article, in what way(s) is quantum mechanics a probabilistic theory?

Quantum mechanics is a probabilistic theory because it is impossible to attempt to compute where a certain electron is or will be at a certain time. Probabilistic theory can be used to compute this through the Schrodinger equation.

**<u>HW07:</u>**

**Questions 9 & 10 refer to the readings: "A Quantum Threat to Special Relativity" & "Is the moon there when nobody looks?"**

**9.** What is meant by the terms ***realism, locality*** & ***completeness***? What are some examples of hidden variables?

To me, realism can be described as the idea that things happen whether someone is there to witness it. For example, if a tree falls in the middle of the woods and there is nothing around to hear it, does it still make a sound? Locality represents an intuition that objects around us can only be directly influenced by other objects in its immediate surrounding. Completeness is a description of the world that is represented by the smallest physical attributes such as particles, electrons, waves, atoms, etc. Completeness describes the complete world as one. A great example of hidden variables is the example referred to in class about 2 socks being put into different boxes, mixed up and sent to opposite sides of the universe. Once you discover the color of one sock, you know the color of the other one... entanglement. These socks are hidden variables until one sock's color is discovered.

**10.** Does ***entanglement*** allow for faster-than-light communication? If so, what kind of information can be communicated? If not, why not?

Entanglement doesn't allow for faster than the speed of light communication because this method of entanglement involves observations that cannot be controlled. For example, in the sock analogy described above, the person/ thing that discovers the color of one of the socks cannot transmit a signal to the other person receiving the other sock before he/she already knows what it is. Therefore no information is actually being transmitted in this process.

**11.** In the two Aspect experiments discussed in class, where the goal was to produce a "single-photon" source, the calcium atoms were excited to the upper level by a two-photon absorption process. Why did the experimenters excite the calcium atoms with a laser pulse of 3.05 eV photons followed by a pulse of 2.13 eV photons, rather than with single photons with the same energy as the two original photons combined (single-photon excitation)?

The experimenters excited the calcium atoms with two seperate pulses because they wanted only one electron to be spontaneously emitted at a given time, and exciteing two photons seperately to get the highest energy level to emit an electron is the only way to emit a photon at 2.93 eV and bring the energy level back down to ground state.

**12.** In your own words, explain what the anti-correlation parameter ($\boldsymbol{\alpha}$) is, both in terms of its mathematical definition, and in terms of what it physically tells us, in the context of single-photon experiments as performed by Aspect. Why didn't Aspect measure $\boldsymbol{\alpha} = 0$ if photons are supposed to be acting like particles?

The anti correlation parameter alpha is equal to the probability for Nc (probability for both photomultiplier A and B are both triggered) to be triggered divided by the probability that photomultiplier A times the probability that photomultiplier B is triggered. If there is a great intensity of photons being emitted then alpha will be high and this represents a wavelike nature of light. If the intensity is small, then alpha will be small and represents a particle like nature of light. Alpha is not zero because Aspect could not fire a single photon. That is the only way it could be zero.

**(Essay)** In class we have discussed correlated measurements performed on systems of two entangled atoms. The assigned reading "The Reality of the Quantum World" discusses correlated measurements performed on entangled photon pairs. In what ways are these systems of entangled photon pairs similar or different from systems of entangled atom pairs? In what sense are the particles in each system entangled (i.e., what properties are correlated for each of the two types of systems)? What types of measurements are performed to determine these properties, and what are probabilities for the possible results of these measurements for both types of systems?

Entanglement is used to describe a system where one measurement effects another measurement. Systems of entangled atom pairs have a few similarities and a few differences to entangled photon pairs. First of all, an atom will always have a magnetic moment around it also known as the spin of the atom. When two atoms are entangled, we know that the two atoms will have opposite spins (for example if you know the spin of one atom is up, and then you know that the spin of the other atom must be down.) The best way to describe entanglement of a photon pair is by using the example of light passing through a polarizer. When a photon passes through the polarizer at one specific angle like 135 degrees, then you know that if you fire another photon into the same polarizer with the same reference angle that you will get the same result (the photon will pass through.) However if you were to rotate that same polarizer 90 degrees then you know that the photons won't pass through it. The main difference between entangled atom pairs and entangled photon pairs is that photon pairs will behave with each other and entangled atom pairs will always oppose each other. A major similarity between these two types of entanglement is that if you know the outcome of one measurement, then you will know the outcome of the other one as well, just as I described above. A good way to measure entanglement of atom pairs is through using a stern-gerlach analyzer and a good way to measure the entanglement of photon pairs is through playing with the polarization of light.

**(Essay)** As discussed in class and in the readings, what do the two single-photon experiments performed by Aspect tell us about the nature of photons? How were the two experiments designed to demonstrate the particle and the wave nature of photons? When answering, don't concern yourself with technical details (such as how the photons were produced); focus instead on how the design of each experimental setup determined which type of photon behavior would be observed. How are the elements of these two experiments combined in a delayed-choice experiment, and what do delayed-choice experiments (along with the two Aspect experiments) tell us about the nature of photons?

When Aspect performed these two single photon experiments, he made one distinct difference in the design of the experiment. In the first experiment, Aspect shot a beam of light through a beam splitter and off a mirror (either A or B) and into a photomultiplier that corresponds to the given mirror. In the second experiment as discussed in class, Aspect added a second beam splitter near the two photomultipliers so that a photon could travel on either the path (the photon can hit either mirror A or mirror B and still hit either one of the two photomultipliers (A or B), but may be interfered with depending on the lengths of the paths. These two different experiments tell us that a photon can either behave like a particle where a photon must travel in one distinct path to get to a particular photomultiplier, as displayed in experiment one, or the photon could behave like a wave, where the photon could take either path to get to a particular photomultiplier as shown in experiment two. In the delayed choice experiment, experiment three from lecture, these first two experiments are in some sense combined. In this experiment, the photon is triggered by a voltage to take either path A or path B or both. The photon would take path A if it is detected by photomultiplier A and if it takes path B then you know that the photon was detected by either photomultiplier 1 or 2. This only happens when there is a voltage applied to PCA, and we know that there won't be any interference involved because we know which path was taken. On the other hand, if you don't apply a voltage, then you know that the photon could have taken either path. In this case we know that the photon was detected by photomultiplier A, and because two different paths are possible, we know that there is possibly interference involved. This final experiment tells us that a photon cannot behave like a wave and like a particle at the same time.

**HW09**:

**3.** How does thinking of electrons as waves rather than particles explain why energy levels in the hydrogen atom are quantized? How does it answer Bohr's question of why electrons don't radiate energy when they are in one of these energy levels?

Thinking of electrons as waves instead of particles explains why energy levels of hydrogen are quantized because waves can't accelerate, they are at a constant velocity which means that they will stay in their given energy level unless they are excited by another electron. An electron can't be in between energy levels, therefore they are quantized. This goes with bohr's theory that electrons cant accelerate; a partical wave that starts propagating around the nucleus must meet up with its original path therefore it cant radiate energy.

**Exam Questions:**

**Exam 2**

**E1. (10 Points)** In the sequence of screenshots shown below (taken from the PhET Quantum Wave Interference simulation), we see: A) a bright spot (representing the probability density for a single electron) emerges from an electron gun; B) passes through both slits; and C) a single electron is detected on the far screen. After many electrons pass through and are detected, a fringe pattern develops (not shown).

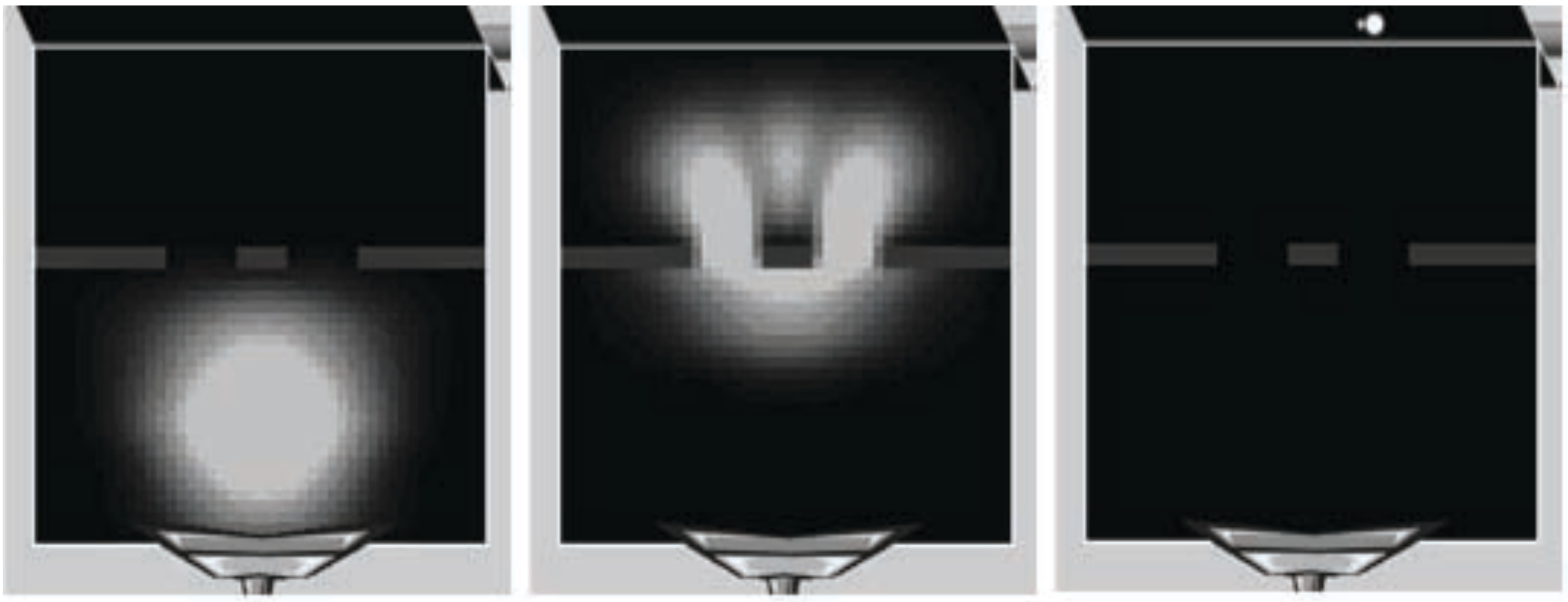

(A) (B) (C)

**Three students discuss the Quantum Wave Interference simulation:**

**Student 1**: The probability density is so large because we don't know the true position of the electron. Since only a single dot at a time appears on the detecting screen, the electron must have been a tiny particle traveling somewhere inside that blob, so that the electron went through one slit or the other on its way to the point where it was detected.

**Student 2**: The blob represents the electron itself, since a free electron is properly described by a wave packet. The electron acts as a wave and will go through both slits and interfere with itself. That's why a distinct interference pattern will show up on the screen after shooting many electrons.

**Student 3**: All we can really know is the probability for where the electron will be detected. Quantum mechanics may predict the likelihood for a measurement outcome, but it really doesn't tell us what the electron is doing between being emitted from the gun and being detected at the screen.

**E1.A (2 Points)** In terms of the interpretations of quantum phenomena we've discussed in class, how would you characterize the perspective represented by Student 1's statement? What assumptions are being made by Student 1 that allows you to identify their perspective on this double-slit experiment?

Student One interprets this sequence of screen shots classically, he obviously is thinking of this problem not quantum mechanically because if he did he would think the electron is going through both slits at the same time although he is thinking of this in terms of the Bohr model a bit. I think this is because he knows that we don't know the true position of the electron which means he is also thinking of it in terms of the uncertainty principle too. He thinks classically because he thinks it can't go through 2 slits at the same time.

**E1.B (6 Points)** For each of the first two statements (made by Students 1 & 2), what rationale or evidence (experimental or otherwise, if any) exists that favors or refutes these two points of view? As for the third statement, is Student 3 saying that Students 1 & 2 are wrong? Why would a practicing physicist choose to agree or disagree with Student 3?

For Student 1, I agree that the prob. density is large because we don't know position of the electron – we never do. I disagree that this can't be represented quantum mechanically. From experiments in the past it is proven that we get fringes (pattern).

For Student 2, I disagree that the electron is the blob because in the brighter part of the blob there is a higher probability that an electron will be detected than in the dimmer part. Although I agree the electron acts as a wave, I disagree that a single electron can be described as a wave packet.

The third student isn't saying the first 2 are wrong. All he is saying is that the interference patterns are a result of probability not classical physics and that both are right. We don't know how we get the results we do so we work with probabilities.

**E1.C (2 Points)** Which student(s) (if any) do you ***personally*** agree with? If you have a different interpretation of what is happening in this experiment, then say what that is. Would it be reasonable or not to agree with ***both*** Student 1 & Student 2? This question is about your personal beliefs, and so there is no "correct" or "incorrect" answer, but you will be graded on making a reasonable effort in explaining why you believe what you do.

I think from what I have learned in this class that Student 3 is correct. Probability can show us patterns but we really don't know what's going on before. It is reasonable to agree with both Student One who thinks classically and Student 2 who thinks quantum mechanically because that allows you to form your own ideas about what is going on but the truth is that we don't know what's going on between emission and the screen.

**E3. (8 POINTS TOTAL)** For the diagrams below depicting Experiments X & Y, M = Mirror, BS=Beam Splitter, PM = Photomultiplier, N = Counter. In each experiment a single-photon source sends photons to the right through the apparatus one at a time.

## EXPERIMENT X

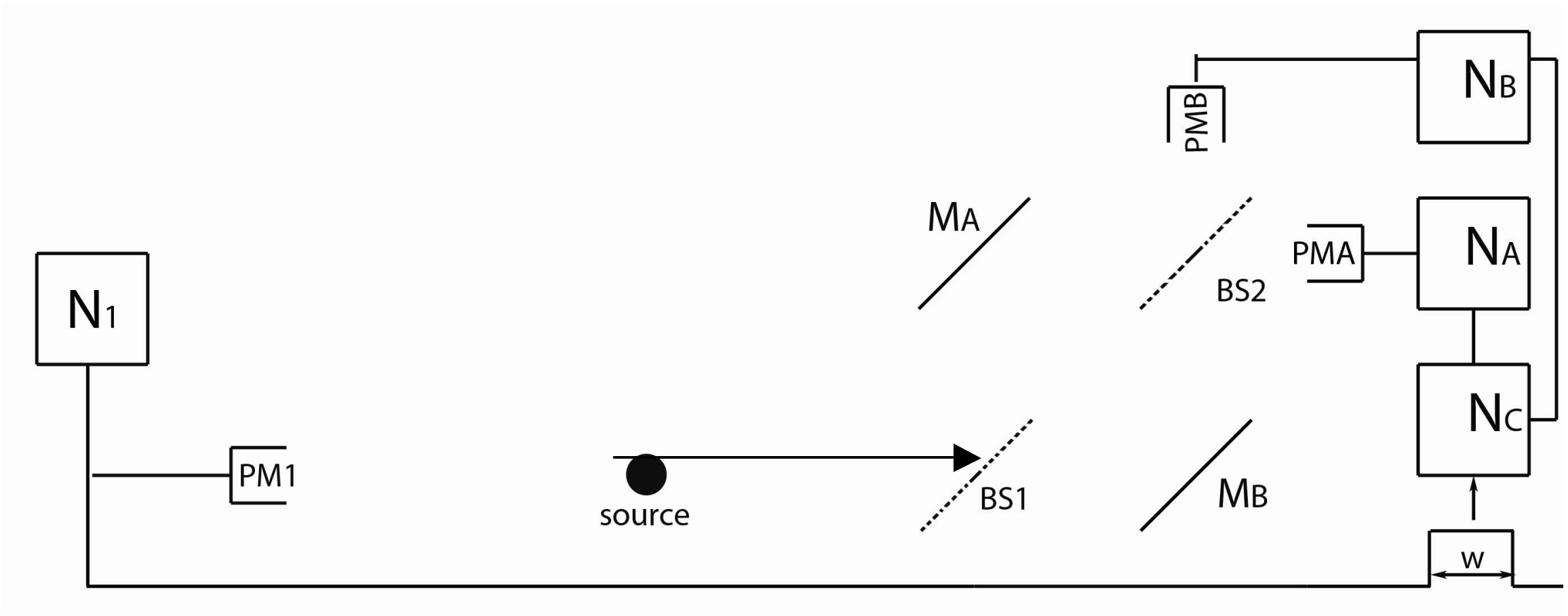

## EXPERIMENT Y

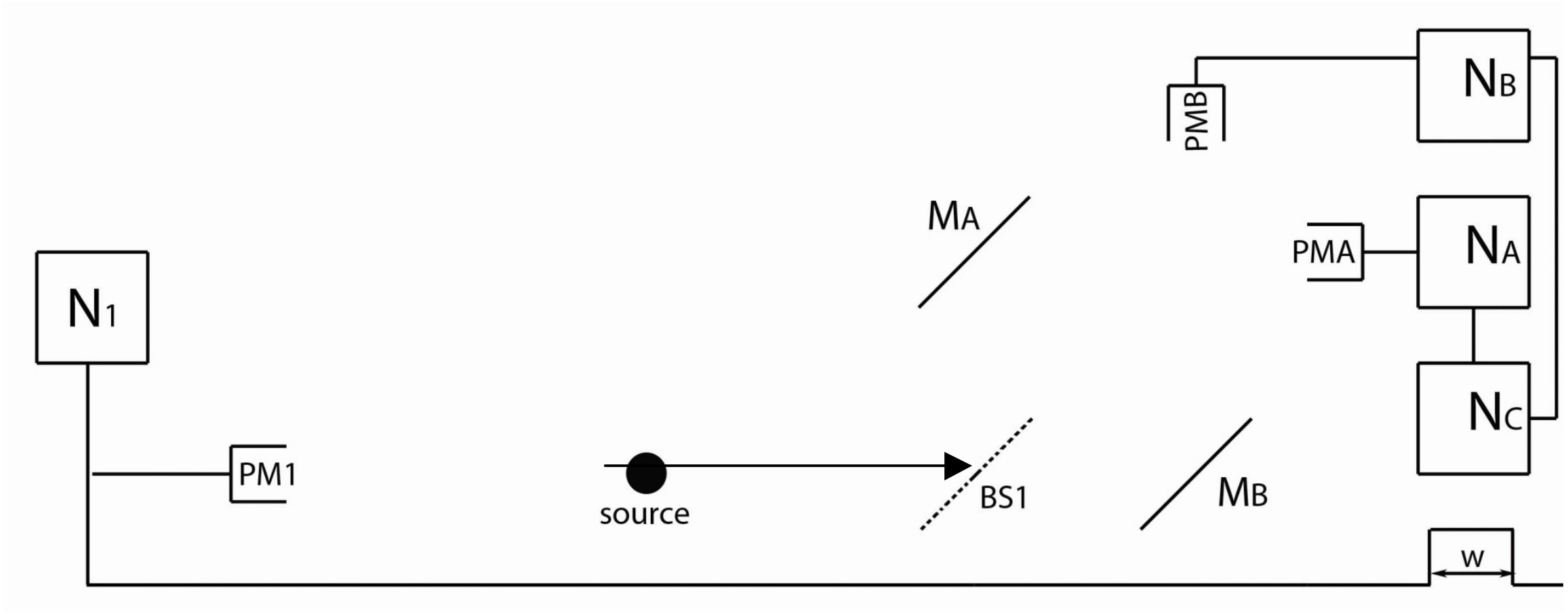

**E3.A (3 Points)** For which experimental setup (X or Y) would you expect photons to exhibit particle-like behavior? Describe in what sense the photon is behaving like a particle during this experiment. What features of the experimental setup allow you to draw this conclusion without actually conducting the experiment?

In setup Y, the photons exhibit particle like behavior because the photon can only have one path to get to a particular photomultiplier. I know this because beam splitter 1 will either allow the photon through or reflect it. If it reflects it will go to PMA, if it is let through it will go to PMB. It can't take Path A to get to PMB thus since there is one path to take, it acts as a particle.

**E3.B (3 Points)** For which experimental setup (X or Y) would you expect photons to exhibit wave-like behavior? Describe in what sense the photon is behaving like a wave during this experiment. What features of the experimental setup allow you to draw this conclusion without actually conducting the experiment?

In setup X the photons exhibit wavelike behavior because the photon can take either Path A or Path B and still get to PMA or PMB, we don't know which path it took, thus since it unpredictable, it acts as a wave. Since it can take either path and still get to either photomultiplier, I know it can be represented as a wave.

**E3.C (2 Points)** Suppose we are conducting Experiment X (the second beam splitter (BS2) is present) when a photon enters the apparatus and encounters the first beam splitter (BS1). Afterwards, while the photon is still travelling through the apparatus (but before it encounters a detector), we suddenly remove the second beam splitter (switch to Experiment Y). Can we determine the probability for the photon to be detected in PMA? If not, why not? If so, what would be that probability? Explain your reasoning.

No, we could not because we don't know which path the photon took, it could have taken Path A in which it would be detected by photomultiplier A or it could have taken Path B and not been detected by PMA. Since it has not been detected yet we can't determine the probability, it's already on a definite path.

**Exam 3**

**E1. (10 Points Total)** A hydrogen atom is in its lowest energy state. Use words, graphs, and diagrams to describe the structure of the Hydrogen atom **in its lowest energy state (ground state)**. Include in your description:

- **(4 Points)** At least two ideas important to any accurate description of a hydrogen atom.
- **(3 Points)** An electron energy level diagram of this atom, including numerical values for the first few energy levels, and indicating the level that the electron is in when it is in its ground state.
- **(3 Points)** A diagram illustrating how to accurately think about the distance of the electron from the nucleus for this atom.

(On these diagrams, be quantitative where possible. Label the axes and include any specific information that can help to characterize hydrogen and its electron in this ground state.)

## [EXPLICITLY AND EXCLUSIVELY USES BOHR MODEL]

## Post-Instruction Survey:

**1.** It is possible for physicists to carefully perform the same measurement and get two very different results that are both correct.

**(Disagree)** Take the example of hidden variables. If you put one red sock and one blue sock into identical boxes and both socks are identical beside their color, and you send them across the universe, then your technically performing the same measurement. When you open one box you find out what color the sock is in that box and it can be either red or blue, two different results. At the same time you also know what is in the other box every time you perform the experiment, in that respect, you are kinda getting the same result.

**2.** The probabilistic nature of quantum mechanics is mostly due to physical limitations of our measurement instruments.

**(Strongly Agree)** The probabilistic nature of quantum mechanics comes from the fact that there are aspects of quantum mechanics that can't be measured due to physical limitations of our measurement instruments. For instance how the uncertainty principle interacts with electrons orbiting a nucleus. Electrons are too small and move too fast for humans to know exactly where an electron is at a certain moment, so we can only perform one measurement at a time. Position and momentum of a particle can't be known at the same time, we can only calculate the probability of finding them there.

**3.** When not being observed, an electron in an atom still exists at a definite (but unknown) position at each moment in time.

**(Strongly Agree)** Every physical thing exists whether it is being observed or not. This is the idea of realism, and I completely agree with it. An electron is a particle therefore I believe that it has a physical manifestation. An electron will definitely still exist at a definite position at every moment in time. This correlates with my answer above.

**4.** I think quantum mechanics is an interesting subject.

**(Strongly Agree)** I found quantum mechanics to be an interesting subject because the concepts around it are not proven. A lot of what is behind quantum mechanics is qualitative which is very different than most physics classes which are quantitative. It is nice to look at a complex subject such as physics from a qualitative manner because for the past two years I've been taking all engineering classes which are all involving math significantly.

## Double-Slit Essay Question:

I agree with Student 1 mostly except for the fact that the electron could be going through both slits at the same time for all we know. I also agree with student 2 because I think that the electron is acting as a wave and again possibly go through both slits at the same time. Therefore I agree more with student 3 because we really don't know what is happening between the moment the electron is shot from the gun and it hits the detection screen.

**Final Essay:**

Although some people think that quantum mechanics is a very boring and difficult subject, I found it to be a very interesting topic to learn about. As an engineer, most of what I study ends up being taught primarily with numbers therefore a lot of the work I have done in the past couple of years has been computational. Honestly, modern physics and quantum mechanics was a very nice break from all of the number crunching because there was a lot more theory involved with understanding topics that are covered in this class. As Carl Wieman says, "Quantum mechanics is the greatest intellectual accomplishment of human race." From what I have learned in this class that quantum mechanics is used in about a third of all the engineering that surrounds us, I would definitely have to agree with Carl that it is a very important achievement by humanity. Since quantum mechanics is so important to how people live their lives, I found myself being able to engage with the class a lot better than most of my classes. A big motivation for me taking this class was that in the past, I have taken primarily math classes, and physics classes and because of all of this background knowledge of these sciences, I thought that physics 3 would be an interesting, yet difficult class to take. Another motivation for me taking this class was that my sister took this class 4 years ago, and she encouraged me to take it so that we can discuss physics not only on the classical level, but on the quantum level as well. My last major motivating factor that enticed me into taking this course is that it is a required course for electrical engineering, which happens to be my major. I guess that that is enough of a motivation to take a class in itself! Coming into this course, i really didn't know much about quantum mechanics at all. All I knew about quantum mechanics was that it had something to do with the study of how light and matter interact with each other. So, the first couple of weeks of this class were exiting because I was learning qualitative things at a rapid rate and i felt like the knowledge that i was gaining was sticking pretty well. Initially, I had a lot of questions about how things differ classically from quantum mechanically, and about the differences between particle-like and wave-like behaviors exhibited a photon. But, over the course of the semester, I felt like I ended up grasping these concepts pretty well.

This class has changed my perception of science and has definitely changed my ideas about physics drastically. Like most people, I have always perceived physics quantitatively, but quantum physics is not so much about that I found out, minus Schrodinger's equation of course.

Rather, most of the topics we covered involving quantum mechanics was primarily conceptual. On top of that, because these topics are so conceptual, most of the ideas behind quantum mechanics are extremely difficult to prove without doing in depth research about a particular topic. Prior to this class I thought of subjects that had to do with concepts that are proven and declared as theorems and laws as sciences, but now that I am almost finished learning about modern physics, I obviously know how theories can be scientific and can lead to very significant legitimate finding. Some of the major topics that we covered in this course that changed my views on physics include the concept of entanglement, hidden variables, locality, realism, and completeness. All of these topics are purely conceptual, yet they explain different aspects of physics not necessarily logically, but in a manner that is comprehensible to human beings. For example, the idea from class that if you but a blue sock and a red sock into different identical boxes and mix them up and send them across the universe, the sock in each box can be looked at as red and blue at the same time from a quantum mechanical point of view, until the true identity of one of the socks is reveled, then we know the true color of both socks. This concept was very intriguing to me, because it defies all of what my brain logically tells me, that the sock is always red **or** blue, but cannot be both. I am glad that i had this change in perception of physics because it lets me think outside the box a little bit better.

The most interesting topics to me were topics that I could see physical applications of. For example, I really enjoyed the sections on lasers because I knew what a laser was, (now I actually know!) before we covered the material, so I found myself paying extra attention to those topics so that i could learn more about them. I really liked the process of how we learned about lasers because we simplified everything down to the requirements that are needed to make a laser work, such as a method of recycling photons(like mirrors), stimulated emission of photons(more atoms in an upper energy level than a lower one, also known as population inversion), and same color of photons that are in phase with each other and go in the same direction. I also really enjoyed the section where we learned about tunneling. The concept that a particle can actually go through a potential barrier was extremely interesting to me. I thought that the long homework problem that related to tunneling was a very cool topic to explore; Even though I know its not actually possible because quantum mechanical concepts like tunneling only apply to microscopic particles, I thought that it was fun to think about something so abstract as a person being able to tunnel through a door. Even if its probability is so incredibly close to zero, we were actually able to put a number to the probability using the tunneling equation. I also thought that the history behind quantum mechanics was fun to learn about. Its crazy how fast humans progress in sciences nowadays. It started out with the Thomson model of an atom which said that the mass of an atom was uniformly distributed through out the atom, then it went to Rutherford's solar system model which said that the majority of the atom is in the middle (nucleus), next it went to Bohr's model of the atom, followed by DeBroglie's model, and then finally Schrodinger's model of the atom(Leon). Schrodinger's model put the concepts together of all of these other models and made a model that actually worked for all cases that a non man-made atom can endure. One last thing that i found pretty interesting were the quotes by famous scientists and mathematicians that were posted on the lecture slides. A lot of those slides were either really informative, or they were hilarious! For example, the Dr. Evil quote about sharks with "freakin" laser beams attached to their heads was a very appropriate quote to put in the laser slides. These are just a few of the topics that we covered in this course that i found interesting, I am also really enjoying what we are learning about as im writing this paper which is nuclear weapons. Pretty much, the most interesting topics to me that we covered were topics that involved applications of quantum mechanics.

Even though there were not too many topics that I found to be uninteresting, there are a few that I wish that we spent less time on. For instance, I thought that we got into the topics aver model vs. theory vs. interpretation etc. were a little bit boring to learn about for more than a few minutes. Also, I didn't really enjoy learning about Stern-Gerlach analyzers because I thought that they were a bit confusing, and i didn't really see the point to learning about them. While learning about Stern-Gerlach analyzers, we did not really discuss anything about the applications of them in depth. We learned how they were involved with entanglement, how one outcome will effect all of the other outcomes, but I really wanted to see a more physical application. Later on I came to found out that Stern-Gerlach analyzers have applications involving things like chemistry. I took the time read a little bit into how Stern-Gerlach deflection patterns are used in supersonic beams, and even though, I found this to be interesting, it was incredibly difficult to comprehend (C Weiser). I think i would have been more involved in learning about this topic we talked about the applications of the Stern-Gerlach analyzer before we learned about how they work. Another topic that I found uninteresting was the math behind the proof of Schrodinger's equation. I found this proof of it to be very confusing, and I lost interest in learning about the the equation very fast, but im glad that we didn't spend as much time on it as past modern physics classes at CU Boulder did. Although there were a few topic that I didn't enjoy too much or that I thought we spent too much time on, there were a lot more topics that I did enjoy.

I think that the structure of this course was designed very well. I think that how this class is set up allows students to access the tools and information thus helping them to succeed in this course. The website for this class is probably the most useful website that I have had throughout college. It was extremely helpful to have all of the lecture slides posted on there for reference, and it was nice to have all the simulations that our teachers provided us to make the learning process a little bit easier, and more enjoyable. Also, I found the help room to be very helpful in allowing students who want to do well on their homework, and who seek further explanation about topics. Without the help room, I would have been a lot more confused about a majority of the topics that we covered in class. On top of that, I found that the clicker questions were helpful as well. The clicker questions keep students who go to lecture but have trouble staying engaged in lecture on their toes and it keeps them alert. I would know, im one of those students... Another good teaching technique that was used in this class was giving out candy to student who interact with the class. Not only does the sugar stimulate our brains for a few minutes, but the candy also acts as a pretty good motivating factor for student to speak their mind. The teaching techniques used in this course were very beneficial and were appreciated greatly.

Overall, I was very satisfied by this class. I feel that I have learned a lot more about physics, and I have developed a different perception about how science works. All of the questions that I had coming into this course were addressed in the first few weeks, and from there on out it was fun to learn about physics in a quantum manner. The motivations that I had for taking this course were addressed as well, as long as I pass this class, which im pretty positive will happen. Learning about quantum mechanics was very interesting, although I wish that we covered more applications of quantum mechanics.

## Student B

### Pre-Instruction Survey:

**1.** It is possible for physicists to carefully perform the same measurement and get two very different results that are both correct.

**(Agree)** I don't know of any examples, but the fact that quantum physics has some things that seem counter-intuitive and contradict classical physics, it seems that this could be a possibility.

**2.** The probabilistic nature of quantum mechanics is mostly due to physical limitations of our measurement instruments.

**(Strongly Agree)** I believe that in the future, we would be able to make more accurate and exact assertions due to technological advances and would not need to rely on probability.

**3.** When not being observed, an electron in an atom still exists at a definite (but unknown) position at each moment in time.

**(Strongly Agree)** An electron is a particle, and every particle has a definite position at each moment in time.

**4.** I think quantum mechanics is an interesting subject.

**(Strongly Agree)** I think that I'm going to learn that what I would think is correct is actually completely incorrect. Plus, it just sounds cool.

**5.** I have heard about quantum mechanics through popular venues (books, films, websites, etc...)

**(Strongly Disagree)** I'm completely out of the "physics loop" and hope to get more into it in this class!

**Homework Problems**

**HW01:**

**13.** What are atoms made of, and how are the parts of the atom configured?

Atoms are made up of protons, electrons and neutrons. There is a dense nucleus in the center of the atom, which contains positively charged protons and neutrally charged neutrons. The nucleus is surrounded by negatively charged electrons.

**21.** The force (F) experienced by a charge (q) in an electric field (E) is given by the equation:

$$\vec{F} = q \cdot \vec{E}$$

In fact, the electric field at any point in space is ***defined*** in terms of the force that would be experienced by a charge placed at that point in space:

$\vec{E} \equiv \dfrac{\vec{F}}{q}$ (Electric field defined as force per unit charge)

How do you (personally) ***interpret*** the concept of an electric field? Is the electric field something that is physically real, even though we can't observe it directly? Or is it a mathematical tool devised by physicists to explain our observations and to make calculations easier? A mix of both, or is it something else entirely? Remember, we are interested in what you actually think – there are no "correct" or "incorrect" answers to this question. Please provide the reasoning behind your response.

I believe that the electric field exists, just as I believe that gravity exists. Just as gravity can cause something to fall, the electric field can exert a force onto a charge. The definition of the electric field at any point in space is the force that would be experienced by a charge placed at that point in space. Similarly, the definition of gravity is the force that something would feel towards the center of the earth at a particular point in space, and just as the force of gravity varies with location, so does the electric field.

**HW02:**

**12.** It is said that the photoelectric effect demonstrates the particle-like nature of light. Explain how this conclusion is reached. That is, what *experimental evidence* is consistent with particle-like behavior for light but not with wave-like behavior? Cite at least two pieces of evidence.

1. Energy depends on color (frequency)
2. Electrons come out of metal immediately (there is no time delay to heat up).

Both of these are contrary to the classical predictions that would be correct if light were just a wave. However, these two realities show that light has particle-like behavior.

**13.** In explaining the observations from the photoelectric effect experiment, did Einstein propose a new theory, a new model, or a new interpretation of the data? In your mind, are there any differences between the three terms (theory, model, interpretation)? If so, what would they be? If not, why would it be OK to use the three terms interchangeably?

Remember, for these types of questions, we are interested in having you express your opinion, so there are no "right" or "wrong" answers. Your response will be graded on the effort you put into it.

I believe that Einstein did not propose a new theory; rather, he dwelled on a previous theory that was not complete. He interpreted the previously available data and came up with explanations for it. This is still a very big achievement, but the original idea did not come from him.

I believe a theory is the first time an idea comes up in somebody's mind. I'm not too sure what a model is. Interpretation is one's belief as to what that theory means, or why it is a reality/false. This is the part that can be different for each person.

**HW03:**

**3.** What was the point of the "Farmer and the Seeds" story (from lecture)?

The purpose of this story was to show us how theories come to exist, and what can be done after a theory is proposed. People can come up with many different theories to explain a certain observation. Then, through experiments, some of those theories are refuted and eliminated. Finally, the results of the experiments can be interpreted differently by different people as well. In general, coming up with a correct theory (explanation for an observation) is not all as obvious as one may think.

**4.** Review the discussion of the terms *theory, model & interpretation* in the solution set from HW02. How would these terms, as **you** understand them, apply to the different elements of the "Farmer and the Seeds" story, and to the schemes you came up with in class? Remember, we are interested in what **you** think. There are no "right" or "wrong" answers to this question.

A theory is an assertion based on observation. Basically, an answer to a question that arises based on observation or experiment. A model is some sort of representation of a procedure or experiment. Interpretation is an explanation of why the theory works. The interpretation involves a lot of analysis from the experiment.

**<u>HW06:</u>**

## Questions 16 – 18 refer to the reading "100 Years of Quantum Mysteries".

**16.** As discussed in this article, what were some of the problems in classical physics that led to the development of quantum theory?

One of the problems was Rutherford's atom model didn't explain why the electrons of an atom don't radiate away here energy and spiral into the nucleus in about a trillionth of a second. Of course this was observed to not happen, but classical physics wasn't able to explain why this is so.

Another example of what classical physics couldn't explain is that according to classical physics, ultraviolet radiation and x-rays should blind you when you look at the heating elements of a stove. This is obviously not the case, but classical physics had no explanation as to why.

**17.** How are the terms *theory* and *interpretation* used in this article?

The word interpretation is used once to explain the Copenhagen interpretation of quantum physics, which is their interpretation of the superposition theory, in order to explain the difference between the theory that atoms exist in superposition in different locations, and the reality that we observe them, which is in only one state. The theory was something they believed in, so they had to interpret it in some way to make it hold, without disproving their observations. Their interpretation was: the act of observing the card (or atom), triggers an abrupt change in its wave function, and from then onward only that part of the wave function survives. This was an answer that is satisfying to some, because at least it explains why we observe the reality that we observe, and it is based on the theory holding as true.

Based on this interpretation, however, others interpreted that the theory is fundamentally wrong, because there was no definite equation to explain the outcomes observed.

Another form of an interpretation is by Hugh Everett, who took the idea of superposition to its extreme as he interpreted that based on the theory of something being able to exist in two superpositions simultaneously, this would also pertain to the world in which we live and that there is a wave function that describes the actions of the universe. But this would also imply that there are what was called "many-worlds" as each component of one's superposition perceives its own world simultaneously.

Therefore, just as we discussed before, a theory is an attempt at explaining a certain observation (or a number of observations), and an interpretation is what tries to explain why that theory is correct.

**18.** Is there any experimental evidence in favor of any of the interpretations discussed in this article? In the cases where there is not, why would a scientist favor one interpretation over another?

The experimental evidence that once we observe an atom, it is no longer in two superpositions and thus collapses into one, is observed. Whether that is the correct interpretation or not (that the reason for this is the observance), we can't be 100% sure, but it does provide an explanation that at least does not refute our observation.

However, the interpretation that everything does indeed exist in both superposition states simultaneously, well, this is less tangible based on our observations. The reason a scientist would favor this interpretation might be that it provides an explanation that is practically impossible to test, and so may be a better philosophical reason to want to accept based on faith than an abstract idea that was tested, but just so bizarre in comparison with what physics was known to be for that particular scientist. In other words, perhaps the latter interpretation explains everything about the theory, but there is really no way to prove it, whereas the former interpretation only explains a particular observation, and is consistent with what we observe, but perhaps it is less general, but more tangible.

**19. This question refers to the reading assignment "Probability"**

According to this article, in what way(s) is quantum mechanics a probabilistic theory?

Quantum mechanics is a probabilistic theory in that it does not measure a definite position of an electron, but rather provides probabilities of having an electron in a particular position. As stated in the article: "one cannot predict where the electron will be – one can only compute the probability that it will of will not be at a given place or places."

**HW07:**

**Questions 9 & 10 refer to the readings: "A Quantum Threat to Special Relativity" & "Is the moon there when nobody looks?"**

**9.** What is meant by the terms ***realism, locality*** & ***completeness***? What are some examples of hidden variables?

Realism is a property in which every measurable quantity exists. In other words, everything is definite, and there is no superposition. The only thing that keeps us from knowing what all the quantities are is our ignorance.

Completeness refers to a theory that can describe everything without leaving anything unknown. By this definition, quantum physics is not complete because when we measure a certain quantity such as the projection of the atom in the Z direction, then we can't know its projection in the X direction.

Locality is the concept of being able to relate all actions to actions that occurred before them. For example, locality can describe a car accident – all the events that lead up to the car accident are clear and relate to one another. Bohr's interpretation of entanglement is not local, because we have no way of explaining how the observation of one atom collapses the wave such that the other atom (which would be miles apart) instantaneously is affected.

**10.** Does ***entanglement*** allow for faster-than-light communication? If so, what kind of information can be communicated? If not, why not?

No, entanglement does not allow for faster-than-light communication. This is because the whole concept of entanglement is based on the fact that the atom pairs start in an indefinite state, and only become a definite state once they are observed. The purpose of communication is to get a meaningful thought from one person to another. If I already know what I want to communicate, then it is not in an indefinite state, so entanglement does not apply. Even if we were to say that I'll create a set of random characters to communicate with someone else, this is the best we can do at mocking entanglement. But still, in that case, the only way to actually communicate those random characters would be for the other person to see them, which would take longer time (to reach the other person) than the speed of light. One thing to keep in mind is that the two people haven't coordinated beforehand, otherwise, it's not counted as communication. The atom pair seems to somehow be able to coordinate beforehand or always be in a "coordinated state" at any instance in time, so the concept, as we know it, of communication, does not apply to the atoms, and hence, we cannot take the property of entanglement and use it in communication.

**11.** In the two Aspect experiments discussed in class, where the goal was to produce a "single-photon" source, the calcium atoms were excited to the upper level by a two-photon absorption process. Why did the experimenters excite the calcium atoms with a laser pulse of 3.05 eV photons followed by a pulse of 2.13 eV photons, rather than with single photons with the same energy as the two original photons combined (single-photon excitation)?

The only way we can conduct the experiment in a way that would provide accurate/meaningful results is such that we have only one photon in the apparatus at a time. If we had used single-photon excitation, in that case, many atoms would get excited and give off photons simultaneously. This would make it hard to see how each individual photon acts. However, if we make it very unlikely that an atom get excited (absorb a photon), then only very few photons will actually be ejected from the excited atoms at the desired energy level (2.25 eV). The best way we can make it very improbable that the atom will get excited to the 5.18 eV state is to depend on a unlikely event as double-excitation. This way, we can have a better chance that there will only be one photon at a time in the apparatus.

**12.** In your own words, explain what the anti-correlation parameter ($\boldsymbol{\alpha}$) is, both in terms of its mathematical definition, and in terms of what it physically tells us, in the context of single-photon experiments as performed by Aspect. Why didn't Aspect measure $\boldsymbol{\alpha} = 0$ if photons are supposed to be acting like particles?

Mathematically, alpha is the probability that both detectors are triggered for a specific photon (a coincidence), divided by the product of the probabilities of each detector to be triggered separately. Physically, it tells us whether or not a coincidence is more likely than not. In the event of a coincidence, the conclusion is that the photon acted like a wave. In the event of each detector being triggered at separate times, the photons act like particles. What this means is that if alpha is greater than one, and if photons act like particles, then alpha is between zero and one. The reason Aspect never measured alpha=0 is because it was never practical to have only one photon at a time in the experiment, and thus have that one photon act like a particle, and then have the numerator of the definition equal to zero. However, if we extrapolate theoretically, we see that alpha should be zero.

**(Essay)** In class we have discussed correlated measurements performed on systems of two entangled atoms. The assigned reading "The Reality of the Quantum World" (available on CU Learn) discusses correlated measurements performed on entangled photon pairs. In what ways are these systems of entangled photon pairs similar or different from systems of entangled atom pairs? In what sense are the particles in each system entangled (i.e., what properties are correlated for each of the two types of systems)? What types of measurements are performed to determine these properties, and what are probabilities for the possible results of these measurements for both types of systems?

They were different in that entangled photon pairs have different properties to measure, and thus require different apparatuses to properly test the theories. In other words, we had to use different experiments because they each have different superposition concepts. With the atom pairs, a Stern-Gerlach analyzer was used because the main correlated property that is associated with atoms is the atomic spin. For the photons, however, we want to know whether the photon behaves like a particle or a wave. The similarity of the two systems of entangled pairs is that in both systems, there is some form of "communication" between the two items. There is some way that one atom is "aware" of the state of the other, and the same for photons. Thus, they both deal with entanglement, but just in different ways.

In the systems of entangled atom pairs, the correlated property between the two atoms was the projection of the atom on a given axis, depending on its angular momentum (atomic spin). The correlated property for the photon pairs is the polarization state of the photon. Though different properties, they each had strictly correlated outcomes- for the atom pair, if one atom had a positive projection on the z-axis, the other had a corresponding negative projection on the z-axis. For the photons, both protons would have the same linear-polarization. Either both photons would pass through the polarized film or neither would.

Measurements that were performed on the atom pairs was the atomic spin, and measurements that were performed on the photon pairs were the paths taken in the apparatus. For the photon experiment, an important factor was the time it took for the photon to start its journey in the apparatus and reach the switch for the different detectors. Basically, they had to turn on the switch after the photon had already started its movement in the apparatus. This was determined by very precise time measurements for the switches.

As for the probabilities, when it comes to atoms, there is a 50-50 (equal) probability for an atom to have a positive or negative orientation on a measured axis. For the photons, with the polarized plates, there was a 50-50 chance for the photon to be in a horizontal state or a vertical state. We notice that whatever the possible states of atoms or photons may be, they must correlate.

**(Essay)** As discussed in class and in the readings, what do the two single-photon experiments performed by Aspect tell us about the nature of photons? How were the two experiments designed to demonstrate the particle and the wave nature of photons? When answering, don't concern yourself with technical details (such as how the photons were produced); focus instead on how the design of each experimental setup determined which type of photon behavior would be observed. How are the elements of these two experiments combined in a delayed-choice experiment, and what do delayed-choice experiments (along with the two Aspect experiments) tell us about the nature of photons?

The two single-photon experiments performed by Aspect tells us that photons sometimes act like particles and sometimes act like waves (and interfere with themselves), but cannot be both a particle and a wave simultaneously. The experiments were designed so that in the first experiment, the photon could only take one path at a time, thus only acting like a particle. The second experiment, however, added an extra beam splitter. This experiment was designed in such a way that when the splitter was fixed in a certain position, there was no way of knowing which path the photon takes. But when we slowly change one of the path lengths by moving one of the beam splitters, interference is observed. Thus, this experiment was designed so we could see wave-like interference as well as just particle-like behavior. The very intriguing result of this experiment is that each photon somehow is "aware" of both paths. So, assuming that the photon can somehow sense that we are conducting experiment one as opposed to experiment two, just to prove whether or not that is the case lead to the third experiment. In other words, if the photon was aware of both paths all along, perhaps it's because it was given a "sneak preview" of the apparatus before being emitted. So in order to fully understand the nature of photons, they had to somehow come up with a way to "trick" the photon into now being able to know or predict the paths that it could possibly take. That's when the delayed-choice element was added to the apparatus. The idea here was to let a photon be emitted and start moving through the apparatus, and then change the path it can take, so it would be caught "off-guard" if it had a preconceived idea about the apparatus before being emitted. All these experiments can combine in one concept- that photons can be waves or particles. Also, there is some kind of phenomena that exists such that the photon is "aware" of both paths. What this means is that maybe the photon is actually in both paths simultaneously (superposition), but collapses to one position when it reaches the detector.

**<u>HW09:</u>**

**3.** How does thinking of electrons as waves rather than particles explain why energy levels in the hydrogen atom are quantized? How does it answer Bohr's question of why electrons don't radiate energy when they are in one of these energy levels?

**[Not answered.]**

**Exam Questions:**

**Exam 2**

**E1. (10 Points)** In the sequence of screenshots shown below (taken from the PhET Quantum Wave Interference simulation), we see: A) a bright spot (representing the probability density for a single electron) emerges from an electron gun; B) passes through both slits; and C) a single electron is detected on the far screen. After many electrons pass through and are detected, a fringe pattern develops (not shown).

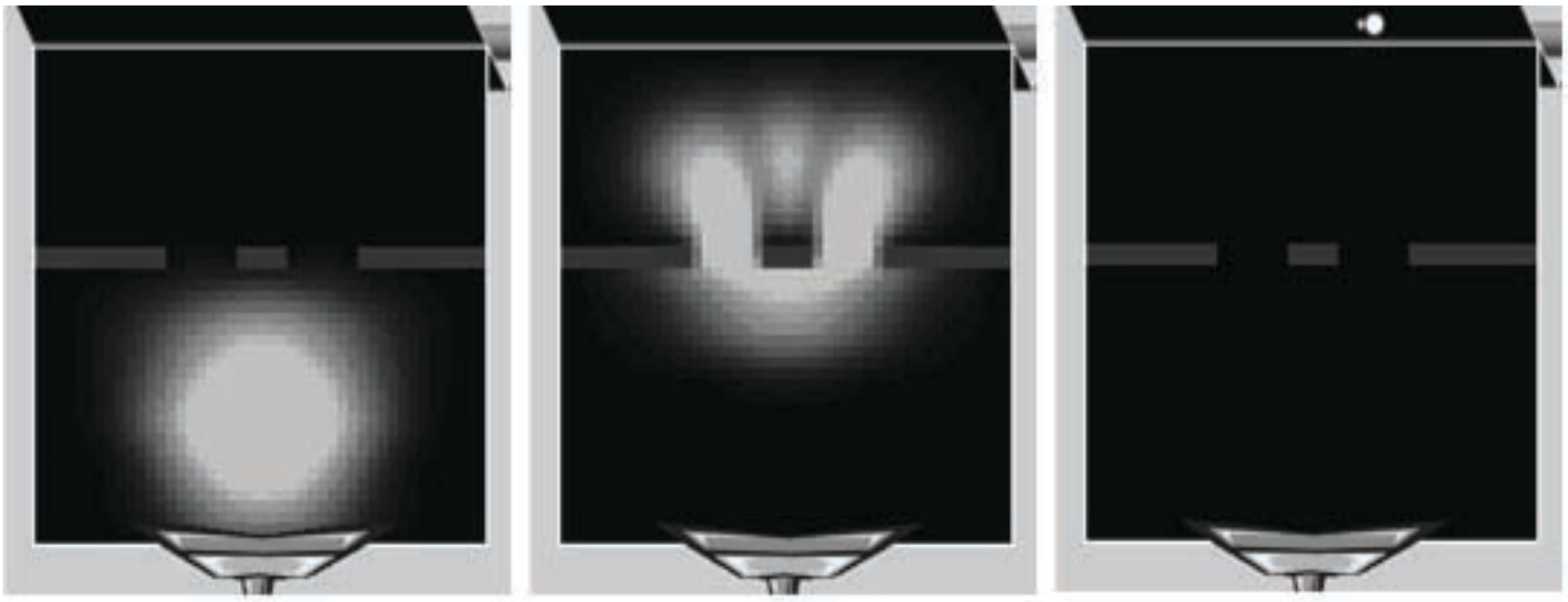

(A) (B) (C)

**Three students discuss the Quantum Wave Interference simulation:**

**Student 1**: The probability density is so large because we don't know the true position of the electron. Since only a single dot at a time appears on the detecting screen, the electron must have been a tiny particle traveling somewhere inside that blob, so that the electron went through one slit or the other on its way to the point where it was detected.

**Student 2**: The blob represents the electron itself, since a free electron is properly described by a wave packet. The electron acts as a wave and will go through both slits and interfere with itself. That's why a distinct interference pattern will show up on the screen after shooting many electrons.

**Student 3**: All we can really know is the probability for where the electron will be detected. Quantum mechanics may predict the likelihood for a measurement outcome, but it really doesn't tell us what the electron is doing between being emitted from the gun and being detected at the screen.

**E1.A (2 Points)** In terms of the interpretations of quantum phenomena we've discussed in class, how would you characterize the perspective represented by Student 1's statement? What assumptions are being made by Student 1 that allows you to identify their perspective on this double-slit experiment?

Student One believes that the electron is indeed just a particle the whole time, but is moving around so fast in a random way that we can't detect it. He does not believe in wave-particle duality of electrons. He does believe that there are hidden variables (i.e., position). He also does not believe that there is a superposition. Overall, he has a realist point of view that the electron has a specific path but we just don't know it.

**E1.B (6 Points)** For each of the first two statements (made by Students 1 & 2), what rationale or evidence (experimental or otherwise, if any) exists that favors or refutes these two points of view? As for the third statement, is Student 3 saying that Students 1 & 2 are wrong? Why would a practicing physicist choose to agree or disagree with Student 3?

Since Student One believes that the electron was traveling within the blob and went through only one slit, he believes that electrons act as particles. This would mean that he would never observe interference. This is not true though because the experiment shows that over a long time, interference is observed. (Even the nickel atoms in a crystal lattice experiment shows this too.) Since Student 2 believes that the electron acts as a wave packet, he suggests that we have a small uncertainty in its position (and large uncertainty in its momentum). However, if we had a small uncertainty in its position, then we could later predict where it would show up on the screen. The double-slit experiment shows this. In other words, the blob doesn't represent the electron, but rather the probability density of the electron to be detected. Experiments show that we don't really know what the electron is doing before we detect it. Student 3 is indeed disagreeing with Students 1 & 2 by saying that Students 1 & 2 can't make some of their claims, as we really just can't tell what the electron is doing between being emitted from the gun and being detected on the screen. He might not be stating that Students 1 & 2 are necessarily wrong, but he says that quantum mechanics can't conclude their conclusions. A practicing physicist would most likely agree with Student 3 because it is consistent with the Aspect experiment for photons as well as probability densities (quantum).

**E1.C (2 Points)** Which student(s) (if any) do you ***personally*** agree with? If you have a different interpretation of what is happening in this experiment, then say what that is. Would it be reasonable or not to agree with ***both*** Student 1 & Student 2? This question is about your personal beliefs, and so there is no "correct" or "incorrect" answer, but you will be graded on making a reasonable effort in explaining why you believe what you do.

I personally believe that the electron acts like a wave until we observe it. This is Dirac's interpretation. Student 1 & Student 2 can't both be right because that would suggest that the electron acts like a wave and particle at the same time, and there is experimental evidence that refutes this.

**E3. (8 POINTS TOTAL)** For the diagrams below depicting Experiments X & Y, M = Mirror, BS=Beam Splitter, PM = Photomultiplier, N = Counter. In each experiment a single-photon source sends photons to the right through the apparatus one at a time.

## EXPERIMENT X

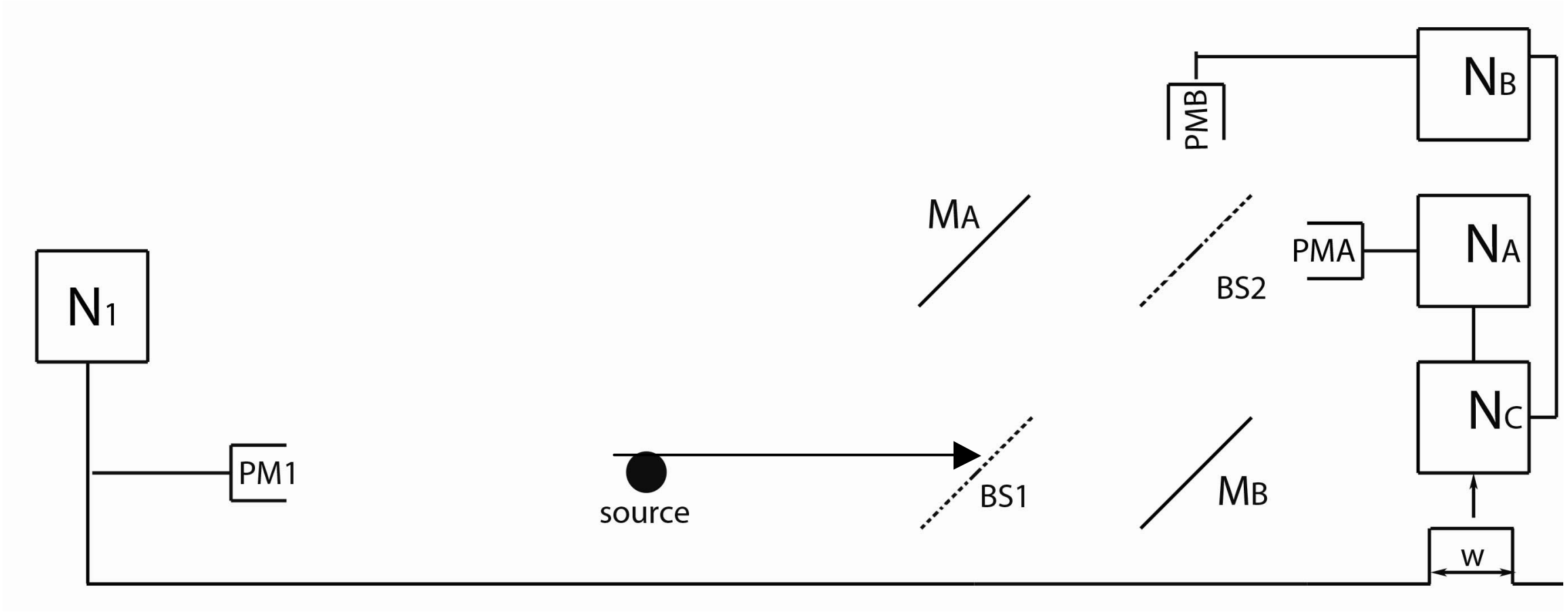

## EXPERIMENT Y

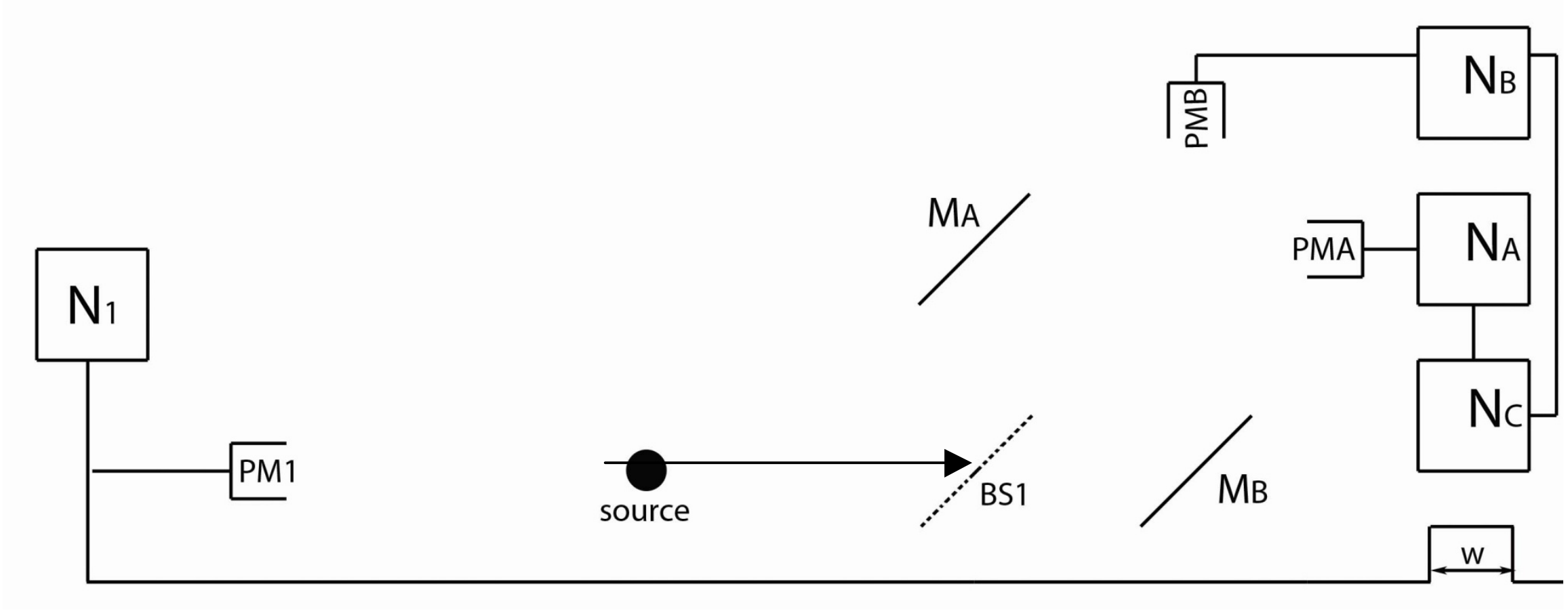

**E3.A (3 Points)** For which experimental setup (X or Y) would you expect photons to exhibit particle-like behavior? Describe in what sense the photon is behaving like a particle during this experiment. What features of the experimental setup allow you to draw this conclusion without actually conducting the experiment?

Particle-like behavior expected in setup Y. Photon's path is predictable depending on the detector in which it was detected. It either gets reflected or transmitted at BS1, thus if detected at PMA, it must have been reflected and if detected at PMB it must have been transmitted. We also know that alpha = zero if there is only one photon in the apparatus during the time constant. This implies that PC (interference) = 0 -> no wave-like behavior -> acts like a particle. There is only one beam splitter, so it will act like a particle (we know this even before conducting experiment).

**E3.B (3 Points)** For which experimental setup (X or Y) would you expect photons to exhibit wave-like behavior? Describe in what sense the photon is behaving like a wave during this experiment. What features of the experimental setup allow you to draw this conclusion without actually conducting the experiment?

Wave-like behavior expected in setup X. Photon behaves like a wave because there is interference if we change the path length (move BS2). Thus it seems to interfere with itself. In this experiment, we can't know which path the photon takes due to the existence of BS2 (it could be detected by either PM and have taken either path). We can also change BS2's location such that all photons are detected in PMA or PMB. Throughout the experiment, it seems that the photon somehow "knows" that there are both paths. The BS2 let's us conclude this before starting the experiment (that it can behave like a wave).

**E3.C (2 Points)** Suppose we are conducting Experiment X (the second beam splitter (BS2) is present) when a photon enters the apparatus and encounters the first beam splitter (BS1). Afterwards, while the photon is still travelling through the apparatus (but before it encounters a detector), we suddenly remove the second beam splitter (switch to Experiment Y). Can we determine the probability for the photon to be detected in PMA? If not, why not? If so, what would be that probability? Explain your reasoning.

This is the delayed-choice experiment. We can indeed predict the path that the photon took if BS2 is not present depending on the detector in which it was detected. Thus, the probability of being detected in PMA would be 0.5. It would act just as if we ran experiment Y and behave like a particle. Put the beam splitter back and it acts like a wave again. There is no "tricking" the photon!

**Exam 3**

**E1. (10 Points Total)** A hydrogen atom is in its lowest energy state. Use words, graphs, and diagrams to describe the structure of the Hydrogen atom **in its lowest energy state (ground state)**. Include in your description:

- **(4 Points)** At least two ideas important to any accurate description of a hydrogen atom.
- **(3 Points)** An electron energy level diagram of this atom, including numerical values for the first few energy levels, and indicating the level that the electron is in when it is in its ground state.
- **(3 Points)** A diagram illustrating how to accurately think about the distance of the electron from the nucleus for this atom.

(On these diagrams, be quantitative where possible. Label the axes and include any specific information that can help to characterize hydrogen and its electron in this ground state.)

**[IMPLICITLY AND EXCLUSIVELY USES SCHRODINGER MODEL]**

## Post-Instruction Survey

**1.** It is possible for physicists to carefully perform the same measurement and get two very different results that are both correct.

**(Strongly Agree)** This is possible especially when it comes to measuring the position of an electron. This is because there is no definite position to begin with. All we can know is the probability of finding the electron in a particular position, but probability does not determine where the electron will be when we measure it.

**2.** The probabilistic nature of quantum mechanics is mostly due to physical limitations of our measurement instruments.

**(Strongly Disagree)** It seems that the probabilistic nature of quantum mechanics is mostly due to the nature of sub-atomic particles rather than the limitations of our measurement instruments. If the particles were in definite states and definite positions to begin with, or even if there were a wave function that could define the exact state of the particles at any time, then one could argue that the problem is our measurement instruments. Perhaps such a formula will exist in the future, but that would mean that the limitation is our knowledge, not our instruments.

**3.** When not being observed, an electron in an atom still exists at a definite (but unknown) position at each moment in time.

**(Disagree)** This thought process only makes sense if one were to view electrons as particles (like billiard balls). However, we know from experimentation that the electron has wave-like properties and can be described in the form of an electron cloud (Schrodinger's model). Thus, we can have an idea of where we are likely to find the electron if we make a measurement, but when we don't make a measurement, the electron should not be acting like a particle. But then again, we can't be 100% sure of what's happening when we aren't measuring...

**4.** I think quantum mechanics is an interesting subject.

**(Strongly Agree)** The fact that there are truths associated with quantum mechanics that still can't be explained is a very interesting concept. I have never been taught something in school that is proven in experiments but still lacks a proper reasoning (such as entanglement). I also think it's very interesting to learn how sub-atomic particles behave so differently than macroscopic particles.

## Double-Slit Essay Question:

I agree with student three because it seems that the electron can act as a wave until we observe it. Even if this isn't the reality, there's nothing we can know about it from when the electron is emitted to when it is detected. However, student one and student two cannot be both correct because the electron cannot act like a wave (student 2) and a particle (student 1) at the same time, because there is experimental evidence that refutes this.

**Final Essay:**

Poem and Reflection Superposition Duality

So I was sitting in Larry's office, deciding on what classes to take,
Physics 3 or organic chemistry was a tough decision I had to make.
While I sat there considering my options, Larry spoke up to help me out,
You'll like the physics class, he says, assuring that I will without a doubt.
He tells me I'll enjoy the lectures as they resemble stand-up comedy,
And learn about modern physics with a hint of educational parody.
So what the heck is quantum physics? I had absolutely no idea!
But it seemed more attractive than the other option as I have a chemistry phobia!
So came the first day of the semester, and I walked into class,
To see a vibrant game-show host with a gift certificate to half-fast.
Seeing the life and energy, I knew that this class, I wouldn't be ditching,
Especially with a passionate professor who uses words like darn-tootin' and bitchin!
So of course I assumed I'd be learning stuff that would intuitively make sense,
But then I learned of core theories that were contradicted by experiments,
Like encountering situations in which it seems that photons straddle the fence,
Is it a wave or a particle? These abstract discussions made me a little bit tense,
As well as question all the physics I had previously been taught and hence,
I would doubt my intuition and maybe not want to always put in my two cents.
But as I reflect upon it all now, at the end of this course,
And recap everything we've learned from Electromagnetic waves to nuclear force,
I developed a new way of thinking, and of course there's no remorse,
That I took such a challenging class, but will admit that it was far from the worst.
I look back now on all the interesting topics that we learned about,
And the full spectrum of quantum physics we came across in our course route,
There are some things that were contradictory that left me dumbfound,
Starting out with light, how it's a wave but doesn't move up and down.
And how imaginary numbers are used to calculate real situations,
Is not a common practice we deal with in all that many occasions.
So I can finally accept that photons have wave-particle duality,
It was quite a fight but eventually, I was able to change my mentality.
But the fact that they exist, yet don't have any mass,
Whatever! I won't argue, I'll just accept it and pass.

And we’re allowed to say that Einstein was wrong? I thought that was sacrilegious!
But then again, on second thought, what isn’t when it comes to quantum physics??
Yes- Einstein was wrong- this calls for a song! Or rather a poem, cause I don’t sing,
Besides, he’s mostly famous for his hair- and sorry Noah, but that’s the most important thing!
Although it’s the truth, I should apologize, and I did not mean for that to hurt,
But who care’s anyways? You’re just as cool when you wear your Hawaiian shirt!
And even more impressive is the impeccable Dr. Evil impersonation,
Yes, after meeting Noah, I believe that physicists will rule the nation!
And he has god policies, like the cheat sheet- without it, this class I would resent,
For all the different entities, we’ve run out of letters to represent!
There’s wavelength, Lambda, Plank’s constant, h, and h-bar has a 2-pi,
S, p, d, and f for orbitals, and then a wave function denoted with psi!
Just thinking of all we have learned, really reminds of this field’s extent,
But enough about the general stuff, now let’s talk about more course content:

I must admit it’s fascinating that a photon gives all its energy to one electron,
And if that’s more than what the work function is, it escapes and then it’s gone!
I also learned about colors, and how they are associated with frequencies,
And how different atoms create different spectra, one such is Balmer’s series.
We then went into energy levels, and how electrons like to be in their lowest state,
But we can shoot them with photons and excite them, and with that in mind we can create,
A powerful concentration of focused energy, also known as a laser,
And we have mastered this art so much that with it we can even create razors,
It’s a tricky combination of population inversion, two mirrors, one less reflective,
But once you get a hang of it you can make a thinner beam that’s more effective,
It’s things like this that show us, without quantum our lives would be quite different,
From the words of professor Wieman, it’s our single biggest accomplishment!

Lots of double-slit experiments, with a wall at the end that would detect,
It was fascinating to see the results and then learn about the photoelectric effect.
Constructive and destructive interference is clearly a property of a wave,
But then the detection wall shows a position, just as a particle would behave.
These curious contradictions in the class are what always kept me alert,

And often I would get confused and to the lecture notes I'd have to revert.
Sometimes I'd forget what agrees with classical views and what's from quantization?
It took a bit to understand them, and then, more practice in memorization.
Contrary to classical belief, frequency is directly related to energy,
Especially when it comes to photos, Plank's constant and frequency create a decent synergy.

After learning some formulas, we went on to learn more conceptual stuff,
And we quickly learned that homework assignments were becoming more and more tough.
Differentiating a theory from interpretation is an important concept indeed,
But really, what the heck does that have to do with a farmer and his seeds?
Anyways, moving on to interpretations, like Many Worlds and Copenhagen,
This is where there was hot debate and many different ideas were undertaken.
Like, how can we really know if Schrödinger cat is dead or alive?
We won't! And since it can't be measured, I'm not taking his side,
And if that's the case, what's the point in even putting up that fight?
It's like bringing up the question: does color exist when there is no light?
And if Many Worlds is indeed correct, so is there really a different me?
Reciting a different poem, far off in some other galaxy?
Well if that's true, you know what I'd do, I'd go out and explore,
And find that other Hamed out there, and get right down to the core,
Of all this philosophical debate that I had no idea exists,
It's so abstract how even amongst the experts nothing is fixed.
But wait- if Copenhagen is right, what happens when I observe my superposition?
One of the two Hameds would collapse! So maybe it wouldn't be such a good decision.
So instead I'll comfort myself to Decoherence, and accept that I exist in only one state,
Then would that mean when I came out of my mom, a tiny air particle determined my fate?
But moving on, there's something even more bizarre called Entanglement.
That's something no one can explain, there's nothing on it- not even a hint!
A phenomena that has bewildered even the most wise,
Is it faster than light communication or just a trick to our eyes?

It’s so wild those top physicists haven’t even attempted a surmise!
So big that whoever can explain the mechanism will win a Nobel prize!

Skip a few lectures to Heisenberg’s Uncertainty principle,
Which by the way, I have to admit, still seems to be quite whimsical.
Fine- our knowledge about the position is inversely proportional to the wave,
But what’s the reason for the weird h-over-2 factor that he gave?
Another thing we learned is that energy of an electron relates to n-squared,
And from the Stern-Gerlach analyzer, we know that atoms are paired,
We also saw the development of the different models of the atom,
How a plum pie was all back then that Mr. Thomson could fathom.
But then we advanced to Rutherford’s solar system perception,
But if that were true, electrons would spiral into the nucleus at inception!
Thankfully Bohr came around to provide the explanation to how so this wasn’t,
His genius explanation of a spiraling electron was: It just doesn’t!
Yes! Very convincing, quite impressive indeed!
But really, de Broglie’s standing wave is what planted the seed,
For Schrödinger’s model that later came, just in time to save the day,
It was brilliantly accurate-yes! But made my life harder in many ways!
Complex differential equations and various electron clouds!
Not quite my cup of tea, but surely his mother was very proud.
So this proved that the position of the electron is based on probability,
Which explained how on the detection sheet there was so much volatility.

Coming up next, was infinite potential wells, tunneling and much more,
And I’m so grateful for Schrödinger to teach me that I can’t run through a door.
Towards the end of the semester, we touched a bit on nuclear combustion,
And what a relief to know that three protons in Uranium keeps the world from total destruction!

Well overall, I must express my thanks to Noah, Charlie, Sam and Dan,
I hope to see you all around campus, indeed to all of you I am a big fan!
Now at our last lecture, class will be over, yes- it is quite tragic,
But one thing I will take away from this course is that physics is cooler than magic!

Oh no! Now I realize that too much time has elapsed-
Because my poem was just measured and now it will collapse!

**Student C**

**Pre-Instruction Survey:**

**1.** It is possible for physicists to carefully perform the same measurement and get two very different results that are both correct.

**(Strongly Agree)** What the two physicists are measuring could be highly unstable and sensitive to multiple external stimulus.

**2.** The probabilistic nature of quantum mechanics is mostly due to physical limitations of our measurement instruments.

**(Neutral)** I don't know what quantum mechanics is yet.

**3.** When not being observed, an electron in an atom still exists at a definite (but unknown) position at each moment in time.

**(Agree)** Because I have been told this since 9th grade.

**4.** I think quantum mechanics is an interesting subject.

**(Neutral)** I don't know yet.

**5.** I have heard about quantum mechanics through popular venues (books, films, websites, etc...)

**(Agree)** I read part of the book In Search Of Schrodinger's Cat by John Gribbin

**Homework Problems**

**<u>HW01</u>:**

**13.** What are atoms made of, and how are the parts of the atom configured?

Atoms are made of a nucleus and electrons that exist at various energy levels around the nucleus. The nucleus is composed of protons and neutrons. The electrons are a large distance outside the nucleus and are significantly smaller than the nucleus.

**21.** The force (F) experienced by a charge (q) in an electric field (E) is given by the equation:

$$\vec{F} = q \cdot \vec{E}$$

In fact, the electric field at any point in space is ***defined*** in terms of the force that would be experienced by a charge placed at that point in space:

$$\vec{E} \equiv \frac{\vec{F}}{q}$$ (Electric field defined as force per unit charge)

How do you (personally) ***interpret*** the concept of an electric field? Is the electric field something that is physically real, even though we can't observe it directly? Or is it a mathematical tool devised by physicists to explain our observations and to make calculations easier? A mix of both, or is it something else entirely? Remember, we are interested in what you actually think – there are no "correct" or "incorrect" answers to this question. Please provide the reasoning behind your response.

An electric field is as real as any other concept that describes a physical phenomena. Mathematics can describe and, under certain conditions, predict the phenomena of the electric field.

**<u>HW02:</u>**

**12.** It is said that the photoelectric effect demonstrates the particle-like nature of light. Explain how this conclusion is reached. That is, what *experimental evidence* is consistent with particle-like behavior for light but not with wave-like behavior? Cite at least two pieces of evidence.

The photoelectric effect demonstrates two examples that both suggest particle-like behavior of light. Photoelectrons are not emitted as soon as the frequency of light shinning on a material goes below a threshold frequency specific to the material. This suggests that light is composed of quantized packets of energy that are absorbed by electrons. One electron can absorb only one packet of light quantum. Photoelectrons have a greater KE as frequency increases. Waves do not output more power if only their frequency changes.

**13.** In explaining the observations from the photoelectric effect experiment, did Einstein propose a new theory, a new model, or a new interpretation of the data? In your mind, are there any differences between the three terms (theory, model, interpretation)? If so, what would they be? If not, why would it be OK to use the three terms interchangeably?

Remember, for these types of questions, we are interested in having you express your opinion, so there are no "right" or "wrong" answers. Your response will be graded on the effort you put into it.

A theory is a system that uses models of the physical world as well as interpretation to explain. A model is typically a representation of a physical phenomena where as an interpretation is based more in opinion. Both interpretation and modeling are not limited to the scope of a theory.

**HW03:**

**3.** What was the point of the "Farmer and the Seeds" story (from lecture)?

The farmer and the seeds was a parable that demonstrated scientific interpretation. Reasoning using tools in our mental toolbox were used to describe a phenomena that was occurring. The only constraints on the interpretation were what tools people were using, their imaginations, and what they were observing.

**4.** Review the discussion of the terms *theory, model & interpretation* in the solution set from HW02. How would these terms, as **you** understand them, apply to the different elements of the "Farmer and the Seeds" story, and to the schemes you came up with in class? Remember, we are interested in what **you** think. There are no "right" or "wrong" answers to this question.

The theory was the scheme, the interpretation and model all wrapped into one. The model was the mathematics and concepts based on what was observed. Interpretation encompassed the scope of the entire process. Everyone interpreted the situation slightly differently.

**HW06:**

**Questions 16 – 18 refer to the reading "100 Years of Quantum Mysteries".**

**16.** As discussed in this article, what were some of the problems in classical physics that led to the development of quantum theory?

Classical physics predicts that an electron orbiting a nucleus would radiate away its energy as electromagnetic energy. But this doesn't happen. The 1900 paper published by Max Plank predicted that certain black body radiators should also be radiating ultraviolet rays. Clearly a new interpretation of the building blocks of reality was needed to explain these weird discontinuities in classical physics.

**17.** How are the terms *theory* and *interpretation* used in this article?

The term interpretation was used repeatedly to talk about how people interpreted Schrodinger's wave equation. Specifically, there was the Copenhagen interpretation, the multi world interpretation, Born's interpretation. All of these were interpretations of what was going on with a theory that had been set forth. Whereas one example of a theory discussed in the paper was Decoherence theory. This theory reveals how tiny interactions can alter outcomes on the quantum level. This theory seems more rigorous. However, this theory is still subject to the same interpretations of multi worlds, and so forth.

**18.** Is there any experimental evidence in favor of any of the interpretations discussed in this article? In the cases where there is not, why would a scientist favor one interpretation over another?

Scientists seem to favor the Copenhagen interpretation. This interpretation matches most directly what we observe, however it has the problem that it means the wave equation has to collapse. How can the atom suddenly collapse?

The multi worlds theory makes sense from a certain perspective. There are multiple ways any outcome can go, and all but one of the outcomes remains. However, this requires the leap that as I am typing this there is also a box with a cat in it that is hearing the sound of a tree falling in the forest who is not actually there but is at the same time….

**19. This question refers to the reading assignment "Probability"**

According to this article, in what way(s) is quantum mechanics a probabilistic theory?

The article stated that to try and determine the exact location of an electron would be "a self defeating experiment which destroys the conditions under which the original question was asked." So to determine the mechanics of tiny quantized things cannot, according this article, be done without talking about the likelihood of the outcome. We only know with what probability the electrons speed and position are. Through tools such as the Schrodinger equation, a decent picture probabilistic can be painted.

**<u>HW07:</u>**

**Questions 9 & 10 refer to the readings: "A Quantum Threat to Special Relativity" & "Is the moon there when nobody looks?"**

**9.** What is meant by the terms ***realism, locality*** & ***completeness***? What are some examples of hidden variables?

Locality: Locality of the two particles that are being separated and measured means that in some way the particles are linked to each other. These two linked particles are then able to influence each other with out traveling faster than the speed of light.

Realism: Realism suggests that no quantum superposition exists. If I see a red sock in the classic two socks in box experiment, the sock was red all along and the other sock was blue all along.

Completeness: If the sum total parts of any experiment is known, the outcome can be predicted. There is completeness to an experiment that can always be predicted. Quantum mechanics suggests otherwise.

Hidden Variables: A hidden variable could influence the outcome of an experiment and explain the non-locality of entangled particles. A tachyon is an example of a hidden variable, it is something that can travel faster than the speed of light.

**10.** Does ***entanglement*** allow for faster-than-light communication? If so, what kind of information can be communicated? If not, why not?

If a stream of entangled particles is being transmitted between two points, the observers at both points would expect to see particles with opposite spins every single time. If the observers suddenly start getting the same spin observed, someone is listening in on the stream. So a definite state cannot be imposed on these entangled transmitted particles, but a line of communication can be made so that no one can listen in on it can be made.

**11.** In the two Aspect experiments discussed in class, where the goal was to produce a "single-photon" source, the calcium atoms were excited to the upper level by a two-photon absorption process. Why did the experimenters excite the calcium atoms with a laser pulse of 3.05 eV photons followed by a pulse of 2.13 eV photons, rather than with single photons with the same energy as the two original photons combined (single-photon excitation)?

Imagine initially the calcium is excited to the 3.05eV using a pulse of photons, then a certain amount of these excited electrons return to the ground state and a small portion of the electrons remain excited. Then a 2.13eV pulse is sent at the calcium atom. There is now a very small number of atoms that will absorb the 2.13eV photons and be excited to an even higher state. This reduces the total number of 2.93eV photons emitted. If a 5.18eV photon were sent initially there would be more 2.93eV photons being emitted than in the double excitation case.

**12.** In your own words, explain what the anti-correlation parameter ($\boldsymbol{\alpha}$) is, both in terms of its mathematical definition, and in terms of what it physically tells us, in the context of single-photon experiments as performed by Aspect. Why didn't Aspect measure $\boldsymbol{\alpha} = 0$ if photons are supposed to be acting like particles?

Mathematically: alpha is the ratio of coincident counts to the probability of the two independent events Na and Nb triggering. Conceptually: When alpha is greater than or equal to 1, wave like behavior should be observed. When alpha is less than 1, non-classical wave behavior is observed. An explanation for why Mr. Aspect didn't observe alpha = 0 is that his measurements were not ideal. There could have been dark counts occurring on the PMTs. There also could have been multiple photons in the pathways approaching the PMTs, there was not necessarily an ideal single photon source.

**(Essay)** In class we have discussed correlated measurements performed on systems of two entangled atoms. The assigned reading "The Reality of the Quantum World" (available on CU Learn) discusses correlated measurements performed on entangled photon pairs. In what ways are these systems of entangled photon pairs similar or different from systems of entangled atom pairs? In what sense are the particles in each system entangled (i.e., what properties are correlated for each of the two types of systems)? What types of measurements are performed to determine these properties, and what are probabilities for the possible results of these measurements for both types of systems?

Both entangled atoms and photons share the similarity of "spooky action at a distance". That is, when we measure one it has gotta collapse the other. Both systems showed very similar relationships of the probability of how the particle would orient itself once observed. One way the two entangled messes differed was the photon's orientation was determined by how the electric field was associated with the photon. The atom's orientation was determined by the spin of the atom, or the magnetic field of the atom, if you will. Both, however, again, exhibit "spooky action at a distance". The observation of the polarization of two entangled photons and the spin of two entangled atoms generates this "spooky action at a distance".

The polarization of entangled photons can be measured as in the Aspect experiment. In this case polarizing glass was used to observe the photons. The spin of entangled atoms can be observed using the classic stern-gerlach analyzer scenario. Two stern-gerlach analyzers are used to measure, and collapse, the superposition of the atoms.

**(Essay)** As discussed in class and in the readings, what do the two single-photon experiments performed by Aspect tell us about the nature of photons? How were the two experiments designed to demonstrate the particle and the wave nature of photons? When answering, don't concern yourself with technical details (such as how the photons were produced); focus instead on how the design of each experimental setup determined which type of photon behavior would be observed. How are the elements of these two experiments combined in a delayed-choice experiment, and what do delayed-choice experiments (along with the two Aspect experiments) tell us about the nature of photons?

Fundamentally, the Aspect experiments tell us that photons act as waves or particles. The other implication of this experiment is some spooky action at a distance is occurring between the photons.

The single beam splitter scenario predicted that the photon like nature of light would cause the photon to enter one of the photomultiplier tubes but not both. The photon has a definite path that it will take.

The double beam splitter scenario predicted that the path of the photon took to enter one of the photomultiplier tube is undeterminable. Not only that, the mirrors can be arranged so that interference of the photons can cause no photons to pass through one pmt and not the other. Both paths are possible and yet only one pmt is firing.

But wait, the story gets better. A delayed choice experiment can be performed that reveals that the photon cannot be both a wave and a particle at once. Not only that, the photon, even once it begins traveling along a path that would suggest particle like behavior, can suddenly be altered so that it exhibits wave like behavior. Can this photon be aware of the actions of humans? The weirdness of quantum mechanics persists.

## HW09:

**3.** How does thinking of electrons as waves rather than particles explain why energy levels in the hydrogen atom are quantized? How does it answer Bohr's question of why electrons don't radiate energy when they are in one of these energy levels?

Any fixed volume, area, or length creates boundary conditions for setting up a standing wave. Within this boundary, only a fundamental frequency and harmonics of that fundamental can establish standing waves in a fixed region. Because each wave has a corresponding amount of momentum, the energy of the electron’s wave is quantized with energy levels corresponding to the fundamental and harmonics.

As a wave, the electron would not radiate away its kinetic energy to an electromagnetic because the wave is propagating at a constant velocity. This implies there is no acceleration of the wave happening.

**Exam Questions:**

**Exam 2**

**E1. (10 Points)** In the sequence of screenshots shown below (taken from the PhET Quantum Wave Interference simulation), we see: A) a bright spot (representing the probability density for a single electron) emerges from an electron gun; B) passes through both slits; and C) a single electron is detected on the far screen. After many electrons pass through and are detected, a fringe pattern develops (not shown).

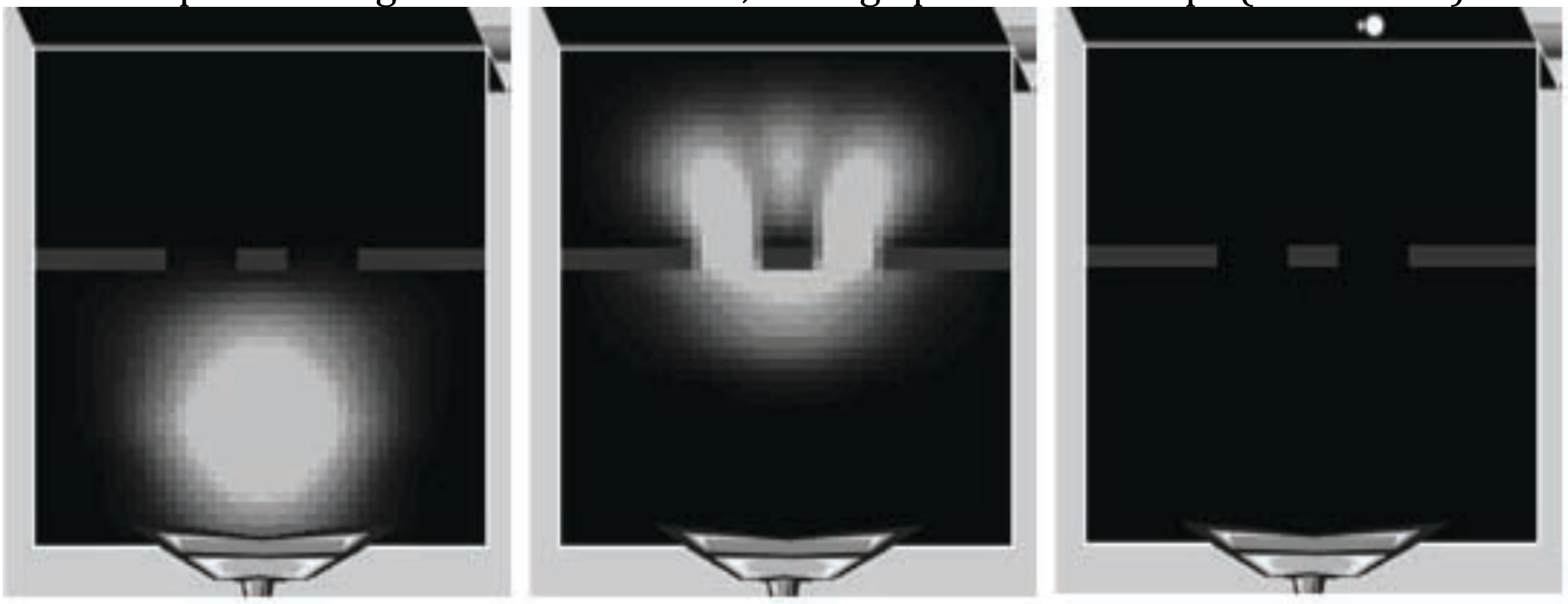

(A) (B) (C)

**Three students discuss the Quantum Wave Interference simulation:**

**Student 1**: The probability density is so large because we don't know the true position of the electron. Since only a single dot at a time appears on the detecting screen, the electron must have been a tiny particle traveling somewhere inside that blob, so that the electron went through one slit or the other on its way to the point where it was detected.

**Student 2**: The blob represents the electron itself, since a free electron is properly described by a wave packet. The electron acts as a wave and will go through both slits and interfere with itself. That's why a distinct interference pattern will show up on the screen after shooting many electrons.

**Student 3**: All we can really know is the probability for where the electron will be detected. Quantum mechanics may predict the likelihood for a measurement outcome, but it really doesn't tell us what the electron is doing between being emitted from the gun and being detected at the screen.

**E1.A (2 Points)** In terms of the interpretations of quantum phenomena we've discussed in class, how would you characterize the perspective represented by Student 1's statement? What assumptions are being made by Student 1 that allows you to identify their perspective on this double-slit experiment?

Student 1 is taking a somewhat realist perspective. They are assuming the electron traveled through one slit or the other. They claim the reality of the situation is the particle-like electron existed in a cloud of probability, and passes through one slit or the other as the cloud moved through the double slits. This explanation does not mention the probability density predicted by the wave equation.

**E1.B (6 Points)** For each of the first two statements (made by Students 1 & 2), what rationale or evidence (experimental or otherwise, if any) exists that favors or refutes these two points of view? As for the third statement, is Student 3 saying that Students 1 & 2 are wrong? Why would a practicing physicist choose to agree or disagree with Student 3?

Student 2 describes the electron as a wave packet. When a double slit experiment is performed, the interference pattern that is observed corresponds to a probability density that can be described by a wave-packet equation. A packet of waves would interfere with itself, creating a probability of the electron to pass through both slits. Also, which slit the electron went through cannot be measured without altering the uncertainty in the momentum.

**E1.C (2 Points)** Which student(s) (if any) do you ***personally*** agree with? If you have a different interpretation of what is happening in this experiment, then say what that is. Would it be reasonable or not to agree with ***both*** Student 1 & Student 2? This question is about your personal beliefs, and so there is no "correct" or "incorrect" answer, but you will be graded on making a reasonable effort in explaining why you believe what you do.

Since electrons show both wave and particle like behavior, it would be reasonable to side with either Student 1 or 2. Student 2 used a more wave-like interpretation, Student 1 used a more particle like interpretation.

I personally visualize the situation as a flow of some fluid that travels through the two slits in waves. It appears through all space as soon as the electron is fired. The electron then rides this chaotic fluid toward the screen and strikes in a location that is somewhat determined by the interference patterns of the fluid. Trying to measure this fluid flow collapses the waves created.

**E3. (8 POINTS TOTAL)** For the diagrams below depicting Experiments X & Y, M = Mirror, BS=Beam Splitter, PM = Photomultiplier, N = Counter. In each experiment a single-photon source sends photons to the right through the apparatus one at a time.

## EXPERIMENT X

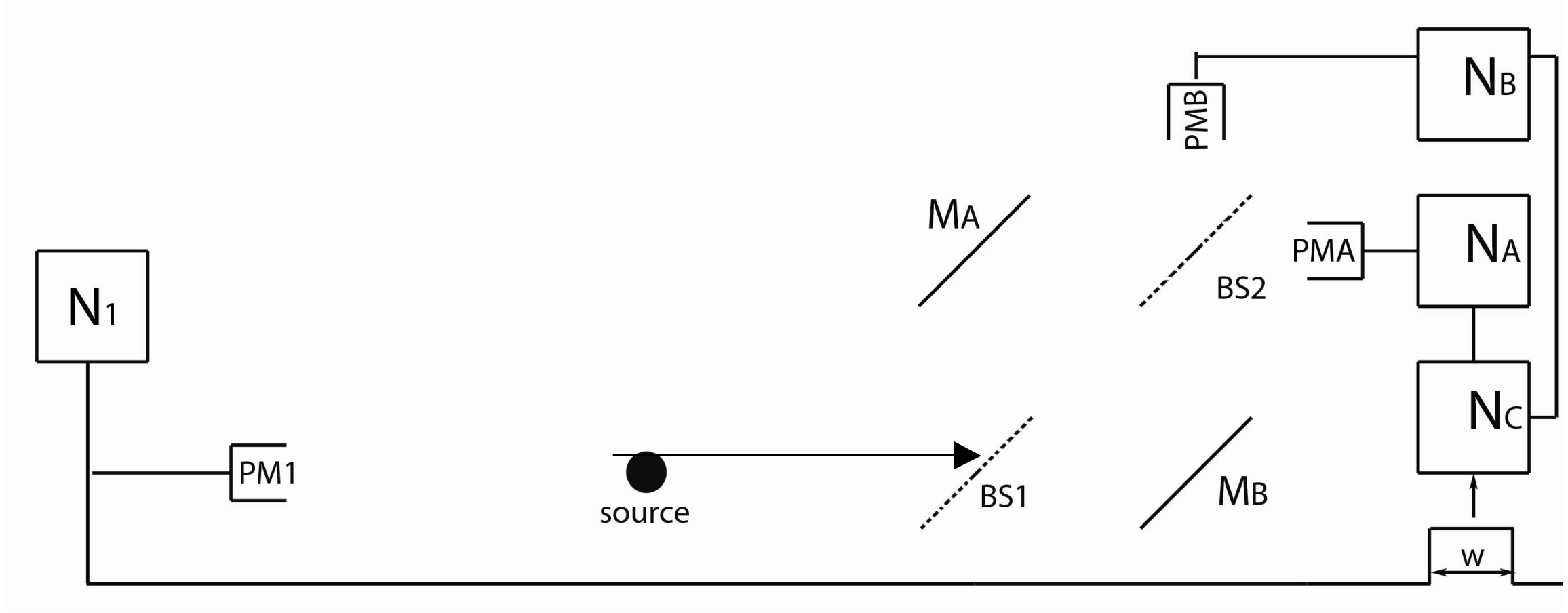

## EXPERIMENT Y

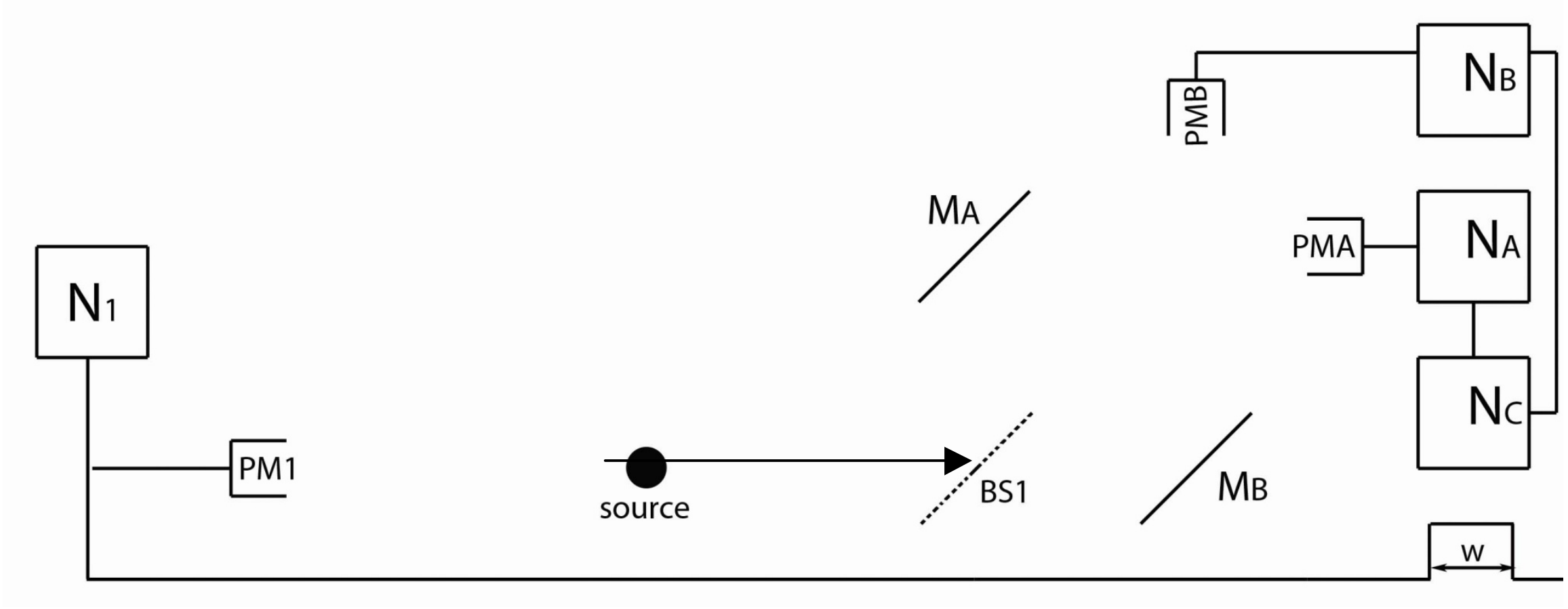

**E3.A (3 Points)** For which experimental setup (X or Y) would you expect photons to exhibit particle-like behavior? Describe in what sense the photon is behaving like a particle during this experiment. What features of the experimental setup allow you to draw this conclusion without actually conducting the experiment?

Experiment Y should show photons acting like a particle. This is due to the fact that which path the photon takes can be determined by which photomultiplier is triggered. If the photon struck mirror B, PMB will fire. If the photon struck mirror A, PMA will fire. If there was truly a single photon in the source only one of the photomultipliers will fire, and each would fire with a 50/50 chance.

**E3.B (3 Points)** For which experimental setup (X or Y) would you expect photons to exhibit wave-like behavior? Describe in what sense the photon is behaving like a wave during this experiment. What features of the experimental setup allow you to draw this conclusion without actually conducting the experiment?

Experiment X should show photons acting like waves. The path the photon took is underterminable. Mirror B could have been hit with a photon and either PMB or PMA could fire. This implies a wave is being propagated through both possible paths. The wave then describes an equal probability of triggering each photomultiplier provided each path is the same length. Interference can happen if the paths are different lengths and cause only one photomultiplier to trigger.

**E3.C (2 Points)** Suppose we are conducting Experiment X (the second beam splitter (BS2) is present) when a photon enters the apparatus and encounters the first beam splitter (BS1). Afterwards, while the photon is still travelling through the apparatus (but before it encounters a detector), we suddenly remove the second beam splitter (switch to Experiment Y). Can we determine the probability for the photon to be detected in PMA? If not, why not? If so, what would be that probability? Explain your reasoning.

First assume that Experiment X is set up so that interference occurs and only PMA is firing. If the photon is still traveling will switch to acting like a particle. The photon will no longer only fire in PMA due to interference, but will instead show particle-like behavior and trigger either PMB or PMA with a 50/50 probability. BS1 results in either path from BS1 being 50/50 probable. Because when BS2 is removed, the path the photon took is now better known and particle like behavior is observed. In other words, once BS2 is removed, PMB firing means MB was hit by a photon and PMA firing means MA was hit by a photon.

**Exam 3**

**E1. (10 Points Total)** A hydrogen atom is in its lowest energy state. Use words, graphs, and diagrams to describe the structure of the Hydrogen atom **in its lowest energy state (ground state)**. Include in your description:

- **(4 Points)** At least two ideas important to any accurate description of a hydrogen atom.
- **(3 Points)** An electron energy level diagram of this atom, including numerical values for the first few energy levels, and indicating the level that the electron is in when it is in its ground state.
- **(3 Points)** A diagram illustrating how to accurately think about the distance of the electron from the nucleus for this atom.

(On these diagrams, be quantitative where possible. Label the axes and include any specific information that can help to characterize hydrogen and its electron in this ground state.)

**[IMPLICITLY AND EXCLUSIVELY USES SCHRODINGER MODEL]**

**Post-Instruction Survey**

**1.** It is possible for physicists to carefully perform the same measurement and get two very different results that are both correct.

**(Strongly Agree)** Two very different results could confirm the same fact. Being correct is nothing more than confirming a fact.

**2.** The probabilistic nature of quantum mechanics is mostly due to physical limitations of our measurement instruments.

**(Neutral)** I have no idea.

**3.** When not being observed, an electron in an atom still exists at a definite (but unknown) position at each moment in time.

**(Neutral)** If an electron orbits a nucleus in a forest and no physicist is there to observe it, does it obey the uncertainty principle?

**4.** I think quantum mechanics is an interesting subject.

**(Strongly Agree)** Quantum mechanics and is strange and interesting and mind stretching. This has been a great course.

**Double-Slit Essay Question:**

Student One is assuming the electron is always a particle. Student Two is assuming that the electron is pretty much a wave and until it gets smooshed by the screen. Student three is sticking to the fact that the electron has a probability of going in certain places on the screen. I think there will always be a more accurate description of observations and quantum mechanics is, for now, an accurate description of reality.

**Final Essay:**

In this present day, products of quantum mechanics can easily be found. Laser printers, plasma screen televisions, and iPhones are just a few of the handful of common electronic devices whose discovery is steeped in quantum mechanics. This paper will discuss the basic quantum mechanics of a component crucial in the operation of a digital camera; a device called a charge-coupled device array (CCD array for short). A CCD array is a grid of doped semi-conducting devices called CCDs. The CCD itself measures the intensity of incident light by storing an electrical charge directly proportional to the amount of photons that strike it. The CCD array is carefully electrically connected such that the array can be used to harvest the stored charge in each CCD. The charges can then be used make a digital picture of the light incident on the array. The CCD array is a huge advantage over film and has almost entirely replaced film for capturing images. It has proven to be one of the many rock star technologies that have been ushered in by the modern digital age.

Let's start by looking at the construction of a discrete CCD. It all begins when a pure semi-conductor is doped to become a p-type semi-conductor. P-type doping is a process in which the valence bands of the atoms in a semi-conductor, such as silicone (Si), are stripped of a few electrons without giving the semi-conductor a net charge so that the semi-conductor becomes more conductive.

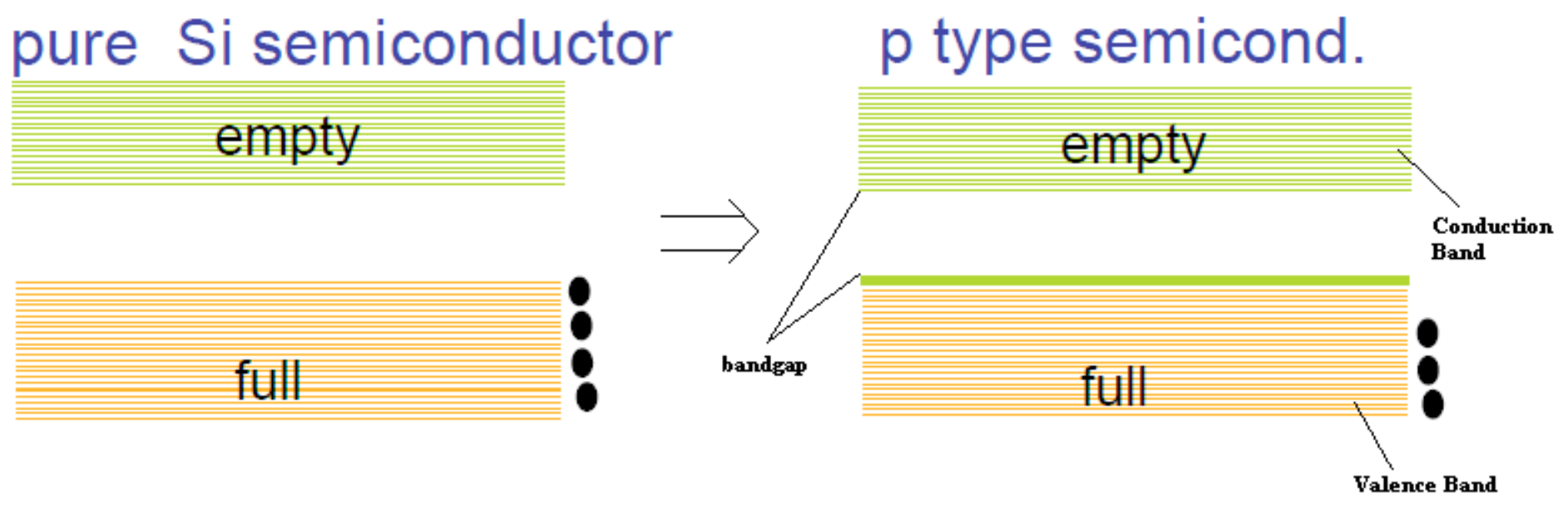

Figure 1: Energy Band Contents Changing Through P-Type Doping[1]

The next step in making a CCD is taking the p-type semi-conductor and placing a transparent oxide (non-conductive) layer on top of it. A typical transparent oxide used is polycrystalline silicon. This p-type semi-conductor and transparent oxide layer are then sandwiched in between two metal plates, a gate and a ground plane, with two metal connections poking out of the plates. The typical size of a CCD used in a digital camera is about a 10 μm cube. The actual process used to produce such a small device is called photolithography and would require another whole paper.

So how does it all work? First, a voltage of about 5V is applied to the metal connections. Second, a photon strikes the CCD through the transparent oxide layer. If the photon has energy greater than the band gap of the p-type semi-conductor, what is called a hole-electron pair is generated in the semi-conductor. The hole-electron pair is then free to move in the valence and conduction band respectively.

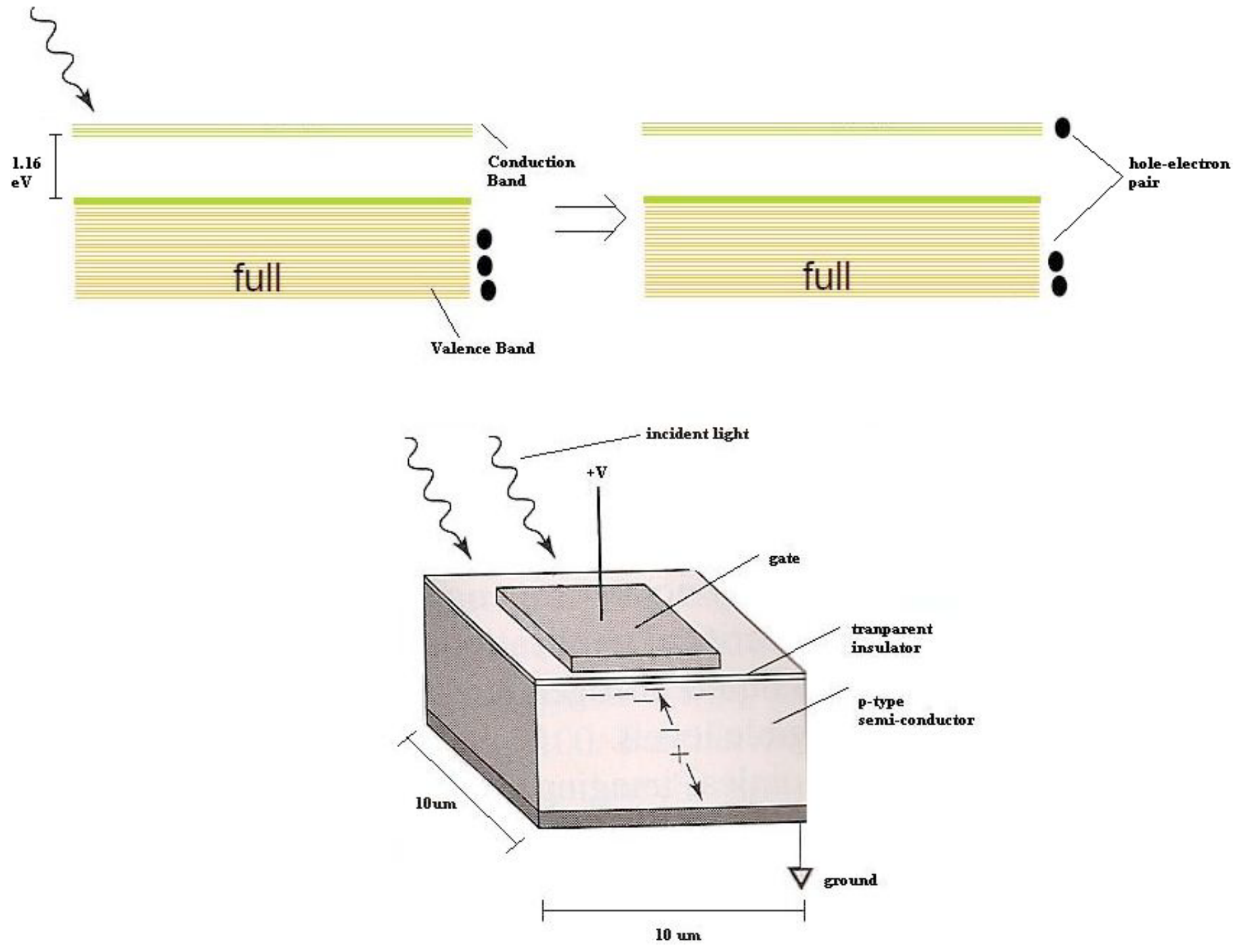

Figure 2: Production Of A Hole-Electron Pair Process.

Thirdly the 5V applied causes the hole to whiz down to the ground plane, through the conduction band, and is filled by the abundance of electrons floating in the ground plane. The 5V causes the electron to be whizzed up through the conduction band to the gate where it is stopped from combining with the gate due the non-conductive oxide layer being in the way.

By doing a little math, it is not hard to show that if Si is used as the semiconductor even the lowest energy light quanta will generate a hole-electron pair. For Si, the band gap energy is about 1.16eV.5 Also, the lowest energy photons are about 700nm in wavelength. Using E = hf , we get

$$E = hf = \frac{hc}{\lambda} = \frac{1242\ eV \cdot nm}{700\ nm} = 1.77\ eV\ >\ 1.16\ eV$$

Even low energy red light will produce a hole-electron pair when striking the CCD. So the number of trapped electrons will be directly proportional to the number of photons hitting the CCD. Since the number of photons in a given area is directly proportional to the intensity, we now have a way, using quantum mechanics, to measuring how "white" or how "black" a 10 μm by 10 μm square is.

So what? Well, if a 500x500 grid of CCDs is made we get a 5mm by 5mm grid. If a lens is placed in front of the 5mm x 5mm grid, there can be a wide image cast onto the small grid. So the brighter the image on part of the grid will mean more charges store in that portion of the grid. More charges stored means a whiter image captured on that portion of the grid. Fewer charges stored means a blacker image. We now have the workings of a quantum mechanical eyeball that can see in black in white. A typical CCD array can be seen in figure 3.

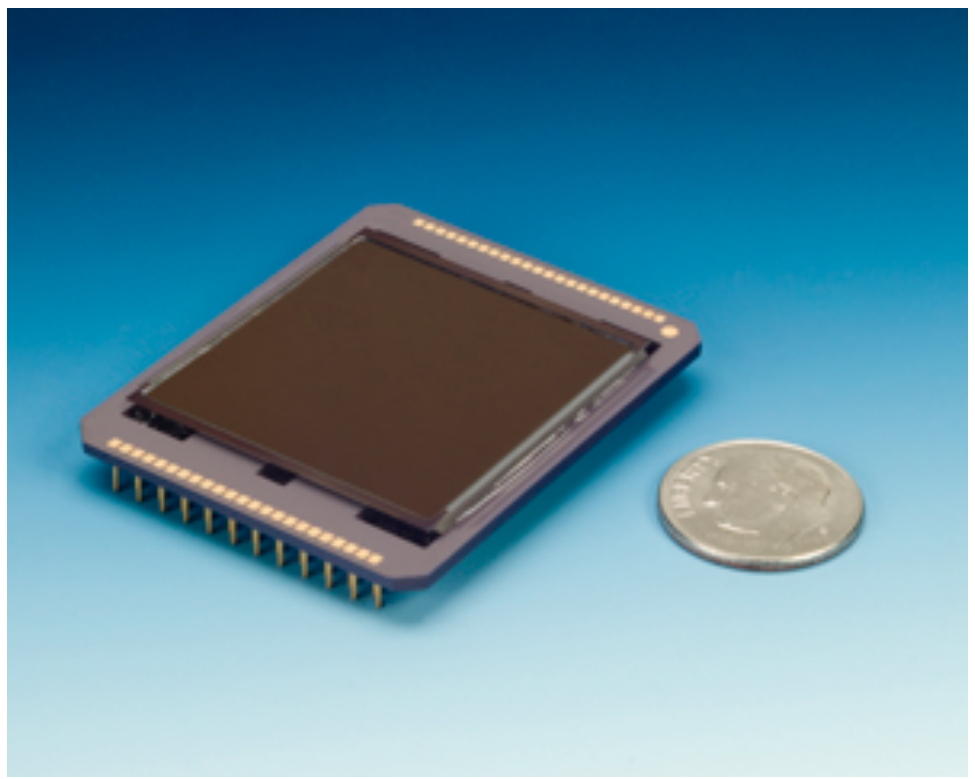

Figure 3: Typical CCD Array

The crux of this grid is how to get the charges out of the grid and converted into something that can be viewed on a computer or printed onto a piece of paper. A simple representation of how the p-type silicone is organized can be seen in figure 4.

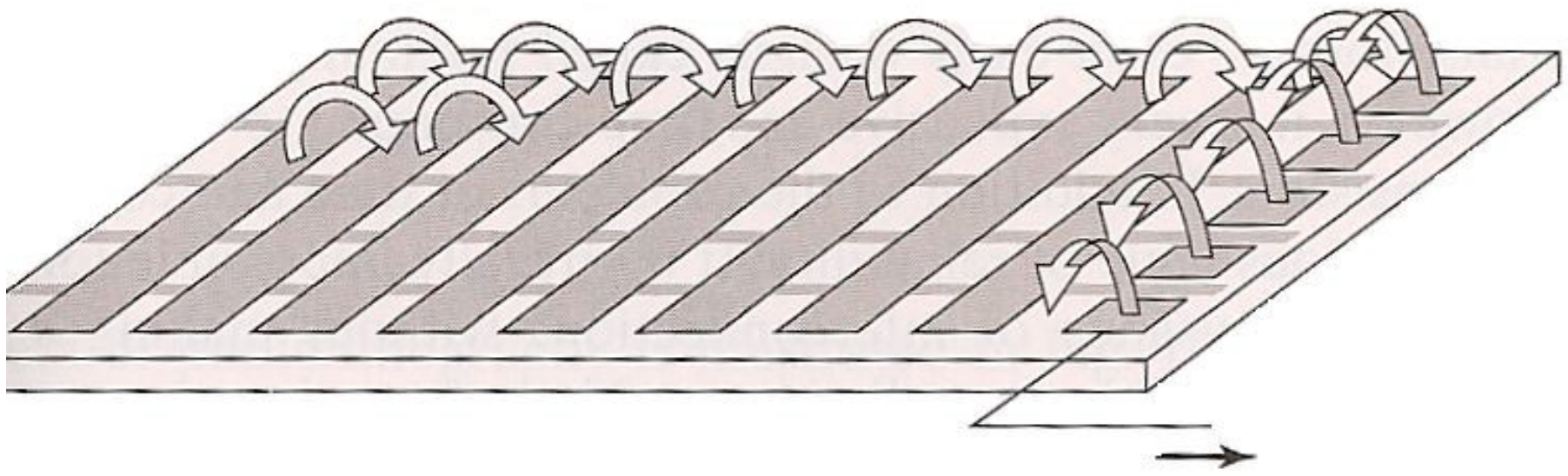

Figure 4: Close Up Picture of the Array

The array is composed of one solid piece of silicone split into rows and columns. The rows of CCDs are separated from each other by rows of Si that have been left undoped. One large ground plane is placed on the bottom of the array as well as a sheet of transparent oxide on top. The array is then split into columns by strips of metal. The charge is pulled off the grid in a conveyor belt like manner. One column of charge is dumped into the end column where the charge contained in each individual CCD can then be sifted out one by one. This processes of dumping out a column of charge and sifting out the rows of the column continues until all the individual pockets of charge are read off the CCD array. The details of how the charge is pulled off the CCD array can vary from array to array.

While it may seem like a disadvantage to only be able to store black and white photographs, CCD arrays can be manufactured so that color images can be captured as well. Essentially the grid of CCDs becomes more complex by filtering and sorting incident light so that a color image can be extracted from the array.9 The main advantage of the CCD over film is sensitivity. A CCD catches about 80% of all incident light where as film captures about 1%. Film captures light nonlinearly. That is to say, how bright a spot is on a piece of film is not directly proportional to the brightness of the image being captured. In a CCD, the number of photons striking is very close to the amount of charge stored in it. The linearity of CCD arrays makes analysis of images produced by CCD arrays simpler than analysis of images produced by film.

The drawback to CCDs over film is film can handle slightly higher temperatures than a CCD. If a CCD made out of Si begins heating up, electrons with kinetic energy higher than 1.16eV will begin popping out of the valence band of the p-type semiconductor and produce a hole-electron pair. This means that the charge stored in the CCD will begin rising even when light is not striking the CCD. This obviously is an undesired phenomenon. While these so called dark counts are generally not a problem for commercial applications, they can become a problem when very precise measurements of light intensity have to be made such as in astronomy applications.

So now you too understand how simple quantum mechanics can describe the operation of an individual CCD and how the charge it stores depends on the intensity of light shinning on it. Using photolithographic techniques p-type silicone, a transparent oxide, and some metal can be arranged to create one of these light intensity-measuring devices. An array of CCDs and a lens can be used to capture a picture electronically. This electronic picture can be extracted from the array by doing a bucket brigade extraction technique whose details involve very little quantum mechanics. Without such a semiconductor array, almost all the video entertainment industry would not be around, astronomy would not be where it is today, and worst of all facebook wouldn’t be littered with countless photographs. Quantum mechanics is, without a doubt, changing the lifestyle of the technologically infused parts of the world. In this case, a quantum mechanical eyeball is the contents of the infusion.

Bibliography

Finkelstein, Noah. “Modern Physics: Quantum Mechanics.” University of Colorado. http://www.colorado.edu/physics/phys2130/phys2130_fa10/Lecture_Notes/class40.pdf

Dubson, Michael A., et al. Modern Physics For Scientists and Engineers. 2nd Edition. Upper Saddle River, NJ: Prentice Hall 2004.

Sedra, Adel S., Smith, Kenneth C. Microelectronic Circuits. 5th Edition. New York, NY: Oxford University Press 2004.

Zeghbroeck, Bart V. “Principles of Semiconductor Devices.” University of Colorado. http://ecee.colorado.edu/~bart/book/book/chapter2/ch2_3.htm#2_3_3_1

Fairchild Imaging. “CCD 3041 2k x 2k Multiport Scientific CCD.” Fairchild Imaging Technology and Design.
http://www.optcorp.com/cart/productimages /FairchildCCD3041sensor-L.jpg

Richmond, Michael. “Introduction to CCDs.” Rochester Institute of Technology. http://spiff.rit.edu/classes/phys445/lectures/ccd1/ccd1.html

Coffman, Valerie. “Photoelectric effect is The reason digital cameras work.” Cornell Center For Materials Research. http://www.ccmr.cornell.edu/education/ask/index.html?quid=711

**Student D**

**Pre-Instruction Survey**

**1.** It is possible for physicists to carefully perform the same measurement and get two very different results that are both correct.

**(Strongly Agree)** It is possible for identical measurements to produce different results if that which is being measured can exist in more than one state at the same time. Thus, one would not know whether the subject of the measurement is the object in one state or the other. Interpreting this question differently, one could comment on the fact that the very act of measuring itself introduces new elements into a system, and thus actually changes the outcome of the measurement.

**2.** The probabilistic nature of quantum mechanics is mostly due to physical limitations of our measurement instruments.

**(Strongly Disagree)** The probabilistic nature of quantum mechanics is a fundamental property of the system. For example: it is impossible to define (not just measure) the position and momentum of an electron at the same instant in time (Heisenberg's uncertainty principle). Thus, the uncertainty exists outside of the instruments used to try to measure those properties. (I would really, really like to learn the math behind these statements!)

**3.** When not being observed, an electron in an atom still exists at a definite (but unknown) position at each moment in time.

**(Agree)** An electron occupies a single definite position at any given point in time. It is only our measurement (and thus knowledge) of that position at any given point in time that is subject to the Heisenberg uncertainty principle, where either the position or the momentum of the electron may be measured to a high level of precision, but not both.

**4.** I think quantum mechanics is an interesting subject.

(Strongly Agree) Quantum mechanics fascinates me precisely because it is so counterintuitive. I want to challenge my perception of the world, and there are few better ways to do that than QM. It is also interesting to me because I am much more used to physics on very large, indeed cosmic scales. It is especially interesting to see how the world of the unimaginably tiny and the world of the unimaginably large interact...

**5.** I have heard about quantum mechanics through popular venues (books, films, websites, etc...)

**(Agree)** In high school, I got a taster of quantum mechanics through generalized physics books, but nothing more in depth. Beyond that, my knowledge of quantum mechanics is limited, and comes primarily from several online lectures by MIT (through itunes U) and several from the University of Madras (posted on youtube).

**Homework Problems**

<u>**HW01:**</u>

**13.** What are atoms made of, and how are the parts of the atom configured?

An atom consists of a number of subatomic particles, whose interactions define the properties of the atom as a whole. Any atom can be conceptualized as a field of electrons surrounding a tiny, very dense nucleus. The nucleus of an atom, which contains the vast majority of the atom's mass, consists of protons and neutrons in various orbital configurations bound together by the strong nuclear force. These nucleons in turn consist of quarks (a proton is composed of two up quarks and one down quark, whereas a neutron is composed of two down quarks and one up quark). The remaining tiny percentage of the atom's mass is in the form of electrons, which exist in the space surrounding the nucleus in orbitals (the regions of space in which those electrons are most likely to be found).

**21.** The force (F) experienced by a charge (q) in an electric field (E) is given by the equation:

$$\vec{F} = q \cdot \vec{E}$$

In fact, the electric field at any point in space is ***defined*** in terms of the force that would be experienced by a charge placed at that point in space:

$$\vec{E} \equiv \frac{\vec{F}}{q}$$ (Electric field defined as force per unit charge)

How do you (personally) ***interpret*** the concept of an electric field? Is the electric field something that is physically real, even though we can't observe it directly? Or is it a mathematical tool devised by physicists to explain our observations and to make calculations easier? A mix of both, or is it something else entirely? Remember, we are interested in what you actually think – there are no "correct" or "incorrect" answers to this question. Please provide the reasoning behind your response.

I interpret an electric field as a mathematical tool that allows us to describe the interactions between charges and certain other entities as a function of those object's positions. An electric field is not real in the physical sense because it cannot be construed as an independent entity. It a description of the relationship between a charge and the force being exerted upon it by either other charges or a varying magnetic field at various points in space and time.

## HW02:

**12.** It is said that the photoelectric effect demonstrates the particle-like nature of light. Explain how this conclusion is reached. That is, what *experimental evidence* is consistent with particle-like behavior for light but not with wave-like behavior? Cite at least two pieces of evidence.

If light has all of the classical properties of a wave, then changing the frequency of light incident on the metal plate while keeping amplitude constant should have no effect on the current observed in the system. However, a particle model of light, where the energy of each quanta is dependent on frequency, explains the linear relationship between the frequency of the incident light and the resulting current in the system. A quantum model of light also explains the observation of a threshold frequency below which no current is observed. If each photon delivers its energy to one electron, whether or not the electron leaves the metal is dependent on the amount of energy delivered in the collision, and thus on the frequency of that photon. Furthermore, observation showed that changing the intensity of the light had no effect on the stopping potential, but if light were a wave, one would expect incident light of a greater intensity to result in electrons with a greater kinetic energy, requiring a higher stopping potential. However, the stopping potential would remain constant if the kinetic energy of the electrons was dependent on the energy of the incident photons, independent of intensity. And finally, the (near) instantaneous start of current is explained by (equally rapid) subatomic collisions, but not by the classical model, where the metal would have to be heated sufficiently by the light wave to produce electrons, and thus current, resulting in a delay.

**13.** In explaining the observations from the photoelectric effect experiment, did Einstein propose a new theory, a new model, or a new interpretation of the data? In your mind, are there any differences between the three terms (theory, model, interpretation)? If so, what would they be? If not, why would it be OK to use the three terms interchangeably?

Remember, for these types of questions, we are interested in having you express your opinion, so there are no "right" or "wrong" answers. Your response will be graded on the effort you put into it.

The theory of light as a particle was not new when Einstein was conducting his work on the photoelectric effect (It had been proposed by Max Planck several years earlier). It was Einstein's correct interpretation of the data that allowed for the phenomena observed to be consistent with the theory proposed years earlier, thus validating the theory.

A theory is, in essence, an explanation of a set of phenomena based upon a large body of experimentation and evidence. A model is the application of that theory to the physical world to predict the results of further experimentation (which in turn acts as a check on the underlying theory). Because theories are in essence abstract constructs, they are reliant on the interpretation of observations. That being said, those interpretations are constrained by the internal logic of the system (a valid interpretation cannot be made that does not stand up to inductive reasoning).

**HW03:**

**3.** What was the point of the "Farmer and the Seeds" story (from lecture)?

The "Farmer and the Seeds" story illustrated the difficulty of determining the veracity of a model applied to a phenomena for which there is limited observational evidence. We came up with several explanations for the observations given to us, but could not favor one over the other because they all predicted the results of our limited observations accurately. Seeing as these models contradicted each other (for example, the number of sprouts cannot both be random and governed by (n(dots)-3)/2)) we were left with a dilemma.

**4.** Review the discussion of the terms *theory, model & interpretation* in the solution set from HW02. How would these terms, as **you** understand them, apply to the different elements of the "Farmer and the Seeds" story, and to the schemes you came up with in class? Remember, we are interested in what **you** think. There are no "right" or "wrong" answers to this question.

We were unable to develop a theory as to why the number of seeds that sprouted obeyed any particular model, mostly because we were a.) unable to favor one model over another given the limited observational evidence. b.) unable to explain the mechanism that caused those models to function. Without that mechanism (why the seeds obeyed any given rule) a theory could not be formed. Finally, we were limited in that we could not possibly have conceived every possible interpretation, and thus were restricted to those interpretations which we already possessed. This in turn effected the creation of models to explain the phenomenon.

**HW06:**

**Questions 16 – 18 refer to the reading "100 Years of Quantum Mysteries".**

**16.** As discussed in this article, what were some of the problems in classical physics that led to the development of quantum theory?

The first and probably most significant problem was the massive discrepancy between the lifetime of a hydrogen atom predicted by classical mechanics and the actual lifetime of a hydrogen atom. Finding an explanation for this discrepancy led to the development of a quantized model of the atom, which in turn led to a search for the underlying principles behind the rules which governed those models. Another major problem facing the classical model was the observation of the spectral lines of hydrogen, which implied quantized energy levels for electrons within the hydrogen atom. Bohr's successful prediction of those lines using his quantized model was a major step forward, but led to further discrepancies (it did not predict the energy levels of larger atoms). Finding explanations for those shortfalls was major driving force in early quantum theory.

**17.** How are the terms *theory* and *interpretation* used in this article?

Much of quantum mechanics originated in the ad-hoc rules put forth to explain observed atomic phenomena. Eventually, models based upon quantitative theory, the Schrodinger wave equations, those rules were explained and refined those rules. The distinction between theory and interpretation in this article comes from the implications of those models. Interpretations are simply a way in which the behavior implied by the equations can be explained qualitatively and in such a way that further useful insights may be developed.

**18.** Is there any experimental evidence in favor of any of the interpretations discussed in this article? In the cases where there is not, why would a scientist favor one interpretation over another?

The Copenhagen interpretation is untestable, as under that interpretation there is no quantitative way to predict when a wave function will collapse. The same applies to the many-worlds interpretation. There is no way for an observer in one of the possible outcomes to observe the result of another. Decoherence could in theory be tested, but obtaining the conditions necessary for tests to occur would be exceptionally difficult. As to why scientists favor one over another, the completeness of the interpretation mathematically, the implications of the interpretation, and the personal "world view" of the scientist all play a role.

**19. This question refers to the reading assignment "Probability"**

According to this article, in what way(s) is quantum mechanics a probabilistic theory?

Quantum mechanics is probabilistic in the sense that we often cannot make deterministic conclusions about properties such as the position of an electron, and when we can, they end up being of little use as they recursively depend upon the manner in which they were measured. Another example is the location of a photon hitting a wall after it passes through two slits. We cannot precisely predict where the photon will "collapse" onto the wall when it hits it before that event actually occurs, but we can ascertain the probability that it will hit any given point.

**HW07:**

**Questions 9 & 10 refer to the readings: "A Quantum Threat to Special Relativity" & "Is the moon there when nobody looks?"**

**9.** What is meant by the terms ***realism, locality*** & ***completeness***? What are some examples of hidden variables?

Realism states that a quantity in a measured system has an objectively real value, even if it isn't known. For example, under a realist interpretation, an atom always has a particular spin, we are simply unable to know that spin before we measure it (it is "hidden"). Locality is the concept that there must always be a causative chain in the real world linking two events, in other words, that one object may only effect another by causing a change in its local surroundings that may eventually propagate to cause a chance in the second object through its local surroundings. Entanglement appears to violate this principle by allowing two particles to influence the state of each other regardless of their physical separation or the material in-between them. For a physical theory to be "Complete" according to the guidelines set by EPR, it must be able to explain the nature and behavior of everything in physical reality. In this sense, quantum mechanics is not complete; if locality is not to be violated quantum mechanics cannot explain all of the physical properties of a system at the most basic level.

**10.** Does ***entanglement*** allow for faster-than-light communication? If so, what kind of information can be communicated? If not, why not?

Einstein's theory of special relativity states that no information can be transmitted faster than the speed of light. Say we have two entangled particles held by two individuals. If they are at opposite ends of the galaxy, and one measures the spin of their particle, they will instantaneously know the spin of the other's particle. However, there is no way to communicate that knowledge to the other faster than the speed of light, and without the knowledge of that initial measurement the measured spin appears completely random to the other party (whether they measure up or down, if they do not know the spin of the other particle, no useful information can be gleaned from the system). Thus, while the action that results in a definite spin state can occur at speeds faster than the speed of light, there is no way to use that action to communicate information, as the result of the initial measurement must still be communicated at speeds equal to or slower than c.

**11.** In the two Aspect experiments discussed in class, where the goal was to produce a "single-photon" source, the calcium atoms were excited to the upper level by a two-photon absorption process. Why did the experimenters excite the calcium atoms with a laser pulse of 3.05 eV photons followed by a pulse of 2.13 eV photons, rather than with single photons with the same energy as the two original photons combined (single-photon excitation)?

Double photon excitation is much less likely to occur than single photon excitation, and thus it is easier to get only a small number of photon pairs that correspond to the energies expected from the double excitation. The goal is to limit the number of excited pairs released in order to make measuring the correlation between the photons in the pair possible.

**12.** In your own words, explain what the anti-correlation parameter (**α**) is, both in terms of its mathematical definition, and in terms of what it physically tells us, in the context of single-photon experiments as performed by Aspect. Why didn't Aspect measure **α** = 0 if photons are supposed to be acting like particles?

Alpha is defined as Pc/(Pa*Pb) where Pa and Pb are the likelihoods that a particular photomultiplier tube in Aspect's experiment is triggered and Pc is the likelihood that the coincidence counter is triggered. In other words, it relates the likelihood of the photons taking either path to the likelihood that both are triggered.

**(Essay)** In class we have discussed correlated measurements performed on systems of two entangled atoms. The assigned reading "The Reality of the Quantum World" (available on CU Learn) discusses correlated measurements performed on entangled photon pairs. In what ways are these systems of entangled photon pairs similar or different from systems of entangled atom pairs? In what sense are the particles in each system entangled (i.e., what properties are correlated for each of the two types of systems)? What types of measurements are performed to determine these properties, and what are probabilities for the possible results of these measurements for both types of systems?

Both entangled pairs rely on the conservation of angular momentum to produce entanglement. In a pair of entangled atoms, angular momentum is conserved as when they are produced they have opposing atomic spin states. In a pair of entangled photons, angular momentum is conserved when the two photons are emitted back-to-back (in precisely opposing directions in a very short interval)).

In the case of atoms, the spin states could be determined by passing each of the atoms though a nonlinear magnetic field in a stern-gerlach analyzer to determine their spin (which would correlate to the channel from which they exited). It was found that the probability of the two atoms having opposing spins in either configuration was equal to 1 (a probability of 0.5 for up/down and a probability of 0.5 for down/up).

The correlation of photons could be determined using opposing photomultiplier tubes linked to a correlator, which would allow the experimenter to determine that the two photons were produced from the same excitation at close to the same instant in time if the two opposing tubes each detected a photon. The probability that the two photons would be detected was close to 1 in the interval of time following the release of the first photon up to twice the lifetime of the intermediate state of the electron excited to produce photon emission within the atom.

**(Essay)** As discussed in class and in the readings, what do the two single-photon experiments performed by Aspect tell us about the nature of photons? How were the two experiments designed to demonstrate the particle and the wave nature of photons? When answering, don't concern yourself with technical details (such as how the photons were produced); focus instead on how the design of each experimental setup determined which type of photon behavior would be observed. How are the elements of these two experiments combined in a delayed-choice experiment, and what do delayed-choice experiments (along with the two Aspect experiments) tell us about the nature of photons?

In the first experiment, two photomultiplier tubes opposed a third photomultiplier tube across a chamber of excited calcium gas. A photon pair would exit in opposite directions from an atom in the gas, where one photon would be detected by the third photomultiplier, and the other would head towards the other two tubes, where it would strike a half-silvered mirror, giving it equal probability of hitting either tube one or tube two. If both of the tubes were triggered at the same time more often than they were triggered separately, the correlation coefficient alpha would be greater than 1, and the results would support the model of light as electromagnetic waves. But if each tube were triggered individually more often than both tubes at once, the coefficient would be less than one, and support the particle model of light. In the first experiment, Aspect found that the coefficient was always less than one, and thus consistent with a particle model of light.

The second experiment used the same setup but introduced a second half-mirror in the path of the photon stream which prevented the collection of any information regarding the specific path of the photon. Initially results were the same as with experiment 1, but when the first mirror was moved back and forth, the number of photons detected in the first or second tubes changed in proportion, causing an interference pattern between the two to be observed, which seemingly indicated that the stream of photons was behaving as a wave (the photons somehow "knew" which path was possible, and adjusted accordingly).

The delayed choice experiment, by changing whether one or two paths were available after the photon had been emitted, demonstrated that depending on the number of paths available, the photon would either act as a particle or as a wave, but never as both. If two paths were available, interference would be observed, but if only one was available, the photon would act as a particle.

Together, the three experiments demonstrated that the properties exhibited by a photon could be either particle-like or wavelike, but the exact nature of those properties was dependent on how the experiment itself was conducted and how observations on the light were made.

## HW09:

**3.** How does thinking of electrons as waves rather than particles explain why energy levels in the hydrogen atom are quantized? How does it answer Bohr's question of why electrons don't radiate energy when they are in one of these energy levels?

Thinking of an electron as a circular standing wave around an atom places boundary conditions on exactly what that wave may be. In order to exist as a circular standing wave in any particular energy level, it must consist of an integral number of wavelengths (C=2*pi*r). Knowing that the De Broglie wavelength of a particle is (lambda=h/p, we can substitute out the wavelength of the electron, yielding C=nh/mv. By considering the attractive force between the electron and the protons in the nucleus, we find that for the electron to be in a circular orbit, v^2 = ke^2/mr. Making a final substitution for v demonstrates that the radii at which the electrons may exist, and thus the energies they may hold, are quantized:

r =n^2h^2/kme^2.

Electrons in those energy levels do not radiate energy because the quantization of the radii at which electrons may exist around the atom forces the angular momentum to be quantized as well (as an integer multiple of h-bar). A quantized angular momentum means that an electron may neither gain nor lose energy while remaining in a single energy level (as the radius (and hence energy, assuming all other aspects of the atom remain constant) of that level has already been shown to be quantized).

**Exam Questions:**

**Exam 2**

**E1. (10 Points)** In the sequence of screenshots shown below (taken from the PhET Quantum Wave Interference simulation), we see: A) a bright spot (representing the probability density for a single electron) emerges from an electron gun; B) passes through both slits; and C) a single electron is detected on the far screen. After many electrons pass through and are detected, a fringe pattern develops (not shown).

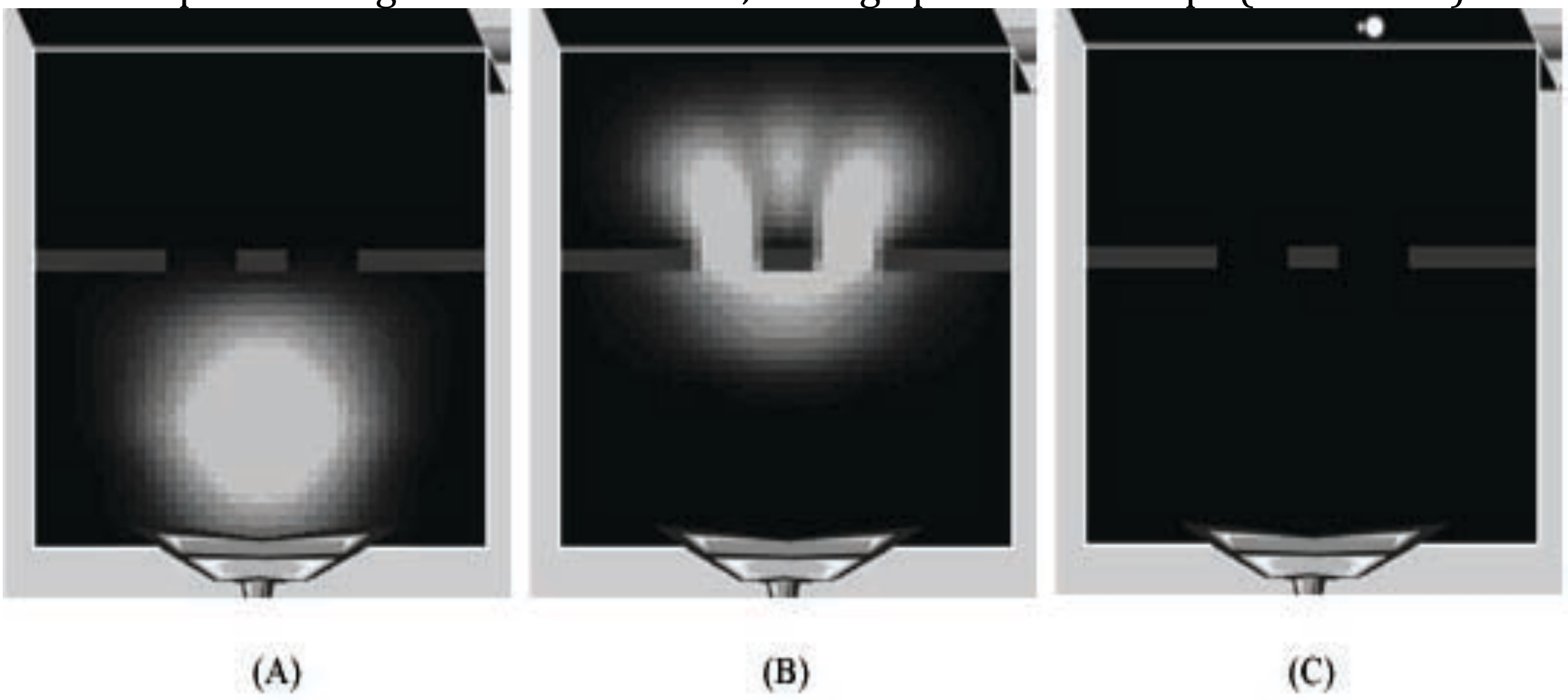

**Three students discuss the Quantum Wave Interference simulation:**

**Student 1**: The probability density is so large because we don't know the true position of the electron. Since only a single dot at a time appears on the detecting screen, the electron must have been a tiny particle traveling somewhere inside that blob, so that the electron went through one slit or the other on its way to the point where it was detected.

**Student 2**: The blob represents the electron itself, since a free electron is properly described by a wave packet. The electron acts as a wave and will go through both slits and interfere with itself. That's why a distinct interference pattern will show up on the screen after shooting many electrons.

**Student 3**: All we can really know is the probability for where the electron will be detected. Quantum mechanics may predict the likelihood for a measurement outcome, but it really doesn't tell us what the electron is doing between being emitted from the gun and being detected at the screen.

**E1.A (2 Points)** In terms of the interpretations of quantum phenomena we've discussed in class, how would you characterize the perspective represented by Student 1's statement? What assumptions are being made by Student 1 that allows you to identify their perspective on this double-slit experiment?

Student 1's statement is consistent with that of someone who holds realism to be true. He/she assumes that: 1) The electron was always a particle with a fixed position in space and time; and 2) The only reason that the probability field is so large is because we are unable to determine its position (a "hidden variable") prior to it striking the screen. Thus, he believes that the properties of the electron are always the same, but we (the observer) are only able to observe those properties under a given set of circumstances (when the particle hits the screen).

**E1.B (6 Points)** For each of the first two statements (made by Students 1 & 2), what rationale or evidence (experimental or otherwise, if any) exists that favors or refutes these two points of view? As for the third statement, is Student 3 saying that Students 1 & 2 are wrong? Why would a practicing physicist choose to agree or disagree with Student 3?

**Rationale/Evidence for Student 1 (aka EPR):**
Realism argument: all objects must have definite properties within the system regardless of observation. Location is real but hidden variable. Makes intuitive sense.

**Against Student 1:**
Idea of definite quantities for all states (Local Realism) does not hold to experiment. Probabilistic provides correct explanation, deterministic does not. Single-photon interference experiments.

**Rationale/Evidence for Student 2 (aka Bohr):**
Electron is a wave function that collapses to a determinate state at plate. Consistent with matter waves argument put forward by deBroglie. Allows for interference with only one electron.

**Against Student 2:**
Fails when applied quantitatively; no mechanism for wave collapse yet developed.

No, Student Three is simply stating the theory behind the interpretations put forth by the first two students. In other words, he is limiting his assessment of the experiment to what can be predicted and explained through existing QM theory. A practicing physicist would tend to agree with Student 3 because his description requires the least assumptions and adheres to what we know as opposed to what we postulate.

**E1.C (2 Points)** Which student(s) (if any) do you ***personally*** agree with? If you have a different interpretation of what is happening in this experiment, then say what that is. Would it be reasonable or not to agree with ***both*** Student 1 & Student 2? This question is about your personal beliefs, and so there is no "correct" or "incorrect" answer, but you will be graded on making a reasonable effort in explaining why you believe what you do.

I personally agree with Student 3. I see no reason to jump to a conclusion regarding the electron's behavior without a quantitative mechanism to explain its behavior between source and the plate. We know from this experiment that an electron exhibits behavior consistent with that of a wave, but we do not know exactly <u>why</u> or <u>how</u> that is so. That being said, I find Student 2's statement a more convenient way to think about the electron's behavior.

**E3. (8 POINTS TOTAL)** For the diagrams below depicting Experiments X & Y, M = Mirror, BS=Beam Splitter, PM = Photomultiplier, N = Counter. In each experiment a single-photon source sends photons to the right through the apparatus one at a time.

## EXPERIMENT X

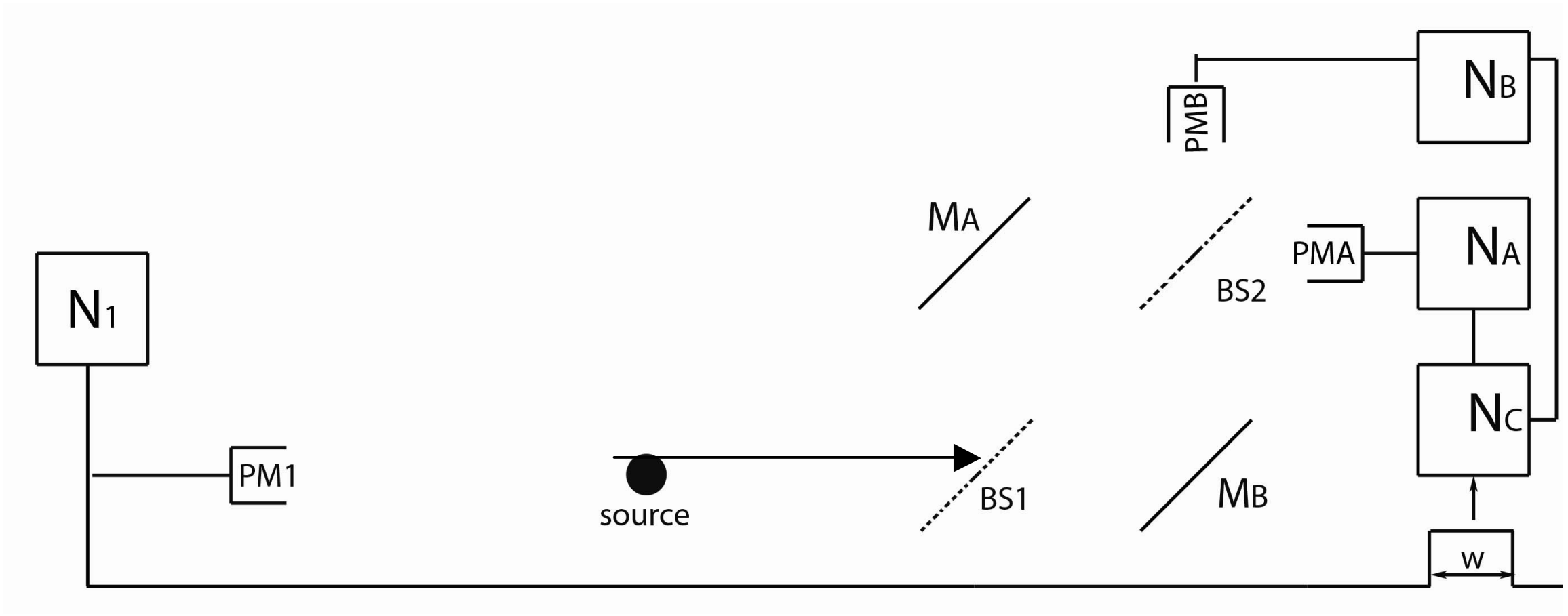

## EXPERIMENT Y

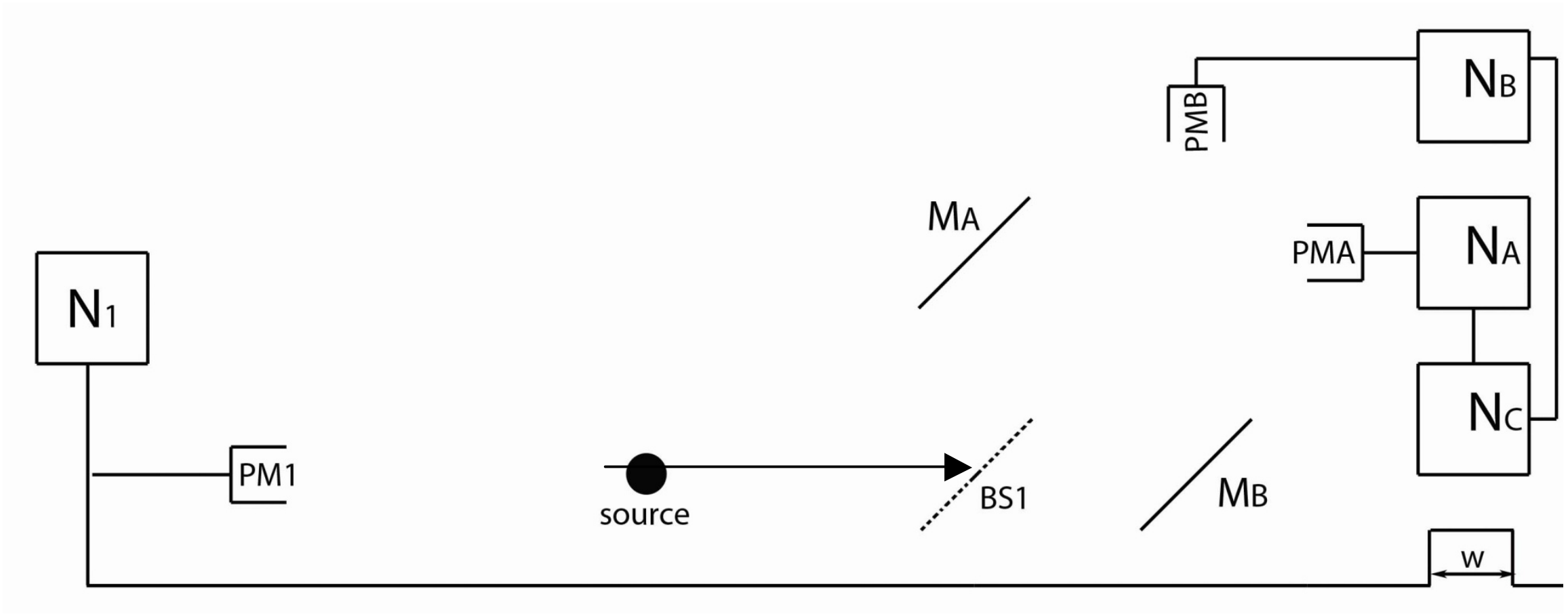

**E3.A (3 Points)** For which experimental setup (X or Y) would you expect photons to exhibit particle-like behavior? Describe in what sense the photon is behaving like a particle during this experiment. What features of the experimental setup allow you to draw this conclusion without actually conducting the experiment?

Experiment Y (Aspect's 1st Experiment)
The photon may take one of 2 paths, but not both, and thus travels along a defined path consistent with the behavior of a particle. The way the experiment is set up, a photon may only take one of:

source->beamsplitter->mirror A->photomultiplier A

source->beamsplitter->mirror B->photomultiplier B

If a photon is to be detected in PM1, its pair must have exited the source in exactly the opposite direction, and by geometry can only take one of the two paths listed above.

**E3.B (3 Points)** For which experimental setup (X or Y) would you expect photons to exhibit wave-like behavior? Describe in what sense the photon is behaving like a wave during this experiment. What features of the experimental setup allow you to draw this conclusion without actually conducting the experiment?

Experiment X (Aspect's 2nd Experiment)
The exact path taken by the photon is rendered indeterminate by the second beamsplitter; we can never know which path the photon actually took to PMA or PMB. If we vary the path length of A or B, <u>and</u> observe interference as a result in the detectors, a logical explanation is that the wave that represents the photon split at beamsplitter 1, and then (due to the difference in phase created by the changed path length) interfered with itself to produce the observed results. The presence of the 2nd beamsplitter essentially randomizes whether a photon traveling along path A or B ends up in PMA or PMB (50% chance of each for fixed path length) thus rendering the path of the photon indeterminate, which allows for the above conclusions to be drawn.

**E3.C (2 Points)** Suppose we are conducting Experiment X (the second beam splitter (BS2) is present) when a photon enters the apparatus and encounters the first beam splitter (BS1). Afterwards, while the photon is still travelling through the apparatus (but before it encounters a detector), we suddenly remove the second beam splitter (switch to Experiment Y). Can we determine the probability for the photon to be detected in PMA? If not, why not? If so, what would be that probability? Explain your reasoning.

Yes, the probability will be 0.5 (same result as Experiment X with equal path lengths, but with a definite path for any given photon).

A photon may exhibit either wave-like or particle-like properties, but not both in the same instant. Removing the 2nd beamsplitter "forces" the photon to exhibit particle-like behavior by making its path definite retroactively (example of a delayed-choice experiment).

# Exam 3

**E1. (10 Points Total)** A hydrogen atom is in its lowest energy state. Use words, graphs, and diagrams to describe the structure of the Hydrogen atom **in its lowest energy state (ground state)**. Include in your description:

- **(4 Points)** At least two ideas important to any accurate description of a hydrogen atom.

- **(3 Points)** An electron energy level diagram of this atom, including numerical values for the first few energy levels, and indicating the level that the electron is in when it is in its ground state.

- **(3 Points)** A diagram illustrating how to accurately think about the distance of the electron from the nucleus for this atom.

(On these diagrams, be quantitative where possible. Label the axes and include any specific information that can help to characterize hydrogen and its electron in this ground state.)

**[<u>DID NOT TAKE EXAM 3</u>]**

**POST-QA-Survey**

**[NOT ANSWERED]**

## Final Essay:

Personal Reflections on My Experiences
in General Physics 3 with Dr. Noah Finkelstein

I wish I could say that I chose to take this class solely because it interested me, but the simple truth is much less exciting. I decided to take this course for two rather mundane reasons: first, to "hedge my bets," as I had not yet decided on a specific major to pursue, and physics 2130 fulfilled requirements for three out of my four top choices (Engineering Physics, Astrophysics, and Physics), and second, because the class suggested for physics majors (PHYS 2170) was unavailable to me as I had not yet declared my major. Although I essentially stumbled into this class, it ended up being an immensely positive experience for me.

Upon entering the class, I was most excited to learn about the various interpretations put forth to explain quantum mechanical phenomena. I already had a fairly strong footing in the actual mathematics of the material, both from my own independent studies and from an exceptional AP Physics course I had taken in my senior year in high school. However, neither of those pursuits had given me a strong grounding in the overarching theoretical principles behind the material, especially when it came to interpreting the experimental data in the more recent work such as Aspect's single photon experiments and electron diffraction. I came in understanding the results of those experiments, but not their implications for the nature of light and matter. This class did a fantastic job of patching those holes in my understanding.

This class was unique for me in that it was the first course in which I have had fun during lectures. I have always enjoyed the science classes I have taken from a purely intellectual standpoint (the material fascinates me, otherwise I wouldn't be there), but this was the first science class I have ever taken where I actually felt relaxed during class and managed to just enjoy the material as it was presented. Usually in the science classes I have taken I am either severely bored at the pace of the material or frustrated in its incompleteness. The lectures in this class were both well-paced and structured in such a way as to allow for the exposition of material that was not directly presented. For example, the discussions of entanglement left me with a very good idea of the basic principles, which I could then use in the online readings to explore the implications it had for special relativity.

For me at least, the most unusual aspect of the class was the complete lack of hostility from my classmates when I asked long and/or slightly off-topic questions. In some of my introductory classes and in most of my classes in high school, questions I posed in class were met with either groans or brusque indifference. There was a strong sense that everyone but me just wanted to get on with the lecture, and that I was somehow a detriment to the class by always asking for more information. I never felt that while attending this class, not even once. Although with my lack of experience, I could find that this is just a pleasant aspect of academic life at the college level, I suspect that it had far more to do with the way lectures were given. I have never been in a class that was so interactive, and yet managed to keep up the pace of the material without erupting into a confused mess.

Out of all of the aspects of this course, I found that I derived by far the most knowledge from the lectures themselves. The way the material was presented allowed the audience to actually get involved with it in a way that made learning the material infinitely easier than it would have been if obtained straight from a text (a process with which I am intimately familiar). The powerpoints struck an excellent balance between the two undesirable extremes of dumb visual aide and lecture text. I found that to be particularly important during the two times I was sick this semester. As opposed to some classes (such as my introduction to astronomy class, where the slides were literally just a slideshow of images without explanation) the slides posted online for class lectures allowed me to glean most of the material I needed to remain current in the class, and the online reading materials and discussions helped me the rest of the way.

After the lectures, one of the things that this class got right where so many other classes falter was in the scope of its assessments. All of the midterms and the final were an accurate reflection of the material covered during the lectures and in the homework and readings. While this may not sound like a massive accomplishment, in my (admittedly very limited) experience I have had only one other science class that managed to make the link between the material taught and the assessments so well. I also enjoyed the online discussions, as they gave me a place to discuss material (such as quantum gravity) that could not be included in the formal curriculum due to time constraints. The remarkably high intellectual level of online discussions surprised me; after years of frustration on sites like physicsforums, it was nice to finally be able to have stimulating discussions with my peers.

I made relatively little use of peer instruction outside of the online discussions, as I have never been particularly good at dealing with other people face-to-face, and the comprehensiveness of the lectures, homework, and discussions left me without a need for extra help. The simulations proved useful when I was just being introduced to subjects, but I found that it was often much more informative to look for the limitations of the simulations; doing so allowed me to explore the nuances of the material in a way that the simulations themselves did not. For example, the simulation of a 1-dimensional potential well made me curious as to how a 3-dimensional well would function, which lead to further exploration of how a wave model of matter explains atomic structure and bonding.

Although this class has not significantly changed my ideas about physics and the practice of science, it has been one of the few courses I have taken that accurately portrays the scientific method of careful observation. The course was exceptional in how it handled conclusions drawn from experimental results, the most memorable example being the refutation of the "hidden variable" interpretation. The class was at its best when discussing the interpretations of experiments and the implications of their results; Aspect's single photon experiments were explained with particular clarity and care.

Almost every topic covered in the class was interesting to me in one way or another, but I especially enjoyed the weeks spent on Schrödinger's equation. Professor Finkelstein did an absolutely fantastic job of explaining the application of Schrödinger without becoming completely lost in the mathematics. But what I appreciated most was the fact that he managed to remain completely true to the material while simplifying it enough for it to be accessible to students without the extensive mathematical background needed to comprehend it fully. Physics classes have often maddened me in how they progress from inaccurate but simple models to slightly less inaccurate models without ever explaining why we use the simplified models in the first place, but in this course, the evolution of atomic models leading up to Schrodinger was explained methodically and exceptionally carefully. Such a focus on how less complex models fail to explain certain aspects of experimental evidence and are subsequently replaced by incrementally more accurate models was to me one of the highlights of the course.

That having been said, the class was not completely without its drawbacks. I found the material towards the very end of the course, specifically the information on nuclear weapons and diodes to be relatively distracting. I would greatly have preferred another two weeks exploring the nuances of applying the Schrodinger wave equation to multi-electron atoms, or even two weeks exploring the rudimentaries of quantum chromodynamics. Of course, a great deal of that has to do with my preference for the more theoretical side of the material, as opposed to the details of its practical applications (this was the primary reason I left Engineering for Astrophysics). However, I understand that physics 2130 tends to be geared towards engineering students as opposed to its counterpart physics 2170, and as such, the time spent on practical applications of quantum mechanics was completely warranted.

Of all of the classes I have taken up to this point in my academic career, this class was struck the best balance between spending time on the origins and theory behind the equations as on their application. It reached an excellent balance between introducing new concepts and ways of thinking and applying those ideas to reach conclusions about experimental results in particular and the physical world in general. It was unique among the classes I have taken in that it managed to seamlessly link the philosophy of science with its application. And finally on a lighter note, I don’t think I will ever forget Dr. Finkelstein pulling out the Darth Vader helmet for the tesla coil demo…

# APPENDIX E

## Collected Excerpts from Student Reflections (Fall 2010)

The following excerpts have been taken from ***each*** (all) of the reflective final essays written by students at the end of the semester. Students were asked to reflect on their experience of learning about quantum mechanics, and to discuss which aspects of the course were most effective in their learning.

**#3:** I would like to begin by taking this opportunity to share what a pleasure it has been taking this course this past semester. Although, admittedly, I had only minimal expectations coming into this class, I was simply amazed by the sheer amount of material I learned this past semester and its applicability to so many of the facets of my major, computer engineering. The structure of the course, the introduction to the mathematical formulations of quantum mechanics and the presentation of theory/interpretation—all—did well to impress and inspire me. [….] I was constantly under the impression that I could succeed at quantum mechanics due to the nature of this class. Whereas in other classes a student may be more concerned with obtaining a good grade versus learning the material, in this class I never felt that there was an issue with achieving both a good grade and a sense of complete understanding.

**#4:** Coming into this class I was excited, I was about to be brought up to date in the physics world. Newton's laws of motion and the basic laws of electromagnetism were interesting enough. It all seemed quite tangible and practical; however it also seemed quite old. It grows tedious to be studying things that were discovered more than a hundred years and sometimes many hundreds of years ago. I wanted something a little more modern. I dreamed of being on the cutting edge of physics, having the tools at my disposal to understand the questions and the advancement of the field. […] Overall it was a great class; I liked the emphasis on interaction and the energy of the lectures. This quickly became my favorite class. I loved the emphasis on how we come to know something and scientific reasoning rather than just knowing it and the integration of historical perspectives into the class

**#12:** Taking this class is probably the best thing that could have happened at this point in my career. As a future engineer, quantum computing is my future. …My future is in computing and micro-processing, if not directly then at least on the end user side. Thus, my future is in quantum computing.

**#13:** Even though quantum mechanics does not directly relate to what I want to pursue in engineering I still think it is important to be able to understand and talk confidently about all aspects in physics. […] Throughout the course of this class, I have found that my favorite area of study ended up being entanglement, probability and Schrodinger's cat…

…The more we got to read the articles assigned, and discuss the theories both in lecture and in the problem solving sessions the more everything began to click. The combination of being able to study the same things in multiple ways with several different points of view helped me the most. …I ended up finding myself wanting to know more about several topics…All in all I'd say that I am walking away from quantum physics with much more then I came in expecting. I was dreading having to take another physics class and I did not think that I would be interested in anything we covered, and I figured it would be mostly over my head. I have heard from previous students what an awful experience physics three can be…, I have found a new passion for physics.

**#19:** In my mind, quantum physics has always been that big scary topic that smart people talk about but nobody I know can really understand. Indeed, the title of the subject has become synonymous with brainy people who can do math and science that is way beyond me. I am a Mechanical Engineering major, and I was compelled to take this course by my advisor who insisted that it would be in my best interest to learn quantum mechanics in order to expand my mind and learn a new way to think about things. In addition, he promised me that if I didn't take it, I wouldn't be able to graduate. With such academic pressure to take the course, I decided to take the plunge and try to tackle the mechanics of things that I cannot see. […] The moment that contributed most to my new understanding about the concepts of quantum mechanics occurred during the discussion of the "many worlds interpretation" and indeterminacy. […] The presentation of the material in this course was as good as it gets. There were so many avenues for learning the material that any student who has failed has only himself to blame. I found the lecture to be immensely beneficial as it was clear that the professor put a lot of time and energy into the preparation each class. During the class, the clicker questions facilitated discussion amongst the students in the class, and this lead me to have to explain concepts to others. I believe that this is one of the best methods of completing the mental model of a new concept as it forces me to fully understand it before vocalizing it. In addition, the peer instructors (LA's) were very helpful in encouraging discussion amongst the students as well as ensuring that students were working in the right direction towards the correct answer while not giving away any shortcuts that would end the learning process. The simulations were very helpful for all students who are more inclined to learn from visuals and need to see concepts to understand them. The only medium of instruction that I found to not be of use was the textbook, from which little was drawn for the course and which I think could be eliminated from future versions of the course. All in all, the success of the class in this semester's PHYS-2130 course can be attributed to the skill and educational knowledge of the professors, and their field of educational research should continue so that all other students can benefit from it as I have.

**#20:** Physics 3 has been a fun course to attend and has changed how I view the world around me. Professor Finklestein did a fantastic job teaching the course and had many traits that I really liked. He was able to keep the whole class engaged in discussion and was able to keep us coming back to class. Professor Finklestein also used many teaching aids, like simulations, that were very helpful this semester in understanding the material. Like most classes, we learned a fair amount this semester and some things were more interesting to learn about than others. […] I personally feel like Physics 3 has changed my ideas about physics and the practice of science an incredible amount. […] The only reason I took Physics 3 this semester is because it is a required course for Electrical and Computer Engineers and because I failed the same course last semester. This fact made believe that I wasn't going to learn much new and that the course was going to be boring. However, I was completely wrong. […] Professor Finklestein made the material fun and made coming to class a pleasure. One thing I found particularly helpful throughout the semester were the simulations. The simulations always helped me really understand whatever we were learning. The nicest thing about the simulations was that it allowed us as students to experiment with the concepts that we are being taught. That way if we are curious about what something will do when the experiment is changed a little bit, we can run the simulation and observe the outcome instead of having to read through an entire chapter and do a bunch of math to understand what would happen. The simulations this semester made the concepts a lot easier to learn and are a very powerful teaching aid. […] I was convinced that retaking the class this semester was going to be extremely boring. I was very wrong.

**#22:** Coming into physics III as a transfer student, I did not know what to expect. After meeting new people and hearing about what they had covered in this class, it appeared clear very early in the semester that this semester's curriculum would be very different. Coming into this class, I knew in the back of my mind that quantum mechanics would be a subject that we would touch. What I did not expect was that we would be spending the entire semester on it. Of course I had questions at the beginning of the semester, and with no previous background in quantum mechanics, my list of questions grew as the semester progressed. […] Instead, this class has enabled me to understand how physics is practiced, and how it works in today's world. A particular example of something that fascinated me was the concept of how lasers work. […] The teaching techniques that left the best impression on me were the well-put together lectures via PowerPoint, as well as the time intensive homework. The PowerPoint lectures were very easy to follow. Because they were clearly written, the basic concepts were easily taken away from each day of lecture. Also the fact that the lectures coincided with the homework very well, made the long homework assignments a little less nerve-racking. What was most helpful in making this class a better experience for me was the lecture and homework reviews…

**#28:** Over the past few months I have learned much about the world in which we live, and the new and exciting areas of study and work. For the last hundred years, physicists have made many great contributions to our knowledge of the universe and the mechanics of its operation. Many of these contributions seem to be outside the scope or validity of classical physics. This course on Quantum Mechanics has really opened my eyes to the current possibilities, and where this exciting field could go in the future. […] Before beginning this class I did not know that physics research was so alive, and the directions it is going in, and its continued importance. This class helped me realize that there will always be a need for pure scientists to research and pioneer new fields of study and thought. This will enable engineers to take these advancements and use them for good. […] Overall I feel that this class will help me in my pursuit of a degree in mechanical engineering. I feel that the more in-depth understanding of how our world operates, from this class, will help me to be a better engineer and contributor to society. While I am not changing my major to physics, I still feel that the class made a significant impact on me and I would definitely like to learn more about many of the topics covered in the class.

**#29:** Truthfully, when I found out that I had to take Modern Physics, I was very apprehensive. Since I switched my major to Biochemistry last year, my advisor told me that algebra based Physics I and II were not sufficient to graduate with a Biochemistry degree. As a result, I had to fulfill my physics requirement by completing PHYS 2130. Physics I and II were challenging enough so I was expecting this class to be just as hard. However, now that the semester is almost completed, I can say that my journey into the quantum realm has been more than fascinating. Modern Physics has imparted a practical knowledge of chemistry and the physical world around us; which most of my other core classes do not accomplish. The level of detail that the curriculum provides will be a stepping stone of knowledge that will help me build my career as a biochemist. Before taking this class I had the slightest clue about bonding/anti-bonding orbitals or electron distributions in various bonds. Though these concepts were not intuitive for me, I can proudly say that learning these complex ideas has not only improved my understanding of chemistry but has also given me further confidence as a biochemist. […] Quantum mechanics has changed my perception of science and the way which I will practice Chemistry now. From day 1, the electron has been one of the main focuses of Quantum Mechanics. […] Another way Quantum Mechanics has taught me more about Chemistry is through the Schrödinger equation for the hydrogen atom. The solutions to this equation for the hydrogen atom give the form of the wave function for atomic orbitals, and the relative energy of various orbitals. Without the Schrödinger equation, I'm not sure how this data could be obtained besides through experiment. These solutions are also relevant when understanding the orbitals of other atoms such as helium, lithium, and carbon, etc. Furthermore, the Schrödinger equation can be extended to transitions between states. Although Chemists had already described the rules for the number of electrons allowed in each shell, they didn't know why! Beforehand, my method of thinking about Chemistry was based on the rules I've been taught. Now I can think about the physical reasons why I am using these guidelines. [….] All of the educators in the help sessions are not only familiar with the material you are studying but can effectively help you either understand it better or use your resources. For example, once when I was stuck on one of the problems on the Quantum Tunneling tutorial, I asked Danny a question and he instantly knew that I should have my quantum tunneling PhET open to find the answer and enhance my understanding-which it did! […] The help room is an unintimidating environment where I can feel free to ask any question, because the students around me are most likely working on the same problem. […] Although I did not initially plan to take this class, it ended up being extremely advantageous to my comprehension of chemistry. From my experience, it should be a required class for Biochemistry majors, regardless of whether they have a deficiency in calculus-based Physics or not. It's comical that on the first day of class, everyone I met was some kind of Physics or Engineering major. They all seemed confused when they asked me what my major was. Little did I know that the Quantum Mechanics material had a tremendous relation to what I study and they were the ones who ended up learning more about the nature of my major, Chemistry. I like to think of Quantum Mechanics as the final puzzle piece to my education.

**#30:** One of the most important motivators for me, when first starting a course, is finding a topic that is fascinating. This came very easily in modern physics because of the different ideas that quantum mechanics offers. The initial spark, before starting the course was my thought of how classical physics will build upon itself in modern times. To my astonishment, I soon found out that we were leaving behind the classical view and moving into a quantum perspective. Of course this completely destroyed my initial questions about the course, but gave me a newly found ideology of physics. The experience of learning quantum mechanics has been excellent in many different ways. […] Many of the tools that this class offered were extremely helpful while learning quantum physics. In particular the study sessions each week gave us real help in figuring out how to approach particular problems. I would like say, thank you to Professor Finkelstein, Charlie, Sam, and Danny for their time spent helping us understand and work through quantum mechanics. Another helpful tool were the simulations that we used to answer problems because they are easy to use, relevant to the topics, and were great visual adds in class. […] All in all I would like to say that I really enjoyed this course and would suggest it to anyone who is interested in reformatting their ideology of atoms.

**#31:** Day by day I am building my dream. When I was a young kid, I would play with Legos, and my dream was to build a Lego spaceship to go to the far reaches of the universe. As I have grown older my dreams have changed. My dream now is to be part of the designing and building of prosthetic limbs to help the disabled. To do this I must become a mechanical engineer. One of the steps to becoming a mechanical engineer and accomplishing my dream is taking Physics 2130. Additionally, I decided to take the class because I enjoy physics, and I thoroughly enjoy building a greater understanding about the world around me. I love learning about concepts that completely change my way of thinking. For example, the section on the wave particle duality of objects such as the electron was eye opening. It is this type of concept that completely defies all common sense and classical physics that makes me feel as if there is an entirely new world waiting to discovered. If such basic definitions as being a wave or being a particle can be broken than what other common notions are waiting to be redefined? **[…]** Throughout this class I feel that my basic idea of physics and logical thinking has been redefined many times. The largest paradigm shifts I have experienced in this course occurred when learning about undeterminable characteristics, entanglement, and tunneling**. [**…] Besides the actual concepts in the course, I greatly enjoyed the way the course was structured and taught. The single best aspect of taking Physics 2130 was the low stress environment. In engineering almost all of my classes are extremely stressful, and due to the stress I do not enjoy them. Too much pressure builds up throughout the year, and too much of my grade depends on the exams. […] I have enjoyed my time taking Physics 2130. I feel that it has helped me move closer to accomplishing my dream of helping build prosthetic limbs and becoming a mechanical engineer. Just as important, I feel that the surprises in quantum mechanics have added more wonder into my world, and for that I am thankful.

**#35:** To be completely honest, I signed up for PHYS 2130 because it is a required course for all electrical & computer engineers at CU. I assumed it would be like the Physics 1 course I dreaded during my freshman year…However, thanks to the dedicated help of Professor Noah Finkelstein, Charles Baily, and our two learning assistants Danny and Sam, this physics class goes down as one of the favorite courses I've taken during my college career. […] One really neat thing about this physics course was it legitimately had an effect on my education overall and the manner in which I practice science. As engineer students we are primarily trained in numerical analysis. Most often in our course work there is only a single correct answer to a problem, however this class has shown me that in practice things are rarely so black and white. PHYS 2130 demonstrated that the process is more important than any numerical value. Thus in my other lab courses I am more willing to accept my experimental results and attempt to explain any abnormalities instead of just redoing the experiment. Thinking back on the semester, I believe the Single Photon Experiment discussions were the most effective in teaching me about the importance of the process. […] In closing, I felt that this course was enjoyable and very effective. As with any course there are some things that can be improved upon such as textbooks and implementation of clickers. However, with PHYS 2130 these things were very minor. What stands out most in my mind is the overwhelming quantity of resources provided to the students. These resources, in addition to a great instruction team, really made the course very effective.

**#39:** As the semester comes to an end, another sixteen credits of classes are on the verge of being added toward my degree. Of the five classes taken, the class being reflected upon is my Modern Physics 2130 class. Modern Physics provided a great learning opportunity and I leave the class soon in knowing that much more than just three credits were acquired. In exploring the impact of the class upon me, my expectations coming into the class will be examined to illustrate my view of physics and science before the class. […] Acquiring knowledge about the quantum mechanical world through Modern Physics caused a shift in the way the scientific world is viewed. Learning about the implications of the double-slit and Aspect experiments along with the electron's wave function literally opened up a whole new world, the quantum world. […] Taking Modern Physics 2130 during the fall semester has provided me with valuable knowledge and a new fascination with physics. The class met and exceeded my expectations coming in, supplying new interest and answering the basics of what the quantum world consists of. Learning about lasers, entanglement, and tunneling was an eye-opening and exciting process, aided greatly by the use of simulations, lecture notes and homework to allow a thorough understanding of all the phenomena. With a quantum view, I see the world in an exciting new way, including informative explanations of the physics behind objects such as lasers and transistors along with a better comprehension of the science behind atoms and the periodic table. The knowledge of quantum mechanics has left me wanting more. While my schedule lacks a physics course next semester, I hope to commit to reading about physics when time allows. I leave Modern Physics 2130 with a keen understanding of the basics surrounding the quantum world and a newfound joy for making sense of the world around me.

**#40:** So naturally this year when I realized I would be taking Physics III – Quantum Mechanics, I was a little intimidated. Little did I know that it would end up being one of my favorite classes ever taken at CU. At the end of the summer when everyone is getting ready to go back to school, a common question is "so what are you taking this semester?" The majority of my friends being business or arts and science majors would respond with the common history, economics or writing classes. So when I would tell them physics III - Quantum Mechanics, it was common to get their eyes to bulge or an "excuse me?" And to be honest I couldn't really tell then what that meant either. It didn't bother me at first but then as class got closer it started to set in. I barely got by one and two, and now I have to take three. But I needed to take the class to get my mechanical engineering degree, so weather I liked it or not, it was going to happen. […] In the end, I had a great quantum mechanics experience. I enjoyed going to class, learned more than I thought was possible and restored my faith in physics. After talking to people who have taken the class in the past, I only wish we could have gone over relativity. It has always seemed like an interesting topic and I just wish we could have discussed it a little. But if we did do relativity, then we might not have gotten to diodes, so I guess it was for the best. I worked hard this semester to finally do well in physics class but I certainly did not do it alone. I did well and learned so much thanks to the hard work of professor Finkelstein, Charlie and Dan. Without them this experience would not have been the same, and for that I am truly grateful.

**#43:** A love hate relationship, best describes my interactions with Physics 3. I had no idea what I was getting my self in to while signing up for the course. My previous knowledge of the class consisted of what my fellow mechanical engineers had informed me. Most of them agreed that the material was really "cool", "trippy" and a lot of the world's ways were explained. The previous semester of physic 3ers had studied relativity, a little bit of Schrödinger's equation and a few topics I cannot recall. The consensus was relativity was really easy but even after completing the course, no one knew what the hell Schrödinger's equation was or how to use it. Schrödinger's equation was described to me as the enormous equation that was written on the board for 3 weeks that no one had any idea what any part of it meant. They obviously did not have Professor Noah Finkelstein or Professor Charlie Baily. I originally had not planned on taking physics 3 due to the fact that mechanical engineers had the option of either take physics 3 or any other science elective, which include organic chemistry or anatomy. Since I am partaking in the bio medical option in mechanical engineering here at CU, I have to take both organic chemistry and anatomy. The fact that I had no requirement for this class and still opted to participate should be flattering. […] The one thing I can say about Physics 3 that I cannot for any other class I've taken so far is that when I tell my dad what I'm learning he doesn't know it all already! My dad got his Chemical Engineering degree from University of Michigan then went on to Medical School at U of M as well. He knows more about everything than any one I've met, which is great but I never get to share new concepts and information with him. When explaining quantum physics though he doesn't already know what I'm going to share him before I do unlike thermo, fluids or any other engineering course. He was clueless about Schrödinger's equation and had no idea that electron tunneling existed! This is one of the greatest feats so far in my life, and I owe this to Noah Finkelstein, Charlie Baily and Danny. […] I also recently just received an internship with an optics lab, XXX. They coat glass to make mirrors for "lasers". When touring the lab I was very proud to say I know how lasers work!! Not every engineering student can say that. I will be working for them all next semester and in to the summer. I chose to write the personal reflection not because it was the "easy" way out but because I wanted to share my great thrill of what I learned in this course. I have had more fun, learned more relevant topics than much of my engineering course combined and for that I thank you!

**#44:** Of all the required classes on my curriculum none were as hated as Modern Physics, I had already labeled it as the most boring and complicated class, the Mt. Everest of classes. The first thing that came to mind was Albert Einstein in front of chalkboard writing out mind numbingly large equations, him frazzled and bordering on insane... I knew this would be me for the next four months. Talking to my roommate about Physics 3 last semester had really scared me especially when he said it was hard for him, I have never seen a class that was hard for him, I was literally quaking in my shoes. As I walked into the class the first day I was expecting a professor as exciting as watching paint dry, it is never my luck to get physic professors that are interesting. What I got though was completely different, he actually understood fun. There will probably never be a better start to a first day than by starting with a game show with an actual prize. This I could tell would be a good year. […] This semester has definitely been a learning experience and even more so my view of physics has changed a lot. At first due to thinking in a classical sense I figured that if two answers are completely different then one must be wrong but I have learned that in quantum mechanics there are actually some things that are best explained with two completely different answers. At the beginning of the semester I found it impossible for particles to be waves and waves to be particles, this is just counter intuitive to everything you experience in life. I found this hard to believe because that is what I have always liked about physics is that it is easy to connect to what I notice and perceive. Though after going through this course Wave-Particle Duality explains a lot about how light and even electrons interact with themselves and other objects. […] This class is definitely on my list favorite classes I have ever had in my life, the atmosphere, the subject matter, the instructors have made this a very positive experience in what I feel has been bland and boring experience previously. With all this new information I feel like I can do some more research and engineer products that will benefit people and all this is thanks to just learning more about the building blocks of this world. This class has been so cool I even contemplated changing majors but I am refraining because I feel my calling is in engineering of some kind.

**#45:** My experience learning about quantum mechanics has been both baffling and enlightening to say the least. Although I had no great expectations of bewilderment, that's just what I got. Not only have my perceptions of science and physicists been adjusted, I have learned a lot about scientific theory and the current extent of knowledge we have of the physical universe. Additionally, some of the most effective employed teaching techniques this semester, in descending order, were peer-instruction, textbook reading, and the concept tests. […] In addition I believe this course has changed my ideas about physics, mainly through the insight offered into scientific theory. It has therefore greatly refined my practice of science as well.

**#47:** Although some people think that quantum mechanics is a very boring and difficult subject, I found it to be a very interesting topic to learn about. As an engineer, most of what I study ends up being taught primarily with numbers therefore a lot of the work I have done in the past couple of years has been computational. Honestly, modern physics and quantum mechanics was a very nice break from all of the number crunching because there was a lot more theory involved with understanding topics that are covered in this class. As Carl Wieman says, "Quantum mechanics is the greatest intellectual accomplishment of human race." From what I have learned in this class that quantum mechanics is used in about a third of all the engineering that surrounds us, I would definitely have to agree with Carl that it is a very important achievement by humanity. Since quantum mechanics is so important to how people live their lives, I found myself being able to engage with the class a lot better than most of my classes. […] This class has changed my perception of science and has definitely changed my ideas about physics drastically. **[…]** I think that the structure of this course was designed very well. I think that how this class is set up allows students to access the tools and information thus helping them to succeed in this course. The website for this class is probably the most useful website that I have had throughout college. It was extremely helpful to have all of the lecture slides posted on there for reference, and it was nice to have all the simulations that our teachers provided us to make the learning process a little bit easier, and more enjoyable. Also, I found the help room to be very helpful in allowing students who want to do well on their homework, and who seek further explanation about topics. Without the help room, I would have been a lot more confused about a majority of the topics that we covered in class. On top of that, I found that the clicker questions were helpful as well. The clicker questions keep students who go to lecture but have trouble staying engaged in lecture on their toes and it keeps them alert. I would know, im one of those students... Another good teaching technique that was used in this class was giving out candy to students who interact with the class. Not only does the sugar stimulate our brains for a few minutes, but the candy also acts as a pretty good motivating factor for student to speak their mind. The teaching techniques used in this course were very beneficial and were appreciated greatly. Overall, I was very satisfied by this class. I feel that I have learned a lot more about physics, and I have developed a different perception about how science works.

**#48:** When deciding whether to switch from Aerospace Engineering to Mechanical Engineering, many things were taken into consideration, but one aspect which I was quite worried about was having to go back and take physics 3 or quantum mechanics, which is not required for aerospace engineers, and I had never had any plans to take. I had made it through the first two required physics courses quite easily. In fact, I enjoyed the content of the first two physics classes, and the content of these is what I am largely interested in pursuing for my career. I understand that when you drop a ball it falls, that when you put two magnets together, they will either attract or repel based on the polarity, and that when you hook a light bulb into a system with a battery (correctly) it will light up. But what makes up light and why does it make a pretty pattern when it is shined through two small slits, I did not know and have never been able to grasp. My knowledge (and many times interest) tends to stop at things I can see and know the effects of, and I have always been a fan of things I understand. However, I am very glad that I did switch and got the opportunity to take this class, because I have learned so much, and never knew that something that I had never really dealt with and had never thought about, or for that matter never knew existed could be so interesting. […] That was all I could think about coming in to this class: that this was going to be a class that I just needed to get through for credit, which I was not going to like it and it was not going to be easy, but if I put enough effort into it I could at least make it to the end. As it turns out, I had spent way too much time worrying about this class. If I could go back and let myself know that this class was going to be a lot of fun, that I was going to learn so much, and that in the end I would actually understand it all, I definitely would, as it would have relieved a lot of unnecessary stress and contemplation. […] The rest of that first lecture also made an impact. I agreed with what the statistics about traditional lectures, that I really don't take away much more than 25% of what is taught in most of my classes, and to try a new format would be a better way to absorb information, with more participation and interaction between all. […] Overall, I have enjoyed this class very much and hope to find more classes that are taught like this one, with professors who really seem to enjoy the subject that they are covering and a more interactive structure as this one uses. I also feel like I need to subscribe to Scientific American magazine to find more articles like that one's assigned as readings in this class, and then take another modern physics class to be able to discuss all of these things. I am very glad I was able to take this class, and am sad to see it ending.

**#50:** "Physics 3 for Engineers" was a required course for me to take through the mechanical engineering program. I've have enjoyed the concepts presented in previous physics courses, but have had trouble being motivated in learning some of the topics not because I found classical physics boring, but because most were taught in very plain, dry and uninteresting ways. Consequently I found it difficult to be interested in the course, though the material itself was of interest of me. Coming into Physics 3 this year was certainly daunting as I expected more of the same difficulty staying interested. Additionally, with "Quantum Mechanics" in the course title, I felt intimidated and unsure if the combination of such foreign topics and an unappealing course structure would allow me to make it through the semester. Fortunately, the teaching style was far from that and the enthusiasm I saw helped me feel comfortable and motivated. […] I feel that quantum mechanics has opened new ways of thinking for me, as in continuing the expansion of my universal perceptions, but also between physics, chemistry and the like. […] Overall, I had a very rewarding and productive semester in physics 3. With every student having different learning styles, I felt that having all the different resources available such as help room, the textbook, the simulations, lecture, etc. catered to any students learning habits. Of course one needed to use all of them effectively to do the work and grasp the concepts, and for that reason I think I improved my ability to learn through mediums that are not my strongest suit. Topics covered in this course, though challenging and often tough to make sense of, were things I am glad I learned for reasons other than that I had to learn them for the class. The innovative thought of the pioneers of quantum theory and the questions they struggled to make sense of relate very closely to the sense of wonder, possibility and mystery that I think everyone feels about the world around us.

**#55:** Entering into Physics III, I had no idea what to expect. Fresh from a rather tedious struggle with Physics II, I honestly began to question my desire to pursue an Engineering Physics major. […] I entered Physics III with a bitter taste in my mouth. Yet, some fragment of my mangled ego compelled me to continue down the path I was on. I have always found physics to be the most intriguing subject, and I was not about to let one class ruin it. I approached Physics III as the deal breaker: if this class was like its predecessor, then maybe mechanical engineering was a more apt major. Almost immediately I knew that this class was not like the others. After some review, we began learning about the photoelectric effect…. After almost every class, I felt the compelling urge to tell anybody who would listen long enough about everything I learned that day. This course definitely changed my ideas about physics. It completely boggled my mind. Things physically impossible were not only achieved but used to create all types of intricate technology. This class was the first class where I was asked about my opinion on something. I cannot express enough how overjoyed I was to compose a paragraph in an engineering class. Usually in classes I am told that the presented material is fact and to learn it only to reproduce something that numerous scientists or mathematicians have already proven. I was presented with the opportunity of creativity, which is a rarity in the engineering world. […] You can be the brightest person in the world, but if you cannot express yourself nor have a fear of doing so, your intelligence become irrelevant. This also helped me as a female engineer. Every so often we encounter male engineers who do not always readily agree with the females. My clique consisted of two prodigious boys. But I quickly learned that I had to be assertive enough to express myself. After getting numerous questions wrong that would have been right by my logic, I decided to reevaluate my strategy. I remember distinctly the first time I refused to change my answer. My two teammates had decided on another, but my refusal meant that none of us would have gotten points. I was scared to stick by my answer, but I was confident I was right. I essentially forced them to change their responses and copy mine. After the answer was unveiled and my correctness confirmed, they were thankful and less hesitant to agree with me from then on. […] Overall this class continually impressed me. Throughout the course, the almost magical results quantum mechanics attained reassured me that I am in the correct major. The teaching style in conjunction with the material made quantum physics attainable. I am not sure if it was the teaching that rejuvenated my passion or the material itself; either way I welcomed back my old friend, physics, with open arms and anticipation. I cannot wait to continue further in quantum physics,

**#56:** Learning quantum mechanics through this course was a very enjoyable and enlightening experience. At first I only took this course because I had to as an engineer and didn't have much of an interest because when I took physics 1 my freshman year, and it made me dread the subject. Learning about incline planes and bullets hitting ballistic gels wasn't exactly the most interesting and fascinating subject. Initially before taking the course my main questions were, what modern physics was even about, what topics will be taught, and will it just be a repeat of physics 1. Once classes began and I attended the first lecture I was pleasantly surprised on how the subject was taught. The class was very engaging and kept me awake because there were a lot of chances for talking with the people around me to discuss a problem, and then use our clickers to answer them. This way of teaching was a much better approach for me than with physics 1. In that class the teacher was just lecturing and deriving equations the whole time. […] The modern physics course this semester was a great class in all honesty. The professor and Charlie were great assets into the learning of the topics presented to us. All I have to say is keep up the great work and there is really nothing needed to change about the course, anyone who is going to take it will have a great time and perhaps, like me, will learn to like physics.

**#59:** As a new transfer student, I had no idea what to expect for class at CU let alone Physics 3. After all the great class discussions and useful help-room hours, the semester is now coming to an end. I enjoyed this course and learned a lot about the quantum physics world and how it applies to engineering topics. The quantum physics world was only a mystery to me before Physics. […] After I receive my Bachelor's degree in Mechanical Engineering, I plan to join the Navy and become a Nuclear Propulsion Officer for 5 years then work as a civilian for an engineering company with a contract to develop products for the military. …As I mentioned earlier, one of my top motivations for taking this course was to understand nuclear reactions and how to apply them to engineering. Learning about neutron-induced fission was my favorite part of the class. Discoveries such as neutron-induced fission and the material covered in this course have given me a new perspective on physics and the practice of science**. […]** The topics that we learned about and the teaching techniques are what made this course material interesting and easier to learn. The topics I found most interesting were either major breakthroughs in quantum mechanics or closely related to engineering applications. I enjoyed learning about the struggle between Einstein and Bohr and their clashing opinions on explaining the behavior of particles. When we first started talking about this struggle, I had hoped we would get into Einstein's Theory of Everything and String Theory but I still feel that I know enough to be able to research and learn more about these topics on my own. I also plan to receive a minor in Electrical Engineering here at CU and the diode topic only helps me in obtaining a more in depth understanding. I was able to learn these interesting topics easier through the various teaching techniques. The class discussions forced us to think more in depth about the topics being presented before us. We were coming up with our own ideas to explain the results of the many experiments learned in class, thinking and discussing topics as if we were the pioneer physicists who got us to where we are in quantum mechanics today. The peer to peer discussions in help room hours allowed us to discuss, learn, and teach these topics amongst ourselves. When we got stuck, we had help available from the professors and learning assistants to give us that extra boost. The class material was supported well by the various teaching techniques which allowed us to get the most out of this course as possible. Overall, I thoroughly enjoyed this course. I felt that Professor Finkelstein, Charlie, Sam, and Danny were great teachers and made the class more enjoyable and interesting. I learned a lot and I'm now eager to find out more about quantum mechanics and were it will take us in the future.

**#60:** This class has been a wonderful, interactive experience with quantum physics that combines the ideas of classical mechanics and E&M along with new ideas and reasoning to explain quantum mechanics in a way that was very easy to understand. […] The first day of class was a strange experience coming in as a freshman and seeing only one person I knew in there. It was intimidating at first, especially since I had questioned my physics knowledge prior to this class. […] Overall, I thought that the class was incredible. I'd recommend it to everyone in engineering, as it gives a good understanding of a very interesting, complex topic while being really fun. I took the class simply because I had to for my major, but as time went on, I would have to say that my interest in the topic has greatly increased, as well as my knowledge of the subject. I'm very glad I took the class, and it has kept me interested in physics overall. It's helped me decide that I'd stay a mechanical engineer with a good understanding of the universe – it'll help with a chemistry understanding of materials as well as an understanding of circuitry and electricity and happenings on the subatomic scale. It's been a wonderful class, explaining things in a way I can easily understand and ask questions about if I don't understand something. It really helped the fact that it was so fast-paced, as I felt comfortable going at the rate the class moved and was amazed to find myself understanding the material so well, even things I didn't fully understand made enough sense that I could apply them using a conceptual thought process. I'm very glad I took this class. It's been a great experience for college so far.

**#63:** As one of the first college classes I had taken I can say I was a bit skeptical about taking quantum mechanics. I had no connections to anyone who has ever taken quantum mechanics or even to people who thought it was easy. What could be easy about Einstein's or Hawking's works, two of the most well known and accepted physicists and theorists? However, there was no need to worry. […] "100 Years of Quantum Mysteries" the idea of quantum cards really changed how I looked at physics. If each physical occurrence, such as neurons firing, is in a superposition state and is "perfectly balanced on its edge and falls down in both directions at the same time", then the outcomes are limitless. It is a factor of the wave function. […] I am pleased with this new information that I have gained from topics I was not aware even existed. Quantum mechanics was an excellent class to transition to from high school. The lecture was amusing and helped to reiterate some of the wording in the book. The weekly homework helped to reinforce the material that would be covered on the test. The help sessions facilitated learning on difficult problems presented on the homework. The staff of this physics class was always available to work through problems using a hands on approach which worked well for me. When all this was coupled with helpful illustrations and simulations, it allowed my mind to fully grasp certain difficult topics. This framework helped to develop good grades from beginning to end and definitely helped my success in the class. I can definitely say that without the help sessions I would be much more hard-pressed to learning the material and submitting the homework by due date. I feel that after this semester I have a full understanding of the material presented by quantum mechanics.

**#64:** I am very excited to have the opportunity to discuss my experience learning quantum mechanics in this course. I have talked to friends and family all semester about how much I enjoy this class – how this class has tied many parts of my electrical engineering degree together. The concepts and material apply to everything I have learned since I started my degree at CU. In addition to the material, I feel like I have found a class that is truly effective at teaching. This class was organized by people who really 'get it' in the sense that they really understand how students learn. I do not know who to give this credit to, whether it is the Professor or the department in general, but I am thrilled with the opportunity to discuss why I think the approach was so effective. […] The point is this: things that we do not experience firsthand can be difficult for us to explain and understand. We do not live on the same order of magnitude as proton interaction and we cannot understand what exactly is happening physically. We need to erase our common sense and explore abstract descriptions because only then can we liberate minds from ourselves. This may be the most important thing I learned in QM. This is a big picture idea, maybe bordering philosophical, and I do not know if that was the intention of the course but it has changed my world. In the second section of the course, I really enjoyed the discussions about realism, hidden variables, the many-worlds interpretation, and everything in between. This plays upon the same type of thing – it really forced me to look at things completely differently. […] In summary, I am trying to escape from the common views of the world and understand it in a new way – or at least accept that all of my common sense accusations do not paint the whole picture. There is a world out there to be explored and understood and I believe that QM is the best medium to understanding it. This class has profoundly impacted me and my sense of the world around me. […] From the first day of class to the last day of class, there was a feeling that someone actually cared that I was learning. It was always apparent the effort that Professor Finkelstein and Charlie were taking. They were in a constant effort to improve the course and make it even better – from Charlie asking me about the reading assignments outside of class, to Professor Finkelstein addressing the constructive criticism and suggestions in front of the whole class, to Charlie spending an entire weekend reading people's drafts in order for them to learn more and get better grades. That is an amazing feeling and actually makes going to class not the classic experience of grades, homework, and stress. One would normally expect a class that is so large to be more difficult to learn in. However, that was simply not the case with this class. I really loved how every day I walk into class there was a topic slide explaining the major ideas of the day. It was nice to sit down each lecture and reflect for a few minutes about what we were about to learn, how that applies to the real world, and the connection with the topics and what we have already learned. In so many of my classes, I do not even know what we are learning on any given day. Everything just seems like random disconnected pieces of information. In this class, a large effort was made to connect everything together and make sense. In two years we are not going to remember how to calculate the probability of an electron tunneling through some potential barrier. However, we will remember the basic concepts of this idea and the important factors that contributed to this probability because we were constantly reminded of the big picture while studying the in depth calculations. […] In conclusion, this class has done many things for me. Firstly, I have learned a tremendous amount of material and I have learned it well. QM is an important class for any engineer and I am very glad my experience in it was so positive. I had friends who hated it so much because their Professor did not care and the structure of the class set them up to fail. In addition to my learning, I have been so impressed by a class that really gets it. I know Professor Finkelstein is involved with researching teaching techniques - well keep it up because you are definitely on to something. Most of it seems obvious but for whatever reason most of my classes completely fail. There was a connection in this class between the teaching unit and the learning unit and I feel like the Professors really *knew* when we were learning and we were not learning. Just taking the time and making the effort to see how the students are doing and not just rambling away material on a chalkboard is an amazing feeling for any student.

**#65:** When I registered for my 2010 fall semester classes, I was not thrilled with my options. I had finished taking all of the classes I was looking forward to and could only enroll in classes I needed to graduate. When fall semester began, I had already resigned myself to a semester of drudgery and boredom. So imagine my surprise when I actually started my classes. My biggest question about physics coming into the class was the one I did not know I needed to ask, "What is quantum mechanics? […] The best part of the class was the lecture. The class was large, but the creative use of candy, clickers, and demonstrations made these lectures interactive and some of my favorites in all of my years at CU. I really liked how the slower pace allowed for some of the more complex ideas of quantum mechanics by repeatedly demonstrating them in different ways. […] Overall this was one of the best classes I have taken. Although I do not think I will be actively pursuing physics in the future, I will take many of the lessons and questions I learned with me.

**#67:** The main reason I ended up taking Physics 3 / Quantum Mechanics was because it was required by my major, electrical engineering. Before entering this class, I had a pretty good understanding of physics, since I took physics 1, circuits 1 through 3, and electromagnetic fields. I even went through some chemistry, all the way through organic. Before I got into class, my main questions were simply what is quantum mechanics? What exactly did Albert Einstein do besides $E=mc^2$? […] This course completely changed my idea of physics and science. I now know that when I came the first day of class I definitely had a classical view of physics in my mind. Since numbers can get infinitely smaller, and were therefore continuous, I assumed that the same thing was true for everything in physics, including that on the atomic level. […] The simulations were probably the most important resource I had throughout the course. It was like controlling dozens of experiments right at my fingertips that I could manipulate in many different ways, so I could explore myself the outcomes of different settings. Also, don't listen to whoever said they didn't want candy given out in class, that was by far the best part! Overall this course has been one of the best, if not the best course I have taken while at the University of Colorado. Both Professors Finkelstein and Baily were terrific in explaining things during lecture… I have learned by far the most in this class and have also learned the most interesting subjects. I am a little disappointed that I took this class in my senior year, like how Professor Baily said on the first day of class that this course made him choose to be a physics major, Quantum Mechanics has given me a tremendous interest in physics and I would love to explore these topics further.

**#70:** When I signed up for this course I expected to do a lot of raw quantum mechanics that involved many complicated differential equations that would not mean anything to me. Since this class was kept primarily on theoretical concepts rather than the hard math, and I feel this has been more useful and allowed me to learn far more than sticking to the math like Schrodinger's Equation. […] This combination of progressive demonstrations and more traditional experimentation allowed a depth of understanding to be gained as it helps to have multiple perspectives on each concept. Through interactively getting the general feeling of the audience about each problem and not stressing correctness but instead understanding, I feel as if a friendlier learning environment was created. Too often is correctness so stressed that it becomes difficult for students to interact with the professor during the lecture through sheer fear of being seen as incorrect in front of so many people. […] I'm glad I took this course and feel I've gained a lot from my experiences in it, while developing my grasp on the world and how it functions.

**#72:** I will provide feedback on the phys2130 class and offer some suggestions, in hopes that this helps my professors, as they helped me tremendously. I will dedicate one section to answer questions posted on the prompt, as well as a section to general class feedback, and the last one for suggestions. This class has taught me a lot, it was very interesting. **[…]** The original motivation that drove me to this class was learning about how the world works. I enjoyed the previous classes to it; however, I thought they were generally flawed and could have been better organized. The conceptual explanation, as well as the time taken to make sure students understood the ideas of physics were not sufficient. Despite having a poor experience I was motivated to learn physics on a deeper level. After having several discussions physics 2130 began to change my perception of reality on a significant level. […] Even though the ideas themselves were very, very interesting, the way professor Finkelstein taught the class was absolutely fascinating. The idea of candy given to the people who speak up is, in my opinion, by far the best thing I've seen in my 13.5 years of school experience. Professor Noah … made [these] the most interesting lectures I have ever listened to. Instead of plugging some numbers in, I was required to understand the concepts and was given a way or a hint on a way to understand them. Not to mention the individual research that I did every time I had to do homework. […] Aside from new ideas I would have to say that professor Finkelstein has to teach every single phys2130 class from now on until he retires. This class is just conceptually too important to have some monotone teacher who really doesn't care either way teach it. At the end of this semester I stand looking back on my experiences during the three semesters that I went to University of Colorado at Boulder, and physics 2130 is by far the best one I've had related to school. The only feeling of dissatisfaction with it is that I wish we covered even more than we did, because it was so good.

**#75:** When I originally registered to take Physics 3, almost a year ago now, I knew nothing about the course; it was simply the next required class for my major. […] This course had an interesting way of challenging my typical views of physics, science, and the world in general. Essentially, it seemed like every day we learned of little ways to disprove classical views of physics. Despite often being confusing, it was a very eye opening experience. Notably, the concept of superposition blew me away. […] An aspect of the class that I really did appreciate was the way it was taught. Unlike most of the lectures I sit through every day, I was rarely bored, and always felt involved. It was almost like I had to reason through things on my own and teach myself, which led to a much deeper understanding of the material. The concept tests were great for challenging and forcing myself to understand everything in detail. They also provided a great opportunity to discuss the information with the students around me. We were able to help each other through the problems, which usually resulted in better learning. Another great set of learning tools was the simulations. Even though I had used similar ones in other physics courses, they had never been nearly as beneficial. For topics like lasers and wave functions, they were invaluable. The way I learn is very visual, so simply hearing a description or seeing a mathematical model is not enough. I am certain that I would never have understood those concepts nearly as well without the assistance of the simulations. In the end, I enjoy the material covered in Physics 3 a lot more now than I did last semester. I think it was mainly a matter of being open to completely new ideas, unlike anything I had learned before. The class was unlike anything I had experienced before, not just based on the curriculum, but the way it was taught, and how I learned it. With the caveat of open mindedness, I would certainly recommend this class to future students.

**#79:** I am an engineer because I love science and math. Physics three was an easy choice for me because I have always wanted to learn more about physics and its applications. My first physics experiences were simply trying to understand the mechanics of the world around me. Senior year of high school was my first opportunity to learn physics in the classroom and really open my eyes. Our physics research and design project for that class was especially memorable. We built an ancient steam powered engine. This first experience motivated me to go into engineering and learn more about physics and how to apply it to mechanical design. When I got to college I did pretty well in physics one and two but these were only basic mechanics and electricity /magnetism. All the while I was looking forward to quantum mechanics because I knew so little about it. My college friends had all said that it was the best class they had ever taken. On top of all that my advisor said that professor Finkelstein was really good. I was really motivated to get the homework done and learn all the concepts well because they were all so interesting. I was thoroughly intrigued by all the non-classical approaches to physics. Following the historical background gave great insight to how these quantum mechanical ideas were conceived and developed. Many different experiments and many different theories were debated and passed on. Different viewpoints still exist today because there is still so much to be discovered about the world. Overall I think the class was run very well. The amount of work was pretty perfectly laid out. It was enough so that you had a lot of practice and kept up to date with the class. At the same time it wasn't too much so that you could get it done in one sitting so you could really focus on the task. The tests were also well written. […] To summarize I had a great learning experience with Finkelstein and the class "Introduction to Quantum Mechanics". All of my expectations were met and my knowledge of the world around me has improved greatly. I hope to pursue physics and other sciences in my future because they are very interesting to me. I cannot see myself becoming a physicist or scientific researcher but I know that I will use my knowledge of physics in my engineering career. All the problems I will solve will require the use of physics because it is the backbone of all things in the real world.

**#82:** Phys 2130 has been one of the most interesting courses of my college career. I could have either taken Organic Chemistry, or Physics 3 to fill the same requirement for mechanical engineering, however, I chose this course because I am much more interested in physics. **[…]** Now, I did not come to learn all this on my own. I enjoyed both Professor Finkelstein's and Charlie's methods of teaching. Both professors have a way of captivating my interest, whether it is the thought-provoking course material, or the […] style of teaching of Professor Finkelstein. Throughout the classes in this course, I was not only listening and learning through lecture, but also engaged in class discussion with my peers during clicker questions, and occasionally outside of class. The homework assignments were of reasonable difficulty, to get me thinking and really understanding, but not too difficult to the point of giving up or just being unable to conceive. We learned some very advanced concepts, were shown enough derivation, and practiced these concepts enough to where I am confident in my knowledge of Physics III. I was only unable to fully comprehend one idea in this course, and Charlie was quick in responding to my emailed question, and helped me overcome this obstacle. Overall, the only topic that we did not cover is relativity, and I am not disappointed. We fit in all the material that we possibly could without compromising depth, and I am happy that I understand the nature of light, the photoelectric effect, the history of atomic model development, Schrodinger's equation, and all the other topics we covered.

**#88:** Deciding between science electives was one of the easiest choices I had to make last semester. I had the choice between physics three or biology, but there was no way that I would choose biology over physics, so physics it was. Physics has always been an interesting subject to me, and the fact that we were finally going to learn about Einstein and his discoveries was something to look forward to. This was the only expectation I had when entering the course since the rumor was, that in this specific course the material covered, changes form professor to professor. The first day of class was a little disappointing because it seemed like all we were going to talk were waves and how light was a wave and a particle at the same time, so how could that possibly be interesting. Now I had the first homework lined up and after working on it for hours, it seemed like physics was going to be a huge pain in the butt with lots of boring lectures and long homework assignments. At this point I had a general bad attitude towards the course, but it didn't take long to change my mind. […] For me the main way to learn the material was doing the homework with a group of people during help room hours. Working with somebody and reasoning through problems, talking about possible solutions, and if stuck talking to LAs. The learning assistances would help me get to the correct answer without actually telling me the answer made all the difference in learning. Many of my classes aren't very interactive and just provide you with new information. Some have clickers, but don't provide much time to discuss the answers, and they don't have a learning assistant to help talk about the possible answers. In our class we have two learning assistances and two professors who come and talk to us for every clicker question which encourages real discussions and reasoning to find the answers. As I mentioned earlier, lecture is very entertaining which makes it much easier to pay attention. Rewarding our ideas and thoughts ..is another positive aspect of lecture. One of the most important things in lecture is that when someone expresses their thoughts and reasoning how they got the answer to the problem, even if this person is wrong, the ideas expressed don't get immediately get shut down but rather discussed and reasoned through why they might not be correct and then we end up with the correct answer in the end. Never getting a straight answer can sometimes be very frustrating, on the other hand this is the most effective way for me to learn. […] This course has changed my thinking about everyday things. A light bulb is no longer a simple light bulb, electrons are exited to a higher energy level and when they jump back down to their ground state they emit photons. Everything is now made of electrons and protons and when walking across the room, scraping off electrons from the carpet and shocking myself on the doorknob makes me think about the work function of the doorknob and my finger. Also if I would walk fast enough there is a slight chance that I could tunnel through the door instead of hitting my face on it. […]

Looking back at the past months I'm glad I had the chance to take modern physics as my science elective and change my thinking about the real world. I must say, for my convenience I will have to keep a balance between the realism and quantum way of thinking, so I believe the moon is still there even if I'm not looking.

**#92:** I heard the topic before but actually I have no idea what is it? So when I know physic department at University of Colorado at Boulder offer Physic 3: Quantum Mechanical so I decided to take this course. I'm unfortunately Computer Science and Business double major but still, I'm really interested in this course so I took it to learn more about Quantum Mechanical phenomena**.** **[**…] Definitely the course has changed my idea about physic and the practice of science. **[…]** It has been really great semester. I learned a lot and expand my understanding about quantum physic. Physic 3 is one of my favorite class this semester. All material covered, teaching technique is really great. Please keep those teaching technique for future and I believe everyone else will love and have a great semester study quantum physics

**#93:** Physics has always fascinated me. Even as a little kid, I was always asking questions about why things happened the way they did in the world around me. **[…]** Though I later switched to Mechanical Engineering and then Electrical/Computer Engineering, Physics remains to be the most interesting subject I have ever studied. When I saw that Physics 3 was a required class for my major, I was very excited and enrolled as soon as I could. Once the semester began, I knew right away that I would enjoy the class. Noah and Charlie were a great team; I loved the atmosphere within the lecture hall. Having professors that are so obviously enthusiastic about a subject really makes a big difference from a student's perspective. Lectures were always interesting and held my attention very well. Being an Electrical Engineering major, most of my classes are largely based on complex mathematics. It was nice to have a class where the focus was to really learn and understand the concepts before being thrown into the math that accompanies them. I appreciated the fact that Noah understood that, with a subject as complex as Quantum Mechanics, it was important to ensure that his students really had a solid understanding of a concept before clouding it with complicated equations and formulae. […] Having so many resources readily available to us was very beneficial. I went to help room almost every week to discuss topics or homework problems I found confusing with other students, LAs, or even the professors. […] I can definitely say that studying Quantum Mechanics has changed the way I think about Physics and science in general. I now look at the world a little differently. For example, when I see color, I don't just see red or blue, I see a material absorbing innumerable photons of varying energies, and emitting photons of a certain energy (or a specific combination of energies) corresponding to the color of light I'm observing. […] When I'm using my new, green (532nm) laser pointer, I can understand what is going on inside this fascinating little device. Within the reflective optical cavity is a gain medium that is excited as it is pumped with energy until a population inversion exists within the cavity. […] Not only did I learn about Quantum Mechanics, I also learned a great deal about science in general. I now know the importance of interpretation and how two well educated and respected individuals can interpret phenomena very differently without one being necessarily wrong. This class taught me to open up my mind and think outside the box.

[…] If anything, studying Quantum Mechanics has only made me more curious about the world around me. I find myself frequently searching the internet for answers to questions I have about physical phenomena. I have learned a different style of thinking that I know will be very valuable in my future. […] It's a beautiful thing for a student to come out of a class with a strengthened thirst for knowledge and learning. I can say with certainty and satisfaction that my thirst for knowledge about the universe I exist in has only been amplified over the course of this past semester; I hope that this thirst will never be quenched.

**#97:** As I approach the final days of my undergraduate college career, I find myself reflecting upon my extensive coursework: Covering topics ranging from economics and marketing to advanced thermodynamics. With this interesting mix of classes all contributing towards my degrees in mechanical engineering, economics, and my certificate in project management, I wonder why I abstained from fully broadening my understanding of the physical sciences until my final semester in college. Physics 2130, has been one of my most interesting and memorable classes at CU, despite its completion being part of a long list of requirements for my graduation in mechanical engineering. **[…]** Specifically, there were a few lectures that were especially powerful in their ability to alter my perception of the physical world. Chronologically, the first of these was the "Local Realism-EPR" Lecture. […] The next lecture in quantum physics that helped to reinforce this paradigm-shift was the lecture on measurement. […] Finally, the lecture on complementarity, which describes the wave-particle duality of protons and electrons as exclusive, also played an integral role in the shift of my mindset of scientific understanding. […] In general, despite having a severe case of the dreaded, senior-itis, I still learned a great amount from Charlie and Prof. Finkelstein in PHYS 2130 and the class proved to be one of my favorite, interesting, and most meaningful classes at the University of Colorado.

**#99:** As a mechanical engineer I am interested in how the world works and the principles that govern it. Even so, I struggled through physics one and two, bogged down by the numerous equations and lackluster topics. Circulating rumors told me that physics three was quite different from the first two courses in that it explored the almost magical phenomena of the world. This renewed my interest in the topic and I was eager to take the course. […] Physics one and two skewed my view of physics. I did not find the "plug and chug" nature of those courses engaging. None of the examples demonstrated seemed applicable to the technology driven world today. Physics 2130 completely altered my opinion of the topic. The class was engaging and really fun. While there were mathematical equations (typical to any physics course), I found the topics engaging and the assigned homework problems and lecture examples relevant to questions that appeared on exams. […] The most effective teaching tool I saw this semester were the simulations. I am a visual learner, so while I attend lecture and listen attentively sometimes the information was in one ear and out the other. The PHET simulations really helped to visualize a number of the concepts. I also appreciated the concept tests. I say appreciated as I did not enjoy them, but they kept me on task and motivated me to do the assigned readings. As I mentioned above, I thought that the assigned homework problems were very beneficial to exam preparation. Dissimilar to the other courses that I am enrolled in, the homework problems were very similar to the material we were tested on. **[…]** Unlike a couple of the other courses that I am enrolled in, I found the lectures for this course very engaging. I enjoyed the candy aspect of the lectures as it motivated kids to speak up and participate in lecture. Moreover, the instructors seemed excited to be there and excited to help us learn. In turn, this made me excited to learn the material.

**#103:** Quantum mechanics is one of the most dynamic sciences today. Much of quantum mechanics has only been around for less than one hundred years, and is still fairly undeveloped. That is what makes this class so interesting to learn about, because quantum mechanics is constantly evolving, keeping it at the cutting edge of scientific theory. […] Although I mainly took this course for educational requirements I was interested in taking a physics class which was nothing like anything I have learned before. This class is far different from any other science course I have taken and is interesting because of that. Any preconceptions that I had about quantum mechanics were completely changed by this course. It made me realize that even today new theories are being explored and changed by experiment. […] This has also changed my view of how science is advanced in today's society. […] In many lectures the professor or TA's simply lecture you, however this class, thankfully, was not like that. Participation was encouraged which was very helpful. The best source of information is not the textbook but the professor, so having such a free flowing lecture allowed the students to access that information in an efficient manner. The clicker questions helped facilitate this learning very much. The clicker questions tested our knowledge which allowed us to see the faults in our arguments. The group clicker questions were even more helpful than any solo question. […] One unique part of this course that I found to be extremely valuable were the online simulations. The simulations allowed us to not only see the physics but to test our own theories and see if they are correct. There is only so much you can do to explain how a laser works. Being able to attempt to make your own laser made it much more interesting and easy to understand. Many times the simulations allowed me to actually see what was going on and to try to replicate what I had seen in class. It is similar to looking at a building and understanding its structure, only to attempt to build that structure on your own. Only then can you realize how little you actually know about something when it is just you doing the experiment.

**#104:**

So I was sitting in Larry's office, deciding on what classes to take,
Physics 3 or organic chemistry was a tough decision I had to make.
While I sat there considering my options, Larry spoke up to help me out,
You'll like the physics class, he says, assuring that I will without a doubt.
He tells me I'll enjoy the lectures …
But as I reflect upon it all now, at the end of this course,
And recap everything we've learned from Electromagnetic waves to nuclear force,
I developed a new way of thinking, and of course there's no remorse,
That I took such a challenging class, but will admit that it was far from the worst.
I look back now on all the interesting topics that we learned about …
After learning some formulas, we went on to learn more conceptual stuff,

And we quickly learned that homework assignments were becoming more and more tough.
Differentiating a theory from interpretation is an important concept indeed,
But really, what the heck does that have to do with a farmer and his seeds?
Anyways, moving on to interpretations, like Many Worlds and Copenhagen,
This is where there was hot debate and many different ideas were undertaken.
Like, how can we really know if Schrödinger cat is dead or alive?

And if Many Worlds is indeed correct, so is there really a different me?
Reciting a different poem, far off in some other galaxy?
Well if that's true, you know what I'd do, I'd go out and explore,
And find that other [me] out there, and get right down to the core,
Of all this philosophical debate that I had no idea exists,
It's so abstract how even amongst the experts nothing is fixed.
But wait- if Copenhagen is right, what happens when I observe my superposition?
One of the two [mes] would collapse! So maybe it wouldn't be such a good decision.
So instead I'll comfort myself to Decoherence, and accept that I exist in only one state,
Then would that mean when I came out of my mom, a tiny air particle determined my fate?
[…]
Now at our last lecture, class will be over, yes- it is quite tragic,
But one thing I will take away from this course is that physics is cooler than magic!

# APPENDIX F

## Selected Student Discussion Threads (Fall 2010)

**INSTRUCTOR POSTINGS HIGHLIGHTED IN RED**

| **SUBJECT: Local Realism/Nonlocality**<br>**SEED QUESTIONS** |
| --- |
| The idea of quantum non locality and special relativity discussion is very interesting, but I did get lost a bit in the actual definition of non locality. |
| I am still confused, if quantum mechanically entangled particles violate either the principle of locality or realism, which one do they violate? |
| First off: modern physics, an entanglement of math, philosophy, and reality. Awesome! So the article talks about some experiment done by Alain Aspect that showed non-locality in action. Can we learn about this experiment? It sounds really interesting. How is it that Bell's inequalities constrain any local theory? What makes them so much more special than the theories proposed before him? |
| So now that we've decided that the world is truly nonlocal, how can we apply nonlocality to our benefit? What are situations where having a nonlocal physical world is better (or worse) than a completely local one? |
| This may not necessarily be a question but I would like to know what Bell said that was so convincing to EPR as to convince them that nonlocality must exist. I feel that knowing two entangled particles always have a definite position and velocity, we just can't measure both at once. Not knowing something does not mean it doesn't exist. In the red sock blue sock experiment from lecture. The socks were always the same color. The fact that we didn't know what color the sock was didn't change the actual color of the sock, and the color of the sock wasn't changing any time after it was put in the box. I guess I am still convinced of locality. |
| From what I understand physical nonlocality is that there is no limit to how far something is and how it can act with something else or that even needs to so does a nonlocal time mean that things anywhere in time can act with each other? |
| How would non-locality actually work? How do two things become entangled and thus lose all position with everything else? |
| Why did the concept of locality and nonlocality originally arise? What problem or question made Einstein and others think about this? |
| I don't understand how two objects can nonlocally influence each other, independently of their space. I understand the analogy used in class with the two boxes with a sock in each one. However, in that case, the boxes were together at some point when the socks were placed into them so space isn't really independent. How could one ever know that two objects actually do influence each other nonlocally? I don't think it is possible. If it is, you probably cannot apply this concept to everything in the physical world. |
| In the article locality and nonlocality are used quite frequently. I understand what it means but I don't understand how it is possible regarding spin up and spin down. I liked the red sock blue sock explanation in class, but how can this be similar to an atom really far away? |
| Can we further discuss “nonlocality” and what it means? |
| I have a hard time grasping the idea of non-locality. As we see the world, everything is caused by something else. This theory would seem to suggest that things can happen, and be related, but not be a direct physical result of the other happening. |
| In the article it discusses the phenomenon of nonlocality. It defines nonlocality as the direct influence of one object on another distant object without actually touching it or any series of entities reaching from here to there. This would imply that stomping the ground in Colorado could cause an earthquake in Australia without affecting any other physical thing. My question is how is this possible? I think stomping the ground would cause some effects at a short distance from the location of the stomping. This could then have an effect on things at greater distances which could then lead to an earthquake somewhere in the world. But if stomping the ground has no effect at short distances then how could it possibly affect somewhere so far away? It only makes sense to me that there must be some kind of chain reaction going on in order for this to be possible but according to nonlocality this is not true. |
| What is the relationship between realism and locality? When local is violated, the world becomes realism? |
| I’m not really sure what nonlocality physics mean? What’s definition of elements of physical reality? |
| Is gravity an example of nonlocality? Say, for instance, and object is dropped from a height above the ground. Intuition tells us that it will fall to the ground. That’s the expected outcome; we know this will happen. We also know that it’s the earth exerting a force on that object to pull it toward the ground, but while the object is in the air, there is nothing “next to” it acting on it. Therefore the principle of locality does not hold in this situation. Are the earth and the dropped object examples of entangled particles? |
| It seems that the entanglement, locality, and nonlocality are the key words in this article, I just barely understand its meaning, still not sure exactly how they are different, and how they relate each other, and how that relate to the quantum world. Seems Einstein keeps arguing about the nonlocality for long time, and he could not accept the concept of being nonlocal, why it bothers him so. Maybe because he is greatest physicist, it doesn’t bother me that much, I guess I don’t understand the concept as well as he does to be able to question it. |
| So if our universe is based off of nonlocality and not locality how come our reality operates sometimes locally and nonlocally? Nothing can happen in zero time even if it appears that the universe operates this way (nonlocally). |
| Date: October 12, 2010 12:59 PM<br>I am personally more convinced with the fact the world could not include realism rather than locality. With realism it just says that everything in the world is in a certain state whether or not it is acted on or observed. I feel like this is exactly what we are observing with quantum mechanics. We see that it is completely random before we observe it, and once we observe it the wave function 'collapses.' |

Date: November 2, 2010 9:48 PM

Where I see your point I disagree. I feel that there is really no way currently to prove that everything is totally random before we measure it. I also feel that the "collapse" of the wave function is a very incomplete answer and very unsatisfying. How does it collapse, and is there some way to predict how it will collapse? Is there a way to force it to collapse to what we want?

Date: November 6, 2010 3:41 PM

I think that's the point of Bell's Theorem (which we talked about a little in class, but didn't really get into). It proves that the outcomes must be random (but correlated), since it shows that no deterministic theory will have the right experimental predictions in all situations.

As for the "collapse" being unsatisfying, welcome to the club!! Many physicists find it unsatisfying, but there aren't a lot of alternative ideas. I think the biggest problem is that, of course, you can never watch the wave function collapse, since it's something that's taking place just as you're making your observation. "Watching" it would mean you would be observing it throughout the entire process, but those kind of "observations" are impossible. If you don't see why, ask yourself what kind of interaction you could have with a particle such that you could observe it collapsing to a point. We don't predict how it will collapse, only the relative probabilities for collapsing to one state instead of another. The outcome is random, and we can't ever force a superposition state to always collapse to the same outcome (otherwise, I'd say it's not in a superposition state, but in a definite state).

Date: October 12, 2010 4:52 PM

Nonlocality states two objects does not have to next to each other in order to interact with other. One object can interact with other object in a distance.

Date: October 12, 2010 6:45 PM

Perhaps a better way of thinking of nonlocality is to contrast it with locality. In locality every action can be broken down into the interaction between matter that is in contact or very close to. For example, a sound generated in your throat creates a vibration that then propagates through the air and then can be heard by another person. Nonlocality says that, in this example, the other person could be on the other side of campus and hear what you said. It does not matter where the entities are in relation to each other, but can still effect each other.

Date: October 18, 2010 7:57 PM

Another way to say this is that there is no "action" connecting two events from happening. With the entangled particles, since there is no way that communication can happen faster then the speed of light, there is no other known mechanism for the atoms to talk to each other. If I threw a book at the wall and made a clock fall down, then the book linked me with the falling of the clock.

Date: October 12, 2010 6:14 PM

I had a question related to this topic about quantum encryption. From what I understand, this can be done by sending a probabilistic signal through a wire or fiber-optic cable. Then, if someone attempts to read the data (man-in-the-middle), the signal at the source will change. That way, whoever is sending the signal will know whether or not someone else is trying to access the signal. Maybe this is a simplistic view, but I do not quite understand how this would be helpful. Wouldn't an intended receiver, just by observing the signal, also cause the signal at the source to take on a definite state?

October 12, 2010 11:11 PM

I was thinking of the same problem. When the intended receiver views the message it collapses into a definite state just as it would if an unintended receiver viewed it. Being that you can't observe something in a superposition state, how can you know if it was really in that state when you got it? I don't know if you could tell if it collapsed when you observed it or when someone else observed it before you.

Date: October 17, 2010 7:53 PM

This is a really interesting problem. Let me see if I can explain how this works without confusing anyone.

First off, assume you have a device that can send polarized photons (the information carriers) along a fiber optic cable. The sender, by convention, is named Alice, and the receiver is named Bob. Alice's device can operate in two modes. In the first mode it can polarize Horizontal (|) and Vertical(-). In the second mode, It can polarize Diagonal (/) or (\). The second mode should be operating 45 degrees off of the first mode.

Now, assume that Alice has a set of binary bits, a secret key. She randomly decides which mode will correspond to which data, but each 90 degree polarization corresponds to certain bits.

If the secret key is : 0 0 1 1 0 0 1 0 1 0
The | - representation is: - - | | - - | - | -
The / \ representation is: / / \ \ / / \ / \ /

At random, the signal changes between sending - | polarization, and / \ polarization. Due to the Heisenberg uncertainty principle, only one polarization can be measured at a time. Bob picks a single polarization (at random) and reads it. About half the data will be correct, and the other half will be random (depending on mode the individual photons were in.) He then, in public, tells Alice which mode he used to read that data. They ignore the other mode. Now, Bob has part of Alice's secret key. They can compare select bits to check for eavesdropping. If Eve (eavesdropper) attempts to read the data, she can also only read in one mode due to Heisenberg. Even if she re-sends the data, she doesn't know the value of the bits in the wrong polarization, they're random. She has aligned both modes to the same mode.

Hopefully that answers the question, at least a little bit. The key to it is, not even the receiver has all the information, but the sender does. The sender can check what the receiver has read, and decide if the line is clean or not.

Date: October 12, 2010 6:45 PM

"Why did the concept of locality and nonlocality originally arise? What problem or question made Einstein and others think about this?"

I think it arose trying to explain the location of electrons around atoms since their location is not known, but you can make a really good guess at their precise location.

Date: October 12, 2010 7:46 PM

My answer to the question "So if our universe is based off of nonlocality and not locality how come our reality operates sometimes locally and nonlocally? Nothing can happen in zero time even if it appears that the universe operates this way (nonlocally)."

I think of it as our universe is a combination of locality and non-locality just as there are several different forces in the universe. For example, if one pushes a block it will move due to the local push force by myself. If that same block is dropped off of a building gravity pulls it down instantly. So gravity is acting non locally and instantly.

Date: October 12, 2010 7:49 PM

I have another question to add to this. Could gravity be a macroscopic demonstration of quantum entanglement? Since everything we know is attracted to each other, if gravity is a demonstration of quantum entanglement, does that mean that all matter is quantum entangled with each other?

> Date: October 17, 2010 11:53 PM
>
> As it stands, gravity is not considered a form of entanglement, since the force itself is limited by general relativity. The force carrier (whatever it turns out to be) is thought to be limited by the speed of light (the space-time curvature withstanding). Entanglement by definition is nonlocal, so it is not bound by the speed of light in this way.

Date: October 12, 2010 8:01 PM

Here is another question. I don't really understand the usage of the term realism with respect to our physics 3 subjects. To use it implies that there is a anti-realism or unrealistic quality that to me does not fit with any of our lecture discussions.
To put it more simply:
1) What does realism truly mean and does that imply there is an anti-realism?
2) If there is a "unrealistic" term then when does it occur?

> Date: October 12, 2010 10:45 PM
>
> I think it depends on what you call "real". I think realism means that you believe that physical objects exist in definite, objectively real states, whether we are observing them or not. Like Einstein said, the moon is always there, even when nobody is looking. He believes that a particle always exists at a definite position, even if we don't know what that position is - so its position is "real", but unknown.
>
> Quantum mechanics says the position of a particle can be described as a superposition of many different positions, which collapses down to one definite value when it interacts with something, like an observer. If you believe THAT (the superposition) is the "real" state of the particle, then maybe we're disagreeing on what it means for something to be real.
>
> As for this class, I think it's Einstein's view that's considered to be "realist". And I guess anything that opposes that view could be looked at as "anti-realist", right?

Date: October 13, 2010 2:29 AM

Realism is the notion that a quantity can be known with certainty before it is measured. The idea is that if the quantity can be known, or in other words predicted, beforehand, then it existed in a definite state beforehand. If it can't be predicted, then it did not exist in a definite state until observed.

Date: October 12, 2010 10:55 PM

Locality, to me, is the idea of physically affecting something via touching it or touching any series of entities reaching from here to that something. It means that you can NOT affect something without somehow coming into direct contact with it, or creating a series of events that each have direct contact with the next event leading to the final event. This is how we experience all things in our daily existence. It is very difficult for us to comprehend the idea of non-locality because coming up with concrete examples is very complicated if not completely impossible. I like to think of it more as a concept in a non-reality. Our reality is what we experience, what we perceive, and how we measure the world around us. The idea of non-locality can really only exist in a parallel world, one that we do not directly experience and can not truly understand in a physical sense. Therefore, it is best to try to understand these terms in a theoretical sense and not with concrete examples.

Date: October 12, 2010 11:00 PM

This is in response to the question: “Can we further discuss “nonlocality” and what it means?”

Nonlocality is the direct influence of one system on another spatially separated system. This violates the principle of locality, which states that an object is directly influenced only by its immediate surroundings. Our local realistic view of the world assumes that systems are separated by time and space. However, quantum nonlocality contradicts these assumptions and proves that there are interconnections between systems at the quantum level. An example of nonlocality is the spin of two electrons that originate from a common source. When the orientation of two electrons from the same source is measured, we find that one electron will spin downwards while the other will spin upwards no matter how far apart the two electrons are. In other words, determining the axis of spin of one electron also determines the spin of the second electron because the second electron will instantly respond to the state of the first electron with no dependency on the distance between them.

Date: October 18, 2010 8:33 PM

How does non-locality work? does the two objects have to be within a finite distance, or they can interact with one or another with no constraints on distance.

Date: October 18, 2010 8:45 PM

From what I understand of non-locality is that one object has a direct influence on another distant object. I think that the example used in lecture explains it best: a source produces a pair of atoms with opposite spins and by measuring the spin of one of the atoms collapses the superposition of states that the second atom is in; therefore by measuring the spin of one atom you know the spin of the other atom regardless of how distant they are from one another.

Date: October 18, 2010 10:44 PM

It seems to me that nonlocality is a very strange, but true phenomenon. The classic experiment is where two particles are shot off in different directions with opposite spins. If we observe the spin of one particle, then we automatically know the spin of the other from superposition. I have two questions: first of all, what is the experiment that proves nonlocality? Second, how are entanglement and nonlocality different?

Date: November 8, 2010 10:46 AM

To try to answer at least part of your questions, entanglement is the idea that these two particles are of different spins, that this is like a property they contain, even if at this moment neither has a definite spin. Nonlocality is the idea that the particles somehow "communicate" or send a signal at a speed that is faster then believed possible, therefore being non local. Entanglement does not say that any signal is sent, making it different then non locality. That's were it gets a little fuzzy for me.

Date: October 18, 2010 10:53 PM

Nonlocality means that suppose there are two objects, and one of the outcome directly affects the other outcome. In the EPR paradox, if Albert observes atom from the plus channel, this outcome affect the result of the other. The other side must be the opposite of what Albert observed.

Date: October 18, 2010 11:21 PM

It seems that the classical and non-quantum universe acts on the theory of locality and realism. Something in the classical realm can not be affected by something that is not anywhere near it unless it is reached by a device. Also, classical and physical things are do not exist only because they are measured. But, it seems that whenever we enter the quantum realm non locality takes place and realism is cannot always describe what we are dealing with. Particles can be affected by other particles with seemingly no interaction and particles are changed when the are measured.

Date: November 2, 2010 9:19 PM

Understanding the concepts of non-locality and realism are all noble goals but what is the real use of these things. It might be my inner (and i guess outer) engineer, but I am having trouble understanding the real world applications of these things. Why do i need to know these things aside from the sheer coolness factor? I keep coming back to this point, but as the compliment to physicists, engineers must always ask "why do we want to use all your crazy ideas?"

Date: November 2, 2010 9:52 PM

I know that some cell phone companies are studying nonlocality and the collapse of the wave function as something that could help improve user privacy. This is due to the collapse of the wave function occurring if someone else actually intercepts the wave function. So nonlocality has the potential to make it almost impossible for someone else to listen in unnoticed.

Date: November 15, 2010 9:59 PM

You do have a very exciting thought, that we might be able to have phone conversations that are actually private. My one concern and what seems like the insurmountable is that as of yet we cannot predict by what mechanism and to what the wave function will collapse. I feel that this is a giant hole in the theory and cannot be considered complete until this is answered.

Date: November 16, 2010 6:02 PM

...but that's why it works for distributing a secure, RANDOM encryption key. We don't know what the outcome of any one measurement will be, but when we do measure it, we know that the other person's measurement yielded the opposite outcome. Also, it's secure because collapsing the wave function puts the photon in a definite state - and we can tell the difference between definite states and entangled superposition states - meaning we could always tell if someone had intercepted the key distribution, because they can't retransmit an entangled particle.